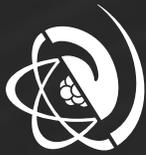

# ORIGINS
## Space Telescope

From first light to life

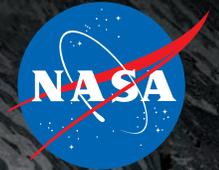

# Mission Concept
# Study Report

**August 2019**

Learn more at
origins.ipac.caltech.edu

**Origins Space Telescope Study Team**

**Origins Space Telescope (*Origins*) Science and Technology Definition Team (STDT) members**

Margaret Meixner (*meixner@stsci.edu*) STScI JHU, Baltimore, MD (*Origins* Study Chair)
Asantha Cooray (*acooray@uci.edu*) University of California at Irvine, Irvine, CA (*Origins* Study Chair)
Dave Leisawitz (*david.t.leisawitz@nasa.gov*) NASA GSFC, Greenbelt, MD (*Origins* Study Scientist)
Johannes Staguhn (*jstaguhn@jhu.edu*) JHU/GSFC, Greenbelt, MD (*Origins* Deputy Study Scientist)
Lee Armus (*lee@ipac.caltech.edu*) California Institute of Technology IPAC , Pasadena, CA
Cara Battersby (*cara.battersby@uconn.edu*) University of Connecticut, Storrs, CT
James "Gerbs" Bauer (*gerbsbauer@gmail.com*) University of Maryland, College Park MD
Edwin "Ted" Bergin (*ebergin@umich.edu*) University of Michigan, Ann Arbor, MI
Charles "Matt" Bradford (*bradford@caltech.edu*) California Institute of Technology JPL, Pasadena, CA
Kimberly Ennico-Smith (*kimberly.ennico@nasa.gov*) NASA Ames Research Center, Moffett Field, CA
Jonathan Fortney (*jfortney@ucsc.edu*) University of California, Santa Cruz, CA
Tiffany Kataria (*tiffany.kataria@jpl.nasa.gov*) California Institute of Technology JPL, Pasadena, CA
Gary Melnick (*gmelnick@cfa.harvard.edu*) Harvard Smithsonian, Boston, MA
Stefanie Milam (*stefanie.n.milam@nasa.gov*) NASA GSFC, Greenbelt, MD
Desika Narayanan (*desika.narayanan@gmail.com*) University of Florida, Gainesville, FL
Deborah Padgett (*Deborah.L.Padgett@jpl.nasa.gov*) California Institute of Technology JPL, Pasadena, CA
Klaus Pontoppidan (*pontoppi@stsci.edu*) STScI JHU, Baltimore, MD
Alexandra Pope (*pope@astro.umass.edu*) University of Massachusetts, Amherst, MA
Thomas Roellig (*thomas.l.roellig@nasa.gov*) NASA Ames Research Center, Moffett Field, CA
Karin Sandstrom (*kmsandstrom@ucsd.edu*) University of California at San Diego, in La Jolla, CA
Kevin Stevenson (*kbs@stsci.edu*) STScI JHU, Baltimore, MD
Kate Su (*ksu@as.arizona.edu*) University of Arizona, Tucson, AZ
Joaquin Vieira (*jvieira@illinois.edu*) University of Illinois, Urbana-Champaign, Illinois
Edward L. "Ned" Wright (*wright@astro.ucla.edu*) UCLA, Los Angeles, CA
Jonas Zmuidzinas (*jonas@caltech.edu*) California Institute of Technology JPL, Pasadena, CA

**Ex-Officio (non-voting) members of *Origins* Science and Technology Definition Team**

Kartik Sheth, NASA HQ, Program Scientist
Dominic Benford, NASA HQ, Deputy Program Scientist
Eric E. Mamajek, Caltech JPL, ExEP Deputy Program Scientist
Susan Neff, NASA GSFC, COR Chief Scientist
Elvire De Beck, Chalmers Institute of Technology, SNSB Liaison
Maryvonne Gerin, LERMA, Obs. De Paris, CNES Liason
Frank Helmich, SRON, Netherlands Institute for Space Research Liaison
Itsuki Sakon, University of Tokyo, JAXA Liaison, MISC instrument lead
Douglas Scott, University of British Columbia, CAS Liaison
Roland Vavrek, ESA/ESAC, ESA Liaison
Martina Wiedner, LERMA, Obs. De Paris, HERO nstrument lead
Sean Carey, Caltech IPAC, Communications Scientist
Denis Burgarella, Laboratoire d'Astrophysique de Marseille, collaborator
Samuel Harvey Moseley, NASA GSFC, instrument scientist

## NASA/GSFC Study Center Engineering Design Team

| Name | Role/Responsibility | Work Email | Org |
|------|---------------------|------------|-----|
| David "Dave" Leisawitz | Study Scientist | david.t.leisawitz@nasa.gov | 605 |
| Johannes Staguhn | Deputy Study Scientist Instrument Scientist | johannes.staguhn@nasa.gov | JHU/665 |
| Ruth Chiang Carter | Study Manager | ruth.c.carter@nasa.gov | 401 |
| Michael "Mike" DiPirro | Technical Lead/Chief Technologist | michael.j.dipirro@nasa.gov | 552 |
| Chi Wu | Mission Systems | chi.k.wu@nasa.gov | 599 |
| Ed Amatucci | Instrument Systems | edward.g.amatucci@nasa.gov | 592 |
| Louis "Lou" G. Fantano | Thermal Systems | louis.g.fantano@nasa.gov | 545 |
| Greg Martins | Mechanical Systems | gregory.e.martins@nasa.gov | 543 |
| Joe Generie | Mechanical Systems | joseph.a.generie@nasa.gov | 543 |
| Joseph " Joe" M. Howard | Optcal Systems | joseph.m.howard@nasa.gov | 551 |
| James Corsetti | Optcal Systems | james.a.corsetti@nasa.gov | 551 |
| Eric Stoneking | ACS | eric.t.stoneking@nasa.gov | 591 |
| Dave Folta | Flight Dynamics | david.c.folta@nasa.gov | 595 |
| Casandra Webster | Flight Dynamics | cassandra.webster@nasa.gov | 595 |
| Jeff Bolognese | Struc Analyst & Modeling | jeffrey.a.bolognese@nasa.gov | 542 |
| Carly Sandin | Materials | carly.sandin@nasa.gov | 541 |
| Daniel Ramspacher | Propulsion | daniel.j.ramspacher@nasa.gov | 597 |
| Alison Rao | Propulsion | alison.rao@nasa.gov | 597 |
| Susanna Petro | Integration and Test | susanna.petro-1@nasa.gov | 568 |
| Kevin Denis | Detectors | kevin.l.denis@nasa.gov | 553 |
| Damon Bradley | Readout Electronics | damon.c.bradley@nasa.gov | 564 |
| Tracee Jamison | Readout Electronics | tracee.l.jamison@nasa.gov | 564 |
| Larry Hilliard | Instrument Electronics | lawrence.m.hilliard@nasa.gov | 555 |
| Steve Tompkins | Ground System | steven.d.tompkins@nasa.gov | 581 |
| Anisa Jamil | Power System | anisa.jamil@nasa.gov | 563 |
| Bob Beaman | Power System | bobby.g.beaman@nasa.gov | Sci. Systems & Application Inc. |
| Porfirio "Porfy" Beltran | Electrical Systems/Avionics | porfirio.beltran@nasa.gov | 565 |
| Cleland P. (Paul) Earle | Electrical Systems/Avionics | cleland.p.earle@nasa.gov | AS & D, Inc. |
| Catherin Lynch | Resource Analyst | joanna.m.rojsirivit@nasa.gov | 401 |
| Chris Derkacz | Schedule | christopher.b.derkacz@nasa.gov | 460 |
| Tom D'Asto | Mechanical Engineer | tom.dasto@ataaerospace.com | ATA Aerospace |
| Ben Gavares | Mechanical Designer | nerses.v.armani@nasa.gov | SGT |

## *Origins* Study Advisory Board

Jon Arenberg (Northrop-Grumman)

John Carlstrom (University of Chicago)

Harry Ferguson (STScI JHU)

Tom Greene (NASA Ames)

George Helou (Caltech IPAC)

Lisa Kaltenegger (Cornell University)

Charles Lawrence (Caltech JPL)

Sarah Lipscy (Ball Aerospace)

John Mather (NASA GSFC)

Samuel Harvey Moseley (NASA GSFC)

George Rieke (University of Arizona)

Marcia Rieke (University of Arizona)

Jean Turner (UCLA)

Meg Urry (Yale University)

**Origins Galaxy Evolution and Cosmology Science Working Group members**

Alexandra Pope (University of Massachusetts Amherst) Working Group Co-Lead
Lee Armus (Caltech IPAC) Working Group Co-Lead
Denis Burgarella (Laboratoire d'Astrophysique de Marseille, Aix-Marseille University)
David Alexander (Durham University)
Matteo Bonato (IRA-INAF)
Caitlin Casey (UT Austin)
Dave Sanders (University of Hawaii)
Kartik Sheth (NASA)
Nick Scoville (Caltech)
Kirsten Larson (Caltech)
Avani Gowardhan (Cornell)
Alex Griffiths (University of Nottingham)
David Leisawitz (NASA GSFC)
Desika Narayanan (University of Florida)
Patrick Ogle (STScI JHU)
Eric Murphy (National Radio Astronomy Observatory)
Phil Mauskopf (ASU)
Anna Sajina (Tufts University)
Douglas Scott (University of British Columbia)
J.D. Smith (University of Toledo)
Duncan Farrah (University of Hawaii)
Irene Shivaei (University of Arizona)
Samir Salim (Indiana University)
Carl Ferkinhoff (Winona State University)
Eiichi Egami (University of Arizona)
Ranga Ram Chary (Caltech IPAC)
Johannes Staguhn (Johns Hopkins University & NASA GSFC)
Matthew Malkan (UCLA)
Justin Spilker (UT Austin)
Alberto Bolatto (University of Maryland at College Park)
Tanio Diaz-Santos (University Diego Portales)
Vassilis Charmandaris (University of Crete)
Susanne Aalto (Chalmers University)
Joaquin Vieira (University of Illinois)
Asantha Cooray (UC Irvine)
C.M. (Matt) Bradford (Caltech JPL)
Phil Appleton (Caltech IPAC)
Eduardo González-Alfonso (Universidad de Alcalá, CfA)
Karina Caputi (University of Groningen)
Allison Kirkpatrick (University of Kansas)
Daniel Dale (University of Wyoming)
Charles Lillie (Lillie Consulting LLC)

### *Origins* Milky Way, ISM, and Nearby Galaxies Science Working Group

Cara Battersby (University of Connecticut), Working Group Co-Lead
Karin Sandstrom (University of California San Diego), Working Group Co-Lead
Angela Adamo (Stockholm University)
Alberto Bolatto (UMD College Park)
Elvire De Beck (Chalmers University of Technology)
William J. Fischer (STScI JHU)
Laura Fissel (National Radio Astronomy Observatory)
Maryvonne Gerin (Observatoire de Paris & CNRS)
Paul Goldsmith (Caltech JPL)
Mark Heyer (University of Massachusetts Amherst)
Charles L H Hull (National Astronomical Observatory of Japan, Joint ALMA Observatory)
Alvaro Labiano (Center for Astrobiology, Madrid)
David Leisawitz (NASA GSFC)
Leslie Looney (University of Illinois)
Jeff Mangum (National Radio Astronomy Observatory)
Mikako Matsuura (Cardiff University)
Peregrine McGehee (College of the Canyons)
Stefanie Milam (NASA GSFC)
Elisabeth Mills (Brandeis University)
Edward Montiel (UC Davis)
Eric Murphy (National Radio Astronomy Observatory)
Paolo Padoan (ICREA & Institute of Cosmos Sciences, University of Barcelona)
Thushara Pillai (Max-Planck-Institut fuer Radioastronomie/Boston University)
Jorge Pineda (Caltech JPL)
Dimitra Rigopoulou (University of Oxford)
Julia Roman-Duval (STScI JHU)
Erik Rosolowsky (University of Alberta)
Sarah Sadavoy (Harvard-Smithsonian Center for Astrophysics)
Marta Sewilo (UMD College Park/NASA GSFC)
Lorant Sjouwerman (National Radio Astronomy Observatory)
J.D. Smith (University of Toledo)
Alessio Traficante (IAPS-INAF, Rome)
Wouter Vlemmings (Chalmers University of Technology)
Doug Johnstone (NRC Canada - Herzberg Astronomy and Astrophysics)
Martina C. Wiedner (Observatoire de Paris & CNRS)
Crystal Brogan (National Radio Astronomy Observatory)
Steve Longmore (Liverpool John Moores University)
Umut Yildiz (Caltech JPL)
Klaus Pontoppidan (STScI JHU)

### *Origins* Solar System Working Group

Stefanie Milam (NASA GSFC) Working Group Co-Lead
James Bauer (UMD College Park) Working Group Co-Lead
Glenn Orton (Caltech JPL)
Timothy A. Livengood (UMD College Park, NASA GSFC)

Leigh N. Fletcher (University of Leicester)
Pablo Santos-Sanz (IAA/CSIC, Spain)
Gordon Bjoraker (NASA GSFC)
Arielle Moullet (SOFIA/USRA)
Mark Gurwell (Harvard-Smithsonian CfA)
Conor Nixon (NASA GSFC)
Kimberly Ennico-Smith (NASA Ames)
Geronimo Villanueva (NASA GSFC)
Martin Cordiner (CUA/NASA GSFC)
Dariusz Lis (Caltech JPL)
Perry Gerakines (NASA GSFC)
William T. Reach (SOFIA/USRA)
Ernesto Palomba (Istituto di Astrofisica e Planetologia Spaziali-INAF)

### *Origins* Disks Working Group

Klaus M. Pontoppidan (STScI JHU) Working Group Co-Lead
Kate Su (University of Arizona) Working Group Co-Lead
Edwin Bergin (University of Michigan)
Susanne Aalto (Chalmers University of Technology, Sweden)
Andrea Banzatti (University of Arizona)
Geoffrey A. Blake (Caltech)
Sean Carey (Caltech IPAC)
Christine Chen (STScI JHU)
Kimberly Ennico-Smith (NASA Ames)
William Fischer (STScI JHU)
Lisseth Gavilan (NASA Ames)
Maryvonne Gerin (LERMA, CNRS, France)
Paul Goldsmith (Caltech JPL)
Joel Green (STScI JHU)
Frank Helmich (SRON & University of Groningen, Netherlands)
Doug Johnstone (NRC, Canada)
Roser Juanola-Parramon (NASA GSFC/STScI JHU)
Grant Kennedy (University of Warwick, UK)
Quentin Kral (LESIA, France)
Lars Kristensen (Center of Astrophysics)
Alexander Krivov (Friedrich Schiller University, Germany)
David Leisawitz (NASA GSFC)
Meredith MacGregor (DTM, Carnegie)
Luca Matra (Center for Astrophysics)
Brenda Matthews (NRC, Canada)
Melissa McClure (Universiteit van Amsterdam, Netherlands)
Gary Melnick (Harvard-Smithsonian Center for Astrophysics)
Stefanie Milam (NASA GSFC)
Karin Öberg (Harvard-Smithsonian Center for Astrophysics)
Debbie Padgett (Caltech JPL)
Mark Wyatt (University of Cambridge, UK)

***Origins* Exoplanets Working Group**
Jonathan Fortney (UC Santa Cuz) Working Group Co-Lead
Tiffany Kataria (Caltech JPL) Working Group Co-Lead
Kevin Stevenson (STScI JHU) Working Group Co-Lead
Jayne Birkby (Univ. of Amsterdam)
Geoffrey A. Blake (Caltech)
David R. Ciardi (Caltech/NExScI)
Pablo Cuartas-Restrepo (Universidad de Antioquia - FACom – SEAP)
William Danchi (NASA GSFC)
Thomas M. Evans (MIT)
Jonathan Fraine (Space Science Institute)
Malcolm Fridlund (Leiden Observatory, Netherlands)
Eric Gaidos (University of Hawaii at Manoa)
Thomas Greene (NASA Ames)
Sam Halverson (MIT)
Jacob Haqq-Misra (Blue Marble Space Institute of Science)
Lisa Kaltenegger (Cornell University)
Stephen Kane (UC Riverside)
Eliza Kempton (UMD College Park)
Ravi Kopparapu (NASA GSFC)
Michael R. Line (Arizona State University)
Eric Mamajek (ex-officio) (Caltech JPL)
Tiffany Meshkat (Caltech IPAC)
Eric Nielsen (Stanford University)
Klaus Pontoppidan (STScI JHU)
Thomas Roellig (NASA Ames)
Kartik Sheth (NASA HQ)
Clara Sousa-Silva (MIT)
Johannes Staguhn (JHU, NASA GSFC)
Keivan Stassun (Vanderbilt University)
Luke Tremblay (Arizona State University)
Miguel de Val-Borro (NASA GSFC)
Eric T. Wolf (UC Boulder)
Robin Wordsworth (Harvard University)
Robert T. Zellem (Caltech JPL)

***Origins* Instrument Study Teams**
**Mid-Infrared Spectrometer and Camera (MISC) Team**
Itsuki Sakon (University of Tokyo, JAXA) MISC Co-Instrument Lead
Thomas Roellig (NASA Ames) MISC Co-Instrument Lead
Kimberly Ennico-Smith (NASA Ames) MISC Science Lead
Keigo Enya (ISAS/JAXA)
Naofumi Fujishiro (Teikyo University)
Shohei Goda (Osaka Univ.)
Thomas Greene (NASA Ames)
Bernard's Helvensteijn (NASA Ames)

Lynn Hofland (NASA Ames)
Masayuki Ido (Osaka University)
Naoto Iida (Kyocera)
Yuji Ikeda (Photocoding) MISC Imager optical and structural design
Satoshi Itoh (Osaka University)
Roy Johnson (NASA Ames)
Masatsugu Kamiura (Kyocera)
Ali Kashani (NASA Ames)
Mitsunobu Kawada (ISAS/JAXA)
Takayuki Kotani (NAOJ)
Takeo Manome (Kyocera)
Taro Matsuo (Osaka University) MISC densified spectrometer design
Naoshi Murakami (Hokkaido University)
Bob McMurray (NASA Ames)
Jun Nishikawa (NAOJ)
Emmett Quigley (NASA Ames)
Yuki Sarugaku (Kyoto Sangyo University)
Hiroshi Shibai (Osaka University)
Takahiro Sumi (Osaka University)
Aoi Takahashi (ISAS/JAXA)
Masayuki Tsuboi (Osaka University)
Takehiko Wada (ISAS/JAXA)
Tomoyasu Yamamuro (OptoCraft)MISC TRA optical and structural design
Kentaro Yanagibashi (Kyocera)

### *Origins* Far-infrared Imager and Polarimeter (FIP) Team
Johannes Staguhn (JHU/NASA GSFC), FIP Instrument Lead
Margaret Meixner (STScI JHU), FIP Science co-Lead
Joaquin Vieira (UIUC), FIP Science co-Lead
Edward Amatucci (NASA GSFC)
Greg Martins (NASA GSFC)
Mike Dipirro (NASA GSFC)
Damon Bradley (NASA GSFC)
Asantha Cooray (UCI)
David Leisawitz (NASA GSFC)
Lee Armus (Caltech IPAC)
Cara Battersby (UConn)
Edwin Bergin (UMich, Ann Arbor)
C. Matt Bradford (Caltech JPL/NASA)
Kimberly Ennico-Smith (NASA Ames)
Gary Melnick (Harvard-Smithsonian Center for Astrophysics)
Stefanie Milam (NASA GSFC)
Desika Narayanan (Univ. of Florida)
Klaus Pontoppidan (STScI JHU)
Alexandra Pope (University of Massachusetts Amherst)
Thomas Roellig (NASA Ames)
Karin Sandstrom (UC San Diego)

Kate Su (U of Az)
Edward Wright (UCLA)
Jonas Zmuidzinas (Caltech)
David Chuss (Villanova)
Eli Dwek (NASA GSFC)
Edward Wollack (NASA GSFC)

## *Origins* Survey Spectrometer (OSS) Team
Matt Bradford (Caltech JPL), OSS Instrument Lead
Lee Armus (Caltech IPAC), OSS Science Lead
Bruce Cameron (Caltech JPL)
Bradley Moore (Caltech JPL)
Edward Amatucci (NASA GSFC)
Damon Bradley (NASA GSFC)
James Corsetti (NASA GSFC)
David Leisawitz (NASA GSFC)
S. Harvey Moseley (NASA GSFC)
 Johannes Staguhn (NASA GSFC)
James Tuttle (Caltech JPL)
Ari Brown (Caltech JPL)
Alexandra Pope (University of Massachusetts)
Margaret Meixner (STScI JHU)
Klaus Pontoppidan (STScI JHU)

## *Origins* HEterodyne Receiver for Origins (HERO) Team
Martina Wiedner (Observatoire de Paris & CNRS, France) HERO Instrument Lead
Maryvonne Gerin (Observatoire de Paris & CNRS, France) HERO Science Lead
André Laurens (CNES) HERO CNES Instrument Support
Susanne Aalto (Chalmers University of Technology, Sweden)
Gabby Aitink-Kroes (SRON Netherlands Institute for Space Research, NL & NOVA, ASTRON, NL)
Marcelino Agúndez (Instituto de Física Fundamental, CSIC, Spain)
Andrey Baryshev (NOVA/Kapteyn Astronomical Institute, NL) Receiver Expert
Victor Belitsky (Chalmers University of Technology, Sweden) Mixer Expert
Olivier Berné (IRAP, CNRS, CNES and Université Paul Sabatier, France)
Wilfried Boland (NOVA, NL)
Bruno Borgo (LESIA, Observatoire de Paris & CNRS, France)
Paola Caselli (Max-Planck-Institute for Extraterrestrial Physics, Germany)
Emmanuel Caux (IRAP, Université de Toulouse, CNRS, CNES, UPS, (France)
Elvire De Beck (Chalmers University of Technology, Sweden)
Ilse De Looze (Ghent University, Belgium)
Yan Delorme (Observatoire de Paris & CNRS, France)
Vincent Desmaris (Chalmers University of Technology, Sweden) Mixer Expert
Anna Maria Di Giorgio (INAF - IAPS, Italy) Instrument Control Expert
Martin Eggens (SRON Netherlands Institute for Space Research, NL)
Brian Ellison (STFC Rutherford Appleton Laboratory, UK)
Malcolm Fridlund (Chalmers University of Technology, Sweden)

Juan Daniel Gallego Puyol (Yebes Observatory, Spain) Amplifier Expert
Emanuele Galli (INAF - IAPS Rome, Italy)
Paul Goldsmith (Caltech JPL)
Christophe GOLDSTEIN (CNES, France)
Antoine Gusdorf (Observatoire de Paris & CNRS, France)
Frank Helmich (SRON Netherlands Institute for Space Research, Groningen, NL & Kapteyn
   Astronomical Institute, Univ. of Groningen, NL & Leiden Observatory, Univ. of Leiden, NL)
   Instrument Scientist
Richard Hills (University of Cambridge, UK)
Michiel Hogerheijde (Leiden Observatory, Leiden University & API, University of Amsterdam, NL)
Leslie Hunt (INAF-Osservatorio di Arcetri, Italy)
Robert Huisman (SRON Netherlands Institute for Space Research, NL)
Willem Jellema (SRON Netherlands Institute for Space Research, NL & Kapteyn Astronomical Insti-
tute, University of Groningen, NL) Optical design
Anders Johansen (Lund Observatory, Sweden)
Carsten Kramer (IRAM, France)
Geert Keizer (SRON Netherlands Institute for Space Research, NL)
Ute Lisenfeld (University of Granada, Spain)
Dariusz Lis (Sorbonne Université, Observatoire de Paris, Université PSL, CNRS, LERMA, France)
Stefanie Milam (NASA GSFC)
Imran Mehdi (Caltech JPL) Local Oscillator Expert
Ramon Navarro (NOVA, ASTRON, NL)
Napoléon Nguyen Tuong (LESIA, Observatoire de Paris & CNRS, France)
Lorenzo Piazzo ( Univ. Roma "Sapienza", Italy)
René Plume (Dept. of Physics & Astronomy, University of Calgary, Canada)
Dimitra Rigopoulou (University of Oxford, U.K.)
Christophe Risacher (Institut de Radioastronomie Millimetrique, France)
Peter Schilke (I. Physikalisches Institut, University of Cologne, Germany)
Russell Shipman (SRON Netherlands Institute for Space Research, NL & Kapteyn Astronomical
   Institute University of Groningen, NL)
Floris van der Tak (SRON & University of Groningen, NL)
Wouter Vlemmings (Chalmers University of Technology, Sweden)
Eva Wirström (Chalmers University of Technology, Sweden)
Friedrich Wyrowski (Max Planck Institute for Radio Astronomy, Germany)

### *Origins* Document and Graphics Preparation Support Team

| | |
|---|---|
| Theophilus Britt Griswold (NASA GSFC) | Tim Pyle (Caltech IPAC) |
| Steve Stuart (NASA GSFC) | Jacob Llamas (Caltech IPAC) |
| Jay Friedlander (NASA GSFC) | Eric Oh (Caltech IPAC) |
| Jena Ballard (NASA GSFC) | Dave Shupe (Caltech IPAC) |
| Gabby Garcia (NASA GSFC) | Alex Lockwood (STScI JHU) |
| J. D. Myers (NASA GSFC) | Leah Hustak (STScI JHU) |
| Robert Hurt (Caltech IPAC) | |
| Janice Lee (Caltech IPAC) | |

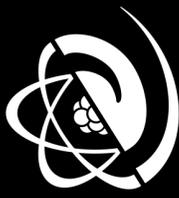
# ORIGINS
## Space Telescope

From
first stars
to life

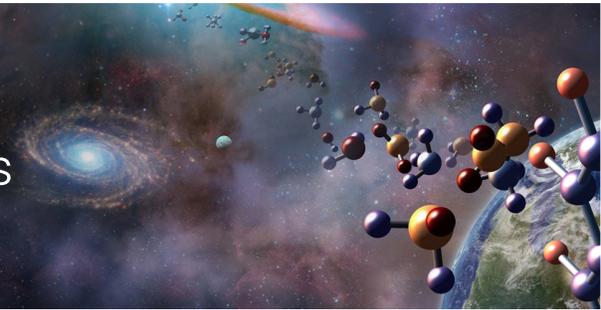

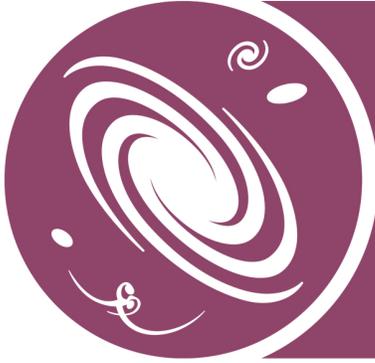

## HOW DOES THE UNIVERSE WORK?

**How do galaxies form stars, make metals, and grow their central supermassive black holes from reionization to today?**

Using sensitive spectroscopic capabilities of a cold telescope in the infrared, Origins will measure properties of star-formation and growing black holes in galaxies across all epochs in the Universe.

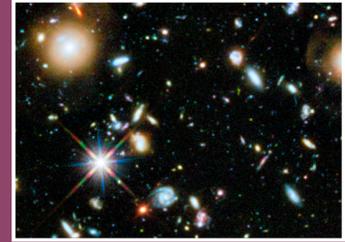

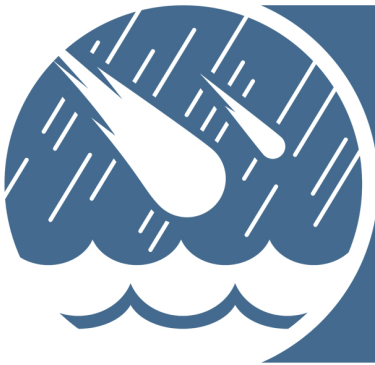

## HOW DID WE GET HERE?

**How do the conditions for habitability develop during the process of planet formation?**

With sensitive and high-resolution far-IR spectroscopy Origins will illuminate the path of water and its abundance to determine the availability of water for habitable planets.

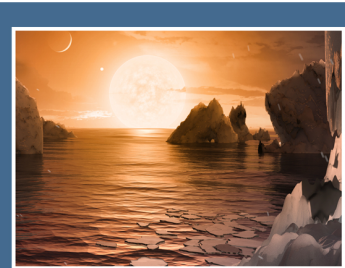

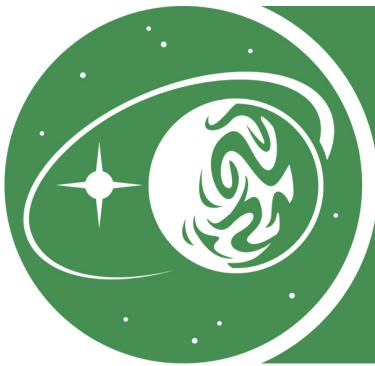

## ARE WE ALONE?

**Do planets orbiting M-dwarf stars support life?**

By obtaining precise mid-infrared transmission and emission spectra, Origins will assess the habitability of nearby exoplanets and search for signs of life.

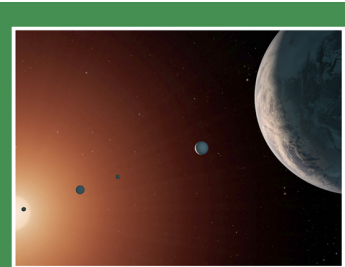

## SCIENCE DRIVERS FOR MISSION DESIGN

## DISCOVERY SPACE OF ORIGINS

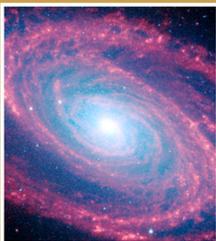
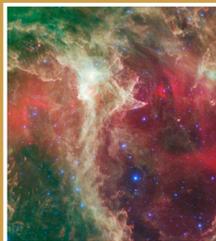
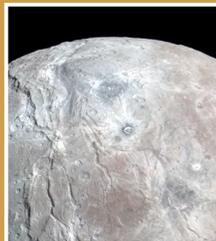

Origins is not only capable of addressing known questions, but has a vast discovery space that will enable astronomers in the 2030s to find new phenomena and address currently unknown questions. All science programs on Origins will be selected by the community via peer review.

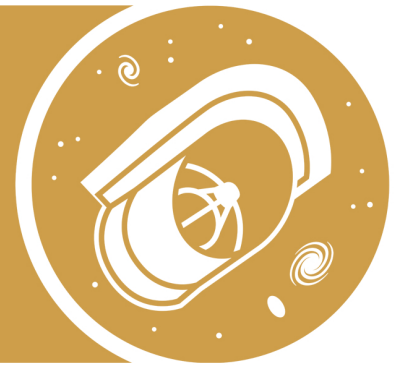

The Origins Space Telescope is a community-led mission concept study sponsored by NASA in preparation for the 2020 Astronomy and Astrophysics Decadal Survey.

**origins.ipac.caltech.edu**

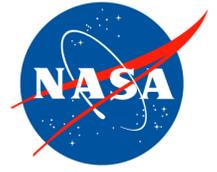

National Aeronautics and Space Administration

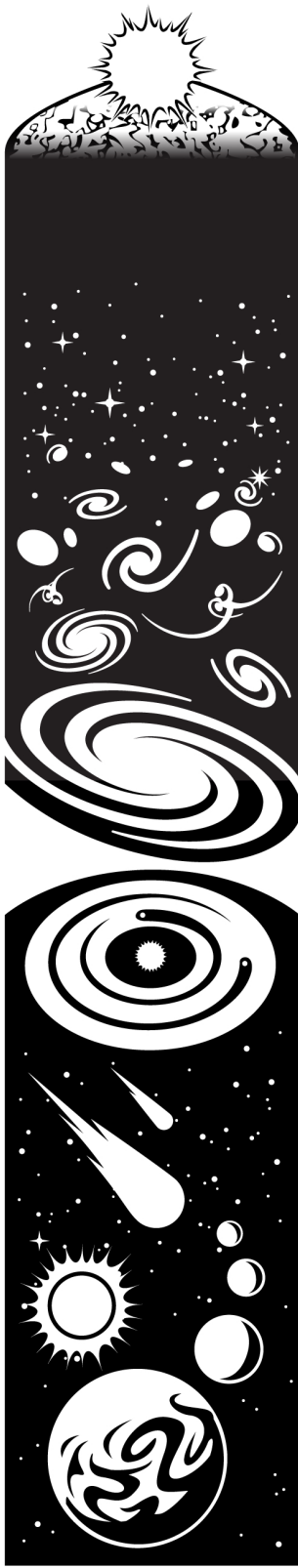

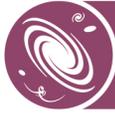

## How do galaxies form stars, make metals, and grow their central supermassive black holes from reionization to today?

Measure the redshifts, star formation rates, and black hole accretion rates in main sequence galaxies since the epoch of reionization, down to a SFR of 1 $M_\odot$/yr at cosmic noon and 10 $M_\odot$/yr at z~6, performing the first unbiased survey of the coevolution of stars and supermassive black holes over cosmic time.

Measure the metal content of galaxies with a sensitivity down to 10% Solar in a galaxy with a stellar mass similar to the Milky-Way at z of 6 as a function of cosmic time, tracing the rise of heavy elements, dust, and organic molecules across redshift, morphology, and environment.

Determine how energetic feedback from AGN and supernovae regulate galaxy growth, quench star formation, and drive galactic ecosystems by measuring galactic outflows as a function of SFR, AGN luminosity, and redshift over the past 10 Gyr.

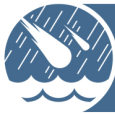

## How do the conditions for habitability develop during the process of planet formation?

Measure the water abundance at all evolutionary stages of star and planet formation and across the range of stellar masses, tracing water vapor and ice at all temperatures between 10 and 1000 K.

Determine the ability of planet-forming disks at all evolutionary stages and around stars of all masses to form planets with masses as low as one Neptune mass using the HD 1-0 line to measure the total disk gas mass.

Determine the cometary contribution to Earth's water by measuring the D/H ratio in over 100 comets in 5 years.

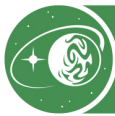

## Do planets orbiting M-dwarf stars support life?

Distinguish between tenuous, clear and cloudy atmospheres on at least 28 temperate, terrestrial planets orbiting M and K dwarfs using $CO_2$ and other spectral features.

Establish the apparent surface temperatures of at least 17 terrestrial exoplanets with the clearest atmospheres and distinguish between boiling and freezing surface water at ≥3 σ confidence (±33 K).

Search for biosignatures on at least 10 planets, highest ranked from Objectives 1 and 2, and if present, detect biosignatures at ≥3.6 σ assuming an Earth-like atmosphere).

**origins.ipac.caltech.edu**

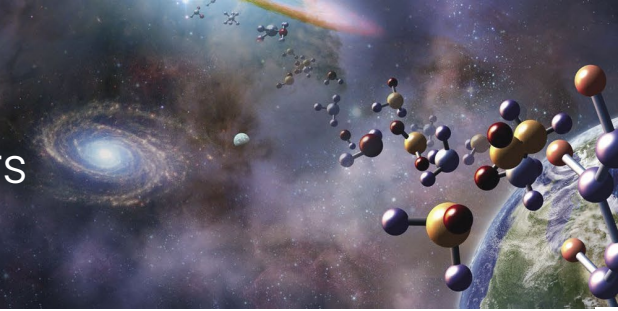

# ORIGINS
## Space Telescope
From first stars to life

| Origins Observatory Level Parameters | |
|---|---|
| **Mission Parameter** | **Value** |
| Telescope: Aperture Diameter/Area | 5.9 m/25 m² |
| Telescope Diffraction Limited at | 30 μm |
| Telescope Temperature | 4.5 K |
| Wavelength Coverage | 2.8–588 μm |
| Maximum Scanning Speed | 60" per second |
| Mass: Dry/Wet (with margin) | 12000 kg/13000 kg |
| Power (with margin) | 4800 W |
| Launch Year | 2035 |
| Launch Vehicle (large vehicle) | SLS or Space-X BFR |
| Orbit | Sun-Earth L2 |
| Propellant lifetime | 10 years, serviceable |

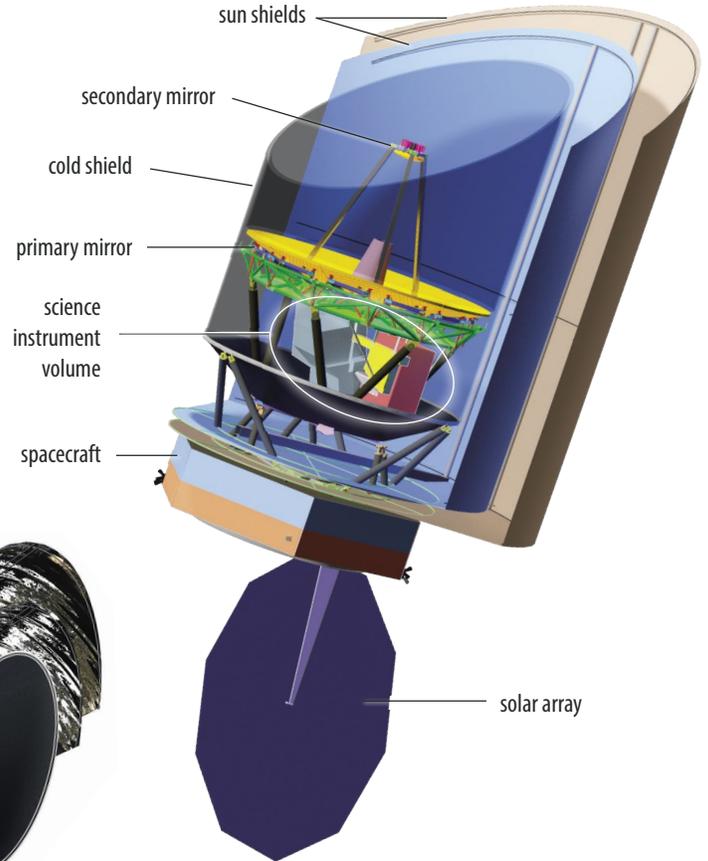

sun shields
secondary mirror
cold shield
primary mirror
science instrument volume
spacecraft
solar array

## Origins Mission Concept

The Origins Space Telescope has a Spitzer-like architecture with simplified deployment for the sun shields. Origins' modular instrument bay facilitates integration and test and allows serviceability. Origins has a low-risk verification program with flight-like tests of critical systems.

The Origins Space Telescope will map wide fields of view, orders of magnitude faster than previous and planned missions, and provide ultra-stable infrared spectroscopy.

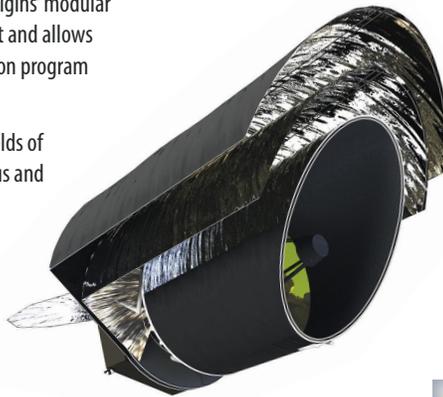

## Origins Detectors

**Detector technology development plan will reach TRL 5 by 2025.**

- Transition-edge Sensor (TES) bolometers:
    operate <50 mK, for FIP, OSS and MISC-T
- Kinetic Inductance Detectors (KIDs) operate <50 mK, for FIP, OSS
- Si: As arrays: operate 7 K, high stability for MISC-T
- HgCdTe arrays: operate 30 K, high stability for MISC-T

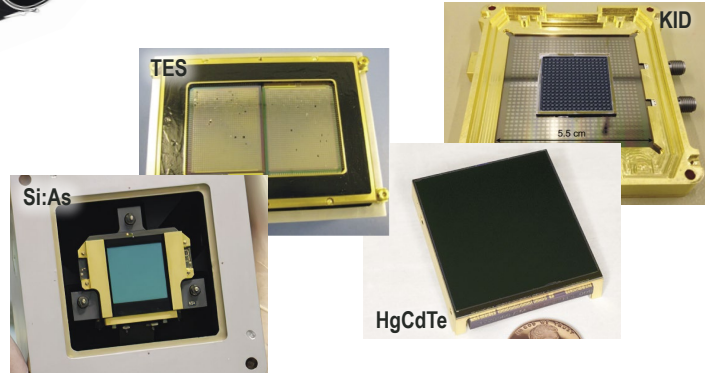

KID
TES
Si:As
HgCdTe

## Origins Mission Schedule

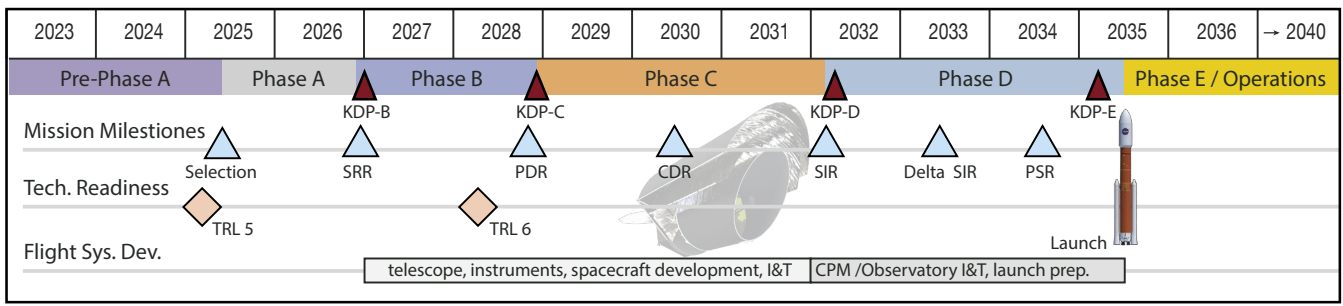

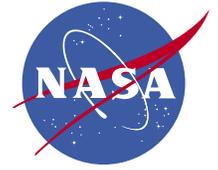

## Origins Instruments Performance

| Instrument/ Observing Mode | Wavelength Coverage (µm) | Field of View (FOV) | Spectral Resolving Power (R=λ/Δλ) | Saturation Limits | Representative Sensitivity 5σ in 1 hr |
|---|---|---|---|---|---|
| *Origins* Survey Spectrometer (OSS) | | | | | |
| Grating | 25–588 µm simultaneously | 6 slits for 6 bands: 2.7´ x 1.4˝ to 14´ x 20˝ | 300 | 5 Jy @ 128 µm | 3.7 x 10⁻²¹ W m⁻² @ 200 µm |
| High Resolution | 25–588 µm with FTS | Slit: 20˝ [2.7˝ to 20˝] | 43,000 x [112 µm/λ] | 5 Jy @ 128 µm | 7.4 x 10⁻²¹ W m⁻² @ 200 µm |
| Ultra-High Resolution | 100–200 µm | One beam: 6.7˝ | 325,000 x [112 µm/λ] | 100 Jy @ 180 µm | 2.8 x 10⁻¹⁹ W m⁻² @ 200 µm |
| Far-IR Imager Polarimeter (FIP) | | | | | |
| Pointed | 50 or 250 µm (selectable) | 50 µm: 3.6´ x 2.5´ 250 µm: 13.5´ x 9´ (109 x 73 pixels) | 3.3 | 50 µm: 1 Jy 250 µm: 5 Jy | 50/250 µm: 0.9/2.5 µJy Confusion limit: 50/250 µm: 120 nJy/1.1 µJy |
| Survey mapping | 50 or 250 µm (selectable) | 60˝ per second scan rate, with above FOVs | 3.3 | 50 µm: 1 Jy 250 µm: 5 Jy | Same as above, confusion limit reached in 50/250 µm: 1.9 hours/2 msec |
| Polarimetry | 50 or 250 µm (selectable) | 50 µm: 3.6´ x 2.5´ 250 µm: 13.5´ x 9´ | 3.3 | 50 µm: 2 Jy 250 µm: 10 Jy | 0.1% in linear and circular polarization, ±1° in pol. Angle |
| Mid-Infrared Spectrometer Camera Transit Spectrometer (MISC-T) | | | | | |
| Ultra-Stable Transit Spectroscopy | 2.8–20 µm in 3 simultaneous bands | 2.8–10.5 µm: 2.5˝ radius 10.5–20 µm: 1.7˝ radius | 2.8-10.5 µm: 50–100 10.5-20 µm: 165–295 | K~3.0 mag 30 Jy @ 3.3µm | Assume K~9.85 mag M-type star, R=50 SNR/sqrt(hr)>12,900 @ 3.3 µm in 60 transits with stability ~5 ppm < 10.5 µm, ~20 ppm ≥ 10.5 µm |

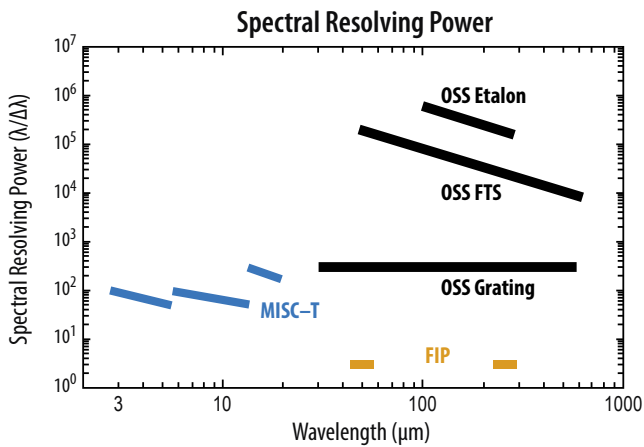

**Spectral Resolving Power**

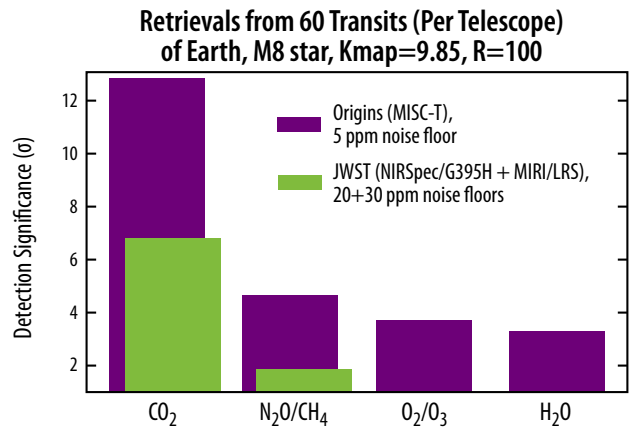

**Retrievals from 60 Transits (Per Telescope) of Earth, M8 star, Kmap=9.85, R=100**

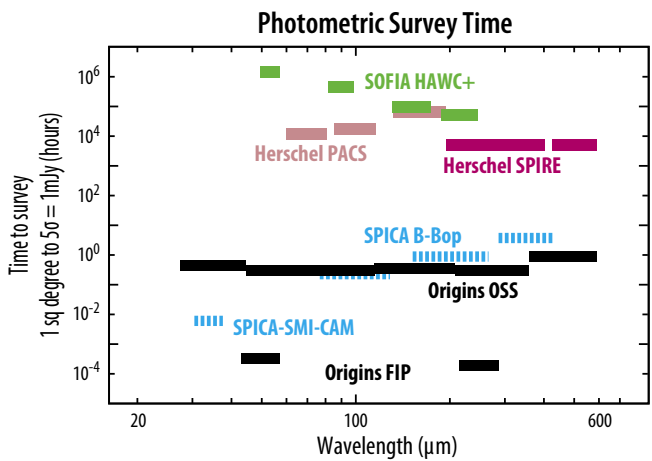

**Photometric Survey Time**

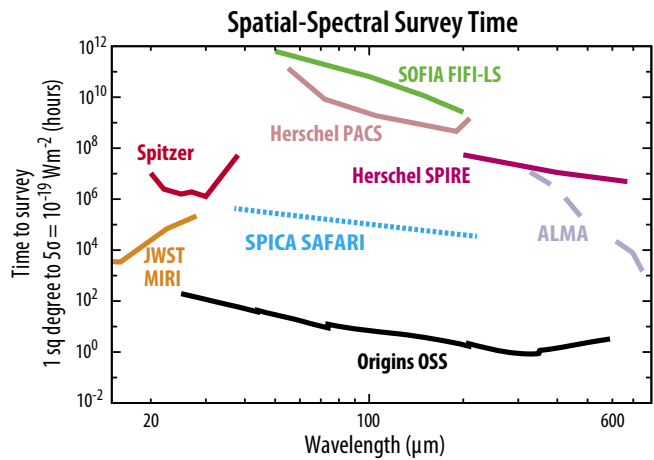

**Spatial-Spectral Survey Time**

## Contents

## EXECUTIVE SUMMARY

> The Origins Space Telescope (*Origins*) traces our cosmic history, from the formation of the first galaxies and the rise of metals to the development of habitable worlds and present-day life. Origins does this through exquisite sensitivity to infrared radiation from ions, atoms, molecules, dust, water vapor and ice, and observations of extra-solar planetary atmospheres, protoplanetary disks, and large-area extragalactic fields. *Origins* operates in the wavelength range 2.8 to 588 µm and is more than 1000 times more sensitive than its predecessors due to its large, cold (4.5 K) telescope and advanced instruments.

*Origins* (Figure ES-1) investigates the creation and dispersal of elements essential to life, the formation of planetary systems and the transport of water to habitable worlds and the atmospheres of exoplanets around nearby K- and M-dwarfs to identify potentially habitable—and even inhabited—worlds. These science priorities are motivated by their profound significance, as well as expected advances from, and limitations of, current and next generation observatories (JWST, WFIRST, ALMA, LSST). The nine key *Origins* scientific objectives (Table ES-1) address NASA's three major astrophysics science goals: **How does the Universe work? How did we get here?** and **Are we alone?**

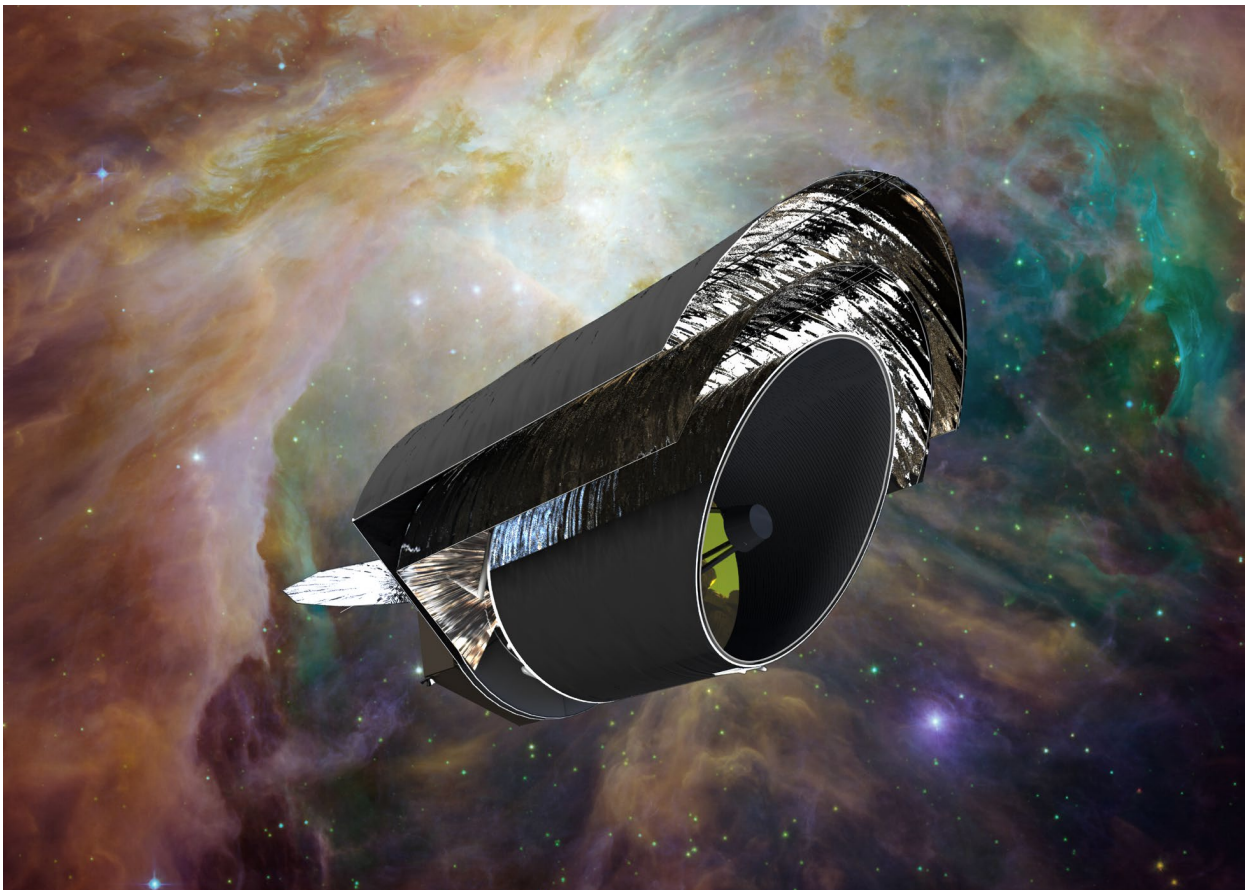

**Figure ES-1:** The *Origins* concept is low-risk and powerful: 1000 times more sensitive than prior far-IR missions. *Origins*, with an aperture diameter of 5.9 m and a suite of powerful instruments, operates with spectral resolving power from 3 to 3x10$^5$ over the wavelength range from 2.8 to 588 µm. The *Origins* design has very few critical deployments and builds upon the technical heritage of *Spitzer*, with passive cooling from a two-layer Sun shield and advanced cryocoolers maintaining the telescope at 4.5 K. *Origins* is agile, enabling >80% observing efficiency, in line with the 90% efficiency achieved with *Herschel*.

**Table ES-1:** Mission Design Scientific Drivers for the *Origins* Space Telescope

| NASA Goal | How Does the Universe Work? | | How Did We Get Here? | | Are We Alone? | |
|---|---|---|---|---|---|---|
| *Origins* Science Goals | 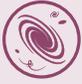 | How do galaxies form stars, make metals, and grow their central supermassive black holes from reionization to today? | 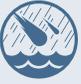 | How do the conditions for habitability develop during the process of planet formation? | 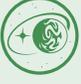 | Do planets orbiting M-dwarf stars support life? |
| *Origins* Scientific Capabilities | Using sensitive spectroscopic capabilities of a cold telescope, *Origins* will measure properties of star-formation and growing black holes in galaxies across all epochs. | | With sensitive, high-resolution spectroscopy, *Origins* will illuminate the path of water and its abundance to determine the availability of water for habitable planets. | | By obtaining precise mid-infrared transmission and emission spectra, *Origins* will assess the habitability of nearby exoplanets and search for signs of life. | |
| *Origins* Scientific Objectives | 1. How does the relative growth of stars and supermassive black holes in galaxies evolve with time? 2. How do galaxies make metals, dust, and organic molecules? 3. How do the relative energetics from supernovae and quasars influence the interstellar medium of galaxies? | | 1. What role does water play in the formation and evolution of habitable planets? 2. How and when do planets form? 3. How were water and life's ingredients delivered to Earth and to exoplanets? | | 1. What fraction of terrestrial planet around M- and K-dwarf stars has tenuous, clear, or cloudy atmospheres? 2. What fraction of terrestrial M-dwarf planets is temperate? 3. What types of temperate, terrestrial, M-dwarf planets support life? | |

*Origins* addresses these questions by achieving its nine key scientific objectives (Table ES-1) in two years. These Objectives drive the instrumental requirements shown in Table ES-2 (a comprehensive Science Traceability Matrix is presented in Section 1.4). The *Origins* design is powerful and versatile, and the infrared emission *Origins* detects is information-rich. *Origins* will enable astronomers in the 2030s to ask new questions not yet imagined, and provide a far-infrared window (Figure ES-2) complementary to planned, next-generation observatories such as LISA, Athena, and ground-based ELTs.

## ES.1 Science Objectives

**How do galaxies form stars, make metals, and grow their central supermassive black holes, from the Epoch of Reionization to today?** Decades of observations have shown that galaxies condensed out of primordial gas, built up their stellar mass, heavy metals, and central supermassive black holes (SMBHs), and evolved into the systems we see today. Yet we still do not understand how this happened. There is a rich interplay between the drivers of galaxy evolution, which can only be understood through new observations in a currently inaccessible wavelength regime, the far-infrared.

Far-infrared observations are required because galaxies are dusty. Dust is a byproduct of star formation that is essential to astrophysical processes, from planetesimal formation in protoplanetary disks, to radiation-driven galactic outflows. Dust also obscures star formation and SMBH growth, since it efficiently absorbs ultraviolet and visible light, rendering the driving processes of galaxy evolution nearly invisible at these wavelengths. However, the dust re-emits this energy in the far-infrared, making optically dim galaxies brilliant infrared sources. Just as importantly, infrared and molecular emission lines, which directly trace star formation, black hole growth, and metal abundances, can escape dusty galaxies, making the mid- and far-infrared the only bands where a direct, unbiased view of galaxy and metal growth is possible. This lesson was reinforced by *Spitzer* and *Herschel*, which provided our first glimpses of dusty galaxies in the infrared during the peak epoch of star formation, when the Universe was only 3 Gyr old. With 1000 times better sensitivity, *Origins* gives us a clear view of galaxy and metal growth across cosmic time, through a deep-and-wide spectroscopic survey in a wavelength regime inaccessible from the ground and that will remain unexplored by JWST.

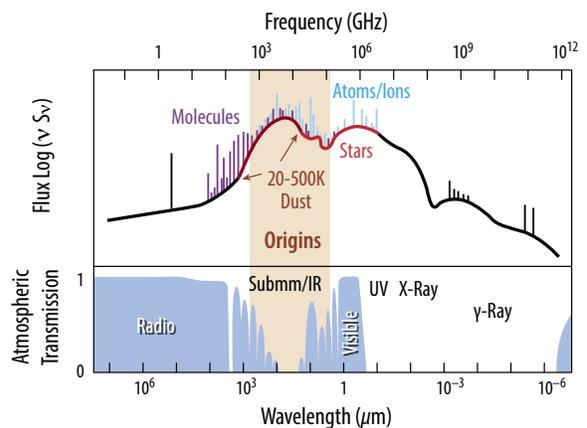

**Figure ES-2:** *Origins* studies the universe at a wavelength range that is inaccessible from the ground. Infrared photons between 2.8 and 588 μm in *Origins'* wavelength range capture emission from stars, molecules, dust, and ions/atoms, enabling a multi-pronged probe into key physical processes in galaxies.

**How do the stars and supermassive black holes in galaxies evolve with time?** *Origins* uses atomic and molecular emission lines and emission from dust grains to measure the density, temperature, and ionization state of the gas where stars are forming and in galactic nuclei. These observations probe the physics of the interstellar medium, characterize the atomic and molecular gas that drives star-formation, measure the buildup of metals from dying stars, and track the growth of SMBHs and their influence, as they drive energetic outflows into the surrounding interstellar medium (Figure ES-3).

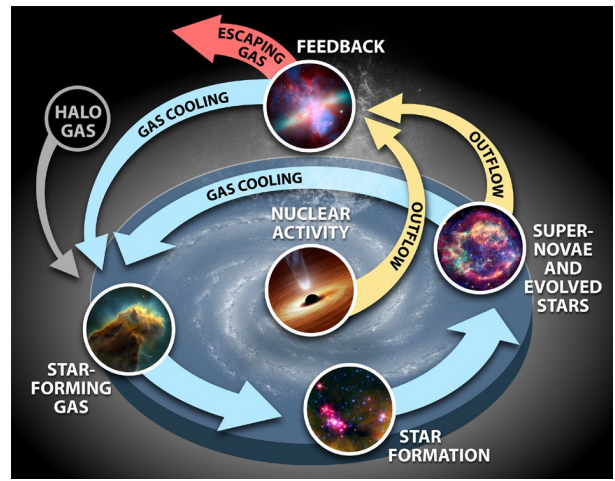

**Figure ES-3:** *Origins* studies the baryon cycle in galactic ecosystems. Energetic processes that shape galaxies and the circumgalactic medium together define this ecosystem. Through its ability to measure the energetics and dynamics of the atomic and molecular gas and dust in and around galaxies, *Origins* can probe nearly all aspects of the galactic ecosystem: star formation and AGN growth; stellar death; AGN- and starburst-driven outflows; and gas cooling and accretion. These measurements provide a complete picture of the lifecycle of galaxies.

**How do galaxies make metals, dust, and organic molecules?** Galaxies are the metal factories of the Universe, and *Origins* studies how heavy elements and dust were made and dispersed throughout the cosmic web over the past 12 billion years. Sensitive metallicity indicators in the infrared can be used to track the growth of heavy elements in even the densest optically-obscured regions inside galaxies.

**How do the relative energetics from supernovae and quasars influence the interstellar medium of galaxies?** Galaxies are made of billions of stars, yet star formation is extremely inefficient on all scales, from single molecular clouds to galaxy clusters. The reason is thought to be 'feedback' from star formation or black hole growth, because supernovae or quasar winds can disrupt star-forming gas. *Origins* studies the role of feedback processes in galaxies over a wide range of environments and redshifts by investigating the processes that drive powerful outflows and surveying the demographics of galactic feedback.

**How do the conditions for habitability develop during the process of planet formation?** Water is essential to all life on our planet. Water provides the liquid medium for life's chemistry and plays an essential biochemical role. However, we do not know how terrestrial planets get their water, since rocky planets with liquid water exist in regions where water in icy, protoplanetary dust would have sublimated and photo-dissociated. With its broad wavelength coverage, *Origins* can detect many water vapor lines that trace the entire range of temperatures found in protoplanetary disks, from the cold snowline to the hot steam line (Figure ES-4). *Origins* can survey all reservoirs of water in more than 1000 planet-forming disks around stars of all masses, including the faint M-dwarfs that likely host most planets in the Galaxy.

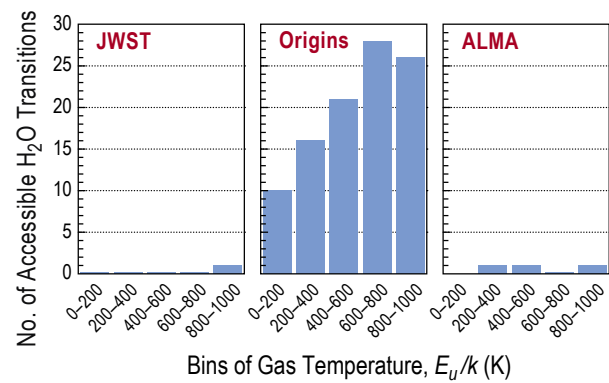

**Figure ES-4:** *Origins* studies more than 100 transitions of water vapor, compared to one and three with JWST and ALMA, respectively. The plot shows the number of $H_2^{16}O$ transitions observable by JWST, *Origins*, and ALMA as a function of the gas temperature for energies above the ground state less than 1000 K. ALMA is limited by atmospheric absorption in its ability to observe water.

Another mystery in planetary formation is the role of hydrogen gas in protoplanetary disks. This gas is the reservoir from which gas giants and planetesimals emerge and the latter are the building blocks of rocky planets. Despite decades of effort, the hydrogen gas mass of protoplanetary disks is essentially unknown because molecular hydrogen, being a symmetric molecule, is largely invisible at the low temperatures of planet-forming gas and the traditional gas mass tracer, CO, is uncertain by factors of 10-100 in disks. *Origins* can observe the HD 112 μm line, which provides a new and robust measure of disk gas mass.

**What role does water play in the formation and evolution of habitable planets?** With its unprecedented sensitivity to weak emission from all forms of water (ice as well as gas), *Origins* deciphers the role of water throughout each phase of planetary system formation (Figure ES-5).

**How and when do planets form?** *Origins* can uniquely use the HD 112 μm emission line, a powerful tool to measure the gas mass of protoplanetary disks to within a factor of 2-3. This precision is one to two orders of magnitude better than alternative tracers and can distinguish between competing models of planet formation and set the timescale for gas giant formation. *Origins*' gas-mass measurements have the potential to provide calibrations for all other observations of protoplanetary disks, including those made with ALMA.

**How were water and life's ingredients delivered to Earth and to exoplanets?** Earth likely formed within the snowline – the distance from a young star where water transitions from a gas to a solid. Thus, the prevailing theory holds that water was delivered to the early Earth via impacts by bodies that formed beyond the snowline.

The evidence for this comes from the high deuterium content of Earth's oceans relative to the protosolar nebula, an excess that is locked when water is formed at a temperature of 10-20 K. In our Solar System, comets and asteroids also carry this signature. Only a handful of comets have been measured to date and they show a range of deuterium abundances. With a larger sample of comets, *Origins* can finally establish if comets were the source of Earth's water.

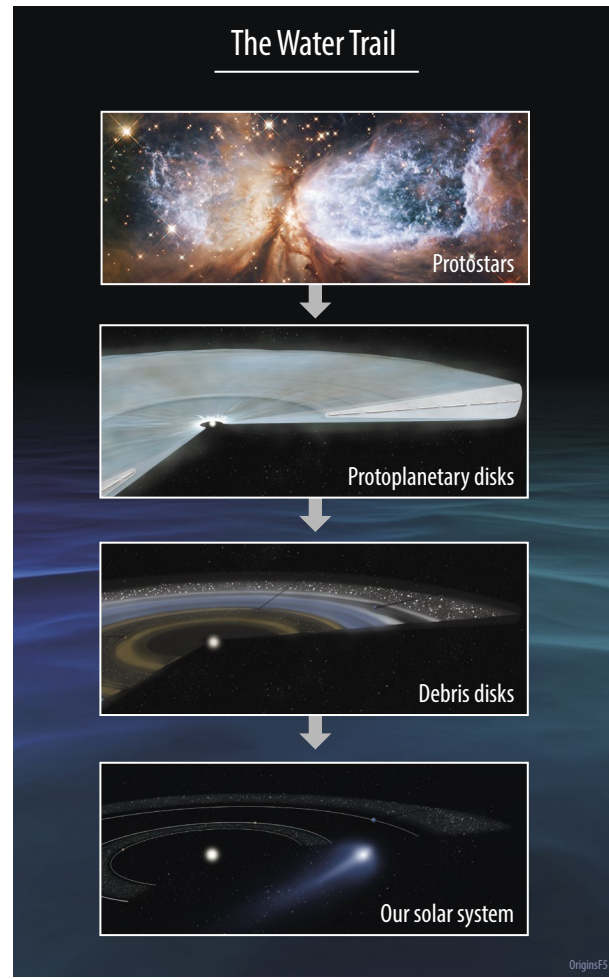

**Figure ES-5:** *Origins* will trace water and gas during all phases of the formation of a planetary system. The trail begins in the "pre-stellar" phase explored by *Herschel*, where a cloud of gas collapses (top) into a still- forming star surrounded by a disk nearly the size of our Solar System and a collapsing envelope of material (2nd from top). Over time, the envelope dissipates, leaving behind a young star and a disk with nascent planets (3rd from top), and eventually leaving behind a new planetary system (bottom). *Origins* will excel at probing the protoplanetary and later phases.

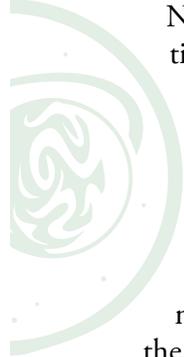

**Do planets orbiting M-dwarf stars support life?** Humankind has long pondered the question, "Are we alone?"

Now scientists and engineers are designing instruments dedicated to answering this question. Our quest to search for life on planets around other stars relies on our ability to measure the chemical composition of their atmospheres and understand the data in the context of models for planet formation and evolution. Using the techniques of transmission and emission spectroscopy, *Origins* will expand upon the legacy of *Hubble* and *Spitzer* – and soon JWST – with a mid-infrared instrument specifically designed to characterize temperate, terrestrial exoplanets. In its search for signs of life, *Origins* will employ a multi-tiered strategy, beginning with a sample of planets with well-determined masses and radii that are transiting nearby M dwarfs, the most abundant stars in the galaxy. With its broad, simultaneous wavelength coverage and unprecedented stability, *Origins* will be uniquely capable of detecting life (Figure ES-6).

**What fraction of terrestrial K- and M-dwarf planets has tenuous, clear, or cloudy atmospheres?** In the first tier of its exoplanet survey, *Origins* will obtain transmission spectra over 2.8–20 μm for temperate, terrestrial planets spanning a broad range of planet sizes, equilibrium temperatures, and orbital distances to distinguish between tenuous, clear, and cloudy atmospheres. Because $CO_2$ absorption features are so large, this tier can include terrestrial planets orbiting stars from late-M to late-K, giving *Origins* a broader perspective in the search for life than JWST.

**What fraction of terrestrial M-dwarf planets is temperate?** For a subset of planets with the clearest atmospheres, *Origins* will measure their thermal emission to determine the temperature structure of their atmospheres. This is critical to assessing climate because it yields an understanding of how incoming stellar and outgoing thermal radiation dictate the heating and cooling of the atmosphere. *Origins* can then determine whether these atmospheric conditions could support liquid water near the surface.

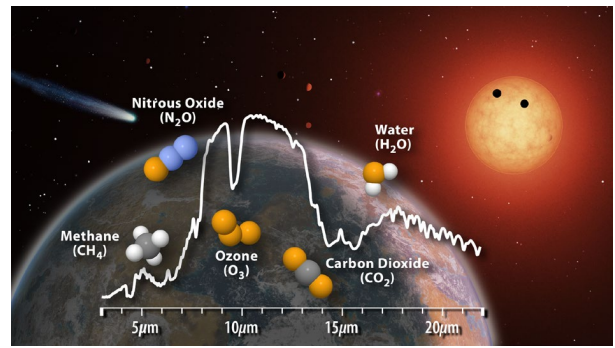

**Figure ES-6:** *Origins* is designed to characterize already-discovered rocky planets that transit M dwarf stars and place critical constraints on their temperatures. By leveraging the mid-infrared wavelengths offered by *Origins*, these atmospheres can be examined for gases that are the most important signatures of life.

**Table ES-2:** Summary of *Origins* Requirements (Full Scientific Traceability Matrix (STM) provided in Table 1-25)

| *Origins* Science Driver | | Technical or Instrument Parameter | | | |
|---|---|---|---|---|---|
| Scientific Goal | Observable | Parameter | Requirement | Design | Scientific Impact with Anticipated Changes |
| 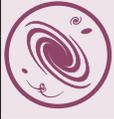 How do galaxies form stars, make metals, and grow their central SMBHs? | Mid- and far- IR rest-frame spectral lines. | Aperture Size | 3.0–5.0 m | 5.9 m | >3.0 m based on the angular resolution needed to study high-z galaxies. >5.0 m driven by z > 6 galaxy studies. |
| | | Aperture Temperature | <6 K | 4.5 K | Sufficiently cold temperature to meet the sensitivity requirements at the longest wavelengths. $T_{tel}$ >6 K impacts the ability to conduct sciences >300 μm. |
| 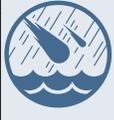 How do the conditions for habitability develop during the process of planet formation? | $H_2{}^{18}O$ $1_{10}-1_{01}$ 547.4–1 μm | $\lambda_{max}$ | >550 μm | 588 μm | $H_2O$ $1_{10}-1_{01}$ is at 538.3-μm; $\lambda_{max}$ <538 μm impacts water sciences. $\lambda_{max}$ <500 μm impacts extragalactic sciences. |
| | $H_2O$ $2_{12}-1_{01}$ 179.5-μm line | $R=\lambda/\Delta\lambda$ | 200,000 | 202,785 | Smaller spectral resolution impacts doppler tomography. |
| | HD 1-0 112-μm line | Spectral line sensitivity | $10^{-20}$ W m$^{-2}$ (1 hr; 5σ) | $5\times10^{-21}$ W m$^{-2}$ (1 hr; 5σ) | Below this sensitivity, *Origins* cannot study sufficient disks at the distance of Orion for gas mass measurements. |
| | | $R=\lambda/\Delta\lambda$ | 40,000 | 43,000 | Lowering R reduces the gas mass accuracy. |
| 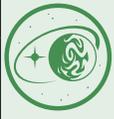 Do planets orbiting M- dwarf stars support life? | $CH_4$ (3.3 & 7.4 μm), $N_2O$ (4.5 & 7.8 μm), $O_3$ (9.7 μm), $CO_2$ (4.3 & 15 μm), $H_2O$ (6.3,17+ μm) | λ min | < 3 μm | 2.8 μm | $CO_2$ at 4.3 μm strongest of all features; $\lambda_{min}$ >5 μm reduces the exoplanet case to surface temperature only. |
| | | Aperture Size | 5.3 m | 5.9 m | An aperture size <5.3 m results in a dramatic drop in the ability to detect $CH_4$ and $N_2O$ – crucial biosignatures - in exoplanet transits over a 5-year mission. |

**What types of temperate, terrestrial M-dwarf planets support life?** *Origins* will be the first observatory with the necessary spectroscopic precision to not only measure habitability indicators ($H_2O$, $CO_2$), but also crucial biosignatures ($O_3$ coupled with $N_2O$ or $CH_4$), which are definitive fingerprints of life on habitable-zone planets. In this observational third tier, *Origins* will obtain additional transit observations for the highest-ranked targets to search for and detect biosignatures with high confidence. The wavelength range afforded by *Origins* will provide access to multiple spectral lines for each molecular species. This will increase the detection significances and prevent potential degeneracies due to overlapping features, averting false positives. This framework robustly detects a variety of potentially habitable planet atmospheres, including the life-bearing Archaean Earth. The entire era of exoplanetary science has shown that Nature's imagination trumps our own, and *Origins'* broad wavelength coverage and precise measurements are guaranteed to give us views into the new and unexpected.

*Origins* **– A Mission for the Astronomical Community:** *Origins* is designed for discovery. While the mission aims to address a specific set of objectives, leading to technical requirements, the outlined science program is intended to be illustrative. *Origins* is a true community observatory, driven by science proposals selected through the usual peer-review process, as used for existing large NASA observatories.

**Unanticipated, Yet Transformative, Discovery Space:** The *Origins*-enabled scientific advances described above are extensions of known phenomena. However, history has shown that order-of-magnitude leaps in sensitivity lead to discoveries of unanticipated phenomena. For example, the sensitivity of IRAS over balloon and airborne infrared telescopes led to the discovery of debris disks, protostars embedded within dark globules, Galactic infrared cirrus, and infrared-bright galaxies, none of which were expected at the time of launch. Likewise, no study anticipated that *Spitzer* would determine the stellar masses of z > 6 galaxies and characterize the TRAPPIST-1 multi-exoplanet system, the coldest known brown dwarfs, measure winds transporting energy in exoplanet atmospheres, and detect dust around white dwarfs produced by shredded asteroids.

*Origins'* sensitivity exceeds that of its predecessor missions by a factor of 1000. Jumps of this magnitude are very rare in astronomy, and have always revolutionized our understanding of the Universe in unforeseen ways. Thus, it is essentially guaranteed that the most transformative discoveries of *Origins* are not even anticipated today.

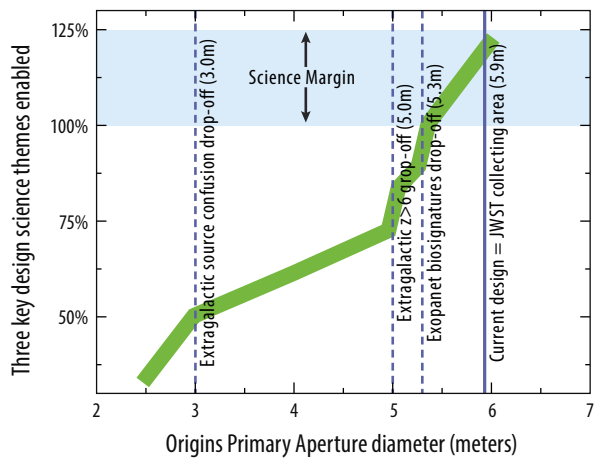

**Figure ES-7:** *Origins'* key science program requires a cold telescope with a primary aperture diameter of 5.3 m. This requirement comes primarily from the exoplanet science case to detect biosignatures in a 5-year mission, given that transit durations are fixed and sensitivity cannot be recovered with a longer single-epoch integration, unlike most other proposed *Origins* observations. The extragalactic study places an aperture size requirement of >5 m, based on the need to detect a statistically significant sample of galaxies at z > 6, to study the formation mechanisms and physical properties of dust and metals during reionization. The minimum primary aperture diameter is 3 m to enable an effective extragalactic and Galactic science program, where source confusion does not compromise the telescope's ability to conduct spectroscopic studies of galaxies at z = 2-3 and the sensitivity is not too poor to study water and gas in proto-planetary disks at the distance of Orion.

## ES.2 Mission Design

***Origins* is >1000 times more sensitive than prior far-IR missions and the design avoids complicated deployments to reduce mission risk.**

The scientific objectives summarized in Table ES-1 are achievable with the low-risk *Origins* design we propose. *Origins* has a *Spitzer*-like architecture (Figure ES-8) and requires only a few simple deployments to transform from launch to operational configuration. With the attributes shown in Table ES-3, the current design carries significant margin between science-driven measurement requirements and estimated performance (Table ES-2) to ensure success of the science mission. The telescope optical system launches in its operational configuration, requiring mirror, barrel, or baffle deployments after launch, but the design allows for mirror segment alignment on orbit to optimize performance. The only deployments are the communication antenna, solar array, telescope cover and sunshields, which are considered low-risk.

This departure from the JWST deployment approach is enabled by the capabilities of new launch vehicles, which are expected to be fully-operational in the mid-2030s. The design is compatible with at least two, and possibly three such launch vehicles. The telescope is a three-mirror anastigmat with an on-axis secondary. It is diffraction-limited at 30 μm and used as a light bucket at shorter wavelengths, where spatial resolution is not a scientific driver, to perform transit spectroscopy for biosignatures in exoplanets. While *Origins* has only 2.8× the collecting area of *Herschel*, cryo-cooling is the dominant factor affecting its extraordinary sensitivity gain (Figure ES-9). To achieve the same sensitivity gain at optical wavelengths, the light-collecting area would have to increase a thousand-fold. Earth's warm atmosphere limits SOFIA's sensitivity, while a relatively warm telescope (~80 K) limited *Herschel*'s.

Three science instruments spanning the wavelength range 2.8 to 588 μm provide the powerful new spectroscopic and imaging capabilities required to achieve the scientific objectives outlined in Section 1 (Table ES-4). The *Origins* Survey Spectrometer (OSS) uses six to take multi-beam spectra simultaneously across the 25 to 588 μm window through long slits enabling deep 3D extra-galactic surveys. When needed, a Fourier transform spectrometer and an etalon provide high and ultra-high spectral resolving power, respectively, especially for studies of water and HD emission lines.

The Far-IR Imager/Polarimeter (FIP) provides imaging and polarimetric measurement capabilities at 50 and 250 μm. Its fast mapping enables rapid follow-up of transient or variable sources and efficient monitoring campaigns. FIP surveys take advantage of *Origins*' agility. Similar to *Herschel*, *Origins* can scan-map the sky at 60'' per second, which is essential, as the FIP 250 μm channel reaches the extragalactic source confusion limit in a few milliseconds. FIP will enable wide area (≥1000 deg²) photometric surveys, leading to large statistical multiwavelength studies of populations of astronomical objects in wide areas, complementing LSST and WFIRST. *Origins* will enable the astronomical community to thoroughly explore the currently unknown, faint, far-infrared Universe.

**Table ES-3:** *Origins* Observatory-level Parameters

| Mission Parameter | Value |
|---|---|
| Telescope: Aperture Diameter/Area | 5.9 m/25 m² |
| Telescope Diffraction Limited at | 30 μm |
| Telescope Temperature | 4.5 K |
| Wavelength Coverage | 2.8–588 μm |
| Maximum Scanning Speed | 60'' per second |
| Mass: Dry/Wet (with margin) | 12000 kg/13000 kg |
| Power (with margin) | 4800 W |
| Launch Year | 2035 |
| Launch Vehicle (large vehicle) | SLS or Space-X BFR |
| Orbit | Sun-Earth L2 |
| Propellant lifetime | 10 years, serviceable |

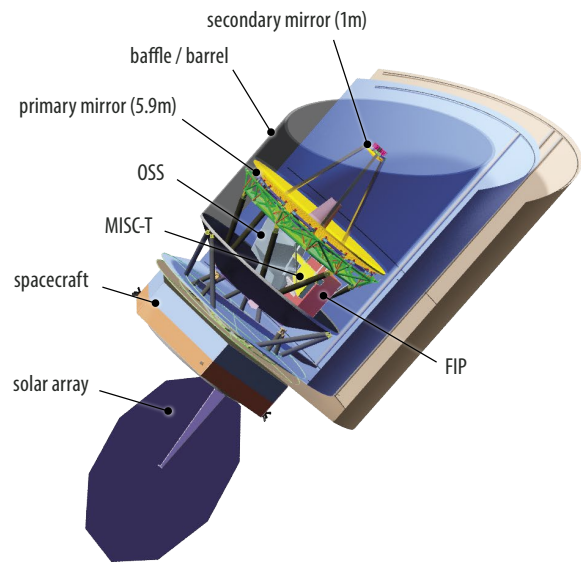

**Figure ES-8:** *Origins* builds on substantial heritage from *Spitzer* to minimize assembly, integration and testing, and deployment risks. A cutaway view shows the *Origins* instrument locations and major elements of the flight system.

The Mid-Infrared Spectrometer and Camera Transit spectrometer (MISC-T) measures R=50 to 300 spectra in the 2.8 to 20 μm range with three bands that operate simultaneously.

MISC-T provides exquisite stability and precision for exoplanet transits (5 ppm between 2.8 to 10 μm). It employs pupil densification to mitigate observatory jitter and relies on a detector stability improvement relative to current state-of-the-art.

To maximize the sensitivity gain, OSS and FIP incorporate next-generation detectors. At least two promising detector technologies already exist. Advanced detectors enable *Origins* to make the first ever fast and wide-area photometric and spatial-spectral surveys in the far-infrared.

**Figure ES-9:** *Origins* taps into a vast, unexplored scientific discovery space, defined by a three-orders-of-magnitude improvement in sensitivity relative to all previously-flown far-infrared observatories. With a temperature of 4.5 K, *Origins*' sensitivity is limited by astronomical background photon noise (lower black curve). SOFIA (220 K), *Herschel* (80 K), and JWST (40 K) are shown in comparison with *Origins* (4.5 K). *Origins*' sensitivity bridges the gap between JWST/MIRI in orbit and ALMA on the ground.

In the mid-infrared, from 2.8 to 20 μm, *Origins* builds on the amazing discoveries anticipated from JWST. JWST is required to deliver extraordinary sensitivity, but transiting exoplanet spectroscopy was not a major design driver. The *Origins* team, however, prioritized exoplanet biosignature detection in the important 2.8 to 10 μm range, and accordingly established 5 ppm as the required system-level stability for MISC-T. The *Origins* Technology Development Plan calls for investment in ultra-stable mid-IR detectors and offers multiple parallel development paths to reduce risk.

**Table ES-4:** Instrument Capabilities Summary

| Instrument/ Observing Mode | Wavelength Coverage (μm) | Field of View (FOV) | Spectral Resolving Power (R=λ/Δλ) | Saturation Limits | Representative Sensitivity 5σ in 1 hr |
|---|---|---|---|---|---|
| *Origins* Survey Spectrometer (OSS) | | | | | |
| Grating | 25–588 μm simultaneously | 6 slits for 6 bands: 2.7″ x 1.4″ to 14′ x 20″ | 300 | 5 Jy @ 128 μm | 3.7 x 10⁻²¹ W m⁻² @ 200 μm |
| High Resolution | 25–588 μm with FTS | Slit: 20″ [2.7″ to 20″] | 43,000 x [112 μm/λ] | 5 Jy @ 128 μm | 7.4 x 10⁻²¹ W m⁻² @ 200 μm |
| Ultra-High Resolution | 100–200 μm | One beam: 6.7″ | 325,000 x [112 μm/λ] | 100 Jy @ 180 μm | 2.8 x 10⁻¹⁹ W m⁻² @ 200 μm |
| Far-IR Imager Polarimeter (FIP) | | | | | |
| Pointed | 50 or 250 μm (selectable) | 50 μm: 3.6′ x 2.5′ 250 μm: 13.5′ x 9′ (109 x 73 pixels) | 3.3 | 50 μm: 1 Jy 250 μm: 5 Jy | 50/250 μm: 0.9/2.5 μJy Confusion limit: 50/250 μm: 120 nJy/1.1 mJy |
| Survey mapping | 50 or 250 μm (selectable) | 60″ per second scan rate, with above FOVs | 3.3 | 50 μm: 1 Jy 250 μm: 5 Jy | Same as above, confusion limit reached in 50/250 μm: 1.9 hours/2 msec |
| Polarimetry | 50 or 250 μm (selectable) | 50 μm: 3.6′ x 2.5′ 250 μm: 13.5′ x 9′ | 3.3 | 50 μm: 2 Jy 250 μm: 10 Jy | 0.1% in linear and circular polarization, ±1° in pol. Angle |
| Mid-Infrared Spectrometer Camera Transit Spectrometer (MISC-T) | | | | | |
| Ultra-Stable Transit Spectroscopy | 2.8–20 μm in 3 simultaneous bands | 2.8–10.5 μm: 2.5″ radius 10.5–20 μm: 1.7″ radius | 2.8-10.5 μm: 50–100 10.5-20 μm: 165–295 | K~3.0 mag 30 Jy @ 3.3μm | Assume K~9.85 mag M-type star, R=50 SNR/sqrt(hr)>12,900 @ 3.3 μm in 60 transits with stability ~5 ppm < 10.5 μm, ~20 ppm > 10.5 μm |

The *Origins* design concept minimizes complexity. *Origins* has a *Spitzer*-like architecture whose thermal performance is well understood. The solar panel, communication antenna, and telescope cover deployment mechanisms have extensive heritage. The two-layer sunshield deployment relies on stored energy in flexible rods to pull the shield material into its desired shape, and telescoping arms to place each shield at the intended distance from the cold shield (see Figure ES-8). This simple deployment sequence can be tested on the ground in existing facilities. The fully-integrated cryogenic payload assembly comprising the telescope, instruments, and cold shield can be tested cryogenically in Chamber A at NASA's Johnson Space Center, following the NASA's favored "test-as-you-fly" approach.

The cryo-thermal system design leverages *Spitzer* experience and technology developed for JWST and Hitomi. Four current-state- of-the-art cryocoolers will cool the telescope to 4.5 K, with 100% margin in heat lift capacity at each temperature stage. The mirrors do not require time-consuming "cryo-null" figuring because the telescope is diffraction-limited at 30 μm. All of the telescope's mirrors and mirror segments can be diamond-turned and rough-polished to the required precision in existing facilities. The JWST primary mirror segment actuator design is adopted to allow the *Origins* primary mirror segments to be adjusted in space in three degrees of freedom (tip, tilt, and piston), enabling final alignment during commissioning.

The next generation of launch vehicles, including NASA's SLS, SpaceX's BFR, and Blue Origin's 7m New Glenn, have much larger payload fairings than the 5-m diameter fairings available today, enabling the launch of a large-diameter telescope that does not need to be folded and deployed. *Origins* operates in a quasi-halo orbit around the Sun-Earth L2 point. The observatory is robotically serviceable, enabling future instrument upgrades and propellant replenishment to extend the mission beyond its 10-year design lifetime.

The *Origins* mission concept study team worked on a wider set of instrument options than presented in the recommended baseline concept. These options are described in Appendix D. They represent possible upscopes which, if adopted, would enhance the mission's scientific capability. These options include the Heterodyne Receiver for *Origins* (HERO), the MISC Camera, expanded FOVs for OSS and FIP, and additional FIP bands (100 and 500 μm). The HERO upscope would provide nine-beam spectral measurements of selectable lines in the 110 to 620 μm bands up to very high spectral resolving power (~$10^7$) and could, for example, vastly extend Event Horizon Telescope observations of supermassive black holes. The MISC Camera would enable mid-IR imaging and spectroscopy (R~300) over the 5 to 28 μm wavelength range. The concept study report also discusses potential descopes (Section 4). In addition to instrument modes, descopes include decreasing the aperture diameter, which impacts the observatory science capabilities as shown in Figure ES-7.

## ES.3 Enabling Technologies

**Detectors, ancillary detection system components, and cryocoolers are the only *Origins* enabling technologies currently below Technology Readiness Level (TRL) 5. The *Origins* Technology Development Plan outlines a path leading to TRL 5 by Phase A start in 2025 and TRL 6 by mission PDR.**

At far-infrared wavelengths, reaching the fundamental sensitivity limits set by the astronomical background requires a cold telescope equipped with sensitive detectors. The noise equivalent power (NEP) required for FIP imaging is $3\times10^{-19}$ W Hz$^{-\frac{1}{2}}$, whereas the NEP needed for OSS for R=300 spectroscopy is $3\times10^{-20}$ W Hz$^{-\frac{1}{2}}$. Transition-edge sensor (TES) bolometers and kinetic inductance detectors (KIDs) both show great promise, and we recommend maturation of both technologies to TRL 5 and down-selection to a single technology at the beginning of Phase A (Figure ES-10). While the noise requirements for MISC-T's

mid- IR detectors are not particularly challenging, 5 ppm stability over several hours must be demonstrated to meet the *Origins* requirement. Again, the *Origins* Technology Development Plan mitigates risk by recommending the parallel maturation of three detector technologies to provide this stability: HgCdTe and Si:As arrays and TES bolometers (Figure ES-10). An ultra-stable calibration source will be needed for testing.

Mechanical cryocoolers that can reach temperatures of 4.5 K have already flown on Hitomi (2016). These coolers, developed by Sumitomo Heavy Industries, had a required lifetime of 5 years compared to *Origins*' 10 years, but meet its performance requirements. Replacing the compressors' suspension system with a flex spring, a relatively straightforward change, is expected to extend the lifetime. Several US companies have also produced TRL 5 cryocoolers or cryocooler components with a projected 10-year lifetime. The TRL 7 JWST/MIRI cryocooler, for example, has a 6 K operating temperature. Sub-Kelvin coolers operating at 50 mK, as needed for the OSS and FIP detectors, were also flown on Hitomi. A Continuous Adiabatic Demagnetization Refrigerator (CADR) with a much higher cooling power (6 μW vs. 0.4 μW for Hitomi) suitable for *Origins* is currently being developed to TRL 6 under a Strategic Astrophysics Technology (SAT) grant. This new SAT CADR also demonstrates self-shielding of magnetic fields to 1 μT, making it compatible with superconducting detectors that demand an ambient field <30 μT. A straightforward extension of this ADR technology allows operation at even lower temperatures (35 mK) with similar cooling power. Lowering the operating temperature is a simple way to improve TES detector sensitivity, should that become necessary during mission formulation.

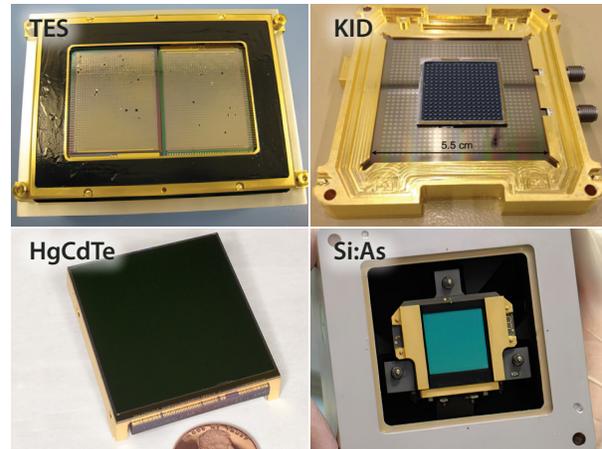

**Figure ES-10:** The robust Origins Technology Development Plan recommends parallel investment in four detector technologies. TES bolometers (HAWC+ array), KIDS (432-pixel device), HgCdTe arrays (JWST/NIRCam), and Si:As (JWST/MIRI).

## ES.4 Schedule and Cost

**The *Origins* team developed a mission design concept, technical approach, technology maturation plan, risk management approach, budget, and a master schedule compatible with NASA guidelines for the Decadal Study and grounded in NASA and industry experience from previous successful large Class A missions.**

*Origins* is a NASA-led mission, managed by a NASA Center, and includes domestic and international partners. The Japanese space agency, JAXA, and a CNES-led European consortium are active participants in the mission concept study, with each contributing an instrument designs. Domestic participants include NASA centers (GSFC, Ames, MSFC), JPL, and industry (Ball Aerospace, Northrop Grumman, Lockheed Martin, Harris Engineering).

The Technical Fact Sheet shows the *Origins* Phase A through E schedule at the highest level. Scheduled milestones and key decision points are consistent with NASA Procedural Requirements (NPR-7120.5) and formulation and development for Class A missions. The schedule supports an April 2035 launch, providing ~10 years from Phase A start to launch, and includes 12.7 months of funded reserve. Much of the design and development work progresses through parallel efforts, and the critical path runs through the most complex instrument (OSS).

*Origins* integration and testing, Phase D, is considerably shorter than that of JWST because *Origins* has very few deployable elements, whereas JWST has many, and we applied lessons learned from the JWST experience. Although the *Origins* primary mirror is segmented, it launches in its operational configuration, and the mirror segments are easier and faster to manufacture than those of JWST. Mirror cryo-null figuring is not required for *Origins*. *Origins* has an isothermal Cryogenic Payload Module (CPM) comprising the cold shield, telescope and instruments, which reduces cool-down and warm-up time. The time needed for testing has also been reduced by selecting non-absorptive, thermally-conducting flight system materials; therefore, staged cooling to avoid condensation on critical components is not required.

The *Origins* lifecycle mission cost is estimated to be in the range $6.7B to $7.3B, including margin, at the 50% and 70% Confidence Levels, respectively. This estimated cost covers mission phases A through E, assuming no foreign contributions, and is given in Base Year 2020 dollars. This cost will evolve until the mission Preliminary Design Review. NASA Goddard Space Flight Center's Cost Estimating and Modeling Analysis (CEMA) office developed this cost estimate using the commercially available PRICE-H parametric cost modeling tool. The cost estimate is based on a detailed master equipment list (MEL) and a detailed Integrated Master Schedule (IMS), both of which are included in the *Origins* Cost Report. The MEL assigns an appropriate Technology Readiness Level (TRL) to each component. The CEMA cost model assumes that all components have matured to at least TRL 5 by the start of Phase A in 2025, and to at least TRL 6 by mission PDR. A separate *Origins* Space Telescope Technology Development Plan describes the maturation of all mission-enabling technologies on this timeline and reports the cost of technology maturation. The mission cost estimate given above includes mission definition and development, the flight segment, the ground segment, and mission and science operations for 5 years. The launch cost ($500M for the SLS launch vehicle, as advised by NASA Headquarters) is also included. NASA GSFC's Resource Analysis Office (RAO) independently estimated the mission cost using different methodology. RAO and CEMA are firewalled from each other, but they both referred to the same MEL and mission schedule. The RAO and CEMA cost estimates agree to within the estimated uncertainty. The *Origins* mission design has not been optimized, and optimization may lead to cost savings. Optimization is planned as a Phase A activity. Japan and several ESA member nations have significant relevant expertise and have demonstrated interest in the *Origins* mission through participation in the study. Foreign contributions are expected to reduce NASA's share of the mission cost.

# 1 - SCIENCE INVESTIGATION

*Origins* answers NASA's top astrophysics questions by piercing the dusty shroud to witness galaxy assembly near cosmic dawn, illuminating the flow of water from cosmic springs to planetary surfaces, and checking the atmospheres of nearby planets for signs of life. This report demonstrates that *Origins* makes breakthrough measurements in each area (Table ES-1) over the course of a two-year notional observing program, although as a public observatory all observing time is awarded through competitive proposals. The science requirements for these breakthroughs have been validated through data-based, end-to-end simulations and detailed feasibility studies that demonstrate the low-risk nature of the studies considered.

*Origins'* broad science goals are met with key objectives that are described in Sections 1.1 (extragalactic), 1.2 (planetary systems) and 1.3 (exoplanets). The full science traceability matrix is shown in Section 1.4. These driving science cases have led to a concept for an observatory with a broad "discovery space." Appendix A highlights *Origins'* "discovery space" with example additional science programs from the astronomical community. History indicates that *Origins'* thousand-fold increase in sensitivity over its predecessors will result in the discovery of new and completely unanticipated phenomena that will enhance and define *Origins'* scientific legacy.

## 1.1 How do galaxies form stars, build up metals, and grow their supermassive black holes from Reionization to today?

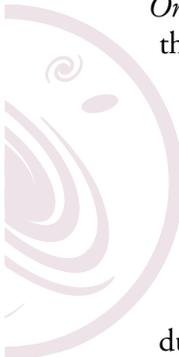

*Origins* creates a 3D map of galaxy formation and evolution by simultaneously measuring the redshift, star formation rate, black hole accretion rate, metallicity, and dust properties in millions of galaxies from the Epoch of Reionization to the present day. *Origins* achieves this science through wide and deep surveys using mid- and far-infrared fine-structure lines and polycyclic aromatic hydrocarbon (PAH) features that trace gas and dust heated by young stars and accreting black holes, and that fully characterize the ISM in the deep survey.

The Milky Way began as a primordial cloud of hydrogen and helium. It has been transformed into stars, compact objects, a central supermassive black hole (SMBH), gas, dust, planets, and life through the action of gravity, star formation, SMBH growth, and the production and dispersal of "metals" (elements heavier than helium). The story of our cosmic origins is intimately tied to these transformative agents, and to understand how they shape galaxies requires a proper accounting of the growth of stellar and SMBH mass over cosmic time. We also need an inventory of the multi-phase interstellar medium (ISM), which supplies the gas and dust to form stars and planets, carries metals far from the stars and supernovae that forged them, and captures radiative and mechanical energy from star formation and SMBH growth. To do this, we must look outside of the Milky Way—and back in time—to survey galaxies at all stages of evolution.

Studying the complex interplay between gas inflow, collapse, feedback from massive stars, and impact of supermassive black holes in galaxies' ecosystems is rooted in the physics of the ISM. The relative fraction of different gas phases (*i.e.,* the molecular, neutral atomic, ionized, and coronal phases) betrays the relative contribution of various channels of energy injection into galaxies. *Origins* is designed to map all of these phases, providing a nearly complete picture of star form ation and black hole growth in galaxies.

The mid- and far-infrared spectral region is rich with spectral features that can be used to diagnose the composition and physical conditions of the gas that forms stars and that accretes onto SMBH, and the dust that reprocesses starlight and catalyzes interstellar chemistry (Table 1-1). Just as importantly, these features can be seen through the thick layers of dust that frequently enshroud star-forming regions and galactic nuclei. *Spitzer* and *Herschel* demonstrated the potential of these features to study gas and dust over almost the entire history of the Universe. With its high sensitivity, fast mapping speed, and wide bandpass, *Origins* realizes that potential through surveys to (i) determine the relative growth rate of stellar and SMBH mass in galaxies from the Epoch of Reionization (EOR, 6 < z < 20) to the present day (Section 1.1.2), (ii) characterize the build-up and dispersal of metals, dust, and complex molecules in the interstellar medium (ISM, Section 1.1.3), and (iii) measure the impact of star formation and SMBH growth on galaxies through supernova- and AGN-driven outflows (Section 1.1.4).

Table 1-2 summarizes the science objectives and Table 1-3 summarizes the measurement requirements to achieve these objectives. Additional science cases targeting nearby and distant galaxies are detailed in Appendix A under Discovery Science. The detailed flow-down of requirements forms the *Origins* science traceability matrix (STM), which is presented in Section 1.4. The primary objectives can be achieved with deep and wide, unbiased 3D spectroscopic surveys covering 0.5 deg$^2$ and 20 deg$^2$, respectively (Section 1.1.4, Table 1-4).

**Table 1-1:** Key Infrared Diagnostic Features

| Species | Wavelength (μm) | Φ[eV] | Diagnostic Utility |
|---|---|---|---|
| **Ionized Atomic Gas** | | | |
| Ne V | 14.3, 24.3 | 97.1 | AGN strength/accretion rate |
| O IV | 25.9 | 54.9 | AGN strength/accretion rate (hot stars) |
| S IV | 10.5 | 34.8 | SB strength/SFR/HII region density, ionization |
| Ne II | 12.3 | 21.6 | '' |
| Ne III | 15.6, 36.0 | 41.0 | '' |
| S III | 18.7, 33.5 | 23.3 | '' |
| Ar III | 21.83 | 27.6 | '' |
| O III | 51.8, 88.4 | 35.1 | '' |
| N III | 57.3 | 29.6 | '' |
| N II | 122, 205 | 14.5 | '' |
| **Neutral Atomic Gas** | | | |
| Si II | 34.8 | 8.2 | Density and temperature probes of photo- dissociated neutral gas at the interface between HII regions and molecular clouds |
| O I | 63.1, 145 | | |
| C II | 158 | 11.3 | |
| C I | 370 | | |
| **Molecular Gas** | | | |
| H$_2$ | 9.66, 12.3, 17.0, 28.2 | | Warm (100–500 K) molecular gas/feedback D/H ratio/ gas mass Column density of cold, dense gas, abundance/ feedback |
| HD | 37, 56, 112 | | |
| OH | 34.6, 53.3, 79.1, 119 | | |
| OH | 98.7, 163 | | |
| H$_2$O | 73.5, 90, 101, 107, 180 | | '' |
| CO | ~2600/J | | High-J, warm/dense molecular gas/feedback |
| **Dust** | | | |
| Silicate | 9.7, 18 | | Optical depth. Hot dust emission in QSOs. PDR tracer. Star formation rate. Grain properties. |
| PAH | 6.7, 7.7, 8.5, 11.3, 17 | | |

---

### Important Definitions
**We adopt the following definitions throughout this section**

- **Starburst Galaxy:** A galaxy producing stars at a rate significantly higher than the average. Usually this enhanced production cannot be sustained by the available gas supply. These galaxies have spectral signatures that indicate heating by young, massive stars.
- **Active Galaxy:** A galaxy with an Active Galactic Nucleus (AGN) harbors a central accreting supermassive black hole. The spectrum of an AGN shows enhanced high-energy emission, excess hot dust, and/or lines of highly ionized atomic species.
- **Galaxy Main Sequence:** The locus of normal galaxies in the star formation rate--stellar mass plane. Most galaxies spend the majority of their lives on the Main Sequence. Starburst, active, or rapidly evolving galaxies can be displaced from the Main Sequence.
- **Metallicity:** The amounts of elements heavier than helium in the interstellar medium of a galaxy, as traced by infrared atomic fine structure lines and the emission of small condensed dust grains. Absolute metallicities are the fractional amount of an element compared to hydrogen. Relative abundances are fractional abundances of two different elements heavier than helium.

**Table 1-2:** Extragalactic Objectives

| NASA Science Goal | How does the Universe work? | |
|---|---|---|
| *Origins* Science Goal | How do galaxies form stars, make metals, and grow their central supermassive black holes from Reionization to today? | |
| *Origins* Scientific Capability | *Origins* will spectroscopically 3D map wide extragalactic fields to simultaneously measure properties of growing supermassive black holes and their galaxy hosts across cosmic time. | |
| **Scientific Objectives Leading to Mission and Instrumental Requirements** | **Objective Goal** | **Technical Statement** |
| | **Objective 1:** How does the relative growth of stars and supermassive black-holes in galaxies evolve with time? | Measure the redshifts, star formation rates, and black hole accretion rates in main-sequence galaxies since the epoch of reionization, down to a SFR of 1 $M_\odot$/yr at cosmic noon and 10 $M_\odot$/yr at z~5, performing the first unbiased survey of the co- evolution of stars and supermassive black holes over cosmic time. |
| | **Objective 2:** How do galaxies make metals, dust, and organic molecules? | Measure the metal content of galaxies with a sensitivity down to 10% Solar in a galaxy with a stellar mass similar to the Milky- Way at z of 6 as a function of cosmic time, tracing the rise of heavy elements, dust, and organic molecules across redshift, morphology, and environment. |
| | **Objective 3:** How do the relative energetics from supernovae and quasars influence the interstellar medium of galaxies? | Determine how energetic feedback from AGN and supernovae regulate galaxy growth, quench star formation, and drive galactic ecosystems by measuring galactic outflows as a function of SFR, AGN luminosity, and redshift over the past 10 Gyr. |

**Table 1-3:** Instrument and Mission Design Drivers

| Technical Parameter | Requirement | Expected Performance | Key Scientific Capability |
|---|---|---|---|
| Maximum Wavelength | > 550 μm | 588 μm | OH 119 absorption line from molecular outflows to z > 4 & [OIII] 88 μm fine structure emission line out to z of 5 for a comparison of far-IR and mid-IR based metallicity indicators |
| Aperture Temperature | < 6 K | 4.5 K | Sensitivity to detect $10^{11} L_\odot$ galaxies at z=6 |
| Line Flux Sensitivity | <$10^{-20}$ W m$^{-2}$ 5σ, 1 hr | 3x$10^{-21}$ W m$^{-2}$ 5σ, 1 hr | Detection of rest frame MIR fine structure lines at z > 6 in ULIRG |
| Spectral Resolution | R > 250 | R=300 in low-res mode for surveys | Detection of narrow fine structure emission lines |

## 1.1.1 The Infrared Toolbox and Origins in the 2030s Context

> *Origins* is designed to detect spectral features that fully characterize star formation, black hole accretion, and the state of the ISM in an unbiased sample of galaxies across cosmic time—something no other telescope can do.

Thermal infrared spectroscopy enables direct measurements of the basic physical properties – density, temperature, pressure, and energetics – of the ionized (T ~ $10^4$ K) and neutral atomic gas, the warm (T ~100-500 K) molecular gas, and the dust in galaxies and the hot ionized gas in the accretion disks of SMBHs.

The key probes of the neutral and ionized gas are fine-structure emission lines from abundant elements, including Oxygen (O), Carbon (C), Neon (Ne), Sulfur (S), Nitrogen (N), Iron (Fe), Argon (Ar), and Silicon (Si). These lines lie between 10 and 400 μm and are easy to identify in galaxy spectra (Figure 1-1). The line centroids measure redshifts, their intensities measure the amount of gas in different phases of the ISM, their ratios indicate the ionization state of the gas (the lines span an order of magnitude in ionization potential), and the line profiles can indicate dusty outflows (Spoon *et al.*, 2009) and measure SMBH masses (Dasyra *et al.*, 2008). The relative contributions of young stars and black holes to the interstellar radiation field can be inferred from the ionization state (Genzel *et al.*, 1998; Lutz *et al.*, 2003; Brandl *et al.*, 2006; Armus *et al.*, 2004, 2007; Farrah *et al.*, 2007).

HII regions are typically studied using optical nebular emission lines, but these lines frequently suffer from heavy extinction by intervening dusty clouds. However, mid- and far-infrared lines, such as those from S, N, and Ar (10-210 μm; Table 1-1), are easy to detect and reveal the physical properties of the ionized gas in even the most heavily obscured HII regions.

The warm molecular gas is directly measured in the mid-infrared via $H_2$ emission lines in the 10-30 μm range while the warm, neutral phases are traced by neutral and singly ionized carbon lines ([*CI*] 370 μm and [*CII*] 158 μm), and neutral oxygen ([*OI*] 63 and 145 μm). These lines provide an essential complement to tracer molecules such as carbon monoxide (CO), which rely on an uncertain calibration and are only visible along sightlines with high column density (typically, $A_v > 1$). Indeed, *Herschel* observations by Pineda *et al.* (2017) have demonstrated that up to half the molecular gas in the Milky Way may be in a "CO-dark" phase and invisible to traditional mm-wave surveys. The metal-poor molecular gas that is found in nearby dwarf galaxies and expected in the early Universe is dominated by a "CO-dark" phase of molecular gas (*e.g.* Cormier *et al.* 2019).

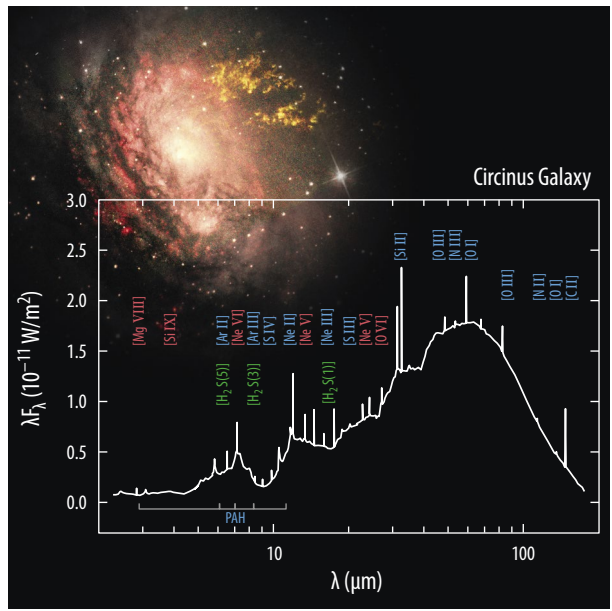

**Figure 1-1:** (Above) *Origins* measures the redshifts, star formation rates, black hole accretion rates, and metal and dust content in galaxies. The infrared spectrum of the nearby active galaxy, Circinus (see inset), is shown using Infrared Space Observatory data (Moorwood, 1999). Emission lines from highly ionized gas heated by the central active nucleus are marked in red, those coming from gas heated by young stars are marked in blue, and those from warm molecular gas are marked in green. PAH molecules are excited by UV photons, and emit broad features in the mid-infrared, through bending and stretching modes.

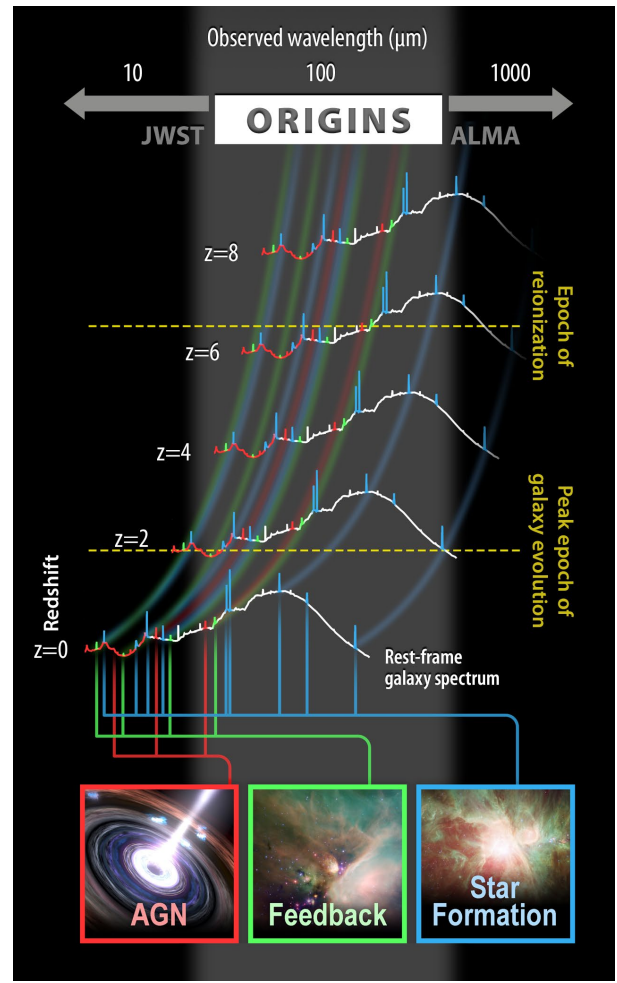

**Figure 1-2:** (Right column) The spectral reach of *Origins* over cosmic time. Schematic representation of how the key spectral diagnostic features of AGN (red), star formation (blue), and energetic feedback (green) move through the wide bandpass of the *Origins* Survey Spectrometer (OSS) with look-back time. *Origins* can measure all of these important processes over the entire history of galaxy evolution, filling in a key gap in wavelength and discovery space between JWST and ALMA.

Meanwhile, small dust grains are measured via broad emission bands from Polycyclic Aromatic Hydrocarbons (PAHs), that are stochastically heated by individual photons and subsequently emit a distinct pattern of broad features in the infrared (3-20 μm). Larger grains are detected via emission and absorption bands from silicate dust and far-infrared continuum emission from grains in thermal equilibrium with their environments (Table 1-1). The PAH features indicate redshift, UV flux (and hence star formation rate), and the presence of an AGN, which destroys them. The temperature and luminosity of the larger grains is related to the star formation rate.

The important lines and dust features lie in a large wavelength gap between JWST and ALMA, which *Origins* fills (Figure 1-2).

At low redshift, JWST/MIRI accesses many of the rest-frame, mid-infrared features prominent in the spectra of AGN and starburst galaxies, but is unable to detect these lines at the moderate redshifts where galaxy evolution is most rapid. For example, JWST can only detect the bright 12.8 μm [NeII] line in galaxies at z < 1.1. Likewise, the 14.3 μm [NeV] line is visible at z < 0.9, and the 25 μm [OIV] line is visible only at z < 0.1. Although JWST can detect the 3.3 μm PAH feature to very high redshifts, it cannot measure the 6-17 μm bands. Thus, the ability to use multiple features to measure the properties of the small dust grains is limited to z < 2-3. JWST is undeniably a powerful observatory that is capable of creating detailed infrared spectral maps of galaxies and, through deep, multi-band imaging surveys, make initial observations of proto-galaxies in the early phases of the Epoch of Reionization. However, JWST is unable to measure the basic properties of the ISM at higher redshift.

At the highest redshifts, ALMA can detect the rest-frame, far-infrared features. For example, ALMA has detected the [OIII] 88 μm line in a z = 9.1 lensed galaxy (Hashimoto *et al.* 2018). The [OIII] line detection has been essential to establishing the redshift, star formation rate, and star formation history of this object, which showed that the galaxy has an intrinsically red color and that most of the stars formed at z ~ 15. The [OIII] line has been a high-redshift workhorse for ALMA, since it directly probes warm, ionized gas in low-metallicity galaxies, and will continue to be used in the next decade of ALMA observations. However, no current observatory is capable of detecting this line below z ~ 7. In short, JWST and ALMA can detect high-z proto-galaxies, but cannot track their evolution due to the far-infrared wavelength gap and the difficulty of conducting wide surveys with either telescope.

No current or planned mid- or far-infrared mission is up to the task. SOFIA covers some of the wavelength gap and can study a handful of bright, nearby galaxies, but these are arguably unlike the sources being detected with ALMA at z > 7. Meanwhile, *Herschel* observed galaxies at z < 1 and built up spectroscopic samples of starburst galaxies and AGN at z < 0.5, but could not access the peak epoch of galaxy evolution because of limited sensitivity (Figure 1-2) . This issue could be partially resolved by SPICA, a proposed joint ESA/JAXA mission with a cold 2.5-m telescope covering the 30-300 micron range. If selected, SPICA can study the ISM in populations of galaxies below z~3, but can only detect a small number of the most luminous sources at higher redshift. In contrast to SOFIA, *Herschel*, and SPICA, *Origins* has a much larger field of view, broader wavelength coverage, and higher sensitivity that enables efficient, unbiased, 3D spectroscopic mapping of galaxies up to z~6 (Table 1-4). These surveys detect many high-redshift, luminous objects and reveal the evolution of normal, rather than only exceptional, galaxies.

A key goal of ESA's *Athena* X-ray observatory, scheduled for launch in the early 2030s, is to study the Universe's earliest super-massive black holes (Aird *et al.*, 2013). X-rays are a direct tracer of active black hole growth; however, because they probe the hot gas near the accretion disk, the soft X-rays can suffer greatly from absorption by gas and dust in the host galaxy. For the most distant galaxies, *Athena* can only observe the extremely hard X-rays that are less susceptible to absorption. Therefore, *Athena* data alone cannot be used to directly measure star formation rates in the host galaxies of the AGN. By comparison, *Origins* can simultaneously measure the star formation and black hole accretion rates in every detected galaxy through the rich infrared diagnostic features, which are nearly immune to the effects of extinction. The powerful combination of *Athena* and *Origins* could provide the clearest picture yet of black hole and stellar growth within galaxies at all epochs.

**Table 1-4:** Extragalactic Surveys for Scientific Objectives

| Proposed Survey | Area and Depth | Applications |
|---|---|---|
| Wide Survey | 20 deg² over 1000 hours | **Maximizes galaxy detections to z=6 (Section 1.1.6; Figure 1-14).** <br> (i.) Metallicity measurements (Objective 2) <br> (ii.) [NeII] SFR and [OIV] measurements (Objective 1) <br> (iii.) AGN identification for feedback studies (Objective 3) |
| Deep Survey | 0.5 deg² over 1000 hours | **Maximizes detections of faint fine-structure lines to z=4 (Section 1.1.5, Figure 1-5)** <br> (i.) Diagnose the multiphase ISM conditions (Objectives 1, 3) <br> (ii.) Absolute metallicities with hydrogen lines (Objective 2) <br> (iii.) OIV for BHAR measurements (Objective 1) |

### 1.1.2 Science Objective 1: The Co-evolution of Stars and Black Holes

> With deep and wide spectroscopic surveys, *Origins* reveals how galaxies and black holes grow together over cosmic time.

#### 1.1.2.1 Infrared Surveys and the Landscape of Galaxy Evolution

The build-up of stellar mass is the most basic metric of galaxy evolution, and has traditionally been studied with deep optical and UV surveys. The deepest infrared surveys to date have revealed that (1) the bulk of star formation in galaxies occurred between redshifts of $1 < z < 3$, with a relatively sharp drop at $z < 1$, and that (2) most of the UV light ever produced by young stars was re-processed by dust into the far infrared (Madau & Dickinson, 2014 and references therein; Figure 1-3). Moreover, recent results show that most star formation is obscured in all galaxies more massive than log M (M$_\odot$) >9.4 at $z < 3$, and that the fraction of obscured star formation, at a given stellar mass, remains nearly constant (50% for low-mass galaxies and nearly 90% for galaxies with log M (M$_\odot$) ~10.4) from $0 < z < 2.5$ (Whitaker *et al.*, 2017). Thus, while UV and optical estimates of the star formation reach well into the EOR they must correct for the number of ionizing photons absorbed by dust. These corrections are uncertain and usually large, but this issue can be resolved by far-IR measurements.

The star formation rate density (SFRD; the number of stars produced per unit time, per co-moving volume) at $z > 3$ remains mostly unconstrained in the IR (Figure 1-3). It may decline quickly toward high redshift, or fall off more gradually relative to the unobscured SF observed in the rest-frame UV (*e.g.*, Madau & Dickinson, 2014; Casey *et al.*, 2018). Constraining the SFRD at high redshift is important to an accurate picture of star formation there. Only a small number of dusty galaxies have been detected near the EOR, but they have been crucial laboratories for understanding star formation and the ISM conditions in the early Universe (*e.g.*, Watson *et al.* 2015, Strandet *et al.*, 2017). With an accurate SFRD, we can address two major issues: the co-evolution of SMBHs and host galaxies and the role of environment on evolution.

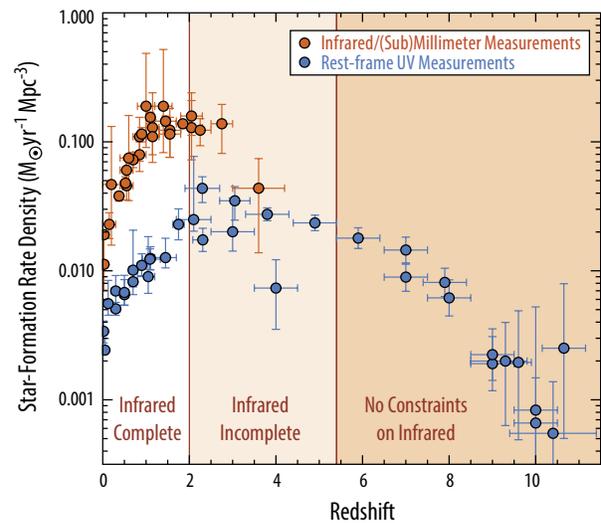

**Figure 1-3:** Star Formation over Cosmic Time. *Origins* provides the first accurate measure of the star formation rate in galaxies at $z > 3$ in the infrared, with sufficient large samples to determine the effects of environment, AGN power, and other physical variables. This figure, adapted from Casey et al. (2018), shows the current state of the art – the lack of infrared data for $z > 2$, and the vast discovery space that can be filled in with *Origins*.

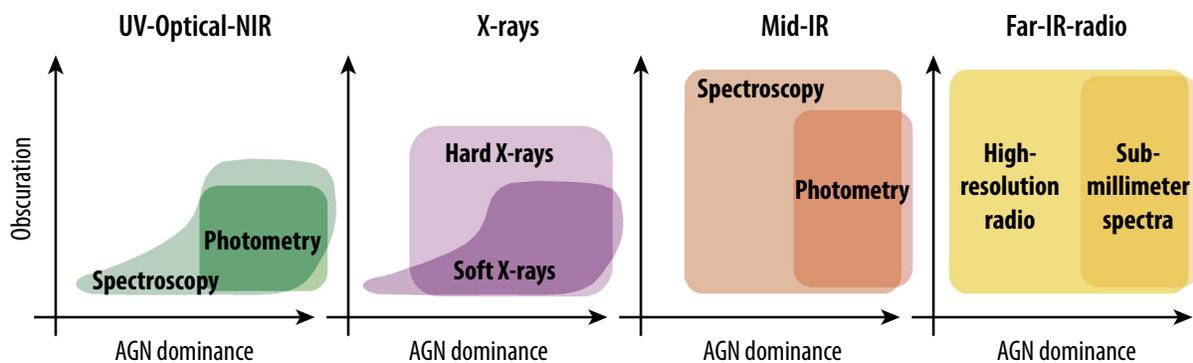

**Figure 1-4:** MIR spectroscopy provides a completely unbiased selection of AGN. These plots show the relative ability of observations in different wavelength regimes to identify obscured AGN. The rich set of IR diagnostic spectral features accessible by *Origins* identifies AGN and measure their power for large, statistical samples as a function of redshift. AGN can also be identified with very high spatial resolution radio observations, but only over extremely limited areas. Figure adapted from Hickox & Alexander (2018).

First, we can learn how SMBHs co-evolve with their host galaxies by comparing the SFRD to the growth rate of SMBHs, as described by the black hole accretion rate density (BHARD, traced by the AGN luminosity). The BHARD shows a similar shape to the SFRD up to z ~ 3 (Madau and Dickinson, 2014), suggesting that galaxy and SMBH growth are linked over much of cosmic time. This correlation is not surprising, since there is a locally established SMBH-stellar mass correlation (Magorrian *et al.*, 1998; Marconi and Hunt, 2003). However, the details of how this relation arises remain unclear: given the vastly different physical scales between star formation and SMBH growth (AGN activity), it is very difficult to self-consistently model both processes. One might expect that any differences between the shapes of the BHARD and SFRD would occur at z>3, when galaxies were rapidly changing, and indeed recent results from X-ray surveys suggest that the BHARD declines more steeply than the SFRD at the highest redshifts (Vito *et al.*, 2018). More data are needed to understand how and why this occurs.

A major obstacle is that existing measurements of the SFRD and BHARD are based on measuring different galaxy populations in different wavelength regimes over different redshift intervals. The average SFRD is often calculated from galaxies selected in UV or IR imaging surveys that rely on photometric redshifts, while the average BHARD is typically estimated from galaxies selected from deep X-ray or optical surveys, then extrapolated to large areas to account for the large fraction (~25-50%) of "missing," Compton-thick objects, the majority of which are not directly identified as such in the X-ray surveys (Hickox & Alexander, 2018). Even in our local neighborhood, many AGN are missing from traditional optically-selected samples (Goulding & Alexander, 2009; Figure 1-5). Piecing this information together to provide an average, global picture of the evolution of galaxies is not only fraught with uncertainties, but also masks the true diversity in the population as a function of lookback time.

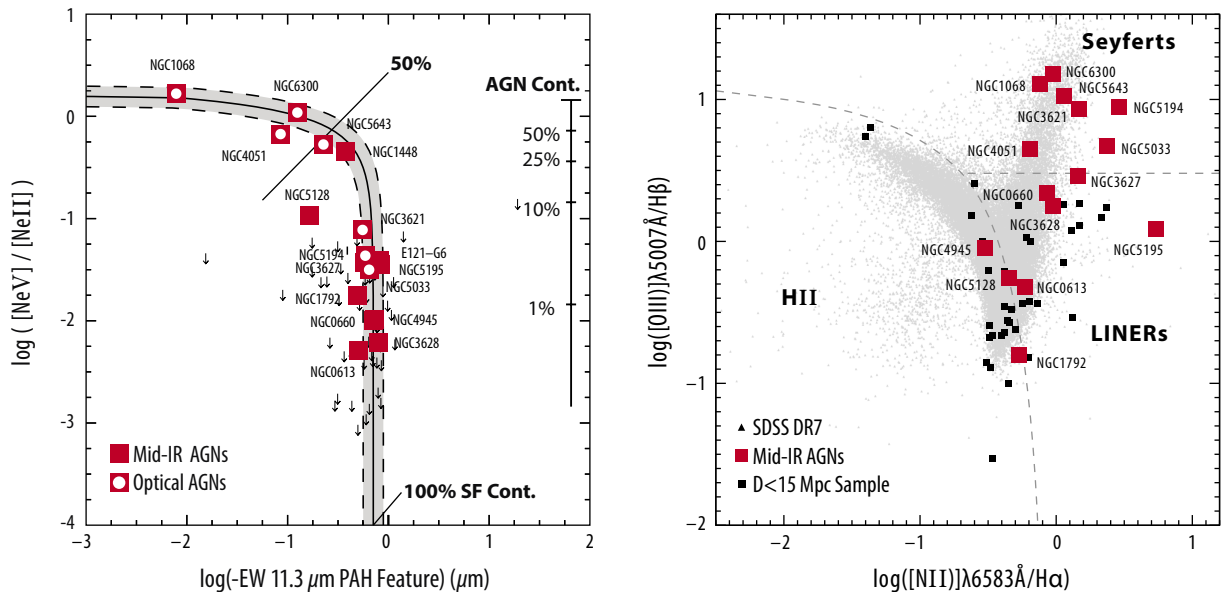

**Figure 1-5:** Diagnostic power of IR fine-structure lines for identifying and quantifying AGN. Traditional optical diagnostic line ratios can fail to detect highly obscured AGN. Here, buried AGN (red squares) are identified in a sample of nearby (D < 15 Mpc) galaxies via their mid-IR diagnostic lines with *Spitzer* IR spectroscopy (left) when optical diagnostics (right) suggest star forming or composite sources. The [NeV]/[NeII] and [OIV]/[NeII] (not shown) MIR line ratios are excellent measures of the AGN contribution, and the high ionization line luminosities can be used to estimate the black hole accretion rate. Figures adapted from Goulding & Alexander (2009).

**Table 1-5:** Extragalactic Science Requirements Flow (Part 1)

| Science Objective 1 |
| --- |
| Measure the redshifts, star formation rates, and black hole accretion rates in main-sequence galaxies since the Epoch of Reionization, down to SFR 1 $M_{\odot}$/yr at z ~ 3 and 10 $M_{\odot}$/yr at z ~ 5, – the first unbiased survey of the co-evolution of stars and SMBHs over cosmic time. |

| Science Observable |
| --- |
| An atlas of galaxy spectra in the infrared with PAH, atomic fine structure, and hydrogen recombination lines; at least 100 galaxies per redshift bin, which results in a million spectra over all redshifts. |

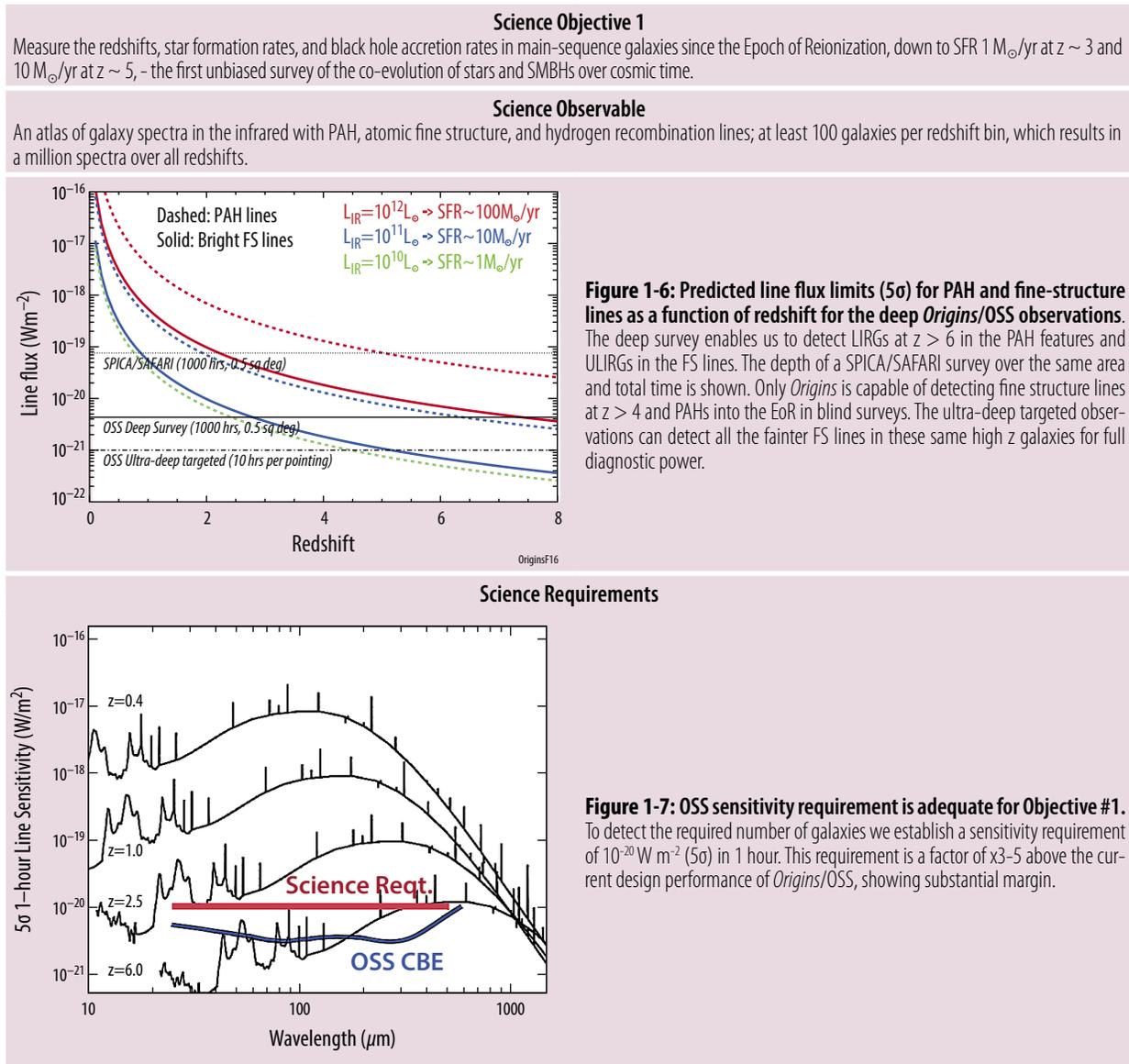

**Figure 1-6: Predicted line flux limits (5σ) for PAH and fine-structure lines as a function of redshift for the deep *Origins*/OSS observations.** The deep survey enables us to detect LIRGs at z > 6 in the PAH features and ULIRGs in the FS lines. The depth of a SPICA/SAFARI survey over the same area and total time is shown. Only *Origins* is capable of detecting fine structure lines at z > 4 and PAHs into the EoR in blind surveys. The ultra-deep targeted observations can detect all the fainter FS lines in these same high z galaxies for full diagnostic power.

**Figure 1-7: OSS sensitivity requirement is adequate for Objective #1.** To detect the required number of galaxies we establish a sensitivity requirement of $10^{-20}$ W m$^{-2}$ (5σ) in 1 hour. This requirement is a factor of x3-5 above the current design performance of *Origins*/OSS, showing substantial margin.

The second issue that an accurate SFRD addresses is the role of environment in galaxy evolution. Environmental density on scales up to at least 10 Mpc is evidently important to galaxy assembly. This environmental effect is demonstrated by, for example, the accelerated growth of galaxies in dense environments and the strong differential clustering of galaxies as a function of luminosity. However, the magnitude and form of environmental influence remains vigorously debated, especially at z > 1. In the local Universe, star-formation rates drop as the environment becomes denser, but the redshift at which this trend reverses is poorly constrained. There is also uncertainty about how the precise environment, *e.g.* the sheets and filaments connecting clusters, influence galaxy evolution, as opposed to simply the average number density of galaxies.

By the 2030s, large area optical surveys (*e.g.*, DES and LSST) will map the buildup of stellar mass as a function of environment. Meanwhile, WFIRST and *Euclid* will determine morphologies over wide fields for galaxies up to z ~ 6, thus establishing how mergers and secular processes depend on redshift, environment, and mass. However, uncertainties in stellar masses and ages of high-redshift galaxies, as derived solely from the optical and near-infrared bands, will propagate to the derived star-formation histories. Likewise, planned *Athena* X-ray surveys (scheduled for launch in the early 2030s) will measure SMBH growth at high redshift through deep and wide surveys (Aird *et al.*, 2013) because the rest-frame hard X-rays that *Athena* will detect are insensitive to absorption. However, *Athena* will miss or be unable to identify Compton-thick AGN, and the X-rays alone cannot be used to measure star formation in the host galaxies. Finally, radio surveys with the SKA or LOFAR will efficiently detect star formation and AGN to z > 6, but cannot directly measure key properties (*e.g.*, ionization state, gas density, or metallicity). To fully understand the influence of environment, mid- and far-infrared tracers will be required.

### 1.1.2.2 The Next Big Step – 3D Surveys with Origins

Accurate measurements of the SFRD and BHARD require infrared spectroscopy, which is sensitive to obscured star formation and provides the most complete identification of growing SMBHs (Hickox & Alexander, 2018). *Origins* can undertake unbiased, 3D extragalactic surveys that simultaneously measure the redshifts, star formation rates, and black hole accretion rates in millions of galaxies from the EOR to today, over survey areas large enough to be immune to the effects of cosmic variance and which identify active galaxies in sufficient numbers to map out the cosmic web. This enormous sample is strictly necessary to understand how star formation and AGN activity depend on redshift, luminosity, galaxy age, and galaxy mass. Hundreds to thousands of spectra are required in each redshift bin from 0 < z < 8, and this drives requirements for *Origins'* spectroscopic surveys. The need to capture high redshift systems also makes *Origins* extremely sensitive at lower redshifts: it measures the star formation rates of galaxies down to 1 $M_\odot$ $yr^{-1}$ – equivalent to the rate at which stars are forming in the Milky Way today – in tens of thousands of galaxies at cosmic noon.

Section 1.1.5 provides yield estimates for a notional two-tier deep and wide *Origins*/OSS survey, and shows how such a survey would firmly establish the relationship between SFR and BHAR over more than 99% of cosmic history. To obtain enough galaxies with precise measurements at z > 4, galaxies detected via features such as [*NeII*] in these surveys would be followed up with targeted observations to detect the fainter fine-structure lines. Combined with a very wide *Origins*/FIP survey (hundreds to thousands of $deg^2$), it would also fill in a missing piece in planned WFIRST, LSST, *Euclid*, SKA, and *Athena* surveys and conclusively measure the impact of cosmic environment on galaxy assembly. Table 1-5 shows the requirement flow, based on the need to detect enough galaxies in each redshift bin. The instrumental sensitivity must be $10^{-20}$ W $m^{-2}$ (5σ) in 1 hour (Figure 1-7). This requirement is a factor of 3-5 times above the current design performance of *Origins*/OSS, resulting in a substantial science margin (Section 1.4).

### 1.1.2.3 How Origins Measures Accurate SFRD and BHARD

Beyond merely detecting enough galaxies at high redshift, *Origins* provides multiple, independent indicators of both the star formation rate and black hole accretion rate in each galaxy, with the ability to decompose the two. This mitigates calibration uncertainties for individual scaling relations and increases the reliability of the inferred SFRD and BHARD. It also provides deeper insight into the underlying astrophysical phenomena than correlated population statistics. The observables include PAH emission features, spectral features that trace heating by HII and photo-dissociation regions or AGN, and the continuum spectral energy distribution.

**PAH Emission Features:** PAH features are excellent probes of the UV luminosity, and therefore the star-formation rate. They are also bright: in starburst galaxies PAHs contribute a few percent of the bolometric luminosity and can be seen to large distances, where they also indicate the redshift. *Spitzer* demonstrated the utility of the PAH bands in hundreds of local and distant galaxies, including luminous galaxies out to z ~ 2 (Yan *et al.*, 2005; Houck *et al.*, 2005; Smith *et al.*, 2007; Brandl *et al.*, 2006; Armus *et al.*, 2004, 2007; Pope *et al.*, 2008; Menendez-Delmestre *et al.*, 2007; Valiante *et al.*, 2007; Huang *et al.*, 2007; Rigby *et al.*, 2008), and in the hyper-luminous galaxy GN$_{20}$ at z = 4.2 (Riechers *et al.*, 2014), which showed that organic molecules exist at high redshift.

Notably, PAH features in low-redshift galaxies become fainter when the gas-phase metallicity drops below ~20% of the Solar value (*e.g.*, Engelbracht *et al.*, 2005). In a purely photometric survey, or one with a small spectroscopic window, this could be a severe limitation for using PAHs to measure star formation. However, *Origins*/OSS will measure multiple PAH features in each source, as well as the dust continuum and the fine-structure star formation indicators in many sources, and can thus use this apparent correlation with metallicity to determine how dust grains evolve in galaxies (Section 1.1.3). For sources at higher redshifts, with fainter luminosities or lower metallicities, we can use locally established relations to correct the PAH-SFR calibration on a source-by-source basis whenever we also have measurements of the metallicity from the fine-structure lines.

PAHs also provide an unbiased ratio of starburst-to-AGN power. Obscured AGN heat the surrounding dust to higher temperatures than in starburst galaxies, with peak rest-frame emission around 10 μm (starbursts typically peak near 50-100 μm). The equivalent width (EQW) of the PAH features indicates the relative proportion of AGN and star formation heating: the typical 6.2 μm PAH EQW for pure starbursts is about 0.5 - 0.7 μm, whereas pure AGN have PAH EQWs ~0.1 μm or lower, as the grains are destroyed by the harsh radiation. The PAH EQW is often used in concert with the atomic fine-structure line ratios in a diagnostic diagram to identify the power sources in very dusty galaxies (*e.g.*, Genzel *et al.*, 1998; Armus *et al.*, 2007; Figure 1-5).

**Atomic Gas Lines:** Low-ionization atomic emission lines, such as [*NeII*] 12.8 μm, [*NeIII*] 15.5 μm, and [*SIII*] 18.7 and 33.5 μm, are produced in the HII regions surrounding young stars and correlate extremely well with the star-formation rates derived from the dust continuum and hydrogen recombination lines (*e.g.*, Ho & Keto, 2007; Inami *et al.*, 2013). Meanwhile, the rest-frame, mid-infrared spectrum contains tracers of highly-ionized gas, such as [*SIV*] 10.5 μm, [*NeV*] 14.3 and 24.3 μm, and [*OIV*] 25.9 μm, with ionization potentials of 35-97 eV. The ratios between these lines and those with much lower ionization potential (*e.g.*, [*NeII*] 12.7 μm) determine the heating from obscured AGN, which emit a harder spectrum than massive stars and thus raise the ionization state of the gas (Figure 1-5).

Previous work has demonstrated the robustness of this approach. While emission lines such as [*NeV*] are seen in the spectra of extremely hot objects in the Milky Way, including planetary nebulae (Bernard-Salas *et al.*, 2001; Pottasch *et al.*, 2001), this line is not seen in the nuclear or integrated spectrum of a galaxy unless an AGN is present. Similarly, [*OIV*] can be excited by OB stars, and this line is seen in starburst galaxies (*e.g.*, Lutz *et al.*, 1998; Smith *et al.*, 2004; Devost *et al.*, 2007). However, AGN produce extremely large [*OIV*]/[*NeII*] line flux ratios (near unity or above), and this ratio is often used as an extinction-free diagnostic of AGN power (*e.g.*, Genzel *et al.*, 1998; Lutz *et al.*, 2003; Armus *et al.*, 2006; Armus *et al.*, 2007). The excess [*OIV*] emission can then be directly attributed to AGN heating, providing an independent measure of the AGN luminosity. [*NeV*] and [*OIV*] have been well calibrated in samples of local galaxies, and can measure the AGN power in even heavily obscured galaxies (Gruppioni *et al.*, 2016).

Observations with *Origins*/OSS are extremely sensitive to obscured and/or low-luminosity AGN. The spectral line limits achieved in the Deep survey proposed (~3x10$^{-21}$ W m$^{-2}$; Section 1.1.5), when applied to the [*OIV*] 25 μm line, correspond to bolometric AGN luminosities of ~10$^9$, 10$^{11}$, and 10$^{12}$ solar luminosities, or equivalent mass accretion rates of 0.001, 0.1, and 1 Solar Mass per year at redshifts of z=1, 3, and 5, respectively (assuming the correlation between line and AGN luminosity in Gruppioni *et al.* 2016, and a 10% efficiency in the conversion of mass to energy). These limits are comparable to the typical AGN luminosities expected in the *Athena* wide field imaging surveys at these redshifts. Unlike *Athena*, *Origins* also simultaneously measures redshifts and star formation rates for every detected source. Together, *Athena* and *Origins* could provide independent estimates of the Compton-thick AGN fraction (*e.g.*, Alexander *et al.* 2008; Lansbury *et al.* 2017) and a nearly complete picture of the population.

The high-resolution mode of *Origins*/OSS can also be used to estimate the mass of the black hole in AGN-dominated systems, regardless of their dust content. As shown by Dasyra *et al.* (2008), the widths of the mid-infrared [*NeV*] and [*OIV*] emission lines are correlated with the mass of the central SMBH (as determined via reverberation mapping). Velocity dispersions of 100-400 km/s are found in AGN with black hole masses of $10^7 - 10^9$ M$_\odot$. These velocities can be easily resolved with follow-up OSS spectroscopy of AGN detected in the deep or wide *Origins* /OSS surveys, giving us an estimate of the black hole masses, even for Compton-thick sources.

Since infrared emission is strongly enhanced in merging galaxies (*e.g.*, Sanders *et al.* 1988), *Origins* surveys detect interacting galaxies across the entire merger sequence, and over a large range in mass ratio, building on results from IRAS, ISO, *Spitzer* and *Herschel* to understand how star formation and black hole growth proceed as galaxies collide. These merging galaxies form the progenitors of gravitational wave sources for the next generation of observatories, such as LISA. Our deep *Origins* survey detects PAH and fine-structure emission lines from galaxies with stellar masses of $10^8 - 10^{10}$ M$_\odot$ at z ~ 4-6 (see Figures 1-6 and 1-17). To the extent that these are merging galaxies with central black hole masses of $10^5 - 10^7$ M$_\odot$, the eventual SMBH mergers should be detectable with LISA.

**Dust Continuum:** Finally, the shape of the mid- and far-IR continuum is sensitive to the relative heating by AGN and star formation (*e.g.*, Kirkpatrick *et al.*, 2015). With a measured redshift, the far-IR continuum emission can be fit using templates to constrain the relative fraction of the bolometric luminosity due to young stars and the AGN. This approach takes advantage of the well-established correlation between the far-IR luminosity and the star formation rate (*e.g.*, Murphy *et al.*, 2011; Kennicutt *et al.*, 1998), which is the basis of existing measurements of the dust-obscured SFRD (Madau & Dickinson, 2014). Meanwhile, the mid-IR continuum indicates the AGN luminosity in systems where it dominates, and the combination of the two leads to the template-based fitting approach. As described above, with *Origins*/OSS these continuum-based estimates of the ongoing star formation in galaxies can be cross-calibrated with those derived from spectroscopic measurements of low-ionization atomic emission lines and small grains. Comparing these techniques can isolate the heating of dust by older stars and help to understand the different timescales probed by different star formation indicators, which helps to interpret dust-continuum star formation rates based on measurements from *Origins*/FIP that can extend the analysis to lower luminosity galaxies over a significant fraction of the sky.

### 1.1.3 Science Objective 2: The Rise of Metals and the First Dust

> Using well-understood PAH and emission-line metrics, *Origins* maps the buildup of metals in the ISM of Milky Way-sized galaxies across cosmic time.

### 1.1.3.1 The Landscape of High Redshift Metallicity and Dust Indicators

The present-day Universe is rich in metals that efficiently cool the ISM, are essential to forming stars and planets, and form the building blocks of life. But it did not start this way — metals built up in the ISM by successive generations of stars, as seen by the fact that high-redshift galaxies contain fewer metals than present-day galaxies of similar mass. Studies at low redshift indicate that the rate at which the gas-phase metallicity increases with time depends on a galaxy's star formation history and halo environment, and extensive studies indicate that the metallicity of a typical galaxy increases by a factor of 5-10 from z = 1 to z = 0. Our knowledge of the metal enrichment beyond z ~ 1 is much more limited (*e.g.*, Sanders *et al.*, 2015), and understanding how metallicity evolves at higher redshift is necessary to understanding galaxy growth, since varying the metal content of galaxies dramatically alters how they form stars and build up structure.

What little we know about metals in the very early Universe (z ≥ 4) comes from quasar absorption-line studies of the circum- and inter-galactic medium (IGM). These studies are important, as they trace the dominant reservoir of metals in the Universe (McQuinn, 2016). However, they provide an incomplete picture. Absorption-line studies probe individual lines of sight to bright background objects like QSOs, and the inferred metallicities rely on assumptions about the structure and ionization state of the absorbing gas. Although the next generation of large, ground based telescopes, like TMT and GMT, will increase the number of sightlines by using high-redshift galaxies as absorption-line probes, these measurements suffer the same limitations. A more fundamental limitation of these IGM absorption measurements is that they do not probe the interiors of galaxies where the heavy elements are actually produced. This limitation is a serious problem, because the mass and contents of the outflows that bring metals into the IGM depend on the stars that form—which depends on the metallicity.

Upcoming and planned facilities will provide the first abundance estimates in samples of early galaxies. Spectroscopic surveys with *Euclid* and WFIRST will provide metallicities of galaxies to z ~ 4, using traditional rest-frame optical nebular line tracers. JWST, with its highly sensitive mid-infrared spectroscopic coverage, will be able to use the same lines to probe galaxy metallicities at faint luminosities to even higher redshifts over limited areas (*e.g.*, Windhorst *et al.*, 2009). These missions will provide an important first glimpse of the relative trends of metal content in samples of galaxies across cosmic time. However, their use of rest-frame optical lines results in two significant limitations.

First, rest-frame optical metallicity surveys can either miss or incorrectly characterize metal content in sources with high dust extinction (*e.g.*, Zahid *et al.*, 2014). This problem is critical, since during the peak of star forming activity in the Universe (z ~ 1-3), most star formation occurred in very dusty galaxies (Elbaz *et al.*, 2018).

Second, absolute abundance measurements are highly uncertain if one does not know the gas temperature (*e.g.*, Kewley & Ellison, 2008). This temperature sensitivity has led to a decades-old problem: there is a factor of 3-5 systematic uncertainty between strong-line abundances calibrated against temperature-sensitive (but faint) auroral lines, and those tied to photo-ionization models. Using these indicators, we cannot reliably ascertain whether the bulk of the Universe has a metal content below or above that of the Sun (*e.g.*, López-Sánchez *et al.*, 2012). This discrepancy is also an issue for understanding feedback in cosmological simulations, which, given the systematic uncertainties, have been forced to treat the absolute normalization of optically-derived gas-phase abundances as a free parameter (*e.g.*, Davé *et al.*, 2011).

Mid- and far-infrared emission lines do not have these limitations, since the key lines arise from transitions that are insensitive to gas temperature and extinction. Figure 1-8 illustrates several such tracers in the rest-frame mid- and far-infrared. The power of infrared diagnostics to trace galaxy metallicities as a function of cosmic time remains largely unexploited, since *Spitzer* and *Herschel* could only use them for nearby galaxies (*e.g.*, Fernández-Ontiveros *et al.*, 2016; Pereira-Santaella *et al.*, 2017). Similar studies of very high-z systems with ground-based facilities like ALMA are moving forward, but can only observe small and biased samples due to the limited atmospheric windows and small sky coverage.

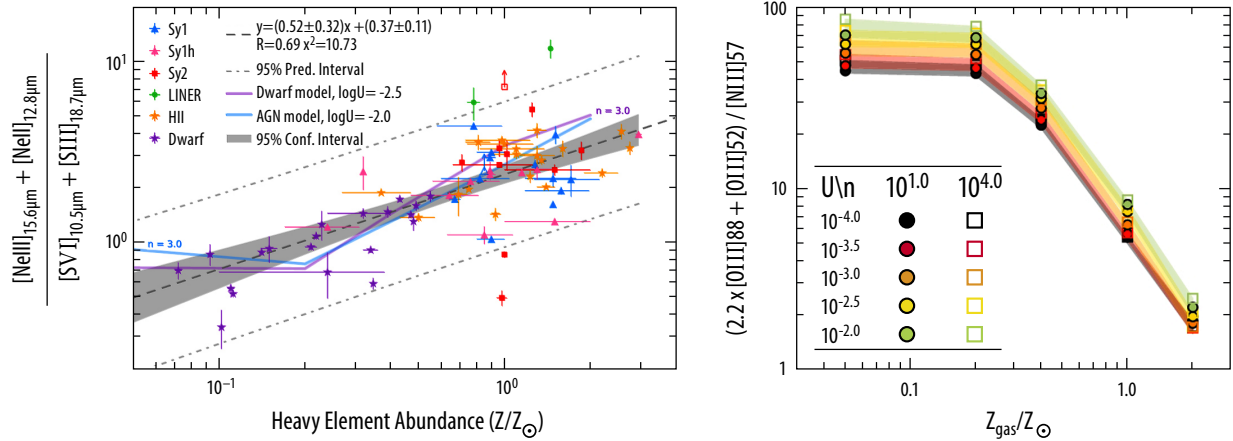

**Figure 1-8:** Key Infrared Metallicity Diagnostic Lines. With a sensitive, broadband spectrometer covering the rest-frame mid and far-infrared features across a large redshift range, *Origins* will measure the gas-phase metallicity in galaxies since EoR, over a wide range in physical conditions. Infrared metallicity line diagnostic diagrams for the bright mid- (left) and far- (right) infrared emission lines from Fernández-Ontiveros et al. (2016) and Pereira-Santaella et al. (2017) are shown. By using these line flux ratios, OSS will trace the buildup of metals in galaxies from 0.1-2 Solar in all environments.

Dust is also a repository of metals (especially C, Si, Fe, and O) that influences star formation. The discovery of large dust reservoirs (~$10^8$ M$_\odot$) in some galaxies and quasar hosts at z > 5 − 7 (*e.g.*, Strandet *et al.*, 2017), only several hundred Myr after the Big Bang, has challenged theories of early dust production in galaxies. Possible explanations for the large amount of dust include ultra-high mass supernovae, rapid ISM growth, or fast-evolving massive AGB stars. Meanwhile, dust is also highly sensitive to the abundance of heavy elements. For example, the strength of PAH emission drops rapidly with decreasing metallicity, and PAH emission is entirely absent for metallicities ⅒ the solar value (*e.g.*, Sandstrom *et al.*, 2012). This absence is attributed to photo-destruction of PAHs in the harsh UV fields of low metallicity galaxies and inhibited formation in dense clouds with lower free carbon content. *Spitzer* permitted a rough calibration of this trend, and JWST will complete the mapping between PAH emission and direct abundance measures in the nearby Universe. These measurements sets the stage for a study of PAHs and metallicity at high redshift.

### 1.1.3.2 The Next Big Step – Galaxy Metallicities and Dust Properties at High Redshift

Measuring the rise of metals and dust from the EOR (z>6) up to cosmic noon (z~3) requires a sensitive survey to measure the metallicity and dust properties in many galaxies. To chart early enrichment requires measuring the metallicity down to 10% of the Solar value in a Milky Way-sized galaxy at z=6, which drives the sensitivity, mapping speed, and bandpass requirements (Table 1-6). Such a survey detects more luminous systems up to z=8 (Figure 1-9), and hundreds of thousands of galaxies overall, producing an SDSS-like survey in the far-IR. This large sample reveals how metal and dust production depends on luminosity, star formation rate, mean stellar age, and environment over most of cosmic time, as thousands of galaxies in each redshift bin are required to isolate these effects. Fortunately, the requirements are similar to those for Objective 1 (Table 1-5), so we envision using the same two-tier notional survey (Section 1.1.5 and Table 1-4), with targeted follow-up on galaxies with bright [*NeII*] emission. Figures 1-9 and 1-10 show the expected yields in galaxy counts and metallicity measurements for a line sensitivity of 4x$10^{-21}$ W m$^{-2}$ (5σ) in 1 hour at 100 μm (Table 1-6).

**Table 1-6:** Extragalactic Science Requirements Flow (Part 2)

| Science Objective 2 |
| --- |
| Measure the metal content of galaxies with a sensitivity down to 10% Solar in a galaxy with a stellar mass similar to the Milky-Way at z=6, as a function of cosmic time, tracing the rise of heavy elements, dust and organic molecules across redshift, morphology and environment. |

| Science Observable |
| --- |
| A catalog of oxygen and nitrogen fine-structure line fluxes from galaxies spanning the luminosity range $10^{11} < L_{IR} < 10^{13} L_\odot$ to measure relative metallicities in nearly $10^5$ galaxies over the redshift range $1 < z < 9$. |

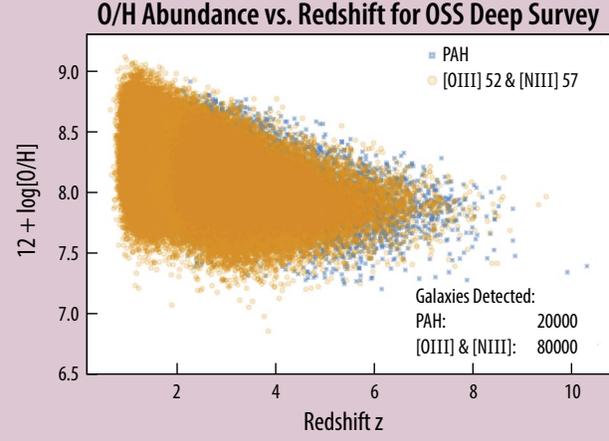

**Figure 1-9: O/H abundance vs redshift detected with *Origins*.** O/H abundance versus redshift for galaxies detected in the OSS Deep Survey. Shown are galaxies detected in either the PAH or [OIII] and [NIII] metallicity probes. PAH detections are based on the detections of the 7.7-μm feature, which enters the OSS band at z=2.2. [OIII] and [NIII] galaxies are simultaneously detected in the [OIII] 52 and [NIII] 57-μm lines.

| Science Requirements |
| --- |
| A line sensitivity of 4x10$^{-21}$ W m$^{-2}$ (5σ) in 1 hr at 100 μm and sufficient beams on the sky to give mapping speeds of ~ 10$^{-4}$ deg$^2$ (10$^{-19}$ W m$^{-2}$)$^{-2}$ sec$^{-1}$. Wavelength coverage from: 25 to 500 μm. |

### 1.1.3.3 How Origins Measures the Metallicity and Dust Properties

**Emission Lines:** The IR lines used to measure abundances (such as from [*OIII*], [*NIII*], and [*SIV*]; see Figure 1-8) have low ionization potentials, are easily excited by stellar UV radiation, and thus trace local (frequently obscured) metal enrichment. In particular, the ratio of the [*OIII*] 52 μm and [*NIII*] 57 μm lines is sensitive to the metallicity, and can be observed in the wide bandwidth of *Origins*/OSS from z~0 nearly to z~9 (Figure 1-9). The wide bandwidth provides a second advantage, which is that the ISM can be fully modeled, eliminating most systematics in the abundance diagnostics. To measure these lines at the highest redshifts, systems with bright lines are targeted with deeper observations to detect the fainter diagnostics.

**Dust Features:** The relative intensities of broad dust features in the rest-frame mid-infrared (3–30 μm) constrain the dust composition and grain size distribution, as well as the ambient radiation field through the sensitivity to ionization state. These features are also easy to detect where PAHs exist, since they contribute up to 20% of the total infrared luminosity of star-forming galaxies (Smith *et al.*, 2007) and are stochastically heated to high temperatures. In contrast, while the thermal dust continuum at longer wavelengths is a useful diagnostic of star formation, at z>5 the larger dust temperature in many galaxies approaches TCMB, rendering differential detection in the Rayleigh-Jeans tail of their dust SEDs increasingly difficult, and thus limiting the ability to measure the grain emissivity. *Origins*/OSS detects the PAH features to z > 8, enabling a study of dust growth since the EOR. For example, due to the very different expected dust compositions and mass yields of low metallicity Pop-II and Pop-III stars, the strength and shape of the rest-frame mid-IR silicate feature may allow us to probe directly the properties of the earliest phases of star formation. Rest-frame UV absorption studies against gamma-ray bursts or background quasars provide complementary information (*e.g.*, on the slope and UV absorption features of the dust extinction curve at these epochs, which constrain grain composition and size), but do not sample the full range of galaxy properties or constrain dust mass.

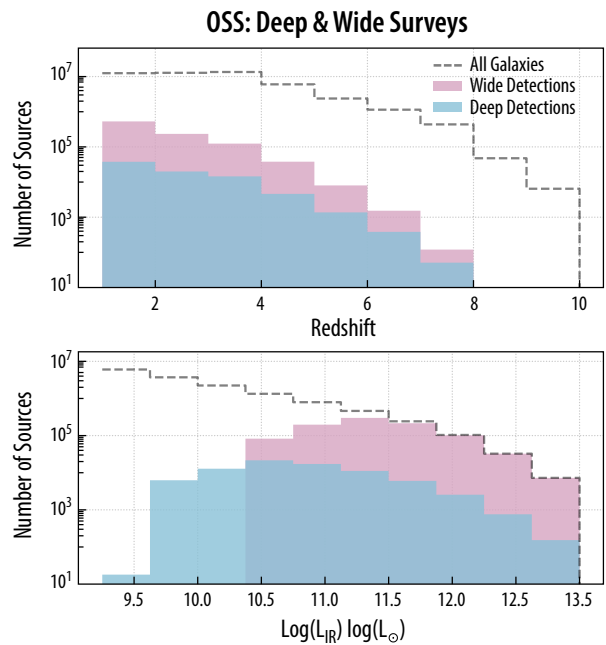

**Figure 1-10:** Metal Content in Galaxies seen by *Origins*. *Origins* will determine the metallicities for an unbiased sample of nearly a million galaxies. These plots show the expected numbers of unique galaxies detected as a function of redshift and IR luminosity in the OSS Deep and Wide Surveys. These galaxies are all detected in the [OIII] 52 and [NIII] 57-μm lines.

The comparable redshift range where PAHs and gas-phase metallicities can be measured enables a complete view of metal enrichment in galaxies.

### 1.1.4 Science Objective 3: Linking Feedback to the Decline of Cosmic Star Formation and the Present Day Mass Function of Galaxies

> *Origins* measures starburst- and AGN-driven outflows over the past 10 billion years.

### 1.1.4.1 The Landscape of Feedback Measurements

Star formation is regulated by stellar winds, supernovae, and AGN, which inject energy and momentum into the ISM (*e.g.*, Croton *et al.* 2006; Governato *et al.*, 2007), heating and disrupting gas that would otherwise collapse to form stars. In extreme cases, this "feedback" can even remove gas from the galaxy; this occurs in starburst galaxies, where thousands of concentrated supernovae combine to drive fast, powerful winds, and also in AGN (Figure 1-11).

There is ample evidence that feedback is important in almost all galaxies. First, the correlation between the SMBH mass and stellar bulge (Magorrian, 1998; Marconi & Hunt, 2003) suggests co-evolution of SMBH and galaxy growth (Section 1.1.2). Since rapid SMBH growth results in luminous quasars, the implication is that growing SMBHs must slow or quench star formation. Second, star formation is very inefficient (Figure 1-12; Kennicutt *et al.*, 1998), with a peak ratio of stellar mass to halo mass of about $\frac{1}{30}$ for a halo mass of $10^{12}$ M$_\odot$ (Moster *et al.*, 2010), which is far below that expected from the cosmic baryon fraction. The efficiency is even lower at low or high masses, which is attributed to supernovae and AGN, respectively, removing gas from galaxies (Baldry *et al.*, 2008; Springel *et al.*, 2005; Henriques *et al.*, 2015), although simulations indicate that stellar and AGN feedback together are important to regulate growth in both regimes (*e.g.*, Hopkins *et al.*, 2014; Henriques *et*

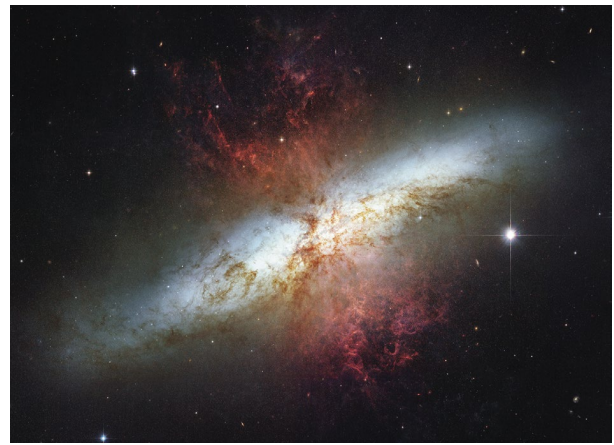

**Figure 1-11:** Galactic Outflows. *Origins* detects and measures galactic outflows in galaxies over the past 10 Gyr of cosmic time. In this image of the nearby starburst galaxy M82, the warm ionized gas (red) traces a familiar hourglass shape extending along the minor axis (stellar disk in blue), driven by the collective effect of young stars and supernovae. Starbursts and AGN can eject energy and metal-enriched gas into the inter-galactic medium, where they can sometimes be seen in emission against dark space, and in absorption against the light of background QSOs.

*al.*, 2018). Third, most of the baryons in the Universe lie in the gas between galaxies, and studies of absorption lines imprinted on the spectra of distant QSOs have shown that some of this gas is metal-enriched, especially in the circumgalactic medium, where it contributes >25% of the mass (Steidel *et al.*, 1994; Adelberger *et al.*, 2003; Rudie *et al.*, 2012; Turner *et al.*, 2014; Werk *et al.*, 2014). Many of these metals are removed from galaxies by galactic winds, which have been studied in nearby starburst galaxies (*e.g.*, Armus *et al.*, 1990; Heckman *et al.*, 1990, 2000; Fischer *et al.*, 2010; Sturm *et al.*, 2011; Martin, 1999; Martin *et al.*, 2012; Veilleux *et al.*, 2005, 2013), and in star forming galaxies at z~1-3 (Kornei *et al.*, 2012; Shapley *et al.*, 2003). AGN can also drive winds that heat and enrich the intergalactic medium, and there is evidence for high-velocity outflows, negative feedback, and/or extremely turbulent interstellar media in some high-redshift quasars and dusty, IR-bright AGN (*e.g.*, Diaz-Santos *et al.*,

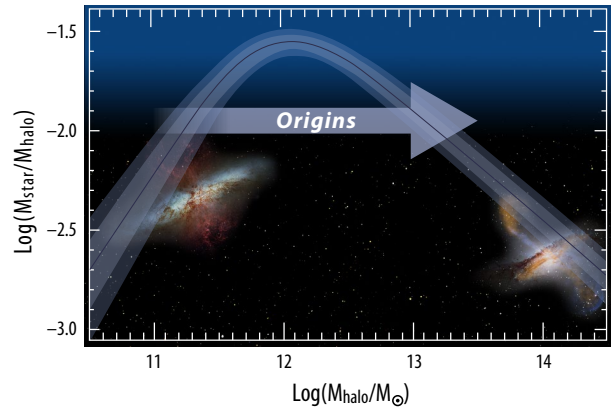

**Figure 1-12:** The inefficiency of star formation. *Origins* measures feedback over a huge range in galaxy mass, when galaxies were rapidly evolving. The plot shows the ratio of stellar to halo mass as a function of halo mass in galaxies (Moster et al., 2010; Erb, 2015). The ratio drops off precipitously above and below a halo mass of about $10^{12}$ $M_\odot$, suggesting extremely inefficient star formation in both low and high mass galaxies. It is believed that stellar feedback dominates at the low-mass end, while AGN dominate at the high-mass end. *Origins* probes both kinds of feedback over at least three orders of magnitude in galaxy mass.

*al.*, 2017). Finally, galaxy formation theories require feedback to reproduce the observed distribution of stellar mass and luminosity (Somerville & Dave, 2015; Hopkins *et al.*, 2014). Feedback can also explain the mass-metallicity relation (Tremonti *et al.*, 2004) and the existence of the main sequence of star-forming galaxies (Elbaz *et al.*, 2010).

Despite the panoply of evidence that feedback does shape galaxies, we still do not understand how it works, which essentially amounts to measuring how energy and momentum from supernovae and AGN couple to the ISM (as well as the circumgalactic and intergalactic media). There is clearly enough energy, as cosmological simulations with different feedback prescriptions can arrive at the observed stellar mass and luminosity distributions, but even slight differences between models predict very different conditions in the ISM (*e.g.*, Hopkins *et al.*, 2013). Measurements of the state of the ISM, and how mass is partitioned in each state, can identify the primary feedback channels as a function of star formation rate, AGN luminosity, galaxy mass, etc.

Perhaps the most important issue relates to galactic outflows (Figure 1-11), which may be the primary mechanism by which galaxies are "quenched." Much, and probably most, of the mass in these outflows is in molecular gas, but we do not know how common they are at z>3, nor whether they are linked to major mergers. For a normal galaxy, we expect about one major merger event per Δz=2, and if fast outflows that can deplete star-forming gas are triggered by these mergers then we would expect to find one fast outflow per ~100 massive galaxies at z=3. Significantly higher rates would imply that major mergers are not the primary trigger, while significantly fewer detections would imply that depletion times are very short or that the fast outflows seen in LIRGs and ULIRGs at low redshift are not important at z>3. In general, observations are required to verify model predictions for scaling relations between the mass outflow rate and star formation rate, galaxy mass, and redshift (*e.g.*, Hayward & Hopkins, 2017).

A related issue is that we do not know the fraction of gas in outflows that can escape the galactic potential, which may also change with redshift. Massive (>100 $M_\odot$ yr$^{-1}$), fast (>1000 km s$^{-1}$) molecular outflows have been detected in ULIRGs at z < 0.5 with *Herschel* (Fischer *et al.*, 2010; Sturm *et al.*, 2011; Gonzalez-Alfonso *et al.*, 2014). Such winds can quench star formation by removing the fuel on timescales of a Gyr (*e.g.*, Feruglio *et al.*, 2010), and the fastest outflows seem to occur in the sources with the most powerful AGN (Veilleux *et al.*, 2013; Cicone *et al.*, 2014; Gowardhan *et al.*, 2018). However, these measurements do not place strong constraints on the fraction of the gas that can fully escape the galaxy; gas that fails to do so may re-accrete relatively quickly.

Determining how feedback operates is a major goal for existing and planned facilities, such as JWST, ALMA, XRISM, and *Athena*. X-ray observations probe the hot phase of the ISM and the effect of AGN feedback in galaxy clusters, and can detect hot outflows to z>1 through absorption or emission line velocities, but it is the cool outflows that can quench star formation. ALMA and JWST can detect these molecular and dusty outflows, including at high redshift with ALMA, but, as described above, they cannot amass the large sample needed to determine how feedback scales with redshift, mass, star formation rate, AGN luminosity, environment, and metallicity.

### 1.1.4.2 The Next Big Step – A Census of Outflows and Shock-heated Gas up to z > 5

An ambitious infrared survey is needed to measure the gas velocity, mass, metal content, energy, and momentum in a large number of galaxies with outflows or large-scale shocks ($\sim 10^3$) up to and beyond cosmic noon (see Figure 1-13). This survey provides the most complete, and unbiased, view of feedback in action, and definitively establishes the frequency and importance of outflows at high redshift. We envision a two-step process in which a wide survey identifies sources with molecular absorption, strong $H_2$ emission, or blueshifted high-ionization atomic lines from AGN-driven outflows (Spoon & Holt, 2009). Detected sources would then be followed up by deep, high-resolution observations. The requirements (Table 1-7) are based on local analogs to systems that are normal at high redshift (corresponding to L*), and the expectation that there are $\sim 5000$ ULIRGs/deg$^2$.

NGC 6240 is a nearby (z=0.025), interacting ULIRG with a powerful starburst and an obscured AGN (Lutz *et al.*, 2003; Armus *et al.*, 2006). It has the most luminous warm $H_2$ emission in the local Universe (Draine & Woods, 1990), a frothy, turbulent ISM heated by interstellar shocks, and a large superwind (Heckman *et al.*, 1987, 1990). The warmest gas (500-1000 K), traced by the S(3) to S(7) rotational transitions, has a mass of about $10^7$ M$_\odot$, and accounts for $\sim 0.1\%$ of the total molecular gas in NGC 6240 and $\sim 0.5\%$ of the molecular gas within the central kpc (Armus *et al.*, 2006; Tacconi *et al.*, 1999). Systems like NGC 6240 are locally rare but are thought to be common at high redshift, and by projecting this galaxy to high redshift we can estimate the sensitivity needed to detect warm gas via the mid-infrared $H_2$ lines. Figure 1-14 shows a simulated *Origins*/ OSS spectrum based on a 1-hour integration for an NGC 6240-like galaxy at z=5; a line sensitivity of $10^{-20}$ W m$^{-2}$ ($5\sigma$) in 1 hour (as defined for Objective 1) is sufficient to detect the $H_2$ lines with sufficient signal to make precise measurements.

Mrk 231 is an infrared-bright AGN with a massive, fast molecular outflow seen by *Herschel* via blueshifted OH and $H_2O$ absorption (Fischer *et al.*, 2010; Gonzalez-Alfonso *et al.*, 2014, 2017), and so serves as a template for absorption measurements at high redshift. The outflow rate is 1000 M$_\odot$ yr$^{-1}$, with a total of ~$10^9$ M$_\odot$ involved in the outflow (Gonzalez-Alfonso *et al.*, 2017). This measurement implies a ratio of mass outflow rate to current star formation rate of about 4, which is comparable to that measured in other local ULIRGs (Gonzalez-Alfonso *et al.*, 2017). Figure 1-15 shows a simulation of a similar system at z=3; to detect these outflows requires a resolution of 1000 km/s in low-resolution mode (detailed study requires 100 km/s in high-resolution mode). The same line sensitivity of $10^{-20}$ W m$^{-2}$ (5σ) in 1 hour is sufficient.

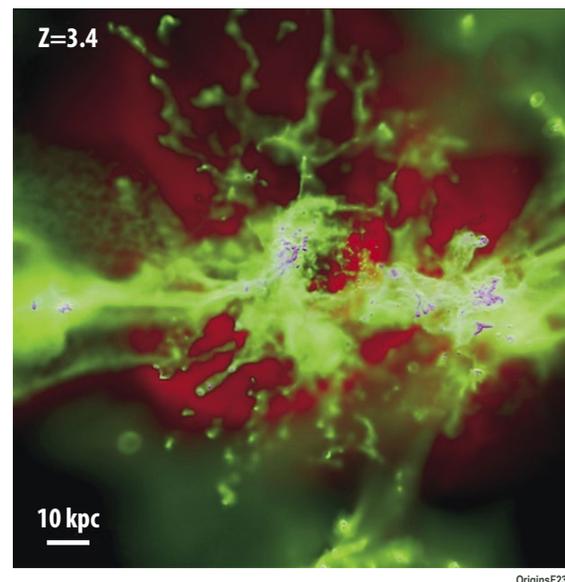

**Figure 1-13:** Feedback at High-z. *Origins* probes the molecular and enriched atomic ISM being ejected by normal galaxies at z > 3. State of the simulation of a Milky Way mass galaxy at z=3.4, showing feedback-driven structure in the cold (magenta), warm (green) and hot (red) gas (Hopkins et al., 2014).

**Table 1-7:** Extragalactic Science Requirements Flow (Part 3)

| Science Objective 3 |
|---|
| Determine how energetic feedback from AGN and supernovae regulate galaxy growth, quench star formation, and drive galactic ecosystems, by measuring galactic outflows as a function of SFR, AGN luminosity and redshift over the past 10 Gyr. |

| Science Observable |
|---|
| A catalog of warm and cold molecular gas velocity and mass outflow rates via mid-IR $H_2$ emission and far-IR OH absorption lines in $10^3$ galaxies. |

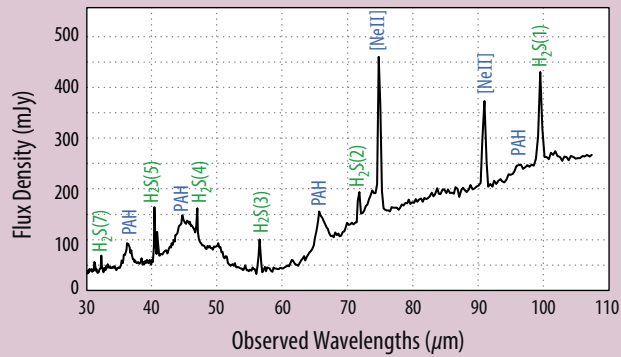

**Figure 1-14:** Simulated rest-frame mid-IR spectrum of the local ULIRG NGC 6240 scaled to LIR=5x$10^{12}$ L$_\odot$ and placed at z=5, as seen with *Origins*/OSS in a 1hr integration. Extremely strong $H_2$ lines from warm molecular gas shock heated by the outflowing galactic wind are seen. Also seen are strong, broad PAH features and the [NeII] and [NeIII] fine structure lines.

| Science Requirements |
|---|

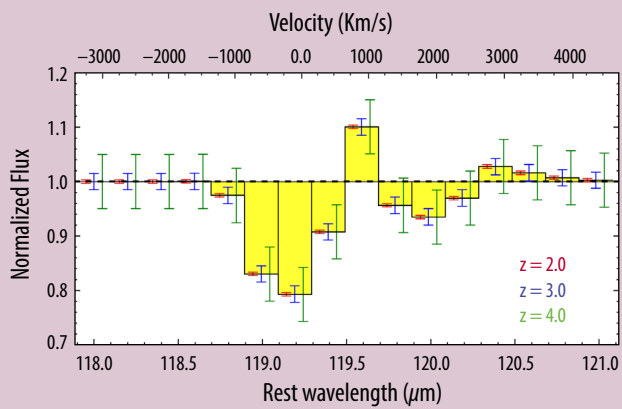

**Figure 1-15:** To measure the velocity and mass outflow rates in the molecular gas in an L* galaxy at z=3, *Origins*/OSS shall be capable of resolving an emission line of 1000 km/s (R=300) in low-resolution mode and 100 km/s (R=3000) in high-resolution FTS mode, with a 5σ, 1hr line flux sensitivity of $10^{-20}$ W m$^{-2}$. *Origins* also must have a long wavelength cutoff of no less than 476 μm (to detect OH 119 μm at z=3).

These requirements also make *Origins* sensitive to very small amounts of warm molecular gas (100 times less than in NGC 6240) at z ~ 1, after the peak of star formation. These galaxies would be prime targets for follow-up with OSS at high resolution to measure the kinematics of the shocked gas. Finally, in nearby galaxies (d < 50 Mpc), where the star-forming disks (D$_{25}$ > 1 arcmin) are resolved in the far-infrared, *Origins* can image outflows and extraplanar gas in atomic and molecular emission lines and through dust. Extended warm gas or dust can easily be seen against the dark background of space, particularly in edge-on galaxies where the full extent of the winds can be traced into the CGM. Resolved spectroscopy of bright lines can be used to estimate how much of the outflowing, metal-enriched material can escape the galactic potential even in normal star-forming galaxies.

### 1.1.4.3 How Origins Measures Outflows

**Molecular Emission:** *Origins* can directly measure the warm molecular gas heated by shocks and in outflows using $H_2$ emission lines, such as the S(3)-S(7) transitions from gas between 500-1000 K. As shown in Figure 1-14, these lines are bright in galaxies with outflows or large-scale shocks. Targeted observations of galaxies with strong lines using the *Origins*/OSS high-resolution mode yield velocities, temperatures, and masses. The *Origins* bandpass enables the use of these lines up to the EOR. In nearby galaxies, the mass of extraplanar, molecular gas can be most reliably measured with $H_2$ and HD lines.

Molecular Absorption: Blueshifted absorption lines indicate nuclear outflows. In unresolved galaxies, these often appear as P-Cygni profiles in $H_2$, $H_2O$, [*OI*], and OH lines. For example, the 79 and 119 μm OH features have been used to estimate the outflow rates and physical conditions of cold, dense molecular gas (Sturm *et al.*, 2011; Gonzalez-Alfonso *et al.*, 2014, 2017), and they can be measured with *Origins* at z~2 in minutes. Lines detected with *Origins*/OSS at low resolution (1000 km/s) can be followed up with the targeted, high-resolution observations to obtain velocities and masses.

**Atomic Emission:** Fine-structure lines from Ne, Fe, and Si (*e.g.*, Figure 1-14) identify galaxies (or regions within galaxies) where shocks dominate the ionization of the gas or have destroyed small dust grains and raised the gas-phase abundance of highly depleted elements. In nearby galaxies, resolved spectroscopy of bright lines produce velocity fields that can be used to estimate how much of the outflowing, metal-enriched material can escape the galactic potential. In particular, the bright [*CII*] 158-μm emission line can be used as an effective tracer of very low column density material (either neutral, molecular, or ionized) predicted by simulations of galactic disk "fountains" (*e.g.*, Walch *et al.*, 2015). *Origins* should easily be able to map extended, faint [*CII*] emission in hundreds of low-redshift galaxies, reaching surface densities of neutral and ionized gas of < 0.1 $M_\odot$ $pc^{-2}$ for the first time.

**Dust Emission Features:** Ratios of the mid-IR PAH emission features constrain the average grain size and ionization state of dust in galaxies, which can indicate processing by fast shocks (*e.g.*, Beirao *et al.*, 2015). Combined with measurements of the $H_2$ emission lines, PAH emission can identify reservoirs of warm (100-500 K) molecular gas heated by slow shocks driven into dense molecular clouds (Ogle *et al.*, 2010; Guillard *et al.*, 2012; Stierwalt *et al.*, 2014). In nearby galaxies, extraplanar PAH and thermal dust continuum emission can also measure the dust content of galactic outflows (*e.g.*, McCormick *et al.*, 2013, Melendez *et al.*, 2015).

### 1.1.5 3D Infrared Surveys with Origins

> With a combination of deep and wide unbiased spectroscopic surveys, wide-area continuum surveys with FIP, and ultra-deep targeted observations with OSS, *Origins* measures star formation and AGN growth, the rise of metals, and feedback in galaxies over cosmic time and across the cosmic web.

Over twenty years ago, the *Hubble* Deep Field Survey revealed the richness of the high-redshift Universe. Since then, advancements in our understanding of galaxy formation and evolution have been anchored in multi-wavelength, extragalactic surveys performed with *Hubble*, *Spitzer*, *Herschel*, and many ground-based observatories. These surveys typically adopt a "wedding-cake" strategy, with several tiers of increasing depth and decreasing area to sample galaxies over a wide range of redshift and luminosity. While these multi-wavelength photometric surveys have revealed much about galaxy evolution, follow-up spectroscopy is often necessary to measure accurate redshifts and the physical properties of the detected galaxies. *Origins* revolutionizes the concept of unbiased, blank-field surveys by covering wide areas with full IR spectroscopy at every point in the field (Figure 1-16).

The science objectives presented in Sections 1.1.2–1.1.4 can be achieved with a multi-tier, 3D survey that takes advantage of the full 25-588 micron OSS wavelength range. The blind surveys are suffi-cient to achieve most of the science goals at z < 4 by themselves, and will also identify galaxies at higher redshift with strong emission or absorption features. Deeper targeted OSS spectra of these galaxies enable important measurements based on the fainter features and precise velocities. Finally, wide-area continuum surveys with FIP extend the census of obscured galaxy growth to the largest scales of the cosmic web. Here we present a nominal survey that allocates 2000 hours to the two-tier OSS survey, 2000 hours to targeted OSS follow-up, and 1000 hours to wide-area FIP surveys.

### 1.1.5.1 Predicted Counts for the 3D OSS Extragalactic Surveys

Tracking the relative growth of stars, metals, and SMBHs (Science Objectives 1 and 2) requires at least 100 galaxies with IR line detections in each redshift bin. This requirement drives the OSS survey design.

To determine the optimal yield from a blind survey, we use the galaxy evolution models from Bethermin *et al.* (2017), coupled with constraints on the line-to-LIR ratios based on observations of galaxies with *Spitzer* and *Herschel* (*e.g.*, Bonato *et al.*, 2019). While this backward evolution model fits all existing number counts and redshift distributions in the infrared and submillimeter, great uncertainty remains in the estimates above z > 4, where existing data are very incomplete. Nevertheless, the number of galaxies estimated from this model is conservative: in the wide survey there are many gravitationally lensed galaxies, which are not accounted for here, and we have accounted for the dependence of PAH emission on metallicity, based on nearby galaxies spanning a range of metallicities.

Based on this model and the requirement of 100 galaxies per redshift bin up to z~7, we explore the yield from a range of surveys from 0.1-1000 deg², each with an integration time of 1000 hours with OSS. Figure 1-17 shows the number of z > 4 and z > 6 galaxies with line detections as a function of area. To maximize the number of z > 4 line detections for the faintest MIR/FIR fine-structure (FS) lines (*e.g.*, [NeV], blue points), we converge on a deep survey of 0.5 deg², which reaches a 5σ depth of 6x10⁻²¹ W m⁻² at 100 μm. This survey yields an unbiased sample of >100 galaxies at z > 4 detected in the full suite of mid-IR and far-IR ISM diagnostics. At higher redshifts, we find that a 20 deg² survey that reaches a 5σ depth of 4x10⁻²⁰ W m⁻² at 100 μm maximizes the number of bright FS lines (*e.g.*, [NeII], green points). Since the PAH lines are bright even when accounting for their decreased flux in low metallicity galaxies (*e.g.*, Shipley *et al.*, 2016), the deep and wide OSS surveys yield ~10⁴ PAH 7.7 μm line detections (red points) at z > 6. We only show estimates for these three lines ([NeV], [NeII], and PAH 7.7 μm), but there are other bright lines and features, so the expected number of line detections for these surveys is much higher. We also note that these estimates are based on 5σ line detections. If OSS mapping speeds are worse by a factor of two, we can still achieve these yields by considering lines detected at 3.5σ.

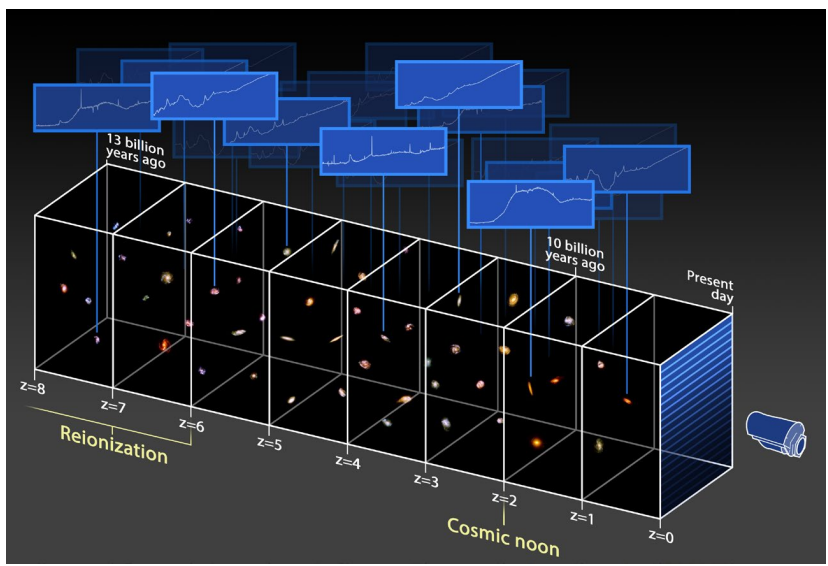

**Figure 1-16:** *Origins* enables three-dimensional (3D) spectroscopic surveys of the Universe of galaxies. The extragalactic science program is based on *Origins'* ability to conduct wide-area spectral surveys at a resolving power R=λ/Δλ of 300 over the full 25-588 μm spectral range. *Origins* can conduct spectroscopic surveys of order 20 deg² in less than 1000 hours, leading to spectra for at least a million galaxies out to z > 8.

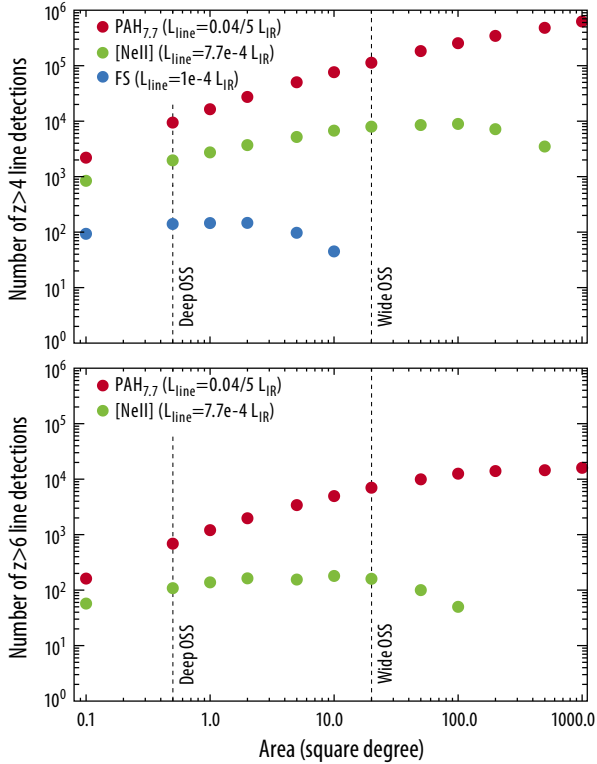

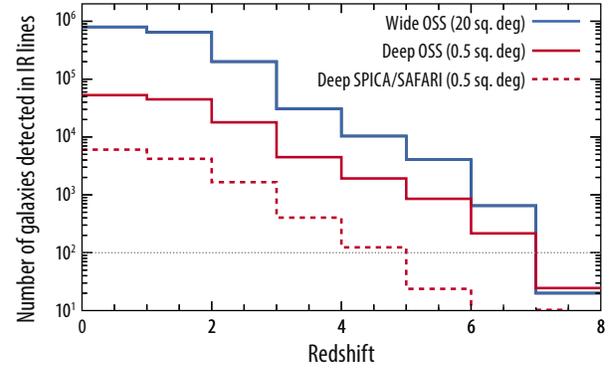

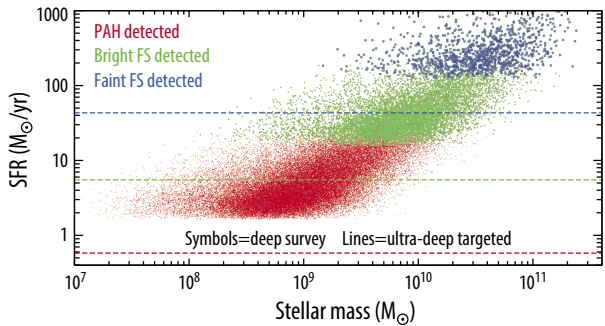

**Figure 1-18:** (Above) Cumulative *Origins*/OSS Galaxy Survey Counts. The redshift distribution of galaxies with detected lines in our two-tier, OSS, blank-field survey. A comparison with SPICA/SAFARI observations over the same amount of time (dashed line) shows only *Origins*/OSS is capable of detecting galaxies into the EoR with sufficient counts (>100 galaxies at 6 < z < 7) to study this important population.

**Figure 1-17:** (Left Top) Expected *Origins*/OSS Line Detections. The expected number of galaxies detected by *Origins* in the proposed Deep and Wide area surveys, in the PAH (red) and bright and faint atomic fine-structure lines (green, blue), as a function of area, at z > 4 (top) and z > 6 (bottom). The deep survey area is selected to maximize z > 4 faint line detections, and the wide tier maximizes the detection of bright lines (like [NeII]) beyond z > 6, into EOR.

**Figure 1-19:** (Left) OSS depths relative to the z ~ 4 galaxy main sequence. The established relation between the SFR and the stellar mass of galaxies at z ~ 4 is shown denoting which galaxies are detected in PAH (red), bright (green) and faint (blue) FS lines in our deep survey and our targeted follow-up (dashed lines). By combining the survey with deeper follow-up at z > 4, we will measure redshifts, SFRs, metallicities and BHARs in >100 typical main-sequence galaxies per redshift bin. Milky Way progenitors are easily detected in the deep survey in the PAH and bright FS lines at z > 4, with fainter sources detected in follow-up with OSS.

The dashed line in Figure 1-18 shows the depth achieved by SPICA/SAFARI (based on their goal capability) for an identical 1000 hour, 0.5 deg² survey. While SPICA is able to study ULIRGs at cosmic noon, it misses the fainter (but typical) star forming galaxies at this epoch, and is not able to perform an unbiased census of galaxies at z > 4. Table 1-8 shows the limiting SFR and stellar mass of the proposed deep OSS surveys at z = 2-6. At z ~ 4, we detect faint, fine structure lines in ULIRGs. To detect these fainter lines in less luminous galaxies, we supplement the unbiased survey with ultra-deep targeted OSS spectra of *e.g.* [NeII] (10 hours per pointing, 5σ depth of 1x10⁻²¹ W m⁻² at 100 μm; dashed horizontal lines in Figure 1-19).

**Table 1-8: Limiting depths (in SFR and stellar mass) of the OSS unbiased survey and targeted high-z follow-up.** While the OSS Deep survey produces an unbiased sample of IR emitting galaxies out to the highest redshifts, additional ultra-deep targeted observations of galaxies selected from the survey are necessary to detect the rich array of fainter FS lines needed to measure the BHAR and conditions in the ISM.

| OSS | Parameters | Limiting SFR ($M_\odot$/yr) | Limiting SFR ($M_\odot$/yr) | Limiting SFR ($M_\odot$/yr) | Limiting $M_\star$ ($10^8 M_\odot$) | Limiting $M_\star$ ($10^8 M_\odot$) | Limiting $M_\star$ ($10^8 M_\odot$) |
|---|---|---|---|---|---|---|---|
| **Observation** | | $\bar{z}$=2 | $\bar{z}$=4 | $\bar{z}$=6 | $\bar{z}$=2 | $\bar{z}$=4 | $\bar{z}$=6 |
| **Deep survey** | 0.5 sq deg 1000 hours | 0.4 | 2 | 6 | 1 | 4 | 10 |
| | | 4 | 20 | 60 | 10 | 30 | 100 |
| | | 30* | 200* | 500* | 100 | 200 | 400 |
| **Targeted high z** | 100 galaxies 1000 hours | n/a | 0.6 | 1.5 | n/a | 1 | 4 |
| | | | 6 | 15 | | 10 | 40 |
| | | | 40* | 100* | | 100 | 200 |

Footnotes: Depths for a given line **PAH**, **bright FS**, **faint FS**, assuming SFR=$10^{-10} \times L_{IR}$ and the relation between SFR and $M_\star$ (Bethermin et al. 2017).
*The SFR will not be measured from these faint lines but this is meant to show what type of galaxy (in terms of SFR).

## 1.1.5.2 A Note on Confusion

The confusion limit has been a fundamental barrier for deep imaging surveys with *Spitzer* and *Herschel*. The improvement in continuum mapping with FIP, at 250 µm for example, is already substantial compared to previous *Herschel*/SPIRE surveys (Figure 1-20). Meanwhile, confusion is much less of a problem for OSS, as the spectra for each galaxy in the field can be used to de-blend galaxies within a given beam. Simulations have shown that this technique recovers redshifts and fluxes for sources up to an order of magnitude below the imaging confusion limit (Raymond *et al.* 2010); since these simulations did not include the bright, mid-IR PAH features, we expect to reliably recover fluxes and redshifts out to the EOR. Instead, confusion occurs when lines from foreground sources overlap in the beam of a high-redshift target. To test whether this confusion is an issue for *Origins*, we used the Bethermin *et al.* (2017) galaxy counts coupled with constraints on the line-to-LIR ratios (Bonato *et al.* 2019) to calculate line confusion at the depths of our proposed OSS extragalactic surveys. Appendix E.1 describes a preliminary study of blind spectral extraction of a simulated *Origins*/OSS 3D data cube of an extragalactic deep field. Figure 1-21 shows the integral line counts per spatial beam and spectral resolution element, compared to the depths of the deep survey in each of the six OSS bands. Spectral confusion is an issue when the number of sources per beam is >¹⁄₁₅ (horizontal dotted line). At the depths of the planned deep survey (dashed vertical lines), spectral line confusion is not expected to limit the ability of these surveys to achieve the primary science objectives.

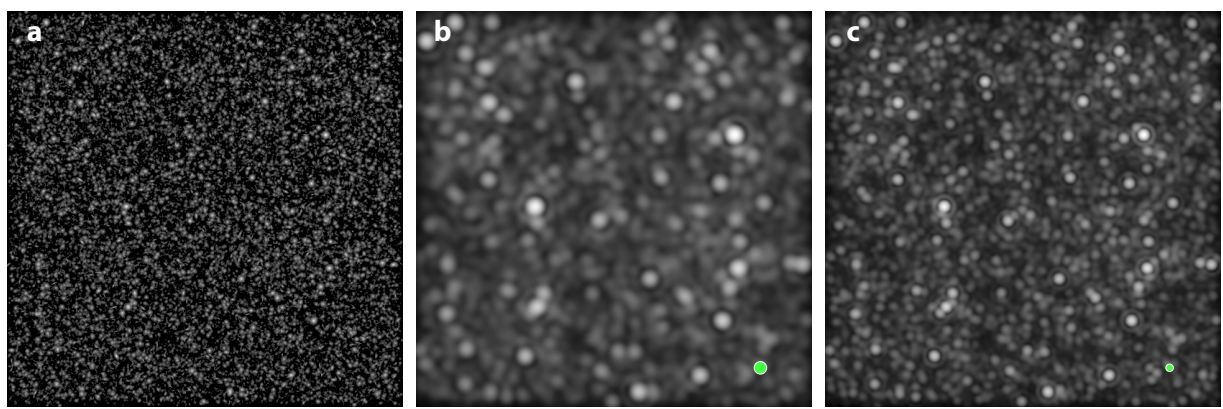

**Figure 1-20:** *Origins* has a deeper confusion limit than Herschel. a) Sky simulation of 9.5 arcmin x 9.5 arcmin at 250 µm, matched to the FoV of the FIP instrument. b) The same map convolved with Herschel/SPIRE 250 PSF. c) The expected *Origins*/FIP 250 µm map over the same area showing the substantial improvement in the source identification and the depth of continuum imaging data relative to previous Herschel/SPIRE surveys. Green circles to the bottom right show the PSF size.

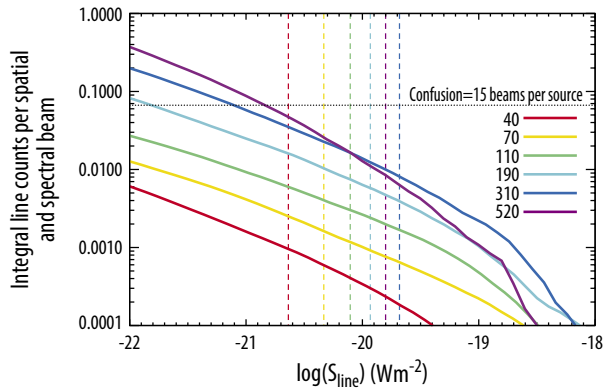

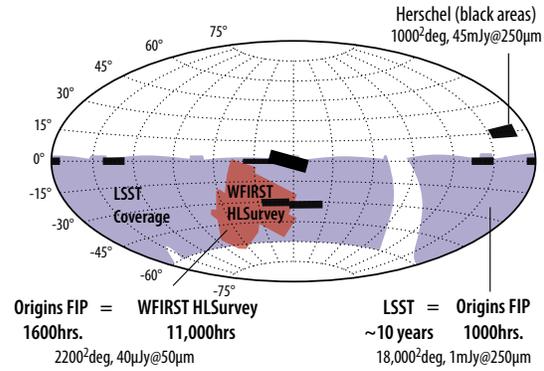

**Figure 1-21:** The Integral Line Counts per spatial beam and spectral resolution element (R=300) show that spectroscopic confusion is not a problem. The integral line counts are shown for each *Origins*/ OSS band in μm (see legend). The detection limit of the deep survey in each band is shown by the vertical dashed lines. The nominal 2D confusion limit of 15 beams per source is shown as the dotted horizontal line. In all bands, the counts for the deep survey are well below the confusion limit.

**Figure 1-22:** *Origins*/FIP can rapidly produce continuum maps over a very large area. This sky projection shows the footprint two of the widest surveys, LSST and WFIRST-HLS, and the time estimate to cover those areas with FIP. The proposed FIP surveys would include a wide, 10,000 deg² at 250 μm and a medium, 500 deg² area at 50 and 250 μm (Table 1-9) that would provide SFRs and dust masses to complement LSST and WFIRST-HLS stellar masses and detect > 99% of the galaxies that will be seen with WFIRST in the near infrared (Figure 1-23).

### 1.1.5.3 FIP Wide-area Surveys

A two-tier continuum survey with FIP covering 1) 500 deg² at 50 and 250 μm, and 2) 10,000 deg² at 250 μm will constrains the role of environment in driving dust-obscured star formation and black hole growth. The integration times and depths reached are listed in Table 1-9 (also Figure 1-22). An ultra-wide field will take a census of rare galaxy populations, such as large galaxies and QSOs at z > 6 during the EOR, and lensed galaxies. A survey at 50 μm can detect more than 99% of the galaxies detected by WFIRST-HLS and provide SFRs and dust masses to complement the stellar mass measurements from the near-infrared. Meanwhile, the medium field probes large-scale structure (including proto-clusters and clusters) as a function of the redshift from 1 < z < 8. Figure 1-23 shows the anticipated redshift and luminosity distribution of galaxies in this survey; FIP surveys are more efficient than μm-wave surveys (*e.g.*, CMB-S$_4$) at detecting dusty galaxies during reionization, which can be followed up by OSS for detailed characterization.

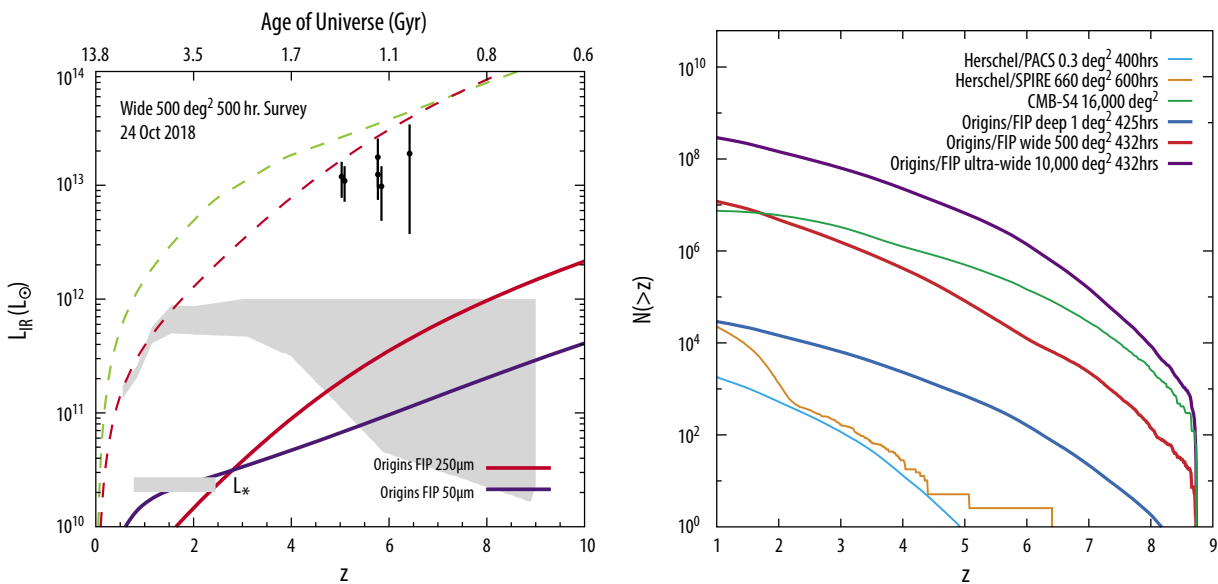

**Figure 1-23:** Redshift distribution and luminosity depth of galaxies in a wide *Origins*/FIP survey (see Table 1-9 for details). An ultra-wide 10,000 deg² survey with *Origins*/FIP at 250 μm detects galaxies out to the epoch of reionization. A deep 50 μm survey covering the WFIRST-HLS footprint detects >99% of all WFIRST-detected galaxies, providing star-formation rate and dust mass measurements that complement stellar mass measurements in the near-infrared with WFIRST.

**Table 1-9:** Continuum Mapping Surveys

| Property | | Medium Field | Ultra-Wide Field |
|---|---|---|---|
| Survey Area | | 500 deg$^2$ | 10,000 deg$^2$ |
| Science Application | | Large-scale structure | Rare sources (lensed galaxies, galaxy clusters and proto-clusters) |
| Survey Time | 50 μm | 415hrs | — |
| | 250 μm | 25hrs | 500 hrs |
| Point Source Depth (5σ) | 50 μm | 40 μJy | — |
| | 250 μm | 1 mJy (confusion limit) | 1 mJy |

### 1.1.5.4 OSS Targeted Observations

Science Objectives 1 and 2 (**Sections 1.1.2-1.1.3**) require precise measurements (*e.g.*, line ratios) of similar quality for >100 galaxies in each redshift bin (Δz=1). The highest redshift galaxies (z > 4-6, depending on the metric) require better sensitivity than in the blind surveys. Galaxies with strong lines (*e.g.*, [*NeII*]) detected in the blind surveys are targeted for deep (10 hours) observations, which increase the sensitivity to a 5σ depth of 1x10$^{-21}$ W m$^{-2}$ at 100 μm. This increase is more than sufficient to match the measurements made at lower redshift.

Science Objective 3 requires high resolution observations of outflows to measure velocities, so sources at z < 4 that show strong and/or blue-shifted H$_2$ lines or blue-shifted OH absorption features are observed in the high-resolution mode. This high resolution enables measurement of the OH absorption profile (separating the doublets), and the centroids and profiles of the rotational H$_2$ and fine-structure lines. To understand the role of feedback in a range of galaxies at z < 4, 100 galaxies from the blind survey are observed at low resolution for 5 hours each to detect faint blueshifted OH absorption and/or broad emission, and an additional 100 galaxies with strong signatures are observed at high resolution (R~3000) for 5 hours each to measure accurate emission line profiles and line ratios as a function of velocity.

### 1.1.5.5 The Legacy of Origins Extragalactic Surveys

The OSS and FIP surveys outlined here produce an unprecedented and unbiased sample of star forming and active galaxies in the infrared. Although they are designed to achieve the science objectives described in **Sections 1.1.2-1.1.4**, they also provide a pool of targets for follow-up with high-resolution OSS observations. The combination of *Origins'* surveys with other wide-area surveys at shorter wavelengths (*e.g.*, LSST or WFIRST) provide measurements of obscured star formation and black hole growth rates, and ISM dust masses (*Origins*), and rest-frame UV emission and stellar mass (LSST/WFIRST) for billions of galaxies. Spatially overlapping observations with JWST, ELT and *Origins* provide measurements of stellar masses as well as star formation histories in addition to SFRs, BHAR, and ISM properties. By creating the first unbiased, 3D maps of the Universe in the infrared, *Origins* creates a spectroscopic database even more powerful than that of the Sloan Digital Sky Survey -- with accurate redshifts, star formation rates, black hole accretion rates, metallicities, and energetic outflow masses for a large sample of galaxies over most of cosmic time.

## 1.2 How do the conditions for habitability develop during the process of planet formation?

Water is essential to life, yet we do not know how the Earth obtained its water, or whether water-rich terrestrial planets are rare or common. *Origins* will measure the mass of gas available to form planets, reveal the trail of water from the earliest stages of planet formation, and test comet populations in the solar system as potential reservoirs for the Earth's oceans.

### 1.2.1 Introduction

With its superlative sensitivity and spectral resolving power, *Origins* follows the trail of water in up to a thousand protoplanetary systems and measures their masses to provide a foundational perspective from which to understand the origins of habitable planets. Water is critical for the emergence and evolution of life on Earth, which also depends on carbon, nitrogen, and other minor elements (Chopra & Lineweaver, 2010). Understanding how the ingredients for life are delivered to exoplanets requires new observations of the water distribution in various stages of the planet formation process. By tracking the abundance and phase (solid or gas) of water throughout this process, we can observe the flow of volatile elements toward their ultimate incorporation into potential biospheres (Marty, 2012). *Origins* will enable us to understand this trail of the water.

Water is not only life-enabling, it is also thought to play a critical role in the growth of planets. Recent Atacama Large Millimeter Array (ALMA) images of intricate dust structures (Andrews *et al.*, 2018) provide dramatic evidence of the importance of the physics of solids in protoplanetary disks. As solid cores are needed to form giant planets, the presence of a massive ice reservoir may profoundly affect the architecture of exoplanetary systems (Ida & Lin, 2004; Raymond *et al.*, 2004) and the prevalence of water worlds (Zeng *et al.*, 2018). The dust seen by ALMA is likely icy, with a mass and volume dominated by water ice, in which case planetesimal formation may only occur beyond the water snowline (Drazkowska & Alibert, 2017). Thus, understanding planet formation requires observations of water ice and gas in protoplanetary disks.

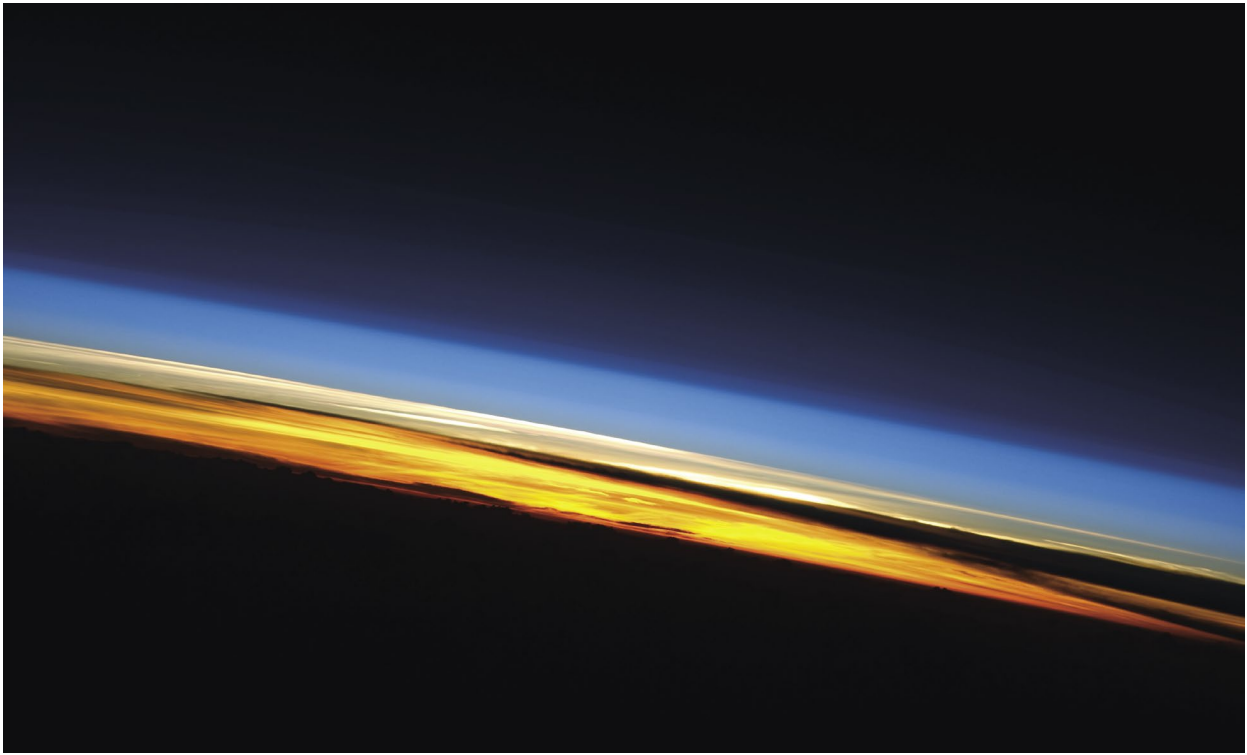

**Figure 1-24:** Earth's atmosphere — seen here in a photograph taken from the International Space Station — is mostly comprised of volatile elements delivered to the Earth, likely after its initial formation. This delivery process started in the interstellar medium, and was shaped by chemistry and dynamics of water and other volatiles in the solar nebula. *Origins* tracks the delivery of water and volatile elements to planets and addresses how the conditions for habitability develop during the process of planet formation.

**Table 1-10:** Galactic Objectives

| | | |
|---|---|---|
| NASA Science Goal | How did we get here? | |
| *Origins* Science Goal | How do the conditions for habitability develop during planet formation? | |
| *Origins* Scientific Capability | Using sensitive and high-resolution far-IR spectroscopy, *Origins* will map the water trail in our Galaxy. | |
| **Scientific Objectives Leading to Mission and Instrumental Requirements** | **Objective Goal** | **Technical Statement** |
| | **Objective 1**: What role does water play in the formation and evolution of habitable planets? | Measure the water abundance at all evolutionary stages of star and planet formation and across the range of stellar masses, tracing water vapor and ice at all temperatures between 10 and 1000 K. |
| | **Objective 2**: How and when do planets form? | Determine the ability of planet-forming disks at all evolutionary stages and around stars of all masses to form planets with masses as low as one Neptune mass using the HD 1−0 line to measure the total disk gas mass. |
| | **Objective 3**: How were water and life's ingredients delivered to Earth and to exoplanets? | Determine the cometary contribution to Earth's water by measuring the D/H ratio in over 100 comets in 5 years. Measure the volatile content in exocomets using [OI] and [CII] in debris disks. |

Water is intimately linked to the composition of exoplanet atmospheres. As the dominant carrier of oxygen, it is a driver of the relative partition of carbon and oxygen between the gas and solid phases, and the C/O ratio. The C/O ratio is used to tie planet and disk composition, spurring a number of groups to derive this ratio in planetary atmospheres (*e.g.*, Madhusudhan, 2012; Konopacky *et al.*, 2013; Line *et al.*, 2014; Dawson & Johnson, 2018) and protoplanetary disks (*e.g.*, Kama *et al.*, 2016; Bergin *et al.*, 2016; Cleeves *et al.*, 2018). Giant planets have C/O ratios set by their birth location and they may, depending on core-envelope mixing and planetesimal accretion, carry this information to later stages (Oberg *et al.*, 2011). While it is thought that many chemical signatures seen in primordial Solar System materials originate in the gas-rich protoplanetary disk phase (Busemann *et al.*, 2006; Mumma & Charnley, 2011; Simon *et al.*, 2011), giant planets, planetesimals, comets and Kuiper belt objects form in a rich chemical environment where primordial water and other volatiles undergo complex chemistry, leading to a wide range of planetary outcomes.

A protoplanetary disk's total gas mass is fundamentally important, as it affects planet formation and evolution. Gas is essential to planetesimal assembly, it determines timescales of giant planet birth, and it factors into a wide range of parameters that affect or characterize planet formation, from chemical abundances (X/H) to the mass effi-ciency of planet formation. The total gas mass is needed to measure the water abundance at all stages in the water trail. Gas masses derived from traditional tracers, such as dust thermal continuum emission and CO isotopologue line emission, are known to have significant (1 - 2 orders of magnitude) systematic uncertainty. *Origins* observes the isotopologue of $H_2$, hydrogen deuteride, HD, to derive much more precise measurements of the disk gas mass.

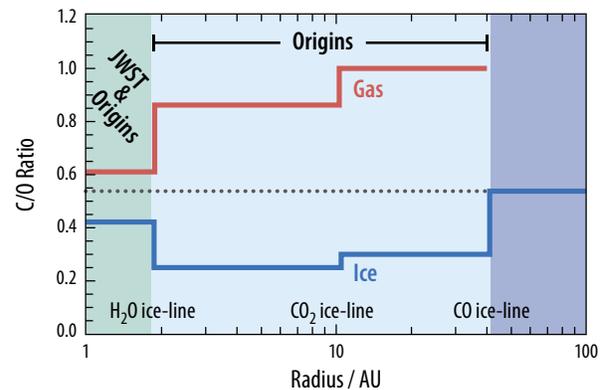

**Figure 1-25:** Chemical expectations of the elemental C/O ratio in gases and ices in a gas-rich protoplanetary disk. Figure adapted from Öberg, Murray-Clay, and Bergin (2011). CO, $CO_2$, and $H_2O$ are the known main carriers of volatile C and O (i.e., not in refractory form, such as silicate grains). In the disk, these main carriers have different sublimation fronts, with CO as the most volatile, evaporating far from the star, then $CO_2$, and finally water. In the disk, the elemental C/O of the gas and ices will shift as these volatile carriers are released from the grain at successively closer distances to the star. Comets born at different distances thus have subtle shifts in the C/O of their volatile ices, while giant planets accrete (at birth) a C/O ratio associated with that distance to the star. A migrating planet might retain the chemical signature of its birth site. This serves as one of the strongest links between composition of a gas giant atmosphere and the birth site to date. J/WST can probe $CO_2$, but *Origins* is needed to probe the oxygen carried by water beyond the water ice line. *Origins* provides access to the volatile oxygen content carried by water throughout the disk and constrains the location of the important water ice-line.

To follow the water trail, *Origins*:

- (a) traces water and other volatiles back to their nebular origins - *Origins* follows the water from the youngest protoplanetary disks to older debris disks, revealing how planetary systems form.
- (b) builds a census of water in planet-forming disks around stars of all masses - *Origins* delivers velocity-resolved spectra of water lines, tracing, for the first time, the full range of gas temperatures in thousands of planet-forming disks around stars of all masses. This will enable scientists to determine the mass and distribution of water as a function of stellar mass, and establish how, and to what degree, planets are seeded with water during their formation. *Origins* thus enables an understanding of the origins of habitable planetary conditions. *Origins* measures the O in the C/O ratio, which is the main posited link between the disk composition and that of exoplanet atmospheres.
- (c) determines the total amount of primordial planet-forming gas around stars of all masses - *Origins* uses the ground-state (J=1-0) HD line at 112 μm – a reliable proxy for molecular hydrogen ($H_2$) – to measure the total gas mass available for the formation of planets around thousands of stars of all masses. The efficiency of planet formation as a function of stellar mass is derived by folding in known exoplanet demographics. *Origins* accurately measures the abundance of water (and all other species).

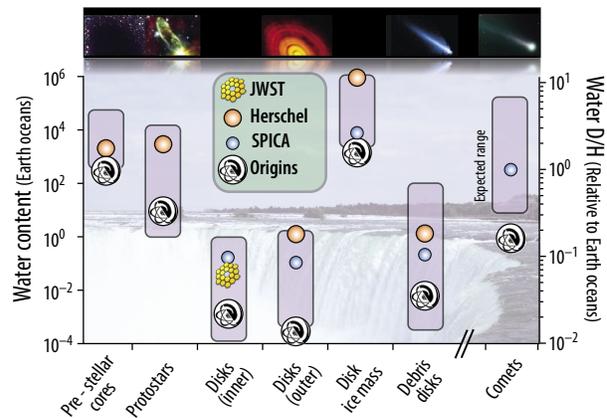

- (d) reveals the volatile content of exo-solar comets by measuring the [*OI*] and [*CII*] fine structure lines in debris disks. The source of this emission is cometary outgassing. These observations determine the volatile composition of planetesimal debris during the final stage of planet assembly.
- (e) traces the origin of water on Earth and in our Solar System by determining the D/H ratio in hundreds of comets, providing, for the first time, a statistically-significant sample of this critical fingerprint for the origin of water on Earth. *Origins* uses low-lying lines of $H_2O$ and HDO to determine the D/H ratio precisely. This large sample will set stringent constraints on cometary delivery of Earth's water, explore the heterogeneity of D/H among comet populations (Oort Cloud, Kuiper Belt), and transform our understanding of the origins of water in the Solar System.

**Figure 1-26:** Overview of the sensitivity of *Origins* to the water mass at all phases and evolutionary stages, illustrating the promise of *Origins* relative to other major facilities with access to the spectrum of water vapor. These facilities are SPICA (which is not yet accepted, and is undergoing ESA Phase A), JWST (which will launch prior to *Origins*), and Herschel (which ended operations in 2013). For each case, the maximum range of expected values is provided based on models and/or observations. *Origins*' main asset is that it is a bigger telescope than Herschel or SPICA, that it is a cool telescope operating at the far-infrared wavelengths, and that it is much more sensitive than Herschel or JWST to cold water. *Origins* also has higher spectral resolving power than JWST or SPICA, thereby allowing for greater sensitivity to spectral lines in sources with low line/continuum contrast (disks) and/or narrow spectral lines (debris disks, comets). Furthermore, as a cooled large aperature telescope, *Origins* has a factor of 1000 gain in sensitivity, enabling it to perform the first statisitically-significant surveys to capture all exoplanetary disk spectral types and evolutionary stages. Appendix E.2 provides a summary of the assumptions that provide the expectation values.

**Table 1-11:** Water Trail Requirements Flow

| Science Objective 1 |
| --- |
| Measure the water abundance at all evolutionary stages of planet formation and across the range of stellar mass tracing water vapor and ice at all temperatures between 10 and 1000 K. |

| Science Observable |
| --- |
| Integrated fluxes and resolved profiles of water emission lines for planet-forming disks within 400 pc. |

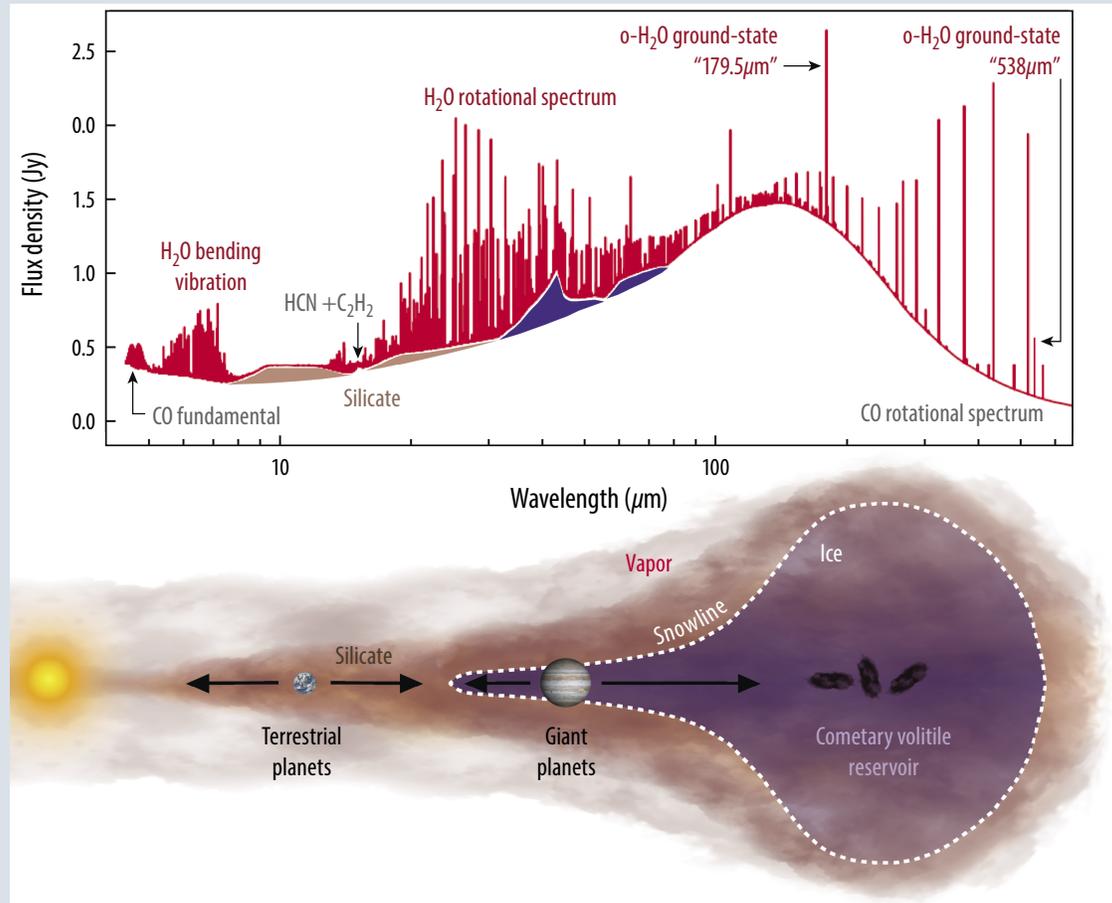

**Figure 1-27:** *Origins* **provides access to critical molecular tracers, including the HD J=1-0 line at 112 µm, and nearly the full H₂O rotational spectrum.** The wavelength range includes warm water lines between 25 and ~100 µm, and the ground state lines at 179.5 µm and 538 µm. *Origins* observers will use these tracers to quantify the gas mass and the location of water in planet-forming disks. The model protoplanetary disk spectrum above is based on Blevins *et al.* 2016, rendered at 4-660 µm and at a uniform 6 km s⁻¹ spectral resolving power, and for a disk distance of 125 pc. Previous spectroscopic observations of disks with *Spitzer* and *Herschel* were under-resolved, and therefore did not show the dramatic line-to-continuum ratios to be revealed by *Origins*.

Also shown is a schematic of the different water regions in a planet-forming disk. The main regions include inner disk warm water vapor, midplane ice, and outer disk cold (photo-evaporated) water vapor. *Origins* will probe the water and gas mass content throughout the disk. (credit: K. Pontoppidan & M. McClure).

| Science Requirements |
| --- |
| To quantify the water content in planet-forming disks, *Origins* must be able to perform high-resolution spectroscopy of water lines between 25 and 580 µm with a line sensitivity of 5x10⁻²¹ W/m² in 1 hour (5σ). The minimum required spectral resolving power is 25,000. Line tomography of water in nearby disks requires spectral resolving powers of at least 200,000 at 179.5 µm and a line sensitivity of 1.5x10⁻¹⁹ W/m² (5σ). |

To illustrate the transformative power of *Origins*, Figure 1-26 shows its mass sensitivity as a function of the expected range of water vapor/ice mass and evolutionary stage. Table 1-10 summarizes the primary science objectives related to the goal to understand the development of habitable conditions. These objectives and the measurements required to accomplish them are described in detail in the following sections.

**Table 1-12:** Summary of key Technical Requirements for *Origins*

| Technical Parameter | Requirement | Expected Performance | Key Scientific Capability |
|---|---|---|---|
| Maximum Wavelength | >540 μm | 590 μm | Ability to detect $H_2O$ $1_{10}-1_{01}$, the coldest water in the most massive disks, at 538.3 μm |
| Aperture Temperature | < 6K | 4.5K | Sufficient sensitivity at 179.5 and 538 μm |
| Line Flux Sensitivity at HD 1−0 112 μm and H20 179.5 μm | $<10^{-20}$ W m$^{-2}$ (5σ, 1hr) | $3 \times 10^{-21}$ W m$^{-2}$ (5σ, 1 hr) | Ability to detect a gas mass down to a Neptune- mass around a solar-mass star. |
| Spectral Resolving Power | 43,000 at 112 μm; 200,000 at 179.5 μm | Line-tomographic imaging of cold water out to 100 AU. | Line-tomographic imaging of cold water out to 100 AU. |

## 1.2.2 Science Objective 1: Water's Role in the Formation and Evolution of Habitable Planets

> The water and volatile content of forming planets relative to their total mass is a critical parameter for habitability. *Origins* will probe the initial conditions for volatile delivery to planets through its unique capability to trace the water during all phases of planet development.

### 1.2.2.1 Tracing the Origins of Life's Ingredients During Planet Formation

The total amount of water in different disk reservoirs – inner, outer, midplane, and surface – is intimately linked to the ability of the disk to provide planetary surfaces with ingredients critical for life (Ciesla & Cuzzi, 2006; Pontoppidan *et al.*, 2014; Bergin *et al.*, 2015). To what extent is the planet-forming material in these reservoirs modified through chemical and dynamical processes, or stellar illumination? And is planet formation, like real estate, all about location? Large uncertainties in existing water and volatile abundances in planet-forming disks hinder our understanding of the pertinent astrophysical processes and the development of habitable conditions.

In the dense interstellar medium water ice forms efficiently on dust grain surfaces at very low temperatures (*i.e.*, ~10 K), reaching abundances of as much as $10^{-4}$ per hydrogen (Boogert *et al.*, 2015). If water is present beyond the snowline at interstellar abundances, then ice mass will be a driver of giant planet formation (Ida & Lin, 2004).

The inner solar system exhibits strong evidence for energetic processing of the planet-forming material, processing potentially strong enough to erase the chemical signature of the interstellar medium (*e.g.*, Connelly *et al.*, 2012).

Water can also be concentrated in specific regions of the disk, in particular around the snowline, through the action of advection (the transport of small solid bodies relative to the gas), and turbulent mixing coupled with freeze-out (the "cold-finger" effect) (Stevenson & Lunine, 1988). These processes can act to enhance the local abundances of ices by several orders of magnitude (Dodson-Robinson *et al.*, 2009; Schoonenberg & Ormel, 2017), potentially changing the carbon-to-oxygen elemental abundance ratio (Öberg *et al.*, 2011), or even catalyzing the formation of planetesimals by gravitational collapse of large concentrations of solids (Ros & Johansen, 2013).

> ### Important Definitions
> **We adopt the following definitions throughout this section**
>
> - **Snowline:** The location in a planet-forming disk where water transitions from vapor to ice.
> - **D/H Ratio:** Ratio of HDO to $H_2O$ measured in comets and planetary bodies.
> - **VSMOW:** Vienna Mean Stanard Ocean Water is a standard benchmark for D/H measurements based on Earth's water.
> - **Unit of Earth Oceans:** a mass standard based on Earth's total water mass, $1.6 \times 10^{24}$ g.

To understand the relative importance of these various processes, it is necessary to survey the water reservoirs, with accurate mass or abundance measurements, in each of many disks. Doing so will provide a true constraint on the O in the C/O ratio for gas and ice reservoirs (*e.g.*, Figure 1-25). *Origins* is the only planned observatory that can set the initial disk composition for comparison to the composition of exoplanetary atmospheres and answer fundamental questions about the role water plays in the formation of terrestrial and giant planets, whether planet-forming chemistry is inherited from the interstellar medium, and the effectiveness of disks in transporting volatiles to the surfaces of warm planets in the habitable zone.

The water trail begins in the interstellar medium and ends with the formation of an ocean on the surface of a habitable planet. *Origins* brings unique and essential information especially to the later stages of habitable planet development, beginning with the protoplanetary disk phase.

### 1.2.2.2 Multi-phase Water in a Planet-forming Disk

A planet-forming, or protoplanetary, disk is defined as the stage during which giant planets fully form, as well as the embryos of terrestrial planets (Williams & Cieza, 2011). During the protoplanetary period, the disk is also thought form a swarm of planetesimals, which evolve into asteroid belts, Kuiper belts, Oort clouds, and other populations (Blum & Wurm, 2008). Planet-forming disks are rich in gas, and the motion of their dust is dictated by interactions with the gas. Surveys of warm dust in young clusters of different ages indicate that the lifetime of the giant planet-forming stage is a few to 10 Myr (Haisch *et al.*, 2001).

Physically and chemically, planet-forming disks have highly heterogeneous internal structures (Dullemond *et al.*, 2014). Disks can generally be divided into distinct vertical and radial regions (Figure 1-27). In the vertical direction, disk surfaces are externally heated by light (X-rays, UV, infrared) from the central star. Due to its high temperature and exposure to high-energy radiation, the surface has a distinct chemistry (Glassgold *et al.*, 2004; Woitke *et al.*, 2009). Extremely young disks with accretion rates in excess of $10^{-6}$ $M_\odot$/yr are dominated instead by internal dissipation of accretion energy (d'Alessio *et al.*, 1999; Kamp *et al.*, 2003). The gas and dust temperatures are much lower in the midplane. The typical midplane is likely characterized by a dense, but vertically thin, disk of large (1-10 mm) ice+dust grains that have settled from higher elevations (Dullemond & Dominik, 2005; Pinte *et al.*, 2016). Detailed images of protoplanetary disks based on ALMA observations show dust midplanes with rings, spirals, and other asymmetric structures (van der Marel *et al.*, 2013; Perez *et al.*, 2014; Flock *et al.*, 2015).

In the radial direction, the inner disk is defined as the region interior to the mid-plane water snow line. The water snow line is roughly the disk radius where the dust has a temperature of ~150 K, corresponding to the gas-solid boundary (Lecar *et al.*, 2006; Min *et al.*, 2011). Depending largely on the stellar and accretion luminosity, the mid-plane snowline is typically located at a stellar-centric radius of 0.5-5 AU. Inside the snowline, all water is in the gas phase, whereas outside the snow line, nearly all water is sequestered as ice. In the outer disk, only a tenuous gas-phase water component is maintained by non-thermal desorption, such as cosmic ray impacts or ultra-violet photo-desorption (Hogerheijde *et al.*, 2011).

Since the inner disk surface is superheated, water vapor there reaches temperatures of 300-1500 K (Woitke *et al.*, 2009). Gas at these temperatures is traced in the mid-infrared. Warm (>300 K) water vapor has been detected in many inner disks by *Spitzer* (Pontoppidan *et al.*, 2010a; Carr & Najita 2011) and *Herschel* (Riviere-Marichalar *et al.*, 2012), and will be explored in greater detail with JWST. Some of this warm water line emission can even be seen with ground-based telescopes or from the stratosphere (Pontoppidan *et al.*, 2010b; Salyk *et al.*, 2015). At these inner disk radii, the dust optical depth toward the midplane is much greater than 1 at all wavelengths, including the ALMA bands, for a minimum mass solar nebula (van der Marel *et al.*, 2016). Similar to the midplane, the disk surface also has a characteristic radius where water transitions from vapor to ice, but because of external heating it is pushed to significantly larger radii than those of the midplane snowline (Meijerink *et al.*, 2009). This surface snowline is typically located at 2-20 AU, depending on the stellar luminosity. Beyond the surface snowline, most water is frozen onto dust grains at all disk elevations. Whereas the midplane snowline is often hidden by optically thick dust, the surface snowline at higher elevations is generally observable. Water lines tracing gas in the critical region between the midplane and the surface snowline at $100 - 300$ K can only be seen at wavelengths beyond ~30 μm, and are thus inaccessible to JWST. Between 30 and 120 μm, dozens of strong water lines trace this gas (Blevins *et al.*, 2016; Banzatti *et al.*, 2017).

Compared to the multitude of water lines tracing warm gas, only a handful of lines trace cold gas. Interior to the surface snowline, cold water vapor at temperatures of 10-100 K is traced by the rotational ground state lines of water (Hogerheijde *et al.*, 2011), such as the well-known ortho 538 μm line ($E_{upper}$=61 K) and the ortho 179.5 μm line ($E_{upper}$=114.4 K) (van Dishoeck *et al.*, 2011). Each of these lines has equivalent para lines, as well as important isotopologue counterparts (*e.g.*, $H_2^{18}O$, HDO). *Origins* will observe the cold and warm gaseous water in protoplanetary disks.

### 1.2.2.3 Water Ice Mass in Planet-forming Disks

*Origins* will be able to measure the total water mass, including water in both gaseous and solid phases. At least half of the solid mass incorporated into planetesimals is thought to be found in the disk midplane in the form of water ice. In the solar nebula, the ratio of water ice mass to the mass of silicates and refractory carbon beyond the snow line (the "ice/rock ratio") was likely at least unity (Lodders *et al.*, 2003; Desch, 2007).

The primary tracer of water ice in disks is the solid-state band in the 43-47 μm region (hereafter the "43 μm feature"), which appears in emission and therefore directly traces the full ice reservoir (Kamp *et al.*, 2018). Many ice features exist at shorter, mid-infrared wavelengths, such as the 3 μm asymmetric stretch feature, but they can only be seen in absorption (*e.g.*, in edge-on disks) since dust grains warm enough to excite them into emission are also much too hot to retain ice. Absorption bands are not reliable bulk mass tracers.

The position and shape of the 43 μm feature is strongly dependent on ice phase (amorphous versus crystalline) and ice temperature, with a center closer to 43 μm for highly crystalline ice (Smith *et al.*, 1994). The 43 μm feature has not been observable by any facility since the Infrared Space Observatory (ISO) mission (Malfait *et al.*, 1998). Another, similar, ice feature centered around 62 μm is wider, shallower, and much more difficult to discern against the bright infrared dust continuum. It was detected in only a few cases with *Herschel*. Using these data, the total ice/rock ratio was estimated to 0.5-4.0, but remains uncertain (McClure *et al.*, 2015; Min *et al.*, 2016). Figure 1-28 shows disk model spectra with different ice/rock ratios compared to the 43 μm ice feature, as observed by ISO.

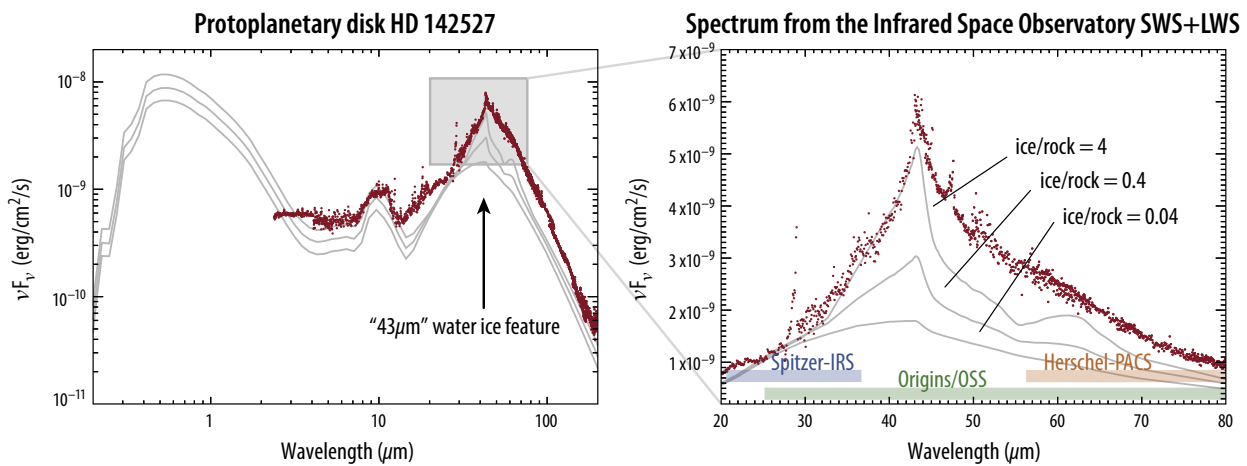

**Figure 1-28:** Disk models with crystalline water ice compared to the Infrared Space Observatory (ISO) spectrum of the planet-forming disk HD 142527. The solar nebula is thought to have had an ice/rock ratio of 1–2, and the ISO spectrum indicates a similar ratio. *Origins* is sensitive to much lower ice masses and much lower-luminosity disks and will observe the 43 and 62 μm water ice features in every disk surveyed for gas-phase water.

**Table 1-13:** Example water and HD transitions for a typical planet-forming disk around a solar-mass young star.

| Transition | Wavelength (μm) | $E_{upper}$ (K) | Strength (W/m² 1 M☉ @ 400 pc) | Width (FWHM km/s) | Line-to- Continuum Ratio |
|---|---|---|---|---|---|
| o-$H_2O$ $7_{52}$–$6_{43}$ | 32.99 | 1524.8 | 6.9 10⁻¹⁸ | 16.7 | 3.4 |
| p-$H_2O$ $6_{60}$–$5_{51}$ | 33.01 | 1503.6 | 6.9 10⁻¹⁸ | 16.7 | 3.4 |
| o-$H_2O$ $5_{32}$–$5_{05}$ | 54.51 | 732.1 | 2.3 10⁻¹⁸ | 15.7 | 2.2 |
| o-$H_2O$ $3_{21}$–$2_{12}$ | 75.38 | 305.2 | 1.1 10⁻¹⁸ | 15.4 | 1.6 |
| o-$H_2O$ $2_{12}$–$1_{01}$ | 179.53 | 114.4 | 2.8 10⁻¹⁹ | 6.4 | 2.4 |
| o-$H_2O$ $1_{10}$–$1_{01}$ | 538.29 | 61.0 | 2.3 10⁻²⁰ | 6.5 | 2.5 |
| HD 1–0 | 112 | 128.5 | 1.9 10⁻²⁰ | 7.7 | 0.1 |

### 1.2.2.4 The Far-infrared Molecular Spectrum of a Planet-forming Disk

Due to the expected ubiquity of water in typical planet-forming disks around solar-mass stars, their infrared spectra (3-600 μm) are likewise expected to be rich in emission lines from water and many other light (*i.e.*, 2-4 atoms) molecular species (Pontoppidan *et al.*, 2010a; Carr & Najita, 2011; Riviere-Marichalar *et al.*, 2012; Fedele *et al.*, 2013; Pascucci *et al.*, 2013; Blevins *et al.*, 2016; Notsu *et al.*, 2016). By comparison, rotational transitions from heavier species with >4-5 atoms tend to appear in the submillimeter range, and are accessible to ALMA (Walsh *et al.*, 2014; Bergner *et al.*, 2018). Figure 1-27 shows a model spectrum of a typical planet-forming disk around a solar-type star at the distance of Ophiuchus (125 pc). While the model is rendered at high spectral resolving power (R=50,000), the water line strengths match those observed by *Spitzer* and *Herschel* at low spectral resolving power. Table 1-13 shows example line fluxes for the representative disk at a distance of Orion (400 pc).

Although typical water abundances in disks around solar-mass stars appear to be consistent with dense cloud values ($[H_2O/H]$~10⁻⁴), *Spitzer* and *Herschel* data also suggest that strong abundance variations may exist for different stellar masses and evolutionary stages (Pontoppidan *et al.*, 2010b; Fedele *et al.*, 2011; Najita *et al.*, 2013). For instance, if the chemical signature of the interstellar medium is erased by chemical processing, water abundances at temperatures below 300 K could be as low as $[H_2O/H]$~10⁻⁶. Such a low abundance corresponds to a water vapor mass as low as 10⁻³ Earth Oceans in the inner disk around a solar-mass star.

### 1.2.2.5 Requirements for Tracing Water in Planet-forming Disks

To understand the trail of water in planet-forming disks, *Origins* must be capable of measuring the mass and distribution of every water reservoir: cold and warm gas, as well as ice in disks at all gas-rich evolutionary phases, and around stars of all masses. In particular, water around lower mass stars must be observed to understand the delivery of volatiles to potentially habitable planets around the M dwarfs. *Origins* will characterize such exoplanets via transit and emission spectroscopy.

**Sample Size:** *Origins* will be capable of detecting all water reservoirs in ~1000 disks. The sample size is driven by a need to place the Solar Nebula into a broader Galactic context. Is water equally common in disks around low- and high-mass stars, and near young OB star clusters? Is our solar system chemically typical or unusual (Pontoppidan *et al.*, 2014; Bergin *et al.*, 2015)?

To answer these questions requires a sample spanning two axes of parameter space: (1) disk dust mass as a proxy for evolutionary stage; and, (2) stellar mass. Dividing each axis into ten logarithmic bins and defining the "minimum statistical sample" per bin as ten disks per bin leads to 10 disk masses x 10 stellar masses x 10 disks per bin = 1000 disks. *Origins* will measure the disk gas mass, while the dust mass for a large sample of disks will likely be well established by 2035 from ALMA observations. Estimates from Evans *et al.* (2009) and others suggest 1000 disks around stars of all masses are not available within the nearest star-forming clusters at ~200 pc. However, the nearest massive star-forming cluster in Orion is known to contain ~3000 protoplanetary disks spanning the full range of stellar masses (Megeath *et al.*, 2012). Consequently, the team selected Orion at a distance of ~400 pc to define the *Origins* reference science program for protoplanetary disk observations.

**Spectral Resolving Power:** The *Origins* planet-forming disk survey must be conducted with spectral resolving powers high enough to satisfy two distinct requirements: (1) maximize the line-to-continuum contrast for water lines tracing gas temperatures between 10 and 1000 K; and (2) resolve selected lines to enable retrieval of tomographic line images (*i.e.,* inversion of the line profile to derive a radial intensity profile assuming a disk in Keplerian rotation; Section 1.2.6.3). The second requirement is defined as the ability to distinguish the double-peaked line profile from a Keplerian disk for the lines of interest.

Figure 1-29 shows how the required resolving powers are defined by illustrating the widths (FWHM) of all water lines in the $25 - 600$ μm wavelength range for a typical disk around a solar-mass star, viewed at an inclination angle of 45 degrees. For a typical disk, the snow line at 1-5 AU is traced by moderately "warm" lines with widths of 10-20 km/s, whereas "hotter" lines may have widths as high as 50 km/s or more. Figure 1-29 also shows that a few lines are significantly narrower, such as the 179.5-μm water ground state line and the HD J=1-0 line at 112 μm. These critical lines trace cold gas beyond the snow line. Matching the minimum resolving power to their typical width maximizes the sensitivity of *Origins* to their detection against the bright dust continuum from the disk. This requirement drives the minimum resolving power to R~45,000 at 112 μm, and R~25,000 at 179.5 μm for the full spectral survey of planet-forming disks. This defines the OSS instrument's Fourier Transform Spectrometer (FTS) intermediate resolution mode.

**Line Sensitivity:** To achieve the science objective, *Origins* must be able to detect all of the relevant water lines, not only in disks around solar-mass stars, but also in disks around the much more common, and much dimmer, low-mass stars, which drive the line sensitivity requirements. For a

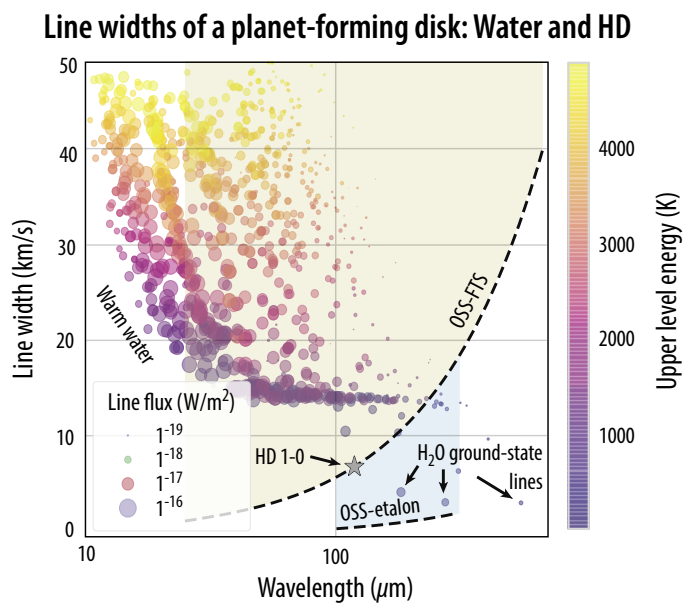

### Line widths of a planet-forming disk: Water and HD

**Figure 1-29:** The resolving power requirement for the OSS-FTS mode is defined by the need to optimize the line-to-continuum ratio for the water ground-state line at 179.5 μm, and the HD J=1-0 line at 112 μm. The resolving power offered by the high-resolution (etalon) mode of OSS is defined by the need to conduct line-tomographic imaging on the water 179 μm and HD 112 μm lines, defined as the ability to separate the two peaks in a double-peaked line formed by a Keplerian disk observed at a 45 degree inclination angle. For cold water, R ~ 200,000 is needed to resolve the 179 μm line.

given abundance, the integrated water line intensity roughly scales with stellar luminosity (Antonellini *et al.*, 2016). Since the water abundances are unknown for most types of disks, we estimate the line flux from a minimum-mass solar nebula (~0.01 M$_\odot$) and scale by the stellar luminosity using the pre-main sequence star tracks of Siess *et al.* (2000). Blevins *et al.* (2016) showed that a typical line flux from the warm water line H$_2$O 3$_{[21]}$-2$_{[12]}$ (75.38 μm) around a solar-mass star is 1.1x10$^{-18}$ W/m$^2$ at 400 pc. The ground-state line at 179.5 μm is weaker, with a typical flux of 2.8x10$^{-19}$ W/m$^2$ (Table 1-13). The luminosity of a 0.1 M$_\odot$ star with an age of 2 Myr is ~20 times lower than that of a solar-mass star, so the luminosity-scaled line fluxes are 5x10$^{-20}$ and 1.4x10$^{-20}$ W/m$^2$, respective-ly. However, disks around lower-mass stars and more evolved disks are expected to be less massive than the minimum-mass solar nebula. Detecting cold water from a disk with ⅒ᵗʰ the mass of the minimum-mass solar nebula around a 0.25 M$_\odot$ star requires detection of a line flux as low as 5x10$^{-21}$ W/m$^2$. We require that *Origins* be able to detect such a line at the 5σ level in 1 hour (Figure 1-30).

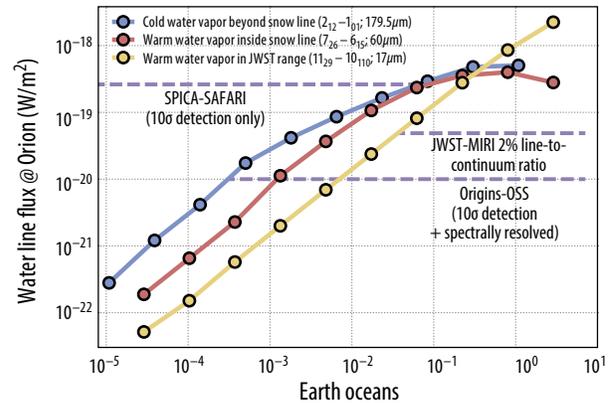

**Figure 1-30:** Dependence of water line flux on the mass of the water vapor reservoir in a disk around a solar-mass star at the distance of Orion. *Origins* is extremely sensitive to water and has the unique capability to spectrally resolve every water line in a typical disk.

### 1.2.2.6 The Last Stage: Delivery of Volatiles During the Debris Disk Phase

During the phase when terrestrial planets are born, optically-thin debris disks emerge. These disks are filled with rocky bodies, or "planetesimals," and dust debris that is continuously replenished through collisions among members of a parent body population stirred by their mutual gravitational interactions. Debris dust is readily detectable and usually discovered through its thermal infrared emission. Because the debris phase can persist for hundreds of Myr, debris disks are the best available tools to search other planetary systems for analogues to major events in the evolution of the Solar System. Such events may include the formation of terrestrial planets and the late heavy bombardments that were triggered by the orbital re-arrangement of the giant planets. Terrestrial worlds evolve after the primordial disk gas has dissipated, and may receive significant contributions of volatiles during their final assembly. The possibility that Earth received its water via impacts during the debris disk stage is a viable hypothesis. *Origins* has the transformative capability to measure the bulk volatile content of outgassing planetesimals during the main phase of terrestrial planet assembly (~10-200 Myr, Chambers, 2013) and during bombardment events (Bottke & Norman, 2017). This is essential for assessing the efficiency of water delivery and the ultimate habitability of planets.

Figure 1-31 shows the evolution of warm circumstellar dust seen with *Spitzer* at 24 μm in debris systems of various ages (Kenyon & Bromley, 2008; Booth *et al.*, 2009). The observations lend support to a model in which the amount of dusty debris gradually diminishes as young planetary systems age, but this long-term trend is interrupted by occasional giant impact events, such as that which formed the Earth's moon, leading to temporary enhancements in the dust emission by 1-2 orders of magnitude.

Volatiles can be released from icy planetesimals either through cometary break-up (Zuckerman & Song, 2012), or following impact events by UV photodesorption from small grains (Grigorieva *et al.*, 2007) or sublimation (Beust *et al.*, 1990). When released into the gas, ices (H$_2$O, CO$_2$, CO, etc.) are rapidly (in ~100 years) dissociated into neutral and ionized atoms, particularly C+ and O. While the short-lived molecules may be difficult to detect, the photodissociation products are measurable with sensitive *Origins* far-infrared spectroscopy and serve as tracers of the volatile contents of extrasolar as-teroids and comets. The [*CII*] 157 μm and [*OI*] 63 μm and 145 μm lines are particularly important, as they can be used to estimate the C/O ratio of the icy planetesimals (Figure 1-25), which currently serves as the strongest link between formation processes and the composition of mature gas giants, ice giants, and terrestrial planet atmospheres (*e.g.*, Öberg, Murray-Clay, and Bergin, 2011).

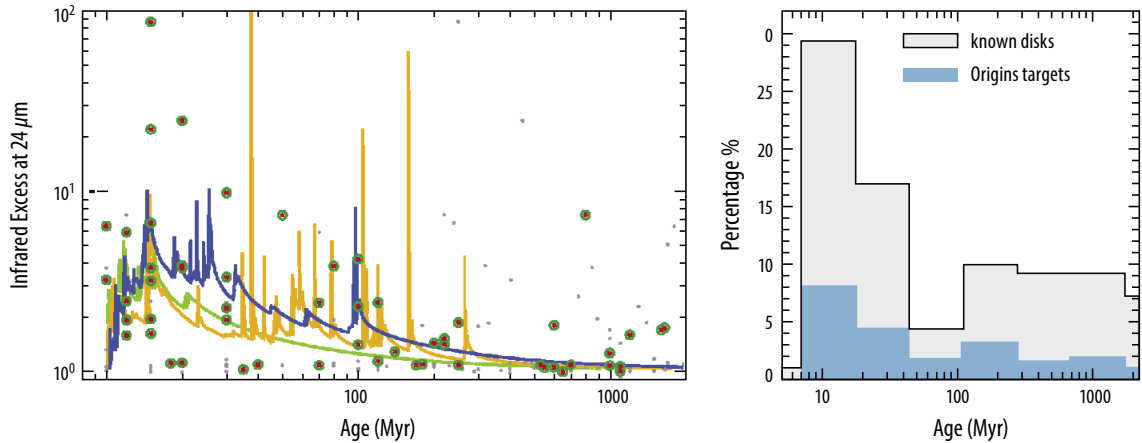

**Figure 1-31:** (Left) Dots and circles represent the amount of warm circumstellar dust as traced by the infrared excess (relative to the stellar spectrum) at 24 μm in debris systems of various ages. The three curves show expected dust levels during the era of terrestrial planet formation, with each curve representing a different realization of the same stochastic process (Genda et al., 2015). Each of the spikes represents a giant impact, leading to the injection of new debris into the system. (Right) The age distribution of the known debris systems characterized by *Spitzer* 24 μm observations (gray area), showing ~40% of the population is younger than 200 Myr. The light blue area shows a potential *Origins* sample (also shown as green/red circles in the left panel).

Limited by their sensitivity, past far-infrared telescopes were only able to detect [*OI*] and [*CII*] in a handful of mostly young and bright debris disks (see reviews by Kral *et al.*, 2017; Hughes *et al.*, 2018). Beta Pictoris is the most completely inventoried system. The C/O ratio in β Pic is thought to exceed the solar value by a factor of 18 (Roberge *et al.*, 2006). However, a detection of the [*OI*] 63-μm emission line with *Herschel* (Brandeker *et al.*, 2016) suggests a lower C/O ratio, with the excess O coming from the photodissociation of H₂O (Kral *et al.*, 2016).

The carbon and oxygen fine structure lines are inaccessible to ALMA and JWST. The limiting factor for *Herschel* is sensitivity. SOFIA also lacks the sensitivity to detect atomic gas in debris disks, but its third-generation instrument HIRMES (Richards *et al.*, 2018, expected completion in late 2020) could potentially extend the [*OI*] detections to another handful of bright, young systems, corresponding to a narrow age range that misses the entire terrestrial planet formation and bombardment phases. The SPICA mission, if selected, could observe relatively bright targets and accomplish a portion of *Origins*' science goals (Kral *et al.*, 2017, Figures 7 and 10). However, its SAFARI instrument would severely under-resolve the lines (Figure 1-32), and they would be overwhelmed by noise from the continuum.

With its high sensitivity and spectral resolving power, *Origins* can detect second-generation gas in debris disks with masses almost two orders of magnitude lower than those detected by *Herschel*. *Origins* spatially resolves the distribution of C and O in nearby systems to distinguish between giant collisions (predicting an asymmetric distribution; Cataldi *et al.*, 2018) and outgassing (expecting an axi-symmetric distribution; Kral *et al.*, 2016). Finally, the distribution and kinematics of atomic gas in debris systems is a powerful tracer of the presence of unseen planets, and complementary to sensitive molecular CO observations by ALMA.

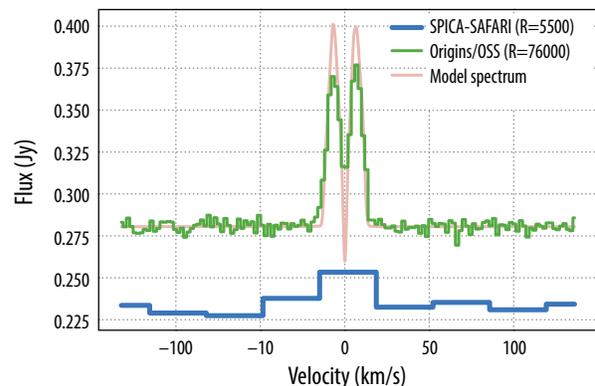

**Figure 1-32:** Spectral resolution is important for detecting fine-structure far-infrared lines in debris disks, as seen in simulated spectra of the [OI] line using a model for Beta Pic, but shifted to a larger distance of 150 pc [Kral et al. 2016]. The noise level is relevant for a 1-hour observation with both *Origins*/OSS and SPICA/SAFARI.

### 1.2.2.7 Requirements for Tracing Volatiles in Debris Disks

**Sample Selection:** Exoplanetary systems of a given age show a wide mass range of dust debris (Holland *et al.*, 2017; Figure 1-31). Therefore, it is essential to draw a sample from known debris disks around a wide range of stellar ages and spectral types. Kral *et al.* (2017) predict the expected line fluxes for ~200 debris disks by assuming that volatiles are collisionally released by a planetesimal population constrained by the dust content measured in the infrared. These models indicate that *Origins* can detect [*OI*] and [*CII*] lines for systems beyond 30 pc within one hour (at 5σ). While most of the debris disks within 30 pc are old, and therefore relatively faint (Kral *et al.*, 2017), a similar detection level can be obtained for the brighter systems (*i.e.*, those younger than 200 Myr). Figure 1-31compares the age distribution of the *Origins* sample (107 systems) to that of the known debris disks. To fully sample the gas content in debris disks, the *Origins* sample comprises ~30-40% of the known debris disks younger than 200 Myr and ~20% of the older disks. This distribution was chosen to better sample earlier stages, which are thought to have a higher frequency of giant impacts. Roughly 25% of the *Origins* debris gas sample has a spatially resolved disk. For these disks *Origins* can employ spectro-imaging to distinguish between giant impacts and outgassing and search for the influence of unseen planets. In the nominal debris disk survey, *Origins* can measure, or sensitively constrain, the water mass in over 100 systems spanning the estimated timescale of the formation of terrestrial worlds (Figure 1-33).

### 1.2.2.8 Inheritance: Water in Protostars

Prior to stellar birth, most water is frozen onto the surfaces of micron-sized dust particles. A central question in the study of Solar System materials, and by by extension, that of all other planetary systems, is how well protoplanetary disks reflect the clouds that formed them. What portion of Earth's water originated in the interstellar medium? If the answer is "all of it," then abundant interstellar ices must be incorporated into planet-forming disks. In contrast, if significant chemical reprocessing occurs when the disk is formed and evolves, then there could be significant chemical abundance variations amongst, and within, planetary systems (see discussion in Cleeves *et al.*, 2014).

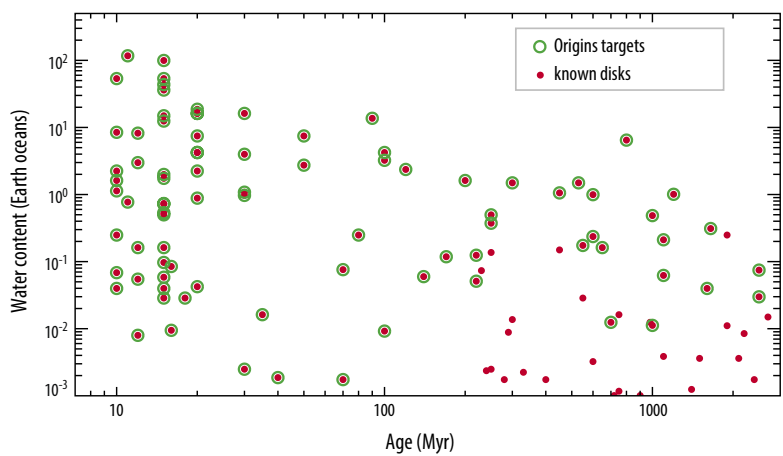

**Figure 1-33:** Expected water content or water mass for the known debris disks based on the model predictions from Kral et al. (2017). *Origins* targets are shown as green/red circles and encompass the entire age range associated with the assembly of terrestrial worlds. *Origins* will trace water via two avenues: directly via $H_2O$, which is not expected to be present due to photo-destruction, and indirectly via O I emissions, as oxygen is expected to be pre-dominantly neutral. This will trace the O originating from evaporating CO and water. *Origins* can readily detect C II and directly determine the O provided by CO. Thus, the water content can be directly estimated by *Origins'* combined C II and O I observations.

Protostars are highly energetic and exhibit a rich chemistry thought to result from the evaporation of ices (Caselli and Ceccarelli, 2012). Figure 1-34 shows a phase diagram for water that includes important chemical transitions. To reprocess or destroy abundant water ice requires energy to release it from its frozen state. Gas-phase chemistry can destroy water evaporating within a newly formed disk, freeing oxygen to react and be sequestered in other species. Hence, interstellar water may be exposed to gas-phase chemical processing during the phase of disk formation and early evolution.

*Herschel* provided a detailed census of water during this critical star formation phase (van Dishoeck *et al.*, 2014 and references therein). Many protostellar sources have an impressive water vapor spectrum extending from the mid- to far-infrared, as shown in Figure 1-35 (*e.g.*, Watson *et al.*, 2007; Herczeg *et al.*, 2012). High spectral resolving power (R > 10⁶) observations with *Herschel*/HIFI demonstrated that this emission is dominated by shocks associated with the outflowing gas (Kristensen *et al.*, 2017). For most *Herschel* observations of water during this phase, the young disk is hidden from view by optically-thick water vapor emission in the stellar outflow. The central portions of the envelope were estimated to be optically thick at far-infrared wavelengths, limiting the observational probes to trace the warmer outer portions of the infalling envelope. However, this is the initial state and is thus an important part of the water story.

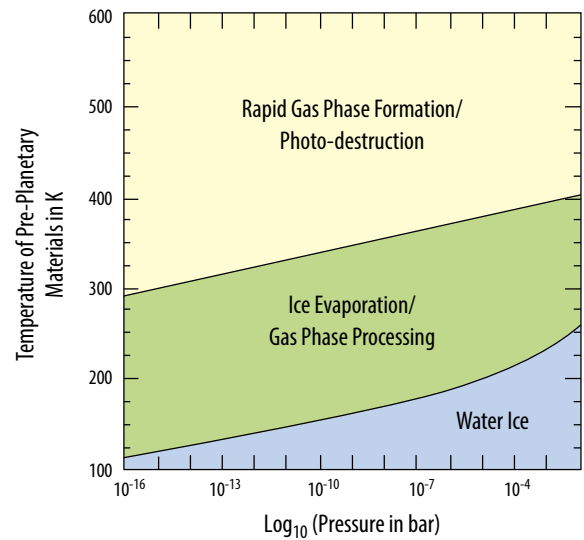

**Figure 1-34:** Water phase and chemistry diagram. The solid line shows water sublimation temperature as a function of pressure. A dense, collapsing envelope surrounding a protostar has a pressure at the low end of the range shown, while the disk midplane at 1 AU corresponds to the upper end of the pressure scale. Above the sublimation temperature, water transitions from ice to vapor, whereupon the water vapor is subject to gas phase processing. Over timescales dependent on the local ionization, this lowers water vapor abundance. Above ~400 K, hydrogen-addition reactions with oxygen raise the water vapor abundance in balance with photodestruction.

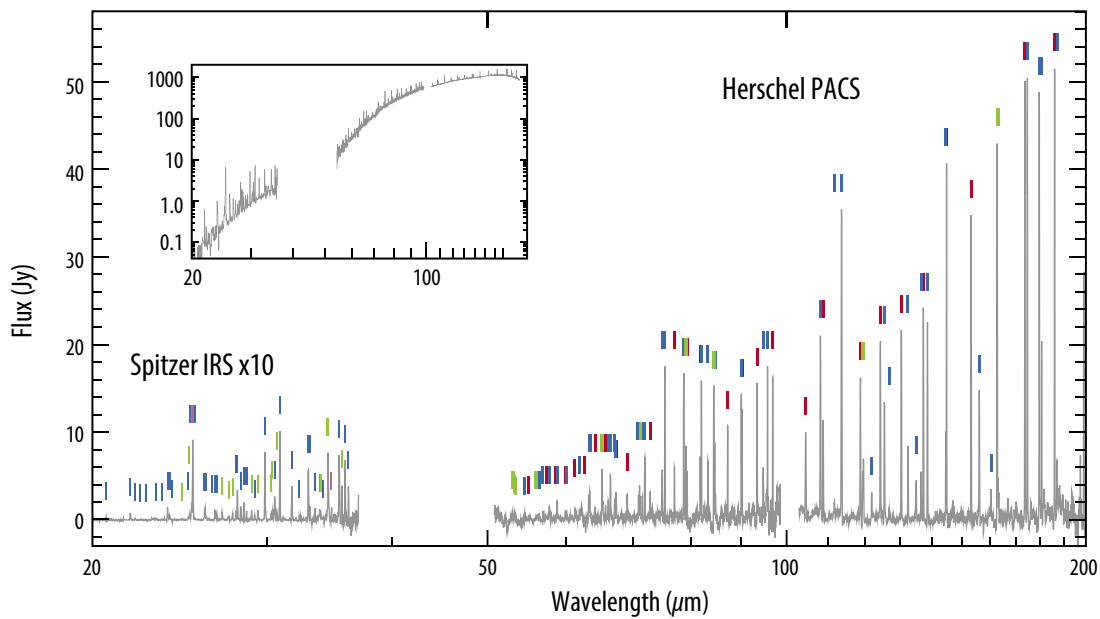

**Figure 1-35:** Herschel/PACS and *Spitzer*/IRS spectra of the rich mid- to far-infrared spectrum of a template protostar, NGC 1333 IRAS 4B (Herczeg et al., 2012). The blue dashes correspond to the water transitions, red to CO, green to OH, and purple to atomic (neutral or ionized) lines. While this spectrum illustrates the wealth of spectral lines in the far-infrared, much of this emission is associated with outflowing gas, and not the warm, infalling envelope or protostellar disk. This gas is not a central part of the water story. However, it is an important aspect of interstellar medium evolution (feedback) and is another area where *Origins* can have significant scientific impact.

Observing the initial accretion of water onto the youngest disks may not be a unique science driver for *Origins* by the mid 2030s. Section 1.2.5 outlines the ability of ALMA (and NOEMA) to observe a handful of transitions of warm $H_2^{18}O$. These lines are observable from the ground in the most emissive protostellar sources, which are much brighter than planet-forming disks. Thus, ALMA will likely observe warm water, on small (<100 AU) scales, toward protostars. However, extensive ALMA surveys of warm water will be very difficult, as they require times when the atmospheric water vapor content is very low, which occurs only under the best (and rarest) of terrestrial weather conditions. Further, ALMA will not have access to the ground state transitions tracing cold water, nor to most of the rotational spectrum of $H_2O$ and its isotopologues. With the spectral resolving power offered by the *Origins* OSS/Etalon, surveys of water isotopologue ($H_2^{18}O$ and $H_2^{17}O$) can disentangle the outflow emission and extract the contribution from the envelope. Thus, *Origins* can study the water content of the warm infalling envelope in a range of sources to complement ground-based studies, provide stringent constraints on the water excitation, and determine its abundance.

### 1.2.2.9 Requirements for Tracing Water in Protostars

The protostellar case does not rise to the level of importance that it drives *Origins* science requirements, but it complements the prioritized goal to understand the development of habitable conditions and benefits from the capabilities incorporated to satisfy other requirements.

**Sample:** The lifetime of the protostellar phase is <1 Myr (Evans *et al.*, 2009). Thus, these objects are rare compared to those in the longer-lived, less obscured planet-forming disk phase, so accumulating a representative statistical sample covering a range of evolutionary stages requires the ability to detect water emission in protostars at greater distances. The most representative sample of young protostars is from the *Spitzer* cores-to-disks program (Evans *et al.*, 2009). This sample contains 50-100 sources with a range in bolometric luminosity $L_{bol}$ = 0.1-50 $L_\odot$ (most sources have $L_{bol}$ = 0.5-5 $L_\odot$). This list can be supplemented with targets from the Gould Belt survey (Dunham *et al.*, 2015), resulting in a sample of ~100 sources in nearby (<500 pc) star-forming clouds that captures all evolutionary stages and a wide range of luminosities.

**Spectral Resolving Power:** *Herschel* spectral measurements of $H_2^{18}O$ toward low-mass protostars suggest line widths ~4 km/s (Visser *et al.*, 2013). Resolving the line therefore requires resolving powers of R > $10^5$. These lines must be at least minimally resolved to maximize the line-to-continuum ratio because protostars are surrounded by dusty envelopes with characteristic temperatures of tens of degrees Kelvin and peak continuum emission between $100 - 300$ µm. With the resolving power R ~$2 \times 10^5$ provided by the OSS etalon *Origins* will detect and characterize water in protostellar envelopes.

**Line Sensitivity:** The sensitivity of *Origins* to the water mass in the warm gas within infalling protostellar envelopes is investigated using the radiative transfer code RADLite (Pontoppidan *et al.*, 2009). The basic model is a central, solar-mass star surrounded by a compact, but massive, disk with a radius of 20 AU, and an extended infalling envelope, following a simple structure from Shu *et al.* (1977), with a total mass of 1 $M_\odot$. Sensitivity to the water mass is tested by varying the abundance of water vapor in the envelope, outside the snow line between X($H_2O$/H)=$10^{-7}$-$10^{-12}$, and assuming a fractional abundance of $H_2^{18}O$ of 1/500. The abundance inside the snowline is kept constant at $10^{-6}$. We find that *Origins* does not lose mass sensitivity by observing the 181 µm transition ($\Delta E/k$ ~ 70 K) as opposed to the line at 547 µm ($\Delta E/k$ ~ 27 K). With a focus on nearby objects to maximize spatial resolution, guest observers will exploit the superlative *Origins* sensitivity to map the rare $H_2^{18}O$ isotopologue, punching through the surrounding envelope to measure the water mass in the inner envelopes of 100 protostars down to 100 Earth Oceans in 100 hours.

### 1.2.3 Science Objective 2: How and When Do Planets Form?

> *Origins* will provide the first complete and accurate census of the planet-forming mass during all stages of planet formation, revealing the timescale for planet formation and providing a foundational reference for all other observations of planet-forming disks.

The protoplanetary disk mass is the most fundamental quantity determining whether planets can form, and on what time scale. Estimates of disk masses are complicated by the fact that molecular hydrogen ($H_2$) has no permanent dipole moment and is also the lightest molecule. This fact results in inherently-weak emission and a high $E_{upper}/k$=512 K for the ground state S(0) line, making $H_2$ unemissive at temperatures that characterize much of the disk mass (*i.e.,* 10-30 K). To counter this difficulty, the thermal continuum emission of the dust grains (Andrews & Williams, 2007), or rotational lines of CO (Williams & Best, 2014), are often used as a proxy for total mass under the assumption they can be calibrated to trace the total gas mass. However, sensitive observations have demonstrated that grains in disks undergo substantial growth, making the determination via dust inherently uncertain (Ricci *et al.*, 2010). Further, recent observations indicate that the CO abundance in disks may be orders of magnitude lower than in molecular clouds due to a combination of sequestration below the CO snow line and chemistry (Favre *et al.*, 2013; Kama *et al.*, 2016). These uncertainties are well known, and inhibit knowledge of the timescale during which gas is available to form giant planets and understanding of the dynamical evolution of the seeds of terrestrial worlds and the resulting chemical composition of planetary embryos.

#### 1.2.3.1 How Origins Uses HD to Measure the Disk Gas Mass

The fundamental (J=1-0) rotation transition of HD at 112 µm has been proposed as a much more robust measure of protoplanetary disk mass. While HD is a different molecular species, and therefore remains an indirect tracer of $H_2$, it may be as close as it is possible to get to a direct total mass tracer. Due to its simple chemistry, HD is widely thought to have an abundance closely matching that of elemental deuterium. The D abundance, in turn, is known from UV absorption spectroscopy of the local interstellar bubble (Linsky, 1998). Using *Herschel*, Bergin *et al.* (2013) accurately measured the molecular gas mass of the TW Hya disk by taking advantage of the fact that the lowest rotational transition of HD is ~10⁶ times more emissive than the lowest transition of $H_2$ for a given gas mass at 20 K. Due to *Herschel*'s limited lifetime, and its limited spectral resolving power and sensitivity, the only other deep HD observations obtained were toward six disks, with the result being two additional detections (McClure *et al.*, 2016). The utility of HD for mass measurements has been confirmed via additional independent analysis by Trapman *et al.* (2017).

**Table 1-14:** Breakdown of methods used to determine protoplanetary disk gas mass.

| Probe | Strengths | Weaknesses | Uncertainty |
|---|---|---|---|
| Dust Thermal Continuum | • Widely detectable at a range of wavelengths from cm to sub-mm via ground-based facilities.<br>• Spatially resolved with ALMA.<br>• Numerous statistical surveys exist. | • Often optically thick.<br>• Large fraction of solids likely hidden in pebbles and planetesimals.<br>• Conversion of dust mass to gas mass is uncertain.<br>• ALMA surveys find x10–100 difference between dust- estimated and CO estimated $H_2$ mass. | x10–100 |
| CO Isotopologue Emission | • Strong calibration from interstellar medium of the CO abundance to $H_2$.<br>• Isotopologues can be observed from the ground.<br>• Optically thin.<br>• Numerous statistical surveys exist. | • ISM calibration is uncertain, as CO can be chemically- processed in the gas into a less volatile form or may be locked in large grains in the disk midplane.<br>• ALMA surveys find 1-2 order of magnitude difference between dust-estimated $H_2$ mass and CO-estimated $H_2$ mass. | x10–100 |
| HD 1–0 112 µm Line | • Isotopologue of $H_2$ with similar chemical properties.<br>• Readily surveyable with *Origins*, but statistics do not exist.<br>• Can determine mass to within a factor of 2–3. | • Requires space-based platform to obtain sensitivity needed for surveys.<br>• Optically thick for most massive disks.<br>• Strong dependence on gas temperature; *Origins* simultaneously constrains temperature.<br>• Beyond local bubble D/H abundance uncertain to 30%. | x2–3 |

Table 1-14 provides a breakdown of the strengths and weaknesses of each indirect probe of the gas mass. HD has two weaknesses. First, the atomic D/H ratio is used to set the calibration between the HD and $H_2$ mass, and this varies by ~30% beyond the local bubble (Linsky *et al.*, 2007). Second, the HD gas mass has a strong dependence on the assumed gas temperature and is proportional to $e^{-128/T}$. As shown in Figure 1-27, *Origins* has access to most of the spectrum of CO and its isotopologues down to the J=5 transition. These lines are commonly used as direct probes of the gas temperature. Combined with easy access to lower-J transitions from ground-based facilities (*e.g.*, ALMA), which further constrain the temperature, *Origins* HD measurements will yield mass estimates good to within a factor of 2-3, a major advance over the orders-of-magnitude uncertainty inherent in alternative methods.

### 1.2.3.2 Requirements for Measuring Disk Masses with HD

**Line Sensitivity:** The predicted flux of the HD J=1-0 (and J=2-1) line is somewhat uncertain, given the small number of existing observations. We use the *Herschel* detections and upper limits in combination with models to produce a relatively conservative prediction. Bergin *et al.* (2015) and McClure *et al.* (2016) reported three detections and four upper limits. Most of these targets are relatively luminous and

**Table 1-15:** Disk Mass requirements Flow

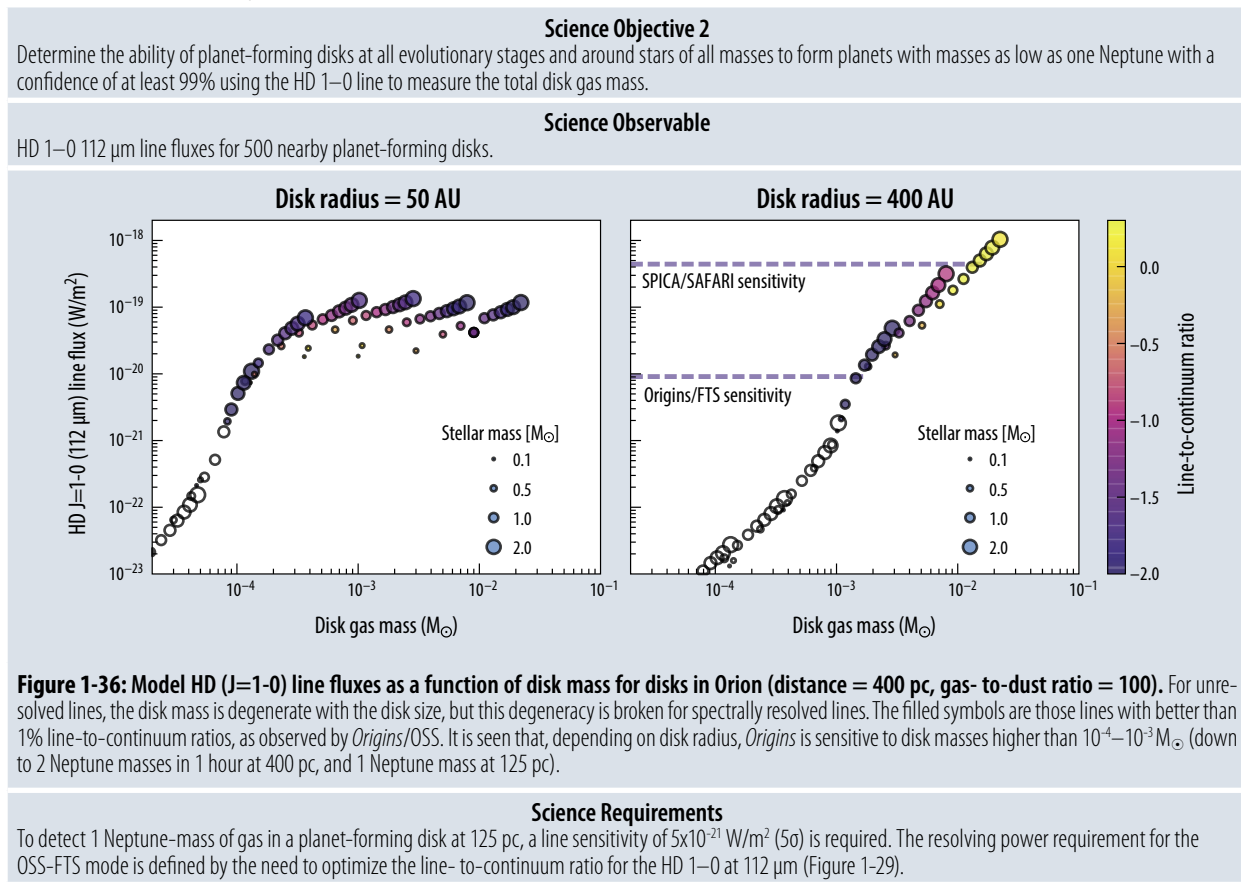

**Science Objective 2**

Determine the ability of planet-forming disks at all evolutionary stages and around stars of all masses to form planets with masses as low as one Neptune with a confidence of at least 99% using the HD 1−0 line to measure the total disk gas mass.

**Science Observable**

HD 1−0 112 μm line fluxes for 500 nearby planet-forming disks.

**Figure 1-36: Model HD (J=1-0) line fluxes as a function of disk mass for disks in Orion (distance = 400 pc, gas- to-dust ratio = 100).** For unresolved lines, the disk mass is degenerate with the disk size, but this degeneracy is broken for spectrally resolved lines. The filled symbols are those lines with better than 1% line-to-continuum ratios, as observed by *Origins*/OSS. It is seen that, depending on disk radius, *Origins* is sensitive to disk masses higher than $10^{-4}$–$10^{-3}$ M$_\odot$ (down to 2 Neptune masses in 1 hour at 400 pc, and 1 Neptune mass at 125 pc).

**Science Requirements**

To detect 1 Neptune-mass of gas in a planet-forming disk at 125 pc, a line sensitivity of 5x10⁻²¹ W/m² (5σ) is required. The resolving power requirement for the OSS-FTS mode is defined by the need to optimize the line- to-continuum ratio for the HD 1−0 at 112 μm (Figure 1-29).

are thought to contain massive disks. Trapman *et al.* (2017) estimate HD J=1-0 line fluxes consistent with the observations. For the purpose of estimating the general sensitivity of the HD lines to the disk mass, an independent grid of spectra is rendered using the team's reference disk model (Figure 1-36).

HD traces total gas mass, and therefore the reservoir used to form gas giants and smaller gas-rich planets. To constrain the ability of a disk to support the formation of such planets, a minimum requirement for an HD measurement is the ability to detect as little as 1 Neptune-mass, or $0.5 \times 10^{-4}$ $M_\odot$. The expected line flux from a small disk (less than 50 AU in radius) with this mass at a distance of 125 pc (*e.g.*, Ophiuchus or Upper Sco) and a line-to-continuum contrast of just under 1% is $5 \times 10^{-21}$ W/m$^2$. The Trapman *et al.*

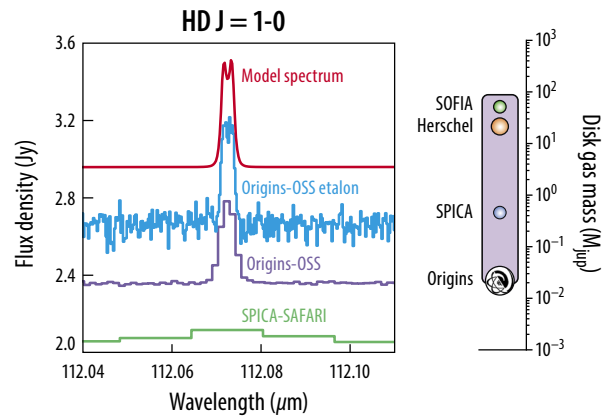

**Figure 1-37:** Comparison of the HD J=1-0 line as observed with different OSS modes and SPICA/SAFARI. SAFARI has lower resolving power than OSS, which will lead to strongly under-resolved lines and low line-to-continuum contrast. The right-hand side illustrates the improvement to the disk gas mass sensitivity offered by *Origins* compared to other missions.

(2017) models are more optimistic and predict a HD J=1-0 line flux of $2 \times 10^{-20}$ W/m$^2$ for a 200 AU disk and $10^{-5}$ $M_\odot$. Assuming a value in between defines the requirement to detect HD J=1-0 lines with fluxes of $1.0 \times 10^{-20}$ W/m$^2$. Since the line-to-continuum ratio of typical HD J=1-0 lines may be low, we allow for a sensitivity reduction due to noise from the source continuum on an unresolved line. This reduction leads to a requirement to obtain 5σ measurements of $5 \times 10^{-21}$ W/m$^2$ integrated flux in a single resolution element to detect 1 Neptune-mass of gas at 125 pc. At this level, the lowest-mass disks are observable in the nearest star-forming regions, and more massive disks are accessible at the distance of Orion.

**Resolving Power:** Detecting HD in the lowest-mass disks requires maximizing the line-to-continuum contrast, which occurs at R~40,000 at 112 µm. Higher resolving power is needed for line tomography, which is essential to derive the mass distribution and break model degeneracies. Figure 1-37 shows a HD J=1-0 line profile for different *Origins*/OSS instrument modes, and the corresponding simulated observation with SPICA/SAFARI. This demonstrates the need for R~200,000 to fully resolve the line as offered by OSS in the etalon mode. As in the case of water, it is expected that detailed line tomographic images of HD will be carried out for disks around solar-mass stars at distances of ~125 pc.

### 1.2.4 Science Objective 3: How Were Water and Life's Ingredients Delivered to Earth and to Exoplanets?

> *Origins* precisely measures the D/H ratio in ≥100 comets, building a sample that determines the cometary contribution to Earth's oceans.

The origin of Earth's water remains an unsolved puzzle. Based on a comparison between the mineralogy of Earth and the asteroid belt, it is widely thought that Earth formed interior to the midplane nebular snowline (Chyba, 1990; van Dishoeck *et al.*, 2014). Hence, embryonic Earth would have contained little to no water. Water is posited to have been delivered later by comets or small bodies from the outer parts of the asteroid belt that were dynamically perturbed, possibly by the giant planets (*e.g.*, Raymond and Cossou, 2014).

The ratio of deuterium to hydrogen (D/H) of Earth's oceans is significantly higher than that of the proto-Sun and the interstellar medium (Bockelee-Morvan *et al.*, 2012; Morbidelli *et al.*, 2000; see

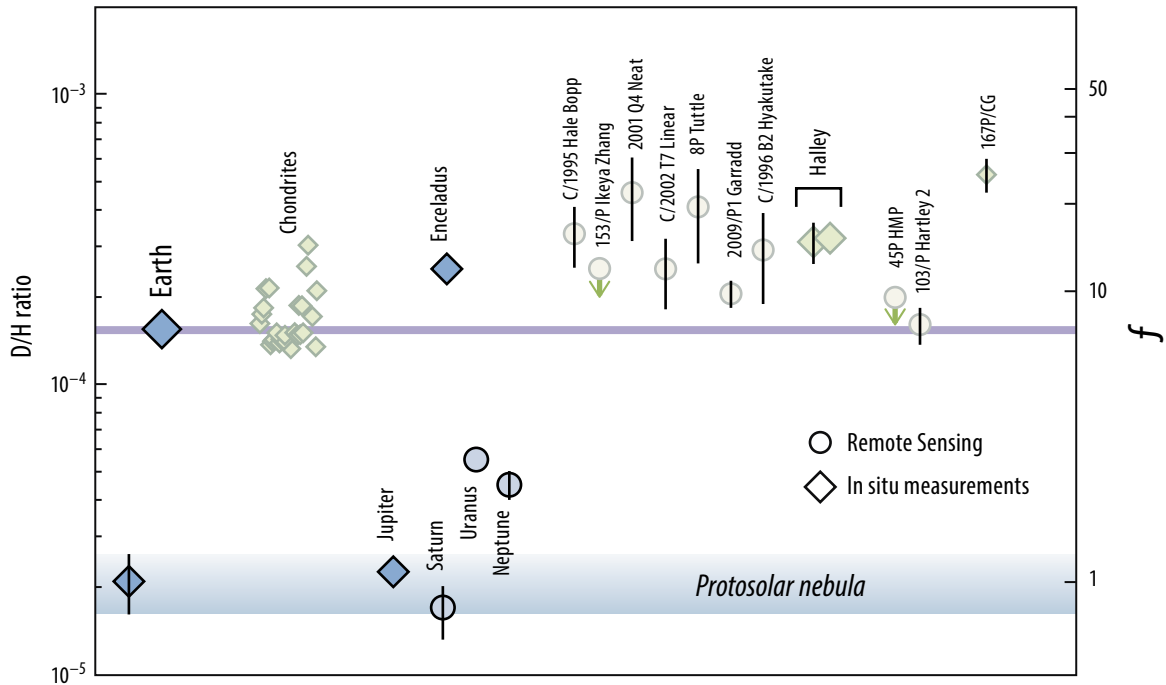

**Figure 1-38:** The current state-of-the-art in D/H measurements in the solar system (Altwegg et al., 2015) is limited by sensitive measurements of $H_2O$ and HDO. D/H ratios are shown on the left and the enrichment factor, f = $[D/H]_{object}/[D/H]_{PSN}$ (scale on the right) normalizes the D/H ratio to its protosolar nebula value. *Origins* provides precise D/H ratio statistics within the Solar System.

Figure 1-38). Similarly, cometary water, as measured in comets Halley, Hale-Bopp, and seven more, is enriched in deuterium to levels at, or above, those of the Earth (Ceccarelli *et al.*, 2014), suggesting that comets may have been an important source of Earth's water. Yet, the relative importance of a cometary carrier compared to, for example, hydrated minerals in rocky bodies (*e.g.*, asteroids traced by chondrites), remains a subject of intense debate, as many comets have D/H ratios well above that of the Earth. For example, *in situ* measurements of the Jupiter-family comet, 67P/Churyumov-Gerasimenko, made in 2014 with ESA's Rosetta mission found the elevated D/H ratio ~3.4 VSMOW (Altwegg *et al.*, 2015; Figure 1-38). Was chondritic material the primary reservoir (Morbidelli *et al.*, 2000; Marty, 2012), or did comets play a significant role?

Recently, *Herschel* observations of the Jupiter-family comet 103P/Hartley 2 (Hartogh *et al.*, 2011) and the Oort cloud comet 2009/P₁ Garradd (Bockelee-Morvan *et al.*, 2012) revealed Earth-like D/H ratios, demonstrating the existence of comets of the right composition to deliver water to the inner solar system. Jupiter-family comets are theorized to arise from the Kuiper belt while the Oort cloud formed near the ice giants.

The deuterium enrichment of the cold interstellar chemistry is reset to the protosolar nebula value (~$10^{-5}$) inside the midplane water snowline, and mixing leads to a positive radial gradient of D/H in the nebula water (*e.g.*, Robert *et al.*, 2000; Albertsson *et al.*, 2014), accounting for the fact that meteoritic material has a lower D/H ratio than Oort cloud comets (Figure 1-38). However, contrary to expectations, 103P/Hartley 2 and 2009/P₁ Garradd have similar D/H ratios (Yang, Ciesla, and Alexander, 2013). A great many more comets will have to be observed to determine if there is heterogeneity in D/H within or between the comet populations, and to trace water delivery to the inner solar system to a particular population. The migration of small bodies, the transport of volatiles, and the origin of Earth's oceans can be understood with the help of a large number of precise D/H measurements.

*Origins* simultaneously observes $H_2^{18}O$ and HDO, with the ability to constrain the ratio to fractions of the D/H ratio seen in the Earth (~$10^{-5}$ on the scale in Figure 1-38). With its superlative sensitivity, *Origins* will be able to detect comets with production rates two orders of magnitude below that of *Herschel*, and below that detectable on the ground. This capability enables a survey of hundreds of comets with sufficient precision to check for variability in D/H within each cometary population.

### 1.2.4.1 Requirements for Measuring the D/H Ratio in Comets

**Spectral Resolving Power:** The minimum required spectral resolving power for comets is driven by the expected line intensity and the line-to-continuum ratio. Figure 1-39 shows the *Herschel*/HIFI detection of HDO and $H_2^{18}O$ toward comet 103P/Hartley 2, illustrating the narrow lines with line widths of 1-2 km/s. Based on simulations, the minimum resolving power to reliably detect cometary lines is R > $10^4$. Far-infrared continuum measurements from *Herschel*/SPIRE toward C/2006 W3 Christensen (Bockelee-Morvan *et al.*, 2010) indicate that the continuum is not a major noise source at this resolution.

**Sample Size:** Given the dispersion seen in D/H from object to object in the small sample of comets measured thus far, interpretations of the existing data must be considered tentative. To date, D/H measurements of five Jupiter-family comets and ten Oort cloud comets are available, and not all of the Oort cloud comets have measurements precise enough to enable cross-comparison. The existing sample size and measurement errors provide too little information to determine whether the Oort cloud and Kuiper belt populations differ in their average D/H ratio.

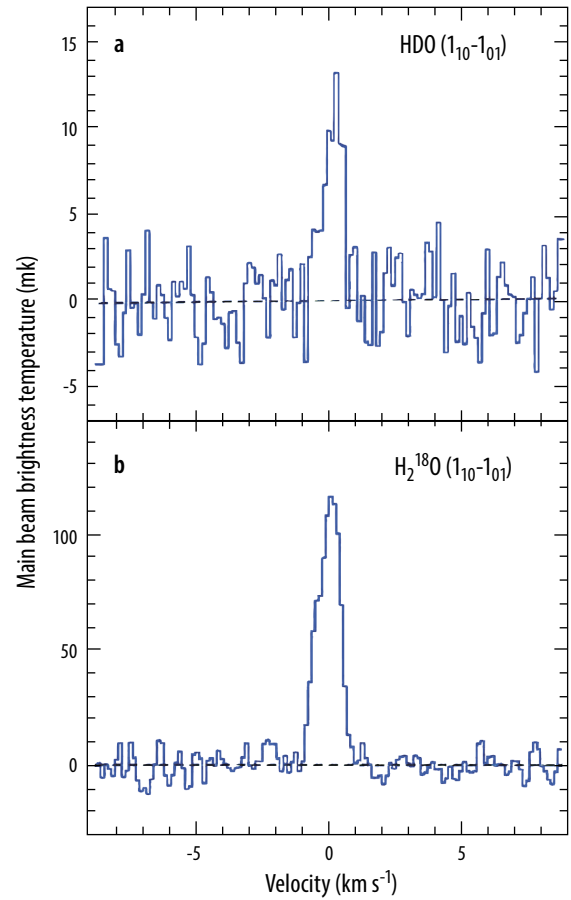

**Figure 1-39:** Herschel/HIFI observations of the Jupiter-family comet 103P/Hartley 2 (Hartogh et al., 2011), where the D/H ~ 1 VSMOW. These data demonstrate the 1-2 km/s linewidths typically measured in comets and demonstrate the need for spectral resolving power ~$4 \times 10^4$ for a comet survey.

A minimum of ~100 comets (50 in each family) is needed to measure the D/H dispersion within each family, to capture potentially rare comets with low D/H values down to 0.5 VSMOW, and to study comets with high dust-to-gas ratios. In a 5-year nominal mission lifetime, *Origins* has access to a wide range of short- and long-period comets. The size of the available sample of new, dynamic (mostly long-period) comets is predicted using JPL's HORIZONS database, taking into account the observatory field-of-regard and a target rate of motion of ≤60 mas/s (see Milam *et al.*, 2016 for comet rate of motion statistics). A query for comets reaching perihelion over the 5-year period from January 2012 through December 2016 yielded ~300 long-period comets and ~200 short-period comets as potential *Origins* targets. The visual magnitude at perihelion of each comet is then converted to an approximate maximum comet activity in number of water molecules per second and adjusted by the geocentric distance. While this method is approximate since some comets may be dust-rich, it provides a reason-able yield estimate. With a minimum requirement of a sample size of 100, the yield is not limited by the number of detectable comets, but rather by the available observing time.

**Sensitivity:** This sample sets the sensitivity requirement, since only a few of the potential *Origins* targets are bright. *Origins* must attain a D/H ratio measurement error of ~$10^{-5}$ and be able to detect weaker comets well below existing detection thresholds (water production rates <$10^{29}$ s$^{-1}$). In terms of the metric of water production relative to geocentric distance, *Origins* must probe comets down to a value of 0.04 for Earth-like D/H (see Figure 1-40), which sets the sensitivity requirement at 3.6 x $10^{-21}$ W/m$^2$ in 1 hour (3σ) for the HDO line at 234.6 μm. Figure 1-40 shows the simulated *Origins*/OSS yield of comets for 5σ D/H ratios of 0.5, 1, and 2 VSMOW. By comparison, *Herschel* sensitivity provided access to tens of bright comets, of which only some were observed, each of which needed at least 5x longer integration time.

### 1.2.5 The Decisive Advantage of Origins for Following the Water Trail

> *Origins* is the only planned or proposed mission that can fully explore the trail of water.

During the next two decades, SOFIA, ALMA, JWST, and, if selected by ESA, SPICA, will each endeavor to study limited segments of the water trail, but even in combination they will be unable to follow the trail of water from the early stages of star and disk formation to the surface of a habitable planet (Figure 1-27). *Origins* provides unique access to key water transitions, with an unprecedented combination of high line sensitivity and high spectral resolving power. The combination of these assets gives *Origins* a decisive advantage for the study of water and the development of habitable conditions during planet formation.

While a trace amount of very hot water (>1000 K) is sometimes found in close proximity to young stars, the formation of water in the ISM and subsequent accretion onto young disks, the assembly of planetesimals, and evolution of young planetary systems all take place at temperatures of 10 to a few hundred Kelvin. As shown in Figure 1-26, the OSS wavelength coverage of 25 to 588 μm contains almost every water transition with upper state energies less than 1000 K, including the important ground-state para and otho H$_2^{16}$O transitions at 179.5 and 538 μm, respectively.

**JWST:** JWST has a long wavelength cut-off just beyond 28 μm, providing access to only one water line with an upper state energy less than 1000 K (the 541 − 414 transition at 25.9 μm with $E_u/k$ = 844 K). Moreover, given JWST/MIRI's modest spectral resolving power (R=1500-4000), it will generally not resolve water lines in young stars and disks. Thus, while JWST will provide integrated line intensities for water at the disk surface within the inner (~1 AU) and hot regions (≥ 500 K) of planet-forming disks, it will probe <1% of the volume of typical disks. Further, the lack of kinematic information will lead to significant model dependencies and uncertain interpretations. Finally, the dust optical depth in the inner disk at the JWST wavelengths is considerable higher than in the far-infrared, effectively shielding the disk midplane from inspection. Complementary *Origins* observations of cooler water lines and kinematic information from *Origins* will be critical to fully understanding the JWST spectra.

**Table 1-16:** D/H in Comets Requirements Flow

| Science Objective 3 |
|---|
| Definitively determine the cometary contribution to Earth's water by measuring the D/H ratio with high precision (>0.5 VSMOW) in over 100 comets in 5 years. |

| Science Observable |
|---|
| HDO and H$_2$0 line fluxes for at least 100 long- and short-period comets. |

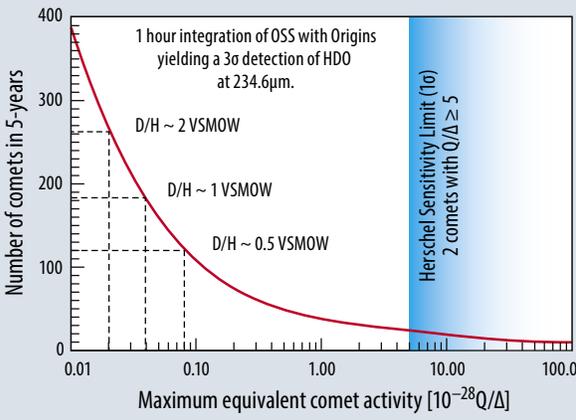

**Figure 1-40:** Simulations for a nominal 5-year lifetime of the population of comets that will approach the inner solar system as a function of their water production rate/Earth-comet distance. Herschel is able to probe the area in blue. *Origins* enables studies of comets two orders of magnitude lower, for D/H ratios similar to Earth/Hartley 2. In a 5-year lifetime, *Origins* could observe over 200 comets with D/H ratios encompassing all known measurements in the solar system.

| Science Requirements |
|---|
| To detect HDO in the fuill sample of ~200 comets, the required line sensitivity at 234.6 μm is 3.6 x 10$^{-21}$ W/m$^2$ in 1 hour (3σ) set by the need to reach a maximum equivalent comet activity of 0.04 and D/H ~ 1 VSMOW. |

**ALMA:** ALMA offers high spectral resolving power and broad wavelength coverage in 10 receiver bands between 315 μm and 8.6 mm. There are ten water transitions with $E_u/k$ ≤1000 K within this wavelength range. However, ALMA must observe through the atmosphere, so even under very favorable atmospheric conditions (Figure 1-41), it is only able to observe three $H_2^{16}O$ transitions – at 1635 μm (183 GHz), 922 μm (325 GHz), and 613 μm (488 GHz) – and only the 922 μm line is sensitive to gas cooler than 200 K. Although ALMA has unobscured line sensitivities similar to that of *Origins*, its 325 GHz line is 300 times weaker than the ground-state water lines and ~1000 times weaker than the warmer far-infrared lines in a planet-forming disk, making *Origins* far more sensitive to water detection and characterization than ALMA. Further, the 183 and 325 GHz lines can mase at the high densities expected in a disk (Neufeld and Melnick, 1991). At lower densities, before maser emission is produced, the upper states still experience a population inversion that makes modeling the emission from these lines unusually dependent on the gas density and temperature. These lines must be observed during the best weather conditions available at the ALMA site, and ideally with ALMA in a long-baseline (high resolution) array configuration. While a few water lines are technically accessible to ALMA, the right conditions rarely align, making ALMA an inefficient tracer of the water trail.

**SOFIA:** SOFIA is equipped with high spectral resolving power instruments and has access to the same wavelength range as *Origins*. Although SOFIA is able to observe above 40,000 ft., where the overhead water burden is greatly reduced relative to ALMA, the residual water features with $E_u/k$ ≤800

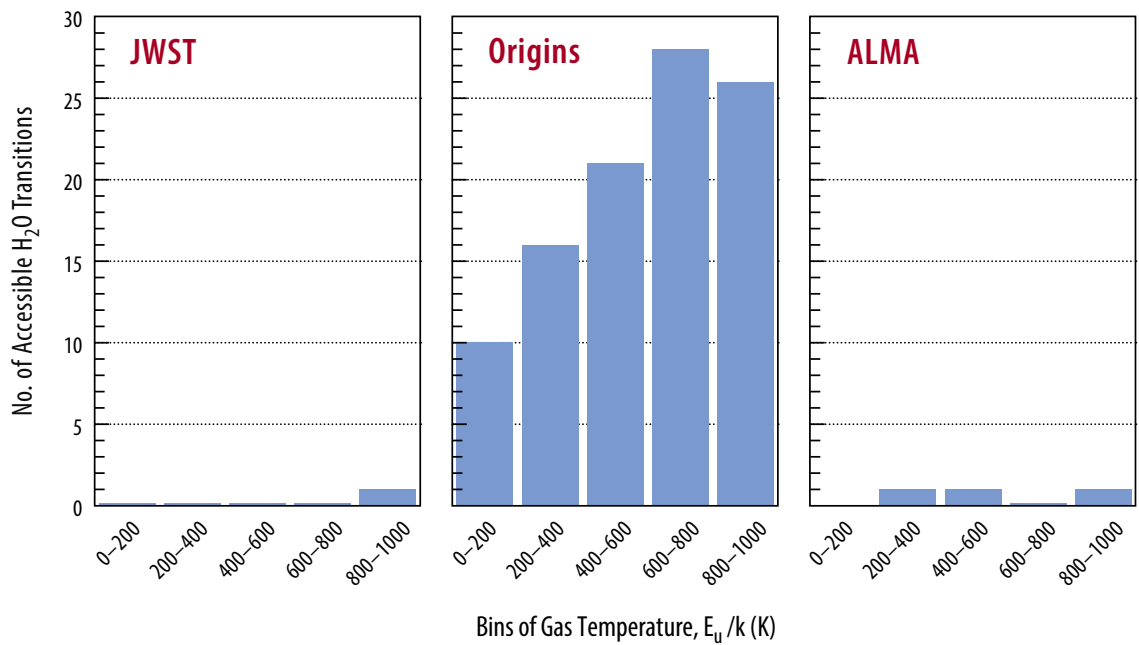

**Figure 1-41:** Number of $H_2^{16}O$ transitions observable by JWST, *Origins*, and ALMA with energies above the ground state less than 1000 K. For ALMA, blockage due to atmospheric absorption is a major factor in assessing the observability of a given water transition, so very good conditions (i.e., 0.5 mm of zenith precipitable water vapor and 90° telescope elevation) were assumed. To be considered observable, the atmospheric transmission at a given wavelength must be >10%.

K remain prohibitively strong in the atmosphere, restricting SOFIA's access to diagnostic $H_2^{16}O$ lines excited at cooler temperatures. In addition, SOFIA's telescope is warm (~250 K) and has an effective collecting area less than 20% that of *Origins*, which results in line sensitivities 2-3 orders of magnitude less than those of *Origins* (Figure ES-9). In particular, SOFIA/HIRMES is expected to obtain high-resolution spectra of water lines that sample warm gas (≥ 300 K) toward tens of the brightest planet-forming disks, and to observe HD at high spectral resolving power to measure the gas mass for a few of the brightest, most massive planet-forming disks. HIRMES will not be able to conduct a comprehensive evolutionary survey of disks.

**SPICA:** If selected by ESA, SPICA will carry spectrometers operating between 12 and 230 μm. Within the limited 12–18 μm range, the SPICA Mid-Infrared Instrument has a spectral resolving power of R~28,000, corresponding to Δv ≥ 11 km s$^{-1}$; however, there are no water lines within this wavelength range with $E_u/k$ ≤1000 K, which limits SPICA to resolving lines emitted from relatively hot gas. At the longer wavelengths, which trace cooler water, SPICA/SAFARI has a spectral resolving power of less than a few thousand, and so under-resolves water lines in disks. Since the line profile contains much of the dynamical information connected to the radial distribution of water across the disk, tomographic imaging (Section 1.2.2.4) is not possible with SPICA beyond 1 AU. Moreover, with lines under-resolved by factors of 10-100 the line-to-continuum ratio is low, limiting the effective line sensitivity. Line dilution is the main reason that *Herschel*/PACS was only able to detect a few water lines from disks. This problem is even greater for measuring disk masses, as the HD J=1-0 line is intrinsically faint with low line-to-continuum ratios (McClure *et al.*, 2016). Like SOFIA, SPICA's cold (8 K) telescope has less than 20% of the effective collecting area of *Origins*, resulting in ~10 times lower line sensitivities than *Origins*. The low spectral resolving power typically leads to another order of magnitude of decrease of effective sensitivity to water lines in planet-forming disks. Therefore, SPICA, as currently equipped, only detects warm water and HD in the brightest disks; a comprehensive disk survey and tomographic imaging are beyond its capabilities.

In the case of protostars, it has been demonstrated by the velocity-resolved water spectra obtained with SWAS, Odin, and *Herschel*/HIFI that water lines often display complex profiles, including in many cases self-absorption due to foreground gas and P-Cygni profiles toward infall sources. Finally, study of Solar System comets requires high spectral resolving power (*i.e.,* Δv ≥ 1.5 km s$^{-1}$) to identify jets, asymmetric water distributions, and temperature variations across the coma.

Only a telescope with the wavelength coverage, sensitivity, and spectral resolving power of *Origins* operating above the Earth's atmosphere can trace the formation and evolution of water from its origins in the interstellar medium to planet-forming disks and, ultimately, Solar System comets. Tracing this Water Trail is key to understanding how Earth and earth-like planets get their water and become habitable.

### 1.3 Do Planets Orbiting M-Dwarf Stars Support Life?

By obtaining precise mid-infrared transmission and emission spectra, *Origins* will assess the habitability of nearby exoplanets and search for signs of life.

### 1.3.1 Introduction

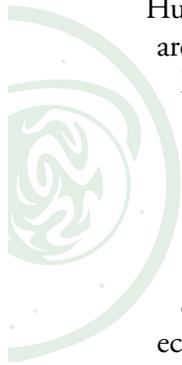

Humankind has long pondered the question, "Are we alone?" Now scientists and engineers are designing instruments dedicated to answering this question. Our quest to search for life on planets around other stars relies on our ability to measure the chemical composition of their atmospheres and to understand the data in the context of models for planet formation and evolution. *Origins* seeks to answer this question with a mid-infrared instrument specifically designed for the characterization of temperate, terrestrial exoplanets.

*Origins* utilizes the transit technique, where planetary atmospheres are observed during primary transit (when a planet passes in front of its host star) and secondary eclipse (when a planet passes behind its host star). While *Origins* conducts atmospheric reconnaissance over a range of transiting planet types, its main focus is the search for habitability indicators and biosignatures in the atmospheres of terrestrial exoplanets in the habitable zone of mid-to-late M dwarf stars, which are a compelling testbed for assessing habitability (Section 1.3.3.2). We employ a multi-tiered strategy in our search for signs of life from a sample of terrestrial exoplanets that are identified before *Origins* would launch.

Before describing how *Origins* characterizes these atmospheres, we first define several terms used throughout the section.

First, we rely on the "classical" definition of the habitable zone as the orbital distance range where flux levels from the parent star allow liquid water to be stable on a planetary surface. The location of the inner edge of the habitable zone is not well understood and depends on a multitude of factors.

Climate models demonstrate that moist greenhouse atmospheres can occur with surface temperatures of $280 - 350$ K (*e.g.*, Wolf & Toon, 2015; Popp *et al.*, 2016; Kopparapu *et al.*, 2016; Kopparapu *et al.*, 2017; Ramirez *et al.*, 2018). Since one *Origins* science objective is to better understand how various factors impact the inner edge of the habitable zone, we initially adopt a relatively lax upper limit of 350 K. Secondary eclipse measurements (Section 1.3.5) determine planets' apparent surface temperatures and identify those that have experienced atmospheric collapse near the inner edge, or a runaway greenhouse near the outer edge. For cooler stars, the habitable zones are closer in than for hotter stars. For example, the orbital distance of the habitable zone of a Sun-like star is roughly 1 AU,

---

**Important Definitions**

We adopt the following definitions for these terms throughout the text and use these to trace our scientific motivation and technical design of *Origins*.

- **Habitable zone:** The region surrounding a parent star where liquid water could exist on a planet's surface (Teq = 200 - 350 K).
- **Temperate planet:** A planet in the habitable zone.
- **Terrestrial planet:** A planet that is rocky, with a radius less than 1.75 times the radius of Earth.
- **Habitability indicator:** A gas-phase species that may suggest the presence of life, but could also be produced without life (*i.e.,* abiotically; $CO_2$, $H_2O$).
- **Biosignature:** A combination of gas-phase species that can only be produced by life (*e.g.*, $O_3+CH_4$, $O_3+N_2O$).

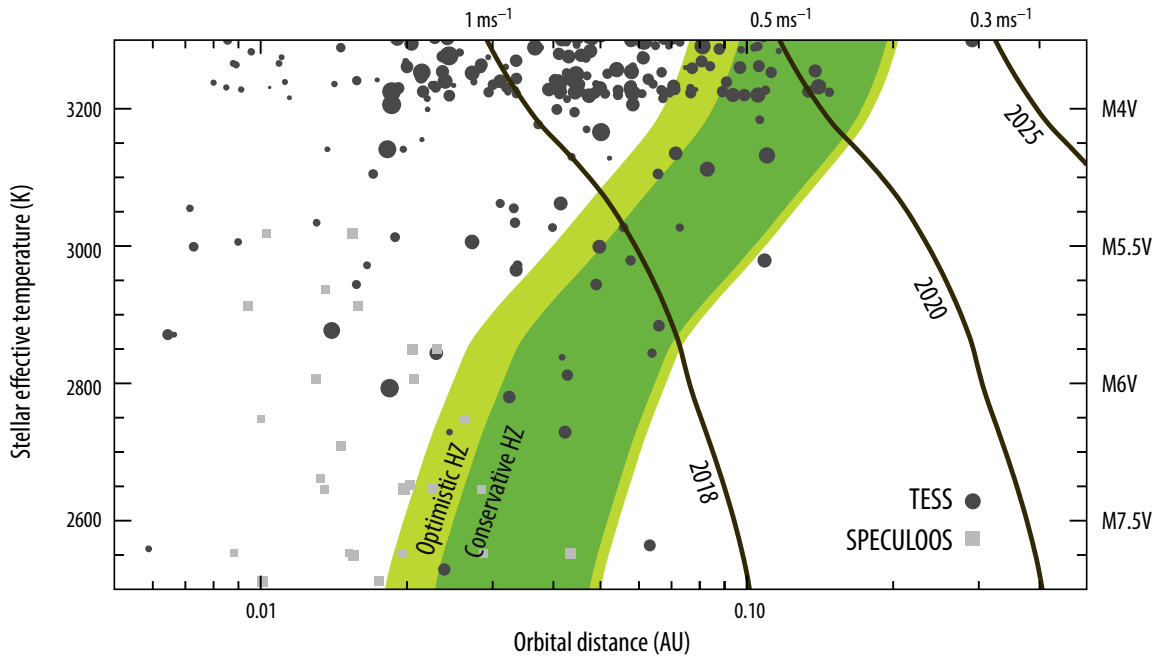

**Figure 1-42:** *Origins* characterizes the atmospheres of terrestrial exoplanets in the habitable zone of M dwarf stars. Shown here is the entire landscape of habitable zone distances (in astronomical units) for mid- to-late M dwarfs, shown as stellar type and stellar effective temperature. Overplotted in black circles and gray squares are estimated TESS and SPECULOOS rocky HZ planet yields from Sullivan et al. (2015) and Delrez et al. (2018), respectively. Green and blue shaded regions indicate the optimistic and conservative habitable zones, respectively, based on estimates from Kopparapu et al. (2014). Black lines indicate estimated radial velocity (RV) semi-amplitude needed to detect a planet in this phase space, annotated with the year it is estimated that RV instruments will reach that precision.

whereas the habitable zone distance for a planet orbiting an M-dwarf is much closer (*i.e.,* ~0.1 AU; Figure 1-42). We define a planet in the habitable zone to be temperate.

Next, we define a terrestrial planet as one that is rocky, and use these terms interchangeably. This definition includes planets like Earth, but also planets that might be scaled-up versions of Earth (*i.e.,* super-Earths). The transition between these terrestrial planets and those that contain a massive gaseous envelope (*i.e.,* mini-Neptunes) has been refined using planet discoveries from Kepler, and is thought to occur between 1.5 and 2.0 $R_{Earth}$ (Wolfgang and Lopez, 2015; Rogers, 2015; Fulton *et al.,* 2017). We adopt 1.75 $R_{Earth}$ as our limit in defining *Origins'* observational program.

Lastly, and most importantly, the terms habitability indicators and biosignature differ across many fields of biological and physical sciences. Here we consider "remote" habitability indicators and biosignatures — that is, the detection of species from distant atmospheres using ground- and space-based telescopes. Additionally, we only refer here to gas-phase species, and not direct measurements of rock mineralogy or fossil morphology. While various studies use "biosignatures" to refer to many types of gas-phase species (Grenfell, 2017), here we adopt the following definitions (Figure 1-43):

A habitability indicator is a gas-phase species that may suggest the presence of life (*e.g.*, Seager & Deming 2010, Youngblood *et al.* 2017 and references therein). We focus primarily here on $CO_2$ and $H_2O$, which can be produced by life but also in large quantities by abiotic processes. These two gases are important as greenhouse gases, as well as potential sources for high $O_2$ concentrations through photolysis.

Biosignatures constitute a combination of gas-phase species such that, when detected together, can only be produced by life and are not readily reproduced by abiotic processes. Classically, these include simultaneous detections of $CH_4 + O_2$ (Lederberg, 1965; Lovelock, 1965) or $O_2 + N_2O$ (Segura *et al.*, 2003). The presence of $O_3$ requires the presence of $O_2$; thus, the two can be used interchangeably in our definition.

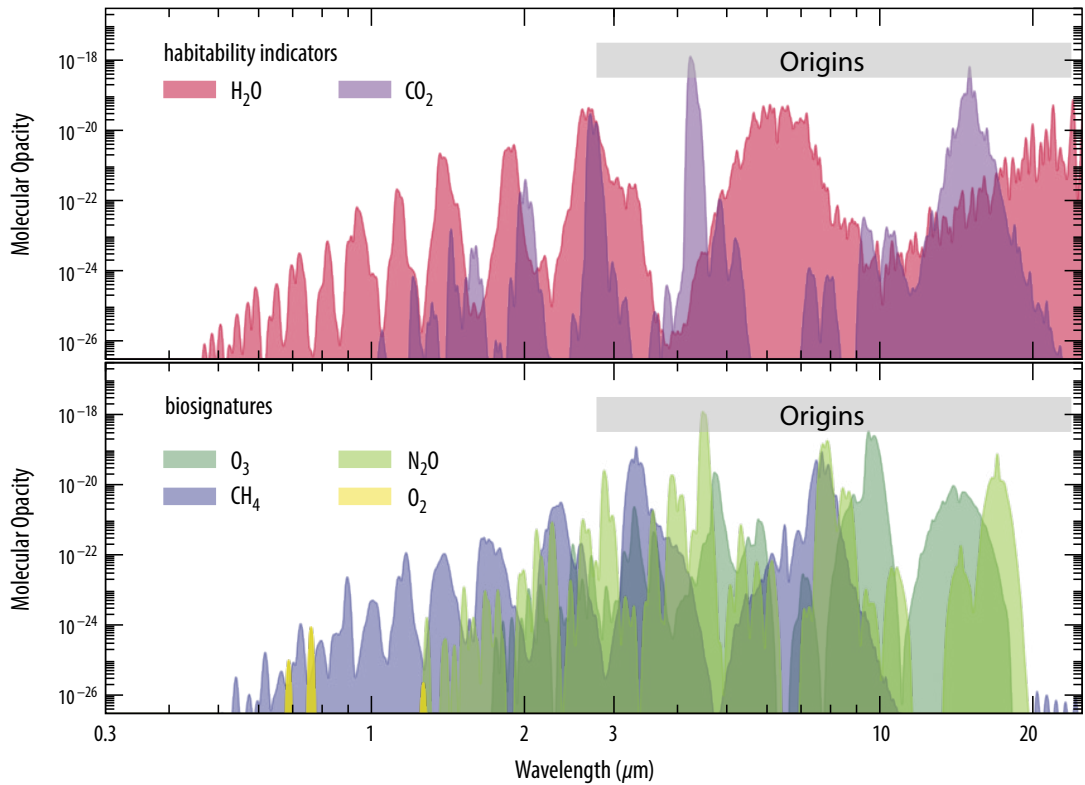

**Figure 1-43:** Molecular opacities of relevant habitability indicator (top) and biosignature (bottom) gases in the mid-infrared. *Origins* is sensitive to multiple bands for each molecular species, which is critical in breaking degeneracies between overlapping spectral signatures.

It is likely that these definitions will change before *Origins* would launch, but the spectral features of many potential biosignatures lie in the *Origins* bandpass. For a comprehensive review of exoplanetary biosignatures see Grenfell *et al.* (2017) and Kaltenegger (2017), as well as many white papers submitted to the National Academies Astrobiology Science Strategy and Astro2020 Science White Paper calls.

Ultimately, detecting and measuring the relative abundances of multiple molecules, including but not limited to habitability indicators and biosignatures, is needed to determine if a terrestrial exoplanet can host life. Table 1-17 summarizes the instrument and mission design drivers needed to accomplish this. *Origins* leverages its broad wavelength coverage in the mid-infrared to search for these molecules and measure atmospheric temperatures, to empirically constrain the types of planets orbiting M and K dwarfs that might develop and harbor life.

### 1.3.2 Transiting Exoplanet Science to Date

> Observations of transiting exoplanets is a proven technique that has amassed a wealth of information about distant worlds. However, many open questions, particularly for terrestrial exoplanets, remain unanswered.

The past 30 years of exoplanet science have shown that planets are common around nearby stars. Now and in the coming years, the exoplanet community seek to understand the atmospheres of these worlds using a range of observational techniques. Built on its experience with *Spitzer*, HST, and in preparation for JWST, the community has developed tremendously successful methods of atmospheric characterization that use the transit technique to probe the radiative, chemical, and advective (*i.e.,* atmospheric dynamics) processes in exoplanet atmospheres (Deming *et al.* 2019, Seager & Deming 2010, Marley *et al.* 2013, 2015; Burrows 2014; Cowan *et al.* 2015; Fischer *et al.* 2015, Heng & Showman 2015, Crossfield 2015, Madhusudhan *et al.* 2016, Deming & Seager 2017, Kreidberg *et al.* 2018, Parmentier & Crossfield 2018, and Fortney 2018). Using these facilities, ~80 transiting exoplanets have been characterized to date, a majority of which are close-in gas giants (so-called "hot Jupiters"). These observational results have provided insights into planet formation and evolution that have fundamentally altered how we understand our own solar system. *Origins* aims to utilize this powerful technique to search for signs of life on temperate, terrestrial planets.

**Table 1-17:** Instrument and Mission Design Drivers

| Technical Parameter | Requirement | Expected Performance | Key Scientific Capability |
|---|---|---|---|
| Min. Wavelength | 3.0 μm | 2.8 μm | Detect $CH_4$ at 3.3 μm |
| Max. Wavelength | 20 μm | 20 μm | Thermal emission; Detect $H_2O$ at 17 + μm |
| Aperture Size | 5.3 m | 5.9 m | Sufficient SNR to detect biosignatures within 5 years |
| Spectroscopic Precision | 5 ppm in 3–10 μm | 5 ppm (2.85–10.5 μm) | Sufficient precision to detect biosignatures at 3.6σ |
| Spectral Resolving Power | >50 in 3–10 μm; >250 @ 17 + μm | 50–100 in 2.8–11 μm 165–295 in 11–20 μm | Resolve spectral features |
| Field of Regard | At least 125 days of visibility per year for all targets | 130 days of visibility at the ecliptic; more at higher latitudes | Sufficient number of transit opportunities to detect biosignatures within a 5-year mission |

### 1.3.2.1 A Primer on Observing Transiting Exoplanets and Characterizing Their Atmospheres

There are three standard viewing geometries for transiting exoplanets: the primary transit, secondary eclipse, and phase curve. These methods typically require orbital inclinations near 90 degrees, which places the observer's line of sight close to the system's orbital plane. Figure 1-44 provides a schematic of a transit and eclipse, as well as simulated spectra for a temperate, terrestrial planet that would be characterized by *Origins*.

The primary transit occurs when the planet passes in front of its host star, thus blocking part of the star's light as seen from Earth. The fractional dip in light is determined by the planet-to-star area ratio and provides information about the size of the planet. With a mass determination from radial velocity measurements, we can learn about a planet's bulk density and begin to constrain its internal structure. The first detection of a transit event was made nearly two decades ago (Charbonneau *et al.* 2000). We obtain the planet's transmission spectrum by measuring the planet's apparent change in size as a function of wavelength. Light from the host star passes through the planet's atmospheric annulus where it interacts with atoms and molecules. The annulus becomes opaque (and the planet appears larger) at wavelengths where these chemical species strong-ly absorb in the planetary atmosphere. Transmission spectroscopy data are sensitive to the relative chemical abundances and the presence of cloud or haze particles within the atmosphere.

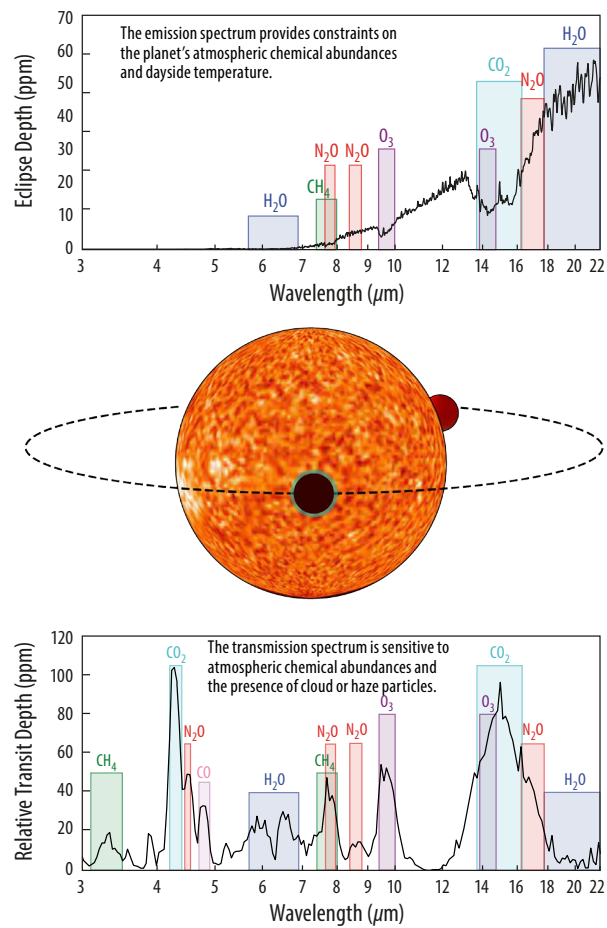

**Figure 1-44:** The geometry and spectra for a typical transiting exoplanet. As a transiting planet passes in front of its host star, its apparent size changes as molecules absorb light at different wavelengths. When the planet passes behind the star, the planet's dayside thermal emission is measured, thus constraining a terrestrial planet's apparent surface temperature.

The secondary eclipse occurs when the planet passes behind its host star, and the planet's light is blocked for the duration of the eclipse. Similar to the primary transit, the fractional dip in light for a secondary eclipse is given by the planet-to-star area ratio, multiplied by the planet-to-star flux ratio. At infrared wavelengths, the secondary eclipse provides constraints on the planet's dayside temperature. The first eclipse measurements were made by Charbonneau *et al.* (2005) and Deming *et al.* (2005). Secondary eclipse data are used to obtain the planet's emission spectrum. The infrared radiation emitted by the atmosphere and surface depends upon the planet's temperature structure and on molecules that imprint their absorption features on the outgoing radiation. The planet appears fainter at wavelengths where chemical species within its atmosphere absorb light; these are observed as absorption features. The planet's atmosphere can also have emission features when observations probe altitudes where there is a strong thermal inversion (*i.e.*, increasing temperature with increasing altitude, as in Earth's stratosphere). Emission spectroscopy is sensitive to the absolute chemical abundances and thermal profile of the atmosphere. Alonso (2018) provides a recent summary of secondary eclipse measurements through 2017.

Phase curve observations typically extend over at least one complete orbit of the target planet around its host star. Some of the first phase curve measurements were made shortly after the first secondary eclipse observations (*e.g.*, Harrington *et al.*, 2006; Knutson *et al.*, 2007). Photometric phase curves in the infrared provide a longitudinally-resolved 1D map of the planet's temperature. Spectroscopic phase curves provide a second dimension, altitude, by probing different pressure levels at different wavelengths. These observations result in phase-resolved emission spectra that place constraints on the planet's atmospheric composition, thermal structure, and circulation (*i.e.*, winds).

Measuring the circulation of heat from dayside to nightside can help determine whether a rocky planet has a substantial atmosphere, and provides a perspective on the planet's long-term climate (Deming and Seager, 2009). Planets with tenuous atmospheres are likely to have strong day-night contrasts, particularly if they are slow rotators, whereas planets like the Earth exhibit only minor temperature fluctuations from dayside to nightside. Section 1.3-10 provides additional information about phase curve observations.

### 1.3.2.2 Recent Milestones in Transiting Exoplanet Science

With over 5,000 confirmed and candidate transiting exoplanets (NASA Exoplanet Archive), scientists are beginning to learn about the frequency of exoplanets as a function of size and orbital separation (*e.g.*, Fulton and Petigura, 2018), thus painting a picture of planetary system architecture. Some of the first observational efforts to characterize transiting exoplanets used facilities such as *Spitzer* and HST to observe individual gas giant atmospheres. The field has since graduated to the realm of small surveys, including for ten planets with HST (Sing *et al.*, 2016), and now ~40 planets (the *Hubble* Panchromatic Comparative Exoplanet Treasury Survey; Sing *et al.*, in prep.)

Work in the giant exoplanet regime has progressed on several fronts, including identifying the molecular absorption features that control the temperature structure of their atmospheres (Fortney *et al.* 2008, Evans *et al.* 2018), and determining whether the abundances of molecules track with those of their parent stars or are instead influenced by disk chemistry (Oberg *et al.*, 2011; Madhusudhan *et al.*, 2013). The influence of high-temperature clouds on observed spectra is also being studied (*e.g,* Pont *et al.*, 2013, Lines *et al.* 2018). With these observational and modeling efforts, scientists have begun to conduct comparative exoplanetology studies (*e.g.*, Sing *et al.* 2016, Kataria *et al.*, 2016) to understand the physics, chemistry, and dynamics that shape giant exoplanet atmospheres. However, a comparative understanding of terrestrial exoplanet atmospheres remains elusive.

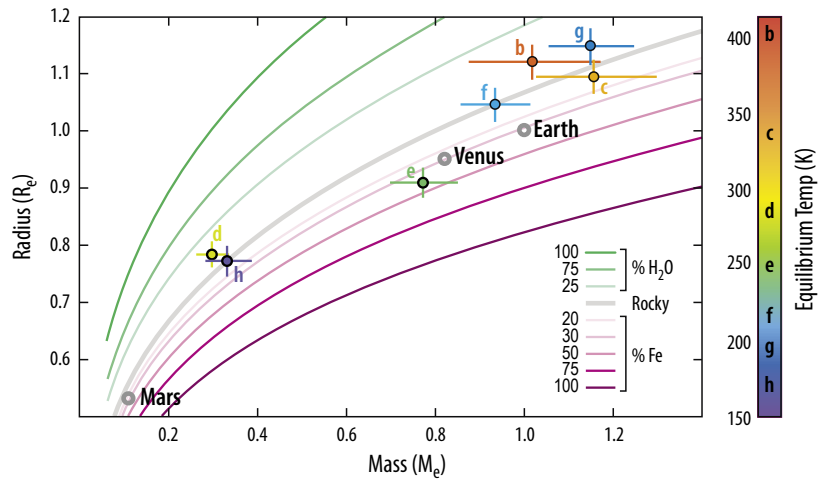

**Figure 1-45:** Mass-radius diagram for the seven TRAPPIST-1 planets, Earth, Mars and Venus. Curves trace idealized compositions of rocky and water-rich interiors. The densities for five of these planets are consistent with having volatile-rich envelopes (i.e., thick atmospheres, oceans, or ice). Adapted from Grimm et al., (2018).

Recent discoveries from the MEarth (Dittmann *et al.*, 2017) and TRAPPIST (Gillon *et al.*, 2016) surveys provide a glimpse into the possible diversity of transiting terrestrial M-dwarf planets. In particular, mass measurements for TRAPPIST-1, a system of seven roughly Earth-sized planets orbiting an M8 star, point to rocky internal structures (Grimm *et al.*, 2018; Figure 1-45), and HST transit observations of the atmospheres of several TRAPPIST-1 planets provide tantalizing hints for high mean-molecular weight (*e.g.*, $H_2O$, $CO_2$) atmospheres, which therefore do not rule out potentially habitable environments (de Wit *et al.*, 2016, 2018). This path of exoplanet detection and follow-up atmospheric characterization will likely be repeated for transiting terrestrial planets found by space-based surveys such as the Transiting Exoplanet Survey Satellite (TESS), an all-sky survey of the nearest and brightest transiting planets (Ricker *et al.* 2015), and ground-based surveys such as the Search for habitable Planets EClipsing ULtra-cOOl Stars (SPECULOOS) survey, the follow-on to the TRAPPIST survey (Gillon 2018).

While JWST is poised to conduct further reconnaissance of many TRAPPIST-1 planets via transmission (*e.g.*, Barstow & Irwin, 2016; Morley *et al.*, 2017), accessing thermal emission will prove challenging for all but the hottest planets in the system, and identification of biosignature gases will remain out of reach. Therefore, a mid-infrared facility like *Origins* is necessary to obtain a complete picture of these and other transiting terrestrial systems, and determine the habitability of these exotic worlds.

### 1.3.2.3 Technical Considerations: Characterization Methods and Spectroscopic Feature Sizes

Close-in transiting planets cannot be spatially imaged with current (or near future) technology. This is because 10-meter-class telescopes do not have sufficient spatial resolution to separate habitable-zone planets from their M dwarf host stars. As a result, time-series observations on the combined (star + planet) light are necessary to learn about the planet. To obtain precise physical constraints, one must observe multiple transits or eclipses to build up sufficient signal-to-noise. Table 1-18 provides a comparison of space-based methods for characterizing temperate, terrestrial planets.

The typical transit depth of a hot Jupiter exoplanet transiting a Sun-like star is ~1%. This is also true for an Earth-size planet transiting a late M dwarf star. This level of precision is sufficient to detect transiting exoplanets (and is often done with relatively-small telescopes from the ground); however, to characterize the atmospheres of these planets requires much more precise instruments. The variation in transit depth (*i.e.*, the transmission spectrum) is typically of order 0.01% (or ~100 ppm). The nominal molecular feature size depends on a number of planetary and stellar parameters, and is equal to $2HR_p/R_s^2$, where $R_p$ and $R_s$ are the radii of the planet and star, respectively, and $H = k_B T/\mu g$ is the atmospheric pressure scale height. Observing planets around smaller stars, such as M dwarfs, yields the most significant gain in molecular feature size, due to its quadratic dependence in the equation above.

**Table 1-18:** Space-Based Method Comparison

| Criterion | Transit Method (M, K Dwarfs) | Direct Imaging (Sun-like Stars) |
|---|---|---|
| A Priori Planet Knowledge | Known population with measured planet radii, masses, and bulk compositions | Unknown planet radii and masses, must search for planets before characterization |
| Expected Temperate, Terrestrial Planet Yields | ~26 planets transiting mid-to-late M dwarfs with $K_{mag} < 11.5$ (includes known, TESS + extended, and SPECULOOS) | 4 meter off-axis design: ~10 planets<br>8 meter on-axis design: ~20 planets<br>8 meter off-axis design: ~40 planets (Stark et al., 2019) |
| Wavelength Range for Atmospheric Characterization | Mid-IR (3–20 μm) provides multiple, strong spectroscopic features of prominent habitability indicators and biosignatures (e.g., $CO_2$, $H_2O$, $CH_4$, $O_3$, and $N_2O$) | Optical and near-UV wavelengths provide access to $O_2$ and $O_3$; near-IR provides $H_2O$ but exact cutoff varies with system distance and aperture size. |

### 1.3.3 Origins and The Search for Life

> With precise mid-infrared transit spectroscopy, *Origins* will measure habitability indicators and biosignatures, including ozone, carbon dioxide, water, nitrous oxide, and methane, in the atmospheres of temperate, terrestrial exoplanets orbiting M-dwarfs.

*Origins* builds upon the legacy of transiting exoplanet science set forth by *Spitzer*, HST, and JWST, by characterizing the atmospheres of transiting exoplanets in the habitable zones of M dwarfs. *Origins* utilizes the Mid-Infrared Spectrometer and Camera Transit spectrometer (MISC-T), which operates from 2.8–20 μm, to conduct a multi-tiered survey of terrestrial M-dwarf planets that will have already been discovered by TESS, SPECULOOS, and other ground- and space-based surveys to:

1. Search for habitability indicators and rule out a tenuous or cloud-dominated atmosphere that would inhibit further observational characterization;
2. Measure the temperatures of planet "surfaces," thus enabling a direct determination of the temperatures required for habitability;
3. For the most promising targets, search for biosignature gases, thus determining the presence of life on another planet.

These comprise the three main Science Objectives (Table 1-19). With this suite of observations, *Origins* provides insights into atmospheric composition, thermal structure, and potential habitability of terrestrial M-dwarf planets that help us assess whether we are alone.

### 1.3.3.1 The Mid-infrared Provides an Ideal Window to Biosignatures

Mid-IR wavelengths are sensitive to the strongest transitions of most molecules, including the habitability indicator gases $H_2O$ and $CO_2$, the biosignature gasses $O_3$, $N_2O$, and $CH_4$ (Figure 1-43, Table 1-20), and other potentially abundant atmospheric constituents, such as ammonia ($NH_3$) and carbon monoxide (CO). The broad mid-IR wavelength coverage *Origins*/MISC-T offers provides the context that comes from the detection of multiple molecular bands. This, in turn, proves critical in definitively measuring atmospheric abundances. Broad wavelength coverage also prepares us for the detection of the unexpected, since exoplanet science to date has shown us that nature's imagination is richer than ours. Beyond the familiar molecules associated with life on modern Earth (*e.g.* $H_2O$, $CH_4$), other biosignature gases have been proposed, *e.g.*, dimethyl sulfide (Domagal-Goldman *et al.*, 2011), phosphine (Sousa-Silva *et al.*, 2019), ammonia (Seager *et al.*, 2013), and many more (Seager *et al.*, 2016). These molecules are expected to be present in exoplanet atmospheres with low abundances but are less susceptible to false positives for life. Additionally, because these gases are spectroscopically active in *Origins*' wavelength range, searches can be carried out at no additional cost to the primary observations.

**Table 1-19:** Exoplanet Objectives

| NASA Science Goal | Are we alone? | |
|---|---|---|
| *Origins* Science Goal | Do planets orbiting M-dwarf stars support life? | |
| *Origins* Scientific Capability | By obtaining precise mid-infrared transmission and emission spectra, *Origins* will assess the habitability of nearby exoplanets and search for signs of life. | |
| | **Objective Goal** | **Technical Statement** |
| Scientific Objectives Leading to Mission and Instrumental Requirements | **Objective 1**: What fraction of terrestrial M and K- dwarf planets have tenuous, clear, or cloudy atmospheres? | Distinguish between tenuous, clear, and cloudy atmospheres on at least 28 temperate, terrestrial planets orbiting M and K dwarfs using $CO_2$ and other spectral features. |
| | **Objective 2**: What fraction of terrestrial M-dwarf planets are temperate? | Establish the apparent surface temperatures of at least 17 terrestrial exoplanets with the clearest atmospheres and distinguish between boiling and freezing surface water at ≥3σ confidence (±33 K). |
| | **Objective 3**: What types of temperate, terrestrial M-dwarf planets support life? | Search for biosignatures on at least 10 planets, highest ranked from Objectives 1 and 2, and if present, detect biosignatures at ≥3.6σ (assuming an Earth-like atmosphere). |

**Table 1-20:** Key terrestrial molecular bands over 3 - 20 µm. Having access to multiple bands of the same feature is critical to breaking degeneracies in molecular abundances.

| Absorber | Wavelength (µm) |
|----------|-----------------|
| $CO_2$ | 4.3, 15 |
| $H_2O$ | 6.2, >17 |
| $CH_4$ | 3.3, 7.7 |
| $N_2O$ | 4.5, 7.8, 8.3, 17 |
| $O_3$ | 9.7, 14.3 |
| $CO$ | 4.8 |

Furthermore, *Origins* provides the best access (in terms of bands that are both strong and well-resolved from the bands of neighboring molecules) to biosignatures relevant to Archaean Earth (see also Figure 1-55). In particular, the simultaneous detection of $CH_4$+$CO_2$, combined with a non-detection of CO, would be robust evidence for an Archaean-like biosphere (Krissansen-Totten *et al.*, 2018). Given that the Archaean comprised a lengthy period of Earth's geologic history, during which oxygen may not have been prevalent, this type of biosphere could be common on terrestrial exoplanets.

The mid-IR also provides information about the planet's thermal emission. Wien's Law ($\lambda_{max}$ in µm = 2898/T) shows that temperate planets are brightest and the contrast with the host star is highest between 7 – 15 µm, which sits in the middle of the MISC-T instrument's bandpass (Figure 1-46).

### 1.3.3.2 M Dwarfs are Compelling Planet-Hosting Stars

M dwarfs are important targets in the search for habitable planets for several reasons. First, thanks to Kepler, the frequency of rocky planets in the habitable zone around M dwarfs is known to be high (~43%, Dressing & Charbonneau 2015). Second, because the habitable zones around M dwarfs are much closer to their stars than G dwarfs like our Sun, the probability of detecting temperate planets in transit is higher—the transit probability for close-in planets is $R_\star/a$, where $R_\star$ is the radius of the star and a is the orbital semi-major axis (Winn, 2010). Third, transits are more frequent around M dwarfs due to the planets' short orbital periods. Fourth, Earth-size planets orbiting M dwarfs feature comparatively large transit depths (~1%) relative to similarly-sized planets orbiting Solar-type stars (~0.01%). Lastly, terrestrial planets in the habitable zones of M dwarfs are the only potentially-habitable planets whose masses can be precisely measured using radial velocity observations with current or near-term facilities. Since both planetary radii and masses will have already been measured by the time of *Origins*' launch, their bulk densities will be known. As a result, all *Origins* planet targets will be vetted at the time of launch and their categorization as rocky planets will be assured. No planet-searching mission phase is required.

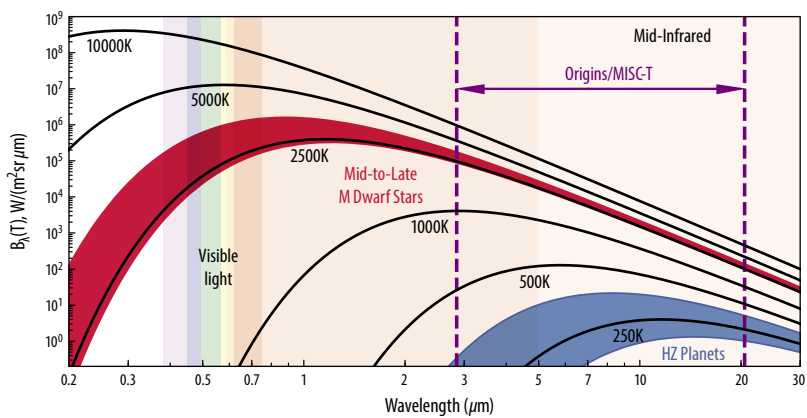

**Figure 1-46:** The spectral radiance of blackbodies at various temperatures (solid curves) highlights *Origins*' opportunity to search for temperate exoplanets in the mid-IR. The peak in emission for mid-to-late M dwarfs (red) occurs near 1 µm and the peak emission for habitable zone planets (blue) is 7 – 15 µm. The planet-star contrast ratio continues to improve at longer wavelengths; however, eventually the overall S/N decreases as the star becomes too faint. The ideal wavelength range to characterize temperate exoplanets is 2.8 – 20 µm (blue vertical dashed lines) -- the same wavelength range as *Origins*' MISC-T instrument. Figure courtesy of Miguel de Val Borro (NASA).

M dwarfs also have characteristics that make assessing the habitability of their planetary systems distinct from that of Sun-like stars. This is both a challenge and an opportunity. It is a challenge in that the formation, evolution, and current environment of the planets may strongly differ from that of the Earth, where our intuition for habitability has necessarily developed. It is a fantastic opportunity, however, because we can learn how a range of different physical processes give rise to planetary atmospheres that may support life.

First, since the habitable zones of M-dwarf planets are much closer to the star than our Sun, the planets are subject to strong tidal effects, including the likelihood of tidal circularization. In some cases, this can result in synchronous rotation, creating permanent day and night sides (Kasting *et al.*, 1993; Joshi & Haberle, 1996; Wordsworth, 2015). Modern climate models for such worlds suggest that this is not a barrier to habitability, as previous concerns of atmospheric freeze-out on the night side would only occur for tenuous atmospheres (Joshi *et al.*, 1997; Joshi, 2003; Yang, Cowan and Abbot, 2013; Kopparapu *et al.*, 2016; Haqq-Misra *et al.*, 2018).

Second, habitable-zone planets orbiting M dwarfs experience higher X-ray to far-ultraviolet (XUV) fluxes and stronger stellar winds relative to those orbiting Sun-like stars. This could lead to increased atmospheric mass loss or surface sterilization, especially at young stellar ages when XUV fluxes are stronger (*e.g.*, Tian, 2009; France *et al.* 2013; Loyd *et al.* 2016; Dong, 2017). However, the masses and bulk densities of most TRAPPIST-1 planets are consistent with having volatile-rich atmospheres (Grimm *et al.* 2018). In addition, early studies of stellar flares using TESS survey data suggest that flaring on M dwarfs is very rarely high enough to erode an ozone layer, and presumably also to strip an atmosphere (Günther *et al.*, 2019). Additionally, life can survive in extreme environments. Extremophiles are organisms that have ecological niches in some of the most inhospitable environments on Earth and have shown that life can survive extreme levels of radiation (*e.g.*, Ordoñez *et al.*, 2009; Kurth *et al.*, 2015), as well as escape periods of high XUV flux by surviving deep under water or below the surface (*e.g.*, Borgonie *et al.*, 2015; Fang *et al.*, 2010).

Finally, M dwarfs and Sun-like stars have different evolution histories at young ages. While the Sun was ~30% fainter at the time of Earth's formation (*i.e.,* the faint young Sun paradox), M dwarfs are brightest during the planet formation era. Thus, M dwarf planets face the potential for a desiccated environment devoid of volatiles and/or significant oxidation of the planetary crust and mantle (Luger & Barnes, 2015; Tian & Ida, 2015). However, similar to volatile delivery on Earth, water on M dwarf planets could be replenished through cometary material.

These and other differences have led to a wealth of studies to understand the composition, climate and potential habitability of rocky M-dwarf planets (see Shields *et al.* 2016 for a recent review), and present questions that only *Origins* can address:

1. What is the abundance of water in the atmospheres of terrestrial M-dwarf planets?
2. What is the atmospheric oxidation state for terrestrial M-dwarf planets, and how does it vary with physical properties such as planet mass and stellar flux?
3. What physical properties (e.g., mass, stellar insolation, climate) set the habitability of terrestrial M-dwarf planets at the inner and outer edges of the habitable zone?

4. Do atmospheric evolution models correctly predict atmospheric composition for sterile M-dwarf planets (planets unlikely to harbor life) such as GJ1132b and the inner TRAP-PIST-1 planets?

**Table 1-21:** Observing Strategy

| Science Objective | # of Planets | Median Observation | Time with *Origins*/MISC-T |
|---|---|---|---|
| # 1 (Section 1.3.4) | 28 | 8 transits | 896 hours |
| # 2 (Section 1.3.5) | 17 | 15 eclipses | 1020 hours |
| # 3 (Section 1.3.6) | 10 | 52 transits | 2080 hours |

5. Are biosignature and habitability indicator gases ($O_3+N_2O$, $O_3+CH_4$, etc.) present in the atmospheres of habitable zone M-dwarf planets (Meadows, 2017; Catling et al., 2018; Wordsworth et al., 2018)?

Answering these questions is essential for putting our own planet in context—how unique is the pathway to habitability? *Origins* enables us to answer these questions for planets with known radii, masses, densities, and orbits across a continuum of temperatures inside and outside the habitable zone.

To achieve *Origins'* three main exoplanet Science Objectives (Table 1-19), **Sections 1.3.4** to **1.3.6** describe a reference ~4000-hour observing program that adopts a "wedding-cake" strategy, wherein each tier addresses a science objective. This strategy is motivated via a trade space study detailed in Section 1.3.3.3. Table 1-21 summarizes this program, allocating 4 hours per transit/eclipse event (3 hours of observation plus 1 hour of settling/overhead).

### 1.3.3.3 Origins Exoplanet Trade-Space Study

In order to determine the ideal wavelength range, number of transits/eclipses, resolving power and aperture size to characterize terrestrial M-dwarf planets and develop well-justified science requirements for *Origins*, members of the *Origins* Exoplanet Science Working Group conducted a comprehensive exoplanet trade-space study (Tremblay *et al.*, in prep), simulating transmission and emission spectra with realistic uncertainties and performing full atmospheric retrievals. For more details see Appendix E.3, here we summarize the results. Trades considered include: wavelength, spectral resolution, mirror size, and number of transits/eclipses. Although there are nearly limitless climates and compositions to consider, for this study the team adopted a TRAPPIST-1e-size planet with an Earth-like atmosphere (Robinson *et al.*, 2011) orbiting an M8 star with a K-band magnitude of 9.85. This magnitude is based on TESS and SPECULOOS simulations. The uncertainties derive from using realistic instrument throughputs and adding (in quadrature) the photon-limited performance to an assumed 5 ppm noise floor. We aimed to determine the nominal instrumental setup in order to robustly detect key habitability indicator and biosignature gases using a fully Bayesian evidence-based method (Trotta, 2008).

This study reached the following conclusions:

1. The $3 - 20$ μm wavelength range is necessary to capture at least two molecular bands per molecule in order to obtain robust detections and break degeneracies. The $3 - 5$ μm range in particular is important for capturing the 3.3 μm band of $CH_4$ and placing stringent upper limits on the CO abundance, a key false positive indicator (Meadows et al. 2018).
2. The longer wavelengths >8 μm, are necessary for constraining the surface conditions and thermal emission of temperate worlds, as this is where atmospheric windows are present;
3. In contrast to stellar spectroscopy, relatively low spectral resolutions (as low as R=50) are required to adequately resolve the molecular bands; there are additional gains in precision with R~100, and diminishing gains thereafter;
4. A minimum aperture size of 5.3 meters ensures that *Origins* will be sensitive to biosignatures at >3.6σ (for the setup described above) within the 5-year mission lifetime; and
5. *Origins* requires ~60 transits to detect the biosignature pair $O_3+N_2O$ at 3.6σ confidence (for the setup described above).

Achieving similar detection significances for other targets requires more or less telescope time. The amount of time was determined by taking ratios of the nominal molecular feature sizes, transit durations, and stellar magnitudes.

### 1.3.4 Science Objective 1: Identify Terrestrial Exoplanets with the Strongest Molecular Features

> *Origins* will distinguish between tenuous, clear, and cloudy atmospheres on at least 28 temperate, terrestrial planets orbiting M and K dwarfs using $CO_2$ and other spectral features.

In the first tier of its exoplanet survey, *Origins* obtains transmission spectra over 3.0–20 μm for temperate, terrestrial planets spanning a broad range of planet sizes, equilibrium temperatures, and orbital distances to distinguish between tenuous, clear, and cloudy atmospheres (Figure 1-47). In particular, *Origins* focuses on the detection of $CO_2$ absorption as well as other spectral features. Because the spectral features due to $CO_2$ absorption are large, for this first tier we can include terrestrial planets orbiting early-M and late-K dwarfs in our observational sample. However, given the small signal sizes expected

**Table 1-22:** Exoplanet Biosignatures Requirements Flow 1

| Science Objective 1 |
|---|
| Distinguish between tenuous, clear, and cloudy atmospheres on at least 28 temperate, terrestrial planets orbiting M and K dwarfs using $CO_2$ and other spectral features. |

| Science Observable |
|---|
| Transmission spectroscopy to detect spectral modulation between 3–10.5 μm for planets selected from terrestrial exoplanets with *a priori* known masses and radii. |

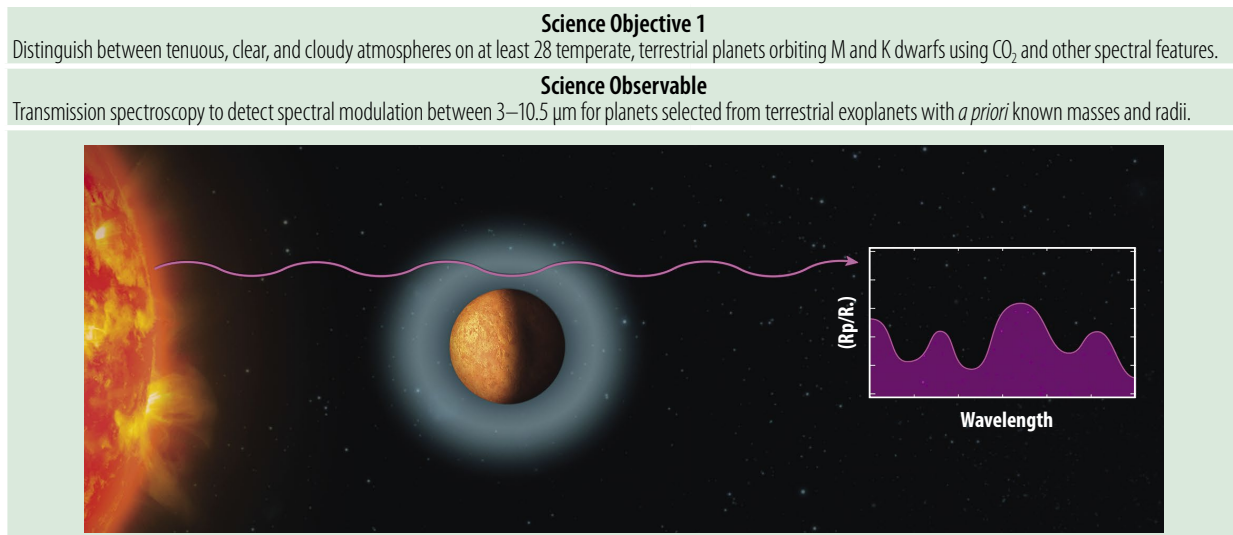

**Figure 1-47:** *Origins* **will conduct transmission spectroscopy to distinguish between tenuous, clear and cloudy atmospheres.** Illustration depicting spectral modulation in observations of an exoplanet atmosphere. During transit, stellar light is transmitted through the planet's annulus; when observed at multiple wavelengths, this yields a transmission spectrum. An atmosphere that is large and uninhibited by clouds at these wavelengths will exhibit large spectral features.

**Science Requirements**

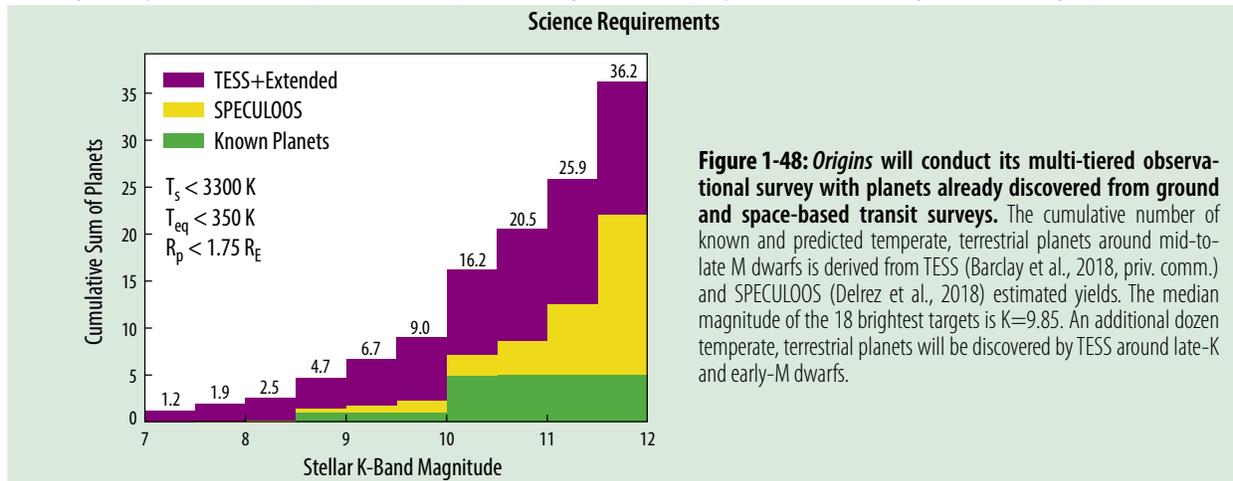

**Figure 1-48:** *Origins* **will conduct its multi-tiered observational survey with planets already discovered from ground and space-based transit surveys.** The cumulative number of known and predicted temperate, terrestrial planets around mid-to-late M dwarfs is derived from TESS (Barclay et al., 2018, priv. comm.) and SPECULOOS (Delrez et al., 2018) estimated yields. The median magnitude of the 18 brightest targets is K=9.85. An additional dozen temperate, terrestrial planets will be discovered by TESS around late-K and early-M dwarfs.

for detecting thermal emission and biosignatures for terrestrial planets, these planets do not feature in the upper observational tiers and ultimately do not drive *Origins*' science and technical requirements. Still, expanding beyond just mid-to-late M-dwarf planets in the first tier allows *Origins* to expand its discovery space, and provide a broader perspective in the search for life.

*Origins* will conduct transmission spectroscopy for at least 28 temperate, terrestrial exoplanets transiting nearby M and K dwarf stars. These planets will already have been identified from transit surveys such as TESS and SPECULOOS. They will also have well-constrained masses from ground-based radial velocity (RV) programs. Nearly all of the ~3,000 confirmed Kepler planets and over 2,200 Kepler candidate planets are in systems that are too distant (and faint) to obtain precise planetary spectra; however, TESS is performing a full-sky survey to identify the nearest transiting exoplanets, many of which will orbit bright host stars amenable to follow-up atmospheric characterization. Based on current yield estimates (Barclay *et al.*, 2018), TESS is expected to discover ~20 temperate, terrestrial exoplanets, many of which will transit cool, bright M dwarf stars. Additional habitable-zone planets could be found by doubling TESS's sector duration from 27 days to 54 days during its proposed extended mission. Habitable-zone planet discoveries are also expected from the ground, including from SPECULOOS, which is expected to identify an additional 14±5 temperate, terrestrial planets orbiting late M to early L dwarfs, which are too faint for detection with the TESS optical bandpass. Many of these systems will be amenable for follow-up studies. Figure 1-48 demonstrates exoplanet yields specifically for mid-to-late M dwarfs. With a cutoff of K=11.5, ~28 of these systems will be suitable for atmospheric characterization.

To observe an average of eight transits per planet, assuming four hours per transit observation, *Origins* would need ~900 hours to detect spectral modulation at >3.6σ in the atmospheres of these 28 exoplanets (Table 1-22). For those planets with measured spectral features, *Origins* can verify that these planets have appreciable atmospheres and, at the terminator, are not predominantly obscured by aerosols, which are likely to hinder attempts to constrain the presence of liquid water on their surfaces and search for biosignatures. Since multiple transit observations are necessary to build up sufficient signal-to-noise, the resultant spectra reveal atmospheric conditions at the terminator averaged over many orbits. These data enable the identification of trends that can be used to inform planet formation models and predict which planets are best suited for an in-depth search for signs of life.

It is possible that this initial atmospheric reconnaissance of habitable-zone planets orbiting mid-to-late M dwarf stars will be conducted with JWST. However, it is unknown whether JWST's instruments will have the precision to do this for the early-M dwarfs, let alone the late-K stars where the $CO_2$ feature size is predicted to be ≤8 ppm. Still, if JWST finds any systems with strong molecular features, this will enable *Origins* to devote more of the 4000-hour reference program to survey additional planets, particularly in the upper tiers, to detect thermal emission and biosignature gases.

### 1.3.5 Science Objective 2: Measure Apparent Surface Temperatures

> *Origins* will establish whether the apparent surface temperatures of at least 14 habitable zone exoplanets with the clearest atmospheres are consistent with liquid water at >95%.

While transmission spectra provide the initial reconnaissance of these worlds, additional observations are necessary to measure the planetary thermal emission to determine the temperature structure of their atmospheres. We select at least 17 habitable zone M-dwarf planets with the clearest atmospheres for targeted observations at secondary eclipse. Using repeated eclipse observations from 2.8 – 20 μm, we build up a high S/N planetary emission spectrum (Figure 1-49). An emission spectrum can be interpreted in terms of the pressure and temperature probed at each wavelength. The wavelengths that probe deepest (an opacity window from ~10.5 to 12 μm, as shown in Figure 1-43), down to the apparent planetary surface, reveal the pressure and temperature there. We can then determine whether these conditions could support liquid water.

**Table 1-23:** Exoplanet Biosignatures Requirements Flow 2

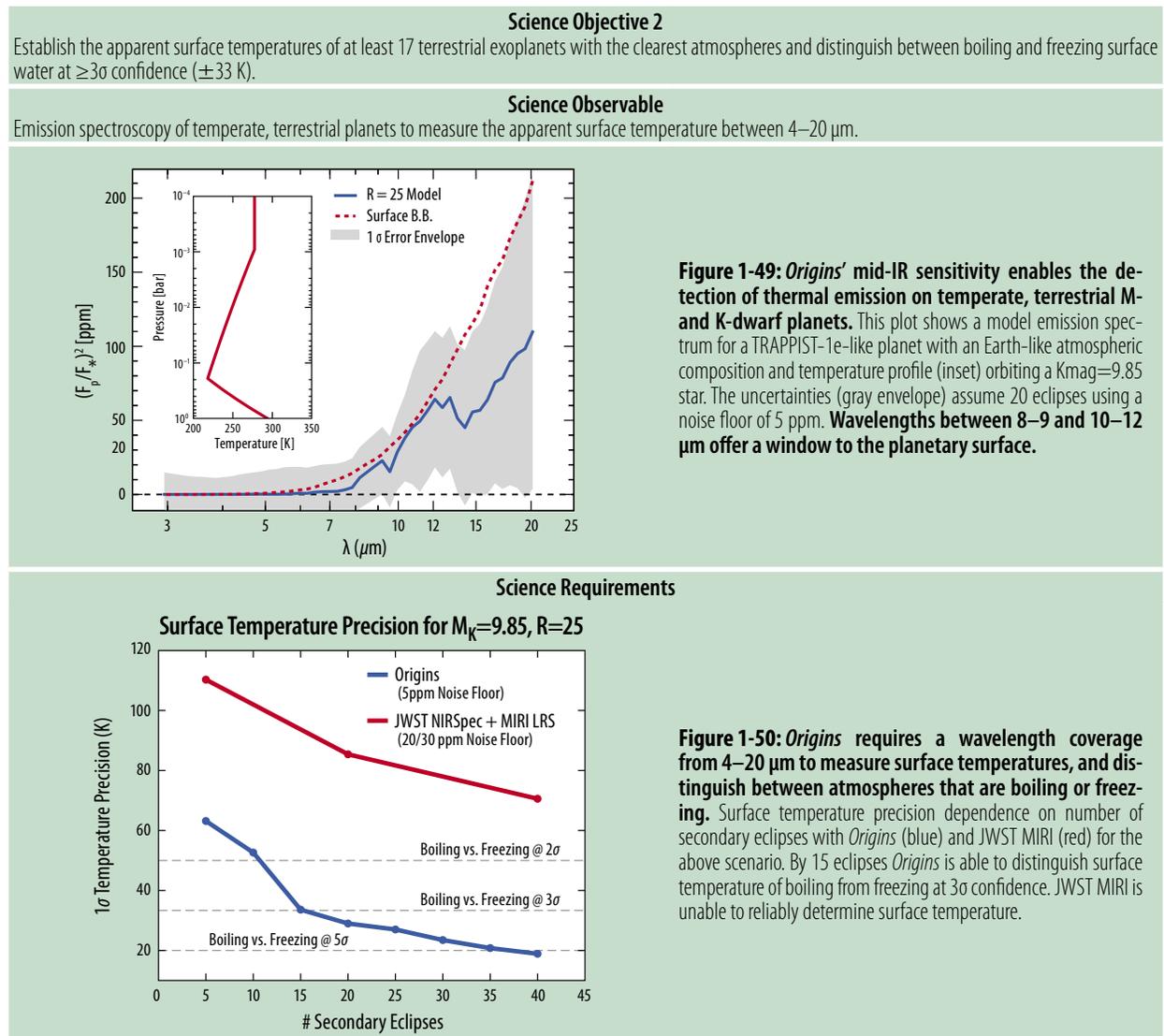

| Science Objective 2 |
| --- |
| Establish the apparent surface temperatures of at least 17 terrestrial exoplanets with the clearest atmospheres and distinguish between boiling and freezing surface water at ≥3σ confidence (±33 K). |

| Science Observable |
| --- |
| Emission spectroscopy of temperate, terrestrial planets to measure the apparent surface temperature between 4–20 μm. |

**Figure 1-49:** *Origins'* mid-IR sensitivity enables the detection of thermal emission on temperate, terrestrial M- and K-dwarf planets. This plot shows a model emission spectrum for a TRAPPIST-1e-like planet with an Earth-like atmospheric composition and temperature profile (inset) orbiting a Kmag=9.85 star. The uncertainties (gray envelope) assume 20 eclipses using a noise floor of 5 ppm. **Wavelengths between 8–9 and 10–12 μm offer a window to the planetary surface.**

**Figure 1-50:** *Origins* requires a wavelength coverage from 4–20 μm to measure surface temperatures, and distinguish between atmospheres that are boiling or freezing. Surface temperature precision dependence on number of secondary eclipses with *Origins* (blue) and JWST MIRI (red) for the above scenario. By 15 eclipses *Origins* is able to distinguish surface temperature of boiling from freezing at 3σ confidence. JWST MIRI is unable to reliably determine surface temperature.

The surface temperature is only part of the picture, however. Thermal emission spectroscopy is the only method that provides information on overall planetary climate. Reconnaissance at optical wavelength provides little guidance. The thermal emission spectrum determines a planet's full vertical temperature structure (*e.g.*, whether it has a stratosphere like Earth) and the abundance of greenhouse gases, such as $H_2O$, $CO_2$, and $CH_4$. The temperature structure is critical to assessing the planet's climate, because it depends on how incoming stellar and outgoing thermal radiation heat and cool the atmosphere.

*Origins* provides the opportunity to map the inner edge of the habitable zone through thermal emission measurements (Figure 1-50). Models suggest a dramatic atmospheric transition at $T_{eq}$=300-350 K, but there is significant uncertainty in this limit due to the role of clouds and transport of water vapor in state-of-the-art 3D models (*e.g.*, Kopparapu *et al.*, 2017). *Origins* can determine where the boundary is because greenhouse atmospheres have significantly higher surface temperatures than temperate worlds.

In addition, emission spectroscopy can yield a thermal "eclipse map" of the planet's dayside. This is achieved with time-resolved eclipse ingress and egress light curves (de Wit *et al.*, 2012; Majeau *et al.*, 2012; Mandel & Agol, 2002). Eclipse maps can show if the planet's hottest point in the atmosphere is offset from "high noon," a feature seen on Earth, which would indicate robust atmospheric circulation. With *Origins*, it is possible that eclipse observations can be used to map terrestrial planets, providing two-dimensional (or three-dimensional, since it is conducted spectroscopically) temperature maps of terrestrial exoplanet daysides for the first time. These maps, as well as spectroscopic phase-curve observations (which might be conducted as *Origins* ancillary science; see Section 1.3.10), can provide important insights to the climates of rocky M-dwarf planets.

### 1.3.6 Science Objective 3: Search for Signs of Life

> *Origins* will be able to detect biosignatures at ≥3.6σ, assuming an Earth-like atmosphere, for at least 10 promising planets identified as part of Objectives 1 and 2.

In the deepest tier of our 4000-hour observational survey, *Origins* will obtain additional transits for at least 10 targets to search for biosignatures ($CH_4 + O_3$ or $N_2O + O_3$), which would be detected at greater than 3.6σ, assuming an Earth-like atmosphere (Figure 1-51). Using this final tier of observations, we will attempt to find life outside the solar system.

**Table 1-24:** Exoplanet Biosignatures Requirements Flow 3

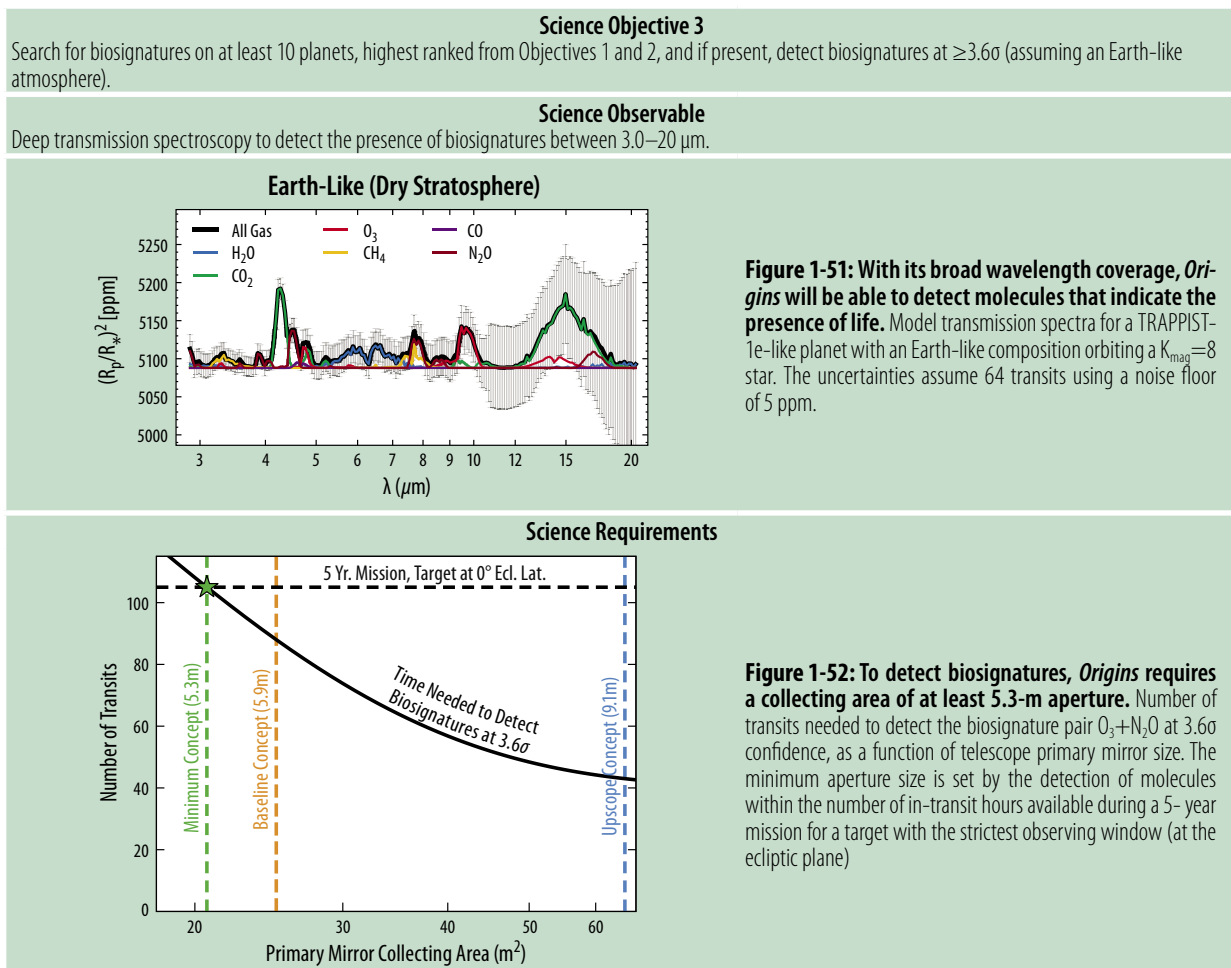

**Science Objective 3**

Search for biosignatures on at least 10 planets, highest ranked from Objectives 1 and 2, and if present, detect biosignatures at ≥3.6σ (assuming an Earth-like atmosphere).

**Science Observable**

Deep transmission spectroscopy to detect the presence of biosignatures between 3.0–20 μm.

**Earth-Like (Dry Stratosphere)**

**Figure 1-51: With its broad wavelength coverage, *Origins* will be able to detect molecules that indicate the presence of life.** Model transmission spectra for a TRAPPIST-1e-like planet with an Earth-like composition orbiting a $K_{mag}$=8 star. The uncertainties assume 64 transits using a noise floor of 5 ppm.

**Science Requirements**

**Figure 1-52: To detect biosignatures, *Origins* requires a collecting area of at least 5.3-m aperture.** Number of transits needed to detect the biosignature pair $O_3+N_2O$ at 3.6σ confidence, as a function of telescope primary mirror size. The minimum aperture size is set by the detection of molecules within the number of in-transit hours available during a 5-year mission for a target with the strictest observing window (at the ecliptic plane)

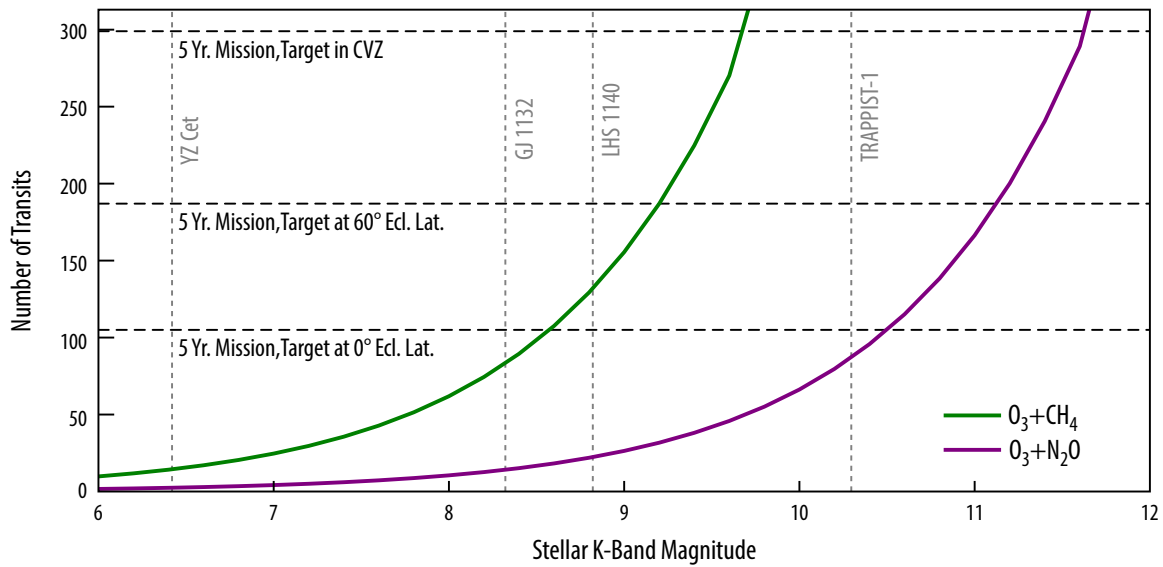

**Figure 1-53:** Number of transits needed for *Origins* to detect biosignatures at 3.6σ confidence (solid curves). For a typical host star brightness of K=9.85, *Origins* would detect the biosignature combination of $O_3+N_2O$ in ~60 transits, assuming an Earth-size planet with an Earth-like atmosphere transiting an M8 star (similar to TRAPPIST-1). Such a feat could be completed in less than 5 years, regardless of the system's ecliptic latitude (black horizontal dashed lines). The biosignature pair $O_3+CH_4$ is more challenging to detect for Earth-like compositions, but the lifetime of $CH_4$ is expected to be longer in planets orbiting M dwarfs. Longer lifetimes lead to higher abundances, which are easier to detect with fewer transits. Note that for this model atmosphere, and assuming a 20/30-ppm NIRSpec/MIRI noise floor, JWST can only detect $CO_2$ at 3.6σ confidence. CVZ = continuous viewing zone.

Because of the broad wavelength coverage of *Origins*/MISC-T, multiple bands from each molecule are accessible in a single observation, increasing the significance of the detection of each molecule, particularly for ozone, methane, and nitrous oxide (Figure 1-43). Having at least two bands per molecular species also prevents degeneracies that can arise due to overlapping spectral features. In the exciting event that high-resolution ground-based optical instruments on ELTs are able to detect $O_2$ in a handful of terrestrial planets orbiting M dwarfs (Snellen *et al.* 2013, Rodler & Lopez-Morales, 2014), *Origins* would still deliver transformational science. Such a ground- based detection would not be decisive, as oxygen alone is not a biosignature (Meadows *et al.* 2018), nor would such a detection yield any information on planetary climate. (See Section 1.3.8 for further discussion.)

While our characterization strategy is based on the ability to detect an atmosphere that resembles the current Earth, the framework is robust enough to allow for the detection of many types of habitable planet atmospheres, including the life-bearing Archaean Earth with high concentrations of $CH_4$ and $CO_2$, before the rise of atmospheric oxygen (Krissansen-Totton *et al.* 2018; see also Figure 1-54). Our broad, simultaneous wavelength coverage can enable the detection of Venus-like, Titan-like or other atmospheric compositions. Additionally, the *Origins* terrestrial exoplanet survey has enough margin to allow for changes to each observational tier. Our total survey amounts to 3996 hours of observing, leaving 4 hours of explicit margin.

However, the implicit margin assumes observing each transit/eclipse for 4 hours, when in fact only 3 hours per transit/eclipse is sufficient for characterization and the remaining hour can be used for telescope settling and overheads. Thus, reducing the number of targets or the sensitivity in one tier frees up a large number of hours. Taken together, the multi-tiered strategy we adopt for *Origins* will allow for the high-confidence identification of life on a terrestrial planet.

### 1.3.7 The Observational Impact of M-Dwarf Host Stars

Observations of transiting planets have shown that stellar activity can affect a measured spectrophoto-metric transit signal (*e.g.*, Pont *et al.*, 2008; Czesla *et al.*, 2009; Agol *et al.*, 2010; Berta *et al.*, 2011; Carter *et al.*, 2011; Désert *et al.*, 2011; Sing *et al.*, 2011; Barstow *et al.*, 2015; Zellem *et al.*, 2015; Rackham *et al.*, 2018). However, there are numerous ways to mitigate these effects, many of which become much less important in the near- to mid-infrared. In the following sections, we outline the M dwarf properties that might affect *Origins* transit and eclipse observations, and the ways in which these will be addressed.

### 1.3.7.1 M-Dwarf Activity

While M-dwarf stars can be comparatively more active than Sun-like stars, stellar activity generally has a larger impact on the observed transit spectrum in the visible than in the infrared (Oshagh *et al.* 2014; Zellem *et al.* 2017; Rackham *et al.* 2017; Morris *et al.* 2018). Indeed, recent studies have confirmed that active stars have a negligible impact on the observed infrared spectrum. For example, recent *Spitzer* observations of GJ 1214b (Fraine *et al.*, 2013) and the TRAPPIST-1 multi-planet system (Morris *et al.*, 2018), both of which transit active M-dwarf stars, find that stellar activity is negligible on the observed exoplanet signal. Similarly, *Spitzer* observations of the warm Neptune HAT-P-11b find that, despite observing multiple spot crossings, stellar activity did not alter the observed planetary signal (Fraine *et al.*, 2014). HST observations of the hot Jupiter HD 189733b, which orbits an active K-dwarf, found that stellar activity could alter the planet's visible and near-infrared spectrum, result-ing in a potential false-positive detection of atmospheric hazes, but that these effects are significantly dampened at longer wavelengths (McCullough *et al.*, 2014). Based on these studies, stellar activity is not expected to significantly impact the performance of exoplanet observations with *Origins*.

For the rare cases of an exceptionally active and bright host star featuring an exoplanet with a large transit depth, Zellem *et al.* (2017) have shown that epoch-to-epoch planetary changes due to stellar variability can be corrected. *Origins'* high-precision measurements can be used to characterize the host star over the entire MISC-T passband during the out-of-transit or in-eclipse portions of the light curve, which probe the light of the star alone. Any changes in the observed planetary transit depth, $\delta$, induced by stellar activity can be corrected (adapted from Zellem *et al.*, 2017) using:

$$\Delta\delta = \delta_{quiescent}\left(\frac{F_{quiescent} - F_{active}}{F_{active}}\right)$$

Furthermore, most temperate planets emit less flux at wavelengths shorter than ~5 μm; therefore, MISC-T's bluest channel can be used to monitor and correct for stellar variability. In addition, per-sistent, non-variable, unocculted activity (spots and plages; Rackham *et al.*, 2017) can be identified by comparing the host star's spectrum to stellar models (*e.g.*, PHOENIX spectra; Husser *et al.*, 2013, Wakeford *et al.*, 2019). Finally, *Origins* has the capability to identify epoch-to-epoch stellar variabil-ity by searching for changes in the host star's spectrum (observed in the $H_2O$ lines at >17 μm using R>250). If stellar variability is observed and found to impact the planet's spectrum, we will use the method described in Zellem *et al.* (2017) to correct for this effect.

### 1.3.7.2 M-Dwarf Star Spot Monitoring

Stars do not have uniformly luminous surfaces and the transit of planets across them produce wavelength-dependent signals that have nothing to do with the presence of planetary atmospheres. These signals are small but can limit the search for atmospheric signatures. Transiting planets crossing cooler spots produce a change in the integrated spectrum of the star (*i.e.,* making it appear transiently hotter). Planets transiting a spotted star (but not the spots) make the star appear transiently cooler (the "out-of-spot" effect). Likewise, the spectral features of molecules such as $H_2O$ or CO that are more pronounced in spots appear less or more prominent in the disk-integrated spectrum of the star during a transit. This raises the possibility that the out-of-spot effect could produce features in a planet transit spectrum that could obscure or even mimic true absorption features by any atmosphere (Rackham *et al.*, 2018).

These effects can be mitigated, or at least identified, by independent photometric or spectroscopic data at other wavelengths, but this depends on knowledge of spot coverage fraction and temperature difference, which cannot be uniquely determined from disk-integrated data. Also, the use of models to extrapolate from other observed wavelengths to the mid-infrared are limited by the fidelity of these models. A more robust approach is to distinguish between variation due to differences between regions on the star, and variation due to an (unknown) planetary atmosphere in the spectrum itself. This means features in the stellar spectrum indicative of temperature differences need to be resolved in the *Origins* wavelength range. Most of the variation between 10-30 μm model spectra of solar-metallicity M dwarfs (Allard *et al.*, 2011) for the 3200 K ("non-spot") and 2900 K ("spot") case can be recovered with a resolution of R>250. At lower resolutions, the spectra cannot be significantly distinguished, while higher resolution yields diminishing returns. The MISC-T instrument has been designed with star spot monitoring in mind, implementing a R>250 spectroscopic channel at 10.5 – 20 μm.

### 1.3.7.3 M-Dwarf Stellar Granulation

Observations of transiting exoplanets can also be affected by so-called stellar granulation— convective cells on the stellar surface that appear as bright granules surrounded by dark inter-granular lanes—which Kepler has shown can be seen as an excess of stochastic noise in ultra-precise light curves. For example, Bastien *et al.* (2013, 2016) demonstrated that, in solar-type stars, the granulation appears as an excess noise of ~15 ppm in the visible-light Kepler bandpass. Because the granulation pattern is variable in both time and space, it constitutes an irreducible systematic noise floor that is imprinted on exoplanet transit light curves and transmission spectra.

Fortunately, the situation becomes substantially better in the infrared, principally because the contrast between the bright granules and the dark inter-granular lanes is greatly reduced. For example, applying a simple Rayleigh-Jeans scaling of the above visible-light granulation noise results in an estimated noise at 10 μm of ~5 ppm. Recent studies that attempt to model the planet transit across stellar surface granules at different wavelengths confirm this estimate. Chiavassa *et al.* (2017) find that the granular noise in the transit light curve leads to an irreducible error in the measured planet radius of ~0.2% for an Earth-size planet orbiting a K-dwarf. Sarkar *et al.* (2018) find that, for the case of GJ 1214b (a super-Earth transiting an $M_4$ dwarf with a transit timescale of ~1 hr), the granulation creates irreducible noise in the spectrum at 8 μm of ~1 ppm.

The Sarkar *et al.* (2018) study used out-of-transit (*i.e.,* full integrated light) simulations and did not consider the transit chord specifically, so the estimate of ~1 ppm is likely to be an underestimate by a factor of a few. Full simulations are underway (Rackham *et al.*, in prep) but the final granulation noise for transmission spectroscopy in the mid-infrared is expected to be in the range of 1-10 ppm, being smallest at the longest wavelengths (*i.e.,* >10 μm) and for the longest transit durations (*i.e.,* >1 hr). Therefore, it is not expected that stellar granulation will provide a large noise source for *Origins* observations of transiting planets.

### 1.3.8 Origins Compared to Other Facilities

> *Origins* offers transformative studies of exoplanet atmospheres unmatched by other telescopes.

With the launch of JWST, the characterization of transiting exoplanets will take another step forward. JWST's larger collecting area, broader wavelength coverage, and higher resolution compared to HST will yield unparalleled constraints in atmospheric composition, thermal structure, and dynamics, thus revolutionizing our understanding of these distant worlds and placing our own solar system into context.

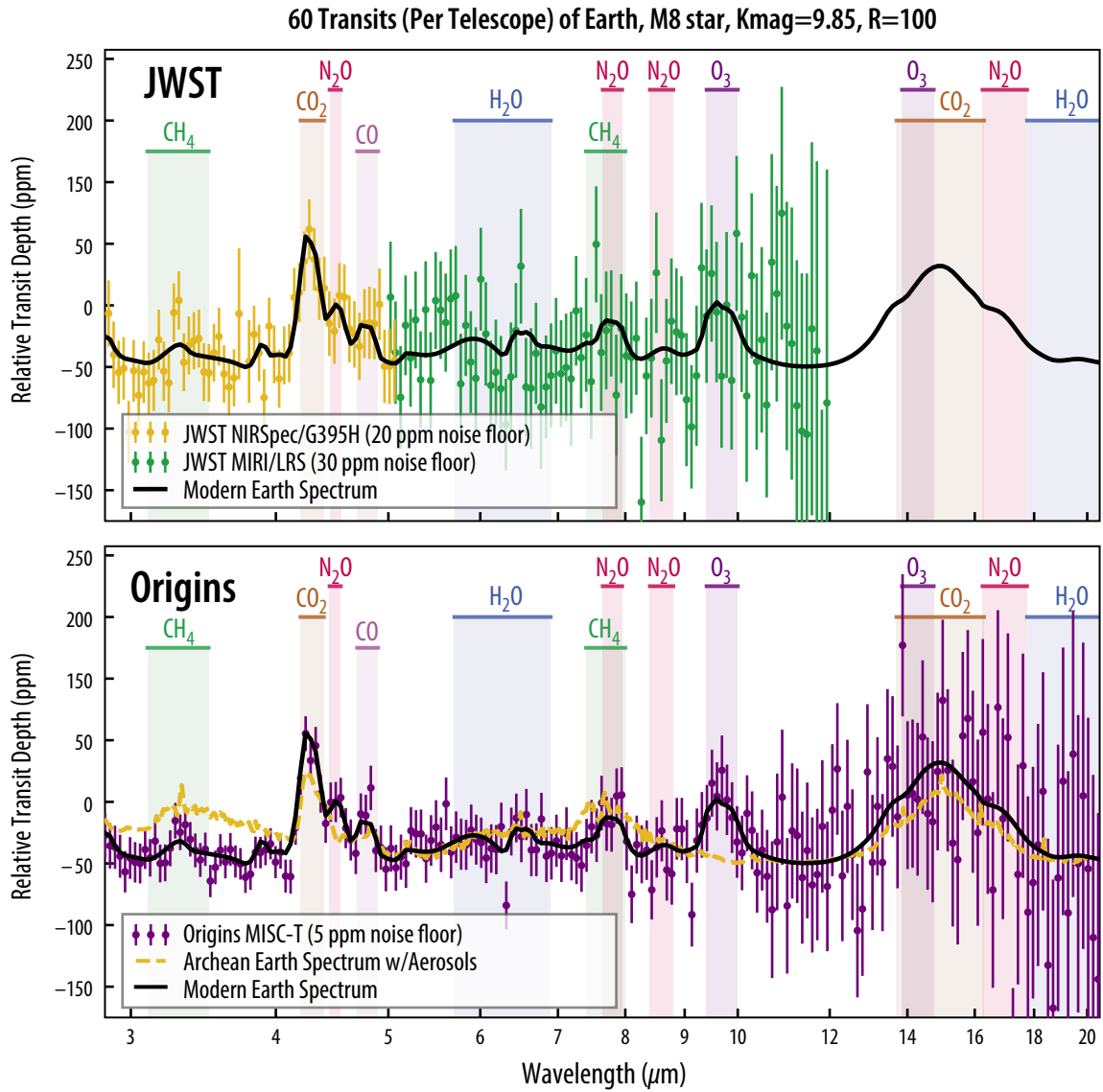

**Figure 1-54:** With its broad wavelength coverage and anticipated noise floor of 5 ppm, *Origins*/MISC-T is designed to detect key habitability indicators and biosignature gases. These plots show simulated transmission spectra for a TRAPPIST-1e-like planet with an Earth-like composition (60 transits, R = 100), comparing JWST and *Origins* measurements based on current best estimates for their respective instrument noise floors (20 − 30 ppm for JWST; 5 ppm for *Origins*).

Meanwhile, *Origins*' broad, simultaneous wavelength coverage from 2.8-20 μm will enable the detection of multiple bands of key biosignature and habitability indicator gases. *Origins*' simultaneous wavelength coverage is vital not only to measure atmospheric abundances, but also planetary temperatures and signs of life. In comparison, JWST will have spectral coverage from 1-14 μm in multiple observing modes, with an anticipated drop in performance longward of 10 μm, limiting its access to these key molecules (Figure 1-54, top panel; see also Batalha *et al.* 2018). It is at these longer wavelengths where *Origins* emission spectroscopy will measure overall planetary climates (Figure 1-49 and Figure 1-50). In addition, *Origins*/MISC-T will baseline a 5-ppm detector noise floor. In comparison, experiences from *Hubble* and *Spitzer* and current predictions point to a noise floor for JWST observations of transiting exoplanets of >20-30 ppm (Greene *et al.* 2016). Given this potential noise floor, the JWST Science Working Group has approved the following statement emphasizing caution regarding the potential characterization of terrestrial exoplanets. "Caution is required in predicting what JWST will and will not be able to do. We know JWST observers will look for atmospheric chemistry and biosignatures of small planets through the transit technique. But we don't know how well JWST will work and we don't know what Nature has given us. In particular, we do not yet know the noise floor for any of the instruments, and we will not know until well into the mission…"

Therefore, *Origins*/MISC-T will be expressly designed not only to detect terrestrial exoplanet atmospheres, but also to spectrally measure the abundances of key molecular gases (Figure 1-54, bottom panel; Figure 1-55). With its broad, simultaneous wavelength coverage and low noise floor, *Origins* will be uniquely capable of detecting life.

ESA's ARIEL space telescope is also focused on transiting planet science, with a wavelength coverage from 2-8 μm, and possibly visible light photometry. The ARIEL science case is distinct from *Origins*, and it does little detailed atmospheric characterization. The ARIEL science case is built upon a lower S/N statistical assessment of atmospheres of 1000 planets, essentially reconnaissance of these worlds to look for general trends in spectra with planetary mass and temperature over a large sample size. Due to its small mirror (80 cm), ARIEL will deliver low S/N spectra compared to JWST and *Origins* per hour of observation. It will focus mostly on hot planets that are Neptune-size and larger and will not touch the habitable planet population.

ESA's SPICA will have essentially no impact on the study of exoplanetary atmospheres, which is not even discussed as a science case in SPICA documentation. At its shortest wavelength (12 μm), parent stars are quite dim, yielding low S/N detections of transits or eclipses given SPICA's relatively small, 2.5 m mirror.

The ground-based Extremely Large Telescopes (ELTs) include science goals of probing habitable zone rocky exoplanets around M stars. If properly instrumented, ELTs can use very high-resolution spectroscopy (R>100,000) to search for $O_2$ in the optical in transit (Snellen *et al.* 2013, Rodler & Lopez-Morales 2014, Serindag & Snellen, 2019), which is the most developed science case. This same science could be extended across the near-infrared for detections of $CH_4$ and $H_2O$. For the nearest few target stars, high-contrast AO and high-resolution could be combined for biosignatures detection, however, thermal background noise limits this ground-based approach to wavelengths shorter than 5 μm (Snellen *et al.* 2015). It is important to keep in mind that high-spectral resolution observations until now have provided only molecular detections, not abundance determinations, due to the loss of continuum information (Brogi & Line 2018). Space-based observations of ELT transiting targets with *Origins* would complement any earlier detections by yielding atmospheric abundances from transmission spectra and a robust view of planetary climate from secondary eclipse spectra.

Another ground-based avenue is thermal emission. Emission from M-star habitable zone planets is within reach for a handful of systems at N-band (10 μm) (Quanz *et al.*, 2014) but will likely come via photometry, rather than spectroscopy. Taken together, the expected yield of characterizable

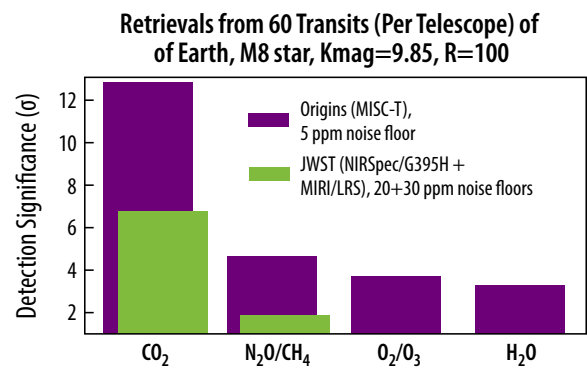

**Figure 1-55:** Detection significance comparison between JWST and *Origins*. Assuming noise floors of 20 and 30 ppm for NIRSpec/GRISM and MIRI/LRS, respectively, JWST will only be sensitive to $CO_2$ at >3.6σ. With a 5- ppm noise floor, *Origins* will be sensitive to $CO_2$ and the biosignature pair $O_3+N_2O$. Figure courtesy Kevin Stevenson (STScI).

M-dwarf exoplanets would be ~10, pending the advancements in instrumentation, as well as requisite observing time. Such visible or thermal IR observations would be entirely complementary to *Origins* transit and emission spectra.

Overall, at the time of *Origins* launch we expect to have some knowledge about M dwarf habitable-zone planets, such as whether a planet has an atmosphere or contains potential biosignatures gases such as $O_2$, but the picture will be incomplete. *Origins* will determine the atmospheric compositions and thermal structures of these potentially habitable planets, thus opening the door for comparative exoplanetology of rocky worlds.

### 1.3.9 Origins' Mid-Infrared Spectrometer and Camera Transit spectrometer (MISC-T)

The instrument that will deliver transformational spectra of exoplanet atmospheres for *Origins* is the Mid-Infrared Spectrometer and Camera Transit spectrometer (MISC-T). MISC-T will operate from 2.8 – 20 μm, with a typical spectral resolution of R~100 across these wavelengths, reaching R~250 at the longest wavelengths. This entire spectral range will be monitored simultaneously. This simultaneous coverage eliminates the significant uncertainty introduced from other observational platforms (including JWST) of stitching observations together across multiple wavelength ranges for observations taken at different times. MISC-T has a required spectro-photometric stability of 5 ppm on timescales of hours to days. Such a low noise floor enables the co-addition of multiple transit observations to enable the detection of spectral signatures in molecules of temperate Earth-sized planets. See Section 3.2 for technical details on the MISC-T instrument.

### 1.3.10 Ancillary Exoplanet Science with Origins

*Origins* will be capable of performing additional exoplanet-related science over a broad range of planet types, such as the characterization of temperate giant planets, understanding the nature of clouds, and phase-curve observations of terrestrial exoplanets.

#### Characterizing Temperate Giant Planets

The last decade of exoplanet science has shown there is a broad continuum of worlds from sub-Earths to super-Jupiters around Sun-like stars. For giant planets with hydrogen-dominated atmospheres, which range from ~10 to ~1000 Earth masses, even after JWST we will be sorely lacking in understanding these atmospheres across a wide range of temperatures. In particular, the best characterized JWST and ARIEL targets will be far hotter (~600 K and above) than the solar system's cold gas giants. *Origins'* wavelength coverage and sensitivity will enable transformational science in the chemistry and structure of temperate giant planets around stars from types A to M. In these cooler atmospheres, the abundances of $CH_4$ and $NH_3$ probe metallicity, non-equilibrium chemistry, and the strength of vertical mixing (Zahnle & Marley, 2014). *Origins* can probe the rich photochemistry expected for these atmospheres, via production of HCN, $C_2H_2$, $C_2H_4$, and $C_2H_6$, from methane as is seen in Jupiter, beyond 5 μm. The population of giant planets *Origins* can probe in thermal emission (down to ~250 K) will be cool enough to, for the first time, serve as a bridge to the solar system's very cold (~100 K) worlds.

#### Understanding the Nature of Clouds in Planetary Atmospheres

*Origins* can also advance our understanding of the physics and chemistry of clouds, which are currently a major source of uncertainty in models of these atmospheres. Since most clouds appear only as Rayleigh or gray scatterers in the optical and near-IR, we have no handle on the composition of clouds whose presence we infer today. However, a number of cloud species known or expected to impact the spectra of these planets over a wide temperature range have Mie-scattering features in the mid-infrared (Wakeford & Sing, 2015). *Origins* would enable identification of the clouds that impact these planets.

## Phase Curves of Terrestrial Exoplanets

Similar to giant exoplanet phase-curve observations conducted with HST and eventually JWST, *Origins*/MISC-T will be able to measure phase-resolved thermal emission (*i.e.,* spectroscopic phase curves) of potentially habitable M-dwarf planets. Because these close-in planets are expected to be tidally locked and hence synchronously rotating, rotation periods are expected to be on the order of days to weeks—much longer than the rotation periods of many solar system planets. These properties strongly influence a planet's atmospheric circulation (*e.g.,* Carone *et al.*, 2018). For slower rotators (P>20 days), the circulation of potentially habitable M-dwarf planets can exhibit weak radial flow from dayside to nightside and the so-called "eyeball" climate state, in which a liquid water ocean only exists near the substellar point (*e.g.,* Pierrehumbert, 2010; Angerhausen *et al.*, 2013). Comparatively faster rotators (P~1 day) can exhibit broad upper atmosphere super-rotation, or super-rotating, high-latitude jets with banding (Carone *et al.*, 2018). If the planet is eccentric, pseudo-locking of its orbit and rotation may result in weak seasonal climatic cycles (*e.g.,* Driscoll & Barnes, 2015). Each of these circulation patterns influence the atmosphere's distribution of heat and the formation and transport of aerosols, all of which shape thermal phase curves.

Therefore, *Origins* phase-curve observations of terrestrial M-dwarf planets can be used to constrain a wealth of properties. First, they can be used to distinguish between planets with and without atmospheres (*e.g.,* water-rich versus water-poor), and also to confirm the presence of dayside clouds (Yang, Cowan and Abbot, 2013). Second, phase-curve observations can constrain the planet's rotation rate—the morphology of thermal phase curves can differentiate synchronous rotators from those in spin-orbit resonances (Wang *et al.*, 2014). This is also applicable to eccentric planets (*e.g.,* Boutle *et al.*, 2017). Third, the amplitude and morphology of phase-curve observations can help distinguish between different climatic states (Wolf, 2017; Kopparapu *et al.*, 2017). Climate modeling results show that, for terrestrial planets around M-dwarfs, there is a clear distinction between runaway and non-runaway greenhouse climate states.

Lastly, phase-curve observations with *Origins* can also be conducted for non-transiting planets. Like transiting planets, they can be used to detect an atmosphere and distinguish between different atmospheric compositions (*i.e.,* dry planets vs. "aquaplanets"), but also between different rotation rates (Turbet *et al.*, 2016). Importantly, thermal phase curves of non-transiting planets can be used to measure the planet's inclination, as has been demonstrated for hot Jupiters (Crossfield *et al.*, 2010).

### 1.3.11 Towards Understanding Habitable Planets with Origins

The *Origins* path to characterizing potentially habitable planets' atmospheres uses the context only transiting planets can provide. *Origins* will probe the atmospheres of catalogued exoplanets with known masses, radii, densities, and orbits. *Origins* will utilize the techniques of atmospheric transmission spectra and emission spectra honed by nearly two decades of practice and optimization, which will be further refined in the JWST era. However, *Origins* is unique because it will provide high precision access to thermal infrared wavelengths. These wavelengths allow a direct constraint of a planet's atmospheric temperature structure and provide multiple bands covering the important molecules of biological interest. The entire era of exoplanetary science has shown that Nature's imagination is richer than our own, so *Origins*' broad wavelength coverage will give us views into the new and unexpected. The 3-20 micron wavelength coverage provided by *Origins* provides the best access to multiple molecules that are habitability indicators or biosignatures (in particular, $O_3$, $CH_4$, $CO_2$, $N_2O$) or allow false positive scenarios to be ruled out (in particular, $CH_4$ and $CO$) for both Modern and Archaean Earth-like atmospheres. Ultimately, The *Origins* path to discovering potentially habitable worlds is one where we can understand the atmospheric composition of the worlds we see.

## 1.4 Origins Science Traceability Matrix

The *Origins* science case described in Sections 1.1, 1.2, and 1.3, defines the *Origins* Baseline Mission Concept (Section 2.0). Executing an observation requires understanding the choreography needed for the observatory and instruments to work together during a measurement. The *Origins* team took a holistic approach to the observatory by understanding not only the specific instruments and telescope needed to make a measurement but also how the observatory works to make the observation effectively and efficiently.

The three *Origins* science themes and their science objectives are captured in the mission's Science Traceability Matrix (STM; Table 1-25). The STM's six columns show the flow-down from NASA Science Goals (Column 1), to the *Origins* Science theme Goal or Question (Column 2), the prioritized Science Objectives (Column 3), to Science Requirements (Column 4), to the Instrument/Telescope Requirements (Column 5), and Mission Requirements (Column 6). The STM rows are the three science themes. Science Requirements (Column 4) has two sub-columns: the Science observable and its corresponding Measurement Requirement. The four Instrument Requirement (Column 5) sub-columns show the parameter, technical requirement needed to achieve the science goal, the instrument (Section 3.0), and the current best estimate (CBE) performance of that instrument based on the study. The Science Goals are traced from left to right (Columns 1 to 5). Mission Requirements (Column 6) presents overarching mission capabilities needed to support a science theme. Table 2-2 in Section 2.2 also captures the Mission Functional requirements, such as telescope collecting area and observatory pointing stability, as these parameters also stem from the science measurement requirements. A detailed list of mission requirements is given in Appendix B.

The science questions, objectives and requirements in Table 1-25 summarize the science cases described in Section 1.1 (extragalactic science), 1.2 (water trail and planetary system formation), and 1.3 (exoplanet science) of this report. We refer the reader to these sections for quantitative explanations of the science requirements and how they drive technical requirements for the instruments. In particular, each of the science sections contains tables showing quantitative support for three science objectives and the flow-down to measurement requirements and technical performance requirements for the mission. For theme 1, extragalactic science objectives (Section 1.1), Tables 1-4, 1-5 and 1-6 delineate the flow from objectives to measurements, and Tables 1-8 and 1-9 show quantitative expectations for the surveys envisioned for the science measurements. For theme 2, water trail and planetary system formation (section 1.2), Tables 1-12, 1-15 and 1-15 show the flow from objectives to requirements. In section 1.3, search for life in planets orbiting M-dwarfs, the science flow is shown in Tables 1-11, 1-23 and 1-23 for science objectives 1 to 3, respectively. The far-infrared polarization requirement is in support of Galactic astronomy (A.4 & A.9), as described in Appendix A, which features example general observer programs. In addition to these science sections, Appendix E provides supporting quantitative calculations and analysis results.

The requirements and corresponding margins between predicted performance and requirements will be refined during Phase A. As part of this process, the *Origins* team will analyze the science impact of a failure to meet specific requirements, and will develop a minimum science case for the mission.

Several *Origins* programs (*e.g.*, Extra-galactic, Section 1.1) require large-area sky surveys, with the consequence that *Origins* needs to accomplish large-area surveys efficiently. To support such surveys, the mission required an observatory survey mode that supports 60 arcsecd per second motion. The optimal performance of far-infrared direct detectors are such that rapid motion of a source over the detectors is required, which necessitates on-sky scanning or motion of a field steering mirror. The instrument and observatory need stability to fulfill the mid-infrared spectroscopy of exoplanet goals. Targets are located all over the sky, which means the field of regard over the year must cover the entire sky. *Origins*' >80% observing efficiency requirement guides attitude control system development and operational scenarios outlined in the science cases.

The science requirements specified in the STM set the minimums necessary to achieve the science goals. However, the *Origins* design is roughly 25% more capable than strictly required (*e.g.* Figure ES-7), leading to what is denoted as "science margin" in the scientific capabilities. The margin can be interpreted in two ways: margin captures the uncertainty in the model predictions or the margin implies that the scientific objectives can be reached with lower integration times than predicted in calculations for the report. In addition, the Baseline Mission provides a good starting point upon *Origins* selection, with the flexibility to explore alternatives during an engineering Phase A study. For example, Figure ES-7 shows the fraction of each science theme that can be accomplished compared with the telescope diameter (collecting area). The Baseline Mission has the same collecting area as JWST, but has a round primary mirror (5.9-m diameter) that fits without deployments in the large launch vehicles under development. This telescope size provides an estimated 25% margin for *Origins*' science driver cases.

**Table 1-25:** *Origins* Science Traceability Matrix: Theme 1, Extragalactic Science

| NASA Science Goals | *Origins* Science Goal/ Question | Science Objectives | Science Requirements | | Instrument Requirements | | | | Mission Requirements | |
|---|---|---|---|---|---|---|---|---|---|---|
| | | | Science Observable | Measurement Requirement | Parameter | Technical Requirement | Instr | CBE Performance | Driver | Parameter |
| How does the Universe work? | How do galaxies form stars, build up metals, and grow their central supermassive black holes from reionization to today? | Objective #1 and #2: Measure the redshifts, star formation rates and black hole accretion rates in main-sequence galaxies since the epoch of reionization, down to a SFR of 1 $M_\odot$/yr at cosmic noon and 10 $M_\odot$/yr at z~5, performing the first unbiased survey of the co-evolution of stars and supermassive black holes over cosmic time. Measure the metal and dust content of at least $10^5$ galaxies out to z = 6 as a function of cosmic time, morphology, and environment, tracing the rise of heavy elements, dust, and organic molecules. | A catalog of $10^6$ galaxies with star-formation rates and the black hole accretion rates measured using mid and far-IR emission lines. | Moderate (R $\geq$ 100) spectral resolution spectral mapping survey spanning 0.5 deg$^2$ and 20 deg$^2$ to deep and shallow depths, respectively. | Wavelength range | 25–500 μm | OSS-grating | 25–588 μm | Point source sensitivity, stability and systematic error control | • To meet Objectives #1- #3 in a 4000hr observing campaign, telescope aperture with diameter > 5.0m. Source confusion requires a diameter > 3.0m. • To meet Objectives #1- #3, a cold aperture with a temperature < 6K. • Down to a line flux sensitivity of $10^{-19}$ W m$^{-2}$ ability to map better than 0.15 deg$^2$/hr and efficient scan mapping at a rate as high as 60 arcsec/sec. • To enable access to all targets of interest, the field of regard shall be 4π sr over the course of the mission. • To enable object matching with near-IR counterparts, the pointing accuracy after ground processing shall be ≤ 0.5 arcsec at 50 μm. • To enable survey sciences, FIP and OSS must allow parallel observations. • The spacecraft shall enable a data transmission of 21 Tbits/day (99% of data collected/day with FIP and OSS surveys in parallel). |
| | | | | | Spectral resolving power R=λ/Δλ | $\geq$ 100 | | 300 | | |
| | | | | | Angular resolution | ≤ 8″ at 100 μm | | 4.3″ at 100 μm | | |
| | | | | | Number of beams (beams capture field-of-view) | 50 at 100 μm | | 60 at 100 μm; 96 at 25 μm; 36 at 500 μm | | |
| | | | | | Spectral line sensitivity | 1.5 x 10$^{-20}$ W m$^{-2}$ at 250 μm (1 hr; 5σ) | | 8 x 10$^{-21}$ W m$^{-2}$ at 250 μm (1 hr; 5σ) | | |
| | | | Multi-tiered survey leveraging a deep 1 deg$^2$ 2 μm imaging from JWST NIRCAM, a ~500 deg$^2$ medium depth survey for large-scale structure overlapping with WFIRST-HLS, and map the full 10,000 deg$^2$ for comparison with LSST and Euclid. | Extragalactic: In a deep integration the ability to resolve the CIB at 50 μm and de-blend the 250 μm map. Galactic: Ability to map star-forming regions, including point sources with flux densities ≤ 0.5 Jy at 50 μm and ≤ 1 Jy at 250 μm. | Wavelengths | 50 and 250 μm | FIP- continuum mapping | 50 and 250 μm | | |
| | | | | | Angular resolution | ≤ 3″ at 50 μm to resolve > 99% CIB | | 2.1″ | | |
| | | | | | Flux sensitivity | 1.75 μJy (5σ) at 50 μm over 1 deg$^2$ in 400 hours. 3.8 μJy (5σ) at 250 μm over 1 deg$^2$ in 25 hours. | | 0.2 μJy (5σ) at 50 μm over 1 deg$^2$ in 400 hours. 0.6 μJy (5σ) at 250 μm over 1 deg$^2$ in 25 hours. | | |
| | | | | | Polarization sensitivity | 1% (3σ) in linear and circular polarization | | 0.1% (3σ), 1 degree in pol angle | | |
| | | | Objective #3: To establish an accurate model of galaxy growth and evolution, measure galactic outflows in at least $10^3$ galaxies as a function of luminosity over the past 10 Gyr to determine the relative role of supernovae and AGN feedback. | Diagnostic mid-IR and far-IR OH 34.6 μm, 53.3 μm, 79.1 μm, 119 μm lines for z = 0 to 3 for 1000 galaxies, with higher resolution follow-up observations for 100 galaxies. | For follow-up observations of 1000 galaxies from Objective #1, ability to spectrally resolve to better than 100 km s$^{-1}$ (assuming S/N = 5) and flux sensitivity to detect $10^{11}$ $L_\odot$ source at a redshift of 3. | Wavelength range | 35 – 476 μm | OSS-FTS | 25–588 μm | | |
| | | | | | Spectral resolving power R=λ/Δλ | $\geq$ 3000 over the wavelength range | | 43,000 x (112 μm/λ) as provided by FTS element | | |
| | | | | | Angular resolution | ≤ 8″ at 100 μm to avoid source confusion at λ<100 μm | | 4.3″ | | |
| | | | | | Spectral line sensitivity | 2 x 10$^{-20}$ W m$^{-2}$ at 119 μm (1 hr; 5σ) | | 8x10$^{-21}$ W m$^{-2}$ at 119 μm (1 hr; 5σ) | | |

**Table 1-25 (cont.):** *Origins* Science Traceability Matrix: Theme 2, Water

| NASA Science Goals | *Origins* Science Goal/ Question | Science Objectives | Science Requirements | | Instrument Requirements | | | | Mission Requirements |
|---|---|---|---|---|---|---|---|---|---|
| | | | Science Observable | Measurement Requirement | Parameter | Technical Requirement | Instr | CBE Performance | Driver | Parameter |
| How did we get here? | How do the conditions for habitability develop during the process of planet formation? | Objective #1: Measure the water mass at all evolutionary stages of star and planet formation and across the range of stellar masses, tracing water vapor and ice at all temperatures between 10 and 1000 K. | $H_2O\ 2_{12}-1_{01}$ 179.5-µm line emission in protoplanetary disks. Also, representative $H_2O$ lines with upper level energies between 114 K and 5000 K in the spectral range 25 to 548 µm (longest is $H_2^{18}O\ 1_{10}-1_{01}$ at 547.4-µm) | Detect 179.5 µm and other representative $H_2O$ emission lines out to 547.7 µm in a 0.7 x MMSN disk around a 0.25 solar mass star at the distance of Orion. | Wavelength range | 25 to 548 µm | OSS-FTS | 25 to 588 µm | Point source sensitivity, stability and systematic error control | • To meet objectives #1–#3 in a total of 4000 hour observing campaign, a telescope collecting area of at least 20 m² (5.0m diameter) to meet the sensitivity requirements.<br>• To meet Objectives #1–#3, a cold aperture with a temperature < 6K.<br>• To meet objective #3 with a sufficient number of comets, a minimum mission lifetime of 5 years.<br>• To meet objective #3 ability to track on Solar system objects with a rate as low as 60 mas/s.<br>• To meet objective #1 with a detection of the ground-state water $H_2O\ 2_{12}-1_{01}$ 179.5-µm line, a cold aperture with a temperature < 6 K.<br>• To enable access to all targets of interest, the field of regard shall be 4π sr over the course of the mission. |
| | | | | | Angular resolution | ≤ 15" to avoid source confusion at Orion | | 7" FWHM at 179.5 µm | | |
| | | | | | Spectral line sensitivity | $1 \times 10^{-20}$ W m⁻² at 179.5 µm (1 hr; 5σ) | | $7 \times 10^{-21}$ W m⁻² at 179.5 µm (1 hr; 5σ) | | |
| | | | | | Spectral resolving power R=λ/Δλ | 25,000 at 179.5 µm | | 27,000 at 179.5 µm | | |
| | | | | Spectrally resolve H2O 179.5 µm emission line enabling Doppler tomographic measurements, with continuum sensitivity for sources ≤ 10 Jy at 179.5 µm. | Wavelength range | 179.5 +/- 10 µm | OSS-etalon | 100 to 200 µm | | |
| | | | | | Continuum Saturation limit | 10 Jy at 179.5 µm | | 100 Jy at 179.5 µm | | |
| | | | | | Spectral resolving power R=λ/Δλ | 200,000 at 179.5 µm | | 202,785 at 179.5 µm | | |
| | | Objective #2: Determine the ability of planet-forming disks at all evolutionary stages and around stars of all masses to form planets with masses as low as one Neptune using the HD 1−0 line to measure the total disk gas mass. | Use the HD J=1- 0 line at 112 µm as a tracer of total gas mass. | Detect HD J = 1−0 line emission at 112 µm in a 0.7 x MMSN disk around a solar mass star at d = 125 pc. Ability to study disks ≤ 10 Jy at 100 µm | Wavelength range | 112 +/- 10 µm | OSS-FTS | Same as above | | |
| | | | | | Spectral line sensitivity | $1 \times 10^{-20}$ W m⁻² at 112 µm (1 hr; 5σ) | | Same as above | | |
| | | | | | Spectral resolving power R=λ/Δλ | 40,000 at 112 µm | | 43,000 at 112 µm | | |
| | | Objective #3: Definitively determine the cometary contribution to Earth's water by measuring the D/H ratio with high precision (<0.1 VSMOW*) in over 200 comets in 5 years.<br>*VSMOW= Vienna Standard Mean Ocean Water | D/H ratio in water via observations of the water ($H_2O$) and deuterated water (HDO) emission lines from periodic and Oort cloud comets. | Detect water $H_2O$ (243.7, 273.2 µm), $H_2^{18}O$ (272.1 µm), and deuterated water HDO (234.6 µm) line emission, in a typical comet. Ability to study D/H ratios for comets ≤ 3 Jy at 234.6 µm. | Wavelength range | 230 - 275 µm | OSS-FTS | Same as above | | |
| | | | | | Angular resolution | ≤ 16" at 234.6 µm | | 10 arcsec at 234.6 µm | | |
| | | | | | Continuum Saturation limit | 3 Jy at 128 µm | | 5 Jy at 128 µm | | |
| | | | | | Spectral resolving power R=λ/Δλ | ≥ 10,000 | | 37,000 at 128 µm | | |

**Table 1-25 (cont.):** *Origins* Science Traceability Matrix: Theme 3, Exoplanet Science

| NASA Science Goals | *Origins* Science Goal/ Question | Science Objectives | Science Requirements | | Instrument Requirements | | | | Mission Requirements | |
|---|---|---|---|---|---|---|---|---|---|---|
| | | | Science Observable | Measurement Requirement | Parameter | Technical Requirement | Instr | CBE Performance | Driver | Parameter |
| Are We Alone? | Do planets orbiting M-dwarf stars support life? | Objective #1: Distinguish between tenuous, clear, and cloudy atmospheres on at least 28 temperate, terrestrial planets orbiting M and K dwarfs using $CO_2$ and other spectral features. | Transmission spectra sensitive to $CO_2$ (4.2 μm), $H_2O$ (6.3 μm), $O_3$ (9.7 μm), and $CH_4$ (3.3 μm). | Simultaneously 3.0–10.5 μm transmission spectra with an average of eight 4- hour transits to achieve ≥3.6σ detection of spectral features, for stars with $K_{mag}$ ≥ 4 to 11.5 | Wavelength range | 3–10.5 μm simultaneous | MISC-Transit Spectrometer | 2.85–20.5 μm | Point source sensitivity, stability and systematic error control | • Pointing jitter and drift shall not compromise the ability of MISC–T to make transit/eclipse spectroscopy measurements. From spacecraft attitude control, jitter must be ≤ 50 mas (rms) at > 1 Hz, and the drift must be ≤ 50 mas (rms) over 1–10 hours. • Including the MISC-T feedback to the field steering mirror, jitter must be ≤ 10 mas (rms) over 1–10 hours. • Spatial Resolution provided by an equivalent field-stop radius ≤ 2.5" at 2.8–10.5 μm and ≤ 1.5" at 10.5–20 μm • To meet objectives #1–#3 in a total of 4000 hour observing campaign, a telescope collecting area of 22.6 m² (5.3m diameter) • To enable access to all targets of interest, the field of regard shall be 4π sr over the course of the mission. |
| | | | | | Spectral resolving power R=λ/Δλ | 50–100 | | 50–100 | | |
| | | | | | Sensitivity | K~9.85 mag M-type star SNR/sqrt(hr) > 10,000 @ 3.3 μm | | K~9.85 mag M-type star SNR/sqrt(hr) > 12,900 @ 3.3 μm | | |
| | | | | | Saturation/Brightness Limit | $K_{mag}$ = 4.0 star | | $K_{mag}$ =2.9 of star Or 30 Jy at 3.3 μm | | |
| | | | | | Spectroscopic precision | ≤ 5 ppm at λ = 3-10.5 μm (R=50) | | < 5 ppm at 10 μm (R=50) | | |
| | | Objective #2: Establish the apparent surface temperatures of at least 17 terrestrial exoplanets with the clearest atmospheres and distinguish between boiling and freezing surface water at ≥3σ confidence (± 33 K). | Emission spectra with sufficient time resolution to resolve ingress and egress. | Simultaneously 4.0–20 μm emission spectra with an average of 15 4-hour eclipses to achieve 33 K 1σ precision on the apparent surface temperatures. | Wavelength range | 11–20 μm simultaneous | | Same as above | | |
| | | | | | Spectral resolving power R=λ/Δλ | R > 250 at λ > 17 μm | | 260–295 over 17–20.5 μm | | |
| | | | | | Spectroscopic precision | ≤ 20 ppm at 15 μm (R = 100) | | ≤ 15 ppm at 15 μm (R = 100) | | |
| | | | | | Eclipse monitoring cadence | ≤ 60.0 sec for $K_{mag}$ = 11.5 star | | Sensitivity and precision above | | |
| | | Objective #3: Search for biosignatures on at least 10 planets, highest ranked from Objectives 1 and 2, and if present, detect biosignatures at ≥3.6σ (assuming an Earth-like atmosphere) | Transmission spectra sensitive to $CH_4$, $N_2O$, and $O_3$. | Simultaneously 3.0–20 μm transmission spectra with an average of 52 4-hour transits to achieve ≥3.6σ detections of biosignatures (if present). | Wavelength range | 3.0–20 μm simultaneous | | Same as above | | |
| | | | | | Spectral resolving power R=λ/Δλ | Same as above | | Same as above | | |
| | | | | | Angular resolution | ≤ 5 ppm at λ = 3–10.5 μm (R=50); ≤ 20 ppm at λ = 10.5–20 μm (R=100) | | ≤ 5 ppm at 10 μm (R=50); ≤ 20 ppm at 20 μm (R=100) | | |
| | | | | | Spectral line sensitivity | ≥125 days of visibility per year for any given target | | 130 days of visibility at the ecliptic; more at higher latitudes | | |

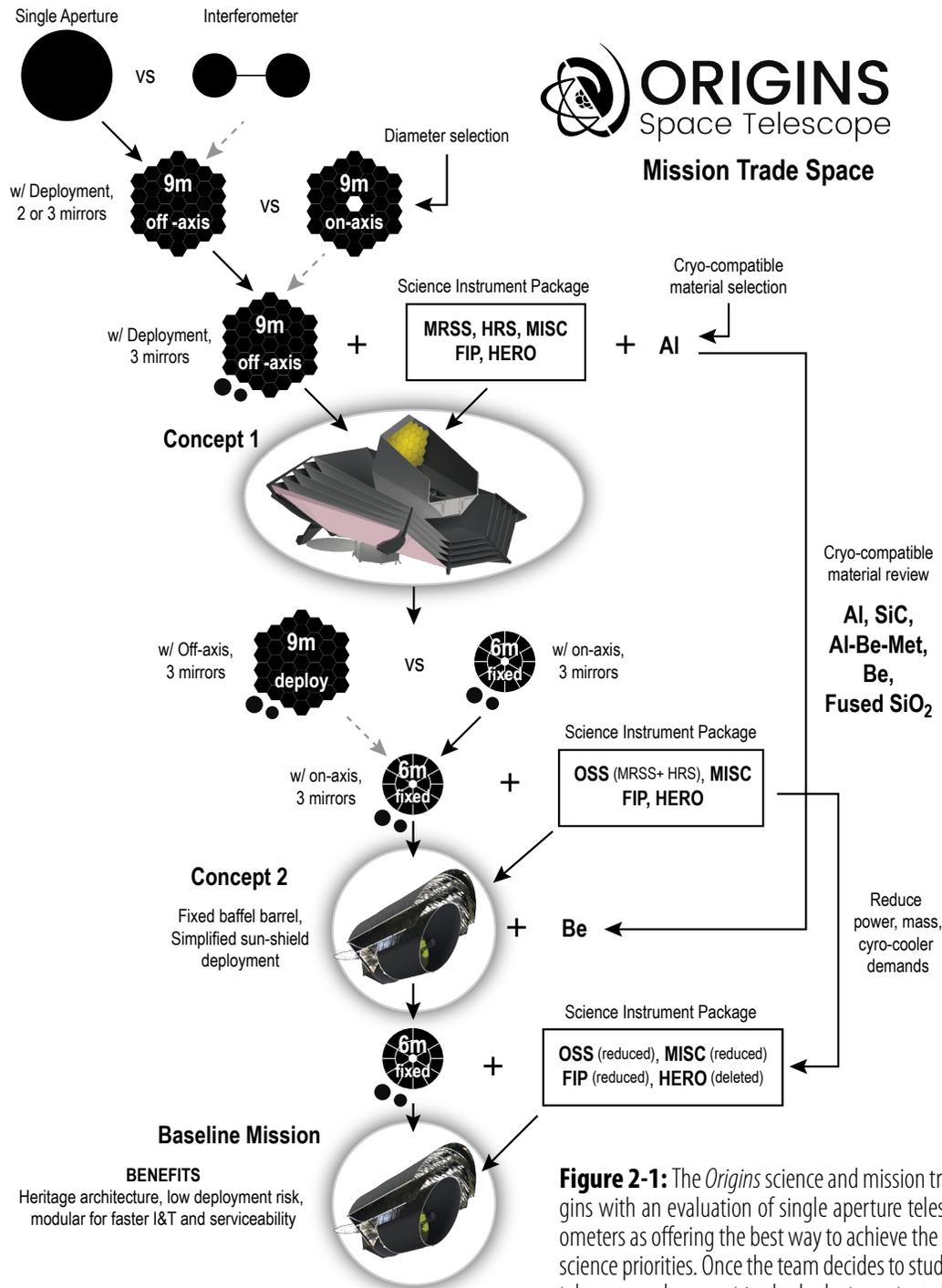

## 2 - BASELINE MISSION OVERVIEW

### 2.1 Trade-Space Path to Baseline Concept

The *Origins* team has conducted a series of science and mission trades to maximize science for the mission cost. Figure 2-1 illustrates major items in the science and mission trade space. Table 2-1 summarizes key trades and the rationale for the decisions made during the course of development of the *Origins* mission concept. These trades are guided by lessons learned with *Spitzer*, *Herschel*, and James

**Figure 2-1:** The *Origins* science and mission trade discussion begins with an evaluation of single aperture telescopes vs. interferometers as offering the best way to achieve the team's preliminary science priorities. Once the team decides to study a single aperture telescope, subsequent trades look at aperture size, packaging and deployment schemes, materials, and instrument selection.

**Table 2-1:** *Origins* Space Telescope major trades and decisions

| Trade | Decision | Rationale |
|---|---|---|
| Single-aperture telescope or interferometer | Single-aperture telescope | The science case requires superlative spectral line sensitivity and high spectral resolving power. Science case does not require sub-arcsecond spatial resolution. |
| JWST-like (deployable) vs. Spitzer-like (non- deployable) aperture | *Spitzer*-like | Minimizes complexity and fits into a launch vehicle that is under development and likely to exist in the anticipated timeframe. |
| Launch vehicle with 5 m diameter fairing vs. larger vehicle | Larger vehicle | Human exploration and competition are driving development of SLS, Space X BFR, and Blue *Origins'* New Glenn. *Origins* would be volume- constrained in a 5 m fairing, requiring complex deployments from a stowed configuration and limiting volume available for instruments, and on-orbit performance could not be fully-verified with ground tests. |
| On-axis vs. off-axis telescope | On-axis | An on-axis three-mirror anastigmat satisfies performance requirements with good Strehl ratio over a large field-of-view, and enables a larger telescope to fit into the launch vehicle without necessitating deployment of the secondary mirror. |
| Telescope temperature 4.5 K vs. warmer | 4.5 K | To satisfy science-driven sensitivity requirements in the far-IR, 4.5 K is preferred, but up to 6K is acceptable. The former can easily be attained with current state of the art cryocooler technology. A telescope warmer than 6 K would not meet sensitivity requirements. |
| Expendable cryogen vs. mechanical cryocoolers | Cryocoolers | Expendable cryogen would require a large, massive cryostat and would be mission life-time-limiting, whereas cryocooler technology has matured and has none of these issues. |
| Materials: beryllium vs. aluminum vs. SiC, carbon fiber, or other | Beryllium was chosen for optical components and instrument structure | Material properties (e.g., low density, high stiffness and strength, low CTE, high thermal conductivity) of Be at cryogenic temperatures are superior to alternatives, outweighing handling challenges. |
| Primary mirror segmentation: hex vs. pie wedges; various segment sizes and numbers | Circular mirror with pie-shaped segments in two annuli | Circular maximizes collecting area in the fairing without requiring mirror deployment. Of four segmentation schemes considered, two are preferred because the segments can be fabricated in existing facilities and have only two optical prescriptions. |
| Far-IR detectors: Transition Edge Sensors (TES) bolometers, Microwave Kinetic Inductance Detectors (MKIDs), or other | Transition-Edge Sensor bolometers | TES bolometers and KID detectors are both viable options. TES bolometers were selected because their room temperature readout system encompasses requirements for readout of alternative detector types, including KIDs, and using only one type of far-IR detector is preferred for costing simplicity. |

Webb Space Telescope (JWST). The *Origins* STDT has driven the process from the outset with scientific proposals representing the science community's wide array of interests. A single-aperture telescope is preferred over a spatial interferometry for extraordinary sensitivity is required by the highest ranked proposals. *Origins* study Concept 1 has the most sensitive telescope with a suite of instruments capable of all the proposed science and is described in the interim report (*Origins* team 2018).

However, Concept 1's greater capability has a high level of complexity. With Concept 2, *Origins* team decides on a design with minimal deployments utilizing the full capacity of the large launchers under development for human space exploration. The *Origins* Baseline Mission Concept results from a descope trade study in which science, technical, and cost information is evaluated in an effort to optimize science per dollar in a mission with a self-imposed cost target.

During the descope phase, the *Origins* team preserved the most compelling science objectives that emerged during the *Origins* initial concept studies. The final recommendation is an *Origins* Baseline Mission Concept that satisfies ~80% of Concept 2's three science theme goals and is capable of all the science outlined in Section 1.

Figure 2-2 shows the nature of the descopes is predominantly instrumentation. The HEterodyne Receiver for *Origins* (HERO), the Mid Infrared Spectrometer and Camera (MISC) imager and two bands of the Far Infrared Imager and Polarimeter (FIP) (100 and 500 μm) are removed. For completeness, this study report describes HERO and the MISC imager concepts in Appendix D and offer them as potential upscopes. In addition to these instrument losses, both FIP and *Origins* Spectrometer for Surveys (OSS) retain only half their pixels from Concept 2 which halves their mapping speed.

The telescope structure is also the instrument mounting structure. This structure is generally driven by launch criteria. Launch criteria requires a certain minimum frequency to be met to allow use of lower loads and skip early coupled loads analysis. Frequency goes as stiffness over mass. To meet the minimum frequency requirements of the SLS launch vehicle a high stiffness to mass material is required.

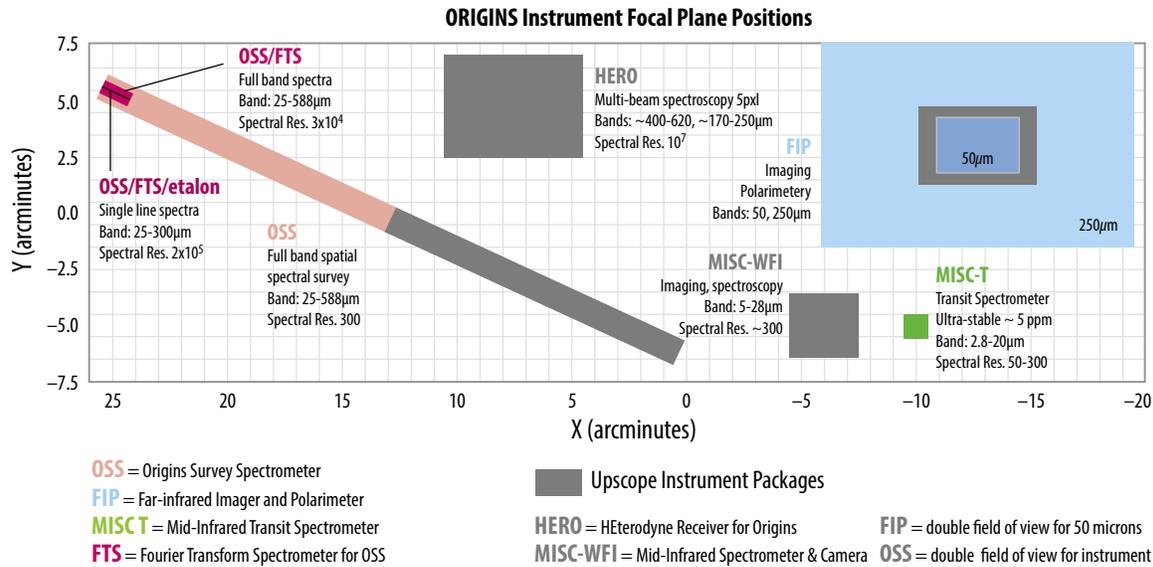

**ORIGINS Instrument Focal Plane Positions**

**Figure 2-2:** The *Origins* Space Telescope focal plane showing all the instruments studied and the result of the descope exercise. Those items shown in grey were descoped and two (HERO & MISC-WFI) of these are described as possible upscopes in Appendix D.

Beryllium was selected based on its high stiffness to mass ration (6 times greater than aluminum alloy) while providing high thermal conductance. Lowering the mass also decreases the support strut cross section which decreases conducted heat load to the cold telescope and instruments. Further discussion of the materials evaluation is provided in Section 2.6.4 and Appendix C.3.6

## 2.2 Mission Requirements

The *Origins* mission objective is to perform 2.8–588 μm mid- and far-IR science observations with unprecedented sensitivity to transform our understanding of the Universe, from formation of the earliest galaxies to the qualities of potentially habitable worlds. *Origins* mission requirements are provided in full in Appendix B. The mission requirements are derived from the STM in Section 1 and standard engineering guidelines and requirements. Table 2-2 illustrates the rationale for the *Origins*, mission requirements, and their flow-down to the flight and ground elements.

**Table 2-2:** *Origins* Mission Requirements, Rationale and Flowdown

| | Mission Requirements | Rationale | Observatory Requirements | Ground System Requirements | Operation Requirements |
|---|---|---|---|---|---|
| **Telescope** | *Origins* payload comprised of OSS, FIP and MISC instruments and the telescope | Enable observations that meet *Origins* mission science requirements in the STM | Fully accommodate telescope and three instruments; meeting all interface requirements | Receive data downlink, process and archive data | • Spacecraft and instrument operation planning and execution<br>• Monitoring health & safety telemetry<br>• Uplink command/data |
| | Telescope diameter ≥ 4 m | Meet driving angular resolution requirement | 5.9 m diameter circular primary mirror | | |
| | Telescope collecting area ≥ 22.6 m² (diameter ≥ 5.3 m) | Meet driving scientific objectives in a total of 4,000 hour observing campaign; meet driving sensitivity requirements | 5.9 m diameter circular primary mirror; 25.4 m² collecting area | | |
| | Telescope temperature ≤ 6 K | Scientific Objectives in Themes 1 (extragalactic) and 2 (planetary habitability) | • 4.5K as the baseline design<br>• Cryocoolers, radiators, sunshields<br>• Thermal decoupling & isolation | | Telemetry monitoring |
| **Pointing and Tracking** | Telescope Field Steering Mirror (FSM) with minimum throw of 1 arcmin | Enables small area mapping | Telescope design; FSM control loop follows tip/tilt commands from ACS | | |
| | Track/scan rate ≥ 60 mas/sec | Map wide survey areas with efficient scan mapping | ACS design | | |
| | Pointing jitter at frequencies > 1 Hz ≤ 50 mas (1σ) and pointing drift ≤ 50 mas up to 10 hours | Pointing drift and jitter not compromise MISC ability to measure transit spectroscopy | • ACS design meet jitter and drift requirements<br>• Telescope FSM to accept feedback from MISC tip-tilt sensor to lower these values to < 10 mas | | |
| | Point to an accuracy of 2 arcsec (1σ) after slew and settling | Enable target acquisition | ACS design | | |
| | Pointing knowledge after ground data processing ≤ 0.5 arcsec | Enable object matching with near-IR counterparts | ACS design | | |
| | 4 π steradian Field of Regard (FoR) over the course of the mission | Enable access to all targets of interest; FoR > 40% of the sky at any time | • Roll +/- 5 deg<br>• Pitch + 85 deg to +150 deg<br>• Yaw +/-180 deg | | Observation planning and execution |
| **Science Operations and Data** | Observing efficiency ≥ 80% | Design goal: Achieve all mission-design scientific objectives within 40% of a 5 year mission; Operations goals: allow a highly efficient GO program with 100% of a 5 year mission. | • ACS design<br>• Minimize disruptions due to momentum unload, orbit maintenance, data downlink<br>• Data storage and downlink<br>• Design robustness<br>• Fault detection and recovery | • Ground station availability<br>• Data transfer from ground station to Science Data Processing facility | Efficient observation planning and execution |
| | OSS and FIP surveying simultaneously; 21 Tbits/day data collection capability | Accommodate driving observation scenario | • Power subsystem design<br>• Onboard data storage capacity & downlink | • Ground station availability<br>• Data transfer from ground station to Science Data Processing facility | Observation planning and execution |
| | 5 year mission life, 10 year consumables | Allow an efficient science observing program for the astronomical community; extra consumables to support an extended mission | • NASA Class A Payload (No credible single point failure, parts, test program . . .)<br>• Reliability<br>• Radiation protection<br>• Micrometeroid protection<br>• Contamination protection<br>• 10 year Propellant<br>• 10 year solar array | • 10 year ground system support<br>• Processing and archiving 10 years worth of data | • 10 year spacecraft and instrument operations<br>• 10 year data operations |
| | Minimize background heat and light | Reduce background noise and enable detection of faint signals | • Sun-Earth L2 orbit<br>• Orbit between 4 deg and 29 deg off Sun-Earth axis<br>• Sunshields<br>• Onboard orbit determination<br>• Compatible with NASA SLS launcher with 8.4 m fairing or/and Space X BFR with 9 m fairing | | Orbit maintenance maneuver planning and execution |
| | ≥ 99% of the *Origins* science observation data processed and archived. | Minimize loss of science data | • Downlink bit error rate<br>• Onboard data storage | • Ground station availability<br>• Ground processing system<br>• Data archiving | • Data downlink operation<br>• Data processing operation |
| | Level Zero production data available 72 hours from time of instrument observation data collection | Data latency requirement | • Downlink data rate | • Ground station availability<br>• Data transfer from ground station to processing facility<br>• Processing algorithms<br>• Processing power | • Data downlink operation<br>– Data processing operation efficiency |

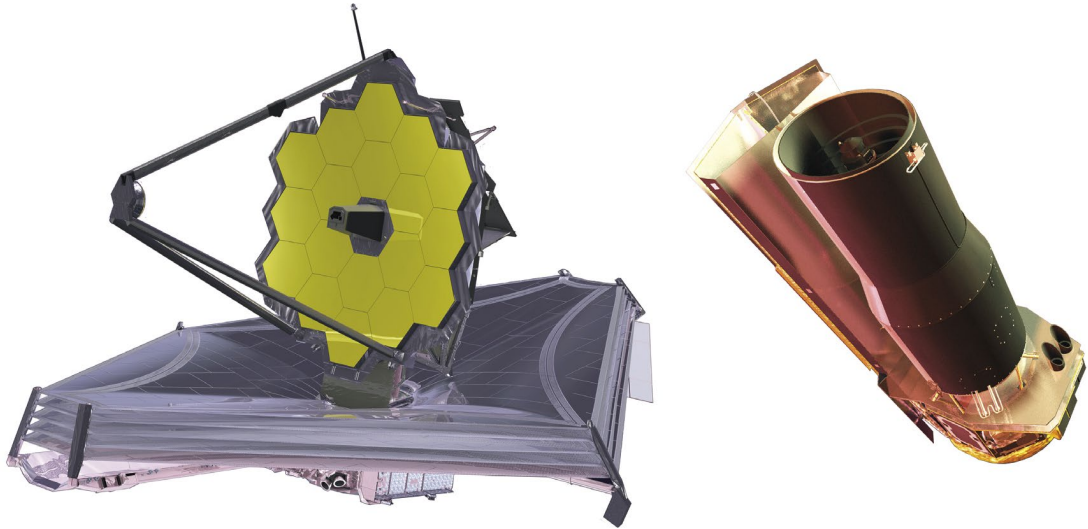

**Figure 2-3:** The *Origins* architecture is more like *Spitzer* (right) than JWST (left).

## 2.3 Observatory Architecture

At 4.5 K, the 5.9-m telescope is sky background-limited over 3 to 250 microns in the mid-to-far infrared wavelength range and only about a factor of 2.7 above the background at 600 microns. The instruments are state-of-the-art spectrometers and imagers in these wavelengths with advanced detectors that take advantage of the low background from the telescope. The thermal system passively and actively cools the telescope and instruments to 4.5 K and below.

The *Origins* design is driven by the cryogenic payload, testability, serviceability (both on the ground and on-orbit), and to minimize deployments. The resulting architecture is more akin to the *Spitzer* Space Telescope (*Spitzer*) than JWST (Figure 2-3). Like *Spitzer*, *Origins* uses two sunshield layers rather than a five-layer JWST design. Eliminating three sunshield layers was possible because the telescope is surrounded by a barrel that is highly reflective toward the sunshield while radiatively cooling in the direction of deep space (Figure 2-4). The sunshield also wraps around the bottom providing protection against the Sun, Earth, and Moon over a 25% larger Field of Regard (FoR) than JWST.

The resulting barrel temperature is 35 K (*Spitzer*'s was 34 K). This protects the colder surfaces within from higher temperature radiation. The *Spitzer* scheme offers several advantages, including: the design is compact, requires few deployments, provides stray light protection, eases trade analyses by making the thermal design separable into subsystems inside and outside the barrel, and thermal performance is easier to verify on the ground. The two-layer sunshield is extremely easy to deploy when compared with JWST, and can be readily deployed, tested, and stowed on the ground. The sunshield shape was determined by a combination of the need to block Sun-, Earth-, and Moon-shine from the aperture and a desire to maintain as small a cross-section to solar pressure as possible. The 35° cut angle in the baffle, barrel, and sunshields keeps the Sun and Earth from shining into the baffle for the complete FoR for the entire orbit around Sun-Earth Lagrange Point 2 (SEL2). The large ~1.2 m gap between the shields and the barrel allows efficient radiation to deep space, as shown in Figure 2-5.

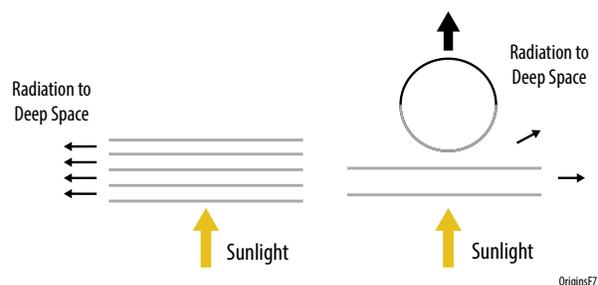

**Figure 2-4:** A two-layer sunshield plus radiator can achieve the same radiative temperature as a five layer sunshield.

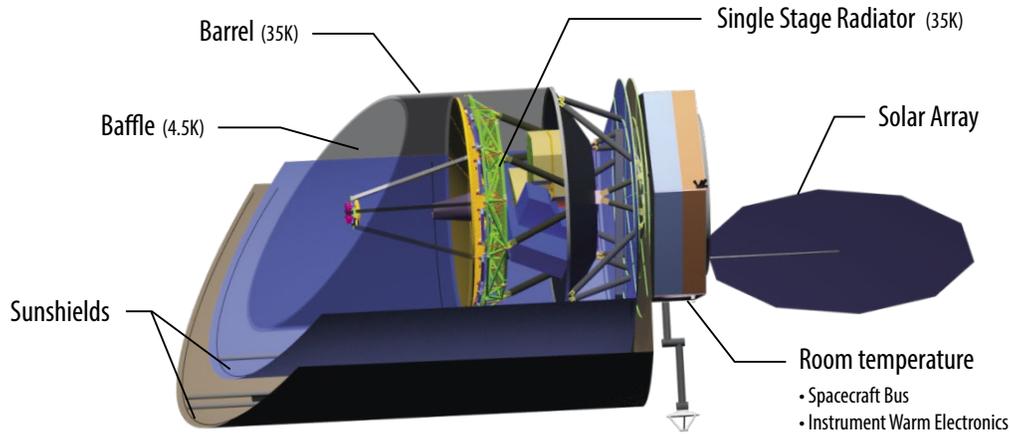

**Figure 2-5:** *Origins* temperature zones are designed to maximize cryo-cooling effiiciency while ensuring a 4.5 K environment for the telescope and instruments. The instruments are housed in the 4.5 K zone beneath the primary mirror. The 35 K radiator is more clearly shown in Figure 2-9.

*Origins* uses staged cooling provided by mechanical cryocoolers with very high reliability and at technology readiness levels (TRLs) >4. In contrast to *Spitzer*, which relied on stored liquid helium (5.5 years with 6 mW of cooling at 5.0 K) to provide cold gas to cool the telescope, *Origins* has multiple long life (>10 years) cryocoolers that together produce 200 mW of cooling at 4.5 K. The multi-stage cryocoolers also provide 400 mW of cooling at the mid-stage temperature of 20 K to intercept conducted and radiated heat within the barrel, and 20 W of cooling at 70 K to intercept heat conducted along the structure supporting the barrel from the warm spacecraft. A simple heat flow budget to the cryocooler stages is provided in Table 2-3. This represents the heat flow at the worst-case attitude of the spacecraft with worst-case End-of-Life (EOL) emissive properties.

While cooling is planned as a shared resource (cryocoolers all operate together, none are dedicated to a particular system), *Origins* has baselined four cryocoolers rather than one larger cryocooler because: cryocooler size is closer to those already developed; if less cryocooling is required, it is easy to remove a cryocooler; individual cooling heads may be positioned in several places on the telescope and instrument mounting surfaces to provide local heat sinks and minimize thermal straps ; and, in the very unlikely event of a failure of the mechanical part of a cryocooler, multiple coolers provide redundancy. The high thermal conductance of the structures make all the 4.5 K cooling nodes isothermal. (Note that the history of cryocoolers in space is that they do not fail, but only get turned off at the end of the mission. Please refer to *Origins* Technology Development Plan, Figure 4).

Table 2-3 shows the calculated heat flows to each of the cryocooler stages. At first glance it looks coincidental that each of the stages has close to the same heat load margin. What is actually taking place is the balancing of heat loads by strategic interception of heat along the struts and harnesses, and the tradeoffs that can be made within the cryocoolers themselves between the fraction of heat lifted between the 20 K and 4.5 K stages. Note that there is a small contribution from radiation to space at 20 K.

**Table 2-3:** The heat flows to each of the three cryocooler stages shows a margin of at least 100% at each temperature. Due to the complex interaction between radiation, conduction and radiation to space, only the totals are given for the 70 K heat loads. Heat from harnesses are not currently intercepted at 20 K, but will be in a future iteration.

| T (K) | Item | Heat Load | Total | Capability | Margin (%) |
|---|---|---|---|---|---|
| 4.5 | Harnesses | 36 mW | 90 mW | 200 mW | 122% |
| | Dissipation | 31 mW | | | |
| | 4 K Bipods | 17 mW | | | |
| | Radiation | 5 mW | | | |
| 20 | Harnesses | 0 mW | 193 mW | 400 mW | 107% |
| | Shield Suspension | 28 mW | | | |
| | 4 K Bipods | 134 mW | | | |
| | Radiation | 33 mW | | | |
| 70 | Harnesses | | 9.38 W | 20 W | 113% |
| | Supports | | | | |
| | Radiation | | | | |

## Enabling Technology

The *Origins* Baseline Mission design emphasized use of high TRL technologies wherever possible. Specifically, JWST-size mirror segments with JWST actuators are planned for the telescope. Note that the reasons for the selection of these actuators are: the electronics will be operated only down to 35 K (within its already qualified range to 20 K), the motors will be tested to 4 K, but there is no reason to think that they will not continue to operate since the differential contraction between 20 K and 4 K is negligible, and the winding resistance will be constant over that temperature range. The range of motion is larger and the resolution is far better than required for performance at the longer wavelengths of *Origins*.

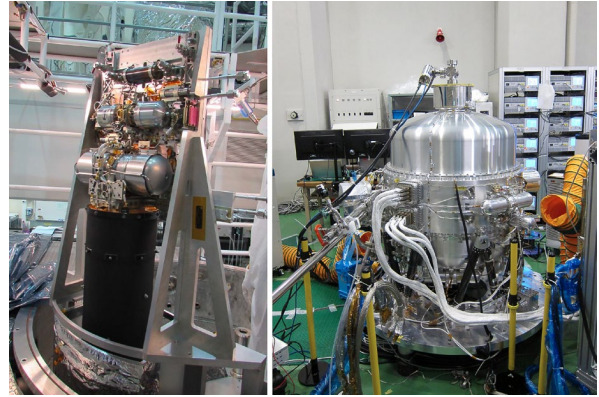

**Figure 2-6:** *Origins* baselines a cryocooler design adapted from present flight-worthy cryocoolers. Two examples of such are the MIRI cryocooler (shown in a test stand at left), or the several JAXA cryocoolers mounted on the Hitomi dewar (right).

Existing technology methods for achieving high resolution far infrared spectroscopy are baselined for the instruments. No deformable mirror is required for high precision measurements of transiting exoplanets. As a result, the enabling technology requirements can be very simply stated as cryocoolers and detectors. Specifically required are 4.5 K cryocoolers for 10 year or greater life, sub-Kelvin coolers producing 0.05 K with a 6 µW heat lift, mid-infrared detectors with a stability of 5ppm over a few hours, and far infrared detector arrays with sensitivities of $\sim3\times10^{-19}$ (imaging) to $3\times10^{-20}$ W/$\sqrt{Hz}$ (spectroscopy) and $\sim10^4$ pixels per instrument channel, with $>10^3$ per array, mosaicking up to $10^4$. The plan to reach those milestones is given in *Origins* Technology Development Plan, a separate volume in this report. This plan allows ample time during Pre-Phase A to mature the technologies to TRL 5, and during Phase A resources to reach TRL 6 before the instrument Preliminary Design Reviews (PDRs).

Mechanical cryocoolers for space missions have been under development in Europe, Japan, and the United States (US) for several decades. In the US, the Advanced Cryocooler Development Program (ACTDP) resulted in three selectable TRL-5 designs from three different vendors for JWST. Although the JWST cooler targeted 6 K and *Origins* requires 4.5 K, there are relatively straightforward development paths for each of several cryocooler types to reach TRL 6 for the *Origins* 4.5 K cooler. The Japanese Aerospace and eXploration Agency (JAXA) has already flown a 4.5 K long-life mechanical cryocooler on the Hitomi mission (Figure 2-6). The Hitomi cooler, with appropriate improvements for a 10-year lifetime, could provide the cooling required for *Origins*. An Adiabatic Demagnetization Refrigerator (ADR) cooling x-ray micocalorimeters to 0.05 K have been flown on Hitomi.

## Other Architectural Features

The two *Origins* observatory elements are the Cryogenic Payload Module (CPM) and Spacecraft Bus Module (SBM) (Figure 2-7). The CPM consists of the 5.9-meter telescope with the inner baffle, outer barrel, and two-layer sunshield, and the Instrument Deck that houses the cold side of the three science instruments. The spacecraft bus consists of the typical bus subsystems, as well as the instrument warm electronics and four cryocoolers.

The instruments are clustered near and upstream of the focal plane, immediately behind the primary mirror structure. The instrument electronics and four cryocooler compressors are located in the spacecraft bus, away from the CPM. *Origins* mitigates long harness lengths by using cold/low-dissipation amplifiers mounted near the detectors.

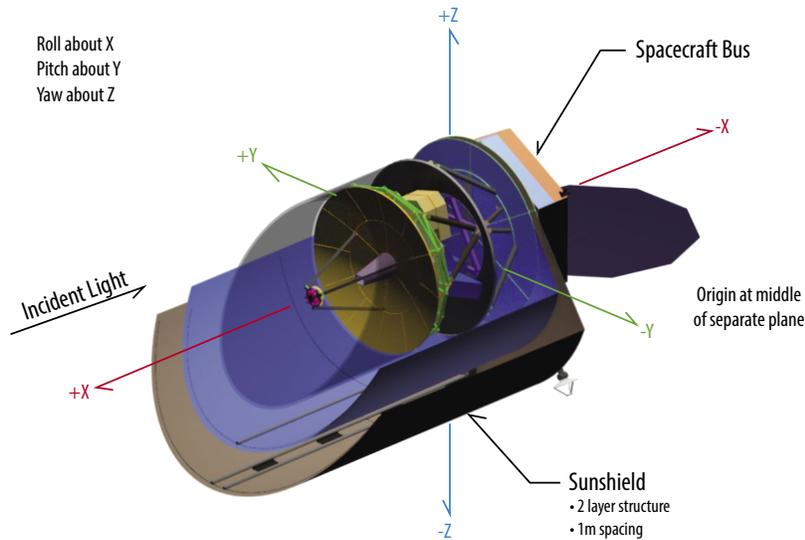

Roll about X
Pitch about Y
Yaw about Z

Incident Light

+Z

+Y

-X

-Y

+X

-Z

Spacecraft Bus

Origin at middle
of separate plane

Sunshield
• 2 layer structure
• 1m spacing

**Figure 2-7:** *Origins* cut-away view showing the observatory coordinate system. The Spacecraft Bus Module is toward the back right in this view and the Cryogenic Payload Module is everything front and left of that including the telescope barrel and sunshields. The *Origins* coordinate system uses +X as the boresight and −Z as the direction to the Sun when the pitch angle is 90 degrees and the roll is 0 degrees.

The spacecraft bus provides the required attitude control, power, thermal control, command and data handling (C&DH), propulsion, and data transmission. The spacecraft is three-axis controlled with a bipropellant propulsion system, and powered by a Direct Energy Transfer (DET) electrical power system with a one-axis gimballed solar array. The design mission lifetime is five years. Propellant is the lifetime limiting expendable, and sufficient propellant is carried to allow for 10 years of operation.

*Origins* is a Class A mission with no credible single point failure. Risk is reduced through design simplification and by minimizing the number of deployments: the only deployments are the solar array, solar shield, and an ejectable aperture cover. Risk is also reduced because the *Origins* observatory can fit into NASA Johnson Space Center (JSC) Chamber A (see Figure 2-4), a large Thermal Vacuum chamber recently refurbished for JWST and outfitted with a helium shroud that has been operated as cold as 11 K. This chamber is suitable for CPM cryogenic testing and vacuum testing of the entire observatory, including sunshield deployment, following the "Test-As-You- Fly" objectives.

*Origins* can be launched on NASA's Space Launch System (SLS) with 8.4-m fairing or Space X's Big Falcon Rocket (BFR) with 9-m fairing. As presently designed *Origins* is just slightly larger in diameter (0.2 m) than the allowed payload diameter of the Blue *Origins* New Glenn fairing. *Origins*' coordinate system is shown in Figure 2-7.

The outside of the barrel is 35 K and is *Origins* radiative boundary. An 11 K shroud has negligible impact on radiative heat transfer vs. the nateral inner solar system effective background of 7.2 K. The heat exchange of a 35 K radiator differs by only 0.8%. The remanent radiation at 7.2 K at 3 mW is <4% of the cryocoolers' capability. Even 11 K radiation would represent an additional 19 mW on the cryocoolers which could easily be accomodated for test. However, a GSE cover will be used in the JSC testing to provide stray light protection for the instruments. This cover will be cooled by a GSE cryocooler of the sort used to cool the MIRI instrument during JWST testing.

To be compatible with the SLS launch vehicle, the *Origins* observatory was stiffened until the normal modes were predicted to meet the requirements in the SLS User Guidelines. From analysis of the *Origins* Baseline design, the primary lateral and axial modes are predicted to be frequencies of 8.09 Hz and 16.7 Hz, respectively. These results meet the SLS User Guidelines allowing simpler structural analysis going forward. To reach these frequency values several design improvements were made: the barrel and baffle were stiffened, supported at the top at launch via the aperture cover, extra struts were added from the spacecraft to the barrel, the bottom of the barrel was made in the shape of a cone, and the support of the propellant tanks was stiffened. These improvements added some mass and had ther-mal impact, but result in a robust structural design without requiring a detailed coupled loads analysis. See Section C.2 in Appendix C for details. A launch lock and deployment were briefly considered to provide stiffness for launch, then release to provide thermal isolation, but a simpler structure that still met the thermal margins was selected instead.

## 2.4 Observatory Thermal Design

To keep telescope emission lower than or comparable to the sky background, the telescope is cooled to ~4.5 K. The cooling system is a combination of passive cooling provided by a two-layer deployed sunshield, a single stage radiator at 35 K, and four high-TRL mechanical cryocoolers in parallel. The cryocooler candidates are currently TRL 4 or 5 and have technology development plans to reach TRL 6 by PDR. The lightweight two-layer sunshield is positioned between the telescope and Sun/Earth/Moon. The layers are reflective in the Sun-ward direction, but effectively black in the perpendicular direction, cooling to deep space. The sunshield layers are designed so the outer sunshield only sees the inner sunshield, and the inner sunshield only sees the outer sunshield and the 35 K barrel. The Sun-facing side of the outer sunshield is coated with Goddard Composite Coating (GCC) for best emissivity to absorptivity. Inside the 35 K barrel, the cryocoolers provide multi-stage intermediate cooling for the structure and wires. Together, the cryocoolers span the 300 K to 4.5 K temperatures of the spacecraft to telescope/instruments with two intermediate cooling intercepts. Notionally, these cooling intercepts are at ~70 K, ~20 K, and 4.5 K. Each of the four cryocoolers can provide 50 mW of cooling at 4.5 K, 100 mW at 20 K and 5W at 70 K. A heat flow map shows the thermal performance of the CPM in Figure 2-8.

The observatory thermal model (Figure 2-9) was developed to simulate *Origins* end-to-end thermal performance. This figure is colored by thermal control coating. The integrated model includes both bus and CPM hardware elements in sufficient detail to accurately simulate key heat loads and heat transfer paths. The thermal model employs 7170 nodes to simulate *Origins* thermal performance.

Multiple stages of cooling intercept heat from warmer temperatures resulting in a modest heat load (<100 mW) at 4.5 K. Furthermore, staged cooling offers relative immunity to external heat load disturbances. Isothermal conditions are provided at each stage by use of thermally-conductive materials (*i.e.,* beryllium, aluminum) with a shiny Vapor Deposited Aluminum (VDA) thermal coating. Heat is easily transported to the cryocooler heat-exchangers with minimal temperature gradient as is seen in Figure 2-10. The thermal architecture is simple enough to be verified by analysis and simple calculation. There are no warm items in the cold part of the observatory that can cause difficult-to-verify sneak paths for radiation.

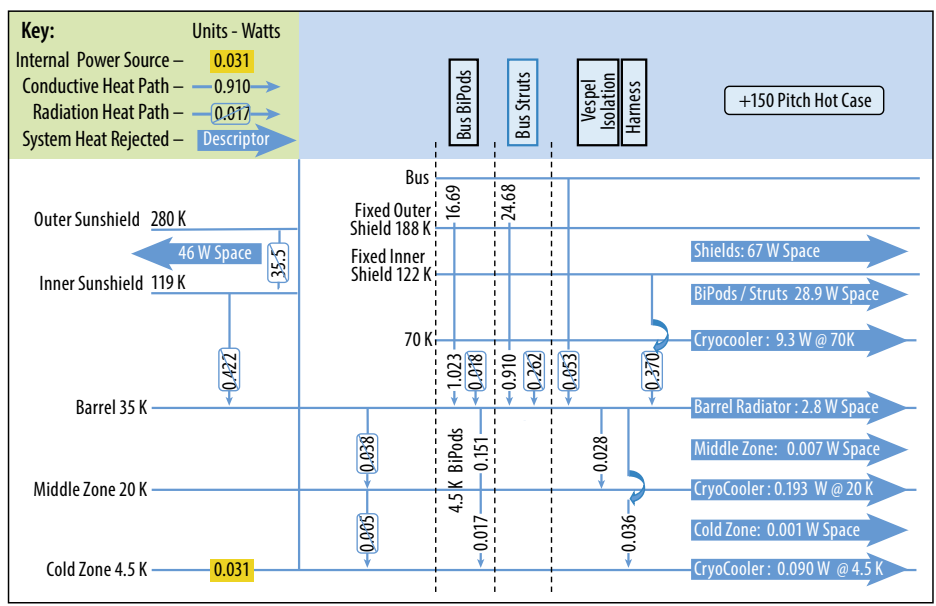

**Figure 2-8:** *Origins* Heat Flow Map of the CPM shows the multiple stages of cooling results in a reasonably small heat load to the cryocooler and radiative cooler stages

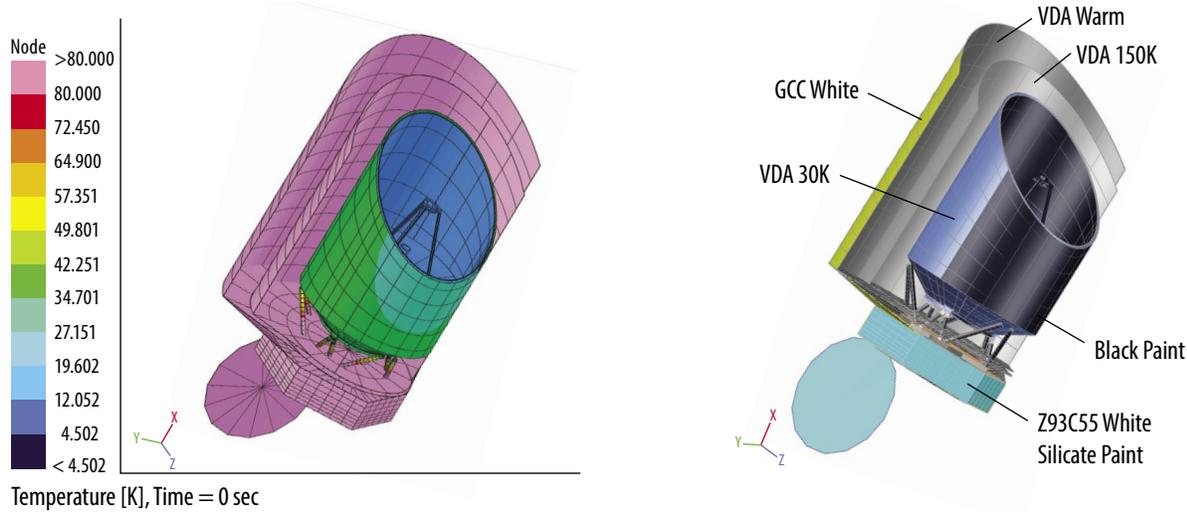

**Figure 2-9:** A Spitzer-like architecture with staged cooling, a two layer sunshield, a deep space radiator, and cryocoolers delivers the required thermal performance with a relatively simple-to-model design.

*Origins* applies a 100 % design margin to each of the cryocooled stages, defined as (capability- heat load)/heat load. This conservative margin is standard practice for the concept design stage.

Combining this margin with the cooling capability at each stage establishes 100 mW at 4.5 K, 200 mW at 20 K, and 10 W at 70 K as required heat load maxima for the baseline design. Table 2-4 summarize thermal model hot and cold design refrigeration performance results for the full Observatory pitch angle range. Data show refrigeration 100% margin requirements are satisfied for all cases.

As a sanity check, a completely independent analysis of a completely different architecture resulted in a similar conclusion: the a 4.5 K telescope can be achieved with a reasonable sized set of 4.5 K cryocoolers. Please refer to Appendix F.

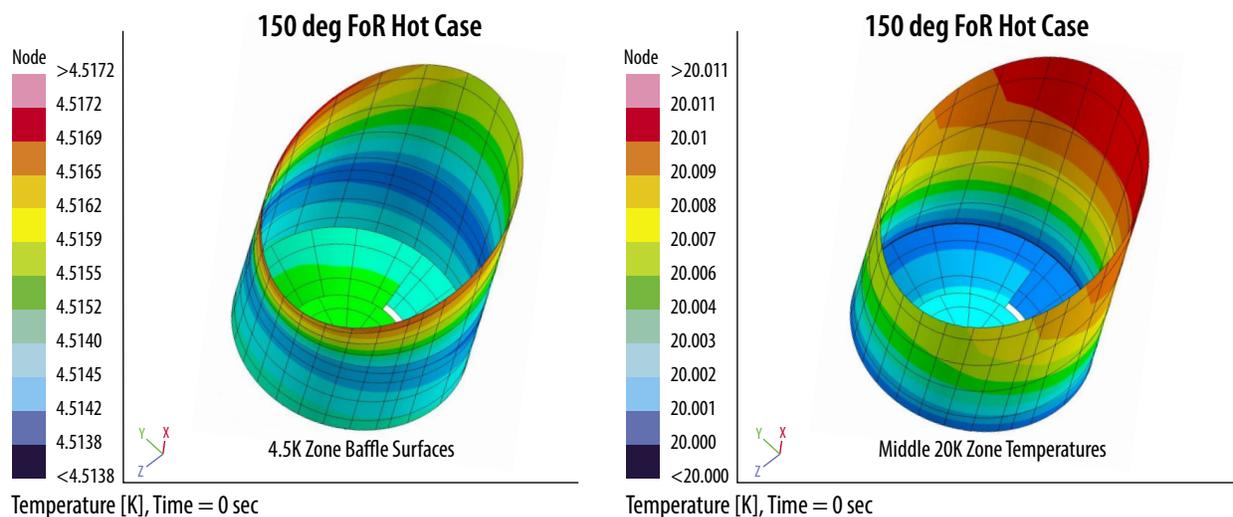

**Figure 2-10:** The 4.5 K "shell" and 20 K shield temperature contours remain nearly isothermal for the worst case, 150 ° pitch attitude.

**Table 2-4:** Thermal modeling demonstrates that the 4.5 K, 20 K, and 70 K refrigeration 100% margin requirements are satisfied for all cases.

| | | Hot Cases | | | Cold Cases | | |
|---|---|---|---|---|---|---|---|
| **Observatory Pitch Angle (deg)** | | 90 | 135 | 150 | 85 | 90 | 135 |
| **CPM Cold Zone Power Dissipation (Watts)** | | W | W | W | W | W | W |
| Instrument Power Dissipation | | 0.031 | 0.031 | 0.031 | 0031 | 0.031 | 0.031 |
| Parasitic Thru Electrical Wires | | 0.014 | 0.014 | 0.014 | 0.014 | 0.014 | 0.014 |
| | Total: | 0.045 | 0.045 | 0.045 | 0.045 | 0.045 | 0.045 |
| **Cold Zone 4.5 K Cryo Cold Head Nodes (100 mW)** | | | | | | | |
| Four Cold Heads 4.5 K Refrigeration | | -0.093 | -0.089 | -0.090 | -0.079 | -0.080 | -0.079 |
| **Allowable Sums Between 0 and -0.100 W** | Total: | -0.093 | -0.089 | -0.090 | -0.079 | -0.080 | -0.079 |
| **Cold Zone 20 K Intercepts (200 mW Available)** | | | | | | | |
| Mid Zone 20K Intercept Power | | -0.050 | -0.056 | -0.060 | -0.034 | -0.034 | -0.041 |
| 20 K BiPod Intercepts (Six Total - One each BiPod) | | -0.113 | -0.125 | -0.133 | -0.082 | -0.082 | -0.096 |
| **Allowable Sums Between 0 and -0.200 W** | Total: | -0.163 | -0.181 | -0.192 | -0.115 | -0.116 | -0.137 |
| **BUS BiPod 70K Intercept Power (10 Watts Available)** | | | | | | | |
| 70 K Central BiPod Intercepts (Six Total - One Each BiPod) | | -4.113 | -6.571 | -7.442 | -2.993 | -3.128 | -5.292 |
| 70 K Outer Strut Intercepts (Six Total - One Each Strut) | | -0.421 | -1.510 | -1.940 | 0.158 | 0.087 | -0.927 |
| **Allowable Sums Between 0 and -10.0 W** | Total: | -4.533 | -8.081 | -9.382 | -2.835 | -3.072 | -6.219 |

Design hot and cold thermal simulations were performed to bound thermal performance for the full Observatory attitude/pitch range. Bounding cold/hot parameters include the solar irradiance at SEL2 (1291 W/m² to 1421 W/m² [*NGST,1998*]), Beginning of Life (BOL)/End of Life (EOL) thermal optical properties, and Bus component power dissipation (20 or 30% margin added to hot case). See Table 2-5. The worst-case results for the sunshield temperatures are shown in Figure 2-11.

The *Origins* thermal design minimizes the conducted and radiated heat load to the CPM and exploits the high and low emittance properties of materials that have been accurately measured at low temperatures. The Sunshield blocks the solar, Earth, and Moon heat loads for the full required observing FoR and minimizes radiated heat loads. Thin wall titanium tubes provide conduction isolation from the spacecraft whose mechanical interface is cold biased and maintained at less than 5°C in the hottest case. Multiple shields between the CPM and spacecraft reduce the radiated heat load from the SBM. The SBM design fully accommodates high power dissipating elements (cryocoolers, OSS and FIP instrument electronic boxes), and SBM temperature sensitive components.

The CPM energy balance includes substantial radiative cooling to enable the cryocooler heat loads to be reasonably small. Parasitic heat loads are diminished by thermally isolating the CPM from heat sources, maximizing its coupling to the space sink, and introducing the mechanical refrigeration to achieve required cold zone temperatures. Radiation heat load inputs are minimized by highly polished VDA thermal coatings. These coatings effective emittance decreases with decreasing temperature as discussed in Section C.1 of Appendix C.

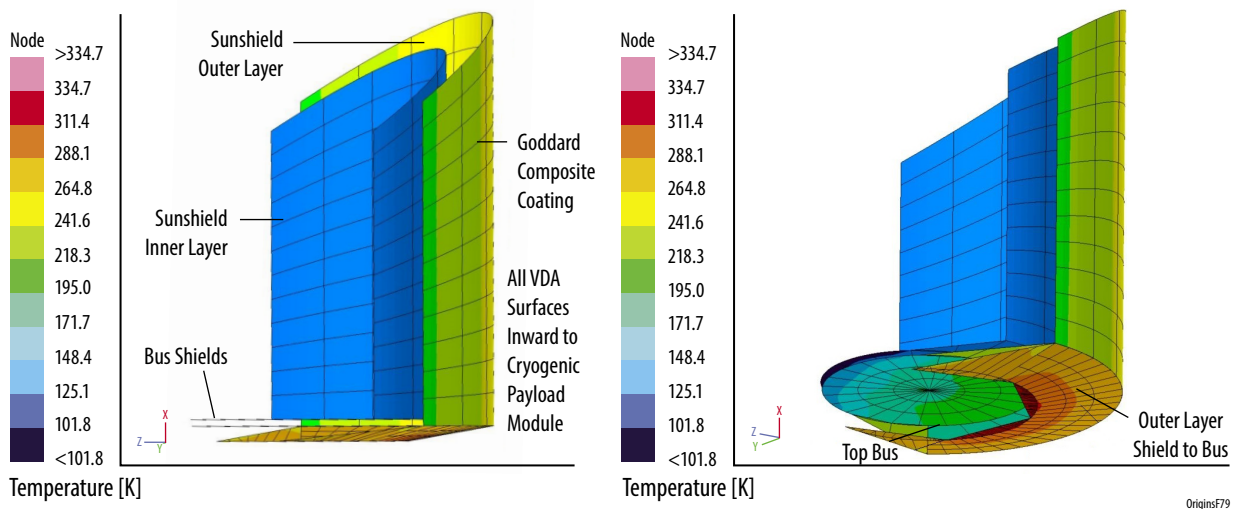

**Figure 2-11:** The Outer Sunshield with GCC maintains room temperature or lower for all FoR over nearly all of its surface.

The CPM is sub-divided into three thermal stages (Barrel 35 K, 20K, and 4.5 K zones) to exploit primary 4.5 K cold head refrigeration as well as the refrigeration capacity of the returning cold gas. An additional stage of cryocooling at 70 K is provided by the same coolers to intercept heat conducted from the warm bus to Barrel by the harnesses, struts and bipods. The Barrel is cooled by a deep space radiator coating about 140 de-grees of the barrel away from the sunshield. The barrel temperature is made nearly isothermal (Figure 2-12) by using high purity aluminum foil coatings on the outer and inner surface. This makes the radiator on the deep-space facing side of the barrel more efficient.

**Table 2-5:** *Origins* uses well-documented thermal optical properties for hot and cold design cases.

| Design Hot and Cold Thermal Optical Properties | | | | |
|---|---|---|---|---|
| | Cold Cases | | Hot Cases | |
| | Beginning of Life | | End of Life | |
| | Alpha | Emittance | Alpha | Emittance |
| **Material** | | | | |
| Kapton 3mil | 0.45 | 0.80 | 0.57 | 0.93 |
| Solar Array Cell | 0.60 | 0.76 | 0.82 | 0.76 |
| GCC White on Kapton | 0.07 | 0.70 | 0.17 | 0.64 |
| Z93C55 White Paint | 0.21 | 0.91 | .037 | 0.88 |
| BIRB Black Paint | 0.99 | 0.88 | 0.99 | 0.88 |
| **Cold Zone VDA Assumptions** | | | | |
| VDA Warm | 0.12 | 0.0250 | 0.12 | 0.0250 |
| VDA_150K | 0.12 | 0.0150 | 0.12 | 0.0150 |
| VDA_100K | 0.12 | 0.0125 | 0.12 | 0.0125 |
| VDA_30K | 0.12 | 0.0100 | 0.12 | 0.0100 |

A 20 K environment is provided inside the Barrel by lining it with a high conductivity aluminum foil sandwiched between layers of double-aluminized Kapton (DAK). This DAK/Al/DAK sandwich was used successfully in the JWST/MIRI thermal shield which also operates at 20 K. Refer to Section 2.4 for more details on the DAK/Al/DAK sandwich mounting to the inside of the Barrel. The 20 K shield is cooled by a 20 K cryocooler stage. The inner-most thermal zone is the 4.5 K zone, which includes the baffle, telescope mirrors, structure, and the volume beneath the Primary Mirror that houses the instruments. Cooling to this zone is provided by the coldest mechanical cryocooler stage at 4.5 K. Note that the actual cryocooler temperature will be 4.3-4.4 K under normal loads. 4.5 K represents an upper value for the cryocooler cold stage under the full load plus margin.

Both the infrared radiation exchange matrix and absorbed solar heat loads were calculated using Monte Carlo methods included within the Thermal Desktop thermal analyses program (Version 6.0).

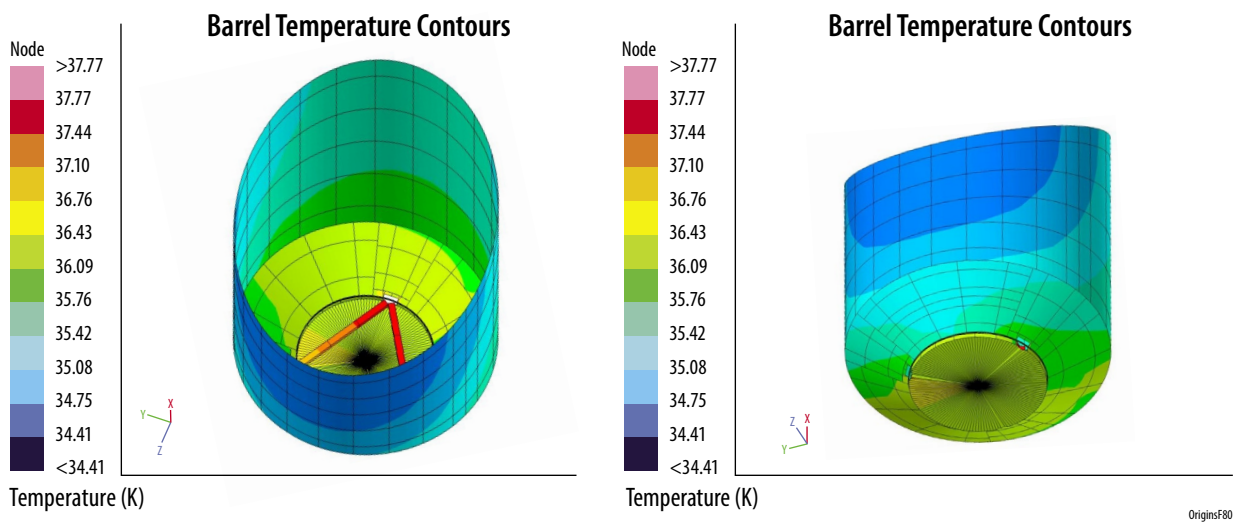

**Figure 2-12:** The Barrel temperature contours show that the barrel temperature remains below 36 K and nearly isothermal in the worst case attitude (150° pitch) over nearly all of its surface.

Hot cases shot 250 thousand rays per surface and cold cases shot 100 thousand. Solution routine accuracy control parameters were set very tight which was necessary to capture milliwatt heat flows. Both the ray trace Energy Cutoff Fraction and Bij/Fij Cutoff Fraction parameters were set to 1E-07. Setting this value small was found to be important for simulation accuracy due to the presence of many low emittance surfaces in the system. The thermal model radiation exchange matrix included roughly 9.7 million radiation couplings.

Steady state numerical solutions were determined using the SINDA thermal network analyzer for each case considered. Boundary nodes were used to simulate the refrigeration sources (4.5K cryocooler heat exchangers, 20 K Bipod Intercepts, and 70 K Bipod Intercepts). SINDA calculates a negative energy value for these nodes which represents the cooling energy required to hold the specified temperature at that location. These are the values reported in Section C.1 of Appendix C. Tight solution control constants were also set for the steady state solution. Results were reviewed for each case and it was verified that the calculated energy balance for each non- boundary thermal model node was better than 10$^{-7}$.

The cryocooler compressor heat rejection thermal design is identical for all four installations to simplify the qualification and test program. Constant conductance heat pipes (CCHP's) transport cryo cooler heat loads to perimeter radiator panels. CCHP's are also used within aluminum honeycomb radiator panels to achieve high radiator efficiencies. The strategic architecture is fully ground testable (telescope facing skyward for test) and minimizes cost and complexity.

Spacecraft thermal design details are shown in Section C.1 of Appendix C.

## 2.5 Baseline Design Configuration

To understand the tradeoffs and engineering discussion to follow, the baseline design configuration is laid out in this section. The *Origins* observatory is comprised of the SBM and CPM. The overall *Origins* height is 12.7 meters and overall diameter is 6.3 meters in the launch configuration (Figure 2-13). *Origins* launch configuration weight is 9,618 kg. This allows *Origins* to fit into the SLS and BFR launch vehicles.

Once on orbit, *Origins* deploys its communication antenna, solar array, telescope cover, and sunshields. After deployments, the overall size of *Origins* is 12.7 meters long and ~9 meters in diameter (Figure 2-14). The colors in the figures simply depict different components for clarity.

The remainder of this section is devoted to a description of the payload, the CPM, the components of which are shown in Figure 2-15.

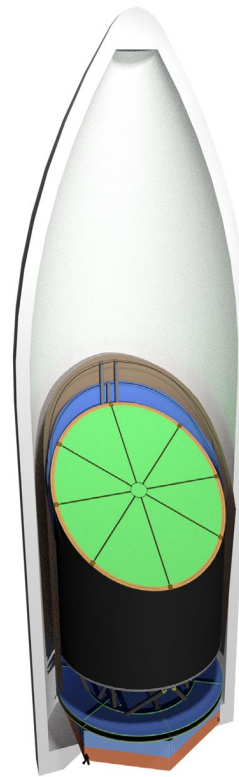

**Figure 2-13:** (left) The *Origins* launch configuration is sized to fit the SLS and BFR launch vehicles. (SLS launch fairing shown).

**Figure 2-14:** (below) The *Origins* on-orbit configuration, with the sunshields deployed, is ~13 m long and ~9 m in diameter

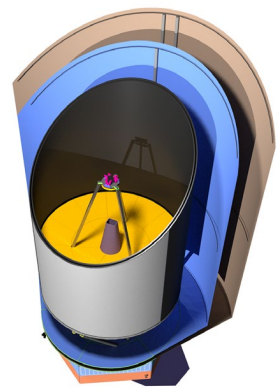

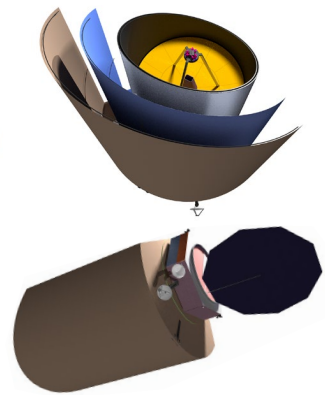

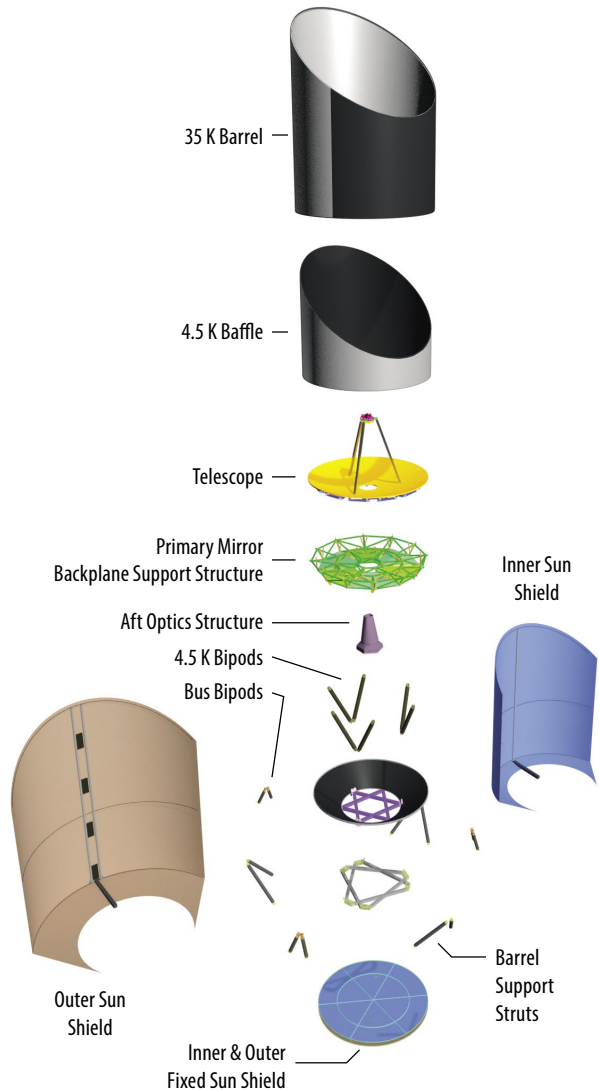

35 K Barrel

4.5 K Baffle

Telescope

Primary Mirror
Backplane Support Structure

Aft Optics Structure

4.5 K Bipods

Bus Bipods

Inner Sun
Shield

Outer Sun
Shield

Inner & Outer
Fixed Sun Shield

Barrel
Support
Struts

**Figure 2-15:** (left) *Origins* CPM components exploded view shows the *Origins* configuration. Instruments are not shown in this diagram.

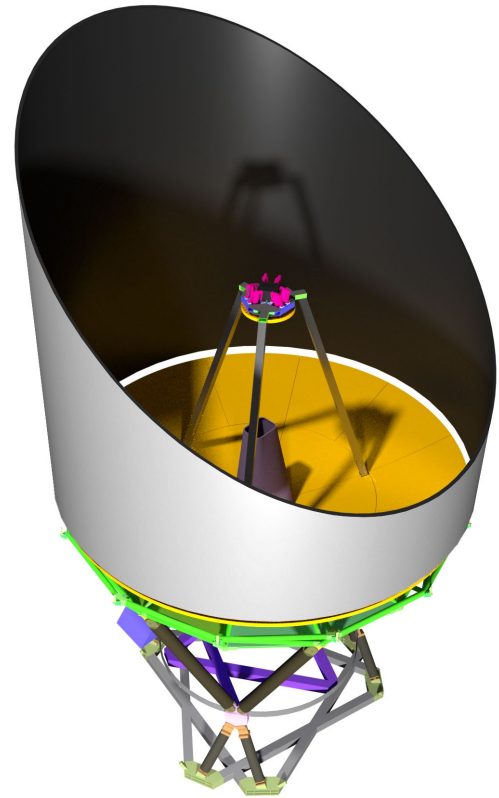

**Figure 2-16:** 4.5 K Baffle is supported from the PMBSS.

## Baffle

The CPM Baffle (Figure 2-16) serves to baffle incoming light and shield the mirrors. For stiffness at launch the Baffle is a 38.1-mm thick aluminum honeycomb core panel. It is coated on the mirror-side with Ball Infrared Black (BIRB) or similar non-reflective coating, and the exterior is shiny aluminum. The Baffle is mounted to the Primary Mirror Backplane Support Structure (PMBSS); however, due to material thermal contraction mismatch between the Baffle and the PMBSS, radially-mounted titanium flexures are used at the interface to allow for deflections during cool down. Thermal straps in parallel with the flexures provide thermal contact through which the baffle is cooled.

## Bipod Support Subassembly

The Bipod Support Subassembly is the CPM's primary support structure. The subassembly is optimized to create the most open access to the instrument volume, to be stiff and strong enough to withstand the launch environment, and to limit thermal heat transfer from the spacecraft bus. The subassembly has three sets of Bus Bipods, a 35 K Deck structure, and three sets of 4.5 K Bipods (Figure 2-17).

The Bus Bipods are made of titanium (Ti-6AL-4V) tubes (165.1-mm diameter, 1.27-mm wall thickness). The tubes end fittings support a hinge at the bottom and a flexure at the top. The hinge and the flexures are positioned perpendicular to the radius of the CPM to allow for thermal deflections due to the large temperature differences at the 35 K Deck. To limit radiative thermal loads along the Bus Bipod tubes, several thin titanium discs are inserted along the inside of each tube to provide thermal radiation blocks, effectively reducing thermal heat transfer along the tube. The Bus Bipods interface with a titanium Bipod Interface Bracket. The 4.5 K Bipods also mount to this bracket. The 4.5 K Bipods are $M_{55}J$ fiber composite tubes with titanium end fittings. M55J fiber composite was selected for the 4.5 K Bipod tubes due to its excellent mechanical and thermal properties. The end fittings are bonded in each end of the composite tube. The end fittings are of a lug-and-clevis design, positioned with the pin normal to the radius of the CPM, so as to avoid imparting loads to the PMBSS while on-orbit and at temperature.

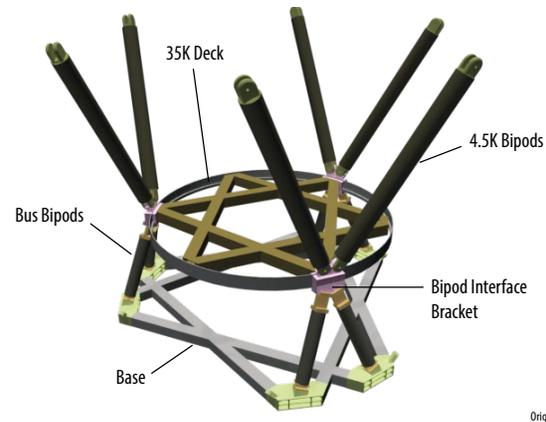

**Figure 2-17:** The *Origins* Bipod Support Subassembly base supports three sets of Bus Bipods, which support the 35-K Deck. The bottoms of the 4.5 K Bipods are attached to the 35 K Deck.

Also mounted to the Bipod Interface Brackets is an aluminum structure, the 35 K Deck. This six-pointed-star shaped structure supports the Barrel and cold-side electronics. The 35 K Deck also provides rigidity between the three sets of bipods.

**Barrel**

The Barrel is another CPM thermal shield comprised of cylindrical and conic sections. It is a 38.1-mm thick aluminum honeycomb core panel with high-purity aluminum face sheets on top of ordinary 6061 Aluminum to provide rigidity and minimize thermal gradients. The bottom of the barrel tapers to the 35 K Deck. Spacecraft bipods support the 35 K deck from the CPM Base as shown in Figure 2-17. Access doors are located around the Barrel for instrument access and integration.

The cone helps provide stiffness to the support of the cylindrical section. Bonded in the cone at the 35 Deck interface is a robust ring to mount to the 35 K Deck. To further stiffen this structure, titanium struts support the Barrel at the cone-cylinder interface from the CPM Base. These struts are 177.8-mm diameter tubes with 2.29-mm wall thickness. The strut end fittings are also titanium and hinged at each end to allow for thermal deflection of the Barrel (Figure 2-18).

**20 K Shield**

Mounted to the inside of the Barrel is a lightweight thermal shield described in Section 2.4 and in Section C.1, Appendix C. The shield is offset from the surface of the Barrel by 63 mm using thin-wall Vespel tubes spaced ~1 meter apart. See Figure 2-19. The tubes are fitted at each end with composite end fittings and are fastened to the Barrel on one end and to the thermal shield on the other.

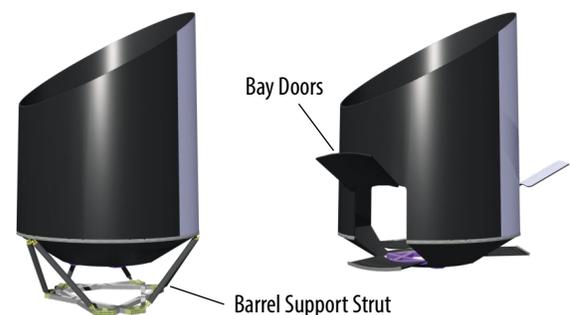

**Figure 2-18:** Barrel Support Struts support the Barrel and connect to the CPM Base, providing stiffness to the assembly. For clarity, the Bus Bipods are not shown. Doors for access to the instruments are notional. The doors, providing access to the instruments, harnesses, and cryocooler lines, may be hinged at the left and right rather than the top and bottom.

## CPM Base

The CPM Base is an aluminum box beam structure similar to the 35 K Deck, but with the points rotated to align with the Bus Bipods. The entire telescope is attached to the CPM Base, which is mounted on top of the spacecraft bus.

## Sunshields

Two sunshields (Inner Sunshield, Outer Sunshield) are deployed from the CPM. See Figure 2-20. The two sunshields have similar designs. The shielding material is DAK, with the Sun-facing side of the outer layer protected by GCC, as described in Section 2.4.

Each sunshield is mounted to the top- side of the CPM Base, with the Inner Sunshield mounted higher than the Outer Sunshield. The horizontal arm of each shield assembly is fixed to the Base, each containing a telescoping arm that is retracted and locked before launch. In the stowed configuration, each shield is retracted inward, allowing the shield material to wrap around the Barrel. Wrapping the Barrel protects the VDA from direct sunlight on-orbit, which could otherwise overheat the structure.

A vertical post is located at the end of horizontal arm of the Inner and Outer Sunshields. These vertical posts are lightweight aluminum tubes extending from the arms vertically to the end of each shield. The Outer Sunshield has an additional, closely-positioned post that can straddle the Inner Sunshield post so the Outer Sunshield can be fully retracted to the Barrel in the stowed condition. Extending laterally on each side from the posts are flexible, small-diameter, hollow, fiber composite rods. The design uses a pair of rods at the bottom, middle, and top of the shields. The rods are curved to maintain the curved shield shape upon deployment. For stowage, these rods are bent under preload to conform to the shape of the Barrel.

The Inner and Outer Sunshield DAK is fixed along the masts and flexible rods. The shielding for the Inner and Outer Sunshields extends radially inward at the bottom of each shield. Each shield is attached to its respective fixed sunshield located along the Bus Bipods. The Outer Sunshield has a second section attached to the top edge of the SBM.

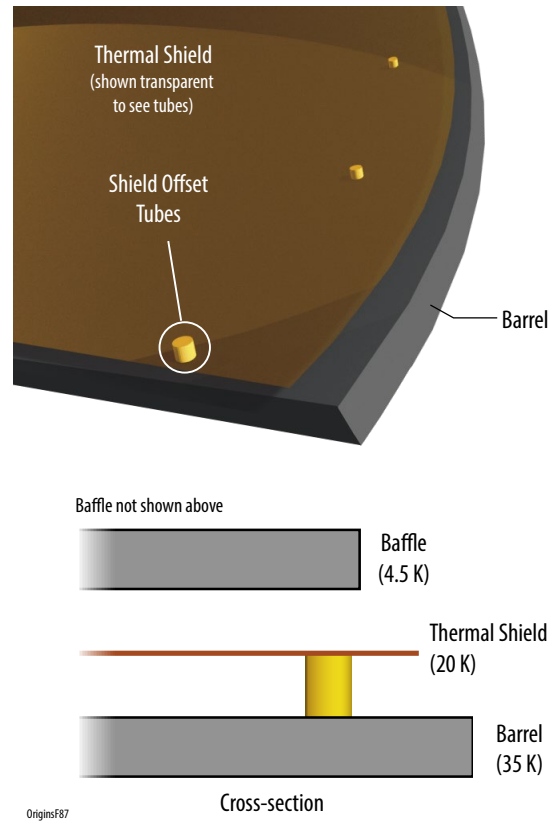

**Figure 2-19:** The Thermal Shield is offset from the Barrel to provide an effective radiative shield. The shield is fastened to thinwalled Vespel® tubes mounted to the inside of the Barrel.

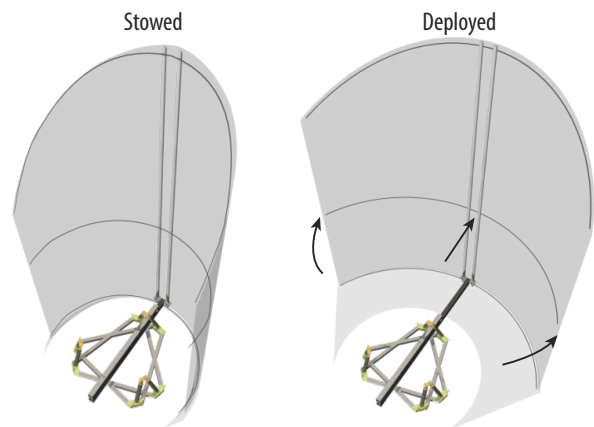

**Figure 2-20:** In the launch configuration, both shields are wrapped tightly around the Barrel. The simple linear deployment mechanism is attached to the Base and released on-orbit, allowing each shield to deploy. Outer Sunshield shown.

The deployment scheme is simple, with very few actuators and moving parts. The deployment can be demonstrated on the ground. The sunshield deployment system uses springs to force out the telescoping arms of each shield assembly and uses the stored energy in the flexible rods to pull the shield material into its final shape. Two restraint systems are used to hold the deployment system in the stowed launch configuration. A launch lock device secures the retracted telescoping arm inside the fixed portion of each main horizontal deployment arm. Kevlar strings tensioned through the hollow flexible rods are used to hold the rods to the shape of the Barrel. To release the sunshields, the strings are released and then retracted with a spring-loaded reel. The arm launch lock releases the telescoping arm section that places each shield at the intended radius from the Barrel.

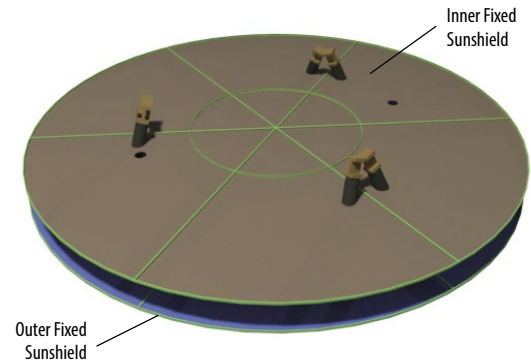

**Figure 2-21:** Fixed Sunshields are mounted along the Bus Bipods at the appropriate position to optimize thermal shielding. The deployed shields are attached to the respective Fixed Sunshield. Thermal closeouts are taped to the bipods and struts (Deployable shields not shown for clarity).

The Inner Fixed Sunshield and Outer Fixed Sunshield close out the Inner and Outer Sunshields, respectively, at the bottom, protecting the cold CPM from the warm SBM. They are mounted to the Bus Bipods at the appropriate height to achieve proper thermal effects. The fixed sunshield structure is comprised of aluminum tube sections that hoop around the bipods and at the major shield diameter with spokes attaching the two hoops (Figure 2-21). The support structure is covered in DAK. The Fixed Shields do not change position and are attached to the Deployable shields making for a continuous thermal shield. A skirt connects the fixed to the deployable parts of each shield. The skirt is sewn and taped to the two segments, pleated and folded for stowage. Please refer to the deployment video (*https://asd.gsfc.nasa.gov/firs/docs/*).

## 2.6 Telescope

### 2.6.1 Optical Layout

As shown in Figure 2-22, the *Origins* telescope is a Three Mirror Anastigmat (TMA)) and is composed of four mirrors: three with optical power (the elliptical primary, hyperbolic secondary, and elliptical tertiary mirrors) and a flat field-steering mirror (FSM). The TMA is the same general optical design form that has been proven on JWST and is therefore well understood and low risk. An initial, off-axis optical design considered for *Origins* resulted in a TMA with a "freeform" surface on all four mirrors to assist in correction of optical aberrations. Later studies found these design aspects to be costly and more challenging to implement, and so the design was modified. Other benefits of choosing the on-axis architecture over the off-axis one include avoiding a secondary mirror that requires deployment on orbit, as well as allowing the instrument volume to fit compactly behind the primary mirror. As such, the final baseline design was constrained to have an on-axis pupil (obstructed primary mirror) with the remaining three mirrors simple on-axis conics. The obstructed primary mirror also made the observatory simpler to package in the fairing as compared to studies regarding a telescope with an unobstructed primary mirror. While a number of aperture shapes were investigated for the purpose of this design study (based on the goal of trying to find the one easiest to package and deploy), ultimately a circular aperture was chosen as it yielded the cleanest point spread function (PSF) while being possible to fit it already fully-deployed within the chosen fairing before launch.

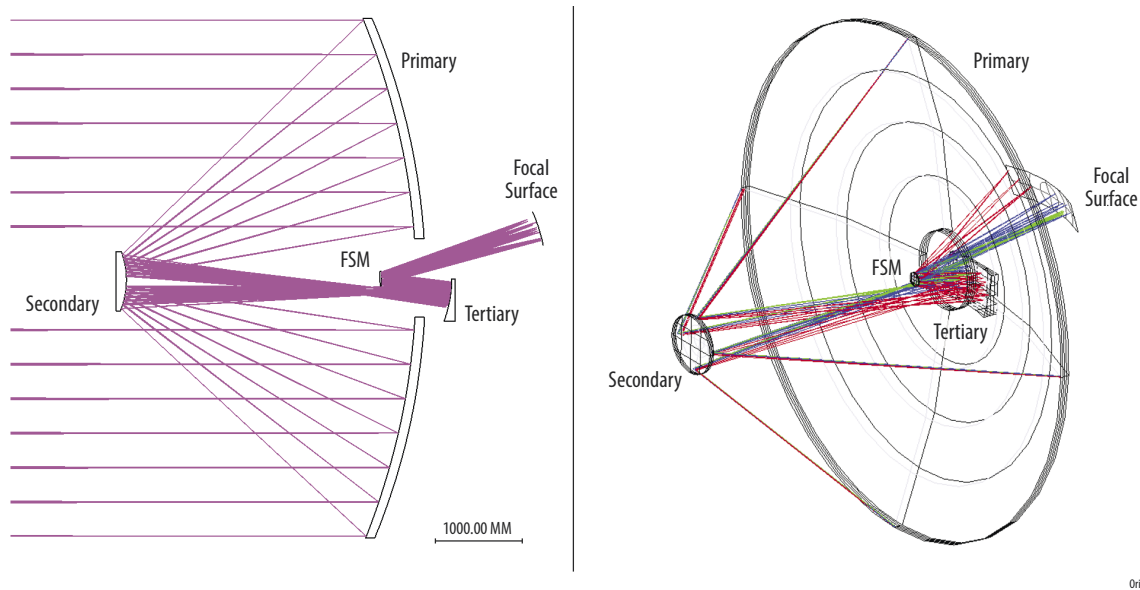

**Figure 2-22:** *Origins* baseline telescope layout shows the locations of the four mirrors and focal surface from both side (left) and perspective (right) views.

The FSM has about the same size, weight, range of motion and speed as the JWST Fine Steering Mirror. It is baselined to use the same actuators as JWST with the addition of superconducting voice coils to limit the dissipation during use.

The primary mirror has 18 segments and is f/0.63. Such a fast primary mirror was desirable for shortening the distance between the vertices of the primary and secondary mirrors (being ~3.33 m), therefore reducing the overall size of the observatory. Between the secondary and tertiary mirrors, the rays form an internal image (Cassegrain focus) that allows the tertiary mirror to image a real exit pupil. This real image will be surrounded by baffling to reject stray light. The FSM is placed at the exit pupil of the telescope, which actively tilts to control the Field Of View (FOV), directing it into each instrument as the observatory slowly drifts over the course of an observation. The telescope image surface is concave, with its center of curvature located at the FSM surface. Placing the FSM at the exit pupil (which is also at the center of curvature of the telescope image surface) prevents defocus from occurring during tilting of the FSM. This effectively makes a locally- telecentric system for each field point, for which each chief ray is normal to the curved image surface.

The optical design of the telescope corresponds to the final prescription of the system on orbit. As such, the mirrors' geometry (including radius of curvature) will be different on the ground as compared to on orbit to account for the effects of gravity sag, CTE, etc. The JWST team had to compensate for between 100 and 200 nm of gravitational sag while figuring the mirror (Daukantas, 2011). As *Origins* is baselined with the same material as JWST and roughly the same diameter, it is likely that these numbers will be similar. Local distortions on the mirror segments due to gravity sag show up in interferometric measurements on an otherwise-nulled optic, as was the case for JWST (Howard, *et al.*, 2007). It will be imperative to predict the form and magnitude of these distortions using a combination of finite element modeling and optical software such as SigFit. Once computed, these results will be compared with observed measurements to give confidence the mirrors will behave on orbit as expected.

The physical dimensions and masses of each of the four mirrors are tabulated in Table 2-6. Each mirror is assumed to be made of beryllium (O30-H) with a density of 1.85 g/cm³. The light- weighting factor for each element follows the convention that a value of 0.75 corresponds to 25% of the solid mass of the final optic. The lateral dimension is different for each shape: diameters for circles, length and width for rect-

**Table 2-6:** The shapes, sizes, masses, and quantity of the mirrors composing the *Origins* optical design are summarized.

| # | Abbrev. | Name | Shape | X [mm] | Y [mm] | Thi. [mm] | Vol [cm²] | Bulk Mass [kg] | LW Frac. [—] | LW Mass [kg] | Quantity [—] | Total Mass [kg] |
|---|---------|------|-------|--------|--------|-----------|-----------|----------------|--------------|--------------|--------------|-----------------|
| 1 | IMS | Inner Mirror Segment | Keystone | See below | | 60.33 | 80859 | 149.589 | 0.86 | 21.060 | 6 | 126.359 |
| 2 | OMS | Outer Mirror Segment | Keystone | See below | | 60.33 | 100137 | 185.253 | 0.86 | 25.790 | 12 | 309.479 |
| 3 | SM | Secondary Mirror | Circular | 700 | — | 120 | 46181 | 85.436 | 0.75 | 21.359 | 1 | 21.359 |
| 4 | TM | Tertiary Mirror | Rectangular | 760 | 480 | 80.6 | 29403 | 54.395 | 0.50 | 27.198 | 1 | 27.198 |
| 5 | FSM | Field Steering Mirror | Circular | 160 | — | 20 | 402 | 0.744 | 0.00 | 0.744 | 1 | 0.744 |
| | | | | | | | | | | | **Total** | **485.139** |

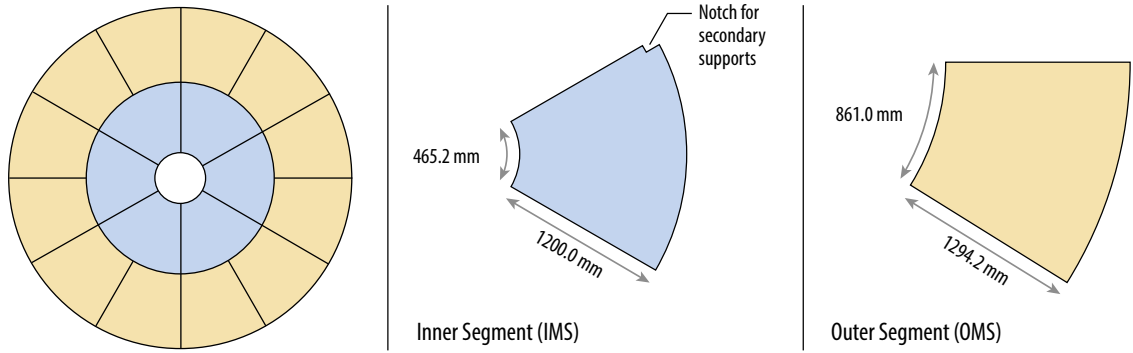

Inner Segment (IMS) — 465.2 mm, 1200.0 mm, Notch for secondary supports

Outer Segment (OMS) — 861.0 mm, 1294.2 mm

OriginsF91

**Figure 2-23:** The twelve *Origins* primary mirror outer segments are identical. The inner segments are of one design, with three of the six mirrored to accommodate a notch for the secondary mirror struts.

angles. Figure 2-23 shows dimensions for the six inner and twelve outer "keystone" segments that make up the primary mirror. Because the primary mirror is on- axis, all of the outer segments are identical. The inner segments have identical optical prescriptions; however, two variants are needed to accommodate a notch for the secondary mirror struts. Three of the inner mirror segments match the outline shown in Figure 2-23, and three segments are reflections of this shape. All four mirrors are protected gold-coated for improved reflectance in the infrared. Each mirror coating is assumed to have 98% reflectance and 2% emissivity which is consistent with measurements made on JWST's protected gold-coated mirrors up to a wavelength of 29µm (Keski-Kuha 2012). Non-JWST measurements on protected gold coatings in the terahertz regime have reported reflectance values around 99% between 100 and 600 µm (Naftaly 2011). Based on this the total throughput for the *Origins* telescope is about 92%.

### 2.6.2 Imaging Performance

*Origins* is required to be diffraction-limited at a wavelength of 30 µm. Two different optical performance metrics are used to evaluate this requirement. The first is root-mean-square wavefront error (RMSWE) over the FOV (with the footprint of each of the three instruments overlaid), as shown in Figure 2-24. The Fourier Transform Spectrometer (FTS) OSS FOV is shown in magenta but is part of the greater OSS FOV allocation. A common standard for diffraction-limited performance is to have less than 0.07λ of wavefront error. For a design wavelength of 30 µm, this corresponds to less than 2.1 µm RMSWE. Figure 2-24 shows that this specification is met across each instrument's FOV. The performance is slightly degraded (to about 0.08λ) in the upper left corner of the plot; however, this is outside of the FOV allocation of OSS: the closest instrument. The second metric is to evaluate Strehl ratio over the FOV as shown in Figure 2-25. A Strehl ratio of greater than 0.8 is defined as a diffraction-limited design. Figure 2-25 shows the requirement is met across the FOV of each instrument. The telescope is designed over a rectangular 15 x 46 arcminute full FOV; however, this FOV is not symmetric being longer in the direction of OSS than FIP. This is due to the fact that during the design process, further FOV was added to OSS after the FIP design had been completed. Rather than redesign FIP at that stage, OSS' FOV was extended in the direction away from the center of the FOV. This asymmetry can be addressed in a later design by just shifting both instruments' designated FOVs by approximately 3 arcminutes.

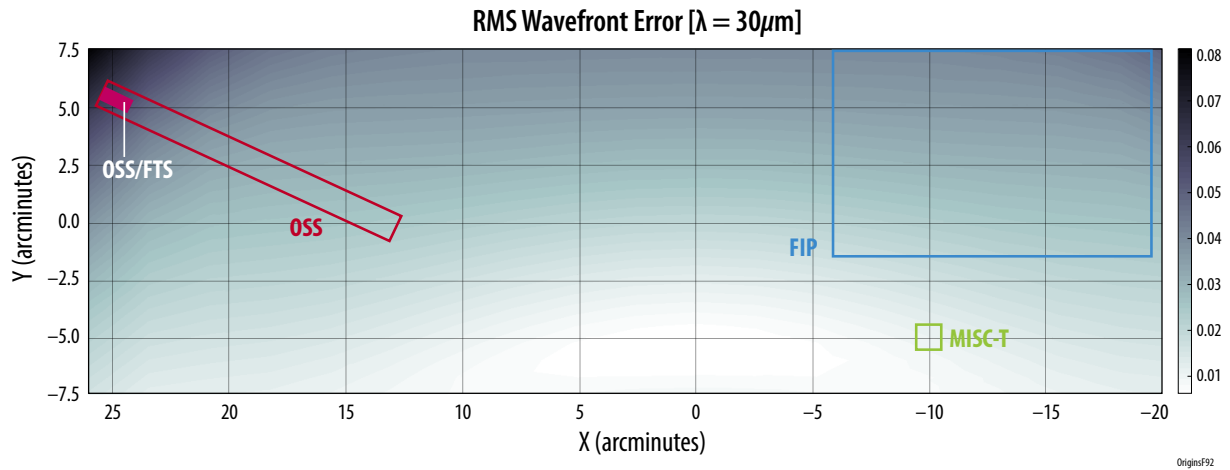

**Figure 2-24:** An evaluation of the baseline *Origins'* RMS wavefront error as a function of FOV shows that the performance requirement (less than 0.07λ) is met across each instruments FOV. Note that the colorbar is set such that light corresponds to better optical performance and dark to poorer optical performance.

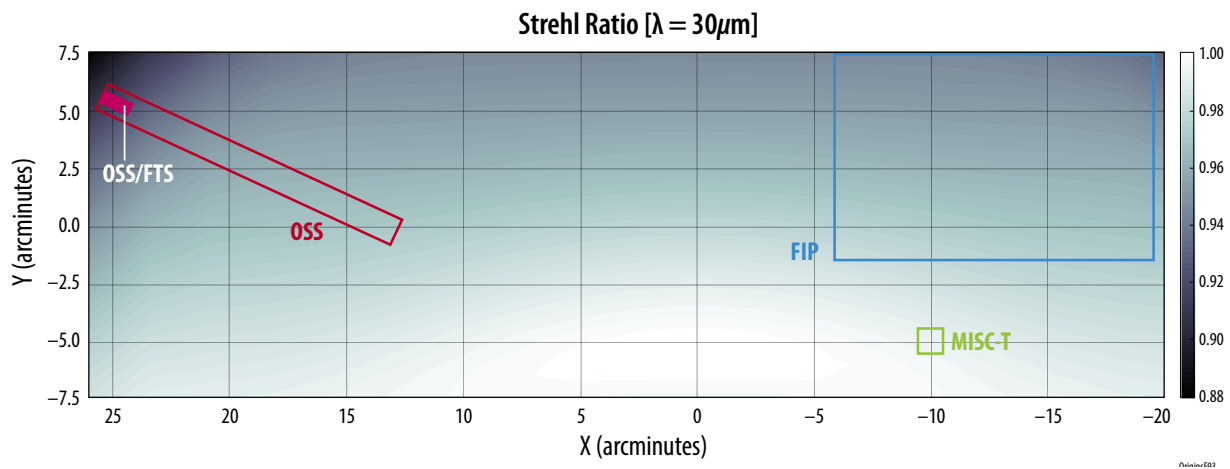

**Figure 2-25:** An evaluation of the baseline *Origins* Strehl ratio as a function of FOV shows that the performance requirement (greater than 0.8 Strehl ratio) is met across each instruments FOV. Note that the colorbar is set such that light corresponds to better optical performance and dark to poorer optical performance.

In both **Figures 2-24** and **2-25**, one arcminute of FOV on the sky corresponds to about 24 mm in length at the focal surface. Space is left between instruments to leave appropriate room for mounting structures. Looking at both figures, there is a large amount of space between the different instruments in the x direction of the FOV. This is to make it possible to more easily accommodate any desired up-scopes to the *Origins* baseline design. These upscopes include: greater FOV for OSS and FIP, another channel for MISC, and the addition of the HERO instrument as shown in Appendix D. In future design iterations, once the total number of instruments and FOV allocations for each instrument is decided, the footprints in Figures 2-24 and 2-25 would be reoptimized to make better use of the central 'sweet spot' of optical performance and reduce the large gap between instruments. To show more detail as compared to Figure 2-22, Figure 2-26 is a CODEV® layout showing rays for each of the three baseline instruments passing through the telescope onto the image surface. Details regarding *Origins'* stray light analysis, wavefront error budget, and alternative telescope architecture studies are contained in Section C.3 of Appendix C.

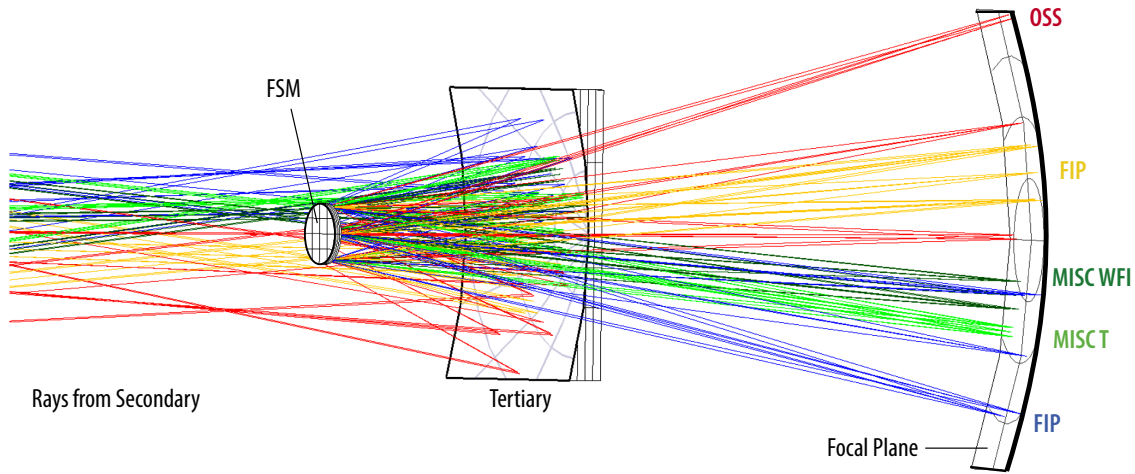

**Figure 2-26:** CODEV® layout showing rays for each of the three baseline instruments passing through the telescope onto the image surface.

### 2.6.3 Telescope Mechanical/Structure

#### Primary Mirror Subassembly

The PMBSS and the Secondary Mirror Support Structure (SMSS) do not, in themselves, guarantee the telescope will be aligned and in focus when cooled in orbit. Actuators for the primary mirror segments and secondary mirror are required for *Origins*. The *Origins* Primary Mirror (PM) Segment Assembly (PMSA) (Figure 2-27) employs proven, JWST-type actuators for tip, tilt, and piston ad-justments on each of the 18 primary mirror segments. The actuator motors are built-to-print from JWST, whereas the mechanism is slightly modified to provide kinematic motion.

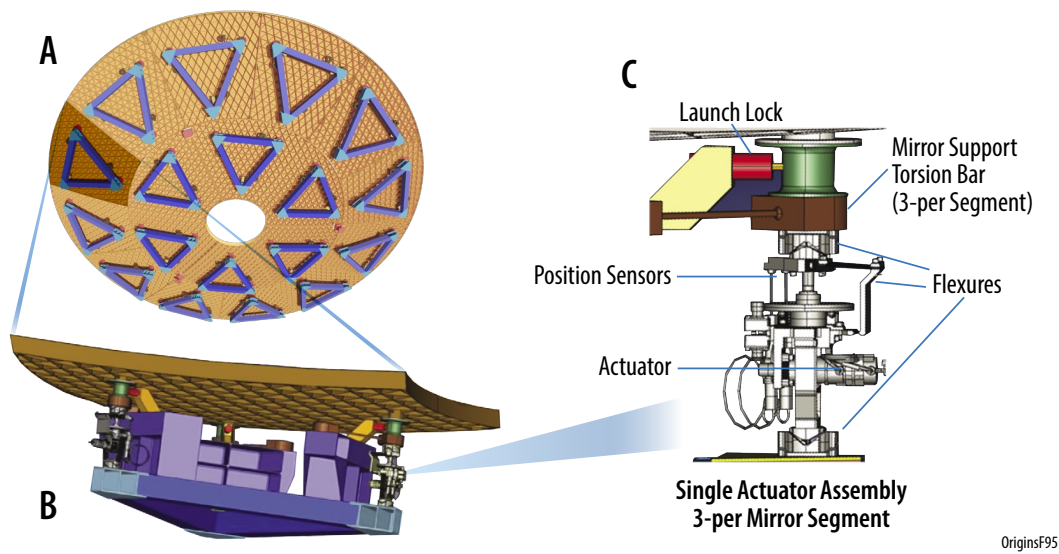

**Figure 2-27:** The (A) *Origins* Primary Mirror Segment Assembly showing a (B) mirror segment and (C) actuating mechanism is based on proven designs. Each segment uses three JWST actuators with flexures to ensure kinematic, reproducible motion.

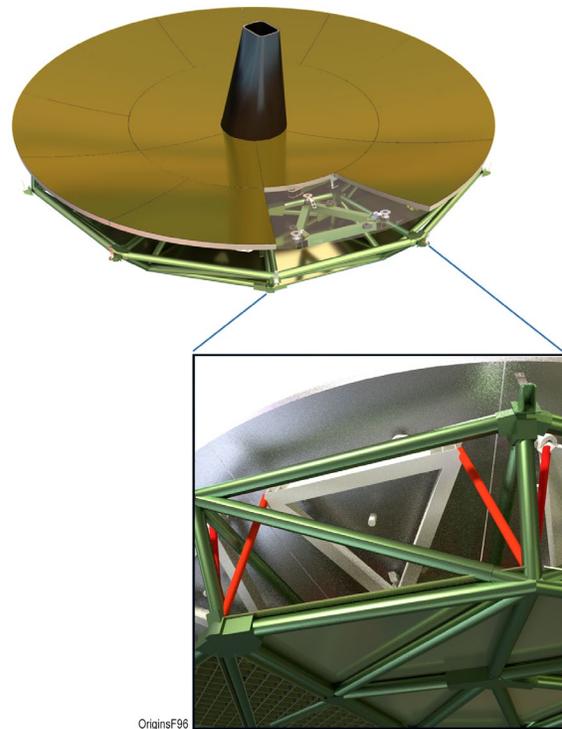

The *Origins* Primary Mirror subassembly is comprised of 18 individual segments. The 12 beryllium outer mirror segments (OMS) are identical. The six beryllium inner mirror segments (IMS) are identical. To meet *Origins* PM fabrication requirements, the team employs analytical optimization for all segments, as well as proven JWST fabrication techniques.

All 18 segments use three JWST actuators (54 actuator subassemblies total) and three beryllium symmetric flexures, yielding a three DOF actuated system. The optical error budget allows the use of only 3 actuators per segment vs. 7 per segment for JWST. Each actuator subassembly system is launch-locked to avoid damage to the flexures during lift-off. The actuator subassembly design allows the required *Origins* PM motion while reducing the need for additional actuators, reducing cost and complexity. Once on orbit, the launch locks are released to allow the *Origins* PMs to tip, tilt, and piston via the three actuators. Alignment will take place using FIP's image of a point source, for instance a quasar. Each actuator subassembly is integrated to a triangular beryllium support structure. The 18 individual PM subassemblies are each integrated to the *Origins* PM Backplane Support Structure at the vertex of the triangular support structure (Figure 2-28).

**Figure 2-28:** The Primary Mirror Segment Assembly Interface support structure recesses into the PMBSS truss elements and is supported by brackets to the lower truss elements.

**Primary Mirror Backplane Support Structure**

The Primary Mirror Backplane Support Structure (PMBSS) supports the PMSAs, Secondary Mirror Subassembly, Baffle, and instruments. The PMBSS is a trussed structure of beryllium tubes and brackets (Figure 2-29). The truss joint brackets serve to connect the truss elements and provide interface-mounting points for the PMSAs.

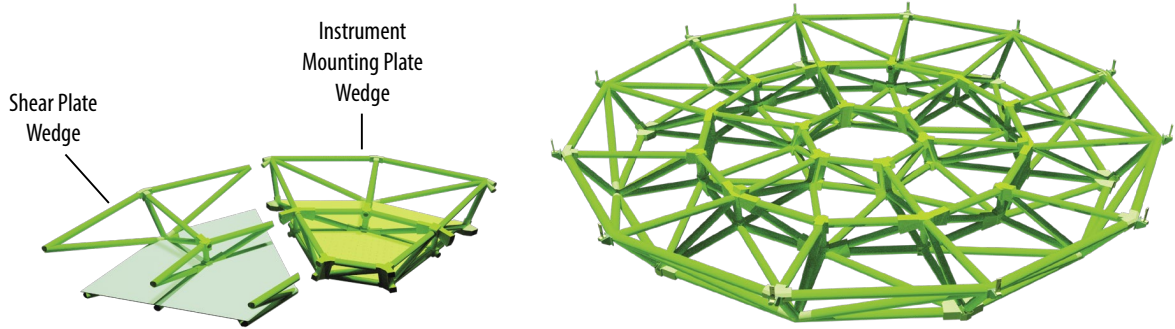

**Figure 2-29:** To optimize weight and strength, the PMBSS is a trussed design. The PMBSS is entirely beryllium to match the Coefficient of Thermal Expansion (CTE) of the Primary Mirror Segment Assembly. Three sets of PMBSS Wedge Subassemblies are fabricated and fastened together to comprise the whole PMBSS.

The design is circumferentially symmetric and can be broken down into three sets of two pie- shaped wedge styles (Figure 2-29). The size of each wedge subassembly allows the components to be brazed together using an existing brazing oven. The six wedges are bolted together to complete the PMBSS overall assembly. Because beryllium is brittle, the design uses through-hole bolted joints instead of tapped threaded holes.

Figure 2-29 shows the Instrument Mounting Plates (IMP) and the PMBSS shear plates. These plates serve to strengthen the assembly and provide interface locations for the instruments. The shear plates are uniformly thick (6.35 mm). The IMPs are 76.2 mm thick and machined to form a uniform isogrid of out-of-plane stiffeners and through-holes for mounting of instruments. These isothermal plates are excellent mounting points for the 4.5 K cryocooler heat exchangers.

### Secondary Mirror Subassembly

The Secondary Mirror Subassembly (inset of Figure 2-30) is comprised of a mirror subassembly mounted on a tripod. The tripod legs are beryllium tubes that protrude through openings in the PMSA assembly and mount directly to the PMBSS similar to the Hershel architecture. The Secondary Mirror Subassembly is actuated and designed similarly to the JWST secondary with six actuators for tip, tilt, piston, translation, and rotation.

### Aft Optics Structure Subassembly

The tertiary mirror and the Field Steering Mirror (FSM) are mounted in the Aft Optics Structure (AOS) (Figure 2-31). The tertiary mirror is a fixed mirror mounted at the end of the AOS. The FSM is an actuated mirror that directs the light into the instrument FOVs.

### Telescope Cover

Contamination control within the CPM is critical, especially above the PMSAs. See Figure 2-13. The Telescope Cover provides contamination control and structural support during launch. It also protects the telescope from exposure to the Sun immediately after launch, and allows a Solar- facing direction for efficient use of the first orbit correction maneuver ~24 hours after launch. After that time, it is deployed and released. The cover is a lightweight but rigid subassembly resembling a bicycle wheel; a central aluminum hub supports radial spokes (aluminum tubes) that connect to a rigid rim. The rim attaches to the Baffle and Barrel through eight equally-spaced launch locks. Between the launch lock housings is an aluminum T-shaped curved beam that sits on the top edge of the Baffle and Barrel. Dowel pins protruding from the top edge of the Baffle and Barrel panels

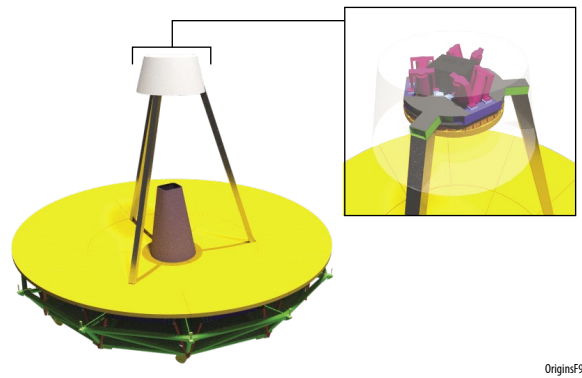

**Figure 2-30:** The Secondary Mirror Subassembly mounts on the end of a tripod connected directly to the PMBSS.

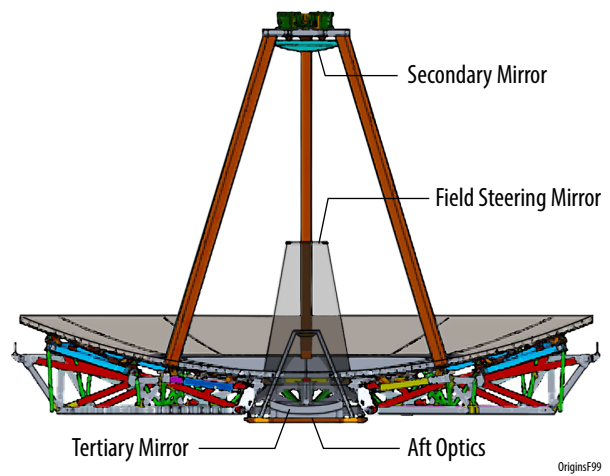

Secondary Mirror

Field Steering Mirror

Tertiary Mirror          Aft Optics

**Figure 2-31:** Aft Optics Structure mounts to the PMBSS and contains the tertiary mirror and Field Steering Mirror.

slip fit through the beam, providing radial support for the Baffle and Barrel cylinders. The entire structure is covered in DAK, with GCC on the outer surface.

The Cover is the final item to be deployed (after the solar array, communication antenna, and sunshields) and is ejected. Eight launch locks are located around the perimeter of the Cover rim. Spring plungers are installed in co-cured edge blocks under each launch lock to assist with Cover deployment. Other than the launch locks, Cover deployment and ejection is very similar to that of the *Spitzer* telescope dust cover.

### 2.6.4 Cryo-compatible Telescope Materials

The *Origins* team completed a materials assessment for the mission's main optical and structural elements. The evaluation team, which included *Origins* team members and industry partner materials experts, developed an iterative process to identify suitable material candidates. The material candidates were chosen based on material performance, spaceflight mission heritage, and knowledge of current manufacturing and processing capabilities.

The evaluation criteria were determined by the driving requirements and included critical material properties and relevant manufacturing challenges. An initial evaluation of primary mirror material candidates yielded five potential options: beryllium, aluminum, AlBeMet®, silicon carbide, and fused silica. The team conducted further trade studies to better understand the trade-offs between material performance and *Origins* requirements. Structural materials were also assessed in a broader framework.

### Driving Requirements

Since the telescope including the backplane is expected to be isothermal, it must be composed of materials with relatively high thermal conductivity, > 4 W/m•K at 4.5 K. It must have low overall mass, namely <900 kg, a large primary with light collecting area > 25 m$^2$ and the materials themselves should have a relatively high TRL ≥ 4.

### Evaluation Criteria

The evaluation criteria were derived from the driving requirements and assessed the candidates' overall material performance, measuring how a material's properties and characteristics serve the application. Evaluation criteria included material properties and considerations, such as: density, stiffness, thermal conductivity, CTE, outgassing, and manufacturability.

### Density

Telescope mass is driven by the size and material density. The size is driven by the science objectives, including the aperture diameter of the primary mirror and number and types of instruments. Density is defined as mass per volume and is one of the most significant material properties for systems launching to space. Materials with a low density and high overall material performance are ideal for meeting *Origins*' overall mass budget.

### Stiffness

Stiffness is another important material property for optical and structural elements. Stiffness, or Young's modulus, is the ability of an object to resist deformation in response to an applied force. High stiffness provides dimensional stability, allowing mirrors and structural components to hold their shape over long periods. Stiffness is also critical for maintaining optical alignments in the system, especially during launch and deployment when a telescope experiences severe vibrations (Edinger, 2005).

Specific stiffness, or specific modulus, is Young's modulus per mass density. Materials with high specific stiffness are generally favored because components have maximum stiffness at minimum weight.

## Thermal Conductivity

A unique objective of *Origins* is achieving a 4.5 K temperature on the cold side of the telescope. Materials with high thermal conductivity can transfer heat faster and cool down uniformly. The primary mirror is one example of where thermal conductivity is critical. Cryocooler heat exchangers are strategically placed on the back of the *Origins* primary mirror structure and need to cool down the entire mirror, including mirror segments further away from the cryocoolers. The connecting elements of the mirror–segment frames and struts–should also have a reasonable thermal conductivity to enable them to reach 4.5 K, and be isothermal under reasonable (~ 10 mW) heat flows at operating temperature.

On the other hand, the warm areas of *Origins* need to be isolated from the cold components to prevent heat transfer. Therefore, some sections of the telescope need to be made from materials with very low or practically no thermal conductivity.

## Coefficient of Thermal Expansion

The CTE is used to determine how a material changes dimensionally as a function of temperature. This is important to analyze when using different materials together, especially for bonds and joints where polymers or metal nodes are used. Using materials with similar CTEs can reduce or eliminate issues with thermomechanical stresses at interfaces or in bonding materials. Some materials, like Invar, have an extremely low CTE and are typically used as fittings to mitigate CTE mismatch between structural components, such as composites tubes, rather than using an adhesive or metal with a large thermal expansion.

## Outgassing

It is crucial for the optics to remain clean before launch and in space, so they can provide clear images and high-quality data during operation. Some materials may outgas in a vacuum environment depending on their material composition and temperatures. Outgassing is a concern for systems in space because this released gas, including water vapor, can potentially condense on cold optical surfaces and distort images and data. During deployment, materials may outgas when in direct path of the sun. The outgassing properties of all organic materials should be evaluated before approval, including adhesives, epoxies, and matrix materials in composites like carbon fiber reinforced polymer (CFRP), polyetheretherketone (PEEK), Ultem®, and Vespel®.

## Manufacturability

When selecting a primary mirror, whether it is a monolithic mirror or a segmented mirror, the material choice can have significant impacts on development. A monolithic mirror made of a material with limited heritage may require a costly facility development program to accommodate the size and material selected, whereas a segmented mirror may require a lengthy segment development program, where the first segment is manufactured and tested, and the remaining segments are manufactured subsequently. For example, JWST underwent an extensive segment development program for the 18 beryllium segments for its 6.5-meter primary mirror. For this reason, the team considered meter-class mirror heritage and manufacturability for point to point hexagonal segments for all material candidates.

Cryogenic testing is also a costly and lengthy process for mirror development. However, *Origins*' relaxed requirements, in comparison to JWST, eliminates the need for testing like cryo- null testing and figuring for mirror shape, which helps reduce the cost and schedule for mirror development.

## Mirror Material Candidates

Mirror materials were evaluated before structural materials due to the inherent complexity of assessing mirror manufacturing. A preliminary list of material candidates is shown in Table 2-7. Candidate mirror materials were systematically evaluated, resulting in five potential options. Major advantages and disadvantages were noted for each material with respect to manufacturability and material performance for *Origins*.

The candidates were chosen using material properties, mirror material heritage, and the knowledge of current manufacturing and processing capabilities. Traditional mirror materials include glasses, ceramic materials, and fused quartz while nontraditional mirror materials include metals, metal alloys, silicon carbide (SiC), and CFRP composites. Nontraditional mirror materials offer opportunities to reduce weight and cost. In the case of *Origins*, nontraditional mirror materials also provide thermal advantages over traditional mirror materials. (Cheng, 2009)

The first assessment was driven by *Origins*' 4.5 K operating temperature. The team first eliminated materials that were a concern for cryogenic temperatures. This included the glasses and glass ceramics: ULE (titania-silicate glass), Zerodur (lithium-aluminosilicate glass-ceramic), and Borosilicate (glass with silica and boron trioxide). However, fused silica was not eliminated because of its prominent heritage as an optical substrate and its potential to perform in cryogenic temperatures. Carbon fiber reinforced polymer (CFRP) was eliminated based on its anticipated poor optical quality and titanium because of its extremely high density. Five mirror material candidates remained: beryllium, aluminum, fused silica, silicon carbide, and aluminum and beryllium metal matrix composite (AlBeMet®).

**Table 2-7:** Candidate mirror materials were systematically evaluated, resulting in five potential options.

| Material | Advantages | Disadvantages |
|---|---|---|
| **Potential Materials** | | |
| Beryllium O-30 | Superior stiffness, extremely lightweight, low CTE | Expensive, brittle, toxic, long machining time |
| Aluminum 6061 | Good structurally, good fabrication time, inexpensive | Heavy, reactive surface, low stiffness |
| AlBeMet® | Good stiffness, low CTE, lightweight | Limited information and heritage, toxic |
| SiC | Excellent stiffness, excellent strength, low CTE | Heavy, expensive, long machining time |
| Fused Silica | Low CTE, lightweight | Low stiffness, low thermal conductivity |
| **Eliminated Materials** | | |
| Titanium | Excellent strength, good thermal performance | Extremely heavy, machining ability |
| ULE | Low CTE, lightweight | Poor thermal performance, low strength |
| Zerodur | Low CTE | Poor thermal performance, low strength |
| Borosilicate | Lightweight | Poor thermal performance, low strength |
| Composite/CFRP | Extremely lightweight, low CTE can be achieved | Surface quality (optical finish), creep |

**Material Trade Matrix**

After the initial evaluation, a material trade matrix was used to identify the top choices among the remaining material candidates. This trade matrix was initially designed for consideration of passive mirror segments for a 9.1 m aperture. The matrix is shown in Table 2-8 and assesses each material on a scale of 1 to 5 for performance, schedule, and cost, and a scale of 1 to 9 for TRL, using NASA standard values. The materials are listed in order of preference, showing fused silica and (SiC) as the top choices.

The results of the primary mirror evaluation show beryllium as the highest performing, highest TRL solution despite programmatic challenges. SiC also has high performance and is another viable option for *Origins*. AlBeMet® offers the opportunity to improve the manufacturability of beryllium, but would ultimately weigh more overall and have a lower stiffness than beryllium. Fused silica has the stron-gest optical heritage among the candidates, which would greatly reduce cost and schedule, but the low thermal conductivity makes it unsuitable for cooling and temperature uniformity. Aluminum has relatively good properties and machinability, but overall lower thermal performance and lower specific stiffness than the top candidates.

**Table 2-8:** A material trade matrix assisted the *Origins* team in identifying the top primary mirror materials.

| Material | Performance | Schedule | Cost | Heritage |
|---|---|---|---|---|
| **Fused Silica** | 4 | 5 | 4 | Glass has max heritage as optic substrate |
| **SiC (multiple)** | 4 | 4 | 3 | Herschel heritage |
| **Beryllium** | 5 | 2 | 2 | JWST heritage |
| **Aluminum 6061** | 2 | 5 | 5 | All-Al telescope was studied and found to be too massive |
| **AlBeMet®** | 4 | 2 | 1 | No meter-class heritage |

### Structural Materials

Ideal structural materials share many of the same characteristics as ideal mirror materials–strong, lightweight, and a near-zero change in thermal expansion. The team considered aluminum, copper, carbon-fiber reinforced polymer (CFRP), titanium, Invar, stainless steel, beryllium, and silicon carbide as the possible structural materials to use on *Origins*. Table 2-9 summarizes the Pros and Cons.

**Table 2-9:** The advantages and disadvantages of each structural material were considered.

| Material | Advantages | Disadvantages |
|---|---|---|
| Aluminum | Good fabrication time, inexpensive , high thermal conductance | Heavy in comparison to CFRP, low stiffness |
| Copper | Excellent thermal properties | Extremely heavy |
| Composite/CFRP | Extremely lightweight, low CTE can be achieved | Creep, outgassing |
| Titanium | Excellent strength, good thermal properties | Extremely heavy, machining ability |
| Invar | Near zero CTE, excellent stiffness and strength | Heavy |
| Stainless steel | Excellent structural material | Extremely heavy |
| Beryllium | Superior stiffness, extremely lightweight, low CTE , high thermal conductance | Expensive, brittle, toxic, long machining time |
| Silicon carbide (SiC) | Excellent stiffness, excellent strength, low CTE | Brittle, expensive, long machining time |

For the 4.5 K structure beryllium was selected. There are several potential issues with the use of beryllium that will be taken into account. Beryllium is somewhat brittle so requires special care in design (sharp corners, proper clearance holes, etc. Beryllium dust is toxic, so fabrication limited to certain places. There is no danger in use after fabrication, though.

The team consulted with several organizations that have used large beryllium structures and they offered the following advice, which we will follow.

These organizations (Lockheed for instance) would not hesitate to use beryllium again after their experiences (NIRCam bench in the case of Lockheed).

The *Origins* plan is to work with vendors early to qualify workmanship and design details.

We will make a wedge of the primary mirror to validate performance after joining of sections together. Finally, the team recognizes the need to resurrect the beryllium processing facilities used by JWST. This will be a pre-Phase A activity.

Standard metals are typically strong and have excellent thermal properties but are generally too heavy to be used in large components. Lightweight materials like beryllium and silicon carbide are ideal, but their brittle nature requires extensive design development for large structural components. CFRP is strong, lightweight, but has low thermal conductance and is not suitable for *Origins* without also providing extensive thermal strapping to isothermalize the structure. Its other weakness is that it will produce water outgassing when warm (>160 K).

For the spacecraft structure, the team selected CFRP whenever possible. *Origins* has critical structures on the warm side (SBM) and the cold side (mirror and instrument support structure and thermally isolating support structures). Therefore, CFRP is an advantageous choice for the SBM, whereas metals are more suited for the cold side structures.

While the observatory is warm before cool down, water vapor will outgas from CFRP. On the cold side, this vapor could condense onto the mirror and instruments, impacting optical throughput. As a result, the team avoided using CFRP as much as practicable on the cold side, opting instead for metals. An exception to this are the 4.5 K bipods which are made of the CFRP M55J to provide the necessary stiffness to thermal conductivity in the range of 4.5 to 35 K. In contrast, high thermal conductance is also critical for the backplane, which will be cooled conductively to 4.5 K, so the team sought a metal with good thermal conductivity down to that temperature. The mirror structures would also ideally be made from the same material as the mirror segments to avoid CTE mismatch between structures on the cold side. Thus, an athermal design of beryllium on the cold side is favorable. However, beryllium grades, such as I-220H, require more design development than more common structural materials like aluminum or composite.

**Contamination Mitigation**

In general, contamination susceptibility of *Origins* will be lower than that of JWST due to the longer wavelengths of *Origins*. Particulate contamination and non-volatile residues will be monitored and mitigated using similar methods to JWST. The *Origins* aperture cover will be in place during transportation, vibration tests, and launch to further prevent contaminants from reaching the telescope or instruments.

Contamination, especially from water, is curtailed by minimizing the use of composites inside the barrel. Vapor emanating from the spacecraft and other warm portions of the observatory has no line-of-sight to the cold optical surfaces and instruments, and will condense on the cold sunshield layer or outside of the barrel.

A model of the on-orbit self-contamination generated by the observatory, including water and hydrazine, will be developed during Phase A using the Distributed Monte Carlo Method Analysis Code (DAC).

## 2.7 Origins Instruments Overview

*Origins'* three powerful instruments that cover wavelengths from 2.8 to 588 µm enable all the science described in Section 1.0. The Science Traceability Matrix (STM, Section 1.4) outlines the science requirements that drive the instrument requirements. The instrument capabilities (Table 2-10) flow from the STM requirements. The *Origins* Survey Spectrometer (OSS, Section 3.1), Mid- Infrared Spectrometer and Camera Transit Spectrometer (MISC-T, Section 3.2) and Far-infrared Imager and Polarimeter (FIP, Section 3.3) are described in detail in later subsections.

**Table 2-10:** Summary of instrument capabilities

| Instrument/ Observing Mode | Wavelength Coverage (µm) | Field of View | Spectral Resolving Power (R=λ/Δλ) | Saturation Limits | Representative sensitivity 5σ in 1 hr |
|---|---|---|---|---|---|
| 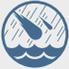 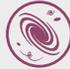 *Origins* Survey Spectrometer (OSS) | | | | | |
| Grating | 6 bands cover 25-588 simultaneously | 6 Slits: 2.7' x 1.4" to 14' x 20" | >300 | 5 Jy @ 128 µm | $3.7 \times 10^{-21}$ W m$^{-2}$ @ 200 µm |
| High Resolution w/Fourier Transform Spectrometer | 25-588 Total range scanned by FTS | Slit: 20" x 2.7' to 20" x 20" | <43,000 x [112 µm/λ] tunable w/FTS scan length | 5 Jy @ 128 µm | $7.4 \times 10^{-21}$ W m$^{-2}$ @ 200 µm |
| Ultra-High- resolution w/ Fabry- Perot | 100-200 Select lines scanned | One beam: 6.7" | 325,000 x [112 µm/λ] | 100 Jy @ 180 µm | $\sim 2.8 \times 10^{-19}$ W m$^{-2}$ @ 200 µm |
| 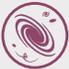 Far-infrared Imager Polarimeter (FIP) | | | | | |
| Pointed | 50 or 250 (selectable) | 50 µm: 3.6' x 2.5' 250 µm: 13.5' x 9' (109x73 pixels) | 3.3 | 50 µm: 1 Jy 250 µm: 5 Jy | 50/250 µm: 0.9/2.5 µJy Confusion limits: 50/250 µm: 120 nJy/ 1.1 mJy |
| Survey mapping | 50 or 250 (selectable) | 60" per second scan rate, with above FOVs | 3.3 | 50 µm: 1 Jy 250 µm: 5 Jy | Same as above, time to reach confusion limit: 50 µm: 1.9 hours 250 µm: 2 msec |
| Polarimetry | 50 or 250 (selectable) | 50 µm: 3.6' x 2.5' 250 µm: 13.5' x 9' | 3.3 | 50 µm: 10 Jy 250 µm: 10 Jy | 0.1% in linear polarization, ±1° in pol. Angle |
| 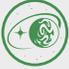 Mid-Infrared Spectrometer Camera Transit Spectrometer (MISC-T) | | | | | |
| Ultra-Stable Spectroscopy | 2.8-20 in 3 simultaneous bands | 2.8–10.5 µm: 2.5 radius 10.5–20 µm: 1.7 radius | 2.8–10.5 µm: 50-100 10.5–20 µm: 165-295 | K~3.0 mag 30 Jy @ 3.3µm | Assume K~9.85 mag M- type star, R=50 SNR/sqrt(hr)>12,900 @ 3.3 µm in 60 transits with stability ~5 ppm < 10.5 µm, ~20 ppm > 10.5 µm |

OSS is a highly capable spectrometer that covers the entire 25 to 588 μm band at moderate (R~300), high (R~4×10⁴), and ultra-high (R~2×10⁵) spectral resolving power. OSS uses six gratings in parallel to take multi-beam spectra simultaneously across the 25 to 588 μm window through long slits. In this grating mode, OSS spatially and spectrally maps up to tens of square degrees of the sky providing 3-D data cubes. When needed, a Fourier transform interferometer and an etalon provide high and ultra-high spectral resolving power, respectively, in a single beam, with insertable elements that redirect the light path. The three OSS spectroscopy modes are packaged into one instrument. To meet its performance requires improved detector sensitivity and larger pixel format size (Section 2.3).

FIP is a simple and robust instrument that provides imaging and polarimetric measurement capabilities at 50 and 250 μm. FIP utilizes *Origins'* fast mapping speed (up to 60 arcsec per second) to map one to thousands of square degrees. FIP's images will be useful for telescope alignment and public relations. FIP's rapid mapping makes photometric variability studies possible for the first time. To meet its performance requires improved detector sensitivity and larger pixel format size (Section 2.3).

MISC-T measures R~50 to 300 spectra in the to 20 μm band with three subsystems that operate simultaneously. MISC-T has no moving parts and is designed to provide exquisite stability and precision (<5 ppm between 2.8 to 10 μm, <20 ppm 11 to 20 μm). The optics design uses densified pupil optics that mitigate for observatory jitter. The improved stability relies on a planned improvement in detector stability including calibration (Section 2.3).

Figure 2-32 shows the location of each instrument under the backplane of the primary mirror assembly. This location is effective for maintaining the instruments (and mirror) at 4.5 K in the cryo-payload module (CPM, Section 2.3).

Table 2-11 summarizes the mass for all three instruments. To determine the mass required for each instrument, the team used best practices (GOLD rules) applied a 30% contingency over the Current Best Estimate (CBE) to arrive at the maximum possible value (MPV) and determine the reserves required. Contingencies at the design level were primarily calculated and based on the TRL of each assembly or component. Table 2-12 lists the power and data rate for all three instruments. The power usage is based on an expected operational model of the instrument for the average power and the peak power is the most stressing case. The science data rate assumes a operational model outlined in the mission operations Section 2.10.

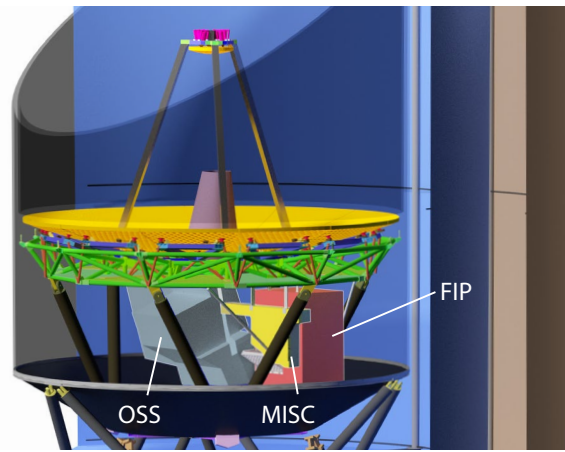

**Figure 2-32:** Each instrument under the backplane of the primary mirror assembly. This location is effective for maintaining the instruments (and mirror) at 4.5 K in the cryo-payload module (CPM, Section 2.3).

**Table 2-11:** Mass estimates for the *Origins* Instruments are derived from best engineering practices.

| Instrument | CBE Mass (kgs) | | | | MEV Mass (kgs) | |
|---|---|---|---|---|---|---|
| | Total Mass | Cold 4.5 K | 35K Region | Warm 263−313K | % Contingency Calculated | Total Mass MEV |
| *Origins* **Survey Spectrometer (OSS)** | 624 | 488 | 21 | 115 | 14% | 712 |
| **Mid-infrared Imager, Spectrometer, Camera (MISC) Transit Spectrometer (T)** | 101 | 76 | 0 | 25 | 15% | 116 |
| **Far-IR Polarimeter/ Imager (FIP)** | 411 | 301 | 15 | 95 | 14% | 469 |
| **Total** | **1136** | **865** | **36** | **235** | **14%** | **1297** |

**Table 2-12:** Power and Science Data Rate estimates for the *Origins* Instruments are derived from best engineering practices.

| | CBE Power (W) (Warm Region Location, Power at 4.5K is negligible) | | | Science Data Rate (Mbps) [2] | Data Interface |
|---|---|---|---|---|---|
| **Instrument** | **Average** | **Peak** | **Standby/ Safehold** | **Average** | |
| ***Origins* Survey Spectrometer (OSS)** | 558 | 945 | 188 | 1.5 Tbps Readout, 146.6 Mbps Science Data | RS422, Spacewire, 1533 |
| **Mid-infrared Imager, Spectrometer, Corornagrapgh (MISC) Mid-Infrared Transit (T) Spectrometer Channel** | 20 | 20 | 20 | 1.86 Avg, 4.10 Max [1] | RS422, Spacewire, 1533 |
| **Far-IR Polarimeter/ Imager (FIP)** | 267 | 350 | 175 | 384 Gbps Readout, 38.4 Mbps Science Data | RS422, Spacewire, 1533 |
| **Total** | **846** | **1316** | **384** | **189** | |

## 2.8 Spacecraft Summary

### 2.8.1 Technical Budgets and Margins

The *Origins* design is a point design with comfortable margins to meet all science requirements. During Phase A, the design will be optimized and the requirements will be revisited and flowed down from top level to lower levels. The *Origins* design contains conservative margins. All observatory components – instrument and spacecraft – have 30% power margin. The Electrical Power subsystem is designed to support the CBE load plus 30% margin at EOL. Instrument and telescope component masses have 38.3% margin over CBE. The four cryocoolers are designed to operate simultaneously, providing 100% performance margin. The input power contingency is subsumed into this factor of 2 margin. Communication links have 3 dB+ margin (see Appendix C for optical communication link analysis). Optical ground receive terminals provide over 50% cloud-free availability on average (*i.e.,* ~100% margin for the ~6 hour daily downlink needed). The 64 Tbit onboard solid-state recorder has enough capacity to store more than 3 days maximum CBE data volume plus 30% margin. The propellant is sized for 10 years, twice the nominal *Origins* lifetime, using the mass MPV. The ACS design meets pointing and slew requirements with conservative margins. Onboard spacecraft processors have over 60% margin.

The current *Origins* baseline mission concept is a point design which serves as a proof of principle; the design has not been optimized. Therefore, there are many potential margins hidden within the unoptimized design. For example, the current propellant tank height of 1.34 m drives the spacecraft bus height and bus structure mass. Because 0.94-m tall tanks could instead hold all of the required propellant, the team will study incorporating this change into the design in Pre- Phase A, which could ultimately provide significant mass savings. Table 2-13 shows the top-level observatory engineering resources.

Table 2-13 is based on MEL summary in Volume 3 of this report. The contingency values were calculated based on TRL levels. Small contingency values in Table 2-13 reflect the low risk levels associated with high-TRL components. In addition to the MEV values, we add contingency in mission

**Table 2-13:** The *Origins* design has 25% mass reserve on top of 11% contingency, and 30% margin for power and data rate

| | Mass | | | | | Power | | | Data Rate | | |
|---|---|---|---|---|---|---|---|---|---|---|---|
| **Item** | **CBE (Kg)** | **Contingency (%)** | **MEV* (Kg)** | **Reserve# (%)** | **Total# (5)** | **CBE (W)** | **Margin (%)** | **Total (W)** | **CBE (Mbps)** | **Margin (%)** | **Total (Mbps)** |
| PAYLOAD | 6,643 | 11% | 7,404 | | | 3,115 | 30%** | 3,510 | 187 | 30% | 243 |
| SPACECRAFT BUS | 2,044 | 8% | 2,209 | | | 977 | 30% | 1,270 | | | |
| OBSERVATORY (Dry) | 8,687 | 11% | 9,613 | 25% | 12,016 | 4,092 | 30% | 4,780 | 187 | 30% | 243 |
| Propellant | 939 | 0% | 939 | | 939 | | | | | | |
| OBSERVATORY (Wet) | 9,625 | | 10,552 | | 12,955 | 4,092 | 30% | 4780 | 187 | 30% | 243 |

**Note:** *MEV – Maximum Expected Value, i.e MEV = CBE + Contingency | # Reserve – On top of mass contingency
**Total** Observatory Dry mass = (CBE dry mass + Contingency)x125% = **138.3%** of Observatory **CBE** dry mass
**30% margin excludes the cryocoolers which has 100% design margin
(The total represents the maximum possible value. summing contingency and reserve.)

implementation and the total contingency meets the requirements for a flight project as per NPR 7120.5E requirements for Class A flight projects. Program Managers assign a Maximum Permissible (or Possible) Value based on their managing the program and launch constraints. These values are the values that are typically in the 25-30% above the CBE not above the MEV values.

The *Origins* science operation profiles impose some challenging requirements on the Attitude Control System (ACS). During MISC inertial pointing for transiting spectroscopy, the pointing stability requirement is 50 mas root-mean-squared (RMS) over 1-10 hours. This requirement drives the need for a field steering mirror control loop to actively attenuate line-of-sight (LOS) jitter, with a MISC-provided wavefront sensor to nullify LOS drift. During OSS inertial pointing operations, the absolute pointing accuracy requirement is 0.15 arcsec RMS. This requirement drives the need for a high-accuracy inertial measurement unit. During OSS and FIP survey operations, scanning rates of up to 60 arcsec/sec are required, with a required absolute pointing accuracy of 2 arcsec RMS. This drives the actuator complement.

### 2.8.2 Attitude Control System

### ACS Driving Requirements

The *Origins* science operation profiles impose some challenging requirements on the Attitude Control System (ACS). During MISC inertial pointing for transiting spectroscopy, the pointing stability requirement is 50 mas root-mean-squared (RMS) over 1-10 hours. This requirement drives the need for a field steering mirror control loop to actively attenuate line-of-sight (LOS) jitter, with a MISC-provided wavefront sensor to nullify LOS drift. During OSS inertial pointing operations, the absolute pointing accuracy requirement is 0.15 arcsec RMS. This requirement drives the need for a high-accuracy inertial measurement unit. During OSS and FIP survey operations, scanning rates of up to 60 arcsec/sec are required, with a required absolute pointing accuracy of 2 arcsec RMS. This drives the actuator complement.

### ACS Architecture

The observatory's size and its range of operational modes require an ACS with fine attitude sensing, high torque authority, and high momentum storage capacity. Commercial off-the-shelf (COTS) star trackers and a high-accuracy inertial measurement unit (IMU) are used to provide onboard attitude determination. For primary actuation, a set of six reaction wheels provide the torque authority needed for survey operations. The accumulated angular momentum is periodically unloaded using thrusters. The reaction wheels are mounted in a pyramid configuration to provide redundancy and graceful degradation. During Phase A, the reaction wheels and redundancy will be revisited to pick the best reaction wheels available in the optimal redundancy configuration.

In conjuction with the reaction wheels, the field steering mirror (FSM) is used to provide fine control of the optical line of sight and to provide some rejection of internal disturbances. The 20- Hz bandwidth FSM control loop follows tip/tilt commands from the ACS control loop assisted by accelerometer and tip/tilt encoder measurements, and feeds back its tip/tilt to the ACS so the ACS can keep it centered. The FSM controller uses measurements from a dedicated wavefront sensor to nullify drift over the hours-long exposures needed for spectroscopy. The *Origins* multi-stage control loop architecture is shown in Figure 2-33.

In addition to active vibration rejection by the FSM control loop, vibrations from the reaction wheels and cryocoolers are further attenuated by mounting them on passive isolators. These isolation mounts are tailored to reject vibrations within a limited frequency range. After isolation, the cryocooler force is 0.1 N per cryocooler. The cryocoolers are synchronized with each other, and have limited and tunable operational speeds, which is helpful in avoiding structural modes on orbit.

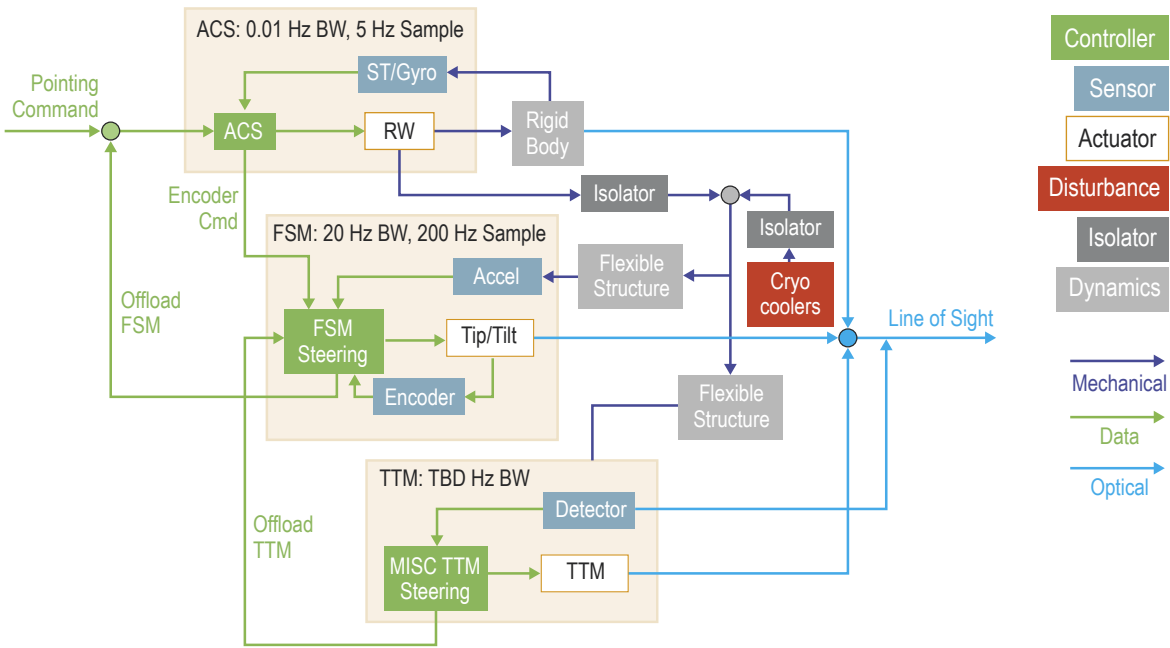

**Figure 2-33:** The *Origins* ACS multi-stage control loop architecture provides fine pointing accuracy and stability by actively suppressing the response of the telescope and instruments to internal disturbances.

## ACS Technology Readiness

The *Origins* ACS concept makes use of a combination of TRL 9 COTS and optical-mechanical components integrated with the instruments. COTS star trackers, IMUs, and reaction wheels with extensive flight heritage in the deep-space environment are available. The field steering mirror controller is part of the telescope. The design assumes the use of the JWST field steering mirror actuators, which can support the desired 20-Hz FSM loop bandwidth. The MISC wavefront sensor is integral to the MISC instrument.

## ACS Pointing Performance

Two cases drive *Origins* pointing performance: inertial pointing and survey operations. The team generated preliminary pointing performance estimates for each case. During inertial pointing, *Origins* pointing performance is driven by sensor noise and internal disturbances propagated through the attitude control loop and flexible structure. Disturbance data from a representative reaction wheel shows tonal components at the rotor spin rate and several harmonics. The cryocoolers are also modeled as tonal disturbances at 50 Hz. A finite-element structural model was not developed for this concept study due to the time and cost. To provide a preliminary estimate of jitter and drift pointing performance, a representative transfer-function structural model was constructed. Twenty flexible modes were placed at frequencies ranging from 0.1 Hz to 300 Hz. To give a conservative upper bound on the excitation of structural flexibility by reaction wheel and cryocooler disturbances, one flexible mode was placed precisely at the cryocooler frequency of 50 Hz, and the wheels were assumed to impart a tonal disturbance precisely at each flexible mode frequency. This model is simpler than modeling multiple wheel harmonics and sweeping wheel speeds through their operational range, but it captures all the primary resonances simultaneously, yielding a reasonable worst-case performance estimate. The team then modeled the ACS control loop and FSM loop. The LOS frequency response was found and decomposed into drift and jitter using the windowing frequency derived from the MISC frame time (10 sec). Figure 2-34 shows the resulting cumulative RMS drift and jitter as a function of frequency. Drift performance (1.1 mas RMS) is dominated by the wavefront sensor noise. Jitter performance (8.5 mas RMS1) is dominated by structural response to tonal disturbances. The cryocooler-induced jitter contribution of 4.5 mas is clearly visible at 50 Hz. Other step increases are due to assumed structural resonances excited by reaction wheel imbalance torques. Structural modes below 10 Hz are effectively suppressed by the FSM loop.

Details of the Kalman filter performance are shown in Section C.5 of Appendix C.

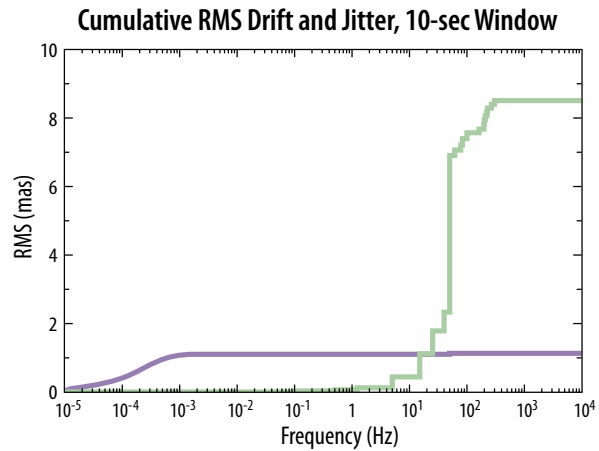

**Figure 2-34:** (left) Drift and jitter performance during inertial pointing complies with requirements under reasonable worst-case assumptions. Wavefront sensor noise drives drift performance, while cryocooler disturbances drive jitter performance.

### ACS Momentum Management

The *Origins* observatory presents a large surface area to the Sun, and the center of pressure is offset from the observatory mass center by ~1.5 m. Preliminary estimates of angular momentum accumulation due to solar radiation pressure torques are ~190 Nms/day. A complement of 250- Nms momentum wheels are baselined in part to meet slew torque requirements, but also to store this angular momentum. This momentum will be unloaded with thrusters at 1-3 day intervals.

Due to wheel torque limits, ~16 min are required to unload one day's worth of momentum. Assuming that momentum unloading is performed using thrusters with a 6-m moment arm between them, the team estimates *Origins* will require 6 kg/year of hydrazine fuel for momentum unloading.

### Safe Mode Attitude Control

The *Origins* observatory will respond to on-orbit anomalies by suspending science operations and transitioning to a safe mode. ACS supports safe mode by acquiring and maintaining a power-positive, thermally safe attitude, with a pitch angle of -20 deg and the solar array gimbal indexed accordingly to point to the Sun. This attitude is in the center of the nominal field of regard, keeping the telescope pointed safely away from the Sun and providing a predictable and benign environment for the thermal and power subsystems. For robustness, ACS uses coarse sun sensors and the IMU for attitude determination in safe mode. Control actuation is performed using the reaction wheels. The yaw angle about the Sun line is modulated to passively unload angular momentum, enabling extended periods without thruster operations.

### 2.8.3 Propulsion System

The *Origins* Propulsion System (PS) is a robust, high heritage design. The regulated bipropellant system provides the Δv and three-axis control required to complete all mission maneuvers including: mid-course correction, orbit insertion and maintenance, and demise. The PS consists of sixteen engines fed by two independent pressure-regulated fluid legs (one for monomethyl hydrazine fuel and one for MON-3 oxidizer) (Figure 2-35). Each leg has its own high-pressure gaseous helium feed section and a low-pressure liquid propellant section. The liquid propellant sections each have a propellant tank with a Propellant Management Device (PMD) and a Propellant Isolation Assembly (PIA). The high-pressure helium feed system consisting of a COPV pressurant tank and regulator-based pressure control assembly (PCA), maintains propellant section pressure. Pressure is monitored in each tank via pressure transducers. The PS has ten fill and drain manual valves (five for each leg) for fluid loading, unloading, and component testing. All PS components are commercial off-the-shelf (COTS) and are TRL 9. The

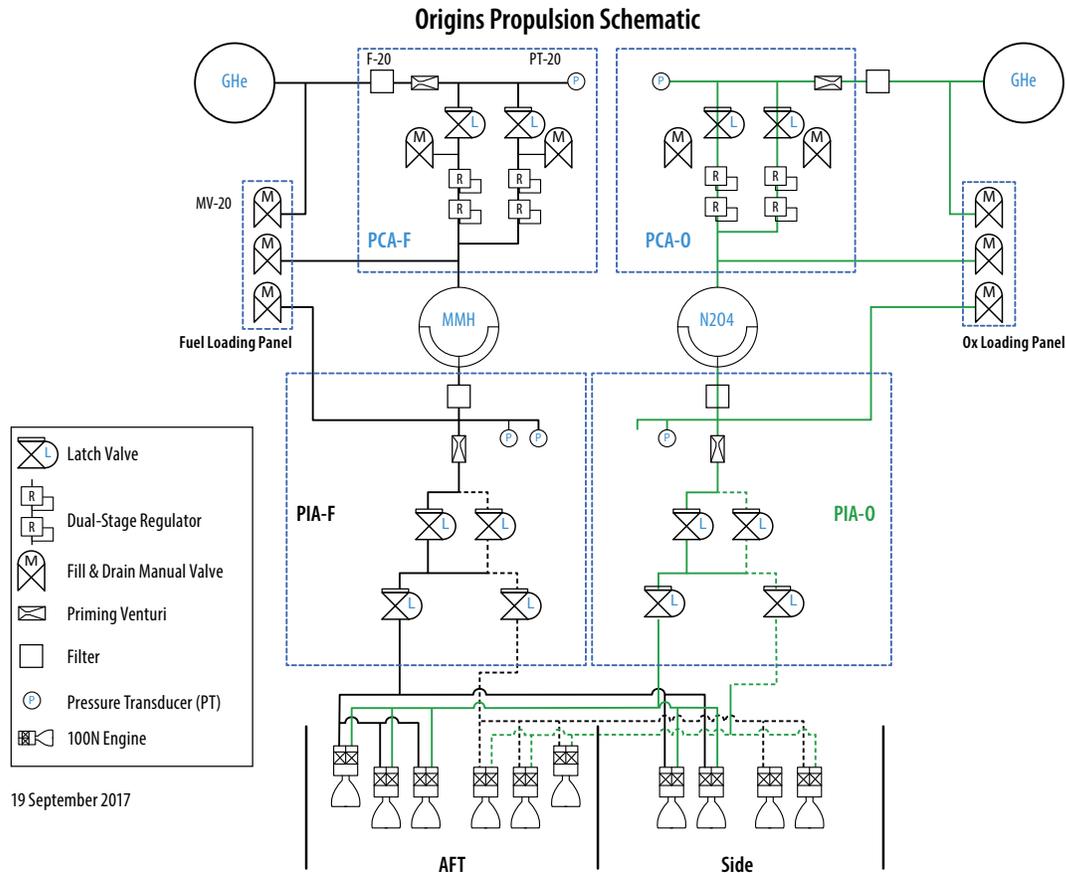

**Origins Propulsion Schematic**

19 September 2017

**Figure 2-35:** The *Origins* regulated bipropellant propulsion system is based on heritage designs from Solar Dynamics Observatory and Europa Clipper.

*Origins* engines are divided into primary and redundant banks that can be independently isolated in the event of a fault. Each PCA has independently-controlled, parallel flow paths, each consisting of a high-pressure latch valve for isolation and fault protection and a dual-staged regulator for pressurant flow control. Each PIA consists of a set of parallel-plumbed low-pressure latch valves and engine-bank isolation low-pressure latch valves used to isolate the primary and redundant banks for fault protection. These latch valves, in series with the single-seat engine, yields the range-safety-required three inhibits from the propellant tank. Venturi surge suppressors are in line between the propellant tank in each leg and the engines to mitigate pressure surges from propellant priming or water hammer events mid-operation. All components are fully cross-strapped electrically. half providing thrust in the -X direction. The engines are canted 30 degrees about the X-axis to provide three-axis control.

The 16 engines are located on the bottom of the spacecraft bus, as shown Figure 2-36 (redundant thrusters are not shown). They are canted 45 degrees from the YZ plane, half providing thrust in the +X direction and the other directions. Propellant is allocated for four mid-course correction (MCC) maneuvers, orbit insertion, 10 years of orbit maintenance (including momentum unloading), and demise. The system is sized to accommodate 930 kg of propellant: 351 kg of fuel and 579 kg of oxidizer. There is also 8.5 kg of pressurant. The propellant is carried in two identical tanks – one for fuel and one for oxidizer – each with a volume of 1.1 m3. The tanks are titanium with a composite overwrapped cylindrical section and contain a titanium Propellant Management Device (PMD). The operating pressure of the fuel and oxidizer feed systems is 2 MPa and the operating pressure of the gaseous helium feed system is 30.6 MPa. The current design engine is the 22-N Aerojet-Rocketdyne R- 6D.

Assuming the nominal oxidizer/fuel mixture ratio for the engine of 1.65 and a 2 MPa psia operating pressure, the specific impulse of the engine is assumed to be a conservative 279 sec. Plume impingement and associated potential contamination effects have been identified as a study to be performed during Phase A.

### 2.8.4 Electrical Power System

The *Origins* Electrical Power System (EPS) is a heritage efficient direct energy transfer (DET) design consisting of solar arrays, electronics, and a battery to collect, store, and distribute electric power (Table 2-14).

*Origins* is in a L2 orbit with no eclipse cycles. The battery will only be used for initial launch needs and cruise phase to the L2. The solar array was sized using a solar constant of 1353 W/m². Solar array degradation over 10 years due to radiation at L2 was determined using the SPENVIS and EQFLUX model runs. 30% contingency is added to all loads except for zero contingency for the cryocoolers which already have 100% design margin (Table 2-13). The solar array power is regulated by the Power Supply Electronics (PSE) using digital (on/off shunts) and linear shunts. Power is distributed via the PSE to the spacecraft loads via both switched and unswitched services with overcurrent protection for switched services and fuses for unswitched services. Battery is connected directly to the power bus. The battery dominated bus nominal voltage is 28 vDC. Due to the high electrical power level the PSE is a standalone box rather than combined with the Command and Data Handling (C&DH) Avionics equipment. The Solar Array orientation needs a flapping type drive to keep it in the sun while the observatory satisfies its pointing requirements. A negligible 5 degree cosine hit is taken on the array to reduce 2-axis array articulation to 1-axis.

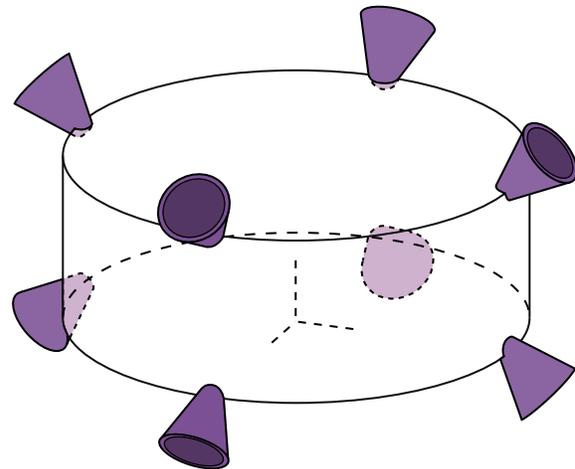

**Figure 2-36:** Current best estimate (CBE) engine locations on the *Origins* spacecraft bus provide delta-V and three- axis control.

**Table 2-14:** EPS Components Dimensions & Mass

| Item | Dimensions (m) | | | # | Area (m²) or Vol (m³) | Total Mass (kg) |
|------|------|------|------|---|------|------|
| Solar Array TJGaAs | | | | 1 | 21.99 | 134.76 |
| Li-Ion Battery 20 Ah 38% DOD | 0.24 | 0.05 | 0.22 | 1 | 0.00 | 6.16 |
| PSE | 0.22 | 0.28 | 1.20 | 1 | 0.07 | 69.4 |
| Solar Array Drive (1 Axis) | | | | 1 | | 24.28 |
| Harness Spacecraft & Solar Array | | | | 1 | | 128.76 |
| | | | | | **Total** | **363.4** |

All the EPS components are at a minimum TRL 7. The topology and hardware elements have heritage to recent spacecraft.

### Solar Array

The solar array uses triple junction Gallium arsenide cells which have an average efficiency of 29.5% operating at 70 °C. Cell power was based on 6x4 cm cells with 14 cells in series. See Figure 2-37. The array has five extra strings for fault tolerance. The array configuration uses the Ultra flex design. It is 6 meters in diameter when deployed.

### Battery

The Lithium Ion battery configuration has eight Saft 20 Ah cells in series. Each cell can be by- passed in case of a failure with corresponding adjustment in the solar array regulator for a lower voltage battery.

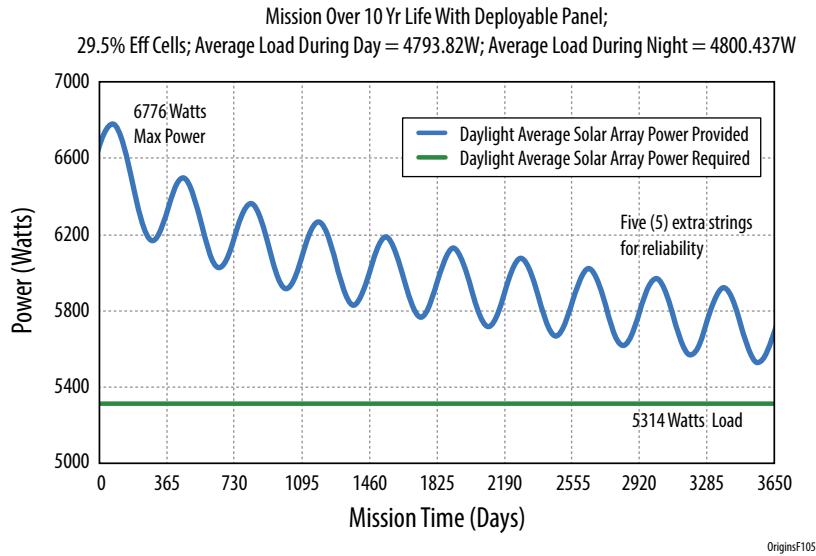

**Figure 2-37:** Solar Array Power and Load vs. Mission Time.

## Cable harness

The cable harness is a nominal 28vDC harness rated for peak power, with a mass estimate of 128.8 kg and power loss estimate of 50.9 watts.

## Summary

The TRL 7 heritage EPS design will fully accommodate the required spacecraft electrical power needs for the full 10-year extended duration of the *Origins* mission with large margin.

### 2.8.5 Communications System

*Origins* transmits ~ 21 Tbits/day science data to the ground via optical communication at 1 Gbps using a two-axis gimballed flight terminal. Commands and telemetry support is provided by heritage S-band transponders with 8w amplifier, omni antennas and a 0.9 m gimballed high gain antenna (Figure 2-38).

Optical communication uses a compact, low-mass and low-power flight terminal to provide orders of magnitude higher downlink rate than RF, is ideally suited for high data volume space missions such as *Origins*. Optical communication is currently the state-of-the-art but undergoing rapid development and deployment that it will be mature and very well established before the start of *Origins* in 2025.

Optical/Ka based European Data Relay System (EDRS) deployment started in 2016 and will be complete in 2020/2021. NASA successfully demonstrated lunar lasercom in 2013-2014. Laser Communication Relay Demonstration (LCRD) is on track for a 2019 launch, and the Deep Space Optical Comm (DSOC) demonstration is slated to fly on NASA Psyche mission in 2022. A US commercial optical satellite constellation, LeoSat, is planned to complete deployment of 78-108 satellites by 2021, ushering in worldwide service.

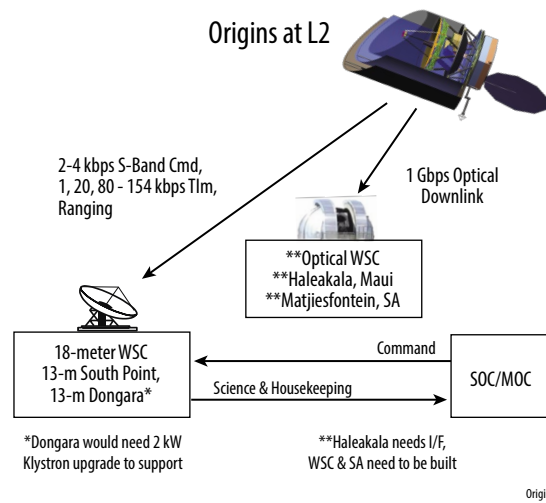

**Figure 2-38:** *Origins* Communication System Consists of Optical Downlink of Science Data and S-Band for Command & Telemetry

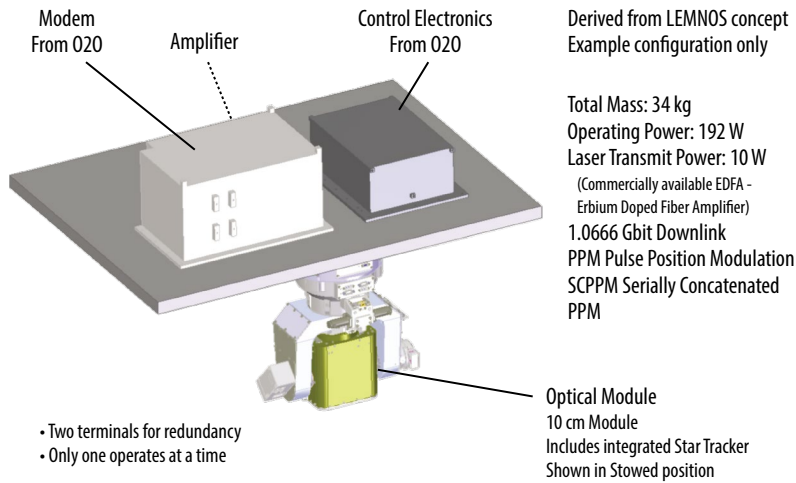

Modem
From O2O

Amplifier

Control Electronics
From O2O

Derived from LEMNOS concept
Example configuration only

Total Mass: 34 kg
Operating Power: 192 W
Laser Transmit Power: 10 W
(Commercially available EDFA -
Erbium Doped Fiber Amplifier)
1.0666 Gbit Downlink
PPM Pulse Position Modulation
SCPPM Serially Concatenated
PPM

Optical Module
10 cm Module
Includes integrated Star Tracker
Shown in Stowed position

• Two terminals for redundancy
• Only one operates at a time

**Figure 2-39:** The *Origins* optical communication system is fully redundant.

The *Origins* optical flight terminal is based on the latest 10 cm design from NASA's Laser- Enhanced Mission Communications Navigation and Operational Services (LEMNOS) program (Figure 2-39). Flight terminal pointing is guided by ground beacon and the system's internal star tracker.

Three geographically-diverse lasercom ground terminals at White Sands, NM, Haleakala, Maui, Hawaii, and Matjiesfontein, South Africa provide cloud-free link availability with ~100% margin for ~6 hours downlink per day. These optical ground terminals and additional ones are expected to be well established and ready for operations by 2030, 11 years from now and 5 years prior to the 2035 *Origins* launch date.

*Origins* uses S-band for housekeeping telemetry, commands, and tracking (Figure 2-40) 1 Kbps downlink and 2 Kbps uplink is achieved using an omni antenna and the 18-m NASA Near Earth Network (NEN) ground station antenna at White Sands, NM with 2 kw klystron. With the 0.9m High Gain Antenna (HGA) downlink increases to 154 kbps and uplink to 256 Kbps. The three NASA Deep Space Network (DSN) stations' 34-m antennas provide S-band emergency backup.

### 2.8.6 Flight Software

The *Origins* Flight Software (OFSW) is a low-risk implementation based on high heritage code and avionics coupled with a robust flight-computing platform. The OFSW has four major components: a small runtime core flight executive (cFE), an expandable catalog/library of reusable core flight software (cFS) components, a process for configuration management and controlled reuse, and an inte-grated development environment (IDE). The cFS allows components to be selected, configured, and deployed into run time systems with all the process artifacts, significantly reducing the development cycle.

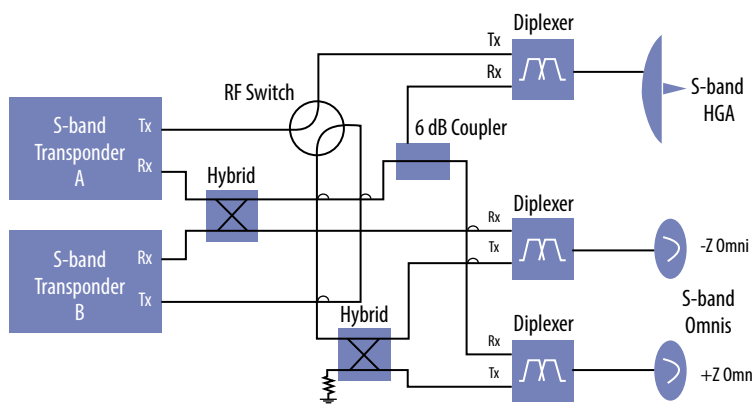

**Figure 2-40:** *Origins* S-Band is a Flight proven Design using Heritage Components.

The cFS was built from multiple successful NASA missions such as RXTE, TRMM, SAMPEX, WMAP and SDO. The TRL 9 cFS is fully tested and considered off-the-shelf products. Only 74% of the overall cFS will need to be parameterized for *Origins*.

**Table 2-15:** OFSW Source Line of Count Reuse Estimates

| CSCI | Estimated Lines of Code | | | | |
|---|---|---|---|---|---|
| | Total | New/ Modified | Autogen | Total Reuse | % Reuse |
| Command & Data Handling Processor (C&DH) | 124239 | 55435 | Partial | 75077 | 60% |
| Attitude Control Electronics Processor (ACE) | 75822 | 7600 | No | 64189 | 85% |
| Solid State Recorder Processor (SSR) | 75866 | 20865 | No | 51001 | 67% |

Only those components unique to the *Origins* system need to be developed as reengineered components. The team conducted a heritage analysis, which determined the reuse source lines of code ranges from 60% for the C&DH processor to 85% for the ACE processor (See Appendix C, Table C.8-2).

The *Origins* OFSW (Figure 2-41) includes all C&DH, GNC, power, thermal, and instrument support software that runs on the

MUSTANG computing platform with the VxWorks real-time operating system. This includes the ACE and SSR software, which runs on two independent MUSTANG computing platforms with the Real-Time Executive for Multiprocessor System operating system. Virtually all of the core C&DH FSW is re-hosted from existing cFS and are common on the C&DH, ACE, and SSR processors (Table 2-15). Further details on the *Origins* Flight Software are in Section C.9 of Appendix C.

## 2.9 Integration and Test

This section describes the element and observatory level Integration & Test (I&T) program for the *Origins* Baseline Concept. The test flow implemented at each level of assembly is discussed as well as the separation of thermal vacuum testing between the hot and cold zones of the observatory. The *Origins* element level I&T consists of the CPM and the SBM. The observatory consists of CPM + SBM. Key facilities, test beds, pathfinders, simulators and ground support equipment that could be used to implement the I&T program and reduce risk are also summarized. All the instruments are delivered to I&T fully qualified and calibrated. The summary level I&T flow is shown in Figure 2-42. Each of these phases is further detailed in the subsequent sections below.

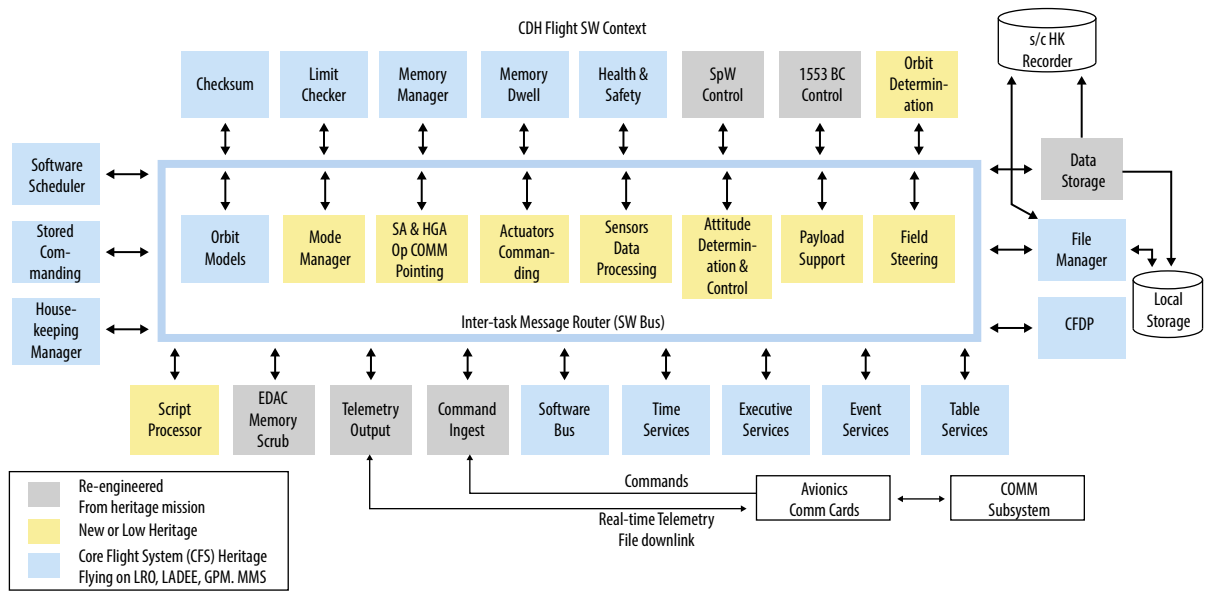

**Figure 2-41:** OFSW Architecture

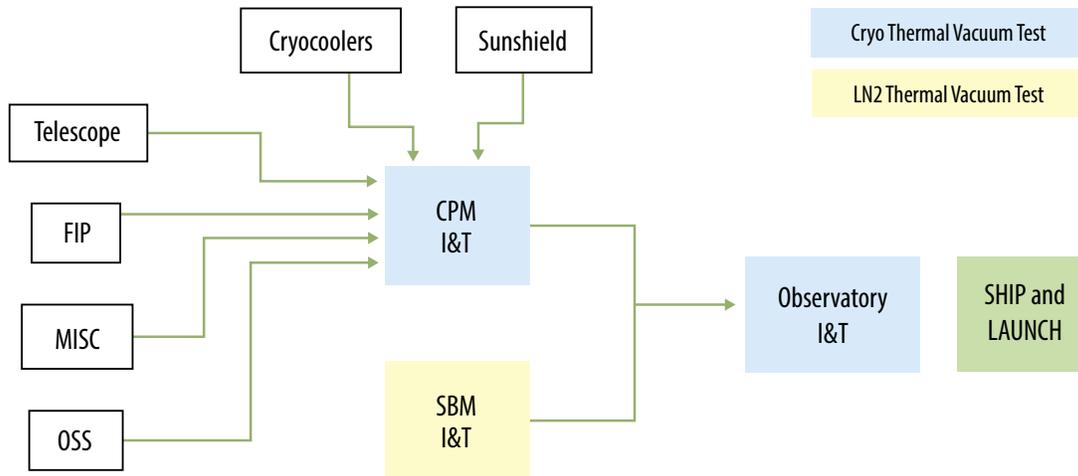

**Figure 2-42:** *Origins* I&T Summary Flow for Cryogenic Payload Module (CPM) and Spacecraft Bus Module (SBM)

### 2.9.1 Telescope I&T

Telescope I&T will start with assembly of the structural subassemblies. One possible venue for mirror assembly is the JWST mirror assembly facility at NASA shown in Figure 2-43.

A rollover fixture is used to orient the structure as required for assembly access. Then the installation of the optics starts. Aft Optics, all primary mirror assemblies and secondary mirror are installed and aligned to the structure. Then instruments mass simulators, cryocoolers, inner optical baffle, telescope harness, outer barrel, sunshield structure and deployment mechanism, electrical boxes and protective cover, are integrated. Details of the assembly process are given in Section C.9 in Appendix C. The completed telescope assembly is then ready for the mechanical test program: Modal Survey, Signature Sine Sweep Frequency Characterization, Protoflight 3- axis Sine Vibration, followed by one more Signature Sine Sweep Frequency Characterization and Protoflight Acoustics. Alignment is performed before and after these tests. Scope of this test program is to verify dynamic stability of mirrors mechanisms. Deployments and barrel will go through acoustics test for the first time. The telescope is then ready for integration with the CPM. The telescope I&T summary flow is shown in Section C.9 in Appendix C.

### 2.9.2 CPM I&T

Instruments, flight cryocoolers, sunshield, telescope and all the associated electronics will be integrate and tested. The telescope has its cover already installed. *Origins* will take advantage of JWST facilities and lessons learned. Chamber A at Johnson Space Center (JSC) will be used for cryogenic vacuum testing. Chamber A was substantially modified for use on JWST including the addition of helium shrouds and an ISO Class 7 environment for preparations inside and outside the chamber. *Origins* schedule will save 50 days out of the total test time in this chamber, during cool down and warm up, using cryocoolers and highly conductive materials. Several large cleanrooms facilities for CPM integration exist throughout the country. The CPM I&T summary flow and test description are shown in Section C.10, Appendix C. The hardware is then prepared for transportation and transported to JSC by a shipping container whose size is compatible with the Super Guppy or Beluga airplanes. Note: It was called Space Telescope Transportation Air, Road and Sea (STTARS) for JWST. Upon arrival at JSC, the flight CPM is removed from the transporter and prepared for installation into the Chamber A (shown in Figure 2-44).

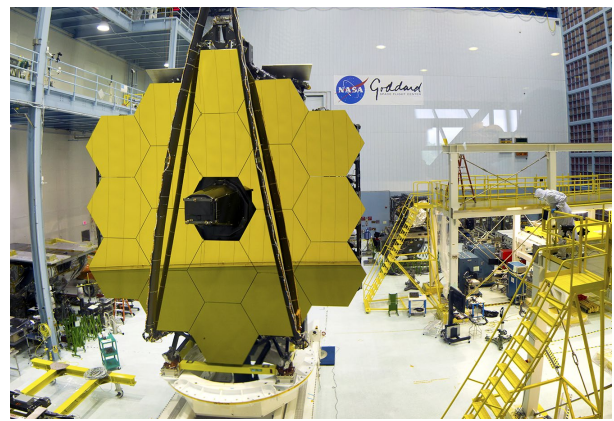

**Figure 2-43:** The JWST telescope was assembled in the Spacecraft Systems Development and Integration Facility at NASA.

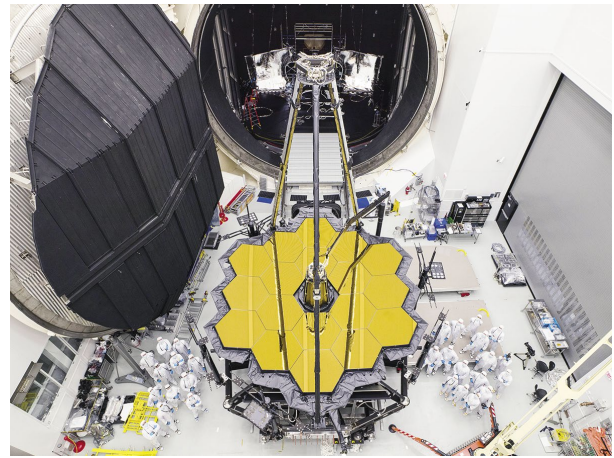

**Figure 2-44:** JSC chamber A supports *Origins* "Test as you fly." The JWST Optical Telescope Element and Integrated Science Instrument Module (OTIS) is shown rolling into the JSC Chamber A. Chamber A and its helium cooled shroud provides a perfect environment for thermal verification of *Origins* with and without the spacecraft.

Preparations include telescope protective cover removal, sensor installation, metrology and ambient functional testing. A 4.5 K GSE cover will be installed to lower the stray light from the 11 K shroud to the instruments. This cover will be cooled by a 4 K GSE cryocooler, similar in concept to the GSE cooler used to cool MIRI during the OTIS cryogenic tests. Once installed into the chamber A the cryogenic testing of the CPM will begin. In general, the objective of the cryogenic vacuum test is to verify the CPM level requirements in the conditions of the expected flight environment, with emphasis on optical measurements that can be performed in this test configuration. This is the first time that the instruments will be fully operational at their flight temperature. The optical tests, described in Section C.3 of Appendix C, will verify the CPM system optical workmanship, and provide optical test data to support the integrated telescope modeling used to predict flight optical performance. After the testing, the CPM is then removed from the chamber and prepared for transportation. The hardware is transported to the integration facility where it will be mated to the SBM to create the *Origins* observatory.

### 2.9.3 Spacecraft Bus Module (SBM) I&T

The SBM I&T could be performed at numerous facilities throughout the country, and consists primarily of the spacecraft bus (which includes the typical spacecraft bus subsystems of power, attitude control, communications, command and data handling), solar array, cryocoolers mass simulators and instrument warm electronics mass simulators. SBM primary structure is delivered to I&T already qualified. The objective of the SBM I&T is to deliver a fully verified and tested element that meets all the functional requirements prior to integration with the CPM at the start of the observatory I&T. The SBM I&T summary flow and test description are shown in Section C.9 in Appendix C.

### 2.9.4 Observatory I&T

The observatory I&T integrates the fully qualified CPM with the fully qualified SBM. Facilities already exists to perform this test campaign. Telescope and instruments warm flight electronics and flight cryocoolers are integrated into SC. Inner and outer sunshield supports and material are integrated into CPM. CPM and SC are integrated. Harness is integrated. Alignment is performed. Full electrical and software compatibility across the interface has been previously validated.

Also, testing at CPM and SBM levels utilized high fidelity simulators to validate those interfaces through ambient and protoflight level environmental testing. The observatory environmental campaign consists of aliveness tests, Electromagnetic Interference/Electromagnetic Compatibilty (EMI)/

(EMC), Protoflight 3- Axis Sine Vibration, Shock and Mass Properties, that can be done with the Variable Center of Gravity Mechanism, already used with JWST and shown in Figure 2-45.

Then the observatory will be shipped to JSC for Protoflight Acoustics. Once this test is completed, the observatory will enter into Chamber A, where a sunshield deployment will be performed in vacuum, followed by a Comprehensive Performance Test (CPT), end to end data test (hardline), special tests (loss of power, etc.). At this point the observatory will leave Chamber A and it will go through functional test, laser communications verification, alignment and leak test. Finally, the observatory is stowed in launch configuration and prepared for transportation to launch site. The observatory I&T flow *Origins* Launch Site Operations are shown in Section C.9 of Appendix C.

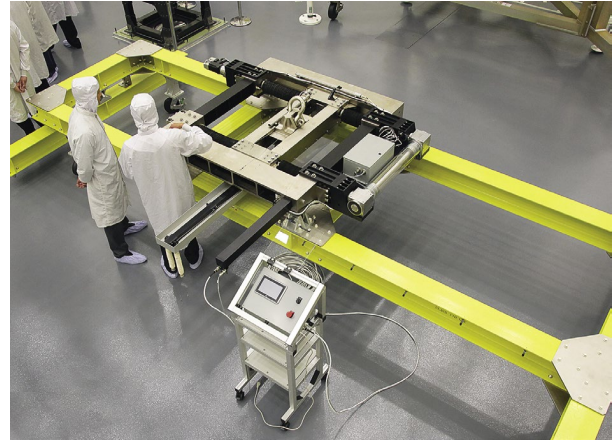

**Figure 2-45:** Variable Center of Gravity Mechanism

### 2.9.5 Reliability and Verification

*Origins* is a Class A mission with no credible single point failure. Risk is reduced through design simplification and by minimizing the number of deployments: the only deployments are the solar array, sunshield, and an aperture cover. Risk is also reduced because the *Origins* observatory can fit into NASA-JSC's Chamber A, a large Thermal Vacuum chamber recently refurbished and used for cryogenic testing on JWST (Figure 2-43). This enables cryogenic testing of the CPM and vacuum testing of the entire observatory, including sunshield deployment, in accordance with the "Test-As-You-Fly" objectives.

### 2.9.6 Verification Testing

#### Observatory or CPM level I&T: Instrument I&T

The vast majority of the OSS, MISC-T, and FIP instrument calibration and verification/validation activities will take place before instrument delivery for observatory integration (See section 3 for details). However, verification of their performance within the integrated observatory will be necessary to ensure that it still performs as expected in end-to-end observatory testing before launch. In particular, the interfaces with the telescope and spacecraft will need to be verified. These include: proper alignment and focus with the telescope (for OSS is verification of the pupil overlap); verify the expected telescope point spread function as measured by FIP; communication between the observatory gyroscope sensors and the telescope fine steering mirror; thermal testing of both cold and warm assemblies; vibration testing with the observatory; and command and data handling with the spacecraft.

#### Telescope Alignment

The FIP instrument can be used to perform phase retrieval holography measurements on astronomical point sources. This method requires to obtain several images of this source at different sub-reflector settings, *i.e.,* measurements of the PSF in several defocused configurations. This method has been demonstrated in ground based sub-millimeter telescopes. Depending on the image size and the dynamic range of the observations, surface accuracies can be measured down to scales significantly smaller than the wavelength at which the observations are taken. Further information is given in Section 2.10.3.

## 2.10 Mission Operations

### 2.10.1 Launch, and Early Operations

Figure 2-46 (also see Table 2-16) illustrates the timeline from launch until normal operations. Separation from the launch vehicle, initial attitude acquisition, and solar array deployment are all critical events. If no ground station is in view, *Origins* will use the Tracking and Data Relay Satellite System (TDRSS) to relay status information to the ground. Once *Origins* is in a nominal configuration, the critical spacecraft subsystems will be checked prior to the first mid-course correction maneuver, which occurs about 1 day after launch. The spacecraft checkout is completed after the first Mid-Course Correction (MCC). The sun shield is deployed after the second MCC, followed by the aperture cover. The cryocoolers will cool the telescope and instruments to 4.5 K within about 7 days after sufficient time for outgassing has elapsed. Warm instrument checkout occurs in parallel with the cool down. Once the payload is at its operational temperature, the telescope is aligned and the instruments are calibrated.

Normal operation starts ~110 days after launch, shortly before L2 orbit insertion maneuver. The *Origins* baseline mission is 5 years and the observatory has sufficient consumables to operate for 10 years. *Origins* is designed to be serviceable (See Section 2.11), so a lifetime beyond 10 years is possible.

**Table 2-16:** *Origins'* post-launch activities enable the observatory to initiate normal operations a few days before arriving at L2

| Time After Launch | Event |
|---|---|
| 30 min | Solar Array Deploy |
| 3 hours | Propulsion Checkout |
| 24 hours | Mid Course Correction (MCC) |
| Day 2 | HGA deploy |
| Day 10 | MCC-2 |
| Day 12 | Sun shield deployment |
| Day 15 | Deploy telescope aperture cover |
| 15–30 | Initial Instrument checkout |
| 16 | Cryocoolers on |
| 23 | Cooldown complete |
| 24–36 | Telescope alignment |
| 37–51 | Instrument calibration |
| 51–90 | Science Commissioning |
| 90+ | Normal Operations |
| 114 | L2 Orbit Insertion |

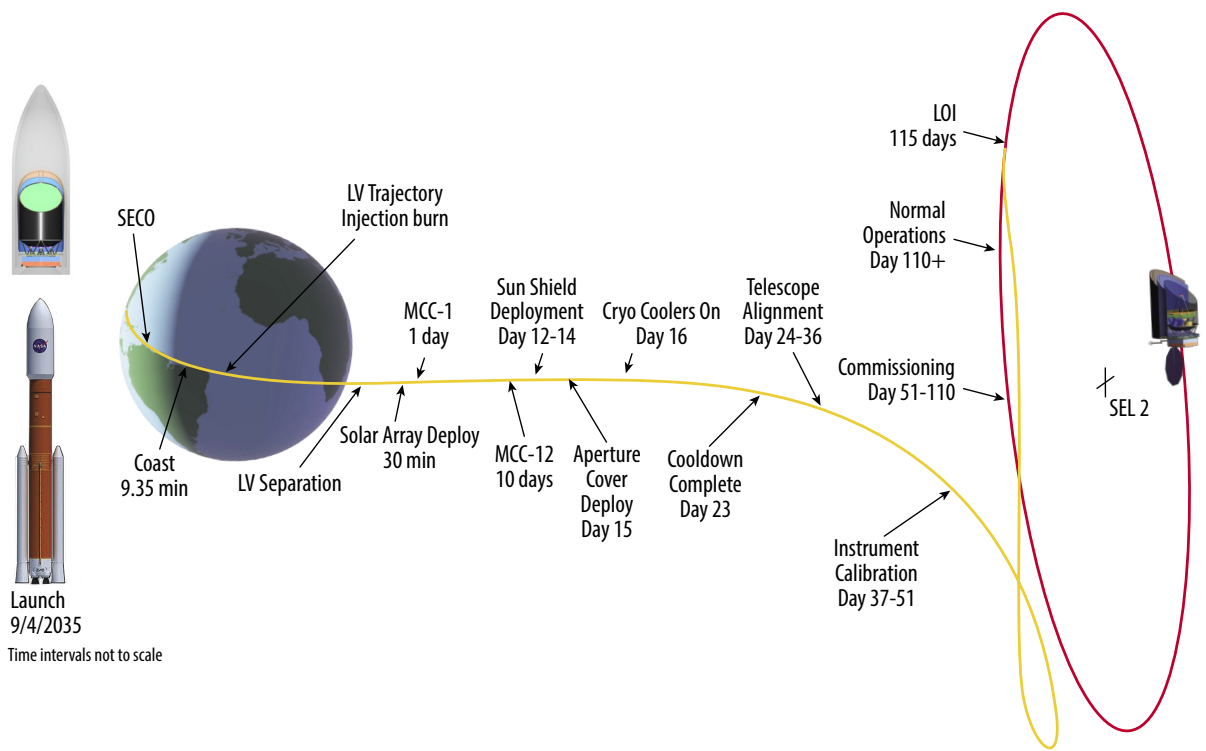

**Figure 2-46:** *Origins'*s post-launch activities enable the observatory to initiate normal operations a few days before arriving at L2.

### 2.10.2 Mission Orbit Design

*Origins* will be placed into an SEL2 orbit to achieve its three science goals. The Sun-Earth system has five Libration Points (Figure 2-47). These points rotate with the Earth as it orbits the Sun, and at these points, the gravitational forces and orbital motion of the spacecraft, Sun, Moon, and Earth interact to create a "stable" environment for the spacecraft.

However, the collinear Libration Points (L1, L2, and L3) are not truly stable. These points are very sensitive to energy, and any perturbation will nudge a spacecraft out of its Libration Point orbit if routine maintenance is not performed. *Origins*, which is planned to make observations at L2, must therefore perform routine station-keeping maneuvers to maintain this orbit. More details of *Origins* orbit and comparisons with other missions are found in Appendix C, Section C.10.

A requirement was also placed on the orbit design so that the observatory does not encounter shadows from the Earth or Moon. Finally, based on the shape of *Origins* and its FoR, an angular constraint was placed on the orbit as well. To ensure no stray light from the Moon reaches the inside of the primary baffle, the angle between the Earth to SEL2 line and the Moon to *Origins* vector must be no larger than 31°. This angle effectively requires *Origins*'s SEL2 orbit to fall within a SEL2-Earth-Vehicle (LEV) angle no larger than 16.7° off the Earth-to-SEL2 line. This, in turn would have required a much larger fuel budget for station keeping, increasing the size and mass of *Origins*. Therefore, the Moon angle was loosened to 44° resulting in an orbit with a LEV angle of 29°, which helps decrease the ΔV Budget. The *Origins* orbit must also maintain an LEV angle of 4° or larger to ensure it does not pass through any Earth or Moon shadows. Figure 2-48 depicts these angular requirements.

From a Flight Dynamics perspective, it is much easier to define the *Origins* mission orbit using LEV angular requirements versus size requirements. For example, JWST's mission orbit is confined to fit within a box that is no bigger than 832,000 km in the RLP Y direction and 500,000 km in the RLP Z direction. By constraining JWST to a box, launch window cases are ultimately thrown out if the mission orbit violates those sizes even by

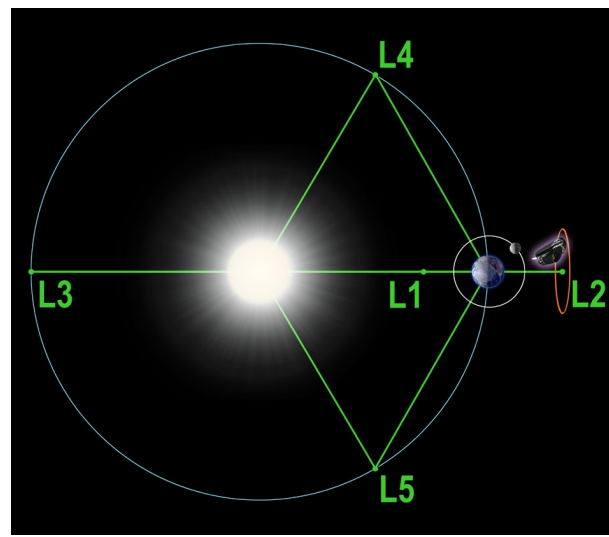

**Figure 2-47:** Our Solar System has five Sun-Earth Libration Points where a spacecraft can be maintained in a relatively-stable environment for performing observations.

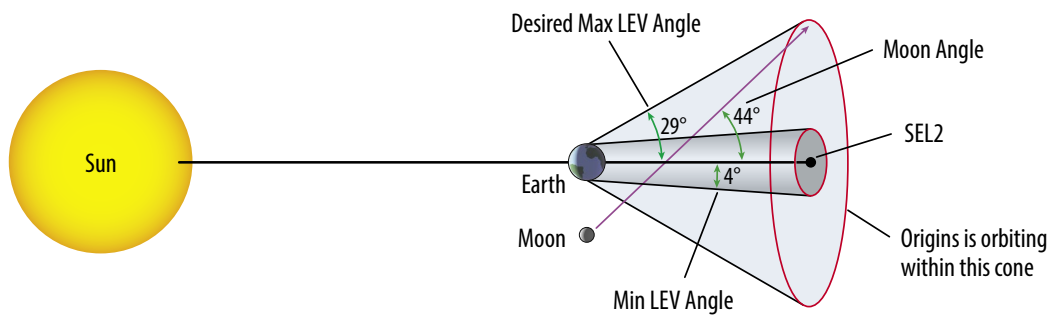

**Figure 2-48:** *Origins* orbits at SEL2 with the appropriate LEV Mission Orbit Angle and Moon Angle to ensure the orbit fulfills all mission requirements.

1 km. By defining *Origins*'s orbit to fit within an LEV angle, it opens the launch window to more opportunities to achieve a desired mission orbit that meets science requirements.

Based on these LEV and moon angle requirements, the resulting orbit, shown in Figure 2-49, was designed for *Origins*. The orbit is shown for 10 years. The observatory's orbital period is 6 months, resulting in two orbits per year. *Origins* inserts into its SEL2 orbit at a LEV of ~8.8° and opens up to a maximum LEV angle of 29.9°. The view in Figure 2-49 is in the RLP YZ frame (as viewed from Earth), and *Origins* orbits counter-clockwise with respect to Earth through its mission lifetime. Further details of the launch and orbit insertion are given in Appendix C, Section C.10.

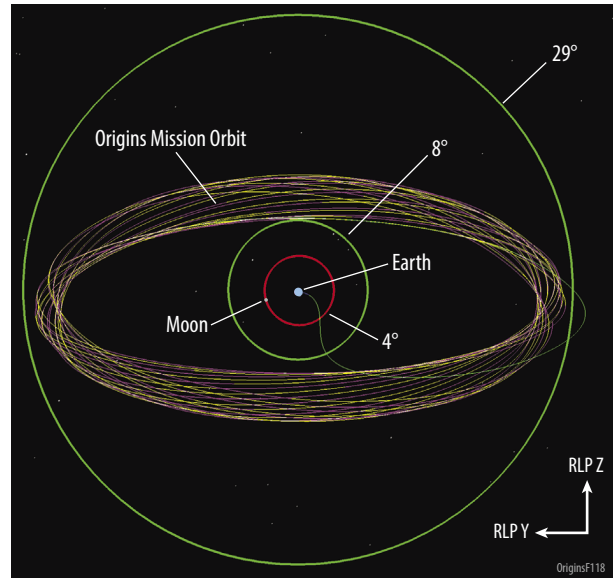

**Figure 2-49:** *Origins* orbits about SEL2 (shown here in the RLP YZ Frame) for 10 years within LEV angles of 8° and 29° (shown in green) meeting all mission requirements.

### 2.10.3 In-Orbit Checkout and Commissioning

Once the sunshields and aperture cover has been deployed, the CPM will begin to cool radiatively. The telescope segment launch locks and any other instrument launch locks will be released. The telescope and instruments will naturally lag behind the barrel during the radiative portion of the cool down. Once the danger of off-gassing of water has passed (~120 K) for the warmest parts of the CPM, the cryocoolers will be turned on. From this point it will take only a few days for the telescope and instruments to reach operating temperature. MISC-T will be able to be turned on. At this point the sub-Kelvin coolers can be started on FIP and OSS, and these two instruments will be then be fully operational.

As an imager, the FIP instrument can be used to perform focus-diverse phase retrieval (FDPR) measurements on astronomical point sources. Phase retrieval refers to a series of algorithmic processes through which one can determine the wavefront error of an optical system through a number of individual measurements. FDPR is one of these methods and requires the acquisition of several images of the input field point with different known values of defocus introduced. This method has been demonstrated in ground based sub-millimeter telescopes. Depending on the image size and the dynamic range of the observations, the accuracy of the wavefront can be measured down to scales significantly smaller than the wavelength at which the observations are taken. FDPR has been used to characterize the wavefronts of the Webb scientific instruments(Aronstein, 2016). Note that a detailed study of this method with its expected performance parameters was not obtained within this study.

### 2.10.4 Ground Systems, Mission and Science Operations

*Origins* mission ground system and mission operations have three drivers – safe and efficient operations, large data volume, and the L2 orbit. As a flagship observatory, all *Origins* operations focus not only on safe operations, but also on ensuring a scientifically-efficient mission, minimizing non-science time to the extent possible. This includes developing and testing procedures for potential anomalies to reduce recovery time and implementing a scheduling system that minimizes slewing time. *Origins* data volume is larger than any previous NASA science mission. The mission's daily science volume is up to 21 Tbits per day (including margin). This large data volume drive the space/ground link optical communications configuration. *Origins*' orbit at $L_2$ drives space/ground link sizing.

## Ground System

The *Origins* ground system (Figure 2-50) is similar to those used by previous NASA Astrophysics flagship missions, including *Hubble* Space Telescope (HST) and JWST. *Origins*' mission and science operations are low risk, take advantage of NASA's development of optical communications, and use established science and mission operations centers.

## Space/Ground Link

Due to *Origins*' large data volume, the team selected optical communications technology to provide science data downlink. The existing optical ground station at White Sands has a 0.6-meter telescope. This station supported the Lunar Laser Communications Demonstration in 2013 and will support the LCRD launching in 2019. *Origins* will use optical terminals with an array of four 0.6 meter telescopes to support its higher data rate and greater distance at SEL2. The *Origins* operations concept assumes using an additional terminal in South Africa. Two stations, one at White Sands and one in South Africa, will provide ~13 hours of visibility to *Origins* daily. At a 1 Gbps download rate, *Origins* can downlink 21 Tbits daily in ~ 6 hours. The ground station sites were selected for their dry climates and geographical/weather diversity. *Origins* onboard storage is sized to store more than three days of data, minimizing the risk of data loss when the weather is unusually cloudy.

*Origins* uses S-band radio frequency (RF) links for commands, housekeeping telemetry, and tracking to supplement the onboard navigation. The baseline design uses existing stations at White Sands and Hawaii. These stations provide daily commanding. *Origins* also uses the uplink to send acknowledgments to close the data delivery protocol to ensure complete and error free science data reception.

Optical communications is a proven technology, but *Origins* requires higher power and a longer lifetime than has currently been flown. NASA's Space Communications and Navigation organization is currently developing flight optical communications terminals for deep space missions and TDRS-like optical relay satellites in geostationary orbit that would meet *Origins* requirements. The Psyche mission is slated to carry the deep space optical communications terminal in 2022. The optical relay satellites are planned to become operational in 2025, well in advance of *Origins'* 2035 launch date."

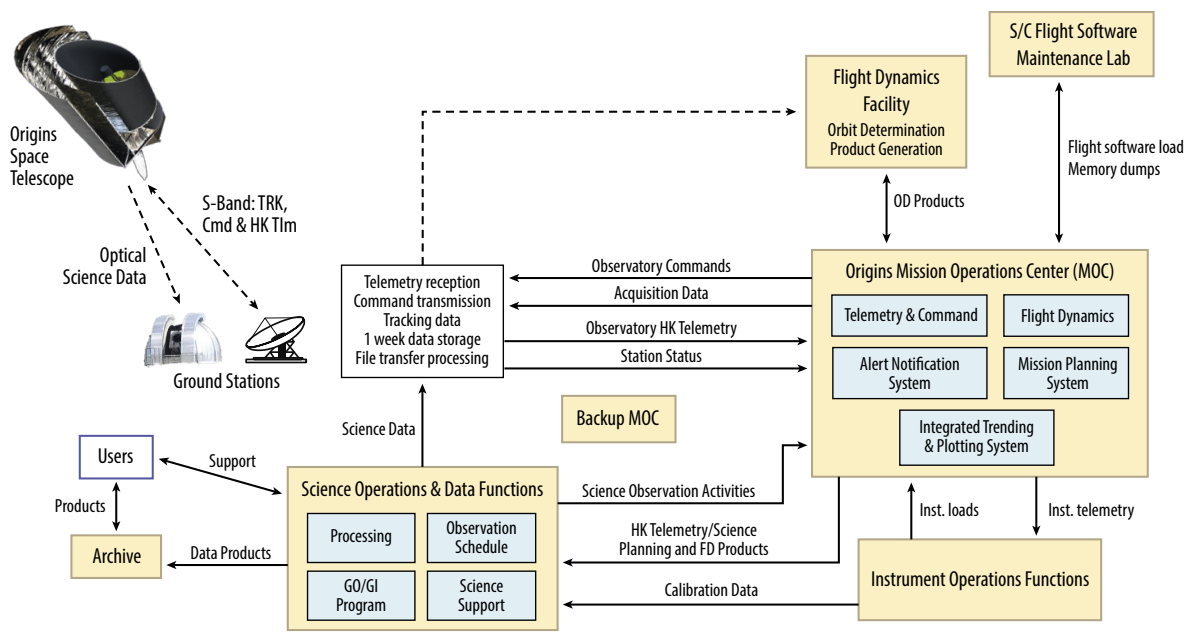

**Figure 2-50:** The *Origins* Ground System is similar to those employed for other large NASA missions and includes the use of optical communication technology advances to downlink the mission's large science data volume.

Ka band RF link is a back up option to optical communications. Data volume reduction via lossless compression is quite likely and will be studied during Phase A.

## Mission Operations Center (MOC)

The MOC is responsible for the safe operation of *Origins* and health of the spacecraft. Its functions include:

- Optical and RF Ground Station Scheduling. *Origins*' daily data volume can vary by a factor of 10 or more up to a maximum of 21 Tbits depending on which instruments are in use. The MOC requests support from the ground stations based on the expected data volume.
- Commanding. The MOC is the sole source of commands uploaded to *Origins*. The MOC integrates instrument and telescope commands from the Science Operations Center (SOC) with spacecraft commands and sends these commands to the ground stations for uplink.
- Monitoring. The MOC monitors housekeeping data and takes appropriate action in the event of a problem. The MOC maintains a comprehensive set of contingency procedures to minimize response time to anomalies.

As a standard practice the *Origins* mission includes a backup MOC that can perform basic functions in the event the primary MOC is unavailable. The primary and backup MOCs are developed and operated by an experienced organization, such as GSFC or experienced spacecraft providers, and are equipped with all necessary people, facilities, tools, and processes. MOC functions benefit from significant reuse of software developed for similar missions. The *Origins* mission operations also benefit from being operated at a location with other missions. This allows for sharing of special expertise, provides resiliency in the event of personnel turnover, and allows *Origins* to benefit from improves processes that are proven on other more risk tolerant missions.

## Science Operations Center

As a large mission observatory, the expectation is that all of *Origins* sciences will be proposal driven by the general observer (GO) astronomical community, with proposals adjudicated by a time-allocation committee. This is similar to existing space observatories, including *Hubble*, *Spitzer*, and Chandra.

With a 5-year lifetime requirement and 10-year goal, *Origins* has ample opportunities to accomplish tremendous science. The *Origins* science usage cases (Table 2-17) shows the mission's three key themes (**Sections 1.1-1.3**) that drive observatory technical requirements. The example science can be accomplished easily within two years of science operations, in addition to leaving significant time for additional programs. A detailed study of science operations, such as a design reference mission, is to be conducted during phase A.

Like the MOC, the *Origins* SOC will be developed and operated by an experienced organization, such as the Space Telescope Science Institute or The Infrared Processing and Analysis Center (IPAC), that will leverage proven software, processes, and staff from similar missions. The SOC supports NASA HQ GO and Guest Investigator (GI) proposal evaluations. The SOC supports GO and GI scientists by providing instrument user's guides, simulators, exposure time tools, and a help desk. Instrument scientists will develop observation templates, monitor instrument performance, and assist in the development of special observations.

The SOC generates the mission's long- and short-term observing plans. This schedule will balance the need for time-sensitive observations, such as those of transiting targets and targets near the ecliptic that are only accessible for ~14 weeks/year, with the need to efficiently manage slewing between targets. The SOC-generated weekly instrument command load is provided to the MOC. The SOC processes science data and produces standard products, adapting existing processing pipeline software augmented with specific capabilities required for *Origins*. It sends data to an archive, such as MAST and/or IRSA. At the beginning of the mission, the SOC also supports telescope alignment and instrument calibration.

**Table 2-17:** Science Use Cases for *Origins'* Main Science Themes

| Observation Program | Science Theme | Instrument Modes | Wavelength (μm) | Typical Map Size or Point Source | Number of Sources/Fields | Total Science Time (hours) |
|---|---|---|---|---|---|---|
| Proto-planetary Disk Spectral Survey | Trail of Water | OSS, FTS spectral scan | 25–600 | Point source | 1000 | 1126 |
| Proto-planetary Disk - Velocity Resolved | Trail of Water | OSS, etalon | 112, 179.5 | Point source | 90 | 122 |
| Debris Disk Spectral Survey | Trail of Water | OSS, FTS spectral scan | 50–180 | Compact source fits into a FOV | 187 | 187 |
| Comets | Trail of Water | OSS, FTS | 128, 179.5, 234.5 | Point source | 100 | 400 |
| OSS DEEP Tier | Galaxy, Black Hole Formation | OSS survey scan | 25–600 | 0.5 deg$^2$ | 1 | 1000 |
| OSS WIDE Tier: 50 sq. deg field | Galaxy, Black Hole Formation | OSS survey scan | 25–600 | 50 deg$^2$ | 1 | 500 |
| OSS Targeted Follow-up, low res | Galaxy, Black Hole Formation | OSS R~300 spectrum | 25–600 | Point source | 400 | 400 |
| OSS Targeted Follow-up, high res | Galaxy, Black Hole Formation | OSS, FTS spectral scan | 25–600 | Point source | 100 | 100 |
| FIP Deep Survey | Galaxy, Black Hole Formation | FIP imaging | 50, 250 | 1 deg$^2$ | 1 | 27 |
| FIP ultra-wide Survey | Galaxy, Black Hole Formation | FIP imaging | 250 | 10,000 deg$^2$ | 1 | 200 |
| Bioindicators : Transit Survey | Are We Alone? | MISC, transit | 3 to 20 | Point source | 12 | 192 |
| Bioindicators: Eclipse Survey | Are We Alone? | MISC, transit | 3 to 20 | Point source | 6 | 384 |
| Biosignatures: Transit Survey | Are We Alone? | MISC, transit | 3 to 20 | Point source | 4 | 1072 |

*Origins* generates almost 1 PByte of raw data per year. Ground data storage is not expected to be a significant challenge due to increases in data storage (including cloud storage), communications, and processing capabilities over time. The *Origins* data volume is about two times the data volume of WFIRST, which launches a decade earlier. Current astrophysics archives at IPAC and MAST hold over 1 and 2.5 PBytes, respectively, and the MAST archive is expected to more than double over the next five years.

Existing scheduling processes should be sufficient for *Origins*. The number of observations is expected to be less than the number for HST, and the SEL2 orbit has fewer constraints, reducing the complexity of scheduling algorithms.

The SOC leads *Origins* s community engagement efforts. This includes support for public affairs, hosting *Origins* conferences, and providing public outreach.

## Other Ground System Elements

Flight Dynamics is responsible for orbit determination and orbit maintenance maneuver planning. To support instrument commissioning and calibration, each instrument has a function that monitors instrument performance. This function maintains instrument Flight SoftWare (FSW), provides updates as required, and is the focal point for resolving instrument anomalies. Spacecraft FSW maintenance teams develop and test any FSW updates required after launch.

## Operations

The *Origins* sunshield is designed to allow a 65° pitch angle and ±5° roll angle, and the observatory can yaw the full 360°. Figure 2-51 shows the FoR. About 75 days per year, the Moon impinges slightly on this FoR. There are no eclipses in this orbit.

Instrument observations range from minutes to days for OSS and FIP mapping. Most observation durations are on the order of hours. OSS and FIP may operate simultaneously; MISC generally operates alone.

*Origins* has no stringent requirements on data latency, and the observatory can store 64 Tbits of science data, more than 3 day's data. Data are downlinked up to 6 hours per day. Data lost due to clouds can be retransmitted later.

The SOC identifies Targets of Opportunity (TOO). When a short-term astronomical event occurs, the MOC and SOC work together to adjust the observatory's schedule and generate commands that allow *Origins* to observe an event in less than 24 hours. TOOs are expected to occur about once per month.

*Origins* science efficiency will be ~89%. The other ~11% of the time it is not taking science (Table 2-18). Non-science activities include:

- Slews. A 30° slew takes ~7 minutes, including settling time. The scheduling system is optimized to minimize slew duration; the efficiency calculation assumes 12 slews per day
- Momentum dumping is required once per day, with a duration of 15 minutes
- Orbit maintenance occurs every 3 weeks.
- Line of sight reacquisition.
- Instrument calibrations.
- Ground contributions to science observation inefficiencies.

A day-in-the life of *Origins* (Figure 2-52) captures a conceptual schedule for activities on the observatory. To maximize science productivity, OSS and FIP operate in parallel, *e.g.* in large extra-galactic surveys. However, parallel operations are not currently anticipated with the MISC-T operations. *Origins* is capable of reaching close to 89% science observing efficiency against the 80% requirement. This efficiency level is comparable to *Herschel*'s science operations and improves over *Spitzer*'s 80% efficiency over its 15 years of science observations.

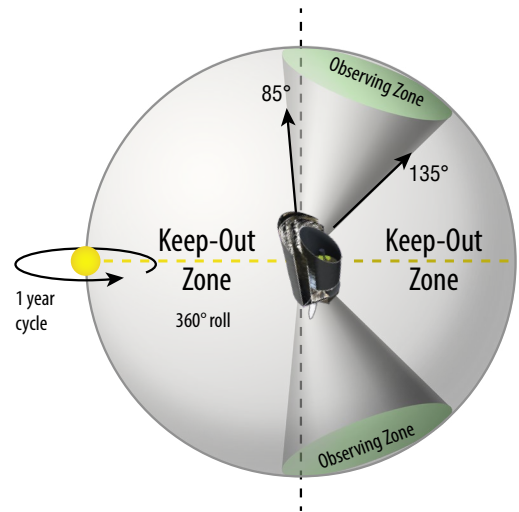

**Figure 2-51:** The mission's FoR allows *Origins* to observe within +85° to +150° (note error in Figure caption) off- pitch off the sun line, yaw 360° around the sun line, and roll ±5°.

**Table 2-18:** *Origins* science observation efficiency is much higher than the 80% requirement.

| Activity | Percent |
|---|---|
| Instruments (e.g., calibrations) | 2.0% |
| Spacecraft Safe Mode | 1.0% |
| Spacecraft Slews (including settling time) | 6.0% |
| Spacecraft Station Keeping | 0.1% |
| Spacecraft Momentum Management | 1.1% |
| Ground Systems | 1.0% |
| Science Operations | 88.8% |
| **Total** | **100.0%** |

Figure 2-53 shows 30 days of operations with the longest expected observation: 16 days of mapping by OSS and FIP. Data volume varies based on instrument operations and observing schedules. When OSS and FIP are operating together for a full day, maximum data volume of 21 Tbits is collected; during transit observations, the data rate is equivalent to the daily science data volume of only 360 Gbits.

## 2.11 On-Orbit Servicing

*Origins*' current mission can be extended and upgraded with on-orbit satellite servicing. The on-orbit servicing will enable flight hardware replacement upgrade capabilities or address degraded performance or non-functioning elements. For example, instruments can be replaced with a next generation instrument to provide enhanced science measurement capabilities, and spacecraft subsystems can be replaced with next generation hardware to provide more efficient power or data management. Additionally, the on-orbit servicing could replenish propellant, extending observatory life by enabling continued station keeping.

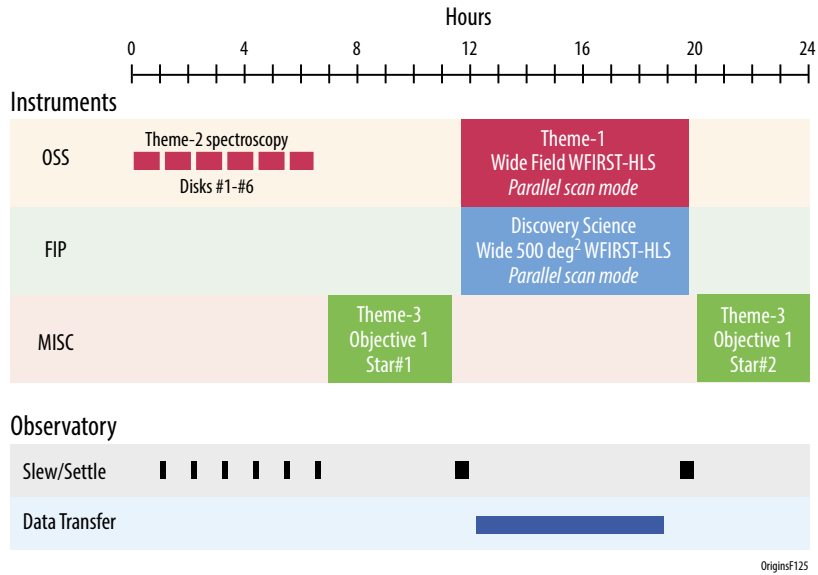

**Figure 2-52:** *Origins* observing plan for a typical day during Year 1. This graphic assumes a science implementation compatible with the described three themes and Discovery Science programs. Actual science targets and observing sequences will be determined by high priority programs allocated based on community proposals to a Time Allocation Committee (TAC), similar to operations of the current great space observatories.

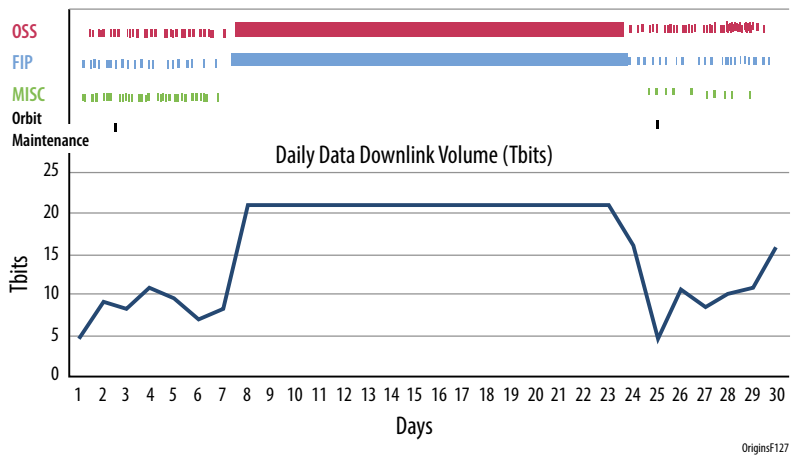

**Figure 2-53:** This example 30-day *Origins* observation plan shows the longest observation anticipated, a 16 day mapping observation by FIP and OSS. Daily data volume varies based on the type of scheduled observations.

*Origins'* operating orbit is SEL2. While on-orbit servicing at SEL2 is yet to be demonstrated, Goddard's Satellite Servicing Projects Division (SSPD) is working to make the servicing of current in-flight missions possible. SSPD missions such as RESTORE-L's mission to service Landsat 7, is helping enable servicing for future missions like WFIRST and *Origins*. For on-orbit servicing, one question that must be answered is where will the servicing occur that will be most beneficial to *Origins* and the Servicer? The two most likely options for servicing are, 1) *Origins* returns to the Earth-Moon regime, 2) the servicer goes to SEL2.

For satellite servicing at SEL2, the Servicer will have all capabilities to reach that orbital regime, rendezvous with *Origins*, and complete all servicing tasks. The servicing vehicle will need a ΔV budget similar to *Origins*, but will have an increased fuel requirement for rendezvous operations with *Origins* and may have less fuel allocated for station-keeping. The servicing vehicle would follow the same trajectory path (approximately 100 days) to reach SEL2 and insert itself into *Origins'* mission orbit for rendezvous. The timing of a direct launch for the outbound transfer is critical since the SEL2 orbit period is 180 days and accounting for any orbital arrival offset would consume additional propellant.

For servicing in the Earth-Moon regime, the roles reverse, and *Origins* would have more responsibilities. The current ΔV budget for *Origins* does not account for returning to the Earth- Moon regime, and so more fuel would have to be included. In order for *Origins* to depart its orbit and insert itself into an Earth-Moon L2 Halo, an additional ~50 m/s would be required (this Delta-V depends on the final *Origins* orbit orientation and amplitude and can be much less). If *Origins* was attempting to be serviced in the planned Near Rectilinear Halo Orbit (NRHO) for the Deep Space Gateway, then the timing of arrival is key and insertion into the NRHO would require an additional ~70 m/s. If a Distant Retrograde Orbit (DRO) is used, an additional 350 m/s would be required. Regardless of the orbit type, the travel time would still be approximately 100 days back to the Earth-Moon regime.

Servicing options will be investigated fully as the *Origins* mission progresses in order to determine the optimal solution in terms of location and ΔV. In general, however, the most efficient (in terms of ΔV) for transferring between the Earth-Moon and SEL2 regimes will be to follow the stable and unstable manifolds that connect the two. An example of these manifolds is shown below (Figure 2-54). In order to minimize the insertion ΔV, an incoming angle was set to 10 degrees to constrain any non-tangential insertions.

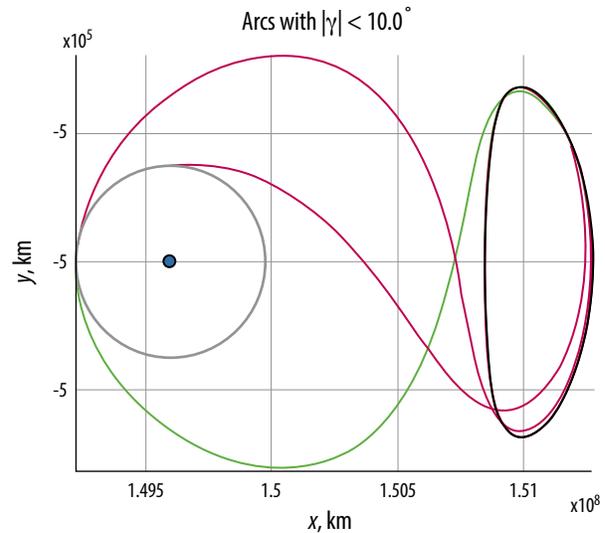

**Figure 2-54:** Stable and Unstable Manifold Transfer between SEL2 to Lunar Orbit, Angle <10°

Regardless of where *Origins* will be serviced, the *Origins* mission includes two types of serviceability accommodations: minor hardware to enable relative navigation and grappling by the servicing vehicle, and modularity to simplify robotic servicing. For example, simple decals with specific markings will be placed on *Origins* to enable the servicing vehicle to determine the relative positions of *Origins* and the servicer.

A grappling fixture can be added to *Origins*, but may not be necessary. Ground testing indicates robotic arms can grapple and control spacecraft at the launch vehicle interface. This will be demonstrated on-orbit with the planned RESTORE-L mission. Major *Origins* flight elements, such as instruments and spacecraft subsystems, are modular. Instruments and spacecraft subsystems are structurally self-contained and have simplified interfaces for power and data. These subsystems are in a mechanical enclosure that can be removed and replaced in one piece. These elements are equipped with blind mate and de-mate connectors and grapple points for a robotic arm.

Serviceable systems are placed radially for visibility and ease of access. These design features will allow a robotic arm to grab and remove or replace these serviceable systems. Table 2-19 shows a representative on-orbit scenario for *Origins*. On-orbit servicing will be further studied during pre-Phase A and Phase A.

**Table 2-19:** On-orbit Servicing Scenario (Example: removal and replacement of an instrument)

| |
|---|
| *Origins* observatory is in a servicing orbit (either SEL2 or Earth-Moon L2), and Origins |
| • Prepares for servicing<br>• Ceases science observations<br>• Science instruments powered down or off<br>• Origins ACS ensures the primary mirror faces away from the Sun<br>• Cryocoolers are turned off and allowed to warm up to ambient temperature (~35 K)<br>• S/A still pointed to sun for power generation<br>• Ready to handover observatory control to a servicing vehicle |
| **Launch Servicing Vehicle and place it in a servicing orbit** |
| **Servicing Vehicle (SV) rendezvous with Origins** |
| **SV captures *Origins* and takes over control** |
| • Origins S/A feathered and ceases power generation<br>• SV provides required power to *Origins*<br>• SV provides data management and transfer<br>• SV provides *Origins* contamination control (e.g., mirror protection)<br>• SV ensures and provides mirror protection from the Sun<br>• SV orients itself and *Origins* in orbit, so that servicing can be performed without damaging *Origins*, including telescope |
| **SV performs *Origins* system checkout for servicing** |
| **SV deploys Robotic Arm One and engages an instrument's grapple fixture** |
| **SV Robotic Arm One removes an instrument** |
| **SV Robotic Arm Two brings a new instrument and inserts it into an instrument slot** |
| **SV checks newly replaced instrument for mechanical fit, power and data connection** |
| **While still mated, SV provides power to *Origins* and *Origins* turns on cryocoolers** |
| **SV disconnects from *Origins*** |
| ***Origins* takes control of its systems: ACS takes over and resumes orbit attitude and control** |
| • If servicing was not performed in *Origins'* operating orbit, *Origins* travels to its operating orbit: *Origins* attitude keeps S/A pointing to the Sun during transit |
| ***Origins* solar array points to sun for power** |
| ***Origins* system self-checkout completed** |
| **Resumes normal operation** |

# 3 - INSTRUMENT DESCRIPTIONS

> *Origins* instrument design studies demonstrate the feasibility of creating instruments that deliver the required science measurements discussed in Section 1. An overview of the instruments (Section 2.7) provides a summary of capabilities and system level resource needs. Each subsection that follows describes the instrument details including operational principle, optical, mechanical, thermal and detector system designs of each instrument: OSS (Section 3.1), MISC-T (Section 3.2) and FIP (Section 3.3).

## 3.1 Origins Survey Spectrometer (OSS)

A highly capable spectrograph enabling breakthrough science, OSS covers the full 25 to 588 micron wavelength range instantaneously at a resolving power (R) of 300 using six logarithmically-spaced grating modules. Each module couples at least 30 and up to 100 spatial beams simultaneously, enabling true (3D) spectral mapping. In addition, OSS provides two high-resolution modes. The first inserts a long-path Fourier-transform interferometer into a portion of the incoming light in advance of the grating backends, enabling R up to 43,000 x [$\lambda/112\ \mu m$], while preserving the grating-based sensitivity for line detection. The second incorporates a scanning etalon in series with the Fourier-transform interferometer to provide R up to 300,000 for the 100-200 micron range. The science drivers for OSS are captured in science Section 1.1, How does the Universe Work? and Section 1.2, How Did We Get Here?, and requirements for OSS are articulated in the STM (Section 1.4). OSS is expected to be a highly used instrument on *Origins*.

## Operation Principle

Figure 3-1 shows the OSS functional approach. At the heart of the instrument are six wide-band grating modules, which combine to span the full 25–588 micron range. These grating modules require no moving parts and will be bolt-and-go structures, similar to the four wide-band grating modules of the *Spitzer* InfraRed Spectrograph (IRS, Houck *et al.*, 2004). In the baseline design, the six slits overlap on the sky so that a point source couples to all six bands simultaneously. The wide-band echelle gratings, particularly the long-wavelength ones which require compact designs, typically only have high blaze efficiency in one linear polarization, so the gratings are used in single-polarization. Light from

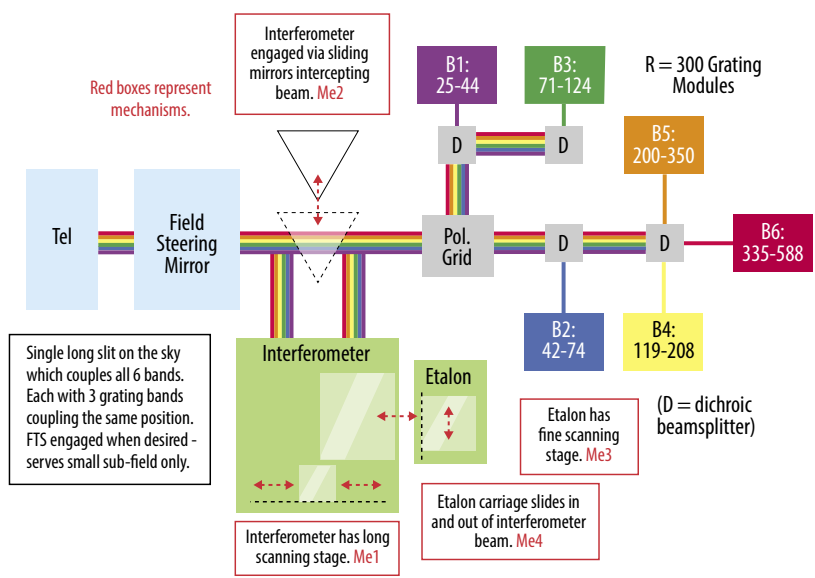

**Figure 3-1:** The OSS Functional Block Diagram shows the six grating modules with their integrated detector arrays, polarizing grid, and dichroic filters (all fixed, with no mechanisms). The beam steering mirror (also known as the field steering mirror or FSM) is part of the telescope. It provides chopping and enables small maps using the grating slits. For high-spectral-resolution measurements over a small sub-field, an interferometer is inserted into the train with a sliding mechanism with two mirrors on it. The interferometer is a Martin-Puplett polarizing Fourier-Transform system with an 8× path multiplier. The baseline design includes an etalon (Fabry-Perot Interferometer) that can be inserted into the interferometer beam to enable very high resolution for Doppler tomography experiments.

**Table 3-1:** OSS R=300 Grating Backends for the Origins Baseline Configuration

| Parameter | Band 1 | Band 2 | Band 3 | Band 4 | Band 5 | Band 6 | $D_{tel}$ scaling |
|---|---|---|---|---|---|---|---|
| λ (μm) | 25–44 | 42–74 | 71–124 | 119–208 | 200–350 | 336–589 | |
| Beam size [″, arcsec] | 1.41 | 2.38 | 4.0 | 6.78 | 11.3 | 19.0 | $\propto D^{-1}$ |
| Slit length [′, arcmin] | 2.7 | 4.0 | 4.7 | 7.9 | 10.7 | 13.6 | $\propto D^{-1}$ |
| Instantaneous FOV [sq deg] | $1.4 \times 10^{-5}$ | $3.5 \times 10^{-5}$ | $7.2 \times 10^{-5}$ | $2.0 \times 10^{-4}$ | $4.6 \times 10^{-4}$ | $9.7 \times 10^{-4}$ | $\propto D^{-2}$ |
| Array size [mm], spectral x spatial | 67x78 | 67x57 | 67x62 | 210x115 | 210x141 | 255x127 | |
| Array format [pixels], spectral x spatial | 168x95 | 168x83 | 168x60 | 168x60 | 168x48 | 140x36 | |
| Pixel pitch [mm], spectral x spatial | 0.40x0.81 | 0.40x0.81 | 0.40x1.0 | 1.25x1.9 | 1.25x2.9 | 1.5x3.5 | |
| **Per-beam sensitivities—includes √2 x penalty for chopping/modulation (with Origins FSM)** | | | | | | | |
| Point source line sensitivity [Wm⁻², 5σ, 1 hr] | $5.0 \times 10^{-21}$ | $3.9 \times 10^{-21}$ | $3.3 \times 10^{-21}$ | $3.7 \times 10^{-21}$ | $3.2 \times 10^{-21}$ | $5.9 \times 10^{-21}$ | $\propto D^{-2}$ |
| Line surface brightness sensitivity [Wm⁻²sr⁻¹, 5σ, 1 hr] | $1.7 \times 10^{-10}$ | $4.7 \times 10^{-11}$ | $1.4 \times 10^{-11}$ | $5.6 \times 10^{-12}$ | $1.7 \times 10^{-12}$ | $1.0 \times 10^{-12}$ | $\propto D^{0}$ |
| Point source R=4 cont. sensitivity [μJy, 5σ, 1 hr] | 2.5 | 3.2 | 4.6 | 12. | 12. | 39. | $\propto D^{-2}$ |
| **Point source mapping speeds—here perfect background subtraction is assumed** | | | | | | | |
| Mapping speed [deg²/(10⁻¹⁹Wm⁻²)²/sec] | $3.2 \times 10^{-6}$ | $1.3 \times 10^{-5}$ | $3.7 \times 10^{-5}$ | $2.6 \times 10^{-4}$ | $2.6 \times 10^{-4}$ | $1.5 \times 10^{-21}$ | $\propto D^{-2}$ |
| Mapping speed [deg²/(μJy)²/sec] | $1.3 \times 10^{-9}$ | $1.9 \times 10^{-9}$ | $1.9 \times 10^{-9}$ | $1.6 \times 10^{-9}$ | $1.6 \times 10^{-9}$ | $3.5 \times 10^{-10}$ | $\propto D^{-2}$ |
| **Intensity mapping sensitivity (noise equivalent intensity / sqrt($N_{beams}$))** | | | | | | | |
| NEI/√($N_{modes}$) [MJy/sr/√sec ], 1σ | 0.57 | 0.29 | .017 | 0.063 | 0.063 | 0.075 | $\propto D^{0}$ |

Numbers are for a 5.9-meter telescope, they can be scaled per the last column, under the assumption that the number of pixels (the optical AΩ or étendu) is fixed.
Notes: Sensitivities assume single-polarization instruments with a product of cold transmission and detector efficiency of 0.25, and an aperture efficiency of 0.72. (Field of view is based on number of beams and a solid angle of π/(4 ln 2) $\theta^2_{FWHM}$, where $\theta_{FWHM}$ is 1.13λ/D, appropriate for an assumed 10dB edge taper.)

the telescope first encounters a polarizing grid; one linear polarization is passed into one arm, one is reflected into a separate arm. One arm then feeds bands 1, 3, and 5 while the other arm feeds bands 2, 4, and 6. Staggering odd- and even-numbered bands allows high-efficiency dichroic filters to separate the bands, and allows the bands to overlap slightly in the two polarizations.

The grating spectrometers are fed with a slit to ensure good sensitivity to individual sources. They are used in point-and-chop mode for deep observations of individual objects; here the observatory FSM modulates the signal from a point source between two positions on the slit. The grating modules also operate as a mapping instrument, in a manner similar to *Herschel*/SPIRE (Griffin *et al.*, 2010), where detectors simply sample the sky as the slit is rastered around. For OSS on *Origins*, this is accomplished either by using the FSM for small fields (on order the telescope FOV), or for large fields by simply scanning the telescope itself as was done for *Herschel*/SPIRE. Table 3-1 shows the high-level instrument parameters, sensitivity, and mapping speed estimates for the base grating system of OSS.

To provide higher spectral resolving power for individual sources, the interferometer is engaged for a small field common to all bands. A sliding carriage moves two mirrors into the beam: the first diverts the beam to the interferometer optics, the second re-inserts it into the beam so it is detected by the grating backends. The interferometer is a Fourier-Transform system designed to provide 7.5 km/s resolution at the 112-micron HD rotational transition. Using the gratings in tandem with the FTS preserves the good underlying line sensitivity of the small per- detector bandwidth (R=300).

Additionally, the interferometer includes an insertable etalon (Fabry-Perot interferometer) that can further improve the resolving power, enabling velocity-resolved measurements of HD and water in protoplanetary disks. This system targets 1 km/s at 112 microns. The etalon cavity operates in high order and is scanned to produce a spectrum.

## Optical Design

**Grating Module Designs:** Each grating module forms a light-tight enclosure that includes the detector array and for which the only opening is a slit. Requirements met by the optical design include:

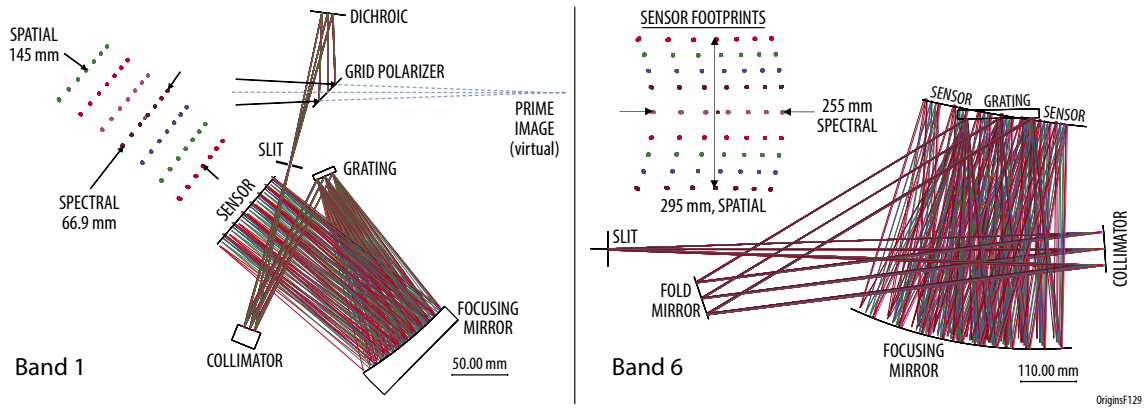

**Figure 3-2:** Grating module designs for Bands 1 (left) and 6 (right), the shortest and longest-wavelength bands in OSS. In both cases the f/14.4 telescope focus is shown, positioned at the input slit for band 6, and the conjugate to it for Band 1. The grating optical designs and focal-plane dimensions shown here represent larger focal planes than are currently baselined (pixel count is limited by readout), and the optics accommodate these, thereby allowing for an upscope. Band 1 in particular can accommodate a large slit – up to 250 beams. The Band 6 focal plane uses the four spatial fields above and below the center, which is obscured by the grating.

- Slit length on sky: at least 100 diffraction-limited beams for bands 1-5, and 75 beams for band 6 (this provides margin – not all of this optical field is used in the baseline design because of readout limitations).
- Intrinsic spectrometer resolving power $\lambda/\Delta\lambda$ of at least 300 over a 1:1.75 bandwidth.
- Strehl >80% (goal >90%) imaging to a focal plane that images a 1.15 $\lambda$/D spatial × R=300 spectral element (band-averaged) onto an area no smaller than 0.25 mm² (to enable multiplexed detector readout).

No constraints are placed on distortion in the focal plane (or 'smile'), the non-linearity of a single spatial position's spectrum as imaged by the system, since the system has a 2D focal plane that can be customized to the optical configuration. The size and mass of the short-wavelength gratings are driven primarily by the pixel pitch and thus the array size, while the long-wavelength modules are driven by the size of the grating. Figure 3-2 shows the shortest and longest of the grating module designs. Notional detector array formats are tabulated for each band in Table 3-1, under the assumption of 1.13 f$\lambda$ sampling spatially by R=300 sampling spectrally. This proof of concept design is not currently optimized. As is typical of wideband grating systems that use a large incidence angle, the systems are highly anamorphic, with spatial f/# much faster (~2×) than the spectral f/#. Pixels are therefore rectangular, elongated in the spatial direction. Additionally, the physical size of an R=300 bin changes across the band, so each focal plane will have at least two spectral pitches across the full band to preserve a given fractional bandwidth per pixel – the tabulated spectral pitches are average values.

**Interferometer Design:** The design of the interferometer (Figure 3-3) is driven by the need to incorporate high spectral resolution at the long OSS wavelengths while maintaining the excellent sensitivity and multi-band capability of the base grating suite. The approach is to intercept and interfere the light from a small field before sending it to the grating modules for detection. This insures that the excellent sensitivity to spectral lines provided by the R=ν/δν = 300 detector bandwidth is preserved; the system can be considered as an extreme implementation of a band-limited Fourier Transform Spectrometer (FTS).

The baseline design has a 30 cm physical throw providing 2.4 meters of optical path difference between the arms of the interferometer. This provides a wavelength-dependent maximum resolving power of 43,000 x [*112 microns / λ*], which provides a good match to expected linewidths of the HD

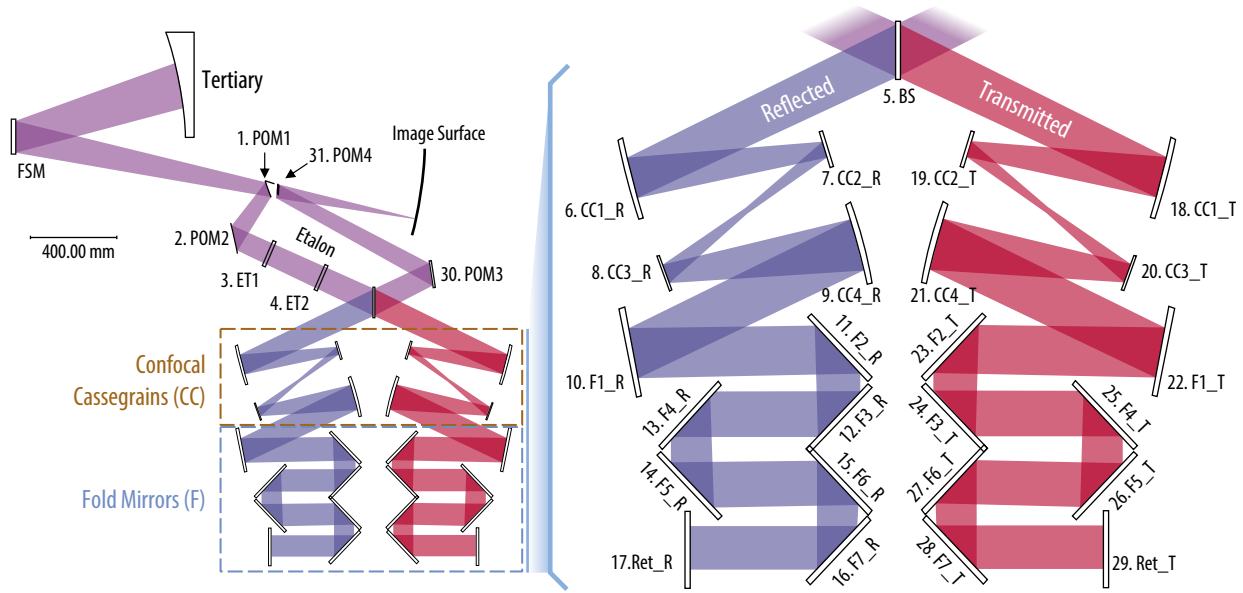

**Figure 3-3:** Optical layout for the OSS interferometer. Light is intercepted in the converging beam from the telescope and is collimated with a diameter of 8 cm. After processing by the interferometer, the light is re-inserted into the original light path from the telescope to the grating modules. The interferometer is engaged with a sliding stage containing POM1 and POM4; it only accesses a small portion of the grating slits as it is designed primarily for single-source or small-field measurements. On the right the mirrors are numbered. The single moving stage carries mirrors 11, 12, 15, 16, 23, 24, 27 and 28.

112 μm transition in protoplanetary disks. Shorter FTS scans can be employed, which would reduce R, but the 1/λ dependence is always present since the single FTS scan is common to all six bands. The collimated beam diameter is 8 cm, which ensures divergence (diffraction) is not a concern for the FTS operation, even at long wavelengths. The long extra optical path between the pickoff and the interferometer retro-reflectors necessitates pupil re-imaging to keep the system compact. This is provided by two sets of confocal off-axis Cassegrain telescopes that image the telescope pupil to the back of the interferometer.

The FTS architecture is a Martin-Puplett polarizing Fourier-transform spectrometer (FTS; Martin & Puplett, 1970; Lambert & Richards, 1978), which is optimal for this application. Relative to other approaches to Fourier transform spectroscopy, this system uses a single input port and single output port. It requires a linearly-polarized detector, but both polarizations can be used with independent detectors. It has unit efficiency in each polarization, unlike a Michelson architecture, which shares power between two ports.

In addition to the FTS, a second, higher-resolution capability enables line-profile Doppler tomography measurements of HD and $H_2O$ lines in the 100–200 μm range. Operation at other wavebands may be possible, which the team will further study in Phase A. This is provided with an etalon (Fabry-Perot interferometer) consisting of a 26-cm long cavity formed by two partially reflecting mirrors that can be inserted into the beam before the FTS beam splitter. With a design finesse of 70, the etalon provides a resolving power of 325,000 (0.9 km/s) at 112 μm, or 200,000 (1.5 km/s) at 179 μm. The etalon uses the same 8 cm collimated beam provided for the FTS; however, to allow for beam walk in the etalon without substantial signal loss, the mirrors must be 14 cm in diameter. The mirrors in the FTS are similarly oversized. Once in the optical train, the cavity spacing can be changed by ~1 mm, providing a wavelength scan of a few orders, which is ample to provide complete coverage. The etalon maintains its parallelism through launch and cool down, so parallelism adjustments are not required in flight.

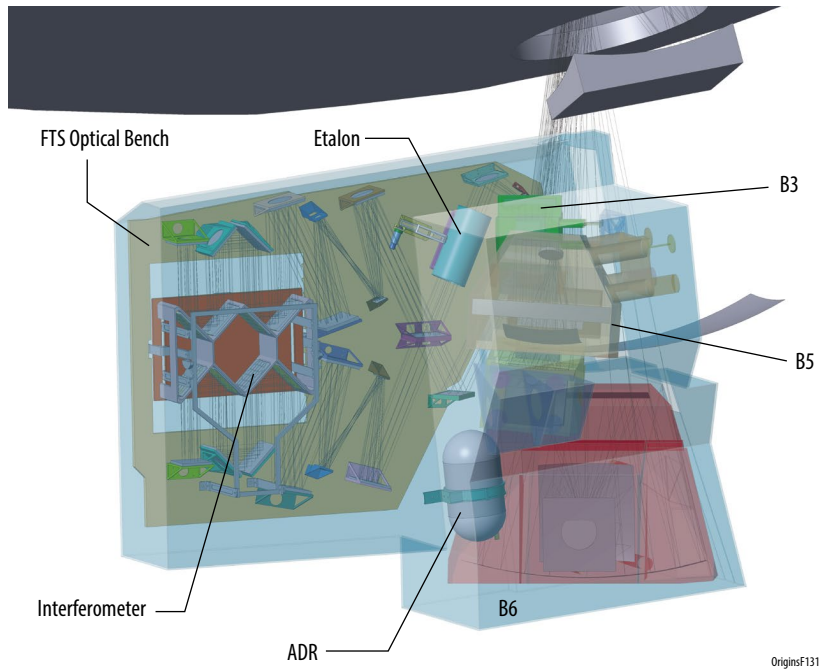

FTS Optical Bench
Etalon
B3
B5
Interferometer
B6
ADR

OriginsF131

**Figure 3-4:** The OSS optical and mechanical design has at its core, the 6 grating models (front right). Light from the telescope comes in from the top (blue rays) and is split into the 6 bands via polarizing grids and dichroics. The FTS spectrometer (back left) picks off light near the entrance of the instrument, passing it through the interferometer and then back through the grating modules. For the highest spectral resolution, an etalon (turquoise tube) is inserted in the light departing from the FTS.

**Full Optical Configuration:** Since all six grating modules must be fed through their slits at the telescope focus, and these slits overlap, packaging is an important consideration, and the solution has been obtained through careful iteration with the observatory mechanical and optical designs. Figure 3-4 shows views of the full configuration with all grating bands, the interferometer and etalon, as well as the field of view of the OSS in the *Origins* focal plane.

## Observing Modes

Table 3-2 summarizes the OSS observing modes. The base R=300 long-slit spectroscopy mode can be used for pointed observations of targets of interest and for mapping. OSS is very efficient for pointed observations since the slits from all six bands overlap so that a complete spectrum is obtained for the full wavelength range simultaneously. For this mode, the *Origins* steering mirror (FSM) will be used to chop a source back and forth with a selectable throw. For point sources, a distance of three long-wavelength beams or 1 arcminute, ensures that the flux in from an unresolved target source is negligible in the off-source position. The chop throw could extend up to the length of the short-wavelength (Band 1) slit -- 2.7 arcminutes, to maintain full efficiency with the source detected in both chop positions. Chop frequencies in this mode are expected to be as low as 0.5 Hz, as the detectors will have good stability to this level. The *Origins* FSM is capable of faster chopping (up to 10 Hz) if needed to overcome other sources of variability.

**Mapping and Detector Requirements:** The OSS slits can also be rastered around the sky to generate maps. In this mode, the scan rate must not be so fast that the short-wavelength beams smear when sampled with the finite bandwidth of a detector. These modes, together with the observatory scan speed therefore set a requirement on the detector time constant t. The team selected 60 arcsec/sec as the maximum observatory scan speed. Operating at this maximum speed then requires a time constant (t) at the short wavelength (25 microns) of 5 ms, which is comfortably within *Origins'* 3 ms requirement. This maximum speed allows fknee to be as high as 17 mHz for recovering the large-scale structure in the intensity mapping experiments.

**Table 3-2:** OSS Observing Mode Summary

| Mode | Band | R | FoV | Modulation | Notes |
|---|---|---|---|---|---|
| Low-Res Pointed | Full | 300 | Full slits | Origins FSM chop along slit | Maximum single-source sensitivity for unresolved lines |
| Low-Res Mapping | Full | 300 | Full slits | Origins FSM or telescope scan | Maximum speed for mapping unresolved lines |
| High-Res Pointed | Full | 43,000 x 112μ/λ | 20 arcsec slit | FTS scan | R is tunable via scan length. Minimum time at max R is 30 min |
| Ultra-high-Res | 100-200 | 300,000 x 112μ/λ | 1 beam | FTS scan | 200 seconds for each position |
| Bright Source | 100-200 | 300 | 20 arcsec slit | Origins FSM chop | Use interferometer + etalon to reduce flux |

**High-Resolution Mode:** A similar relationship exists in the use of the FTS. When the FTS is operating with a constant scan speed of the optical path difference (OPD) in the interferometer $v_{OPD}$, (which is 8 times the speed of the carriage carrying the moving mirrors), the narrow spectral bandwidth of each grating channel corresponds to a narrow audio-band signal according to $f_{audio} = v_{OPD}/\lambda$, so each R=300 detector need only carry signal in a narrow (R= 300) band about the central fringe rate. This provides some natural immunity against systematics and instabilities. In particular, absolute stability is not required over the full scan time. However, to limit the signal loss to 70% at the shortest wavelength, the fringe rate should be less than $(2\pi\tau)^{-1}$, thus the limiting scan speed is $v_{OPD}$,max $= \lambda_{min}/(2\pi\tau)$, which is 1.3 mm/sec using the 3 ms time constant. This scan rate results in a fringe rate at 580 microns of 2.2 Hz, well below the knee. A scan of the full 2.4 m OPD then requires 30 minutes, if the shortest wavelengths are to be recovered. If the recovery of the high- resolution information at the shortest wavelengths is not desired, then the scan could be quicker.

Beam divergence in the interferometer, combined with practical size limitations of the optical design, limit the field of view through the interferometer. The high-res mode will couple a sub-slit of approximately 20 arcseconds through the interferometer, and will be capable of providing the full resolving power over this small slit. For the longest wavelengths this corresponds to a single beam, but for all shorter-wavelength bands, multiple spatial pixels will be available for background subtraction. At the longest wavelengths, the background is dominated by the stable CMB, and with the large beam, the variability induced by pointing drifts while looking at a point source will be small.

**Ultra-high-resolution mode:** When using the etalon in the nominal design with Finesse=70 and cavity spacing of 26 cm, the order number will be several thousand (*e.g.* 4600 at HD 112 microns). Multiple orders will thus be coupled simultaneously to a single R=300 grating channel, and the FTS will be required to sort these orders. Each step in a high-res etalon spectrum will therefore require an order-sorting FTS scan. This scan, however, need only be long enough to clearly resolve the orders from one another, that is it requires R=2 x $m_{FP}$ or 10,000 at HD 112 microns. This is about ¼ of the total FTS range. Additionally, the etalon is not required to operate at wavelengths shorter than 100 microns, so the fringe rate constraint described above only applies to 100 microns. Thus $v_{OPD}$,max for the order sorting operation is 3.2 mm/sec, and the order-sorting scan requires ~330 seconds. A 20-position etalon scan will thus require 1.1 hours, and a 140-position etalon scan covering a full free spectral range (the space between orders) to produce a Nyquist-sampled complete spectrum requires 140 positions, or nearly 8 hours. A subject for phase-A study is the potential to reduce this time with a custom FTS scan which uses ~60 discrete positions instead of a continuous scan. If chosen properly, 60 FTS positions would enable discrimination among the ~30 etalon orders on each grating channel.

## Detectors

To make optimal use of the low-background space environment, the OSS detectors should be more sensitive than those fielded to date. Furthermore, the OSS has a total of 60,000 pixels in the six arrays. The OSS requirements could be met with multiple detector approaches, but all of them require some development; this is described more fully in the *Origins* Technology Development Plan and summarized in Section 2.3 of this report. For this concept design, we have adopted superconducting TES bolometers as the baseline.

## Readout Electronics

For the OSS point design, the team adopted a conservative readout system approach that can accommodate any of the viable detector approaches (Figure 3-5). The team also baselined resonator frequencies at 4–8 GHz and a resonator spacing of 2 MHz. Each readout circuit therefore can process 2,000 pixels, and when partitioned into integer readout circuits for the six arrays, the total number of circuits required for all six is 32. We use the new Xilinx RF-system on chip (SoC) (Farley *et al.*, 2017), which integrates the analog to digital converters (ADC) with programmable digital signal processing logic (floating point gate array (FPGA)-like) in a single chip based on 16-nm-gate complementary metal-oxide semiconductor (CMOS) transistors. A single chip provides eight channels, each capable of processing 2 GHz at 12 bits depth, and the associated waveform generation capability, it represents approximately 8 x (2 / 0.5) = 32 times the processing capability of the fiducial existing FPGA-based readout. Abaco released a board with this chip in late 2018. Initial power dissipation estimates for this board are 50 W at full utilization.

OSS system design uses this power dissipation fiducial of 50 W per 16 GHz of information bandwidth for the ADC and digital spectrometer logic. This is a conservative estimate – while neither of these existing systems are flight-qualified, mixed-signal application-specific integrated circuits (ASICs), which integrate ADCs and signal processing, have been developed for flight implementation of similar applications (Hsiao *et al.*, 2015), and these systems typically have lower power dissipation than their programmable-logic counterparts.

## Data Rate

The OSS data rate is determined by the pixel count (60,000), sampling rate, and required bit depth. The detectors will be sampled at 500 Hz to support an electrical bandwidth of 250 Hz, a comfortable factor of 5 above the 3dB frequency of the detector with its 3 ms time constant. This fast sampling will be useful in particular for identifying cosmic ray events which can have a rise time much faster than the nominal detector speed of response. 12 bits is sufficient to capture the detector's dynamic range under which it is photon noise limited, if a logarithmic scaling is used. The data rate is thus 360 Mbits/sec. In the high-res modes, since only a small fraction of the array is used, the rate could be smaller, but the team expects that the grating data may be valuable for other purposes during these observations (*e.g.*, for monitoring background and taking potentially serendipitous measurements along the slit).

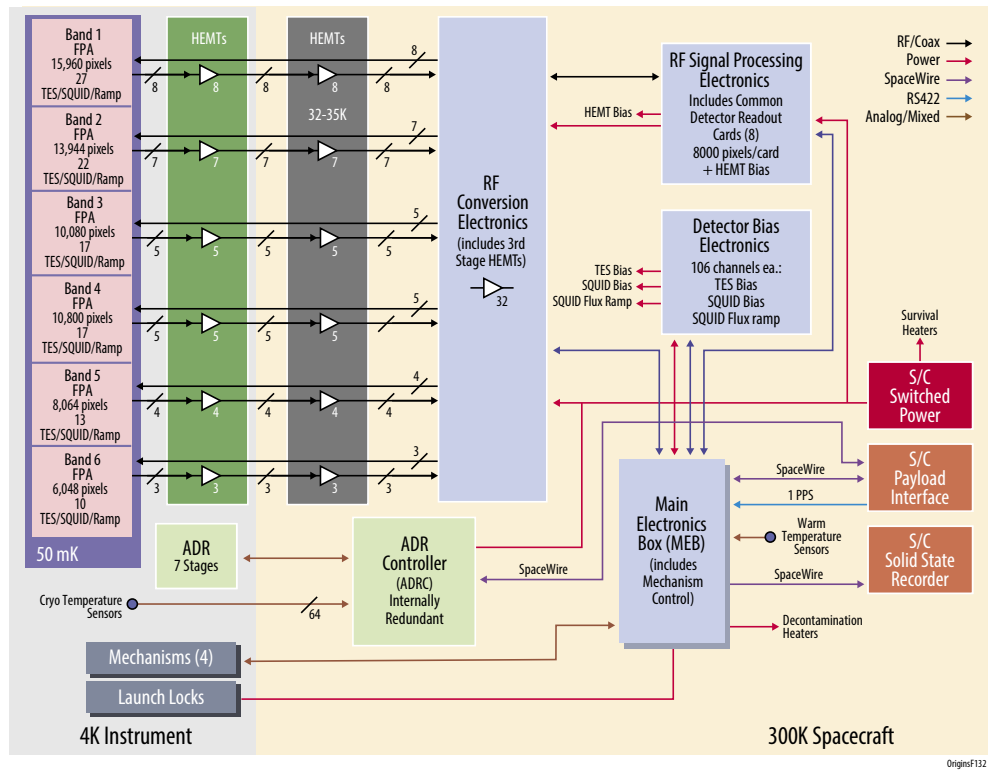

**Figure 3-5:** The readout circuit consists of 32 circuits, each carrying 4 GHz of information bandwidth. Detectors are coupled to micro-resonators, 2000 of which can be read out in each line. On the warm spacecraft side, electronics create the waveform that interacts with the array, and then digitizes and analyzes the return waveform to extract phase shifts that encode the optical power on the detector. This basic approach accommodates all viable detector technologies under development for Origins.

**Thermal and Mechanical Design and Resource Requirements**

The interferometer optics operate at the native observatory-provided 4.5 K, but the grating modules as well as the band separation elements (polarizing grid and dichroic filters) are all cooled further to ensure that their thermal emission does not impact sensitivity. In particular, the potentially lossy grid and dichroic filters before the grating slits (termed the band separation optics) must be cooled to below the microwave background temperature to ensure that they do not add background. The gratings modules themselves need to be below 0.9 K (at least at the long wavelengths) to insure that the integrated power through the bandpass filter with the full native detector etendu ($A\Omega$) does not saturate the detector or degrade the noise equivalent power. These requirements are met with a continuous adiabatic demagnetization refrigerator (CADR) which stages from the observatory-provided 4.5 K system. A schematic is provided in Figure 3-6. The CADR is adapted from the high-heritage designs of the Hitomi ADR and technology advances (Shirron *et al.* 2000; Tuttle *et al.* 2017; Shirron *et al.* 2016); it uses 7 salt pills to provide 3 continuously-cooled stages: at 1.5 K (for the polarizing grid and dichroic filters), 0.7 K (for the grating spectrometer enclosures), and 0.05 K (for the focal plane arrays themselves). A key advantage of the ADR approach is the high Carnot efficiency, and the CADR heat rejection is only about half of the 4.5 K budget even though it is designed to provide 100% lift margin at all of its actively cooled stages. Further information is provided in the *Origins* Technology Development Plan.

A suspension system has been designed together with the cooling system; it is shown schematically in Figure 3-6. The band separation optics and the grating optics modules are suspended kinematically from 4.5 K with bipods made of titanium 15-3-3-3, a low-thermal- conductivity alloy. While composites offer lower thermal conductance at these temperatures, the use of metal eliminates a potential source of water contamination curing cool down. For the grating optics modules cooled by the 0.7 K stage, the titanium supports couple directly from 4.5 K mechanically, but heat is intercepted at the strut midpoints by the 1.6 K stage. Strut dimensions are tuned for each band's mass, insuring a resonant frequency that satisfies a standard mass acceleration curve. Inside each grating module, the focal plane assemblies are also mounted kinematically with Ti-15-3-3-3 bipods, in this case designed to provide resonant frequencies of at least 100 Hz.

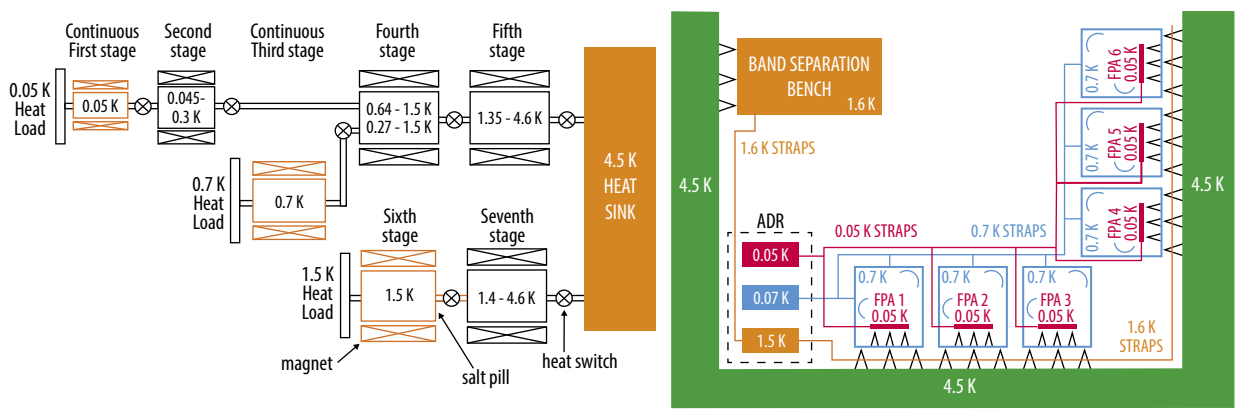

**Figure 3-6:** Left: The seven-stage continuous ADR is actually a five-stage ADR, similar to that used in the FIP instrument, and a two stage ADR. The ADRs are controlled from a single ADR Controller (ADRC) on board the warm side of the spacecraft. Right: Low-thermal-conductivity titanium bipods mount the spectrometer boxes from the optical bench, and the focal plane assemblies inside the spectrometers. These comfortably support launch loads, yet enable the 50 mK cryogenic engine; no special provisions are required to survive launch. Though numerous, the cooling straps from the ADR are the size of large gauge copper wires and weigh a combined total of 1 kilogram.

**Table 3-3:** Heat loads and available lift on OSS Cryogenic stages. A margin of at least 100% over the calculated heat load is maintained at each temperature stage.

| T (K) | Load | Heat Load | Total | Capability | Margin (%) |
|---|---|---|---|---|---|
| 0.05 (Focal planes) | Harnesses conducted | 0.65 µW | 2.67 µW | 6.0 µW | 125 |
| | Dissipation | 0.67 µW | | | |
| | Suspension conducted | 1.35 µW | | | |
| 0.7 (grating modules) | Harnesses conducted | 6.5 µW | 111 µW | 292 µW | 163 |
| | Suspension conducted | 104 µW | | | |
| 1.5 (bench & intercepts) | Harnesses conducted | 0.04 mW | 2.11 mW | 4.25 mW | 101 |
| | Bench Supports conducted | 0.96 mW | | | |
| | FPA Supports | 1.11 mW | | | |
| 4.5 | Amplifier dissipation | 11.8 mW | 25.8 mW | 63 mW | 144 |
| | Mechanisms | 1 mW | | | |
| | ADR heat rejection | 13 mW | | | |

Table 3-3 summarizes the calculated heat load and available lift with the cooler for each of the three OSS cooling stages. The dominant load on the sub-4K OSS-cooled stages is the conducted loads through the mechanical suspension; wiring parasitics are negligible because superconducting cables are used. The 4.5 K situation is more challenging relative to the adopted allocation from the observatory, as this stage must accept the loads from the non-superconducting harnesses, and 4.5 K houses the first stage low-noise amplifier. The 0.38 mW per amplifier x 31 amplifiers corresponds to commercially-available devices operating at up to 8 GHz, but the team expects improvements in amplifier power dissipation will be possible by the time *Origins* is built. A straightforward descope to ease this aspect is to not operate all six bands simultaneously. (Parasitic conduction in harnesses are book-kept at observatory level, but are much smaller than the amplifier dissipation.)

The design includes thermal straps made of pure annealed copper for all stages, sized for the load and distance required. These copper straps, along with the indium used for thermal connections, total approximately 8 kg for OSS.

**Mechanical Approach:** Each of the OSS grating modules and the interferometer bench are discrete opto-mechanical elements that will be fabricated and tested independently and then mounted kinematically in the full instrument. They can be internally aligned and are subjected to environmental testing as units before being integrated into the full instrument. Aluminum spectrometer optical elements (mirrors, gratings) are the baseline, but the structural benches are baselined as beryllium to increase stiffness and reduce mass.

**Mechanisms:** OSS has four mechanisms, tabulated in Table 3-4 with their top-level requirements. Me1 is the FTS interferometer scan mechanism that moves the rooftop mirror carriage. Me2 is for the interferometer insertion; it simply moves the carriage containing the pickoff and re-insertion mirrors into the beam. Position accuracy is not a strong constraint, but this carriage must have an orientation accurate to 20 arcseconds to ensure the interferometer pupil matches the telescope pupil. Me3 scans the etalon, it needs only cover a couple of free spectral ranges at the longest wavelength (FSR = $\lambda_{max}/2$), so a total of 600 microns. However, it must provide translation knowledge to 0.2 microns, ~¼ of a resolution element at 112 microns. Furthermore, it must maintain parallelism to 0.3 arcseconds to insure that the cavity finesse is not impacted by the parallelism. Finally, Me4 allows the etalon assembly to be inserted and removed from the interferometer beam; here the positional accuracy is not critical, but the etalon should be oriented to within 10 arcseconds once inserted to avoid beam walk-off in the etalon.

The carriage for the FTS is a flex-band type design with multiple arms to achieve a linear travel of 300 mm. This is scaled up by a factor of 3 from the unit in Figure 3-7. This design has essentially no friction. A superconducting linear servo motor has the required precision and low losses allowing for a dissipation of less than 1 mW. For the etalon, a piezoelectric inchworm actuator will be used. Position, to the accuracy required, will be read out with either a modified Kaman DIT or a modified Mad City Labs NanoAlign-3. The launch locks will be actuated using typical non-explosive actuators such as the NEA 9100. For the etalon and FTS insertion mechanisms, which are used infrequently, a four-bar linkage connected to a brushless DC motor is used.

**Instrument Control**

Control software for the OSS is straightforward. The system-level schematic is shown in Figure 3-8. A main electronics box (MEB) includes a number of specialized control boards and a LEON-3 CPU. The system includes all critical features:

- Mode management (allowing reboots with software updates, standby, calibration, and the various science modes)
- Instrument support (command processing, data collection, and support for firmware updates)
- Mechanism control. FTS and etalon scan are operated with closed-loop control with encoder feedback at 1 kHz. Interferometer insertion and etalon insertion are open loop commanded, with positions verified with microswitches.
- Power distribution.
- Some degree of on-board autonomy (limit checking, some processing command sequences, failure corrections)

The flight software will be a real-time system based on proven JPL or GSFC in-house or commercial off-the-shelf (COTS) systems. No technical risks have been identified. The system will have an estimated 60,000 total lines of code, of which 85% will be reuse from existing flight instruments. The system has an estimated 50% processor resource margin with a LEON-3 CPU.

**Table 3-4:** OSS Mechanisms

| # | Full name | Function | Accuracy | Qty |
|---|---|---|---|---|
| Me1 | FTS Scanning | Move 20kg mass: a) 100mm at 10mm/sec; b) 300mm at 300micron/sec | 0.8 micron knowledge, 1.6 micron control, angular runout of 1.6 arcsec. | 1 |
| Me1.1 | Me1 Launch Lock | Lock FTS for launch | N/A | 1 |
| Me2 | FTS Insertion | Move FTS pickoff mirrors | Oriented to within 20 arcsec | 1 |
| Me3 | Etalon Fine Scan | Move one etalon mirror 0.5 micron every 10 sec. Total throw of 600 microns. | 0.2 microns knowledge, 0.4 microns control. Angular runout of 0.3 arcsec | 1 |
| Me3.1 | Me3 Launch Lock | Lock Etalon Mirror for launch | N/A | 1 |
| Me4 | Etalon Insertion | Insert Etalon Assembly | Translation control to 2 mm, oriented to 10 arcsec | 1 |
| Me4.1 | Me4 launch lock | Lock Etalon Assembly for launch | N/A | 1 |

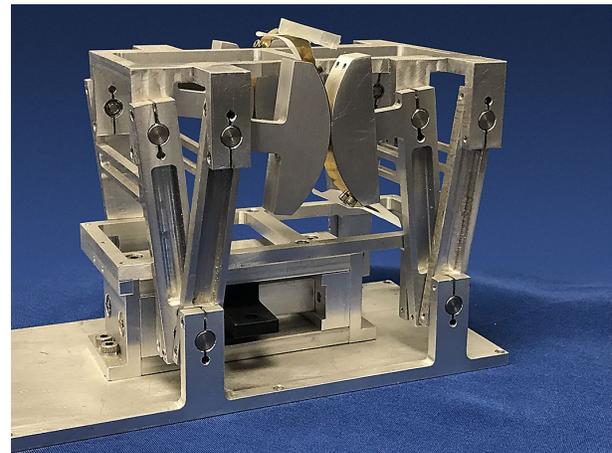

**Figure 3-7:** Flex type mechanism for use with the OSS FTS. This mechanism is free of friction, can be operated in any orientation and has a stroke of 10 mm. For OSS this would be scaled up by a factor of 3.

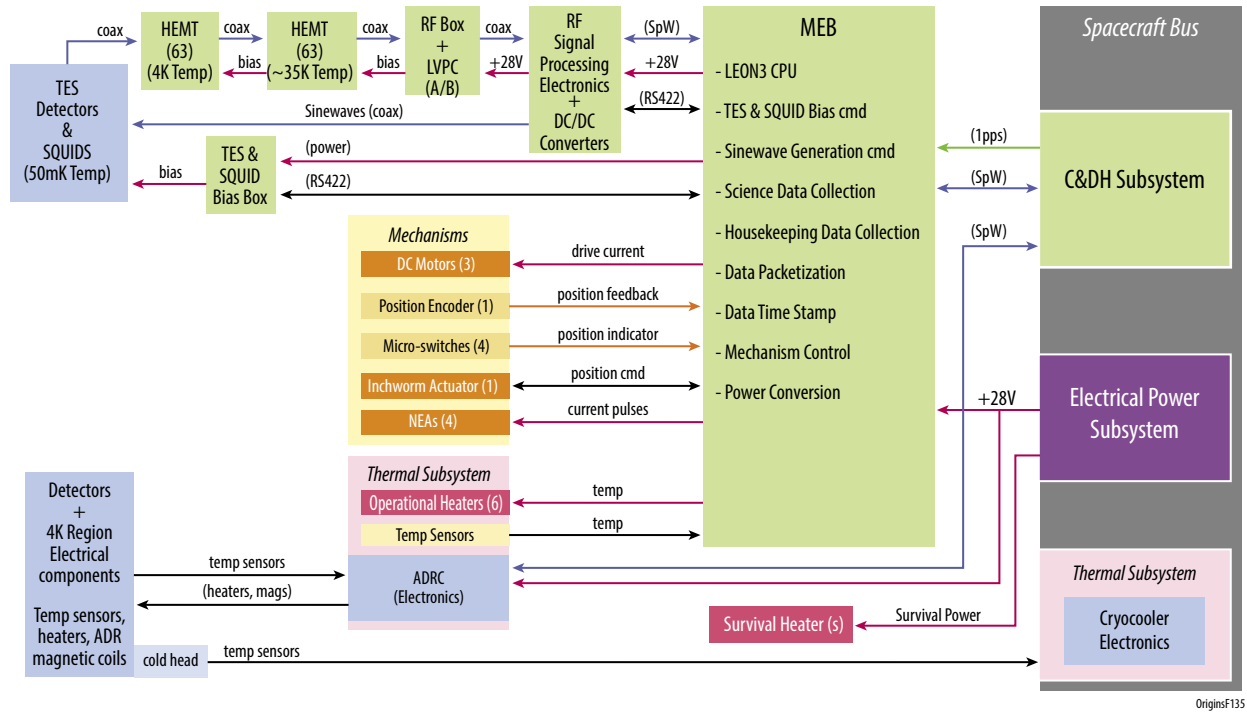

**Figure 3-8:** Block diagram of OSS instrument control. The mode complexity is comparable to instruments that have flown, such as *Herschel* SPIRE, and software re-use is estimated at 85%.

## Risk Management Approach

While the OSS has multiple operating modes, each mode has some degree of heritage, and much of the science can be done in the base grating-only mode. The team has identified the most important risk areas and approaches to their mitigation.

1. At present, the most important risks for OSS are in the detector system, in particular the sensitivity and yield at large array format, and the response to cosmic ray events. These aspects are addressed with a dedicated technology development plan provided in a separate section of this report.

2. Stray light, magnetic susceptibility, and radiofrequency (RF) power are known potential problems with cryogenic superconducting detectors. The stray light risk is mitigated with the design; the grating spectrometer assemblies are light tight with only the slit as entrance points; they can be tested independently with radiative sources (loads) outside the slit to verify that no excess power is coupled to the detectors beyond what is expected in-band from the spectrometer optics. The magnetic susceptibility risk, while often a problem in ground- based and suborbital instruments, and important to understand early in the design cycle, is also straightforward to assess and mitigate for OSS. The ambient magnetic field at L2 is much smaller than Earth's field, so the most important field sources are those on board the Origins spacecraft (e.g., the ADR magnets), but these will be shielded. Their effect can be bounded cleanly because field strength decays as $1/distance^3$. The susceptibility of the focal plane packages can be separately measured with AC-modulated Helm-hotz coils in the lab; and the susceptibility can be shown to be 1-2 orders of magnitude lower that what would impact scientific performance. The RF power may be most challenging among these if bolometers are used; RF power can couple to bolometers directly, generating an analog of stray light, and the RF environment can depend on details of the wiring harness and other observatory-level aspects which are hard to test prior to instrument integration. Here again the grating module boxes help provide a form of mitigation-- they will be used as Faraday cages, with filtered electrical feed-throughs for all bias and readout lines. Additionally, the team will carry out early tests with flight-like cables, electronics, and other identified Origins RF sources (or suitable proxies), at the individual OSS grating module level in Phase C, prior to integration.

3. A final class of system-level performance risk is the failure of mechanisms. This risk will be mitigated by using high-heritage approaches, as well as early design, prototyping, and qualification including cryogenic testing. The FTS insertion carriage will be designed to fail in the 'in' position, so that the interferometer can be used even if the mechanism fails. By the same principle, the etalon insertion mechanism is design to fail in the 'out' position, to leave the FTS path clear.

**Predicted Performance and Margin**

Figure 3-9 shows the expected OSS performance in the far-IR and submillimeter. We assume 72% aperture efficiency to a point source, and 25% instrument transmission (in a single polarization). Galaxy spectra assuming L =$10^{12}$ and L$_\odot$ at various redshifts are overplotted using light curves with continuum smoothed to R=300. The OSS instrument is the suite of R=300 grating with the slit lengths as designed (Table 3-1). Detectors are assumed to operate with NEP = $3\times10^{-20}$ W Hz$^{-\frac{1}{2}}$, a figure which has been demonstrated in the lab. The SPICA / SAFARI-G curve refers to the currently-planned configuration: a 2.5-meter telescope with a suite of R=300 grating spectrometer modules with 4 spatial beams, and detectors with NEP=$2\times10^{-19}$ WHz$^{-\frac{1}{2}}$, the dashed line shows how the SPICA platform could be extended in wavelength beyond the 230 μm requirement. ALMA sensitivity refers to a R=1000 (300 km/s) bin, and the survey speed incorporates the number of tunings of the 16 GHz total bandwidth to cover a 1:1.5 fractional band.

The relationship between the detector performance and the estimated on-sky sensitivity are similar to what has been obtained in ground-based grating spectrometers Z-Spec (Bradford+, 2009) and ZEUS (Stacey+ 2007; Ferkinhoff+ 2012), providing confidence in the estimation.

The OSS instrument sensitivity model includes margin in 3 ways: 1) The sensitivity model includes a factor of 1.7 above the calculated sensitivities values assuming a total slit-to-detector efficiency of 25%, this can allow for a combination of lower efficiency or poorer detector sensitivity. 2) The point-source staring sensitivity (which employs chopping of the *Origins* FSM) includes a factor of $\sqrt{(2)}$ penalty. This would be required if the source is only observed half the time or if the background is only measured with a single spatial mode. But since the baseline approach is to chop the source along the slit and measure the background with the full slit, the team expects that this factor will not apply; thus, it represents margin. 3) Because the system is strongly background limited with the baseline NEP of $3\times10^{-20}$ W/$\sqrt{\text{Hz}}$, some degradation in detector sensitivity can be accommodated without greatly impacting sensitivity. Finally, the most important source of margin is applied to all of the *Origins* science cases: they are based on performance that is 2x poorer than the calculated sensitivity, which itself includes the factors described above. See Figure 3-9 for sensitivity vs. wavelength.

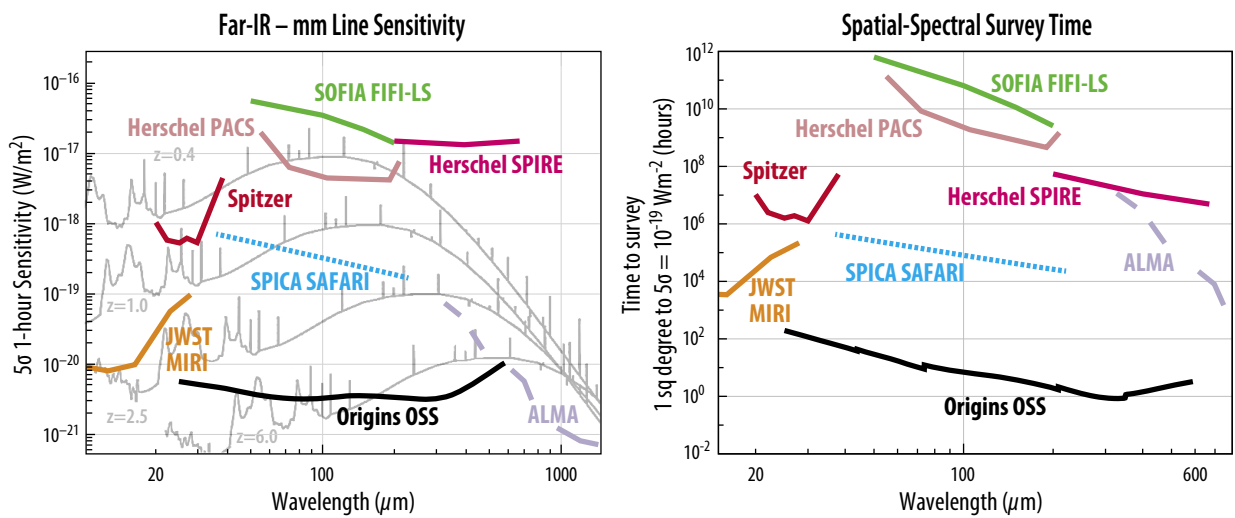

**Figure 3-9:** OSS's spectroscopic sensitivity is superior to prior missions (*Herschel*) by more than 1000 and to the proposed SPICA/SAFARI by more than 100. Its mapping speed is more than 10⁹ times faster than *Herschel* and 10⁴ times faster than SPICA/SAFARI. Left shows the sensitivity in W/m⁻² for a single pointed observation, assuming a factor of sqrt(2) penalty relative to staring for background subtraction (a conservative assumption). The right panel shows the time required for a blind spatial-spectral survey reaching a depth of 10⁻¹⁹ W/m² over a square degree, including the number of spatial beams and the instantaneous bandwidth (here assuming no penalty for background subtraction).

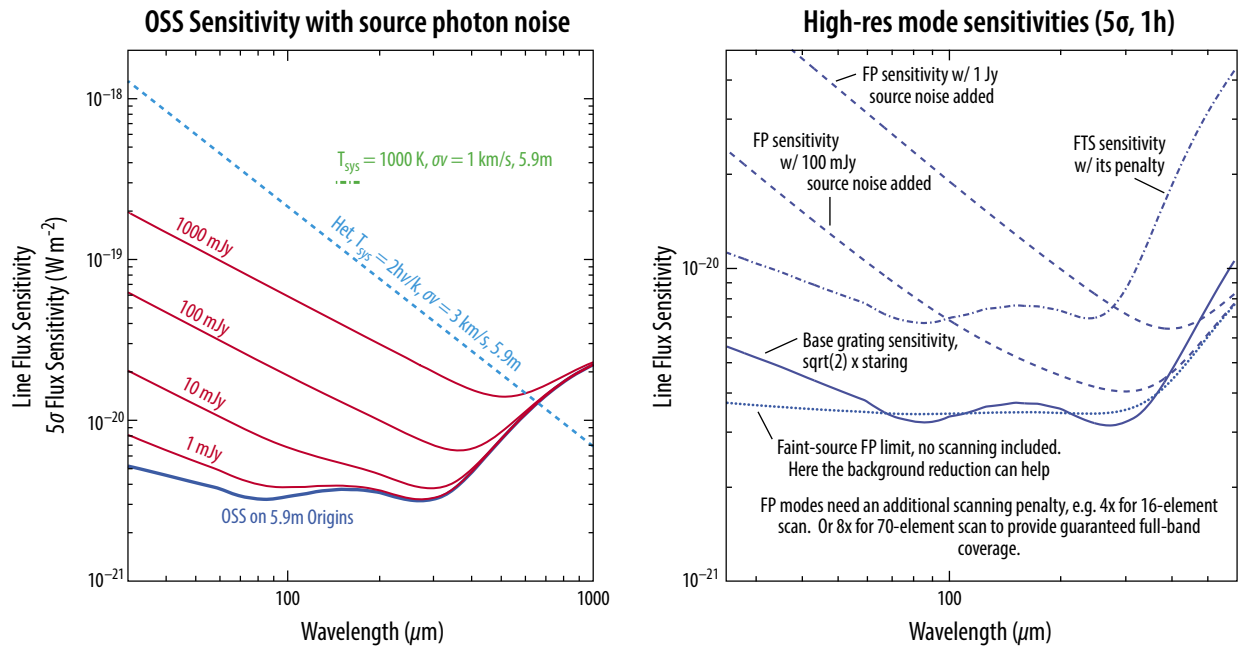

**Figure 3-10:** Sensitivity of OSS for bright sources (left) and in the high-resolution modes (right). The etalon (FP) curves refer to the required scanning time to be incorporated

Further applications of the sensitivity model are shown in Figure 3-10. Bright sources add shot noise and degrade the sensitivity. For the high-resolution modes, the estimated sensitivities include the additional signal loss (52%) and loading from the 4.5-K FTS optics, and another 49% transmission through the etalon, accounting for all the ohmic loss in the mirrors.

## Alignment, Integration, Testing, and Calibration

**Optical Alignment:** The grating modules can be aligned readily either by the vendor or by instrument team members. A good example of this approach is the aluminum grating spectrometer modules of the *Spitzer* IRS, which had tighter tolerances due to the shorter wavelengths. The slit is naturally imaged to the focal plane, and in fact, the slit can be moved by several beams with respect to the rest of the optics to obtain a desired slit position relative to a specific positioning of the grating box. This will be an important aspect in the alignment of the six grating modules with one another, and with the interferometer. This full alignment of the grating modules with one another will require a telescope simulator which can deliver a point-source-like converging beam to the OSS.

The OSS requires that the grating modules be positioned so that their slits overlap (field alignment), and that they all have their pupil aligned with the telescope pupil (pupil alignment). The first consists of simply moving the grating module (or, for small adjustments, moving the slit openings) so that the slits align in position, using a point-source simulator at various field positions. The second requires tipping/tilting the grating modules about the slit positions so that they point to a common pupil in the simulator. These tests will have to be performed with a cryogenic mounting of the grating modules.

Alignment of the interferometer begins with a self-alignment ensuring that the collimated beam remains centered as it traverses the mirrors and retro-reflectors, and that the two sides of the interferometer recombine with good overlap. This should be possible to do warm if good practices are used in the machining and assembly of the FTS bench (uniform materials, thermal stress relieving). Aligning the interferometer with the grating system requires positioning the interferometer insertion mirrors to provide an extremely well-collimated beam to permit the etalon operation; here the distance along the optical axis from the telescope is important, and this may require a cryogenic test. The interferometer must then have its pupil aligned with the telescope pupil; this consists of tilting the interferometer about the insertion mirror. Finally, the re-injection mirror must then be adjusted to ensure the interferometer's pupil aligns with those of the gratings. The team expects precision machining of the interferometer insertion/re-injection mirror pair as a single carriage will effectively eliminate this alignment test.

With this full system aligned in the flight truss which is brought as a unit to the telescope, a full alignment check of all elements should not be required on the observatory; simply verifying pupil overlap (also known as pupil alignment) of the grating modules with the observatory should be sufficient.

**Test Facilities:** OSS will require cryogenic test facilities, but they will be straightforward and pose no particular technical challenges. The detector arrays and individual grating modules will require a refrigerator with 2-K cooling of volumes on order 40 x 40 x 50 cm, with cold fingers allowing the arrays to cool to 50 mK. This will verify basic sensitivity and stray-light rejection at the individual module level. Such systems are available commercially; examples include the closed-cycle dilution refrigerators made by BlueFors, Oxford, and Leiden. For the multi-spectrometer alignment, a larger cryogenic test is required, encompassing ~1.5 meters to accommodate the optical bench and at least three spectrometers simultaneously. However, for this test, the full structure only needs to be cooled to 20 K, as this is sufficient to incur all CTE-related alignment effects. Cold fingers are required at ~250 mK, to allow the detectors to operate in a high-background mode and verify alignment. Full sensitivity is not required. These tests can be carried out in one of the existing thermal-vac chambers at JPL or GSFC.

**Calibration:** Detector electrical properties are measured to at least 1% accuracy on the ground, initially with dark measurements. Next, the full end-to-end spectrometer modules are calibrated absolutely with absolute efficiency measurements (IV curves for bolometers, photon noise statistics if KIDs are used instead) under a range of optical loading conditions in a dedicated low-background spectrometer testbed that can house each of the grating modules. Accurate optical loading with a slit-filling load is readily provided by an actively-controlled cryogenic black body module and this method is expected to provide end-to-end instrument efficiency to better than 2%, based on past experience with similar ground-based spectrometers, and with the SPIRE detector program.

As for *Herschel*, radiometric scientific calibration is ultimately tied to astronomical sources. Given its experience with *Herschel*, the team does not believe an on-board calibration source is needed, and therefore the baseline OSS design does not include one. However, the OSS team will further study this question in Phase A.

A key aspect unique to the grating spectrometer is the need for accurate measurement of the spectral response of every detector. This may reveal low-level grating ghosts and other peculiarities that do not impact sensitivity but could be important in deep observations of sources with a large spectral dynamic range (very bright lines and faint lines). OSS will measure spectral profiles with a dynamic range of at least 3000, and a resolving power of at least 20,000 x (100 $\mu/\lambda$) using a long-path FTS, with the spectral scale anchored by spectral standards such as absorption-cell measurements and a far-IR laser. Once on orbit, the FTS on board OSS is a benefit, as it will provide long-term tracking of the grating system spectral response in the unlikely event there is a shift in the spectral response.

**Heritage and Maturity**

The heritage of the OSS detector system, and its path to flight readiness, is detailed in the *Origins* Technology Development Plan. This section details heritage of the other OSS elements, including the gratings, Fourier transform interferometer, and etalon. The grating modules are straightforward bolted aluminum systems using diamond-machined mirrors and gratings. They have no moving parts. Similar approaches have been employed on the *Spitzer* infrared spectrograph (Figure 3-11 (left)) at even shorter (more demanding) wavelengths.

FTSs have been used at these wavelengths in space in multiple instruments. For example, *Herschel/* SPIRE used an FTS (Figure 3-11 (right)). SPIRE also used multiple sets of rooftop mirrors on the moving stages to provide pathlength amplification.

Scanning Fabry-Perot etalons have been used successfully on ESA's Infrared Space Observatory (ISO) mission; the short-wavelength spectrometer (SWS) and the long-wavelength spectrometer (LWS) on ISO had scanning etalons that were operated in front of the diffraction gratings. These etalons had higher finesse (F = cavity quality factor Q~200 for ISO), and operated at shorter wavelengths than the OSS system, for which F=70, and thus they had much more demanding requirements on mirror flatness and parallelism. OSS does require a larger clear aperture than ISO, but this is comparable to that used on the ground-based submillimeter instrument SPIFI (Bradford *et al.*, 2002, 2003; Oberst *et al.*, 2011). Finally, new high-resolution far-IR etalons are now being built for the HIRMES instrument on SOFIA (Figure 3-12).

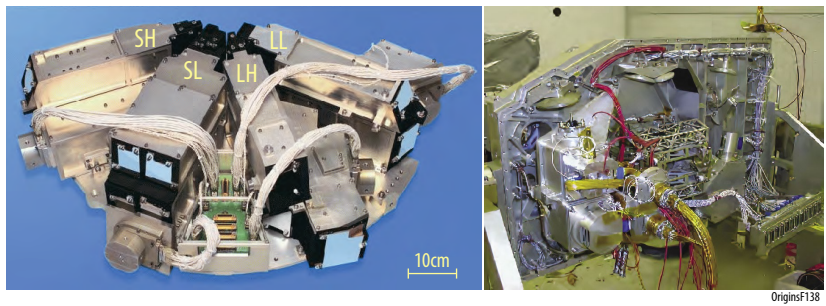

**Figure 3-11:** Previous missions provide heritage for the basic approach of the OSS elements. The bolted aluminum OSS grating spectrometer uses the same approach as the Spitzer infrared spectrograph from Houck et al., 2004. Its four wideband modules are shown at left. The Fourier-transform module path-folding approach and mechanism use the same approach as in the Herschel SPIRE instrument shown at right (Griffin et al., 2010).

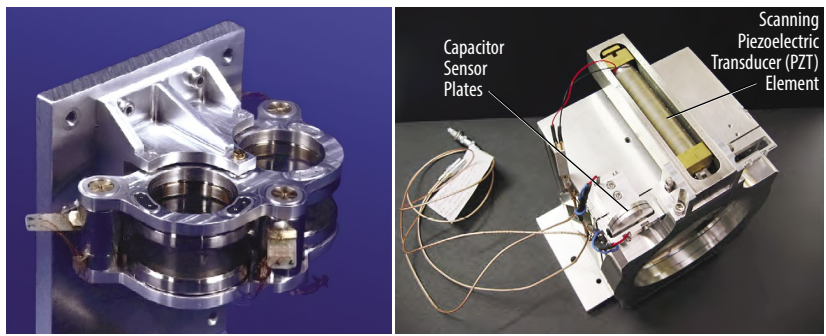

**Figure 3-12:** Heritage of scanning etalons for far-IR astrophysics. Left shows the dual etalon used in the Infrared Space Observatory (ISO) (credit Max Planck Institute for Extraterrestrial Physics, Garching).

**Footnote:** [http://www.mpe.mpg.de/ir/ISO/LesHouches98/tutorial/SWS/sws_image_fp.html]. Right shows an etalon developed at Cornell for the HIRMES instrument on SOFIA. It has demonstrated R=100,000 at 112 μm, using an order of 2,000 and a cavity finesse of 50. OSS will use a similar finesse, but larger cavity spacing.

**Enabling Technology:** The only aspect of OSS that requires technology development is the detector system. The detector technology options, and a plan for development is described in detail in the *Origins* Technology Development Plan and is briefly described in Section 2.3.

## 3.2 Mid-Infrared Spectrometer (Camera) transit spectrometer (MISC-T)

MISC-T provides spectroscopy simultaneously over 2.8 to 20 μm with exquisite stability and precision (<5 ppm between 2.8 to 10 μm, <20 ppm 11 to 20 μm). An additional MISC camera and-spectrometer is studied and described in Appendix D.2 as an upscope option but, for historical reasons, MISC remains as the instrument acronym. The science drivers for MISC-T are captured by the exoplanet case, Are we alone? (Section 1.3), and are articulated in the STM (Section 1.4). Table 3-5 summarizes basic MISC-T measurement capabilities.

**Table 3-5:** The MISC-T instrument fact sheet describing the parameters of the baseline instrument design

| Parameter | MISC Transit Spectrometer |
|---|---|
| Observing modes | MISC Ultra Stable Spectroscopy |
| Spectral Range | 2.8 − 20μm (MISC-T-S; 2.8-5.5μm, MISC-T-M; 5.5-11μm, MISC-T-L; 11-20μm ) |
| Resolving power | R=50 − 100 in 2.8 − 5.5μm (MISC-T-S) R=50 − 100 in 5.5 − 11μm (MISC-T-M) R=165 − 295 in 11 −20μm (MISC-T-L) |
| Angular resolution | Cannot attain spatially resolved information within the field of view |
| Field of View | Determined by the field stop size 3."0 in radius (MISC-T-S) 3."0 in radius (MISC-T-M) 2."0 in radius (MISC-T-L) |
| Detectors | A 2kx2k HgCdTe detector array (30K) for MISC-T-S A 2kx2k HgCdTe detector array (30K) for MISC-T-M A 2kx2k Si:As detector array with a calibration source for MISC-T-L A 256x256 HgCdTe detector array (30K) for MISC-T Tip-Tilt sensor |
| Saturation limit | 29.8 Jy at 3.3μm 27.5 Jy at 6.3μm 4.4 Jy at 14μm calculated for the shortest readout time; assuming partial readout, 10μsec per pixel per read, two reads per pixel to sample up the ramp |

### MISC-T Operation Principle

The MISC-T provides the highest ever spectro-photometric relative stability for observations at 2.8 - 20 μm. To reach this performance requires system optimization. Figure 3-13 shows an error budget allocation for stability. Detectors have the largest allocation. System optimization involves not only the MISC-T optical design and detector/readout electronics choices, but also observatory-level factors, system issues such as the pointing performance and the nature and variability of the *Origins* telescope aberrations. Various instrument design concepts were considered during the present study. The instrument design will be finalized in *Origins* Phase A, but for proof-of-concept and costing purposes our team designed MISC-T with a novel densified pupil optical design and HgCdTe and Si:As detector arrays. The densified pupil design divides up the light from a point-source in the instrument pupil plane up into a number of individual spectrometer optical channels and is described in more detail in Matsuo, *et al.* (2016). This design is relatively insensitive to minor line-of-sight pointing drifts and telescope aberrations, and the detectors do not require a sub-Kelvin refrigerator.

It does have the disadvantage of reduced sensitivity to faint sources compared to the traditional spectrometer design, but for this science program with the relatively bright host star targets spectro-photometric stability is more important than faint-source sensitivity. In order to demonstrate the potential of this novel optical design, a testbed system has been constructed using the densified pupil design and a 1024x1024 Si:As IBC array. Performance testing has just started and so far the design has met all of its design requirements (T. Matsuo, private communication).

Note that an alternative instrument design concept was considered in which TES detectors were fed by a more conventional spectrometer, with Winston cones used to concentrate the rather poor image at these short wavelengths. As this design concept has not been analyzed as thoroughly for overall system performance, especially with respect to errors introduced by telescope image quality and pointing variation, it was not chosen as the baseline, but further work on this alternate concept will be carried into the Phase A Concept Study.

Origins Transit Stability Components (in ppm unless otherwise noted) for a 9.8 K magnitude M-star, 85 transits of 4 hours duration each.

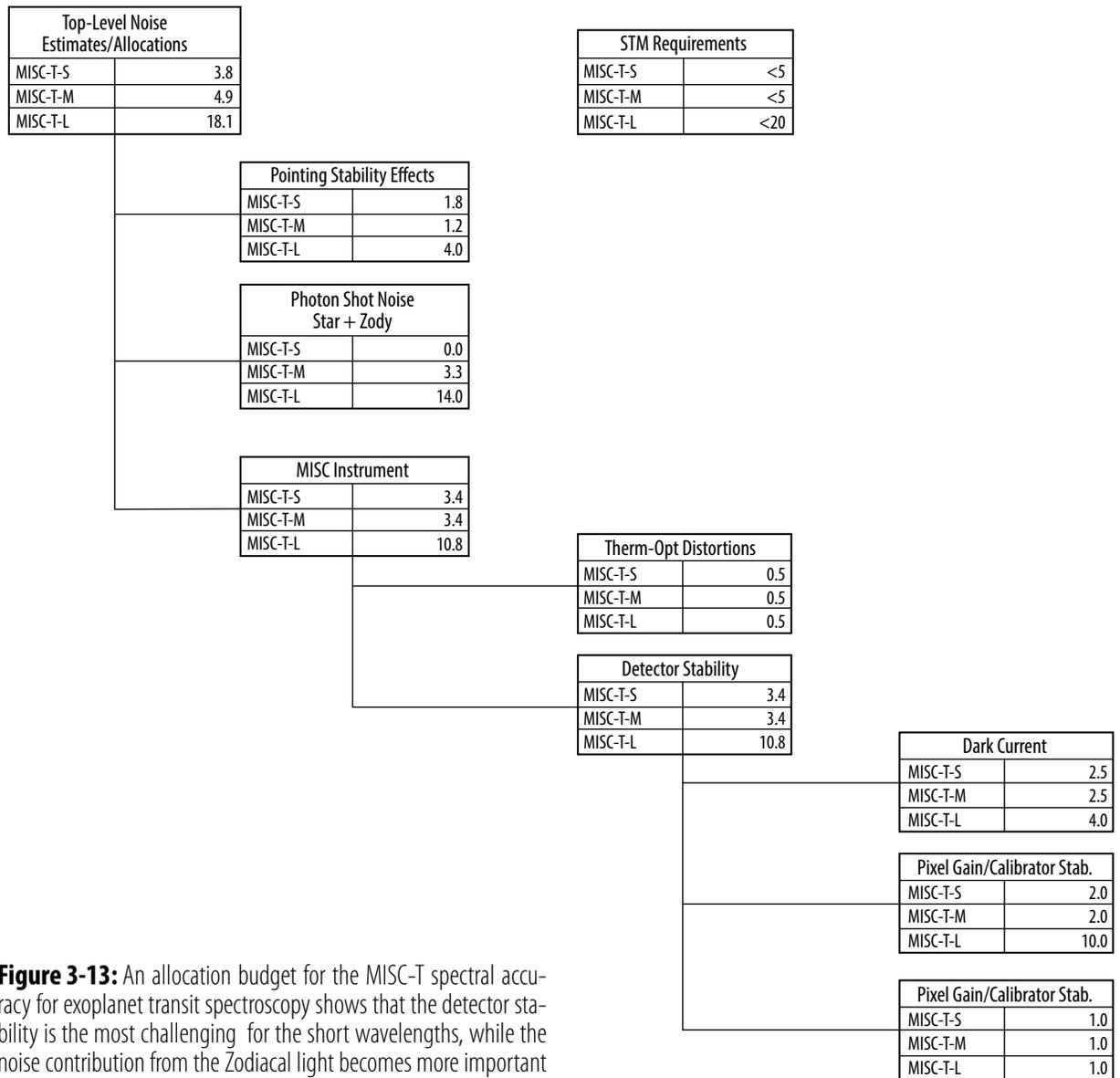

| Top-Level Noise Estimates/Allocations | |
|---|---|
| MISC-T-S | 3.8 |
| MISC-T-M | 4.9 |
| MISC-T-L | 18.1 |

| STM Requirements | |
|---|---|
| MISC-T-S | <5 |
| MISC-T-M | <5 |
| MISC-T-L | <20 |

| Pointing Stability Effects | |
|---|---|
| MISC-T-S | 1.8 |
| MISC-T-M | 1.2 |
| MISC-T-L | 4.0 |

| Photon Shot Noise Star + Zody | |
|---|---|
| MISC-T-S | 0.0 |
| MISC-T-M | 3.3 |
| MISC-T-L | 14.0 |

| MISC Instrument | |
|---|---|
| MISC-T-S | 3.4 |
| MISC-T-M | 3.4 |
| MISC-T-L | 10.8 |

| Therm-Opt Distortions | |
|---|---|
| MISC-T-S | 0.5 |
| MISC-T-M | 0.5 |
| MISC-T-L | 0.5 |

| Detector Stability | |
|---|---|
| MISC-T-S | 3.4 |
| MISC-T-M | 3.4 |
| MISC-T-L | 10.8 |

| Dark Current | |
|---|---|
| MISC-T-S | 2.5 |
| MISC-T-M | 2.5 |
| MISC-T-L | 4.0 |

| Pixel Gain/Calibrator Stab. | |
|---|---|
| MISC-T-S | 2.0 |
| MISC-T-M | 2.0 |
| MISC-T-L | 10.0 |

| Pixel Gain/Calibrator Stab. | |
|---|---|
| MISC-T-S | 1.0 |
| MISC-T-M | 1.0 |
| MISC-T-L | 1.0 |

**Figure 3-13:** An allocation budget for the MISC-T spectral accuracy for exoplanet transit spectroscopy shows that the detector stability is the most challenging for the short wavelengths, while the noise contribution from the Zodiacal light becomes more important in the longest channel.

OriginsF140

Densified pupil spectroscopy is a new optical system approach for transit spectroscopy (Matsuo *et al.*, 2016, 2018) that will greatly improve spectro-photometric performance in the presence of optical disturbances. With pupil desensization, the MISC-T science image will not be disturbed by modest telescope pointing jitter (*e.g.*, 10 mas) or telescope aberrations. The large number of science pixels minimizes the effects of intra- and inter-pixel sensitivity variations. Reference pixels also provide potential detector baseline common-mode gain fluctuation calibration using a cold photon shield mask. As part of the *Origins* technology development plan, the team will refine an optimal pixel-to-pixel calibration technique using an ultra-stable internal calibration source. The MISC-T block diagram is shown in Figure 3-14.

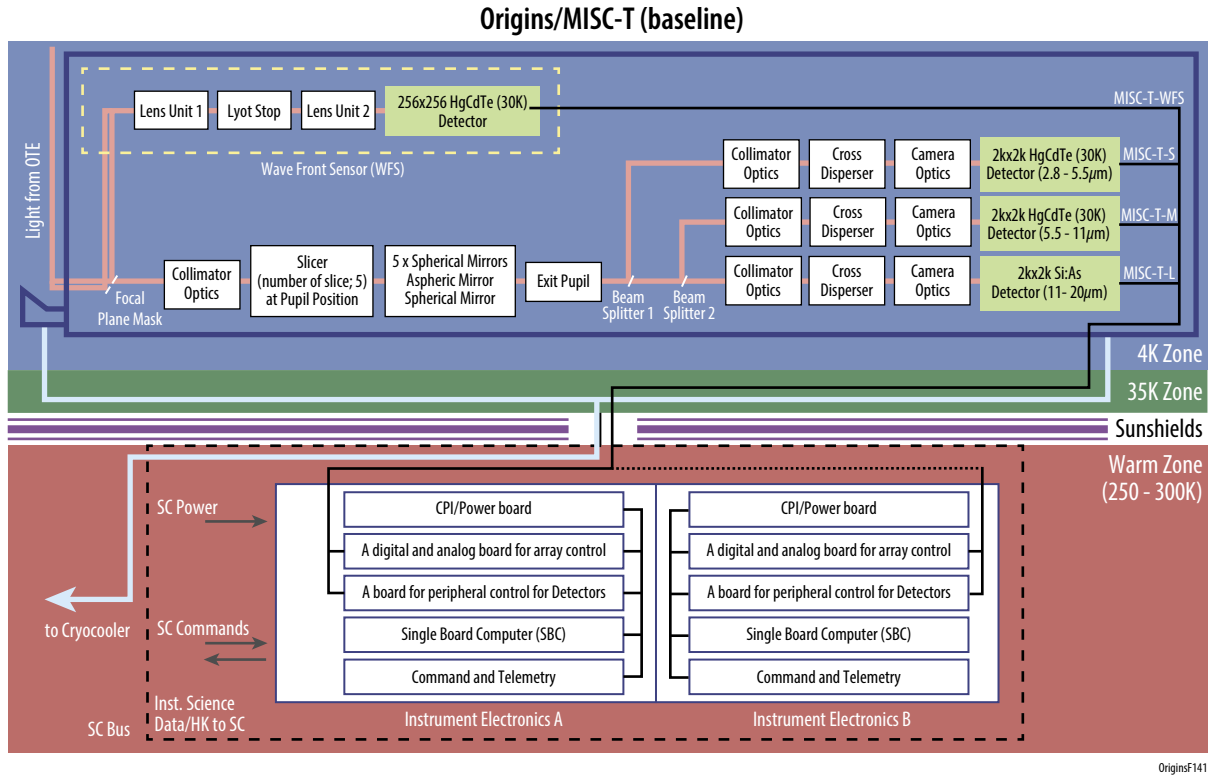

**Figure 3-14:** The MISC-T block diagram showing the optical paths, detector focal planes, electronics, and temperature regimes.

The MISC-T instrument is primarily dedicated to pointed observations to detect bio-signatures in habitable worlds in primary and secondary exoplanet transits. No simultaneous operation between MISC-T and other instruments is expected, therefore the high operating temperature for the HgCdTe detectors will not be a source of noise to the other instruments. MISC-T has three channels (T-S, T-M, T-L) that share the same FOV by means of beam splitters, and all channels are operated simultaneously to cover the full spectral range from 2.8–20 µm at once. A Lyot-coronagraph-based Tip-Tilt sensor (TTS) located in the instrument fore-optics uses the light reflected by a field stop, which corresponds to 0.3% of the light from the target, to send fine pointing information to the FSM in the telescope. The TTS measures the long-term drift using the point source image formed on the focal plane through a Lyot stop and feeds this information back to the observatory to mitigate the long-term drift down to 2 mas over a timescale of a few hours.

**Optical Design**

The MISC-T optical design is based on densified pupil spectroscopy (Matsuo *et al.*, 2016). Conventional spectrographs (*e.g.*, long slit) use focal plane division: light from multiple positions in the focal plane is dispersed into multiple spectra. In contrast, the densified pupil spectrograph uses pupil-plane division and produces spectra from multiple sub-apertures of the pupil plane. As mentioned in Section 3.4, it should be emphasized that in spite of the novel arrangement of optics in the densified pupil design, the optical components themselves are completely ordinary, and as a result we do not feel that this spectrometer design required mention in the *Origins* technology development plan. The performance of this design is currently being demonstrated in tests at NASA Ames Research Center. The densified pupil spectrograph is composed of a pupil division/densification system followed by a normal spectrograph system (**Figures 3-15** and **3-16**).

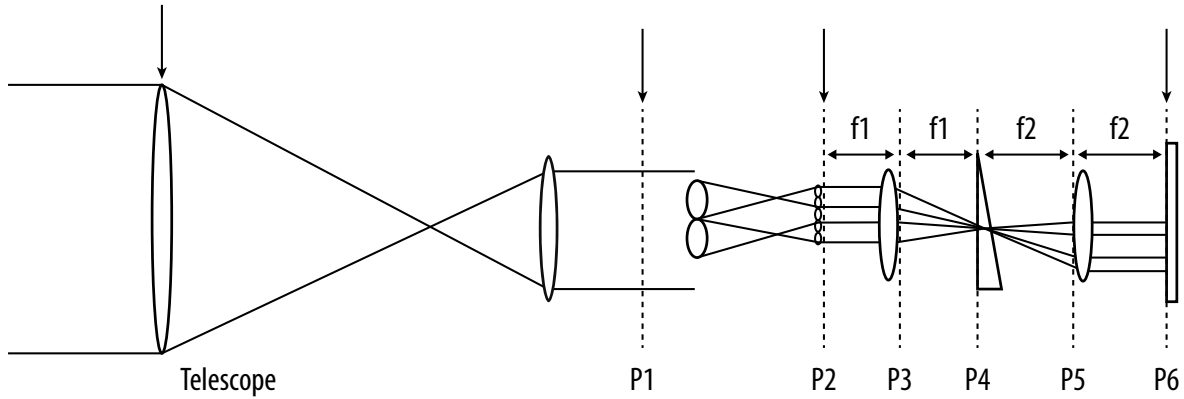

**Figure 3-15:** A schematic diagram of the densified pupil concept. Light from the telescope is collimated into a conjugate pupil plane P1. This plane is divided into sub-apertures (two of which are shown in this figure) and the light is then further focused by lenslets or mirrors into an array of optical paths, each with its own spectral dispersion path, in a conventional spectrometer configuration (P2-P5) with the spectra then imaged onto the detector (P6).

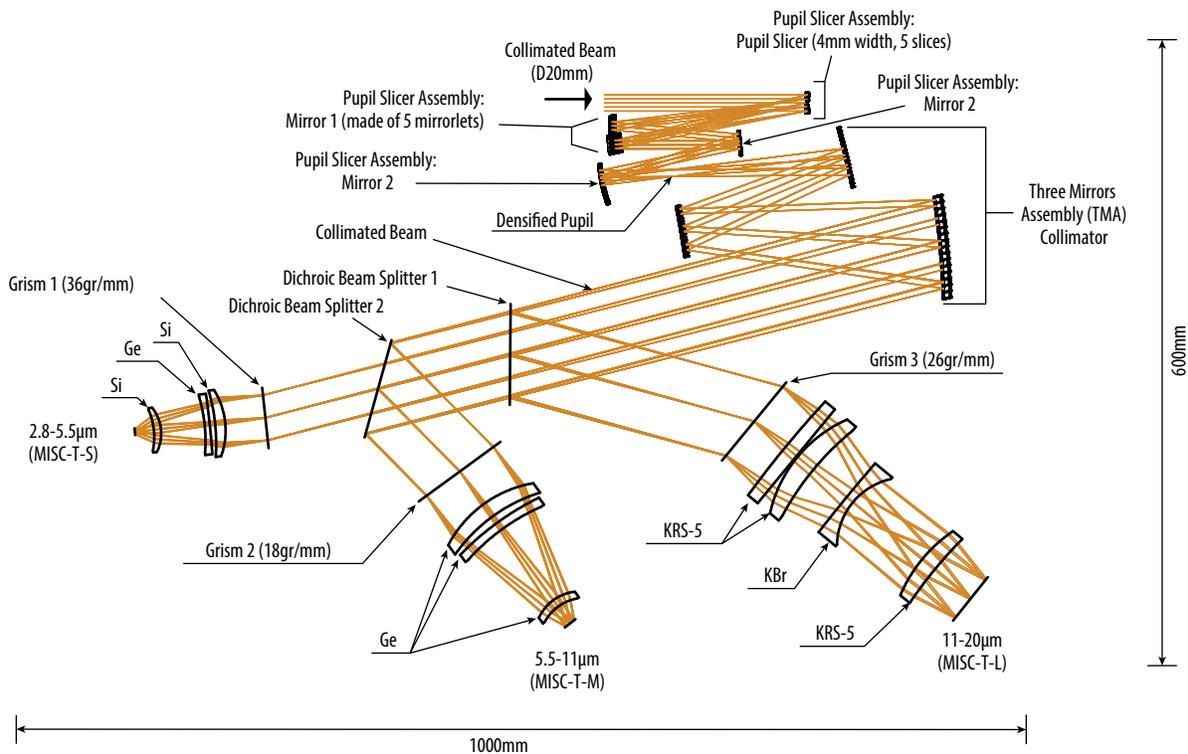

**Figure 3-16:** The MISC-T optical design. The three spectrometers are fed by a collimated beam that emerges from the tip-tilt sensor focal plane mask optics. The dimensions of the instrument in mm are indicated.

The pupil densification plays a key role in performing highly stable spectroscopy. Because of the pupil division into five slices in the case of MISC-T, and densification of the divided pupil, each of the densified sub-pupils acts as a point source and each corresponding beam is collimated by a triple mirror assembly (TMA) collimator as shown in Figure 3-16. The pupil is sliced into five segments, as shown in Figure 3-17. The first dichroic beam splitter reflects the light of 11–20 µm, which is fed to the spectrograph of T-L, and transmits the light of <11 μm. The second dichroic beam splitter reflects the light of 5.5—11 μm, which is fed to the spectrograph of T-M, and transmits the light of <5.5 μm, which is fed to the spectrograph of T-S. Field stops with radii of 3."0, 3."0, and 2."0 are installed in the spectrograph T-S, T-M, and T-L channels, respectively, to reduce the zodiacal background photons. The field stop aperture size determines the field-of- view (FOV) size, and no spatially-resolved information within the FOV is obtained. Each spectrograph employs a transmission grating-prism (grism) to disperse the light in one dimension as a function of the wavelength on a detector plane. Each of the five spectra shares 271 pixels in the dispersion direction by 27 pixels in the cross-dispersion direction for T-S, 551 pixels in the dispersion direction by 54 pixels in the cross-dispersion direction for T-M, and 1586 pixels in the dispersion direction by 65 pixels in the cross-dispersion direction for T-L. The layout of these spectra onto the detector arrays is shown in Figure 3-18. The linear dispersion (Dl) is set to 0.054 μm, 0.108 μm, and 0.074 μm for T-S, T-M, and T-L, respectively.

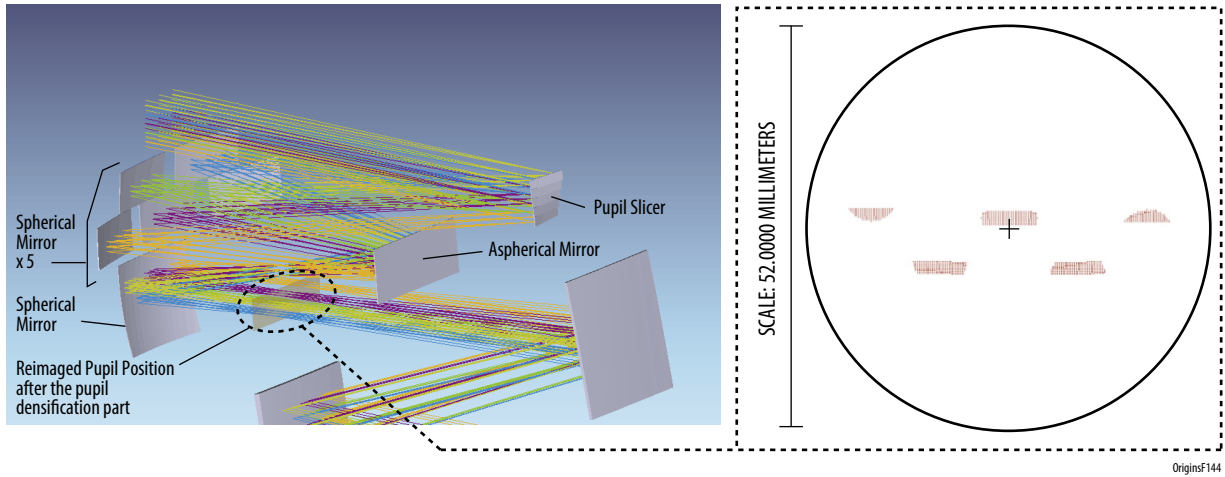

**Figure 3-17:** The telescope pupil is reimaged and sliced as shown at left, with the slices separated as shown in the image on the right.

The TTS corrects any long- term drift in the telescope pointing. The team conducted a trade study between a Lyot- Coronagraph-based field stop with a dedicated TTS array vs. a guider camera with a dichroic beam splitter to extract 2-2.8 μm light of the target. The study showed a clear advantage in the Lyot-Coronagraph method in terms of MISC-T optical throughput, as well as in simplicity and cost. The TTS uses only 0.3% of the light from the target host star, which is reflected by the focal plane mask in the fore-optics; the majority (~98%) of the light from the target passes through a ~5" aperture in the focal plane mask and is lead to the T-S/M/L spectrometer channels. The long-term drift is measured by examining the position of the PSF centroid of the image produced on the detector plane of the TTS through the Lyot stop. Assuming that the actual Lyot stop size is the same size as the *Origins* pupil and has an accuracy of 1%, the TTS will work for M-type target stars brighter than K=10 mag. The TTS feeds back this information to the observatory and the long-term drift can be mitigated down to 2 mas for M-type stars brighter than K=10 mag. (Also see Section 2.8.2) Figure 3-19 shows the MISC-T TTS schematic.

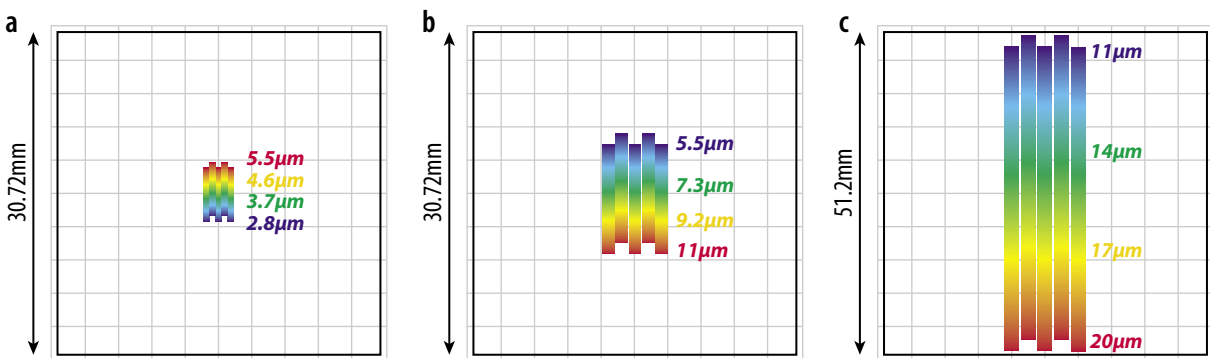

**Figure 3-18:** The spectra from the five pupil slices are overlaid on the short, mid, and long focal planes as shown in panels a-c, respectively.

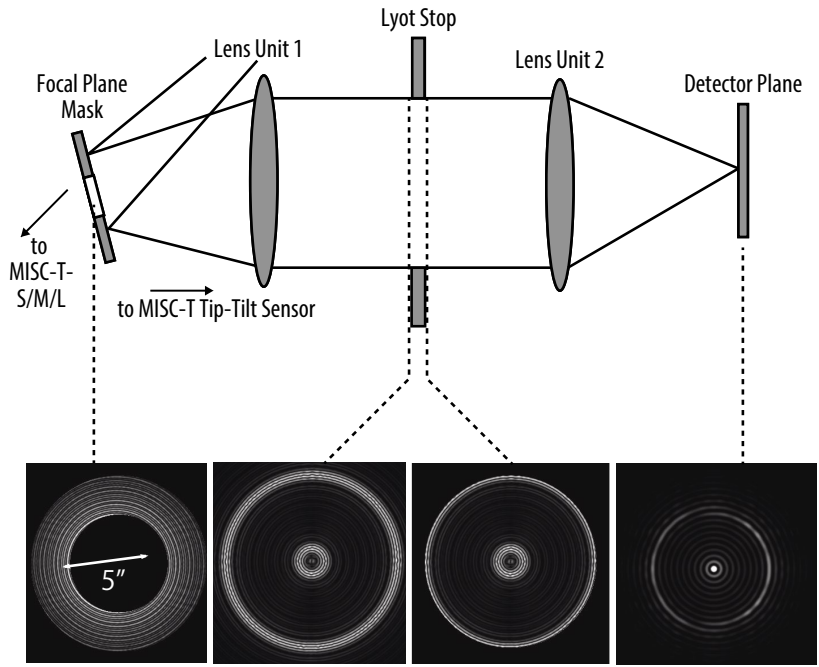

**Figure 3-19:** A schematic view of the MISC-T TTS showing the results of a simulation of a PSF image A) at the Lyot-Coronagraph focal plane mask, B) just before the Lyot stop, C) just after the Lyot stop, and D) at the detector plane.

## MISC-T Observing Modes

MISC-T has a single observing mode: AOT01 (Table 3-6). MISC-T-S, T-M, and T-L share the same FOV by means of beam splitters and are simultaneously operated to produce a complete 2.8–20 μm spectrum at once.

**Table 3-6:** The MISC-T observing modes (baseline) consist of only one mode with data rates that vary depending on the brightness of the target star.

| AOT | Mode | Data rate |
|---|---|---|
| AOT01 | MISC-T Transit Spectroscopy | 1.59Mbps(max)—3.73Mbps(nominal) [MISC-T]<br>0.37Mbps [MISC-T TTS] |

Note 1: Max data rate is calculated for the shortest exposure time $t_{exp}$=4s
Note 2: MISC-T TTS assumes to read and downlink 32 x 32 pixels around the PSF centroid at 20Hz

## MISC-T Detectors

The MISC-T's baseline employs two types of detectors: 1) HgCdTe that is bonded to a Si readout multiplexer, and 2) Si Blocked Impurity Band (BIB) design that is bonded to a Si readout multiplexer. Each of the MISC-T-S and T-M optical paths contains a 2k x 2k HgCdTe detector array operated at 30 K and the MISC-T TTS employs a 256 x 256 HgCdTe detector array operated at ~30 K. HgCdTe detectors developed for the NEOCam mission have successfully demonstrated performance out to 10 μm (Dorn *et al.*, 2016). The MISC-T-L optical path contains a 2k x 2k Si:As detector array bonded to a Si readout multiplexer operating at ~8 K to provide good detective quantum efficiency over the 10.5-20 μm wavelength range. Ultra-stable internal calibration sources are employed to help achieve the required detector stability. These sources are composed of a combination of an infrared source black-body combined with a shorter-wavelength photocell detector used for temperature control. The photocell measures the source output in its blackbody Wien limit, allowing very sensitive feedback of the calibration source temperature, while the infrared wavelengths used for the infrared detector calibration are in the blackbody Rayleigh-Jeans limit.

## MISC-T Read-out Electronics

Amplification of the signals from the detectors is a straightforward reuse of similar technology from previous space applications of HgCdTe and Si:As BIB detectors, including *Spitzer*, JWST, WISE and Near Earth Object Camera (NEOCam). This method employs dedicated satellite chips located in close proximity to the detectors operating at cryogenic temperatures. The cable run from the cold satellite chips near the detectors and the warm electronics in the *Origins* design is shorter than it is in JWST, so the difficulties that were experienced with the long cable runs in the latter mission should be much reduced. MISC-T incorporates flexible readout patterns and a variety of readout strategies, including double-correlated and Fowler sampling techniques. Due to the extreme detector stability requirements, the readout electronics will need to be carefully designed and temperature-stabilized.

## MISC-T Data Rates

MISC-T data rates assume all three MISC-T spectrometer channels are operating simultaneously (Table 3-6).

## MISC-T Thermal and Mechanical Design and Resource Requirements

MISC-T instrument components are located in two thermal zones. The instrument optics and detectors are located in a cold zone at the same ~4.5 K temperature as the telescope. The detectors have a weak thermal link to this cold temperature so they can be heated slightly and held at a stable operating temperature. This operating temperature is ~30 K for the HgCdTe detectors and ~7 K for the Si:As detector. Baffling in the cold instrument prevents IR light leakage from the warmer detectors from reaching the Si:As detector. This was successfully demonstrated by WISE, which had 32 K detectors and 8 K detectors located in the same instrument. The drive electronics that handle instrument data collection and communication with the spacecraft are located in the ~300 K warm zone.

The MISC-T 3D solid model is shown in Figure 3-20. To reduce MISC-T instrument mass, Beryllium was assumed as a baseline material for the mirrors and mirror support structures, as well as the base plate. MISC-T does not have any moving parts or mechanisms. Table 3-7 lists MISC-T resource requirements (volume, mass, power, data rate).

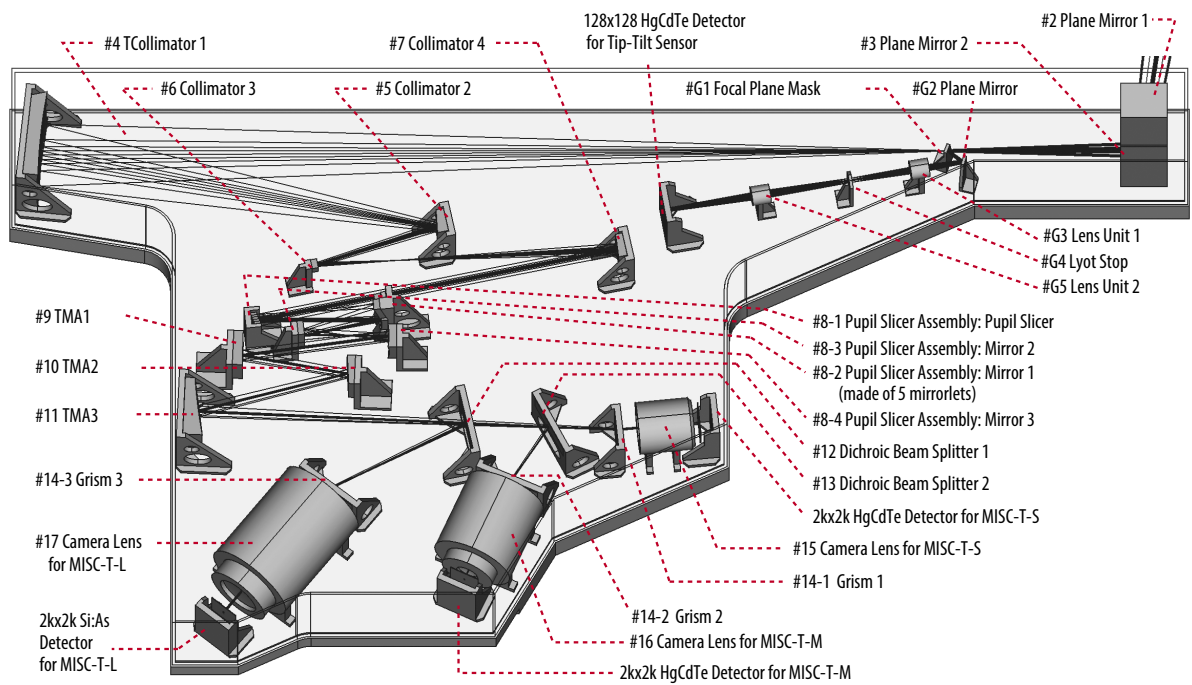

**Figure 3-20:** A 3D solid model of the MISC-T instrument indicating the optical path of the light within the instrument and the location of the mechanical components on the optical baseplate.

**Table 3-7:** MISC-T instrument resource requirements are reasonable.

| Instrument Resource Type | Resource and Units | | MISC-T (Baseline) | | Total |
|---|---|---|---|---|---|
| | | | MISC-T | TTS | |
| Information Related to the Electrical Subsystem | Total Max Data Rate | Mbps | 1.49 | 0.37 | 1.86 |
| | Total Average Data Rate | Mbps | 3.73 | 0.37 | 4.10 |
| Information Related to the Mechanical Model | Volume (Cold Component) | m³ (m x m x m) | 0.15 (2.0 x 0.25 x 0.30) Foreoptics + Tip-Tilt Sensor 0.12 (1.0 x 0.8 x 0.15) Collimator + Pupil Mapping and Densification Part + Spectrographs | | 0.27 |
| Information Related to the Thermal Model | Mass (Cold Component) | kg | 68.87 | | 68.87 |
| | Mass (Warm Component) | kg | 16.00 | | 15.00 |
| | Mass (Other; e.g., Harnessing) | kg | 16.36 | | 16.36 |
| | Total Mass | kg | 101.23 | | 101.23 |
| | Total Peak Power (Cold Part) | W | 0.159 | 0.003 | 0.162 |
| | Total Peak Power (Warm Part) | W | 10 | 3 | 13 |
| | Total Average Power (Cold Part) | W | 0.009 | 0.003 | 0.012 |
| | Total Average Power (Warm Part) | W | 10 | 3 | 13 |
| | Total Standby Power (Cold Part) | W | 0.009 | 0.003 | 0.012 |
| | Total Standby Power (Warm Part) | W | 10 | 3 | 13 |
| | Average Power Dissipation (Detectors) | W | 0.00804 | 0.00268 | 0.01072 |
| | Average Power Dissipation (Heater) | W | 0.009 | 0.003 | 0.012 |

## MISC-T Instrument Control

The MISC-T instrument includes redundant dual-string warm electronics boxes (WEB A and WEB B for MISC-T). Each electronics box contains a single board computer (SBC) and three boards: 1) a CPI/Power board, 2) a digital and analog board for array control (two 2kx2k HgCdTe arrays for T-S and T-M, one 2kx2k Si:As array for T-L, one 256x256 HgCdTe array for T TTS), and 3) a board for peripheral control of the detectors (two 2kx2k HgCdTe arrays for T-S and T-M, one 2kx2k Si:As array for T-L, one 256x256 HgCdTe array for T TTS). The SBC calculates the centroid position of the PSF on the TTS detector array and sends this information back to the observatory. The FSM in the observatory corrects for the pointing accuracy and mitigates the long-term drift down to a few mas over a timescale of a few hours. The detector biasing and readout electronics will need to be very stable so that the detectors themselves can achieve their relative photometric stability requirements, which will necessitate careful temperature control of the electronics boxes.

## MISC-T Risk Management Approach

As an instrument on a NASA Class A mission, the MISC-T instrument has a fully redundant, dual-string, cross-strapped design. The MISC-T instrument has a warm electronics box each with two redundant sets of warm electronics, as shown in the block diagram (Figure 3-21). Within an electronics box the single board computers and the rest of the warm electronics are cross-strapped as shown in Figure 3-21. Although the focal plane arrays are not dual redundant, their multiplexer readouts are arranged so that sections of the arrays fail gracefully, as was done for the JWST instrument arrays.

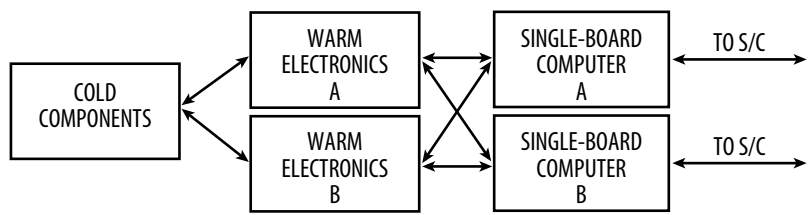

**Figure 3-21:** The cross-strapping design used in the MISC-T modules' warm electronics provides dual-string fault tolerance.

## MISC-T Predicted Performance

MISC-T performance is calculated based on the following assumptions:

**Throughput:** The throughput curves as a function of the wavelength are calculated for T-S, T-M, and T-L, taking account of the reflectance of mirrors (four mirrors for the telescope optics, three mirrors for fore-optics, three collimator mirrors, seven mirrors in the densified pupil spectrograph), the transmittance/reflectance of the dichroic beam splitters, grating efficiency, transmittance of the AR-coated lenses, detector quantum efficiency, and contamination/slit loss. Diffraction losses in the instrument were minimized with oversized optics (at least 12 l/D). The final throughput and detector quantum efficiency is plotted in Figure 3-22.

**Zodiacal Light:** In addition to the light from the host star, there will be additional infrared light coming from the Zodiacal background. As the field stops have to be relatively large (3.0 arc-sec radius for T-S and T-M, 2.0 arc-sec radius for T-L) to capture the majority of the poorly-focused short-wavelength light from the 30 μm diffraction-limited telescope, the shot noise from this extra light can be a significant source of signal stability errors at some wavelengths. The Glasse Zodiacal model (Glasse *et al.*, 2015) is employed for the low background case outside the ecliptic plane while the Wright Zodiacal model for the ecliptic plane (l=0) (Wright 1998) is employed for the high background case. The surface brightness of the zodiacal light, including scattered light and emissive components at representative wavelengths, is summarized in Table 3-8.

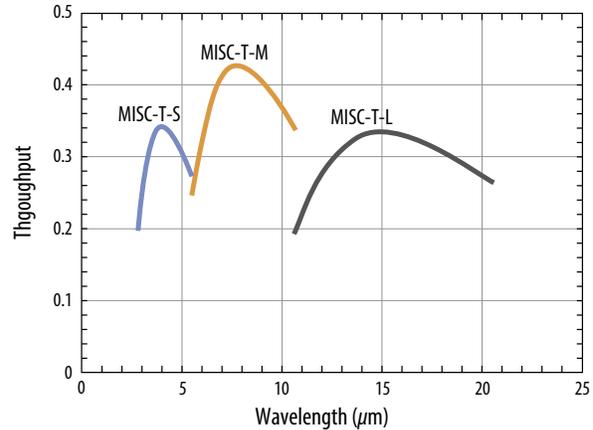

**Figure 3-22:** The throughput as a function of wavelength used for MISC-T performance calculations.

**Table 3-8:** The surface brightness of the zodiacal light models used for the MISC-T performance calculations.

| Wavelength (μm) | Low-background Case (MJy/sr) (Glasse model) | High-background Case (MJy/sr) (Wright model; ecliptic plane) |
|---|---|---|
| 3 | 0.1 | |
| 5 | 0.35 | 0.92 |
| 6 | 1 | |
| 8 | 3 | |
| 10 | 8.5 | 22.76 |
| 11 | 11 | |
| 14 | 14 | 46.58 |
| 20 | 17 | 68.46 |

**Pointing Jitter and Long-term Drift:** The major source of high-frequency pointing jitter is due to the cryocooler and is expected to be <50 mas, with a current best estimate of 9 mas RMS (Section 2.8.2). Low frequency pointing drift has a much stronger effect on spectral stability. The Lyot-Coronagraph- based Tip-Tilt sensor described above is used for measuring the long-term drift so that it can be compensated with the telescope FSM, if necessary. Together the performance is predicted to be 1 mas drift over a timescale of a few hours.

**Detector Performance:** A 2k x 2k HgCdTe detector array operated at 30 K is used in each of the MISC-T-S and T-M channels and a 2k x 2k Si:As detector array operated at ~7 K is used in MISC-T-L. The detector dark current is assumed to be 1.0 e-sec-1pixel-1 for the 2k × 2k HgCdTe detector and 0.2 e-sec-1pixel-1 for the 2k×2k Si:As detector.

Based on these assumptions, the team simulated the light curves of late-M type stars at different K-band magnitudes and calculated the spectro-photometeric accuracy that is realized at representative wavelengths for each type of late-M type star. The results calculated for 3 and 14 μm assuming 84 transits and for a spectral resolution of R=50 are shown in Figure 3-23. Noise sources due to photon shot noise, dark current, zodiacal light, jitter, and drift are included as shown. Aside from dark cur-rent, other detector and readout-electronics noise sources are not included. As an example, a bright star will have more integrated photons in 84 transits, so that the dark current noise contribution from the detectors will be relatively smaller. The flat contribution of the jitter and drift noise to the signal-to-noise over different host star brightnesses may seem odd at first, but in fact this term is a constant multiplicative factor over a range of stellar magnitudes.

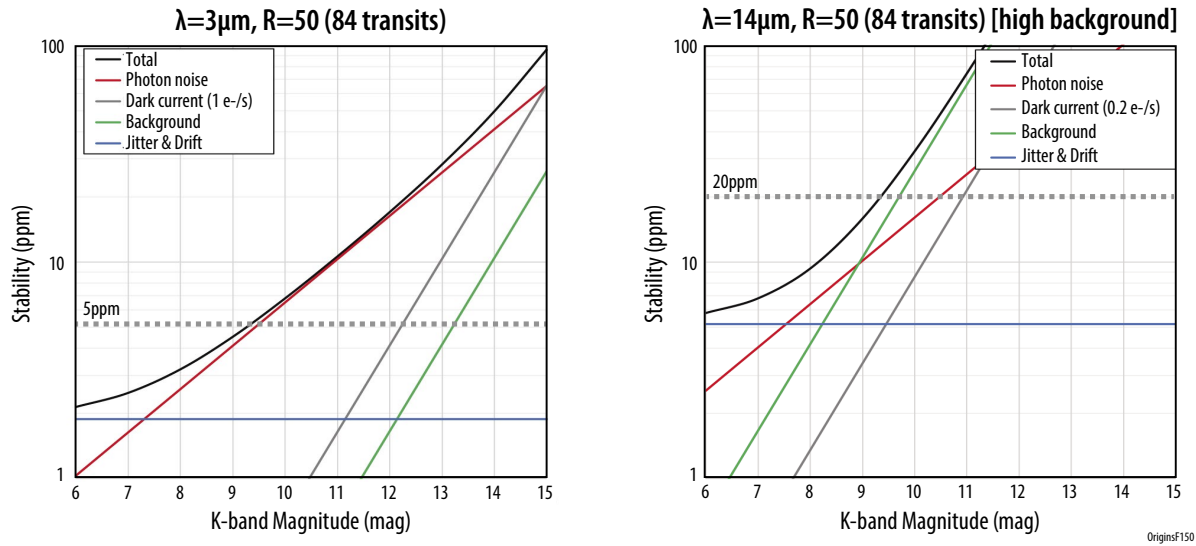

**Figure 3-23:** The current best estimate of the spectro-photometric stability achieved by MISC-T. Left: 3 μm with R=50 is below 5 ppm for stars brighter than magnitude 9.5. Right: 14 μm with R=50 (high Zodical background case; Wright 1998) is below 20 ppm for stars brighter than magnitude 9.3. See the text for further explanations.

**MISC-T Alignment, Integration, Test, and Calibration**

The MISC-T instrument will require its own unique facilities for testing and calibration at the instrument level before it is delivered for integration with the observatory. Aside from the normal vibration, acoustic, and EMI environmental testing, the MISC-T instrument will need a thermal-vacuum test chamber that can be cooled to liquid helium temperatures that is large enough to enclose the complete cold section of the MISC-T instrument and is IR light-tight enough that stray light within the MISC-T detectors' sensitivity wavelengths is no greater than the infrared background that would be expected on-orbit. This test chamber should allow the introduction of IR signals at the expected levels of the on-orbit target sources and the ability to verify the spectral response of the spectrometer. Finally, the test facility will require an ultra-stable light source that can verify the ~few ppm pixel-pixel detector stability requirements needed for the exoplanet transit measurements. This test facility will need to allow connection between the cold MISC-T instrument assembly in the test chamber and the warm MISC-T drive electronics over flight-like cables.

The warm MISC-T instrument drive electronics will not require any particular test facilities other than the usual test facilities used to conduct vibration, acoustic, EMI, and thermal vacuum tests. A spacecraft simulator will be required to test command and data transfers between the MISC-T and spacecraft electronics.

It will be necessary to develop specialized test hardware, but although the scope of this development will be significant, it is straightforward, with no new technology development required. MISC-T instrument I&T is expected to take 12 months, assuming all the external test support hardware and software is ready in time by the start of testing.

## MISC-T On-orbit Checkout and Calibration

In the beginning of the performance verification phase, the following items will be tested:

- [*PV-01*] checkout of the condition of all boards in the warm electronics boxes
- [*PV-02*] checkout of the condition of the detectors (*e.g.*, clock patterns for all observing modes)
- [*PV-03*] checkout of the condition of the internal calibration sources
- [*PV-04*] checkout of the optical alignment
- [*PV-05*] obtaining flat fielding images for each detector array
- [*PV-06*] on-orbit absolute wavelength and flux calibration of the MISC-T spectroscopic modes
  During the nominal observational phase, the following calibrations are updated as needed:
  - [*P1-01*] dark current measurements (before and after each pointed observation)
  - [*P1-02*] annealing the detector arrays to remove hot pixels (beginning of every MISC-T obs. campaign)
  - [*P1-03*] checkout the stability of sensitivity by observing identical stars (once every few months) SOFIA and JWST may provide a new list of standard targets for flux/wavelength calibration. A stable internal calibration source is used to check the stability of MISC-T.

## MISC-T Heritage and Maturity

MISC-T wavelength coverage and capabilities partly overlap with those of JWST/MIRI, Stratospheric Observatory for Infrared Astronomy (SOFIA) instruments: Faint Object infrared Camera for the SOFIA Telescope (FORCAST), Echelon-Cross-Echelle Spectrograph (EXES), Space Infrared Telescope for Cosmology and Astrophysics (SPICA, a mission under study for ESA/JAXA) instrument studies: SPICA Mid-infrared Instrument (SMI, current), Mid-infrared Camera and Spectrometer (MCS), SPICA Coronagraph Instrument (SCI), Tokyo Atacama Observatory (TAO) Multi-field Imager for gazing at the UnKnown Universe (MIMIZUKU), *Spitzer*/IRS, and Thirty Meter Telescope (TMT) Mid-Infrared Camera High-disperser & IFU spectrograph (MICHI), and NEO-Cam. An example of the long wavelength HgCdTe detectors that have been developed for the NEO-Cam mission is shown in Figure 3-24. A summary of the MISC-T technology heritage is detailed in Table 3-9.

**Table 3-9:** MISC-T component heritage is listed below.

| Description | Subsystem/ Component | Heritage |
|---|---|---|
| Be Mirrors and Structures (including base plate) | Component | NIRCam instrument on JWST |
| 2kx2k Si:As with a calibration source | Component | JWST/MIRI, SPICA/SMI |
| 2kx2k HgCdTe | Component | NEOCam |
| Beam Splitter, Short-wave Cut Filters (multi-layer intereference filters) | Component | AKARI/IRC, SPICA/MCS |
| Lyot Stop (for Tip-Tilt sensor) | Component | SPICA/SCI and others |

## Enabling Technology

Stable detectors are critical to MISC-T's overall performance (Figure 3-13). Detectors that meet the stability requirements do not exist today and development is necessary. The *Origins* technology development plan, summarized in section 2.3, outlines the strategy to select detectors that meet requirements from three technologies: HgCdTe and Si:As detectors and TES bolometers. On board calibration sources to monitor and correct stability performance are also part of the development approach.

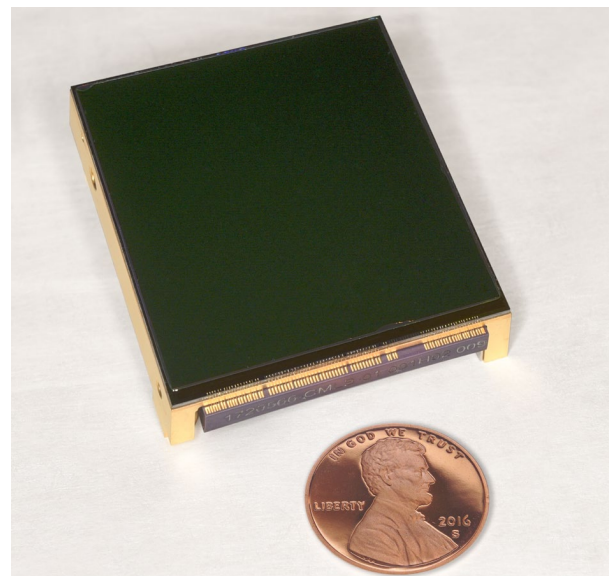

**Figure 3-24:** Long-wavelength HgCdTe detectors have been developed for NASA's NEOCam mission.

**Table 3-10:** FIP instrument capabilities

| Instrument/ Observing Mode | Wavelength Coverage (μm) | Angular Resolution | Field of View | Spectral Resolving Power (R=λ/Δλ) | Saturation Limits | Representative sensitivity 5σ in 1 hr |
|---|---|---|---|---|---|---|
| Pointed | 50 or 250 (selectable) | 50/250μm: 2.5" /12.5" | 50 μm: 3.6' x 2.5' 250 μm: 13.5' x 9' (109 x 73 pixels) | 3.3 | 50 μm: 1 Jy 250 μm: 5 Jy | 50/250 μm: 0.9/2.5 μJy confusion limits 50/250 μm: 120 nJy/ 1.1 mJy |
| Survey mapping | 50 or 250 (selectable) | 50/250μm: 2.5" /12.5" | 60" per second scan rate, with above FOVs | 3.3 | 50 μm: 1 Jy 250 μm: 5 Jy | Same as above, time to reach confusion limit: 50 μm: 1.9 hours 250 μm: 2 millisec |
| Polarimetry Pointed or mapping | 50 or 250 (selectable) | 50/250μm: 2.5" /12.5" | 50 μm: 3.6' x 2.5' 250 μm: 13.5' x 9' | 3.3 | 50 μm: 10 Jy 250 μm: 10 Jy | 0.1% in linear polarization, ±1° in pol. angle |

## 3.3 Far-infrared Imager and Polarimeter (FIP)

A simple and robust instrument, FIP delivers images at 50 μm or 250 μm or polarization at these wavelengths (Table 3-10). The science drivers for FIP are captured by the extra-galactic case, How does the Universe work? (Section 1.1), and are articulated in the STM (Section 1.4). In addition to the science, FIP plays a critical functional role in aligning the mirrors during on orbit commissioning.

### FIP Operation Principles

The FIP functional block diagram (Figure 3-25) shows all components required for the FIP design from input radiation (in pink) through to the detector. FIP consists of Pick-off-Mirror (POM), Fore Optics, Half-Wave Plates on mechanisms, a barrel mechanism with two sets of optics, leading through a filter to the detector plane. The ADR Assembly controls the required mK temperatures for the detector. Launch locks, mechanical structure, and pre-amp electronics and harnessing complete the design.

In pointed or survey imaging and polarimetry mode, FIP observes one of the two bands, either 50 or 250 μm. A single detector with a 109×73 pixel format (~8000 pixels) is used for all the measurements. The wavelength is selected by the optics/filter combination that can be switched with the barrel mechanism. In polarization mode, a half-wave plate modulator and an analyzer (polarizing grid) is brought into the beam. In this set up, the relevant stokes parameters I,Q, and U are being measured. Note that there is no scientific need to observe the circular polarization V.

FIP provides diffraction-limited images (λ/23 μm) [arcsec] and covers an instantaneous FOV of 3.6× x 2.5× at 50 μm and 13.5× x 9× at 250 μm. FIP has an optimized design to meet the maximum achievable optical efficiency in each band, while requiring only a moderate FOV from the telescope. Either the

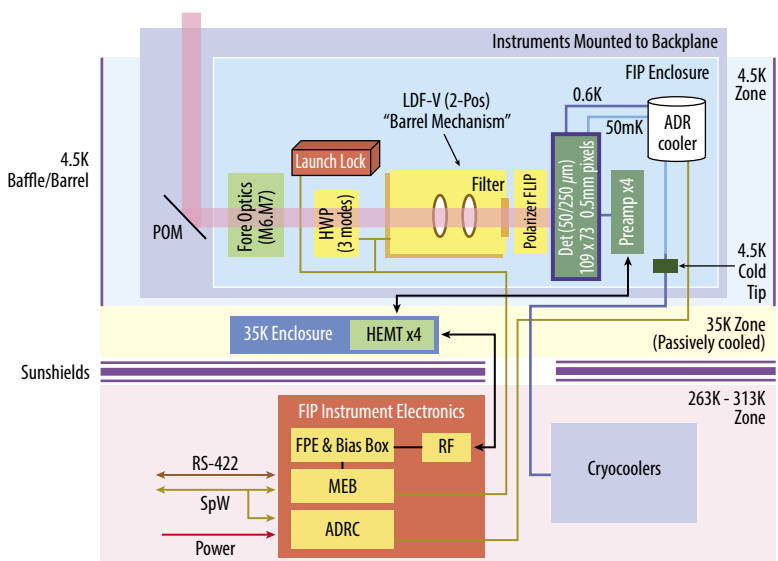

**Figure 3-25:** FIP has a simple design with one detector selectable by filter to 50 or 250 μm. This Functional Block Diagram of FIP shows how the power dissipation in the cold 4.5 K section is minimized by placing the readout electronics in the warm part of the observatory.

FSM (pointed) or the telescope (survey mapping) moves at all times during FIP observations in order to link the detector response of all pixels within the 1/f time of the pixels during observations. For single pointing observations, only the FSM moves. Several observing patterns can be chosen, for example, a Lissajous Pattern for small maps, or a chessboard pattern for large surveys (Kovacs, 2008). For survey mapping, the observatory sweeps the telescope pointing in raster patterns to map significant areas of the sky, hundreds to thousands of square degrees. This FIP operational principle has the significant benefit that the telescope does not need "to settle" at a certain position on the sky. For reconstruction of an image, only the knowledge of where FIP has been pointed at any moment in time is needed. For *Origins*, the star trackers deliver this knowledge.

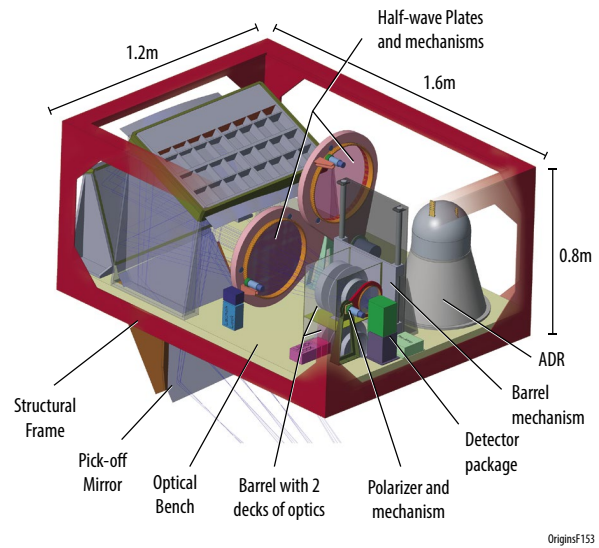

**Figure 3-26:** The FIP mechanical structure and components. The labels point out individual components described in the text.

### FIP Optical Design

The FIP optical system design allows for a compact configuration (Figure 3-26). Light is directed into the FIP instrument box by a pickoff mirror located near the telescope focal surface. Once inside the instrument box, a Dragone collimator comprising a pair of free-form (polynomial surface) mirrors (CM1, CM2), is used to collimate the light and correct aberrations (Figure 3-27). While free-form mirrors would be more difficult to construct than a traditional conic given the same operating wavelength, we do not think they would be a technical challenge to manufacture given that the long wavelength affords much looser surface figure and roughness tolerances relative to optical surfaces.

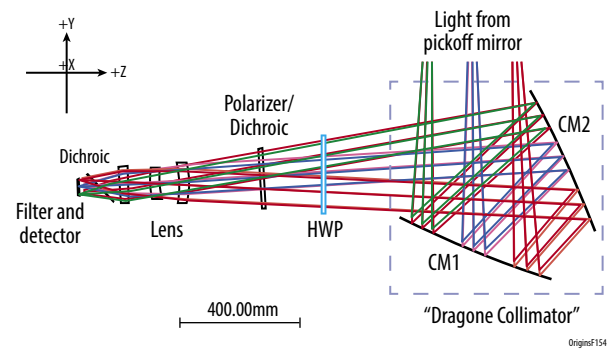

**Figure 3-27:** The FIP optical system minimizes the number of optical components. This configuration shows the layout for a polarization measurement because the HWP is in the beam.

Depending on the mode of operation, the collimated beam either traverses or bypasses a half-wave plate (HWP) before passing through an open window or a polarizer. For each of the two bands, the rays are imaged onto the detector array by a dedicated lens system, which provides the same beam-sampling factor for each, 50 and 250 μm. Like the HWP and polarizer, the lens system switches in and out during operation depending on the waveband. The system uses an f/2 or f/8 lens for the short and long bands, respectively. The HWP, polarizer, and lenses are all switched in and out through mechanisms that are described in the corresponding section.

### FIP Observing Modes

FIP observing modes (Table 3-10) include continuum and polarimetry observations. Either type of observation can be done in pointed or survey mapping modes. Pointed observations are performed by the observatory pointing at the source target, while the FSM performs a small Lissajous scan that moves the target sources on the central part of the detector array, to provide the required pixel to pixel crosslinks. The second observing mode is survey mapping: the observatory moves in a raster form over the sky, while the FSM can be used to create more crosslinks between pixels in one 1/f time, if the raster scan cannot accommodate this criterium.

## FIP Detector Subsystems

The FIP requirements could be met with multiple detector approaches, and all of them require some development. Section 2.3 summarizes the path to create FIP/OSS detectors that is described in detail in the *Origins* Technology Development Plan. For this concept design, we have adopted superconducting TES bolometers as the baseline.

The imaging array for FIP consists of ~8,000 TES elements (pixels), configured in a rectangular shape (109×73). The TES signals are densely multiplexed by microwave superconducting quantum-interference devices (μWave SQUIDs) multiplexers (Irwin *et al.*, 2006). These are followed by cryogenic HEMT amplifiers and room temperature readout electronics, described later, to facilitate readout, signal processing, and subsequent astrophysical imaging. FIP uses membrane-suspended, close-packed TES.

Since each microwave multiplexed TES has its own microwave resonator, this is not the most efficient bandwidth packing approach, but the one with the highest TRL. Since roughly 2,000 resonators fit within a 4 GHz processing bandwidth, and 8,000 pixels need to be read out, a total of four RF signals need to be connected to corresponding HEMT amplifiers and transmitted downstream for signal processing. This architecture is the same as that used for OSS described in Section 3.1.

## FIP Readout Electronics

The FIP RF signals can all be sampled and processed simultaneously by a single RFSoC. The detector feedback and other current signals needed for the detector readout can also be controlled by the RFSoC in combination with digital electronics on the same board (Figure 3-28).

## FIP Data Rates

The data rate is determined by the pixel count (8,000), and the sampling rate is determined by the speed of the apparent sky image of the detector, which is restricted by the detector time constant of a few ms. Finally, the dynamic range of the detector determines the required bit depth. The detectors are sampled at up to 300 Hz to maintain a 150 Hz scientific bandwidth, and 12 bits is sufficient to capture the detector's dynamic range under which it is photon noise limited if a logarithmic scaling is used. The uncompressed data rate is thus equal to or less than 29 Mbits/sec. Depending on the observed sources, the lossless compression rate should be between two and four (the latter for a deep survey on the sky with many weak sources), bringing it down to between 7 and 15 Mbits/s.

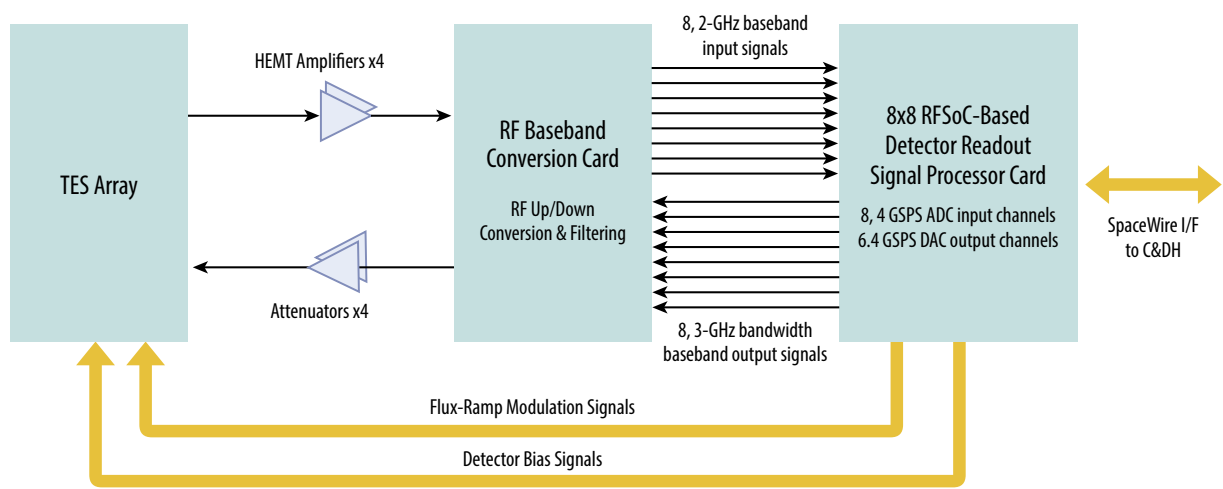

**Figure 3-28:** Detector readout scheme for FIP: FIP and RFSoC samples and processes signals simultaneously.

## FIP Thermal and Mechanical Design

**Thermal Architecture:** The FIP thermal system shares design features with the OSS instrument thermal design described in Section 3.1. Both instruments benefit from significant overlap and synergy in technology maturation and design methods.

The FIP optics and mechanisms operate at 4.5 K with cooling provided by mechanical cryocoolers that are managed at the observatory level. The cooling power of the four cryocoolers can then be shared among the three instruments and telescope (Section 2.4). The dissipation at 4.5 K comes from three main sources within FIP: the HEMTs, the Continuous Adiabatic Demagnetization Refrigerator (CADR), and the mechanisms. The design includes thermal straps made of pure annealed copper for all stages, sized for the load and distance required. These copper straps, along with the indium used for thermal connections, totals approximately 4 kg for FIP. The majority of this mass is for conducting the heat dissipated at 4.5 K to a central heat sink on the instrument structure that is, in turn, connected to a heat sink on the instrument mounting plate to which the 4.5 K cryocooler heat exchanger is also connected. This scheme leads to temperature gradients that are less than 5 mK across the 4.5 K portions of the instrument.

A five-stage CADR is planned for detector cooling (Figure 3-29). Heat to the detectors is intercepted at two locations: an outer guard that includes a cooled filter window, and the array itself, which also includes the microwave SQUID multiplexer (Figure 3-30). The five-stage continuous ADR (CADR) is being advanced from TRL 4 to TRL 6 through a grant from the Strategic Astrophysics Technology

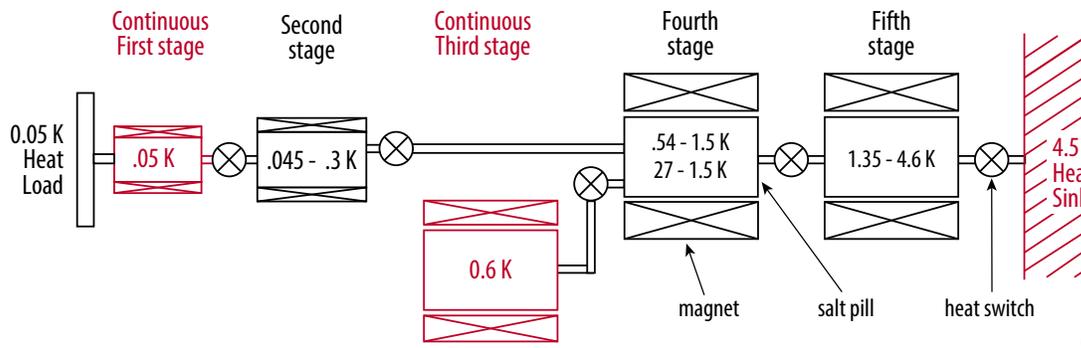

**Figure 3-29:** The FIP CADR provides continuous cooling at two stages: 50 mK (with 0.4 μK stability) and 0.7 K (with 0.1 mK stability), and deposits its heat at a 4.5 K heat sink.

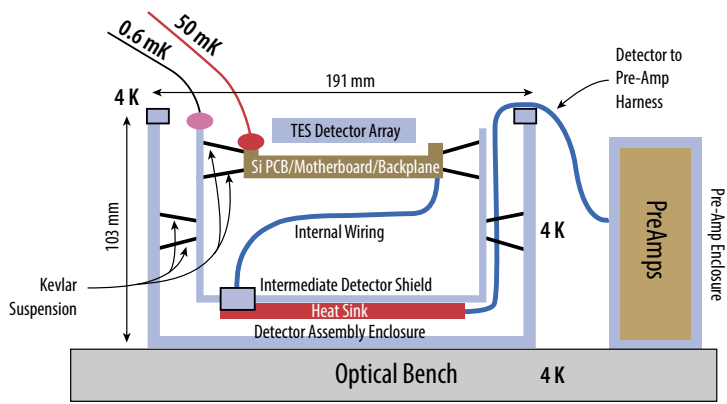

**Figure 3-30:** The FIP Focal Plane Array is cooled to 0.05 K and suspended within a 0.7 K enclosure, providing thermal shielding from the 4.5K stage. The Kevlar suspension, which is common to most instruments with low temperature detectors, provides for excellent stiffness while having a very low heat conductivity.

**Table 3-11:** Calculations of the heat loads and available lift on the FIP cryogenic stages show margins of at least 200% over the calculated heat load maintained at each temperature stage.

| T (K) | Item | Heat Load | Total | Capability * | Margin (%) |
|---|---|---|---|---|---|
| | Harnesses | 0.16 μW | | | |
| 0.05 | Dissipation | 0.17 μW | 0.56 μW | 6.0 μW | 971 |
| | Suspension | 0.23 μW | | | |
| | Harnesses | 7 μW | | | |
| 0.7 | Suspension | 76 μW | 83 μW | 292 μW | 252 |
| | HEMTs | 3.1 mW | | | |
| 4.5 | Mechanisms | 3 mW | 10.1 mW | 64 mW | 533 |
| | CADR | 4 mW | | | |

*Excess capability enables a FIP upscope without resizing the CADR.

Program. This CADR will be capable of providing 6 μW of cooling continuously at 50 mK, which is more than an order of magnitude better than the unit flown on Hitomi.

Table 3-11 summarizes the calculated heat load and available lift with the cooler for each of the three FIP cooling stages. Heat loads at 50 mK are roughly evenly divided between harnesses, a rigid strut suspension, and internal dissipation within the detectors and SQUID multiplexer. The first stage HEMT amplifiers are located at the 4.5 K stage. The 0.38 mW per HEMT x eight HEMTs in the table corresponds to commercially-available amplifiers operating up to 8 GHz, with low noise and 10 dB of gain. The team is currently performing studies to verify this amplifier noise is as advertised. Parasitic conduction in harnesses is book-kept at the observatory level, but is much smaller than the amplifier dissipation.

**Mechanical Architecture:** The FIP structure is shown in Figure 3-26. Mechanical parts include exterior flexure mounts, a housing frame, thin enclosure, optical bench, and brackets that hold the three main mirrors. The design also includes mechanical components associated with the internal optics and detector assemblies. The material for all the mechanical components is beryllium for its light weight and thermal conductivity properties.

**FIP Mechanisms:** FIP has six mechanisms listed with their requirements and operational concepts in Table 3-12.

**Table 3-12:** FIP mechanisms

| Mechanism | Requirement | Design Concept |
|---|---|---|
| Half-wave Plate Flip (2) | Three positions, open center; Element diameter: 150 mm Life: 5 yr, 22 k cycles + test; Duty cycle: low | Superconducting stepper motor drives crank between toggle positions, rotating element housings around axis parallel to optical axis; stop motor at center position to achieve open condition |
| Half-wave Plate Rotation (2) | Rotate HWP at 60 rpm Read position to +/-0.1 deg; Life: 107 revolutions; Duty cycle: high | Superconducting brushless DC motor and resolver to rotate HWP and sense its position |
| Polarizer Flip | Two positions; Element diameter: 150 mm; Life: 5 yr, 22 k cycles + test; Duty cycle: low | Superconducting stepper motor drives crank between toggle positions, rotating element housings around axis parallel to optical axis |
| Lens Flip | Two positions; Element diameter: 150 mm; Elements move simultaneously; Life: 5 yr, 22 k cycles + test; Duty cycle: low | Superconducting stepper motor drives crank between toggle positions, translating linear carriage |

## FIP Instrument Control

Figure 3-31 shows the electrical instrument control system block diagram. The main electronics box (MEB) includes specialized control boards and a LEON3 CPU. System features include:

- Mode management (standby, calibration, execution of observing modes stare and mapping in Total Power and Polarimetry modes)
- Instrument support (command processing, data collection, and support for firmware updates)
- Mechanism control

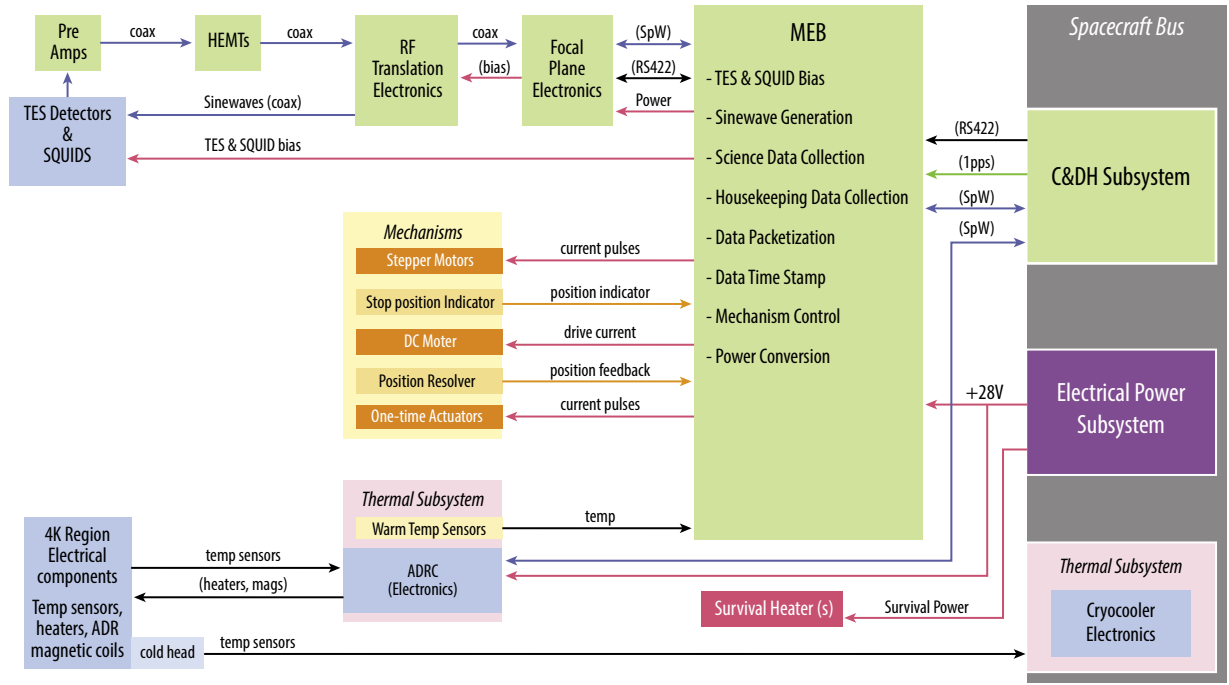

**Figure 3-31:** This FIP Block diagram shows how the instrument is controlled.

- Power distribution, including detector biasing
- Some degree of on-board autonomy (limit checking, some processing command sequences, failure corrections

**FIP Risk Management Approach**

FIP has the same three risk areas and mitigations as OSS (Section 3.1): the detector, straylight/magnetic susceptibility/ radio frequency power and mechanisms. See OSS Section 3.1 for details.

**FIP Predicted Performance**

With the detector NEP=$3 \times 10^{-19}$ W/$\sqrt{\text{Hz}}$, FIP is dominated by astronomical backgrounds. The corresponding point source sensitivity is 0.9 μJy (5σ, 1 hr) at 50 μm, 2.5 μJy (5σ, 1 hr) at 250 μm. Note that the confusion limit at 250 μm does not allow an integration to 2.5 μJy. The proposed surveys with FIP at this wavelength take the confusion limit into account. Additional performance parameters are provided in Table 3-10. Figure 3-32 demonstrates the unprecedented mapping efficiency FIP will provide.

Combined with the fast mapping capability of *Origins*, FIP can survey thousands of square degrees efficiently with significant advantages in continuum mapping speed over OSS, as shown in Figure 3-32. In survey mapping or pointed modes, FIP quickly reaches the confusion limit in the long wavelength band (in 32 ms at 250 μm), but allows integrations of unprecedented depth and takes hours to reach the confusion limit in the 50-μm band.

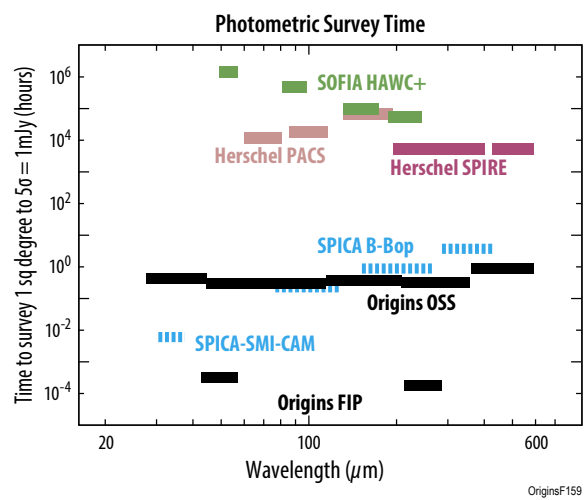

**Figure 3-32:** A comparison of photometric survey times for a number of space-based FIR instruments. Note that Origins/FIP will be the fastest mapping far-IR survey instrument ever flown in space.

### FIP Alignment, Integration, Test, and Calibration

The FIP instrument will require its own unique facilities for testing and calibration at the instrument level before it is delivered for integration with the observatory. With its outer space background limited FIR detectors the major requirement for the optical tests will be to provide an extreme dark environment for the tests with test sources (FIR laser) being attenuated to provide power levels suitable for the detectors. Gain and optical performance will be measured, while the detector readout and data acquisition.

Aside from the normal vibration, acoustic, and EMI environmental testing, the FIP instrument will need a thermal- vacuum test chamber that can be cooled to liquid helium temperatures that is large enough to enclose the complete cold section of the instrument and is FIR light-tight enough that stray light within the detectors' sensitivity wavelengths is no greater than the infrared background that would be expected on-orbit. This test chamber should allow the introduction of FIR signals at the expected levels of the on-orbit target sources and the ability to verify the optical design of the instrument. The facility will need to allow connection between the cold instrument assembly in the test chamber and the warm readout electronics over flight-like cables.

The warm electronics for FIP will not require any particular test facilities other than the usual test facilities used to conduct vibration, acoustic, EMI, and thermal vacuum tests. A spacecraft simulator will be required to test command and data transfers between the FIP and spacecraft electronics. It will be necessary to develop specialized test hardware, but although the scope of this development will be significant, it is well defined, with no new technology development required. FIP instrument I&T is expected to take 12 months, assuming all the external test support hardware and software is ready in time by the start of testing.

Although the vast majority of the FIP instrument calibration and verification/validation activities will take place before instrument delivery for observatory integration, verification of its performance within the integrated observatory will be necessary to ensure that it still performs as expected in end-to-end observatory testing before launch (see Section 2.9).

### FIP On-orbit Checkout and Calibration

In the beginning of the performance verification phase, the following items will be tested:

- [PV-01] checkout of the condition of all boards in the warm electronics boxes
- [PV-02] checkout of the condition of the detectors (*e.g.*, clock patterns for all observing modes)
- [PV-04] checkout of the optical alignment
- [PV-04] obtaining flat fielding images for the detector array
- [PV-05] on-orbit absolute flux calibration
- [PV-06] on-orbit calibration of instrument polarization

During the nominal observational phase, the following calibrations are updated as needed:

- [PV-04] obtaining flat fielding images for the detector array
- [P1-05] orbit absolute flux calibration

### FIP Heritage and Maturity

FIP is a mature instrument design except for the detectors (See discussion in the separate volume: Technology Development Plan). The FIP design, in particular its optical layout, is similar to the SOFIA/HAWC+ instrument (Harper *et al.*, 2018, Staguhn *et al.*, 2016), which has operated routinely for scieence observations since December 2016. HAWC+ demonstrated the success of Backshort Under Grid (BUG) TES kilopixel arrays planned for use in FIP/*Origins*. The HAWC+ on SOFIA (PI C.D. Dowell; Harper *et al.*, 2018) uses three of GSFC's 40x32 pixel TES-based BUG detector arrays.

They provide background limited observations with kilopixel arrays (Staguhn *et al.*, 2016) from a continuum camera on an airborne observatory (to meet SOFIA sensitivity and background requirements, the detectors were designed to have a NEP of $1 \times 10^{-16}$ W/$\sqrt{\text{Hz}}$).

Commissioning of the camera began in Spring of 2016 and science flights performed beginning in December 2016.

First polarimetric data of astronomical sources were obtained during those flights, which were part of the guaranteed time for the instrument team. Figure 3-33 shows the map of the magnetic field as measured with HAWC+ in the Orion Molecular cloud at 89 µm.

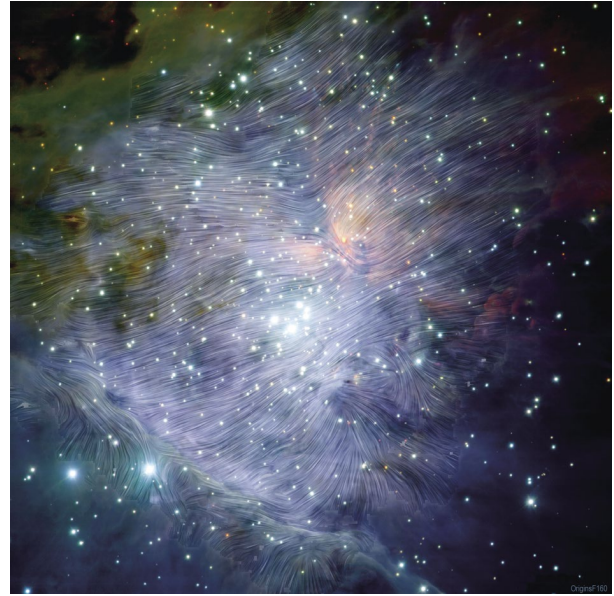

**Figure 3-33:** FIP has heritage from the SOFIA instrument HAWC+. Shown here is a HAWC+ polarimetry image of OMC at 89 µm. The vectors indicate the direction (not strength) of the magnetic field orientation.

Significant amounts of extended emission are also visible in the underlying total power map, another demonstration for the stability of the TES detector arrays used in these instruments.

FIP is a mature instrument design and in most aspects is less complex than HAWC+. With 8,000 pixels, the detector array is only roughly a factor of 3 larger than each of the two focal plane arrays in HAWC+. In light of the fact that the targeted array architectures are tileable, this factor of 3 in pixel count is achievable now. In terms of detector sensitivity on a single pixel level, the required sensitivity has been demonstrated in the Lab (Suzuki *et al.* 2016). The dynamic range requirement might force using two TES per pixel, which could either be read out in parallel or series, the latter demonstrated at GSFC by D. Benford (private communication).

### FIP Enabling Technology

FIP's major enabling technology are the superconducting detectors, which provide background limited performance, combined with wide field capabilities provided by a pixel count of 8,000. The efficient readout of these detectors is provided by µWave SQUID multiplexers, developed at NIST, Boulder (Irwin *et al.*, 2006).

Sub-Kelvin coolers are the other main enabling technology for FIP. The FIP detectors, like the OSS detectors, are cooled to 0.05 K by an Adiabatic Demagnetization Refrigerator (ADR). The path to achieving these technologies is described in the *Origins* Technology Development Plan and summarized in Section 2.3.

# 4 - MANAGEMENT, SCHEDULE, RISKS, COST, AND DESCOPE

> The *Origins* mission will be responsive to community science priorities established in the 2020 Decadal Survey and will be developed by NASA in collaboration with international and industry partners.

## 4.1 Study Team

The *Origins* Decadal mission concept study included two intertwined components: community outreach and an investigation that established the scientific motivation for the mission, and an engineering study conducted with enough attention to detail to demonstrate mission executability. A community-based Science and Technology Definition Team (STDT) developed science goals and measurement requirements. The engineering study, centered at NASA's Goddard Space Flight Center (GSFC), was conducted with input from study partners at NASA's Ames Research Center (ARC), NASA's Jet Propulsion Laboratory (JPL), NASA's Marshall Space Flight Center (MSFC), the Infrared Processing and Analysis Center (IPAC), the Japan Aerospace Exploration Agency (JAXA), the Centre National d'Etudes Spatiales (CNES), a Ball Aerospace-led industry consortium comprising Ball, Northrop-Grumman, and Harris Corporation, and a separate partnership with Lockheed Martin (LM).

Strong multi-institutional partnerships were were sustained throughout the *Origins* mission concept study and were very fruitful. GSFC and JPL partnered to design the OSS instrument. JAXA partnered with NASA Ames to design the MISC instrument. With US involvement, a CNES-led European consortium designed the HERO instrument, an optional upscope to the baseline *Origins* mission concept. GSFC conducted trade studies in collaboration with industry that culminated in the selection of a mission architecture, optical system configuration, optical system error budget, and material selection. Collectively, the partnering institutions brought a great deal of relevant expertise and experience to the study. Scientists around the world participated in discussions of the science case for *Origins*.

The *Origins* study team (Figure 4-1) was charged with delivering this report to NASA Headquarters (HQ) with the understanding that it would be submitted by NASA to the National Academies for the 2020 Decadal Survey in Astronomy and Astrophysics. The STDT Chairs (Meixner, Cooray) are princi-pally responsible for delivering the report and presenting it to the Decadal Survey Committee. As the designated Study Center, GSFC (Carter, Leisawitz, DiPirro, Staguhn) led the engineering and coordination effort. This diverse and experienced team was essential to conducting a study driven by the exploration needs of the scientific community and developing a mission concept grounded in engineering rigor.

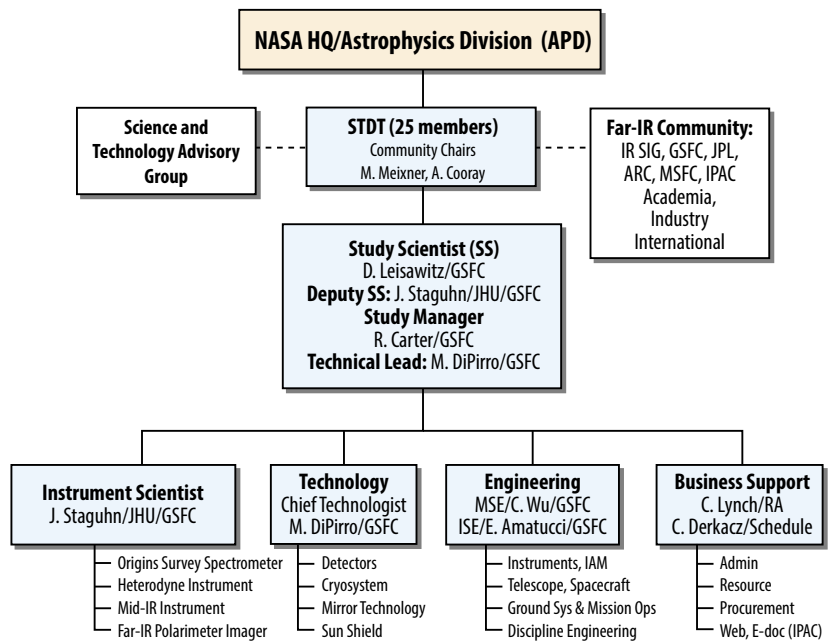

**Figure 4-1:** The *Origins* Study Team brings together scientific and technology leaders and engineers with many years of spaceflight mission experience to ensure that *Origins* represents the best interests of the astrophysics community.

The *Origins* management team (Study Office, community chairs, NASA HQ) and Mission Concept Working Group of engineers and instrument leads (led by Roellig), met weekly via telecon/WebEx to craft the mission concept described in this reprt. To ensure connection with the astronomical community, *Origins* has an external advisory, advocacy and communication group (led by Battersby, Kataria, and Narayanan), and five science working groups representing *Origins*' major topics: galaxy and black hole formation and evolution (led by Pope, Vieira, Armus), formation of planetary systems (led by Bergen, Pontoppidan, and Su), exoplanets (led by Stevenson, Kaltenegger, Fortney, and Kataria), Milkyway and nearby galaxies (led by Sandstrom and Battersby), and our Solar System (led by Milam and Bauer).

Interested community members participated in weekly telecons and public meetings held every 2 to 4 months. These meetings provided essential discussion and led to meaningful mission decisions. The Infrared Science Interest Group (IR-SIG) also served as a valuable conduit for information flow between the general community and the study team.

Section 4.2 presents a notional organization for the *Origins* implementing team based on the fruitful collaboration that resulted in the successful completion of the Decadal mission study. We envision a NASA-led *Origins* mission in which final partner roles and responsibilities for execution will be defined by NASA, JAXA, and ESA during Phase A based on the results of a technology readiness review, further mission studies, and programmatic considerations.

## 4.2 Mission Design Team and Organization

The *Origins* mission is designed to be fully compliant with all requirements specified in NPR 7120.5, Program and Project Management Process and Requirements. The *Origins* mission concept study team formulated a mission design concept, technical approach, technology development plan, risk management approach, budget, and a master schedule compatible with NASA guidelines for the Decadal Study and grounded in experience from previous successful large Class A missions.

*Origins* will be NASA-led, managed by a NASA Center, and include both domestic and international partners. Based on participation in the *Origins* mission concept study, we envisage partnerships with Japan, ESA and its member nations, and Canada, as well as domestic industrial partners and involvement from multiple NASA Centers. Similar partnerships led to the successful infrared missions IRAS, *Herschel* and *Spitzer*.

Notionally, in addition to leading the mission, NASA would be responsible for delivering the spacecraft and cryocoolers, project management and mission systems engineering, and mission integration and testing (I&T). In addition, NASA will solicit proposals to develop the *Origins* instruments, one or more of which may be contributed by a foreign partner. NASA will also provide *Origins* launch services.

Modularity in the *Origins* design will facilitate development-sharing between the partners. GSFC and JPL have the capacity and expertise to lead the telescope development. Instrument provider selection will occur through coordinated multi-Agency Announcements of Opportunity (AO).

NASA will provide overall *Origins* management and integration of the mission elements and competitively select an industry partner as prime contractor for telescope development, spacecraft development and observatory integration. All teams will apply effective management and systems oversight lessons learned at all levels of the partnership and across all international borders and institutional boundaries. To ensure clear communication, systems engineers from across the mission flight and ground segments will serve on the mission systems engineering team and work together to define interfaces and perform system-level analyses.

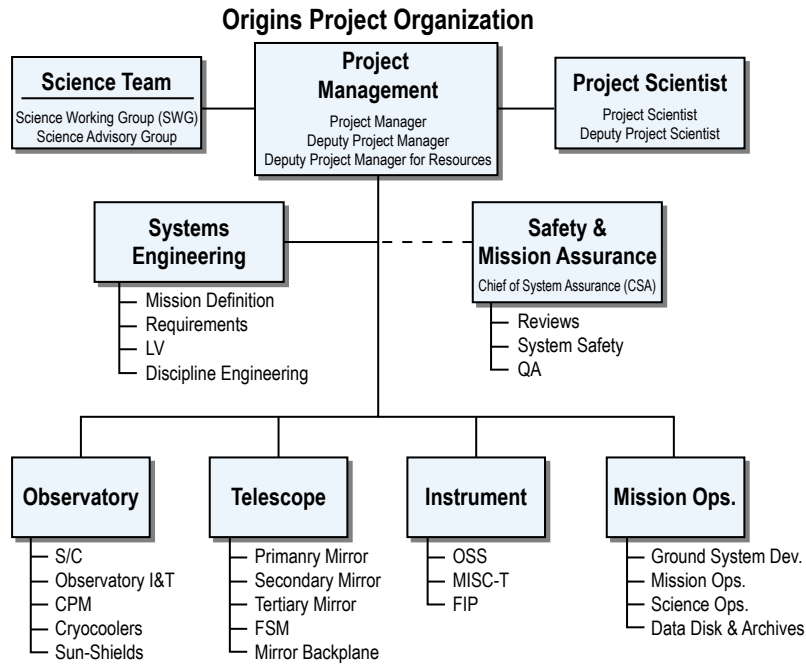

**Figure 4-2:** *Origins* mission implementation organization (notional) is designed to be compliant with NASA project management and systems engineering, and it will be led by NASA, allowing both domestic and international partners.

Engineers and managers across the mission, including all partner organizations, also participate in requirement, design, and other key reviews at the element and mission system level. Resources are applied early in the program to manage issues related to International Traffic in Arms Regulations (ITAR), Export Administration Regulations (EAR), and to establish Technical Assistance Agreements (TAAs). The Program Management Council includes key stakeholders across the partner project teams and expedites issue-resolution across institutional boundaries.

The *Origins* Mission Implementation Organization (Figure 4-2) includes both domestic and international partners under the direction of NASA-led management. It is designed to be compliant with all elements in NPR 7120.5.

### 4.2.1 Mission Partnership Opportunities

As proven to be beneficial during the mission study, the *Origins* team plans to seek and solicit mission partners from both domestic and international organizations. The major criteria for mission partnership is to enhance science products and reduce mission costs to NASA. During Phase A, the *Origins* Project management, in consultation with the Science Team, will determine mission partners, and submit to NASA HQ for final decision. The organizations that participated in the mission study are candidates for partnership in the *Origins* mission, but NASA has made no commitment to include any particular organization.

Although the JPL-led OSS and GSFC-led FIP study teams have already identified US team members who can provide all critical OSS and FIP subsystems, there are many opportunities for partnership with international agencies. The most important is the detector system (Section 2.3) and items detailed in the *Origins* Technology Development Plan. The ultra-sensitive detectors for OSS have similar requirements to those being developed in Europe. A number of groups in the Netherlands and the United Kingdom are developing kinetic inductance detectors and ultra-sensitive transition-edge sensed (TES) bolometers, as well as the readout systems needed to operate these devices. OSS team members are closely coupled to these groups, and in some cases are already collaborating. A potential partnership could be for one of the European groups to provide detector fabrication and delivery, after a common development period early in the project. A similar arrangement could be made for the read-out electronics; French groups developed the readout electronics for the *Herschel*/SPIRE and Planck/HFI bolometers, even though the detectors came from the US and were integrated into the instrument in the UK (SPIRE) and France (Planck/HFI). JAXA has developed a flight-worthy 4.5 K cryocooler with a performance in line with *Origins*' needs. While this cooler has a specified lifetime of 5 rather than 10 years, an incremental development of this cryocooler could be a valuable contribution.

Other possibilities include developing specific optical components. The etalon in OSS would be an excellent partnership opportunity since it is relatively modular and will require independent testing in advance of delivery. A group at SRON in Holland is already developing new multi-layer mirrors that can make high-finesse etalons, and groups at Cardiff (UK) have experience with mirror materials and narrow-band etalon-based filters. These groups have delivered flight hardware, so would be well-positioned to contribute to OSS. A team at JAXA is developing a rotating half-wave plate for LiteBird that has similar requirements to those of FIP.

International partnership offers an opportunity to benefit from a broader experience base and the heritage of space-qualified components to overcome technical challenges, as seen in past international space IR missions. Throughout the *Origins* Mission Concept study, JAXA led the study of the MISC-T instrument with NASA Ames Research Center (NASA/ARC) as a partner. The MISC-T instrument team is developing a testbed for a prototype of the densified pupil spectrograph at NASA/ARC, aiming to achieve TRL 5 by 2022. Laboratoire d'Astrophysique de Marseille is making a contribution to the study of the MISC-T instrument from science and technical aspects.

### 4.2.2 Management Processes and Plans

The *Origins* management process is focused on mission success and a strong commitment to continuously manage risk and stay within the allocated project resources. The *Origins* team has formulated a mission design concept, technical approach, technology integration plan, risk management approach, budget, and a master schedule compatible with the 2020 Decadal Study guidelines. The *Origins* project will maintain appropriate levels of internal and external surveillance across all phases of the project. The Mission Systems Engineer (MSE) and System Assurance Manager (SAM) are responsible for fulfilling the Independent Technical Authority (ITA) responsibilities and will maintain the project's technical standards and requirements in accordance with the NPR 7120.5.

The *Origins* Project Plan, from Phase A through Phase E, will be formulated based on implemented plans from previous successful NASA-led large Class A missions. The Project Plan will include requirements, mission constraints, mission success criteria, descope options, integrated master schedule, mission technical and cost budgets, and a detailed implementation plan. Project planning addresses policies, guidelines, work breakdown structure (WBS), detailed schedules, receivables and deliverable agreements, document trees, decision-making priorities, configuration management, and quality assurance. Understanding that communication is a key element of successful missions, the *Origins* team will communicate by means of teleconferences, regular technical interchange meetings (TIMs), and monthly meetings. The core management team will review *Origins*' status on a daily basis and consult with the partners' key management teams weekly to track progress.

The *Origins* project will conduct all reviews required by NASA senior management. In addition to required reviews, all mission elements will be subjected to peer reviews. The objective of these reviews is to ensure that the flight and ground systems fulfill their mission requirements. An appropriate decision-making process will be devised to identify the flow of responsibility and delegate authority to the lowest possible level. Controlling documents and agreements that identify responsibility and commitments will help guide the decision-making process.

### 4.2.3 Margins and Reserves: Release Strategy

The ability to allocate authority to commit and use reserve and margin is a key management tool. Used in a proper and timely manner, allocated reserve and margins can correct deviations from the development and oprations plan, and reduce or eliminate risks associated with future planned activities. Reserves are allocated according to risks and, if needed, will be released after key milestones. *Origins* will publish reserve and margin status electronically, allowing external stakeholder review.

### 4.3 Risk Management

*Origins* will implement a comprehensive risk management plan, which will rely heaviliy on concpets set forth in NPG 7120.5. The plan will provide the structure within which the mission team will identify, manage, mitigate, track, and control risks to achieve misson success under a fixed budget and schedule. All members of the project team will routinely evaluate, identify, and address risks. The risk management process will include, but not be limited to, maintaining adequate engineering resource margins during all phases of the mission development and implementing a robust design to mitigate functional degradation or failure. In addition, the team will assign a risk manager (RM) to oversee continuous risk management (CRM) practices.

The *Origins* risk management approach will incorporate the following elements:

- **Risk Identification:** Risks will be identified from a variety of sources, including spacecraft, telescope and instrument subsystem leads and engineers and project managers. The project also will rely on the lessons learned from previous missions. Special attention will be given to capturing risks before they are realized as issues. Risk statements will be entered into a CRM database.
- **Risk Analysis:** Each risk will be evaluated to determine the impact, probability of occurrence, and timeframe. Each risk will be examined to determine its relationship to other identified risks. The PM leads a risk management group that ultimately assigns attributes to each risk. This level of awareness enables the PM to allocate the appropriate resources to mitigating the risk.
- **Risk Planning:** Risk information will be translated into planning decisions and actions. In addition, each risk will be assigned an owner to ensure that someone with appropriate abilities manages it.
- **Risk Tracking:** Risk information and metrics that the team defined during planning will be captured, tracked, and analyzed for trends. The risk owner is responsible for the risk until it has been closed.
- **Risk Control:** The PM will evaluate tracking results to determine how to proceed with risks. She or he will decide whether to close risks, continue the current mitigation plan, or invoke the contingency plan.
- **Communication:** The team will encourage an atmosphere of free exchange to ensure that everyone voices all concerns, even those that only are perceived as risks. This process also will be used to convey risk information between all levels of the project team.
- **Lessons Learned:** Taking advantage of lessons learned from similar missions ultimately will save *Origins* time and money and lead to mission success. During Phase A, project leaders will establish with all partners a clear understanding of what is expected of them. Project managers will: oversee spacecraft, telescope and instrument vendors to avoid misunderstandings about requirements and interface changes; and define requirements and interfaces between the instrument and spacecraft in Phase A—before contract negotiations—and sign contracts after the completion of Phase A. Thereafter, the Project will absorb required changes to the spacecraft and instruments without affecting interfaces. Project management will incorporate into the program plan the appropriate number and level of project and subsystem reviews, and develop an appropriate mission-specific management structure and WBS. At all times, Project managers will seek creative solutions to programmatic issues.

### 4.3.1 *Origins* Mission Top Risk Items

The top risk items in this section are based on the *Origins* Baseline mission concept design. The *Origins* mission design minimizes risk by building on extensive heritage from previously flown missions, existing spacecraft bus designs, and flight-proven telescope designs (*e.g.*, the *Spitzer*-like configuration). Most of the flight system elements are in operational configuration at launch, and the sun shields require only a simple two-step deployment. Throughout this pre-Decadal study, the *Origins* team identified risks and adopted prudent mitigations. The risk mitigation strategy includes TRL advancement of the key detector and cryocooler technology. The *Origins* technology roadmap outlines the technology advancement to TRL 5 and TRL 6 during pre-Phase A and Phase A. *Origins* top risks and mitigation approaches are shown in Table 4-1. An overall risk assessment was performed based on the GSFC Risk Matrix Standard Scale (Figure 4-3), and a risk rank was assigned according to the standard scale. Figure 4-4 shows the risk ranks in a 5x5 Risk Matrix.

**Table 4-1:** *Origins* Top Risks

| Risk # | Rank (LxC) | Risk Rating | System | Risk Statement | Mitigation |
|---|---|---|---|---|---|
| 1 | 3 (1x5) | Yellow | Telescope-cover | If telescope cover fails to deploy, then telescope will fail to collect light resulting in complete loss of science | One-time, simple deployment incorporates redundant electronics for actuators. Extensive ground test program under test-as-you-fly conditions. G-negated operation at observatory level will be demonstrated. |
| 2 | 2 (2x4) | Yellow | Observatory Sun Shields | If sun shields fail to deploy, then cryocoolers may not be able to cool telescope and instruments to 4.5 K, resulting in partial or complete loss of science | One-time simple deployment incorporates redundant electronics for actuators. Will be operated at least twice on the ground under test-as-you-fly conditions. |
| 3 | 1 (3x4) | Yellow | Far-IR Detectors | If far-IR detectors cannot meet requirements at TRL 6 by PDR, then mission will be delayed or science may be impacted. | Far-IR detector development plan includes multiple technology development efforts in parallel to TRL 5. Plan to downselect at Phase A and then advance to TRL 6 by PDR. Technology development is planned for 4 years prior to Phase A and two years during Phase A. Detector technology will be funded and managed strategically according to the *Origins* Technology Development Plan. |
| 4 | 1 (3x4) | Yellow | Mid-IR Detectors | If 5 ppm stability cannot be achieved by PDR (TRL 6), then mission will be delayed or science may be impacted. | Mid-IR detector development plan includes multiple technology development efforts in parallel to TRL 5. Plan to downselect at Phase A and then advance to TRL 6 by PDR. Technology development is planned for 4 years prior to Phase A and two years during Phase A. Detector technology will be funded and managed strategically according to the *Origins* Technology Development Plan. |
| 5 | 2 (2x4) | Yellow | Cryocooler Performance | If cryocoolers fail to provide required 4.5 K environment for telescope and instruments, then far-IR science will be severely impacted, resulting in diminished science | The *Origins* cryocooler system has 100% margin in required cooling power. Also, the cryocooler electronics are fully redundant. The cryocooler vendors are qualified to TRL 4 or higher. We have 100% margin on cryocoolers and electronics are redundant. |
| 6 | 4 (2x3) | Yellow | Telescope - Beryllium Mirror | If Be mirror fabrication is delayed, then the overall mission development schedule will be impacted. | In pre-Phase A and Phase A, work early with manufacturers, develop a wedge test for section of backplane and mirrors before PDR, and also have alternate materials selected as needed. |
| 7 | 4 (2x3) | Yellow | OSS mechanisms | If the OSS mechanisms fail, then instrument functionality will be impaired resulting in reduced science return. | Mechanisms are launched in fail-safe positions: FTS is in the beam (most of the FoV is not affected), Etalon is out of the beam (everything except highest resolution mode is not affected). Even without any mechanism operation the basic R=300 resolution is preserved. |
| 8 | 4 (2x3) | Yellow | Launch Vehicle availability | If neither SLS nor Space X BFR launch vehicle is available, mission launch will be delayed or may have to redesign the observatory to fit into another launch vehicle. | During Phase A, determine launch vehicle availability. If both SLS and SpaceX BFR are not available, observatory design is compatible with Blue Origins New Glenn with minor modifications. |
| 9 | 4 (2x3) | Yellow | FIP mechanisms | If the FIP mechanisms fail, then instrument functionality will be impaired and science return will be decreased. | Mechanisms are launched in fail-safe positions: Rotaing half-wave plates are out of FoV, and 50 micron band is selected, enabling the most crucial science. |
| 10 | 2 (2x4) | Yellow | Field Steering Mirror (FSM) | If the FSM mechanism fails, then instrument functionality will be impaired and science return will be decreased. | FSM uses redundant actuators and electronics. Heritage is from JWST. Either Tip or Tilt could be used alone to meet the requirements for the Far IR instruments. |

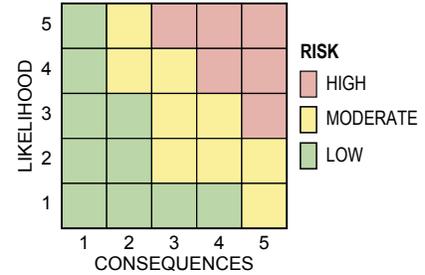

**GSFC Risk Matrix Standard Scale**

| Likelihood | Safety (Estimated likelihood of safety event occurrence) | Technical (Estimated likelihood of not meeting performance requirements) | Cost/Schedule (Estimated likelihood of not meeting cost or schedule commitment) |
|---|---|---|---|
| 5  Very High | $(P_{SE} > 10^{-1})$ | $(P_T > 50\%)$ | $(P_{CS} > 75\%)$ |
| 4  High | $(10^{-2} < P_{SE} \le 10^{-1})$ | $(25\% < P_{CS} \le 50\%)$ | $(50\% < P_{CS} \le 75\%)$ |
| 3  Moderate | $(10^{-3} < P_{SE} \le 10^{-2})$ | $(15\% < P_T \le 25\%)$ | $(25\% < P_{CS} \le 50\%)$ |
| 2  Low | $(10^{-5} < P_{SE} \le 10^{-3})$ | $(2\% < P_T \le 15\%)$ | $(10\% < P_{CS} \le 25\%)$ |
| 1  Very Low | $(10^{-6} < P_{SE} \le 10^{-5})$ | $(0.1\% < P_T \le 2\%)$ | $(2\% < P_{CS} \le 10\%)$ |

**RISK**
HIGH
MODERATE
LOW

| Consequence Categories | | | | | |
|---|---|---|---|---|---|
| **Risk** | **1 Very Low** | **2 Low** | **3 Moderate** | **4 High** | **5 Very High** |
| **Safety** | Negligible or No impact. | Could cause the need for only minor first aid treatment. | May cause minor injury or occupational illness or minor property damage | May cause severe injury or occupational illness or major property damage | May cause death or permanently disabling injury or destruction of property |
| **Technical** | No impact to full mission success criteria | Minor impact to full mission success criteria | Moderate impact to full mission success criteria. Minimum mission success criteria is achievable with margin. | Major impact to full mission success criteria. Minimum mission success criteria is achievable | Minimum mission success criteria is not achievable |
| **Schedule** | Negligible or no schedule impact | Minor impact to schedule milestones; accommodates within reserves; no impact to critical path | Impact to schedule milestones; accommodates within reserves; moderate impact to critical path | Major impact to schedule milestones; major impact to critical path | Cannot meet schedule and program milestones |
| **Cost** | <2% increase over allocated and negligible impact on reserve | Between 2% and 5% increase over allocated and can handle with reserve | Between 5% and 7% increase over allocated and can not handle with reserve | Between 7% and 10% increase over allocated, and/ or exceeds proper reserves | >10% increase over allocated, and/or cannot handle with reserves |

Code 300
GPR 7120.4D
OriginsF165

**Figure 4-3:** *Origins* mission risk assessments were based on the NASA/GSFC Risk Standard Scale.

Development of the far-IR detectors to TRL 5 by mid-2020 is an important part of the *Origins* risk mitigation strategy. Details of the detector technology development are outlined in the *Origins* Technology Roadmap. Either detector type can provide the required far-IR measurements, and *Origins* needs only one of the detectors to reach TRL 5 by 2025.

Cryocooler technology development also represents a risk to OST. Sub-Kelvin cooler maturation is already funded through the SAT program, and the cooler is on track to reach at least TRL 5 by 2021. The Hitomi mission has already demonstrated 4.5 K cryocooler technology, but the cooler lifetime estimate is <10 years. A combination of redesign and survival testing will mitigate the 4.5 K cryocooler risk.

Should *Origins* be recommended for development by the Decadal Survey, NASA will dedicate funding to technology maturation in the early 2020s above the level available through typical Reserch Opportunities in Space and Earth Sciences (ROSES) programs. All *Origins* technology-related mission risks will be retired prior to mission PDR through the maturation efforts described in the *Origins* Technology Development Plan.

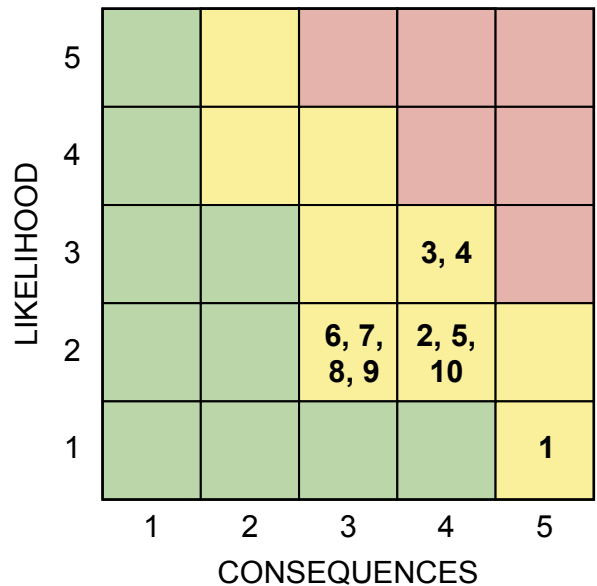

**Figure 4-4:** *Origins* top risks are all yellow.

New risks will inevitably arise during mission implementation, and they will be addressed using the risk management approach outlined above.

## 4.4 Mission Development Schedule

Figure 4-5 shows the *Origins* mission development schedule from Phase A through Phase E. Milestones and key decision points in the *Origins* schedule are consistent with NASA Procedural Requirements NPR-7120.5. Senior WFIRST and JWST managers examined the schedule and considered it reasonable for a strategic science mission with three instruments and a telescope requiring cryogenic testing and verification. The duration of each Phase A through D is comparable to the corresponding Formulation and Development times of previous large missions of similar complexity. The schedule supports an April 2035 launch following ~10 years for development. The project plan provides 12.7 months of funded schedule reserve along the critical path, exceeding by 1.9 months the required reserve according to GPR 7120.7B "Funded Schedule Margin and Budget Margin for Flight Projects" for Phases C and D (total duration 6.25 years). The schedule allows time for travel to and from special integration and test facilities. Much of the design and development work progresses through parallel efforts, and the critical path runs through the most complicated instrument, the *Origins* Survey Spectrometer. The plan includes 5 years of mission operations after launch, and an option to extend the science mission to 10 years.

The *Origins* integration and test phase, Phase D, is considerably shorter than that of JWST because *Origins* has very few deployable elements, whereas JWST has many, and we applied lessons learned from the JWST experience. Although the *Origins* primary mirror is segmented, it launches in its operational configuration, and the mirror segments are easier and faster to manufacture than those of JWST. For example, mirror cryo-null figuring is not required for *Origins*. *Origins* has a Cryogenic Payload Module (CPM) comprising the telescope and instruments. Its design reduces cool-down and warm-up time. The cooling time in particular is greatly reduced because the CPM is actively cooled. Such rapid cooling was seen in the JWST/MIRI test cycle; cooling progressed slowly while radiative, but rapidly as soon as the MIRI instrument cryocooler was activated. Test time is also reduced because we have selected non-absorptive, thermally conducting flight system materials, alleviating the need to orchestrate cooling to avoid condensation on critical components.

## 4.5 Mission Cost and Cost Estimating Methodology

This section summarizes the detailed cost analysis results presented in the *Origins Space Telescope Cost Report*. The estimated costs fully encompass the *Origins* baseline mission's scope, covering all aspects of the *Origins* Work Breakdown Structure (WBS) from Phase A through Phase E. A separate *Origins Space Telescope Technology Development Plan* describes the maturation of all mission-enabling technologies and reports the cost of pre-Phase A technology development. Although future negotiations between NASA and its foreign counterparts ESA and JAXA will likely lead to partnership agreements and cost savings to NASA, no foreign partner contributions are assumed in the cost estimates presented below. This section summarizes the estimated lifecycle cost of the mission (Phases A through E). Phase F (Decommissioning) is not addressed here, as *Origins* carries consumables beyond those needed for five years of science operations and the option exists for an extended science operations phase.

### 4.5.1 NASA GSFC's Cost Estimating, Modeling and Analysis (CEMA) Office Estimate

NASA Goddard Space Flight Center's (GSFC's) Cost Estimating, Modeling and Analysis (CEMA) Office, under the direction of the GSFC Office of the Chief Financial Officer (CFO), parametrically derived *Origins* life-cycle mission cost estimates. GSFC's Resource Analysis Office (RAO) independently derived a mission cost estimate using a top-down parametric model. The CEMA and RAO estimates are consistent to within the estimated uncertainties.

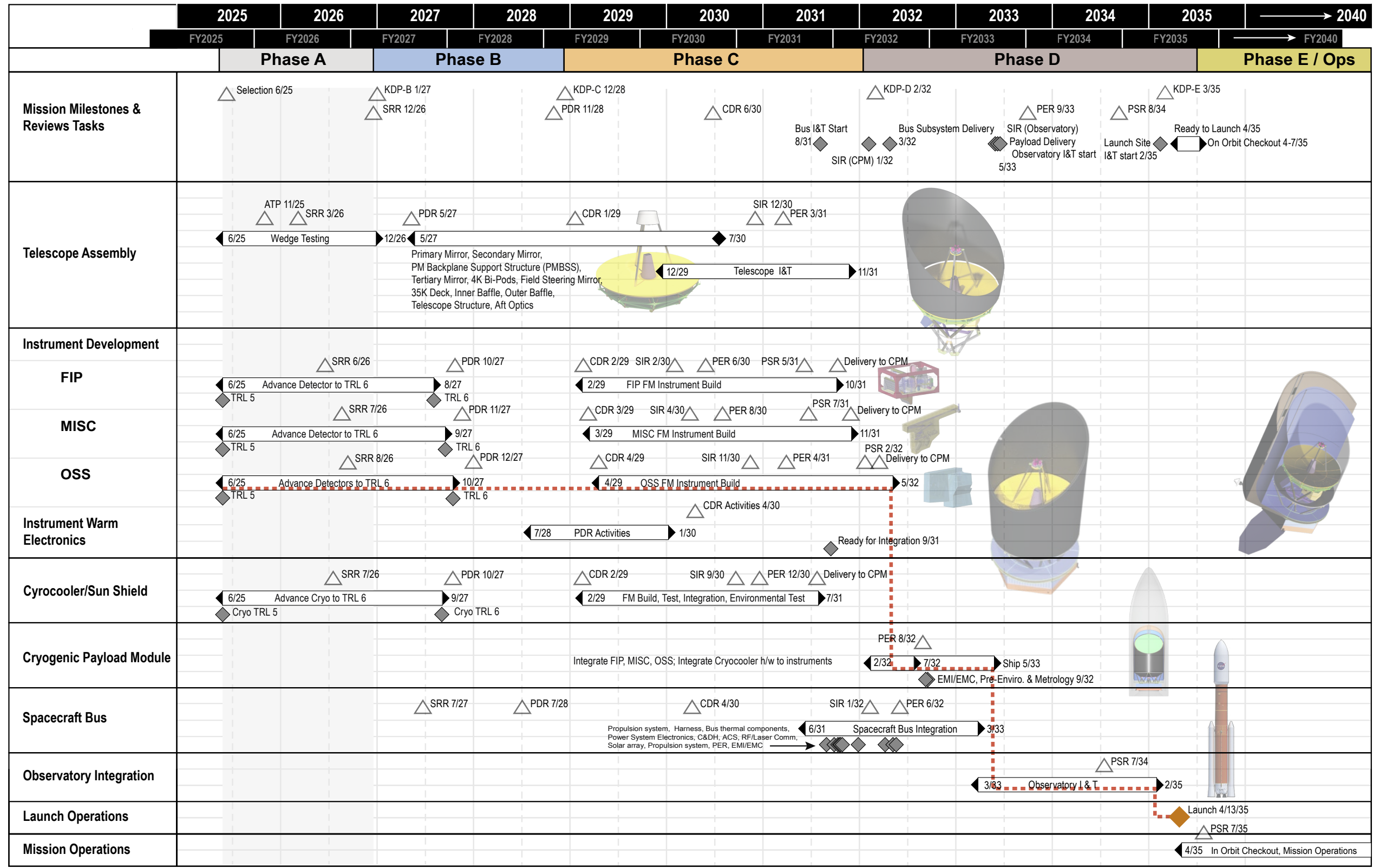

**Figure 4-5:** The *Origins* mission development schedule is derived from the successful schedules of similar scale missions.

Based on directions from NASA HQ SMD, the *Origins* team assumes:

- Mission start, Phase A, is in 2025;
- All mission-enabling technologies will be advanced to TRL 5 at the beginning of Phase A and to TRL 6 by mission PDR;
- The *Origins Technology Development Plan* describes how the required technologies will be matured to TRL 5 prior to Phase A, and fully accounts for the cost of this effort. Pre-Phase A technology maturation costs are not included in the mission cost estimates presented here.
- The cost of technology maturation from TRL 5 to TRL 6 is included in the Phase A and Phase B mission cost estimates given below and maturation to TRL-6 will be completed by PDR.
- SLS Launch Vehicle cost is $500M in real year (RY) dollars.

The CEMA-estimated total cost of the *Origins Space Telescope* mission, Phases A through E, is $6.7B (CY2020) at the 50% Confidence Level (CL) and $7.3B (CY2020) at the 70% CL. Table 4-2 shows the WBS cost breakdown in CY2020 dollars and Real Year (RY) dollars. The mission costs in RY dollars are $8.7B at the 50% Confidence Level (CL) and $10.0B at the 70% CL, which includes the launch vehicle cost of $500M RY. Figure 4-6 shows the funding profile in RY dollars at the 50% and 70% confidence levels, but does not include the launch vehicle and services. We acknowledge that NASA's future funding for a large astrophysics mission may be insufficient to cover the funding profile shown in Figure 4-6 during the peak funding years and that a flatter funding profile would lead to a greater overall mission cost.

**Table 4-2:** *Origins* Mission-level Cost Estimate

| | | | ACEIT Summary (Sum is statistical, not arithmetic) | | | | |
|---|---|---|---|---|---|---|---|
| WBS | Description | Phase A | 50% CL Phases B/C/D | 70% CL Phases B/C/D | 50% CL Phase E | 70% CL Phase E | |
| 1.0 | Project Management | 17 | 281 | 329 | | | |
| 2.0 | Systems Engineering | 11 | 281 | 329 | | | |
| 3.0 | Safety and Mission Assurance | 2 | 161 | 188 | | | |
| 4.0 | Science/Technology | 106 | 132 | 132 | | | |
| 5.0 | Payload | 21 | 2,677 | 3,149 | | | Hardware Elements (Parametrically Estimated) |
| 5.1 | Mid-Infrared Spectrometer Camera (MISC) | | 437 | 516 | | | |
| 5.2 | Origins Survey Spectrometer (OSS) | | 595 | 707 | | | |
| 5.3 | Far-Infrared Imager Polarimeter (FIP) | | 495 | 594 | | | |
| 5.4 | Telescope | | 890 | 1,031 | | | |
| 5.5 | Cryogenic Payload Assy Integration and Test (I&T) | | 256 | 296 | | | |
| 6.0 | Spacecraft | 21 | 1,318 | 1,521 | | | |
| 6.1 | Spacecraft (without Sun Shield and Cryocoolers) | | 864 | 997 | | | |
| 6.2 | Cryocoolers | | 220 | 265 | | | |
| 6.3 | Sun Shield | | 47 | 53 | | | |
| 6.4 | Spacecraft IT& (Sunshield and Cryocoolers) | | 215 | 250 | | | |
| 7.0 | Mission Operations System (MOS) | 15 | 122 | 142 | 480 | 563 | |
| 9.0 | Ground System(s) | 13 | 281 | 329 | | | |
| 10.0 | Systems Integration and Test | 6 | 281 | 329 | | | |
| 8.0 | Launch Vehicle/Services ROM | | 500 | 500 | | | |
| | Total Phase A (BY20) | 211 | | | | | |
| | Total, Phases B/C/D (BY20) | | 6,037 | 6,481 | | | |
| | Total, Phase E (BY20) | | | | 480 | 563 | |

| | | | | | | TOTAL, PHASES A-E ($B) | |
|---|---|---|---|---|---|---|---|
| Subtotal ($M) | Low Range Total (50% CL), Phases A-E (BY20) | 211 | 6,037 | | 480 | | **50% CL** | $6.7 (BY20) |
| | Low Range Total (50% CL), Phases A-E (RY) | 247 | 7,700 | | 753 | | | $8.7 (RY) |
| | High Range Total (70% CL), Phases A-E (BY20) | 211 | | 6,481 | | 563 | **70% CL** | $7.3 (BY20) |
| | High Range Total (70% CL), Phases A-E (RY) | 247 | | 8,899 | | 881 | | $10.0 (RY) |

### 4.5.2 CEMA's Costing Methodology

All costing is consistent with the latest versions of NASA's Cost Estimating Handbook (CEH) and the NASA Space Flight Program and Project Management Requirements (NPR 7120.5). The parametric point-design cost models are based on a technical description of the *Origins* science mission implementation, as reflected in Master Equipment Lists (MELs), the mission integrated master schedule, and hardware heritage descriptions.

The PRICE Estimating Suite (PES) 2016 (16.0 build 5932.2) is used to estimate hardware costs. In PES, cost factors applicable to a Class A mission are adopted, the appropriate NASA New Start Inflation Index issued by the NASA Headquarters Office of the Chief Financial Officer (OCFO) is used, and the global input parameters are tuned to reflect the NASA business environment. The output data are in constant year 2018 dollars, and the top-level cost results are translated into constant year 2020 dollars. The constant year estimated cost is spread over time using the *Origins* mission schedule and the appropriate inflation index, to derive the cost profiles in real dollars (Figure 4-6). Inflation-adjusted historical actual costs are used in rare cases where the PRICE-H cost modeling tool lacks an appropriate Cost Estimating Relationship (CER). Thus, for example, some key components of the Adiabatic Demagnetization Refrigerators (ADRs) are based on actual Hitomi mission ADR costs. The FIP and OSS life-cycle detector costs are estimated by Subject Matter Experts.

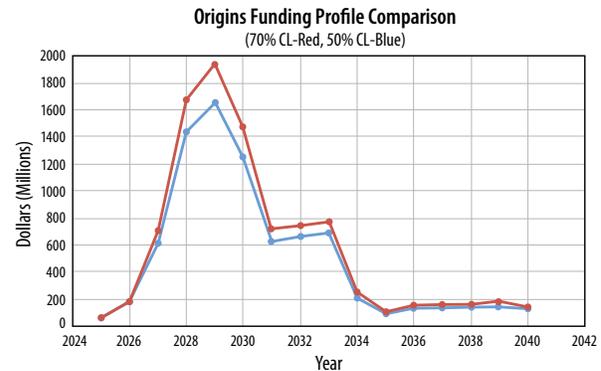

**Figure 4-6:** *Origins* funding profile in RY$ at the 50% CL (blue) and at 70% CL (red). The Launch Vehicle cost is not included here.

The *Origins* decadal mission concept study team developed detailed MELs and a detailed schedule, which served as inputs to the parametric cost models. Preliminary MELs for the *Origins* Telescope and the OSS and FIP instruments were products of separate studies conducted in the GSFC Instrument Design Lab (IDL) and were subsequently refined by the dedicated *Origins* study team. Likewise, the Spacecraft MEL originated in GSFC's Mission Design Lab (MDL) and was further refined by the study team. JAXA designed the MISC-T instrument and provided its MEL. The *Origins* point-design configurations upon which the MELs were based (*i.e.,* the baseline mission concept) represents the team's best effort to satisfy instrument and mission requirements and demonstrate the technical feasibility of the mission, but it is not yet a fully optimized design. Further work in the formulation mission phases will allow the design to be optimized for science return on investment.

As a NASA Class A mission, *Origins* satisfies requirements with respect to safety and mission assurance. These requirements affect subsystem redundancy throughout the spacecraft and instruments, the depth and extent of the qualification test program (engineering models, engineering test and qualification units), and the pedigree of electric, electronic and electromechanical (EEE) parts. The instruments, telescope and spacecraft bus use Class S EEE parts (as defined in the PRICE-H parametric cost tool) qualified to survive for a minimum of 5 years on-orbit. The mission cost estimate assumes consumables sufficient to operate for 10 years, allowing for the possibility of an extended mission.

Because we separately account for the cost of technology maturation, the *Origins* hardware parametric cost models for Phases B, C and D assume that the minimum component TRL is 6. Dedicated technology development funding is expected prior to mission implementation in pre-Phase A and Phase A.

Parametric point-design cost estimates for the hardware elements WBS 5 (Payload) and WBS 6 (Spacecraft) were derived first, and then "wrap factors" on the combined total of WBS 5 and WBS 6 were applied to derive the estimates for other WBS elements. In particular, wrap factors were applied to derive cost estimates for WBS 1 (Management); 2 (Systems Engineering); 3 (Safety and Mission Assurance); 4 (Science); 7 (Mission Operations); 9 (Ground System); and 10 (Systems Integration and Test). The result is a mission-level point-design estimate for WBS 1 through 10.

CEMA performed a cost risk analysis, which is intended to address probabilistically the risk that the parameters that characterize a mission may differ from the CBE selections entered into the parametric model. After developing the *Origins* point-design estimates for payload (WBS 5) and spacecraft (WBS 6), the CEMA Office ran Monte-Carlo simulation uncertainty analyses of WBS 5 and WBS 6 in the PRICE-H tool. Probability distributions were specified for mass and complexity model inputs. The mass uncertainty distribution uses the Current Best Estimate (CBE) as the minimum mass, CBE + specified Contingency for the Maximum Expected Value (MEV) (most likely), and MEV + specified Margin (25%) for the Maximum Possible Value (MPV). A mass contingency value is specified for each component in the MEL. The model's multiple complexity parameters are derived from the TRLs given in the MEL, which reflect component heritage. As with the mass inputs, CEMA uses a triangular probability distribution to reflect low, most likely and high values for the complexity parameters and treats complexity conservatively by adopting an increasing probability of greater complexity.

The CEMA Office performed a mission-level cost risk analysis with Automated Cost Estimating Integrated Tools (ACEIT) to generate a mission-level cumulative cost distribution function ("S-curve"), which captures the inherent risk of the flight hardware, as modeled with PRICE H, and applies it to the other WBS elements, effectively tying overall mission risk to flight hardware risk. In the ACEIT analysis, no risk was applied to Science (WBS 4), Launch Vehicle/Services (WBS 8) and Phase A because these cost elements are considered to be minimally impacted by flight hardware risk. For WBS 4, the point-design estimate with 25% reserve was passed through. For WBS 8, Launch Vehicle and Services, the cost provided by NASA HQ was passed through without reserve. For Phase A, the estimate was passed through without reserve.

The Phase-A cost estimate and the distribution of Phase A funds across the WBS elements (Table 4-2) was based on grass-roots understanding of the efforts required to conduct scientific and engineering analyses, complete technology development to TRL 6 by PDR, and manage the overall Phase A effort. Planned Phase A activities are outlined in Section 4.7.

Figure 4-7 shows the CEMA-derived S-Curve for *Origins*. The final cost estimates by WBS for both the low and high cost cases (Table 4-2) were spread over time using the *Origins* mission implementation schedule and calculated using the appropriate inflation index to arrive at Real Year dollars (RY$), as shown in Figure 4-6.

### 4.5.3 Independent Estimate of the Mission Cost by GSFC's Resource Analysis Office (RAO)

At the present stage of *Origins* mission design maturity, it is common practice to use the industry-standard parametric modeling tools employed by CEMA. Accordingly, we adopt the cost estimates given in Table 4-2 and Figure 4-6 as the best available estimates of the mission cost.

However, to add confidence to the CEMA estimates, GSFC's Resource Analysis Office (RAO) performed an independent assessment. RAO was chartered in 1976 by GSFC Center Management to provide independent, non-advocate assessments of cost and schedule risk for space flight missions. RAO relies on a database comprised of historical cost, technical, and programmatic data collected and normalized internally. For each major WBS element, RAO uses its database to develop top-down statistical models to predict cost and schedule based on mission characteristics. These models are built on actual data and represent cost and schedule for factors both internal and external to project control. (The CEMA-supplied cost phasing profiles shown in Figure 4-6 disregard external factors.)

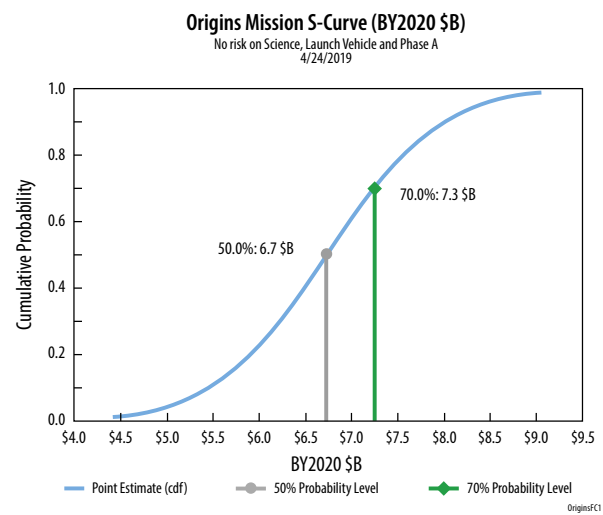

**Figure 4-7:** The CEMA-generated *Origins* Mission-Level S-Curve gives the likelihood that the cost will not exceed a particular value. Cost estimates at 50% CL and 70% CL in Base Year 2020 dollars are shown.

RAO presented 50% CL and 70% CL estimates for three cost and schedule scenarios to indicate the cost and schedule risk at the start of Phase A:

- **"Phase A Ready"** - According to the Management Plan for Large Mission Concept Studies (February 11, 2019), "The final study deliverable shall include: …Roadmap for maturation to both TRL-5 by the start of Phase-A and TRL-6 by the mission PDR." Thus, any project which has substantial recent heritage for all hardware that meets these standards would be afforded a baseline cost and schedule risk.
- **"New Engineering"** – This scenario pertains to a project with existing critical technology which cannot be descoped or utilized in a different way and has no new component technology. The cost and schedule risk reflects the additional effort to modify the hardware for new design, implementation, or engineering.
- **"New Technology"** – This scenario represents state of the art technology which does not exist and requires substantial maturation. The cost and schedule risk in this case accounts for the substantial amount of technology development.

Since the *Origins* Technology Development Plan elevates the maturity of all mission-enabling technology to TRL 5 by the start of Phase A and to TRL 6 by mission PDR, we disregard the "New Technology" scenario. In the spectrum of RAO scenarios, *Origins* lies between "Phase A Ready" and "New Engineering."

Figure 4-8 shows that the CEMA Office's cost estimates are bracketed by RAO estimates corresponding to the two pertinent scenarios at both 50% CL and 70% CL. This general agreement lends confidence to the estimated mission cost.

The *Origins* mission design will be optimized during Pre-Phase A and Phase A. The *Origins* decadal mission concept study team identifies several potential design changes that could yield cost savings, suggesting that the current cost estimate is conservative. For example, the current design can accommodate four, but includes only three, science instruments. A more accurate cost estimate will be derived during Phase A. Japan and several ESA member nations have significant relevant expertise and have demonstrated interest in the *Origins* mission. Foreign contributions are expected to reduce NASA's share of the mission cost.

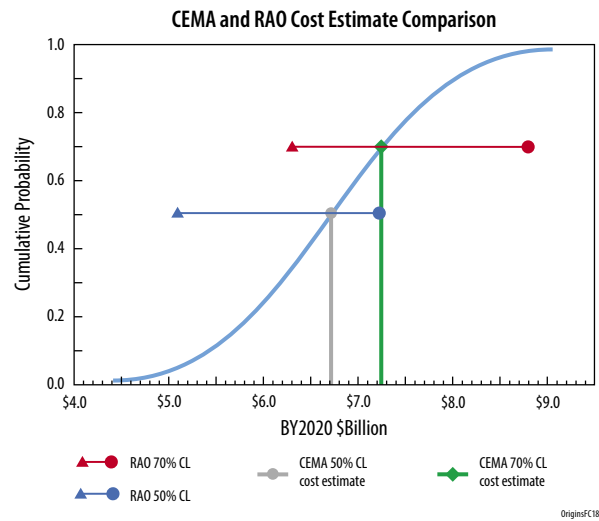

**Figure 4-8:** The independently derived CEMA and RAO cost estimates are consistent.

### 4.6 Mission Descopes and Trade Analysis for Origins Space Telescope Mission

The *Origins* baseline mission concept design results from a detailed review of mission descopes from the *Origins* highly capable mission concept 2 (Section 2). Some of the descopes, notably HERO and MISC-imager, are described in Appendix D on "upscope options" so that reviewers understand scientifically valuable capabilities that could be added to the baseline concept if so desired, but at a cost. Additional descope options were considered during the detailed review but rejected by the STDT as cutting too deeply into the science program for a large NASA mission designed to serve the entire astronomical community, or saving too little in cost for the science lost. Descope options rejected by the STDT are listed in Table 4-3. (A rejected option means that the capability is preserved in the baseline mission concept.)

Table 4-3 shows the possible descopes, the applicable wavelength and spectral resolving power, the science impact and the estimated cost savings. The science impact column documents the science objectives and programs that would have been eliminated if that descope option had been selected. The estimated cost savings are shown as a percentage of total mission cost and can thus be scaled to the mission cost calculated by the Study Center (Section 4.5) or an independent review committee. The estimated cost savings include the ripple effect that removal of an item has on other mission components.

Table 4-3 shows potential mission trade space that can be reconsidered before Phase A. A telescope size descope would be necessary before Phase A. Instrument descopes could be reconsidered during Phase A. A discontinuity exists in the telescope size vs. cost curve corresponding to the size

**Table 4-3:** Potential cost-saving descopes considered but rejected by the *Origins* STDT.

| Descope Option | λ (μm) | Spectral Resolving Power (R=λ/Δλ) | Science Impact | Cost Saving as a Percentage (%) of the total mission cost |
|---|---|---|---|---|
| Remove OSS Etalon (highest resolution spectroscopy capability) | 100 to 200 | 325000 at 112 μm | Reduces Theme 2 (Water and Habitability), Objective 1 (water mass in all evolutionary stages) by limiting Doppler tomographic measurements of water in planet forming disks. | 0.2% |
| Remove OSS Spectrometer Module Band 6 (reduces longest wavelength to 335 μm) | 336 to 589 | 300 | Reduces Theme 2 (Water and Habitability), Objective 1 (water mass in all evolutionary stages) by reducing accessible water lines. Reduces Theme 1 (Extragalactic), Objective 2 (rise of metals) by reducing [OIII] 88 μm line to z <3 (Partial recovery via mid-IR metallicity indicators that are not impacted). | 1.1% |
| Remove OSS Spectrometer Module Bands 6 and 5 (further reduction to longest wavelength, leading to maximum wavelength of 200 μm) | 200 to 350 | 300 | Above, plus Theme 2 (Water and Habitability), Objective 3 (Comet D/H) by reducing accessible HDO lines. Negatively impacts Theme 1 (Extragalactic), Objectives 1 to 3 with restrictions on fine-structure lines, including [CII] at z > 0.5. | 1.9% |
| Remove continuum and polarization mapping provided by the FIP Instrument | 50, 250 | 3.3 | Impacts Theme 1 (Extragalactic), Objective 1 with the removal of rapid mapping of large area extragalactic surveys and discovery space sciences involving small bodies in the Solar system, and large-scale Galactic magnetic field studies via polarization, for example. Partial recovery of imaging capability only (not polarization) by using OSS for continuum mapping, but at a factor of 102–103 slower mapping speed. Loss of functional role of FIP in alignment of the telescope. Loss of public relation images. | 15.5% |
| Remove MISC Transit short channel | 2.8–5.5 | ~50–100 | Negatively impacts Theme 3 (Exoplanet), Objective 1 (habitability indicators) and Objective 3 (biosignatures) by limiting access to $CH_4$, $N_2O$, $CO_2$, CO lines. | 3.5% |
| Remove MISC Transit Long channel | 11.0–20.0 | ~165–295 | Negatively impacts Theme 3 (Exoplanet), Objective 2 (surface temperature) by limiting the ability to conduct emission spectroscopy. | 2.2% |
| Reduce Telescope aperture diameter down to 5.3m from 5.9m | N/A | N/A | All three themes and nine mission design science objectives are preserved (Fig. E5.1-6), but the scientific case has no margin relative to the mission design. | 4.6% |
| Reduce Telescope aperture diameter to 5.0m from 5.9m | N/A | N/A | Negatively impacts Theme 3 (Exoplanet), Objective 3 (biosignatures) by limiting the signal-to-noise ratio of biosignature spectral features, given the finite number of transits visible in a 5-year design mission (Fig 1.3–11). Negatively impacts Theme 1 (Extragalactic), Objective 1 (star- formation/ blackhole accretion) by limiting z > 6 galaxy detections. | 6.9% |
| Reduce Telescope aperture diameter to 3.9m from 5.9m | N/A | N/A | Above and an additional negative impact on Theme 2 (water and habitability), Objectives 1–2 (water and disk mass with HD) by limiting the number of accessible targets to a number well below those needed to meet the statistics necessary for scientific objectives. | 22.6% |
| Reduce Telescope aperture diameter to 3.0m from 5.9m | N/A | N/A | Above and a further negative impact on Theme 1 (Extragalactic), Objective 1–3 via increasing source confusion for reliable source detection. Considered a minimum for any science capabilities over prior missions. | 29.5% |

(3.9 m diameter) below which a smaller launch vehicle could be used.

A warmer telescope temperature was considered and assessed to yield no net cost savings but to have science impact and thus not included in the table of descope options. The *Origins* baseline mission has a telescope and instrument bay temperature of 4.5 K, which makes emission from the telescope and reflective optics in the instruments negligible when compared to photon noise from the dark sky over all but the longest wavelengths of interest (see Figure 4-9). A hypothetical relaxation of the temperature to 6 K would impact Theme 1 (Extragalactic) and Theme 2 (Water) Objectives 1-3 by decreasing the sensitivity at wavelengths above 300 μm, with a factor of 3-6 sensitivity reduction at 300-588 μm. This would exhaust the science margin in the baseline design, and a slightly warmer temperature would erode *Origins*' science measurement capability.

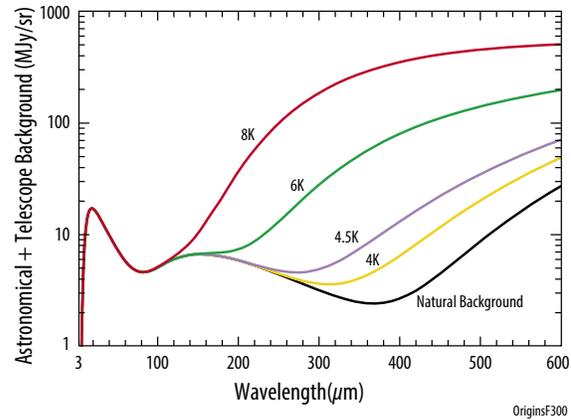

**Figure 4-9:** Increasing the *Origins* telescope temperature even slightly relative to 4.5 K pushes the telescope into a regime in which it dominates natural astronomical background photon noise and severely compromises *Origins* science.

The JWST cryocooler could satisfy the *Origins* cooling power requirements, but only at 6 K rather than 4.5 K. It could be argued that a 6 K descope would further increase the cooler TRL. However, the cryo-thermal design is a highly integrated and optimized system. Thus, increasing the telescope and instrument bay temperature has ripple effects on other elements in the system. In addition to the increased thermal emission from the telescope and instrument optics (about a factor of 10 at wavelengths ≥ 200 μm), the two far IR instruments would be negatively impacted. The far IR detectors require sub-Kelvin cooling. The highest temperature from which sub-Kelvin coolers have been demonstrated is 5 K (Hitomi). There is an ongoing SAT investigation to show ADR operation from 10 K, but this requires more stages of cooling and is less technologically mature overall than the adopted approach with a 4.5 K upper stage. There will also be other instrument effects, such as being unable to use superconducting motors at this higher temperature. The savings realized by not investing in mechanical cryocooler development would lead to a need for more investment in sub-Kelvin coolers, and no net savings would accrue to *Origins*. Finally, as a proof of feasibility for raising the US-provided 4.5 K cryocoolers to TRL 6, the Hitomi cryocoolers provided by Sumitomo Heavy Industries achieved 4.5 K, with the cooling power but not the lifetime (5 year expected life rather than 10) required for *Origins*. This cryocooler is a possible JAXA contribution to *Origins*.

## 4.7 Summary of Phase A Study Topics

*Origins* Phase A activities are scheduled to run for 1 ½ years beginning on 2 June 2025, by which date all of the *Origins* enabling technologies will have matured to TRL 5. During Phase A all of the technologies will substantially progress toward their TRL 6 milestones, and the *Origins* team will conduct additional scientific analyses, refine designs and test components of the science instruments, conduct engineering studies and tests, and address mission-level and programmatic topics. Completion of all Phase A tasks will prepare *Origins* for Phase B and a successful mission. Below is a preliminary list of Phase A study topics by category. This list is not comprehensive, but rather a compilation of the topics identified to date and discussed in this report.

## Science

- The science requirements and corresponding margins between predicted performance and requirements will be refined during Phase A in our Science Traceability Matrix (STM)
- The science impact of failure to meet specific requirements will be assessed, and the team will develop minimum success criteria for the mission.
- The *Origins* Design Reference Mission will be optimized considering all changes to instruments and payload refinements as well as spacecraft performance.

## Science Instruments

- Selection of far-IR and mid-IR detectors for the science instruments
- Refinement of the FTS and etalon and their use in OSS
- FIP rotating half-wave plate development
- On-board calibration capability for MISC-T
- Upscope considerations and analysis, particularly HERO and MISC-camera

## Technology

- Detector selection and advance toward TRL 6
- 4.5 K cryocooler selection and advance toward TRL 6
- Qualification testing of ancillary detection system components

## Engineering

- Thermal model refinement
- Optical wedge test structure analysis
- Sunshield deployment refinement
- Sunshield outer coating analysis and test
- Sunshield analysis to survive launch
- Aperture cover refinement and analysis
- Stray light analysis refinement
- Assess the efficacy of focus-diverse phase retrieval based on FIP 50 µm images
- Model on-orbit self-contamination
- Propellant impingement analysis
- Launch vehicle selection and loads analysis for launch

## Mission

- Detailed study of ground systems and data throughput for Design Reference Mission; consider lossless compression
- Beryllium manufacturing planning
- Detector Fabrication Facilities planning
- Launch vehicle planning
- Study serviceability in greater detail, including instrument replacement and refueling to extend mission life beyond 10 years
- Organizational planning and partner selection
- Documentation and guidance documents as required
- Grass roots cost analysis to complement existing current cost models
- Detailed risk analysis
- Descope planning

## APPENDIX A - DISCOVERY SCIENCE

Starting the first day of scientific operations, *Origins* is available to all astronomers via competitively reviewed proposals. The team anticipates the community will use the observatory for a wide range of purposes, including many that cannot be anticipated 15 years in advance of the mission beginning science operations in the mid-2030s. Current estimates are that many groundbreaking science programs can be achieved with roughly 100 to 400 observing hours. The wide range of scientific capabilities spans observations of targets in the Solar System – such as planets, satellites, and Kuiper Belt Objects – to the formation of the first galaxies during the epoch of reionization.

Enabled by its large, cold telescope and improved detectors, *Origins* provides a tremendous discovery space that facilitates breakthrough discoveries in all areas of astronomy through its dramatic increase in spectral line sensitivity. *Origins*' instrument suite provides a wide range of spectral resolution capabilities – from broadband imaging, to medium resolution spectroscopy, to high-resolution spectroscopy.

The *Origins* study has been driven by community science from the beginning. The top level-1 science objectives (Table ES-1) and discovery space programs (Table A-1) are a subset of the 46 science cases submitted by community members in the form of 2-page proposals in response to a 2016 open call for science programs with a future space observatory. These 46 science cases guided the development of *Origins* design choices and are shown in full in the interim report (*Origins* Space Telescope Interim Study Report 2018, see , see *https://asd.gsfc.nasa.gov/firs/docs/*) on Concept 1. Section A.13 describes how the top 25 proposals from this initial submission can be accomplished with the recommended Baseline *Origins* Concept.

The three primary science themes used to derive requirements for the *Origins* concept (**Sections 1.1-1.3**) could be accomplished within the first two years of operations, assuming the required 80% observing efficiency. However, in practice, *Origins'* actual scientific observations will be based on competitively-reviewed proposals selected from the astronomical community. Such proposals will guide the observatory science operations throughout its mission lifetime requirement of 5 years, with a goal of 10 years. The strength of a large, all-purpose space observatory like *Origins* is that it will expand the science of the time during which it operates, not only the science outlined in the mission's early planning phases. This section highlights potential discovery science programs (Table A-1).

**Table A-1:** *Origins* Discovery Science Highlights

| Section | Title | Instrument/Mode |
|---------|-------|-----------------|
| A.1 | Detection of warm molecular hydrogen from reionization | OSS spectroscopy of lensing galaxy clusters in search of $H_2$ emission lines from first galaxies during and prior to reionization. (270 hours for 3 clusters) |
| A.2 | Mapping Galaxy outflows in the nearby universe | Spectro-imaging with OSS of 158 µm [CII] line out to z < 1 to study gas outflows in 1000 nearby galaxies. (100 hours for 1000 galaxies) |
| A.3 | Wide-field mapping of molecular hydrogen in local group dwarf galaxies | OSS 28 µm spectro-mapping observations of the $H_2$ line to trace dark gas not seen in CO in all local group, star-forming, dwarf galaxies. (100 hours) |
| A.4 | Magnetic fields at galactic scales | FIP polarization mapping of magnetic fields in nearby galaxies. (270 hours for 300 nearby galaxies) |
| A.5 | Follow-up and characterization of LISA and LIGO gravitational-wave sources and other time-domain sciences | FIP fast-scanning mapping observations in search of electromagnetic signals of LISA and LIGO gravitational-wave events; transient identification and follow-up observations with OSS. (Varies, for LISA~75 mins to map 10 deg²) |
| A.6 | Time-domain sciences: proto-star variability as a probe of protoplanetary disk physics and stellar assembly | FIP photometric repeated monitoring observations to study accretion and formation of proto-planetary disks in the inner regions near the star. (100 hours for 500 proto-star monitoring campaign) |
| A.7 | Small bodies in the trans-Neptunian region: constraints on early solar system cometary source region evolution | FIP photometric mapping observations to measure the thermal emission from KBOs in the outer Solar System. (Repeated mapping of 1000 deg² over 400 hours) |
| A.8 | Origins studies of water ice in non-disk sources | Focused OSS studies of water vapor and water ice features. (50 hours to study 200 Galactic ice sources) |
| A.9 | Studying magnetized, turbulent molecular clouds | FIP polarization mapping observations of Galactic star-forming clouds and regions. (200 hours to map 14 star-forming clouds over 50 deg² areas/each.) |
| A.10 | Below the surface: A deep dive into the environment and kinematics of low luminosity protostars | OSS spectroscopic observations to study the accretion of and address issues related to luminosities of proto-stars. (100 hours for 1000 stars) |
| A.11 | Putting the Solar System in context: the frequency of true Kuiper-belt analogues | FIP 50 µm imaging for exo-KBOs around nearby stars. (50 hours for 50 exo-KBO targets) |
| A.12 | Giant planet atmospheres: templates for brown dwarfs and exoplanets | OSS spectroscopic observations of giant planet atmospheres. (100 hours for 4 giant planets) |

## A.1 Detection of Warm Molecular Hydrogen in the Dark Era of Galaxy Formation

Deep spectroscopic observations used to search for warm molecular hydrogen and other key fine-structure cooling lines in gas heated as it collapses inside massive dark matter halos at $7.5 < z < 8.5$ provides a direct measure of the mechanical heating of molecular gas as the first galaxies form. The observations, aimed toward rich clusters, make use of gravitational lensing amplification to search for faint molecular hydrogen line emission from distant galaxies. *Author: Phil Appleton (IPAC)*

**Introduction:** When probing galaxies in the early Universe, *Origins* offers a unique advantage, enabling new observations of key cooling lines of low-metallicity gas in the rest-frame mid-infrared, such as fine-structure lines (*e.g.*, *[Si II]* at 34.8 μm, *[Fe II]* at 26/35.4 μm) and molecular hydrogen lines (*e.g.*, $H_2$ 0-0 S(1) and 0-0 S(3) at 17 and 9.7 μm). $H_2$ lines in particular have a singular importance because they allow *Origins* to identify and study galaxies forming out of extremely metal-poor (or even metal-free) gas in which cooling via fine-structure lines is no longer effective. In this type of low-metallicity environment ($< 10^{-4} – 10^{-3}$ $Z_\odot$), gas cooling is dominated by Lyα and $H_2$ pure-rotational lines, provided that galaxies are massive enough (*e.g.*, $M_{halo} > 10^8$ $M_\odot$ at z ~ 10) to maintain HI atomic cooling (*i.e.,* virial temperature > $10^4$ K).

Since Lyα emission can be easily extinguished by increasingly neutral intergalactic medium (IGM) at high redshift (z>6), $H_2$ emission lines will serve as a powerful (and maybe the only) probe for such first-generation galaxies forming in an extremely metal-poor environment.

For example, the model by Omukai and Kitayama (2003) predicts luminosities of $1.5 \times 10^7$ and $3.3 \times 10^7$ $L_\odot$ for the $H_2$ 0-0 S(1) and 0-0 S(3) lines produced by a giant Pop III forming galaxy at z=8 with $M_{halo} = 10^{11} M_\odot$ the optimal halo mass for maximizing $H_2$ line luminosities. Subsequent studies are in line with this level of $H_2$ line luminosities for most $H_2$-luminous cases (*e.g.*, Mizusawa *et al.*, 2005; Gong *et al.*, 2013). Recently, Liu & Bromm *et al.* (in preparation) have simulated gas collapsing in different forms of massive halos (Warm and Cold Dark Matter). Their models include stellar feedback at high redshifts (7.5-15), which turns on and off periodically with a duty cycle of roughly 100-200 million years.

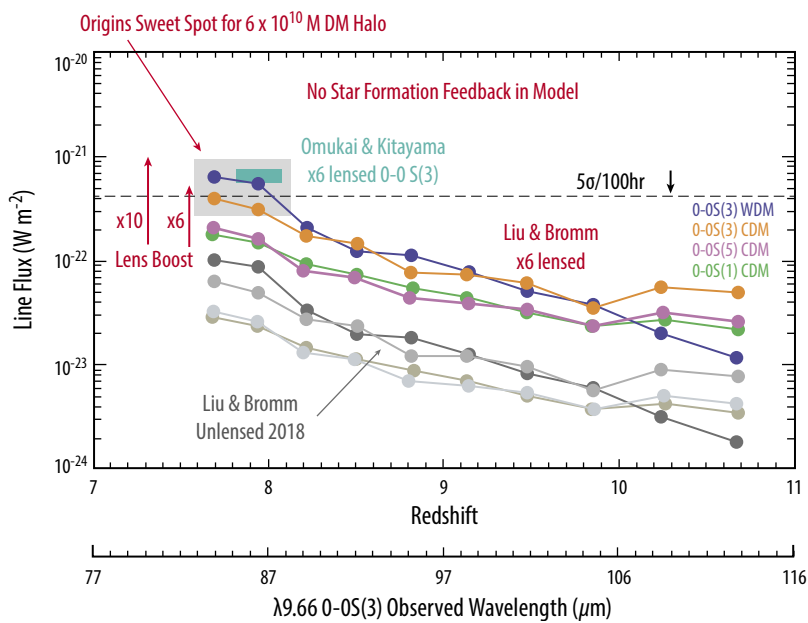

**Figure A-1:** *Origins*/OSS is capable of detecting the molecular hydrogen emission related to primordial gas cooling to form first galaxies. Predictions for 6x lensed (colored lines) and un-lensed (grey lines) for $H_2$ line emission as a function of redshift for a DM halo of mass $6.9 \times 10^{10}$ $M_\odot$, extrapolated from the models by Lui & Bromm et al. (in prepration) and the models of Omukai & Kitayama (2003) (pale blue horizontal line). It also shows the 5σ/100hr sensitivity threshold for OSS for the 80-120 μm (Band 3) detectors (black horizontal line). Detection is only possible in the lower-z range. *Origins* adopts a fiducial observation time of 30 hours per pointing, which includes many beams across the target cluster.

Most of the gas cools through virial shocks, emitting energy in the low-lying (rest frame) mid-IR rotational $H_2$ lines, which are shifted into the far-IR at high-z (6-15). This is an ideal wavelength range for *Origins*/OSS. In this regime, cooling is through $H_2$ and several other important mid-IR lines. The expected line luminosity is at the level of $3.3 \times 10^7$ $L_\odot$ at z=8 (the 0-0 S(3) case). This corresponds to a line flux of $1.6 \times 10^{-22}$ W m$^{-2}$. This flux limit for direct detection is roughly a factor of 5 below the *Origins* 100-hr 5$\sigma$ detection limit (Figure A-1), although the team proposes taking advantage of gravitational lens amplification as a way of increasing the potential of detecting these weak signals.

The $H_2$ luminosities adopted for this feasibility calculation are probably uncertain by at least factor of 10 in either direction. For example, inclusion of magnetic fields in the model leads to generation of MHD C-shocks that are less compressive, generating cooler post-shock gas that shifts the radiating power to the lower-lying rotational lines (especially $H_2$ 0-0S(1)17 um and 0-0S(3)9.7 um). Such models explain well the extreme $H_2$ luminosities of some low-redshift objects, such as Stephan's Quintet (Appleton *et al.*, 2006; Guillard *et al.*, 2009; Lesaffre *et al.*, 2013; Appleton *et al.*, 2017), and the inclusion of MHD shocks may enhance the detectability of high-z $H_2$ lines with *Origins*. Such effects are not included in the predictions here. Alternatively, some forms of stellar feedback can have the opposite effect, heating (and dissociating) the $H_2$ and shifting the cooling power to higher-order pure rotational and ro-vibrational transitions, which emit at shorter wavelengths.

**Proposed Observations:** The *Origins* targets make use of massive clusters of galaxies at redshifts of 0.5-1.5 as powerful gravitational lenses to enhance the faint $H_2$ emission. Figure A-1 shows the predicted model line fluxes expected from the recent numerical models of Lui & Bromm *et al.* (in preparation) for a massive halo of mass $6.9 \times 10^{10}$ $M_\odot$ and specific rotational lines of molecular hydrogen as a function of redshift. The colored curves highlight the expected fluxes from several rotational $H_2$ lines with a lensing magnification factor of 6. As shown, several of the halo models predict emission line strengths that fall within the 5$\sigma$ detection threshold at z = 8 for *Origins*/OSS. The 0-0S(3) 9.7-$\mu$m line is the strongest line in these models, for warm and cold dark matter. For comparison, Figure A-1 also shows a prediction for z = 8 from earlier Omukai and Kitayama (2003) models. It is clear these models predict similar values (within a factor of 2) for the strength of emission in the 0-0S(3) line.

At redshift z = 8, one dark matter halo of mass between 4 and 8 x $10^{10}$ $M_\odot$ is expected to be present within a co-moving volume of 130 Mpc$^3$ (Reed *et al.*, 2003). This volume can be sampled by mapping the lensing caustics of one rich cluster with an amplification factor of 6. To obtain adequate statistics, at minimum, a default program aimed at molecular hydrogen should map three lensings clusters, each with three multi-beam pointings of OSS. To reach the line flux limit each pointing requires 30 hrs of observations, and with three pointings per cluster, leading to a program of 90 hrs per cluster. For three clusters, the total request is 270 hours targeting the 80-120 $\mu$m range for the 0-0S (3) line. If the 0-0 S (1) 17 $\mu$m line is stronger (due to lower temperatures than predicted by the current models), one might expect even stronger lines at 155 $\mu$m. If at least two lines are detected, this would enable a unique line ID and redshift determination. Alternatively, the [*SiII*] 34.8-$\mu$m and [*FeII*] 26/35.4-$\mu$m lines are likely to be as strong or stronger than the $H_2$ lines (Santoro & Shull, 2006), even in low metallicity gas, and these could be used to isolate the redshift of the $H_2$. For low-redshift systems, like Stephan's Quintet, the [*FeII*] and [*SiII*] lines are detected in low-velocity (5-10 km s$^{-1}$ C-shocks) at about the same level as the $H_2$ lines (Appleton *et al.*, 2006; Cluver *et al.*, 2010).

**Scientific Importance:** In 2020-2030, facilities such as JWST, WFIRST, and *Euclid* will probe the rest-frame UV/optical/near-IR photons from the earliest cosmic epochs. Proposed *Origins* observations will provide a crucial and unique missing piece of the galaxy formation puzzle, particularly emission from the darkest era before the onset of major star formation, where gas collapses to form proto-galaxies in the most massive halos at z = 8. *Origins* will be capable of potentially detecting multiple rotational transitions of $H_2$ as well as other key fine-structure lines, like [*SiII*]34.8 μm and [*FeII*]26/35.4 μm, another important coolant in primordial collapse of gas in massive (4-10 x $10^{10}$ $M_\odot$) DM halos.

With these observations, *Origins* will address the critical question of how much mechanical energy is being dissipated through shocks and turbulence in the formation of the first galaxies. These observations provide a completely different, but complementary, view of the formation of the first galaxies by providing unique insight into the molecular gas phase that precedes the first major star formation in galaxies mapped out by JWST.

### A.2 Mapping Galaxy Outflows in the Nearby Universe

> An *Origins*/OSS spectroscopic mapping program to observe the dust continuum and far-IR cooling line emission (particularly [*CII*] 158 μm) from outflowing material around nearby galaxies provides a comprehensive view of outflow characteristics across a statistically-meaningful sample of nearby galaxies and provides critical constraints on models of star formation and galaxy evolution. *Author: Alberto Bolatto (Maryland)*

**Introduction:** The effect of star formation and supermassive black hole-related feedback on galactic scales is one of the key open questions in astrophysics. Feedback is a fundamental ingredient in galaxy evolution, yet it is poorly understood and so constitutes a major unknown in galaxy evolution models. One thing feedback is thought to control is galaxy properties, such as galaxy mass function, quenched fraction, existence of a galaxy main sequence, and enrichment of the interstellar and circumgalactic medium. A key *Origins* goal is to characterize the mechanisms of feedback and quantify the amount and type of material ejected across the spectrum of galaxy masses and types in the local universe.

**Scientific Importance:** Because galaxy outflows consist of multiple phases of gas (ionized, neutral, molecular), it is critical to observe tracers that span these phases. The far-IR provides a unique window to probe the energetics and masses of the major outflow constituents from feedback-driven outflows using the dust continuum and far-IR fine structure lines. The [*CII*] 158-μm line especially is a valuable outflow tracer since it arises from all phases of the gas and is detectable even from low column densities of outflowing material. Figure A-2 shows predictions from numerical galaxy simulations with the expected surface brightness of the [*CII*] 158-μm line.

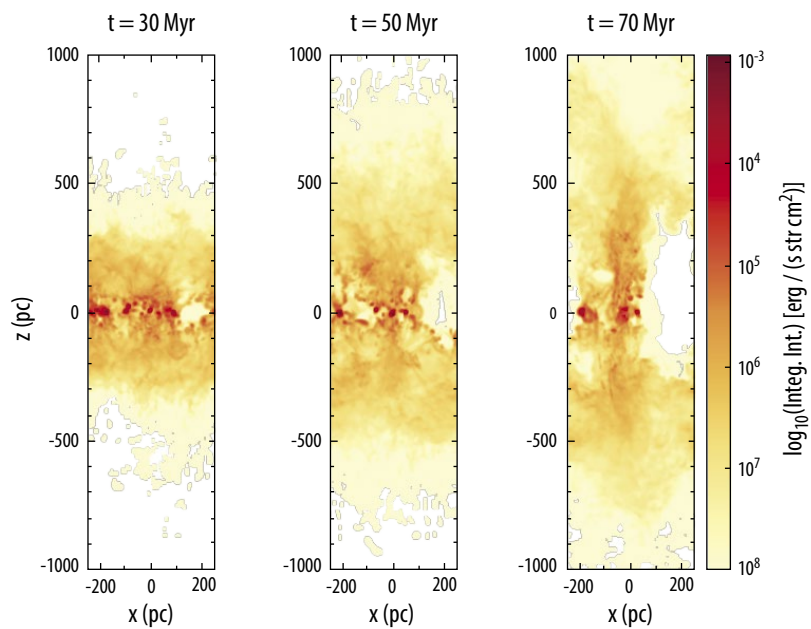

**Figure A-2:** Extraplanar [CII] 158 μm emission as a probe of outflow activity. Simulations of stellar feedback in a galaxy disk, showing star formation driving galactic outflows visible in [CII] emission. The calculations use the SILCC simulations and include radiative transfer to compute [CII] emission (Franeck et al., in prep; Walch et al., 2015). The *Origins*/OSS instrument can efficiently map the areas around nearby galaxies to the depth needed to detect this faint extraplanar [CII] emission.

**Proposed Observations:** To characterize the outflowing material from galaxies, the team uses *Origins*/OSS to perform low resolution (R=300) spectral mapping of extraplanar [*CII*] and dust emission. OSS Band 4 covers the [*CII*] 158-μm line. The faint, off-plane emission in [*CII*] has integrated intensities of $\sim 10^{-8}$ erg/s/cm$^2$/sr. To strongly detect that emission requires sensitivity an order of magnitude better. With *Origins*' beam area at 158 μm, the 1σ sensitivity is therefore $\sim 10^{-21}$ W/m$^2$. Using a Galactic carbon abundance (C/H = $1.6 \times 10^{-4}$), this sensitivity limit corresponds to N(H)$\sim 4 \times 10^{18}$ cm$^2$, assuming $n_H \sim 1$ cm$^{-3}$ in the high-temperature limit (Goldsmith *et al.*, 2012).

With *Origins*/OSS a 3,000-hour survey can cover 10 deg$^2$ to a 1σ sensitivity of $1 \times 10^{-21}$ W/m$^2$ at Band 4. To map a 1-arcmin$^2$ strip across $\sim 1000$ nearby galaxies would require a total area coverage of $\sim 0.3$ deg$^2$. This area can be mapped to the required depth in $\sim 100$ hours of OSS observing time.

The OSS instrument also simultaneously detects dust continuum emission. At the proposed line sensitivity, the 5σ continuum depth is 12.9 μJy at Band 4. Assuming the Planck Collaboration (2014) value of the 160 μm dust emissivity N(H)/$\tau_{160} \sim 1.1 \times 10^{25}$ cm$^{-2}$ measured in the diffuse Galactic ISM, a column density of N(H)$\sim 10^{18}$ cm$^{-2}$ appropriate for these outflows would have $\tau_{160} \sim 10^{-7}$. A blackbody at 30 K with this optical depth would emit $5 \times 10^{-22}$ W/m$^2$/Hz/sr. Thus, the equivalent signal in continuum would be 50 μJy, assuming the *Origins* beam at that wavelength. Therefore, OSS observations should easily detect the dust continuum from the outflow.

**Results:** A 100-hour program with *Origins*/OSS can survey [*CII*] and the dust continuum among the many other tracer observables from 25-588 μm for $\sim 1000$ galaxies. These *Origins* observations conclusively address the frequency of outflows, mass outflow rate, and characteristics of outflowing material in a statistically meaningful sample of z$\sim 0$ galaxies.

## A.3 Wide-field Mapping of Molecular Hydrogen Emission in Local Group Dwarf Galaxies

> With *Origins*/OSS spectral mapping, it is possible to observe the 28 μm rotational line of H$_2$ across all star forming dwarf galaxies of the Local Group (excluding the Magellanic Clouds). These observations provide critical insight into the properties of molecular gas in conditions where most of the H$_2$ is not traced by carbon monoxide. *Origins*/OSS surpasses JWST's mapping speed at 28 μm by many orders of magnitude, making large area maps of H$_2$ emission feasible for the first time. *Author Karin Sandstrom (UCSD)*

**Introduction:** H$_2$ is the most abundant molecule in the ISM, but its rotational levels require relatively high temperatures to excite (E$_u$/k$\sim 510$ K for the first rotational level 0-0 S(0) at 28 μm). Because of this, most H$_2$ emission arises from regions with ample far-UV radiation (*e.g.*, photodissociation regions) on the boundaries of molecular clouds or in regions where shocks contribute significantly to heating molecular gas. Observations of higher rotational and vibrational transitions of H$_2$ can be done from the ground or with mid-IR observatories like JWST, but these lines trace a much smaller fraction of the warm molecular gas component. Most studies of the rotational H$_2$ ladder find that the 0-0 S(0) line is dominated by relatively cool gas at temperatures around 100-300 K (*e.g.*, Roussel *et al.*, 2007).

The H$_2$ 28 μm line is therefore a powerful diagnostic of phases of molecular gas not detected by standard tracers of cold gas, like the mm-rotational lines of CO. In particular, in the relatively low extinction boundaries of molecular clouds, CO may not exist due to photodissociation, leading to layers of "CO-dark $H_2$" that require other observational tracers. The CO-dark layer is predicted to become the dominant reservoir of $H_2$ at low metallicity due to the lack of dust shielding (Wolfire *et al.*, 2010; Glover & Clark, 2012). In these conditions, where far-UV photons are present and the gas is molecular, the 28-μm line will trace a larger component of the gas reservoir. Other potential tracers include the [*CII*] 158-μm line, which is the dominant reservoir of carbon not in the CO form. The full diagnostic power of the $H_2$ 28-μm line as a tracer of CO-dark $H_2$ has yet to be exploited due to the limited mapping capabilities of previous (*e.g.*, *Spitzer*) and planned (*e.g.*, JWST) facilities.

**Scientific Importance:** Although JWST covers the 28-μm spectral region with MIRI IFU, its decreased sensitivity at these longer mid-IR wavelengths means that observing $H_2$, even in nearby galaxies, remains challenging into the 2030s. *Origins*/OSS provides orders-of-magnitude increase in sensitivity and mapping speed. Combining these effects, OSS (in R=300 mode) maps the 28-μm line to a given sensitivity 10,000 times faster than JWST. In addition, OSS in R=300 mode delivers a full spectrum from 25-588 μm, containing a wide array of lines that trace ISM gas in multiple phases as well as dust continuum emission, particularly the [*CII*] 158-μm line. OSS $H_2$ 28 μm mapping is highly complementary to ALMA observations of CO in nearby galaxies.

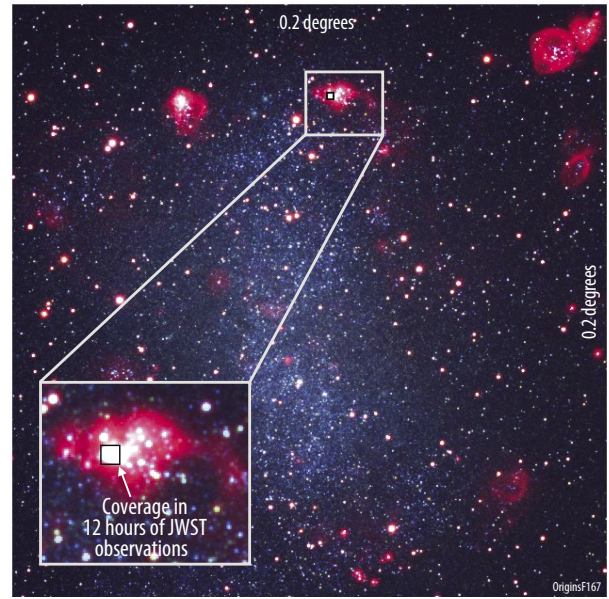

**Figure A-3:** The unprecedented sensitivity and mapping speed of *Origins*/OSS enables wide-field maps of $H_2$ 28 μm rotational emission for the first time. This figure shows the coverage of an 0.04 deg² map of the Local Group dwarf NGC 6822 (color is U, V, and R band imaging from Hunter et al., 2012), which would require 12 hours of OSS observations to obtain the required sensitivity. The area JWST could observe to the same depth in 12 hours is shown in the inset.

**Proposed Observations:** With a 100-hour mapping program, *Origins*/OSS can observe the $H_2$ 28-μm line across all star-forming Local Group dwarf galaxies (Figure A-3). These low metallicity environments are prime examples of regions where most of the $H_2$ is in a CO-dark phase (Rubio *et al.*, 2015; Schruba *et al.*, 2017). The team used example *Spitzer*-IRS observations of the Small Magellanic Cloud to predict the necessary sensitivity to observe the $H_2$ emission in such conditions. These data show typical $H_2$ 0-0 S(0) intensities of ~$1\times10^{-9}$ W/m²/sr in star forming regions (Sandstrom *et al.*, 2012, Jameson *et al.* 2018). With *Origins'* resolution, detecting such regions would require 5σ line sensitivity of ~$2\times10^{-20}$ W/m². In 100 hours, *Origins*/OSS could map 0.3 deg² to this depth, which would be sufficient to cover the full star-forming extent of all Local Group dwarf galaxies (*e.g.*, NGC 6822, WLM, Sextans A, Sextans B, IC 10, and IC 1613, ~0.26 deg² total), excluding the Magellanic Clouds. These *Origins*/OSS observations, in combination with ground-based CO mapping, would enable unprecedented characterization of the state of the warm and cold $H_2$ gas in low metallicity environments dominated by CO-dark molecular phases.

## A.4 Magnetic Fields at Galactic Scales

An imaging polarimetric program of 270 nearby galaxies results in a 25-fold increase in the number of resolved galaxies with detectable polarized dust emission. These observations characterize magnetic field morphology and strength for a large variety of galactic types, ages, and masses. *Origins*/FIP can solve the mystery of the role magnetic fields play in galaxy formation at scales of a few to tens of kpc. *Author: Enrique Lopez-Rodriguez (Ames)*

**Introduction:** Over the past few decades, astronomers have detected the presence of magnetic fields in galaxies at all range of scales, finding them to be ubiquitous on extragalactic sources. These major studies have been performed using optical and radio observations (for a review see Kronberg, 1994; Beck, 2015). Specifically, magnetic field properties have been inferred through the polarization that arises from extinguished light in the interstellar medium (ISM, Figure A-4-left), and Faraday rotation by synchrotron emission (Figure A-4-right). Polarized extinction traces ordered fields (Scarrott *et al.*, 1991), while synchrotron unpolarized emission traces turbulent fields (Fletcher *et al.*, 2011). Magnetic fields provide additional pressure to the ISM and intergalactic gas, couple cosmic rays to the non-relativistic gas, and in specific ratios between the magnetic pressure to the dynamical and thermal energy of the gas, can dominate galaxy dynamics.

The strongest ordered magnetic fields are detected between the optical arms of spiral galaxies (*e.g.*, Figure A-4-left). However, there is only an inconclusive theory to explain these results. For example, if the mean field dynamo in the interarm region is weaker due to outflows driven by star-formation (Chamandy *et al.*, 2015), the magnetic arms are a step in the evolution of galaxy formation (Dobbs & Baba, 2014) or the supernova shock fronts inject and amplify turbulent fields in the interarm regions (Moss *et al.*, 2013). Although concerted efforts (*e.g.*, Li *et al.*, 2015) have been made to explain the role of the magnetic field in the ISM, they are usually ignored in galaxy formation/evolution, and a lot of simplifications are made in magneto-hydrodynamical (MHD) models (*e.g.*, Pakmor & Springel, 2013). In addition, existing MHD simulations explaining the role of the magnetic field in galaxies are more advanced (*e.g.*, Ruszkowski *et al.*, 2017) than any observations currently possible. The observing capability to explain the role of magnetic fields in galaxy evolution is necessary and is fulfilled by *Origins*.

**Scientific Importance:** Although it is known, through observations, that magnetic fields are ubiquitous in galaxies and, through models, that they play a role in galaxy formation and evolution, major open questions still remain, including:

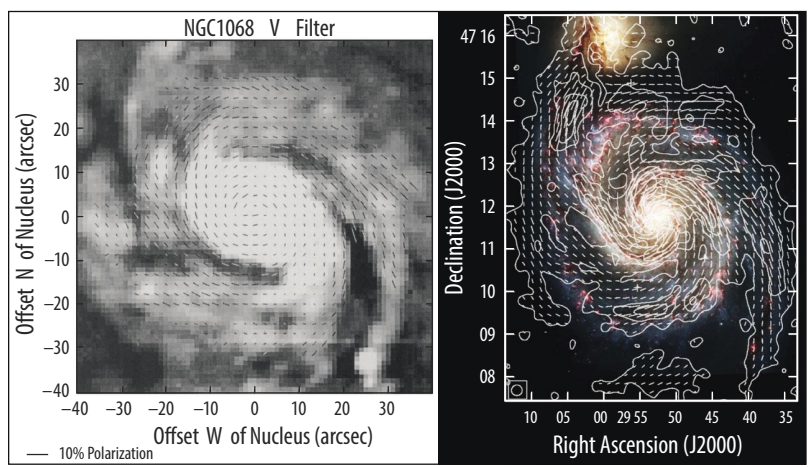

**Figure A-4:** Polarization mapping traces the large-scale magnetic fields of galaxies. Left: The kpc-scale magnetic field of the spiral galaxy, NGC 1068, inferred using optical polarization (Scarrott et al., 1991). The measured polarization at kpc-scales follow the spiral arms, while the central few hundred pc shows a centrosymmetric polarization pattern dominated by dust scattering and no information about the magnetic field can be inferred. Right: The kpc-scale magnetic field of the spiral galaxy, M51, inferred using 6 cm wavelength with the VLA and Effelsberg telescopes (Fletcher et al., 2011). The measured polarization at kpc-scales follows the interarm along the spiral arms.

- How well does magnetic field morphology align with spiral arms?
- How do galactic outflows drive magnetic fields into the extragalactic environment?
- Can this magnetic field magnetize the intergalactic medium?
- What is the origin of the magnetic arms?
- Is there a correlation between small-scale magnetic field amplification in star-forming regions and large-scale fields?
- Can the galactic magnetic field affect the general gas flow of the galaxy?

SOFIA with its new FIR polarimeter, HAWC+, is currently the only facility that can provide new insight with empirical evidence to answer these open questions. The first step (Jones *et al.*, 2019; Lopez-Rodriguez *et al.*, 2019) was a study performed using HAWC+ polarimetric observations of the spiral galaxy NGC 1068, and the starburst M82 at 59 and 83 μm (Figure A-5). For NGC 1068, HAWC+ observed an ordered magnetic field confined to the plane of the galaxy and co-spatial with the spiral arms on scales of 10 kpc. For M82, HAWC+ observed a magnetic field structure perpendicular to the plane of the galaxy on the sky on scales of 3 kpc, and co-spatial with the galactic outflow driven by star forming regions at the core of the galaxy. Although both galaxies have unique classifications, the results show the feasibility of finding magnetic fields in galaxies using FIR polarimetric techniques. However, due to its low sensitivity (several Jy) and moderate angular resolution (>4"), SOFIA/HAWC+ can only perform these observations for a small sample (tens of galaxies). A comprehensive study over a variety of galaxy types, ages, and masses can only be performed by *Origins*.

**Proposed Observations:** *Origins*/FIP's unprecedented sensitivity (~sub-μJy at 50 μm) and angular resolution (2.5 arcseconds at 50 μm) enables it to observe hundreds of galaxies, with the core and host galaxy resolved at a scale of hundreds of pc. Using these data, the team can distinguish the contribution of the star forming regions, active nucleus, and host components (*i.e.,* core, bulge, disk, arms) of the galaxies, which is required to understand how the gas flows and magnetic fields interact.

*Origins* is sensitive to dust emission in the host galaxy and star forming regions with temperatures in the range of 10-100 K. As shown in Figure A-5, the cold dust of host galaxies is detectable at 53-100 μm. *Origins*/FIP will be sensitive to polarimetric observations of 270 galaxies within 100 Mpc (<26 kpc/arcmin) and sizes ~2-45 arcmin. The minimum size of the galaxy at 2 arcmin is based on the resolvability at 250 μm. A galaxy size of 2 arcmin ensures a well resolved galaxy with 10-beam sizes at 250 μm, which allows >100 independent polarization vectors per galaxy. This angular resolution is enough to estimate the magnetic field strengths (Chandrasekhar & Fermi, 1953) as a function of distance to the galactic center. All nearby galaxies within 100 Mpc that have been observed with *Herschel* can be resolved at scales of hundred of pc and observed in the *Origins*/FIP polarimetric mode. Based on the *Origins*/FIP sensitivities and aiming to perform a multi-wavelength polarimetric study within 50-500 μm, each galaxy requires a 3σ detection in the degree of polarization, with typical degree of polarization in the galaxy of 1%, with a polarization uncertainty of 0.3%. This observation will require an average of 20 minutes on-source time at 50-100 μm, and 50 minutes on-source time at 250-500 μm per galaxy; therefore, the required time for this study is ~330 hours on-source time. Each galaxy requires a 3σ detection in the degree of polarization, with typical degree of polarization in the galaxy of 1%, with a polarization uncertainty of 0.3%. *Origins*/FIP provides a definitive study on the structure of the magnetic field in regular, active, and merging galaxies. *Origins*/FIP also reveals groundbreaking insights into galaxy evolution by providing empirical evidence of galactic magnetic fields.

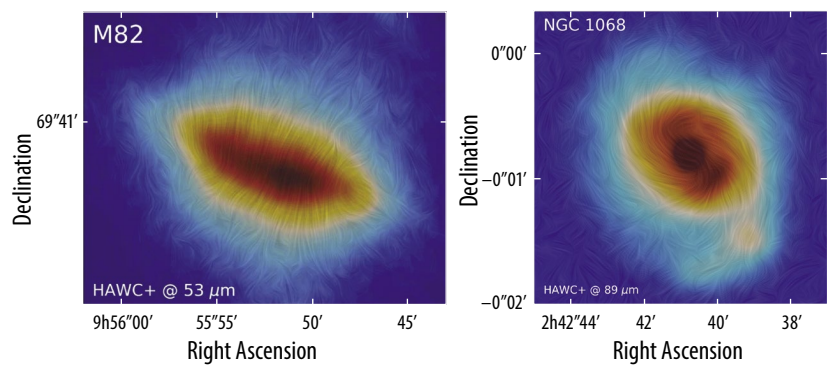

**Figure A-5:** SOFIA/HAWC+ polarimetric observations of M82 (left) and NGC 1068 (right). Total intensity (color scale) and magnetic field morphology inferred from the 90° rotated position angle of polarization is shown. Galactic outflows in M82 (Jones et al., 2019) are dragging the magnetic field out of the galactic plane enriching the intergalactic medium with dust at scales of several kpc. For NGC 1068 (Lopez-Rodriguez et al., 2019), the magnetic field is confined on the galactic plane and follows the spiral arms and gas flow at scaled of tens of kpc.

### A.5 Follow-up and Characterization of LISA and LIGO Gravitational-wave Sources and Time-domain Science

> Rapid mapping (at 60 arcseconds/second scan rate) of relatively large areas corresponding to LISA, LIGO, and other transient error boxes of gravitational wave events, supernova explosions, and time-domain explosive transients identify some key targets-of-opportunity to improve our understanding of the dynamical universe. The infrared also allows spectroscopy of buried sources, generally obscured by dust in the UV and optical, using capabilities that only *Origins* has in the mid 2030s. *Authors: Douglas Scott (UBC), Ryan Lau (Caltech), Mansi Kasliwal (Caltech)*

#### A.5.1 Gravitational Wave Source Follow-up

The Laser Interferometer Space Antenna (LISA; Amaro-Seoane *et al.*, 2017) mission is expected to launch in the early 2030s and have a 5-10 year mission lifetime. Similarly, from the ground, continued operation of gravitational-wave observatories, such as LIGO, is expected. It is likely that *Origins* is the only mid-to-far-IR mission operating at the same time as LISA. The most exciting science expected to come from LISA is detection of the gravitational waves (GWs) from merging supermassive black holes (SMBHs) in distant galaxies, covering the mass range $10^3$-$10^7$ M$_\odot$ over all redshifts. LISA also detects extreme ratio inspirals of compact objects with SMBHs, as well as binary stars in the Milky Way (and potentially new classes of sources). These GW events involve mergers of compact objects and blackholes that are in the few to tens of solar masses.

For all GWs, one of the key scientific goals is to identify the source in electromagnetic (EM) radiation to perform detailed observations during a merger event. EM radiation detection leads not only to redshift, but also to an improved understanding of the physics related to the merger. Finally, with the redshift established from EM radiation, there are a number of applications related to combined GW and EM observations, including measurements of cosmological distances independent of a distance ladder resulting in measurements of cosmological parameters and dark energy, among others.

For LISA, which focuses on gravitational wave observations from space, the most interesting sources are expected to be the rare SMBHs. Pre-dictions for rates are very uncertain, but it is anticipated there are several such events per year, and that most are detected days (or in some cases, months) before the final merging event. Characterization of the first few SMBH merger candidates from LISA is surely a major endeavor for many of the world's observatories. Since no such merging event has yet been observed, predictions for the EM counterparts have a huge degree of uncertainty (*e.g.*, Dotti *et al.*, 2006; Haiman *et al.*, 2009; Schnittman & Krolik, 2008; Tanaka & Menou, 2010) and have mostly focused on optical and X-ray observations. There are no useful predictions related to the expected IR signatures, given the lack of facilities to perform such measurements to date. Finally, there are also no model predictions of what is expected as an EM precursor, or whether it is a simultaneous burst or an afterglow. However, super-massive blackhole mergers are expected in the cores of merging galaxies (or AGNs). Based on observations so far, there is a high probability of any emission from the centers of such galaxies to be obscured at optical wavelengths. Just as for the study of obscured quasars and dusty starforming galaxies at high redshifts, wavelengths longer than the near-IR hold the promise of being able to directly image the merging region, as the long-wavelength infrared radiation penetrates the dust. Even if some optical light gets out, *Origins'* wavelength range probes deeper into the central accretion region. And even if there is a substantial UV or optical flash within the central regions of the galaxy during or simultaneous with the GW burst, such radiation is expected to heat the gas and result in an afterglow in the far-infrared.

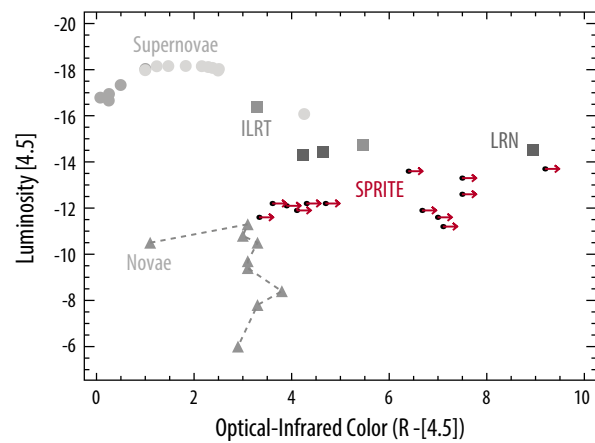

**Figure A-6:** Luminosity-color rendition of the phase space of IR explosions showing optical-infrared color on the x-axis. Since SPRITEs are not detected in the optical, they are shown as limits (red arrows). SPRITEs are much redder than novae (gray triangles) and supernovae (gray circles) and the limits are comparable to late-stage stellar mergers (LRN; black squares) and electron-capture supernovae (ILRT; gray squares). Figure from Kasliwal et al. (2017).

Identifying the counterpart galaxy and obtaining a redshift is critical to answering most of the fundamental astrophysics and cosmology questions that immediately come from the detection of SMBH GW sources: tracking H(z) through standard sirens; understanding the relationship between accretion and star-formation; constraining physics through time delays of photons and gravitons; etc.

***Origins* Capabilities:** LISA's positional uncertainty for SMBH inspirals is expected to be ~10 deg$^2$ in the best cases. It is unclear how much this might be narrowed down using other telescopes, and hence it is desirable for *Origins* to quickly map a field of this size. *Origins*/FIP could cover 10 deg$^2$ at 50 or 250 µm in 75 min, reaching 5σ detection limits at the level of the confusion limit. Comparison with existing long-wavelength maps of the sky would enable newly bright sources to be recognized. Or *Origins*/FIP sky surveys could be designed to make maps at various time intervals, allowing a way to compare over the lifetime of the mission overlapping with LISA. The substantial advantage of *Origins* over other facilities is its ability to scan large 10 deg$^2$ areas rapidly. Once a candidate source is discovered, low-resolution spectroscopy can be performed using OSS. *Origins* is likely have better line sensitivity over the range 5-300 µm than any other facility operating in the LISA era.

## A.5.2 Time-domain Science

This decade is witnessing a renaissance in time-domain astronomy. While a suite of facilities is uncovering optical and radio transients, *Spitzer* has opened up the dynamic infrared sky. Since 2014, *Spitzer* has been conducting the first systematic search in the mid-IR for luminous extragalactic transients; for example, SPitzer InfraRed Intensive Transients Survey (SPIRITS; Kasliwal *et al.*, 2017). SPIRITS has monitored 190 nearby (<20 Mpc) galaxies using warm *Spitzer*/IRAC 3.6 and 4.6 µm bandpasses with cadence baselines ranging from 1 week to 6 months, and has uncovered 157+ explosive transients and 2000+ eruptive stellar variables, including a new class of almost 100 IR "gap" transients with no optical counterparts (Figure A-6).

Although *Spitzer* pioneered time-domain exploration of the IR sky, its wavelength coverage and sensitivity of its 0.85-m mirror limit its ability to fully characterize newly discovered transients. *Origins'* capabilities allow a mid- to far-IR time-domain survey that not only finds, but also characterizes, newly discovered transients. While JWST has substantially improved spatial resolution and sensitivity over *Spitzer*, observatory overheads make a time-domain survey infeasible.

**Proposed Observations:** With *Origins* making use of FIP, OSS, and MISC imaging – an option under the upscopes for mid-IR observations – a key goal is to expand the volume of a mid-IR transient survey, spatially resolve newly identified transients in crowded regions (using MISC-imaging upscope option), and perform follow-up low-resolution mid- and far-IR spectroscopy to identify the nature of a transient. SPIRITS achieves a depth of 20 mag (Vega) at 3.6 μm with a spatial resolution of 2". This corresponds to an absolute magnitude of -8.5 mag, consistent with nova-like transients, out to 5 Mpc. *Origins* can detect nova-like transients in the mid-IR out to a distance of 40 Mpc in 0.5-hr of exposure time. Furthermore, the MISC-imaging upscope option provides a factor of 10 improvement in spatial resolution (~0.2"), which is crucial for resolving transients and variables in crowded regions. Most importantly, the mid-IR spectroscopic capabilities of *Origins*/MISC allow for follow-up characterization of newly discovered transients. Assuming a 4 x 0.5 hr observation per galaxy, *Origins*/MISC could conduct a survey of 100 galaxies in ~200 hr.

### A.6 Time-domain Science: Protostar Variability as a Probe of Protoplanetary Disk Physics and Stellar Assembly

A 100 hour far-IR monitoring survey of ~500 nearby, deeply-embedded protostars on time-scales of weeks to years measures changes in the rate of mass accretion onto the forming star and constrains the physical processes responsible for the redistribution of mass within the inner planet-forming region of the disk. Statistical analysis of the amplitudes and timescales associated with protostellar variability over the lifetime of *Origins* provides unique and quantifiable insight into the importance of large episodic accretion events in determining the final mass of the forming star. *Authors: Doug Johnstone (NRC-Herzberg, University of Victoria), Will Fischer (STScI), Michael M. Dunham (State University of New York at Fredonia)*

**Introduction:** Stars form within the dense molecular cores found in star-forming regions. The material that eventually becomes a star flows inward from the natal core under the influence of gravity (*e.g.*, Shu, 1977; Shu *et al.*, 1987; Masunaga & Inutuka, 2000). Circumstellar disks form early during the evolution of the protostar, acting as a reservoir for the infalling material and channeling it toward the forming star (*e.g.*, Hartmann *et al.*, 1997). Instabilities in the accretion flow within the disk (see review by Armitage, 2015) are evidenced by time-variability in the luminosity of the protostar, which for low-mass stars is dominated by the gravitational release of the binding energy of the protostellar accretion (*e.g.*, Johnstone *et al.*, 2013; Audard *et al.*, 2014; Hartmann *et al.*, 2016). Luminosity change, and therefore accretion variability, is probed most directly at the far-IR peak of the protostellar spectral energy distribution.

Rare, extreme accretion events, whereby the luminosity increases by orders of magnitude, have been observed in a handful of young stars (*e.g.*, FUors: Herbig, 1977; EXors: Herbig, 2008) and have been postulated as significant in terms of the overall mass assembly of stars (*e.g.*, Kenyon *et al.*, 1990; Dunham *et al.*, 2010). Fischer *et al.* (in prep; see also Safron *et al.*, 2015) compared *Spitzer* and WISE mid-IR images of 319 Orion protostars over a 6.5 yr baseline, discovering a few new large outbursts, including HOPS 383 (Figure A-7, Top), and estimating that all protostars undergo a >1 mag brightness burst about every 1000 years. This interval, however, has large uncertainties, and hundreds of protostars must be observed over longer timespans to better constrain the typical extreme outburst behavior.

The JCMT Transient Survey (Herczeg *et al.*, 2017) monitors only 50 bright protostars across eight nearby star-forming regions with a monthly cadence, and yet has uncovered robust evidence of variability at sub-mm (850 μm) wavelengths (Figure A-7; Mairs *et al.*, 2017; Yoo *et al.*, 2017; Johnstone *et al.*, 2018). In the sub-mm, the observed brightness variation scales more closely with the temperature change in the envelope rather than the direct change in the source luminosity, and thus the signal is sig-nificantly diminished versus observations in the far-IR. Nevertheless, the preliminary result from the first eighteen months of the JCMT survey is that 10% of protostars show sub-mm secular variability above 5% over the course of a year (Johnstone *et al.*, 2018).

**Scientific Importance:** Spectacular ALMA observations of disks around young stars show that disk features– rings, gaps, spirals – appear early (*e.g.*, ALMA Partnership *et al.*, 2015; Perez *et al.*, 2016). Thus, the conditions necessary for planet formation are also present. Even with ALMA, however, it is impossible to probe the key physical properties within the inner disk (*e.g.*, viscosity, ionization fraction) that mediate mass flow and structure formation (*e.g.*, Armitage, 2015; Hartmann *et al.*, 2016).

The *Origins*/FIP protostellar variability survey measures changes in the rate of accretion luminosity onto ~500 protostars (ten times the JCMT sample, with an additional order of magnitude sensitivity to fractional variability) across weeks to years. Given the direct relationship of accretion luminosity with mass flow through the disk onto the forming star, and associating variability timescales with orbital times within the disk, this survey provides unique and powerful insight into planet formation conditions within the inner, several AU, disk.

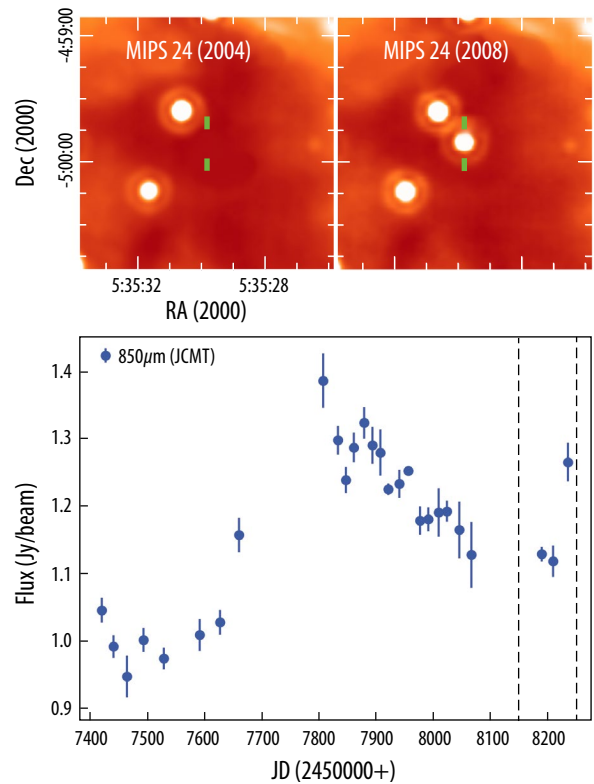

**Figure A-7:** Variability of deeply embedded protostars. Top: Spitzer 24 μm images demonstrating the outburst of HOPS 383 in Orion (Safron et al. 2015). Bottom: Sub-millimeter light curve of the quasi-periodic EC 53 in Serpens Main (Yoo et al., 2017; Johnstone et al., 2018). *Origins*/FIP can monitor 500 protostars in a 100 hr campaign.

Furthermore, the manner in which stars are assembled has a lasting affect on their appearance and evolution, especially if material is accreted in an episodic burst (*e.g.*, Hartmann *et al.*, 1997; Baraffe & Chabrier, 2010; Baraffe *et al.*, 2017). The large sample of protostars and multi-year timeline of *Origins*' survey quantifies, for the first time, the statistical accretion history of protostars and allows direct observation, and detailed characterization and follow-up, of elusive and rare, yet extreme, explosive events.

**Proposed Observations:** Over the 5-year mission, *Origins*/FIP monitors ~500 protostars. The gregarious nature of nearby star-forming regions is such that the optimal imaging strategy allows one to observe on the order of a square degree field per an individual target. Using the fast scanning mode, square degree fields require 30 min per FIP band. To determine dust temperature and to account for possible optical depth effects requires observing two SED measurement points (50 and 250 μm) per epoch, or 1 hour on-sky per field. FIP is sufficiently sensitive such that, although requiring the high background detectors, a sensitivity better than 0.4 mJy per epoch – yielding a S/N > 100 – is achieved, allowing an unprecedented analysis of flux variability down to percent levels.

Two coupled monitoring campaigns are proposed. Eight fields, comprising ~500 protostars, are observed twice yearly over the 5-year mission – requiring ten epochs each and a total of 80 hours. Additionally, two of these eight fields, comprising 178 protostars, are observed for two months at a higher cadence (*i.e.,* ten weekly epochs each, totaling 20 hours). With these observations, which are only possible from space, *Origins* uniquely and conclusively addresses the critical questions of protostellar variability, disk stability, early planet formation, and proto-stellar mass assembly.

## A.7 Small Bodies in the Trans-Neptunian Region: Constraints on Early Solar System Cometary Source Region Evolution

To place meaningful constraints on the fundamental models of solar system evolution, the team proposes to accurately measure the sizes and constrain the thermal properties of Solar System bodies beyond Neptune. Determining the quantity of material in these outermost populations places constraints on models of Solar System formation and history. Characterizing the source regions of comets provides understanding about the transport of volatiles throughout the Solar System's evolution. Thermal measurements yield sizes that, when coupled with optical measurements (for the larger/brighter objects), yield reflectances and compositional information. *Origins*/FIP observations (50 and 250 μm) provide the capability for measuring the sizes of thousands of these bodies, providing essential new information about our Solar system. *Authors: James Bauer (UMD), Kimberly Ennico (NASA/Ames), Amy Lovell (Agnes Scott College), Stefanie Milam (NASA/GSFC)*

**Introduction:** Small Solar System bodies that reside between 30 and 50 AU are referred to as Trans Neptunian Objects (TNOs). TNOs comprise the majority of small bodies within the Solar System and are a collection of dynamically-variegated subpopulations, including Centaurs and Scattered-Disk Objects (SDOs), as well as "cold" (low-inclination and eccentricity) and "hot" (high eccentricity) classical Kuiper Belt populations (KBOs; Gladman *et al.*, 2008).

These minor planets are the reservoir of short-period comets that routinely visit our inner Solar System, clouding the distinction between asteroids and comets (Figure A-8). They are primordial material, unmodified by the evolution of the Solar System, and are the sources of volatile materials to the inner solar system. TNO and outer solar system small body size distributions, down to scales of tens of km or about a few km (Dones *et al.*, 2015 and ref. therein), provide tests of the formation scales of planetesimals and evolutionary models of the early Solar System. FIR surveys uniquely enable derivation of many size measurements for these objects. Such surveys can place constraints on the total amount of material in the Solar System that resides beyond Neptune's orbit, which in turn constrains models of Solar System formation and evolution, and the collisional history of the small body populations in this region.

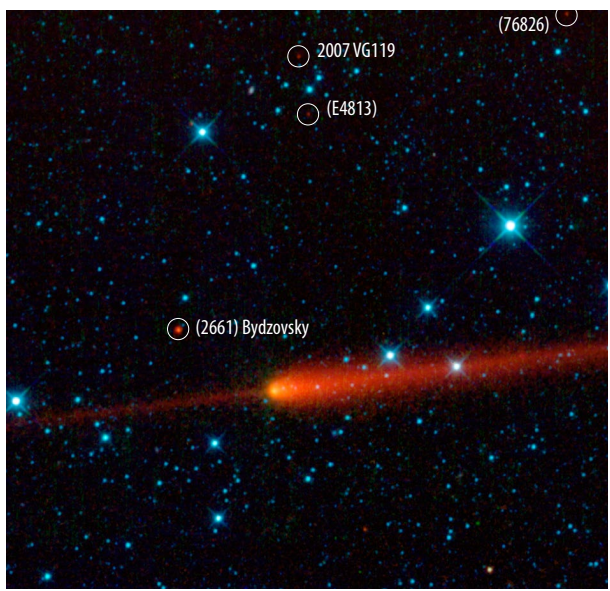

**Figure A-8:** A Wide-field Infrared Survey Explorer (WISE) image of 65P/Gunn at 3.4, 4.6, 12, and, 24 μm. The image includes the thermal signal from several asteroids with sizes ranging from ~4 to 20 km, the thermal signal from comet dust, and the lingering dust from previous perihelion approaches, which appears in this image to proceed the comet. Just as WISE had wavelength sensitivities that spanned the thermal peaks of inner solar system objects, *Origins* (and particularly the FIP bandpasses) spans the thermal emission peaks of outer solar system objects out to hundreds of AU [PIA13115, NASA/JPL-Caltech/UCLA].

**Scientific Importance:** Small bodies can typically vary in their surface reflectivity by factors of 5 or more, while surveys that detect emitted light provide reliable sizes from the flux (Mainzer *et al.*, 2011). New Horizons recently provided a particularly appropriate reminder of this when it imaged a system of TNOs with surface reflectances that ranged from a few percent (Charon) to several tens of percent (Pluto). Previous optical surveys provided alternate size frequency distributions based on inferences of reflectivity, indicative of competing evolution histories for these bodies (Dones *et al.*, 2015), especially at the smaller (TNO diameters <100 km) size scale. Objects at TNO distances are best detected at wavelengths near or below 100 μm. The thermal radiation of objects from 30 to 100 AU from the Sun, with temperatures ~30-50 K, provide sufficient flux at 100 μm to probe the most interesting size scales of the outer solar system populations (Figure A-9). Shorter (~50 μm) and longer (~250 μm) wavelengths better constrain the sizes and temperatures of the objects observed; the two bands 50 μm and 250 μm are provided by the FIP instrument in the baseline design.

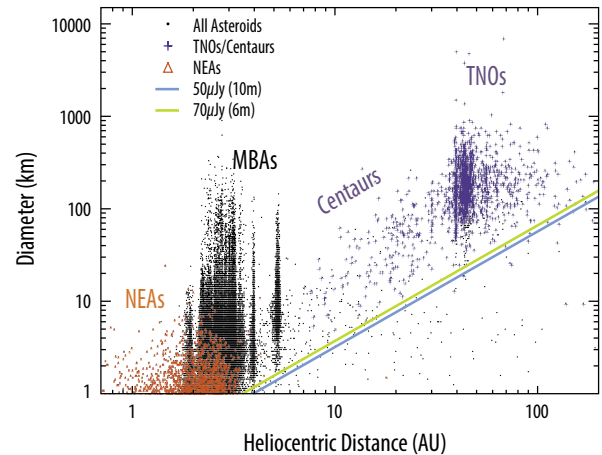

**Figure A-9:** The diameter vs. heliocentric distance of all small bodies in the Solar System. The data from MPC database (2017) includes sizes assuming a 10% surface reflectance in thermal equilibrium with insolation. The lines show the diameters corresponding to continuum flux limits needed to measure these bodies at 100 μm for a 6m (cyan line) and 10 m (green line) telescope. Most of the objects falling below these two lines were first measured at smaller heliocentric distances.

Most surveys have placed order-of-magnitude constraints on larger TNOs, with solar-system absolute magnitudes (H)<9 and sizes >100 km. Beyond 100 AU, the TNO population may flare out as well, with a larger dispersion in inclination, and an increase in the surface density of objects. Studying TNOs, and interlopers into near-TNO distances from even further populations in the Oort cloud, can thus inform a new picture of early Solar System history and how its composition has evolved over the time since it was formed.

**Proposed Observations:** *Origins* images a large area of the sky multiple times to capture large numbers of TNOs and related outer Solar System objects. With sensitivity down to ~60 μJy at 50 μm, a FIP scanning survey in a reasonable integration time would reach large numbers (many thousands) and small (~25 km; Figure A-9) sizes. A second survey using the 250-μm band down to a mJy would provide temperature, and therefore improve size measurements and distance constraints. The survey needs to cover the area several times so motion, ~2 arcseconds per hour (mostly parallax from the Earth's motion), can be detected and confirmed. With coverage of four integrations over the course of several days (~7), an orbital distance can be identified. A repeat observation a few (~4) weeks later can establish an orbit for each object. The multiple passes required to quantify sky motion also mitigates confusion noise. The total time to cover ~1000 deg² of sky is ~400 hours.

**Summary:** No survey has been done on this scale, and no platform would be as effective as *Origins*, as equipped with the FIP instrument. The mission's imaging capabilities at higher spatial resolution coupled with its broad-band wavelength coverage facilitate the large-area survey at greater depth and provide accurate diameters for a statistically-relevant, larger sample than any previous mission, including JWST. *Origins* imaging additionally provides de-coupling of TNO thermal signal from background confusion sources and astrometric measurements to identify and differentiate TNOs from other solar-system populations.

## A.8 Origins Studies of Far-infrared Water-ice Features in Non-disk Sources

*Origins*/OSS can search for the presence of water ice in all sources planned for observation during the Water Trail program, which spans a range of cloud evolutionary stages from starless cores to protoplanetary disks. There is evidence that a significant amount of water (and oxygen) is locked in water ice in clouds at all evolutionary stages. By observing the unique far-infrared water-ice features, *Origins* can not only map the spatial distribution of water ice, thus complementing *Origins*' observations of gas-phase water, but also measure its thermal history through the various evolutionary stages leading to planet formation. *Author: Gary Melnick (CfA)*

**Introduction:** Interstellar dust grains make up less than 1% of the interstellar medium by mass, yet their surfaces provide important sites on which gas-phase atomic and molecular species can settle, migrate, and react, fostering the formation of interstellar ices. As this process continues, layers of ice can accumulate on grain surfaces, altering the infrared absorption and emission properties of these grains and leaving a clear spectral signature of an ice mantle (Figure A-10). Infrared absorption spectroscopy suggests ices are particularly abundant within cold, dense gas, including the mid-plane of planet-forming disks (Section 1.2). After $H_2$, water ice is one of the most abundant molecules within these regions, locking up as much as 30% of the available atomic oxygen.

Water ice formed during the YSO phase, and earlier, can have a profound effect upon the thermal balance in nebular gas, as gas-phase water is predicted to be an important gas coolant – any increase in the ice abundance reduces gas cooling. Moreover, there is increasing evidence that, rather than being a distinct and separate region of ice formation, some water ice in protoplanetary disks is inherited from the nebulae out of which they form (Visser *et al.*, 2009; Cleeves *et al.*, 2014). This suggests that tracing the evolution of water ice from dense clouds to disks is necessary to fully understanding the trail of water to planets.

**Scientific Importance:** Observations of the amorphous and crystalline water ice emission features at 43, 47, and 63 μm (Figure A-10) can advance the study of the ice content in star- and planet-forming regions in several important ways. First, unlike absorption spectroscopy (Figure A-11), which provides the ice content along a narrow line of sight, emission line studies allow extended spatial mapping of the ice distribution. Second, such maps contain information about the thermal history of volatiles and indicate regions of significant ice processing (Poteet *et al.*, 2011, 2013). Amorphous water ice undergoes an irreversible transition to a crystalline state when heated above 90 K for extended periods, and is near-instantaneous when dust is heated above 110 K. Such periods of heating may also facilitate the formation of complex organic molecules within the ice. Third, by observing at far-infrared wavelengths, where dust extinction is greatly reduced relative to shorter wavelengths, it is possible to determine ice masses in the densest parts of pre-stellar cores (*i.e.*, the inner 1000 AU), a region largely inaccessible to JWST.

Significant column densities of crystalline water ice have been discovered toward the Herbig-Haro object HH7 (Molinari *et al.*, 1999) and the bipolar circumstellar outflow from the proto-planetary nebula Frosty Leo (Omont *et al.*, 1990) by the Infrared Space Observatory (ISO) and *Kuiper*, respectively. If *Origins* reveals this to be a common phenomenon, it suggests a scenario in which non-dissociative shocks produce copious amounts of gas-phase water and UV radiation from co-existent dissociative shocks heats grains upon which gas-phase water settles. It also indicates that gas-phase water cooling of shock heated gas may be greatly over estimated.

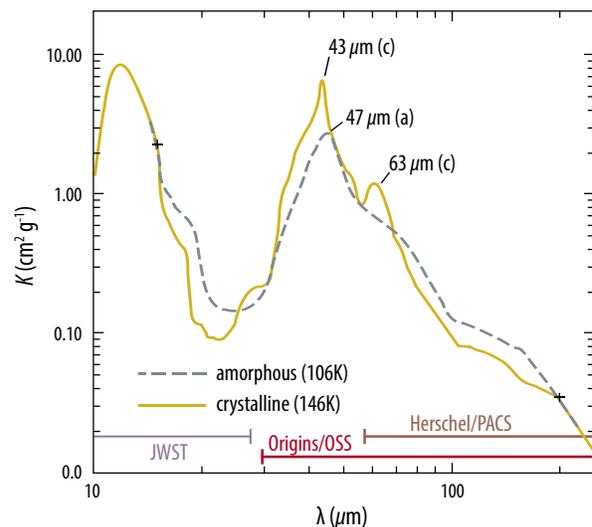

**Figure A-10:** *Origins* can detect the presence of water-ice, as well as study the thermal history of grain mantles by observations of the 43, 47, and 63 μm features. The plot shows the water-ice opacities (McClure et al., 2015). Characteristic wavelengths of the amorphous (dashed line) and crystalline (solid line) ice emission maxima are noted with a and c, respectively.

Water ice delivered to proto-planetary disks is generally believed to be in amorphous form with very little annealed to crystallinity upon delivery (Visser *et al.*, 2009). *Origins* observations of the amorphous and crystalline forms of water ice toward pre-stellar nebulae and planet-forming disks provides an important test of this prediction.

The feature at 43 μm is the strongest and narrowest. Since dust warm enough to excite other ice features into emission at shorter wavelengths would cause any ice present to rapidly sublimate, the far-infrared features available to *Origins* are unique emission tracers of water ice. Toward regions with extended, strong continuum emission, such as Orion, it should be possible to observe these ice features in absorption, enabling a detailed study of ice processing over large areas.

The important 43 and 47 μm features fall in a gap between the long wavelength cut-off of the *Spitzer* and JWST spectrometers and the short wavelength cut-off of the *Herschel* spectrometers. While ISO had access to this wavelength range and the HIRMES instrument on SOFIA can observe these features, *Origins*' significantly greater sensitivity enables it to study water ice toward a wider variety of sources with sufficient SNR to detect weaker features, enabling ice investigations not possible with any other telescope.

**Proposed Observations:** Based on existing ice emission measurements, *Origins* can obtain very high SNR (*i.e.,* SNR >> 100) ice spectra using the OSS instrument in grating mode in less than 1 min of integration time. In a 50-hour program, 200 Galactic sources at various evolutionary stages can be observed, assuming a slew-and-settle time of less than 15 minutes per source.

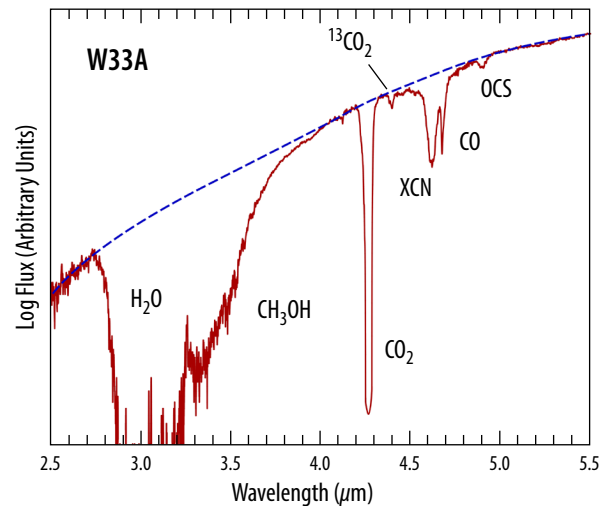

**Figure A-11:** Near infrared spectrum of the star forming region W33A obtained by ISO (Sloan et al., 2003). The blue dashed curve shows the dust spectrum expected without ice, while the red spectrum shows changes to the dust spectrum caused by the buildup of various ices on the grain surface. The composition of the ices between 2.5 and 5 μm is noted above.

## A.9 Studying Magnetized, Turbulent Galactic Clouds

This ~200 hour program using FIP aims to understand the role of magnetic fields and turbulence in star formation, connecting galactic-scale physics to protostellar cores. This program targets 14 molecular clouds of up to 50 deg² in size, yielding a ~10-fold increase in the number of molecular clouds that have been mapped in their entirety and, crucially, for the first time simultaneously providing angular resolution sufficient to map the inferred magnetic field strength at a thousand positions across each cloud. These observations trace the evolution of magnetically-supported neutral hydrogen clouds embedded in the turbulent Galactic environment through the formation of dense molecular filaments; fragmentation into protostellar cores; and then the recycling of gas, kinetic energy, and magnetic fields ejected back into the interstellar medium. *Authors: Laura Fissel (NRAO), Charles L. H. Hull (NAOJ/ALMA)*

**Introduction:** Magnetic fields and turbulence are key regulators of the star-formation process at all spatial scales (Crutcher *et al.*, 2012; Elmegreen *et al.*, 2004). Magnetic fields shape turbulence on the largest scales in the diffuse interstellar medium (ISM) and are a critical factor in the formation and evolution of protostellar cores, envelopes, disks, outflows, and jets. Turbulence itself carries a significant fraction of the energy in our galaxy, and represents the direct coupling of matter from galactic scales to the scales of protostellar envelopes. Over the last two decades a vast amount of work has gone into characterizing magnetic fields and turbulence in the ISM from galaxy → cloud → protostellar core scales, using interferometers plus ground-, balloon-, and space-based single-dish instruments. The *Planck* satellite has recently mapped dust polarization across the entire Milky Way at a resolution of 10' (*Planck* Collaboration, 2015); however, the images lack the spatial resolution to study magnetic fields and turbulence across entire clouds at the ~10" resolution that *Origins* achieves.

More recently, the BLAST balloon-borne telescope has made higher-resolution observations of entire clouds; see Figure A-12, which shows an image of the inferred magnetic field in the Vela C molecular cloud (Fissel *et al.*, 2016). This image has an angular resolution of 2.5', 4x better than *Planck*. However, FIP has 15× BLAST's resolution at 250 µm, implying that while BLAST made over a thousand independent measurements of polarization toward Vela C, an equivalent FIP map would have hundreds of thousands of independent measurements. FIP is also >100× more sensitive than BLAST. FIP's sensitivity and angular resolution changes the landscape of polarization studies of molecular clouds by allowing a dramatic increase in observational statistics, and consequently better comparisons with numerical simulations of star formation, not only across different molecular clouds (because of *Origins'* excellent sensitivity), but inside individual clouds (because of *Origins'* high resolution).

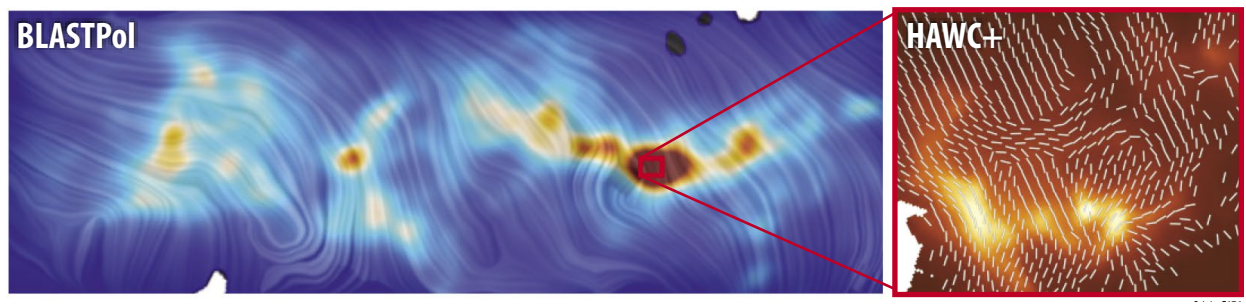

**Figure A-12:** A map of inferred magnetic field lines (texture) and total intensity dust emission (color scale) in the Vela C molecular cloud, from Fissel et al., 2016. This map, made with BLASTPol (the Balloon-borne Large Aperture Submillimeter Telescope for Polarimetry) is the most detailed magnetic field map ever made for a GMC forming high-mass stars. The *Origins* GO project will generate such maps for 14 molecular clouds of this size (min) or 10x larger (max). Inset panel: Observations of magnetic field orientation (line segments) made with the HAWC+ instrument on SOFIA at 89 µm towards the brightest region of Vela C (Fissel et al., in prep). FIP will have 15x the resolution of BLAST at 250 µm (similar to the HAWC+ inset), but with >4000x the sensitivity of HAWC+, yielding maps of magnetic fields in molecular clouds with hundreds of thousands to millions of independent measurements.

**Scientific Importance:** Many crucial questions still remain unanswered, in particular regarding understanding the structure of magnetic fields and the drivers and dissipation of magnetohydrodynamic (MHD) turbulence. *Origins* dramatically advances the field by enabling observations of dust polarization across many orders of magnitude in spatial scale in the ISM. A high-resolution, high-sensitivity, and very wide-field map of the magnetic field orientation has never been possible, and analyzed alongside spectral-line information from ground-based telescopes (or OSS and/or HERO on *Origins*), would reveal the role of the magnetic fields in the turbulent, shocked ISM of molecular clouds where stars are forming.

**Proposed Observations:** To make major breakthroughs in our understanding of the roles of magnetic fields and turbulence in star formation, the team plans to use FIP to perform high-resolution, high-sensitivity, multi-wavelength, wide-field observations of dust polarization in the FIR (at the peak of the dust SED, where dust is generally optically thin). Unlike observations of background starlight polarization in the optical and NIR, these FIP observations allow access to magnetized turbulent flows on all scales ranging from the diffuse neutral hydrogen envelope surrounding molecular clouds to the scales of dense star-forming cores.

Our observations enable estimates of the MHD turbulence power spectrum and magnetic field strength as a function of gas density (Houde *et al.*, 2009). In nearby clouds, *Origins'* high resolution can be used to measure the decorrelation scale of magnetized turbulence (Houde *et al.*, 2009). These observations can also be combined with ground- or space-based maps of CO and other spectral lines to identify turbulent shocks and to determine both energy injection scales and energy losses in the neutral ISM. The combination of spectral-line data and the thermal dust polarization data allow *Origins* to probe material of different densities and temperatures, and help determine where in the medium the emission is originating, thus enabling us to effectively produce tomographic maps of the magnetic field in the turbulent ISM over entire molecular clouds. Furthermore, observations at multiple wavelengths in the FIR (50 and 250 μm) enable continued testing theories of dust-grain composition and alignment (Andersson *et al.*, 2015).

### A.10 Below the Surface: A Deep Dive into the Environment and Kinematics of Low Luminosity Protostars

A survey program of very low luminosity protostars on key emission lines between 25-588 μm using *Origins*/OSS at R~50,000 spectroscopy enables a probe of the nature of the luminosity function – whether low luminosity protostars are low mass cores destined for brown dwarf systems or more massive systems in currently-quiescent states. The outcome of this study addresses the longstanding Luminosity Problem – the vast observed population of apparent low accretion rate protostellar systems compared with collapse models. The sensitivity and spectral resolution uniquely enables *Origins* to characterize the outflows and envelopes of these sources. *Authors: Joel Green (STScI), Yao-Lun Yang (UT Austin), Neal Evans (UT Austin)*

**Introduction:** What is the origin of very low mass stars and brown dwarfs? Is a faint or deeply embedded prestellar core a sign of a low mass star forming, or a quiescent phase of pre-stellar evolution? Predicting how embedded protostellar systems evolve into protoplanetary systems is challenging due to their uncertain final core mass and other properties. It is generally understood that protostellar luminosities in molecular clouds are considerably affected by their current accretion rates, but the luminosities across large samples are, on average, far too small to account for the mass to form a system ("the Luminosity Problem"). It has been hypothesized that most stars are in a "low" phase and only a few are in a high state. The enigmatic Very Low Luminosity objects (VeLLOs) may be protostars in a low accretion state, or they may be candidates for the origin of the low mass end of the IMF (Dun-ham *et al.*, 2008; Kim *et al.*, 2016). To distinguish these outcomes, requires tracing their accretion/outflow history by comparing current outflow rates (derived from FIR high energy emission lines) with historical rates (derived from low energy FIR lines). NIR accretion indicators (measuring the current luminosity) vary on timescales too short for accurate estimation, and are often extinguished in very young sources.

**Scientific Importance:** VeLLOs are too faint, so most of their spectral properties cannot be captured with *Herschel*, and measurements using SOFIA are impossible (*e.g.*, L1014; Figure A-13). Even in bright sources detectable by *Herschel*, PACS's low spectral resolution means the origins of emission lines from gas are confused. Much greater sensitivity, and higher resolving power, like that offered by *Origins*, is required to detect the outflows in low-luminosity systems and separate outflow, infall, and disk components. With *Origins*' sensitivity, VeLLOs can be studied in detail similar to high luminosity sources (*e.g.*, BHR71; Figure A-14), which exhibit rich emission spectra. Additionally, *Origins*' higher spectral resolution will enable detection of dense gas tracers and less abundant isotopologues, and disentangle emission from disks, envelopes, outflows, and the surrounding medium.

Emission lines at far-infrared wavelengths probe a longer timescale compared to the features observed at near- and mid-infrared, where the 30-meter class telescopes and JWST have the best capability. *Origins* covers many molecular transitions, as well as a few ionized lines, including the mid- to high-energy CO emission, water, OH, HCO+, [*OI*], [*NII*], [*CII*], and [*CI*], at impressively high spectral resolution. The CO emission comes from the entrained gas in the envelope due to the outflows and the shocked gas inside the outflow cavities, as revealed by high-resolution *Herschel*/HIFI observations. The entrained gas and shocked gas represent the outflow activity at long and short timescales, respectively. The shocked gas directly probes the shocks due to the current activities of outflows and jets, whereas the outflows drag and heat up the gas around the outflow cavity over time, resulting in the entrained gas at large spatial scale. FIR spectra provide a suite of tracers that probe the outflow activity at multiple timescales.

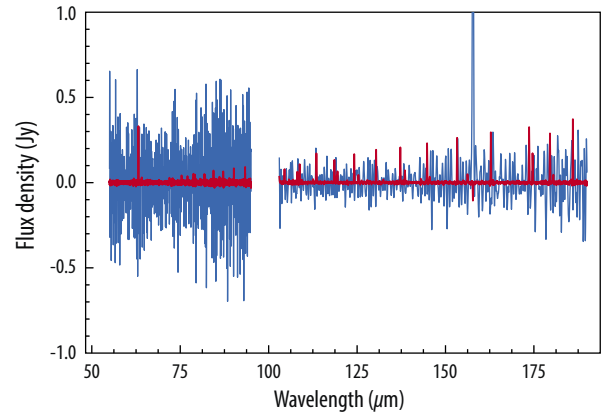

**Figure A-13:** Modeled far-IR spectrum (red) of VeLLO L1014. The model is scaled from the emission line spectrum of BHR71 Herschel-PACS data (BHR71 − 13.5 L$_\odot$; L1014 − 0.1 L$_\odot$, both at 200 pc; Green et al., 2013; Yang et al., 2017, Yang et al., 2018), compared with the actual PACS spectrum of L1014 (blue). The strong [CII] emission line at 157 µm is contaminated by background in the PACS spectrum.

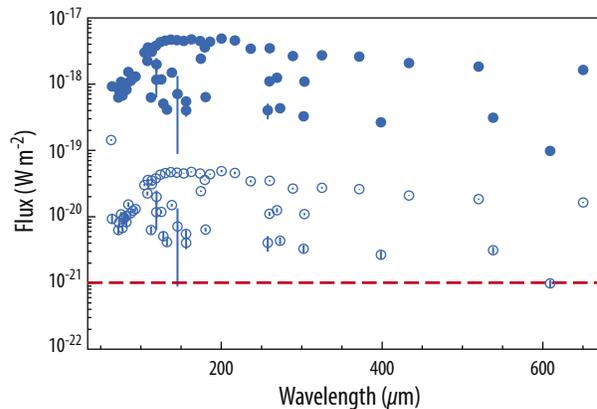

**Figure A-14:** 50-650 µm line fluxes for BHR71. Fluxes are scaled to the luminosity of L1014 at a distance of 200 pc (filled blue circles) and at 2000 pc (open blue circles). The *Origins*/OSS detection threshold (dashed red line) assumes the line emission is unresolved.

Combined with the NIR observations, OSS enables a new picture of the star formation history of protostars with a single shot. Which systems evolve into solar analogues and what are the key environmental factors? *Origins*/OSS addresses the critical question: What is the nature of protostellar evolution that sets the conditions in the protoplanetary disk, which in turn controls planet formation?

**Proposed Observations:** *Origins*/OSS can address this problem with deep integrations on ~100 low luminosity protostars in a variety of cloud environments at much greater range in distance. In the 50-650 μm range, only ~3 VeLLOs were observed, at low S/N. Yang *et al.* (2018) report a best detection threshold of $7\times10^{-18}$ W/m$^2$ with PACS and $3\times10^{-17}$ W/m$^2$ with SPIRE (1σ), at 1-hour integration depth per band (totaling five bands between 50 and 650 μm). *Origins* can achieve the same S/N at spectral resolutions 10-1000 times better, at 100 times greater depth (5σ confidence) in about 6 minutes + overheads, for sources like L1014, and less than an hour per source for analogues in regions up to ten times more distant.

## A.11 Putting the Solar System in Context: Frequency of True Kuiper-belt Analogues

> A small-map imaging program using the *Origins*/FIP 50 μm channel to search for faint emission around nearby stars with dust levels as low as in the Solar System's Kuiper belt provides a definitive measurement of the frequency of true Kuiper-belt analogues and a complete inventory of small body distributions in exoplanetary systems. *Author: Kate Su (Arizona)*

**Introduction:** The Solar System's asteroids and comets (*i.e.,* our "debris disk") are an integral part of our system. These small bodies assisted in the origin of life by delivering water to potentially-habitable worlds and have also upset life's evolution with occasional extinction events. Their locations and orbital structure provide strong constraints on our Solar System's history, a story that includes Neptune's outward migration, capture of Jupiter's Trojans, and the inward bombardment of the terrestrial planets. However, this story lacks context because the true extrasolar analogues of our Solar System's Asteroid and Kuiper belts remain invisible.

While all other stars must host debris disks at some level, only the brightest 20% are currently detectable; the remaining 80% are unknown, as is how this relates to our Solar System. A decade from now, on-going and planned planet-detection missions (TESS, Gaia, WFIRST, etc.) will have detected or set stringent limits on planets around most nearby stars, but the limits on small body populations will remain as poor as they are now (Figure A-15). True Kuiper-belt (KB) analogues, defined as ~45 AU planetesimal belts with peak emission at ~60 μm (*e.g.*, Vitense *et al.*, 2012), can only be detected by resolved imaging, as they are too faint (1%) relative to their host stars to detect as an IR excess. *Origins'* finer resolution and 2 orders of magnitude improvement in sensitivity compared to *Herschel* enable a complete census of true KB analogues among nearby stars, completing the planetary system inventory around nearby stars and placing our system into context.

**Scientific Importance:** A reference KB model (Vitense *et al.*, 2012) has a typical SED peaking at 60 μm from thermal dust emission with steeply decreasing brightness toward submillimeter/

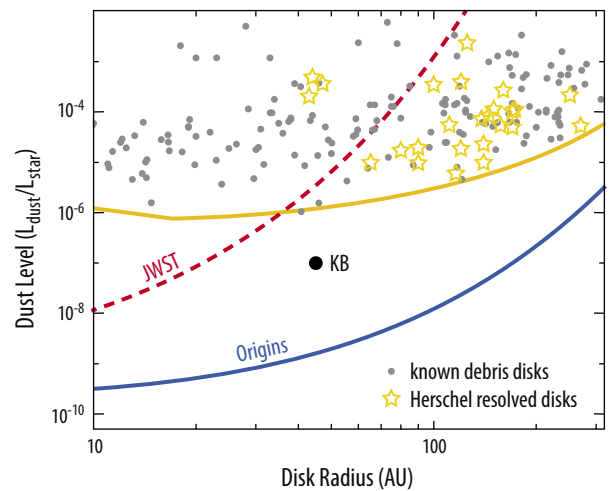

**Figure A-15:** *Origins'* unprecedented far-IR resolution and sensitivity enable a complete census of true Kuiper-belt (KB) analogues among nearby stars, placing our own into context. Points show the known debris systems of circumstellar dust based on measurements by *Herschel* and *Spitzer*. The expected raw sensitivities are shown as the color lines based on unresolved photometry with JWST (22 μm), *Herschel* (100 μm), and *Origins* (50 μm), without systematic noise from stars and calibration. The dust level in the KB is shown as a black dot. In addition to raw sensitivity, resolving the dust structure from the host star is the only robust way to detect faint KB analogues (i.e., setting the number of nearby stars that can fully utilize *Origins'* superb sensitivity; Figure A-16).

millimeter wavelengths; therefore, detecting and resolving exo-KBs is easiest at far-IR wavelengths. Although ALMA has adequate resolution to resolve circumstellar dust emission from the host star, the true KB disk surface brightness would be more than 200 times fainter at 1.3 mm than at 50 μm, making it extremely challenging for nearby stars. SOFIA and SPICA do not have adequate sensitivity and resolution in the far-IR to meet this scientific challenge. The mid-IR wavelength coverage from JWST can only probe warm dust analogous to the Asteroid Belt. By 2030, a complete picture of the planet population around nearby stars down to Saturn mass is anticipated from ongoing and planned planet-hunting surveys and an inventory of warm asteroid-like dust from JWST. At that point, the only missing piece would be a complete census on the frequency of true KB analogues, a survey only *Origins* enables.

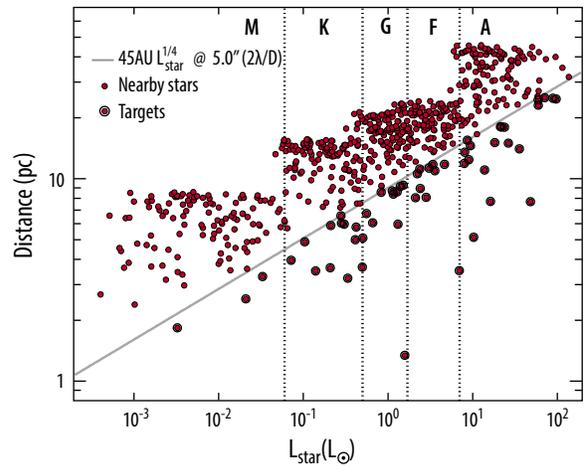

**Figure A-16:** Known stars within 50 pc from the Sun. *Origins'* targets are those that have resolvable KBs by two resolution elements (i.e., stars below the gray line) assuming the expected resolution (FWHM = 2.5") using the FIP 50 μm channel under the *Origins* baseline concept. A total of 56 nearby stars, with 33 in the spectral type of FGK, are suitable for this survey.

**Proposed Observations:** To provide a complete census of true KB analogues, the team uses the *Origins*/FIP 50 μm channel to obtain single pointing, small maps around 56 nearby stars (Figure A-16) where KBs can be spatially resolved at 50 μm by two resolution elements (assuming FWHM = 2.5"). A KB is about ten times fainter than the star's Airy disk at $2\lambda/D$. Thus, to detect disks at $10\sigma$, the PSF wings must be stable at <1% levels, which is equivalent to a contrast better than $10^{-4}$ ($1\sigma$). Contrast can be significantly improved by detecting the disk at more than at $2\lambda/D$ using very high S/N images without a coronagraph. The photospheric flux densities of these targets range from 20-250 μJy at 50 μm. The goal is to obtain S/N of 10,000 on the target's photosphere, which is needed so photon noise from the PSF does not dominate over the faint disk. Assuming a point source sensitivity of 1 μJy at $5\sigma$ in 1 hour, the required S/N on the photosphere can be achieved easily in less than 1 hour per target (the expected confusion limit is 0.12 μJy at 50 μm with an integration of 1.9 hour). Therefore, the survey can be completed in a total of 50 hours under the capability of the *Origins* baseline mission. This survey provides a complete census of true KB analogues around nearby stars, putting our Solar System into context.

## A.12 Giant Planet Atmospheres: Templates for Brown Dwarfs and Exoplanets

The four giant planets of our Solar System represent the closest and best examples of a whole class of gaseous, substellar objects commonplace in our Universe. Their ever-changing atmospheres provide the interface between their dynamic interiors and the external magnetospheric environment. *Origins* far-infrared observations (25-588 μm) provide the capability for comparative planetology of the four giants, as well as long-term monitoring of atmospheric cycles to connect the JWST and *Origins* eras. *Authors: Leigh Fletcher (University of Leicester), Glenn Orton (JPL), Imke de Pater (Berkeley), Arielle Moullet (SOFIA/USRA), Stefanie Milam (NASA's GSFC)*

**Introduction:** Brown dwarfs and directly-imaged exoplanets demonstrate rotational variability of their light curves, related to poorly-understood cloud formation and thermal contrasts on these distant, unresolved worlds. The four giants of our Solar System provide an ideal template for studying the sources of spatial and temporal variability, in an effort to understand how dynamic activity (banded structures, discrete vortices, vertical mixing) varies as a function of planetary metallicity, irradiation, and other driving properties. Although our local giants have been studied for decades at visible and near-IR wavelengths, the required spatial resolutions in the far-infrared for robust meteorological studies have been beyond the reach of previous space telescopes and Earth-based observatories. This wavelength range is key, providing the temperatures (from collision-induced $H_2$-He opacity), humidity (from rotational $NH_3$ lines), wind shears, atmospheric stability, middle-atmosphere composition (from stratospheric emission features of $H_2O$, CO, $CH_4$, HCN, etc.), and aerosol structure within these planetary atmospheres (Figure A-17). In short, the far-infrared reveals the environmental conditions underpinning the color and cloud changes observed in the visible, probing deeper pressures in the troposphere and higher altitudes in the stratosphere (via FTS and ultra-high resolution spectroscopy) than possible with JWST/MIRI. Furthermore, a key requirement for a step-change in giant planet atmospheric characterization is the need for high-temporal-cadence observations under invariant conditions. This time-domain science is required over multiple timescales: (i) short-term to identify small-scale changes in the atmospheric properties due to moist convective processes and waves; (ii) intermediate-term to understand large-scale changes to the belt/zone structure; and (iii) long-term to monitor seasonal and non-seasonal changes to these worlds.

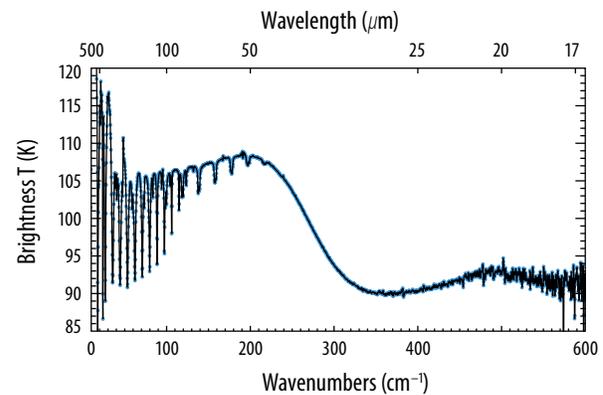

**Figure A-17:** Far-IR spectrum of Saturn, as measured by *Cassini*/ CIRS, showing the rotational lines at the longest wavelengths and the smooth collision-induced continuum that allows temperature, windshear, aerosol, para-$H_2$ and helium sounding. Comparable, repeatable, and spatially-resolved measurements are desired across the discs of each giant planet (from Fletcher et al., 2015).

**Scientific Importance:** Understanding these temporal variations, and specifically the thermal, gaseous, and aerosol changes that underpin them, provides a ground-truth catalog of giant planet variability as a resource to the exoplanetary community. Combined with the proposed decade-long observing record from JWST, *Origins* would add to an unparalleled record of atmospheric variability on all four worlds.

**Proposed Observations:** Regular, global scale far-IR mapping of the four giants using *Origins*/OSS (i) low-resolution (R~300) 25-588 μm grating spectroscopy; and (ii) moderate resolution (R~43,000) Fourier transform spectroscopy of selected stratospheric emission lines in the 110-588 μm range. *Origins* must be able to track the four giant planets and fit the full discs (45" for Jupiter) into the FOV (or use efficient mosaicking built into the observing strategy). Saturation limits must be sufficient (unlike JWST/MIRI) to observe the full 25-588 μm spectrum of each target, potentially via the use of neutral density filters. This requires a dynamic range of blackbody brightness temperatures of 110-220 K (Jupiter), 80-170 K (Saturn), 50-140 K (Uranus), and 50-160 K (Neptune). Sensitivity limits should be improved beyond MIRI to permit rapid imaging of the ice giants Uranus and Neptune. Even if the spatial resolution is insufficient to resolve distant Neptune (2.2"), disc-averaged spectroscopy can provide insights into global temperature/composition and its rotational variability (*e.g.*, Orton *et al.*, 2014). Assuming these conditions, the scientific products would include:

- Global temperature maps from the troposphere (via $H_2$-He collision induced continuum at 25-200 μm) and stratosphere (from $CH_4$, CO, and HCN emissions in the far-IR). Temperatures are used to derive 3D wind patterns and key meteorological tracers, such as the vorticity distributions, as well as assess the global energy balance of all four worlds.

- Distributions of tropospheric aerosols and condensation clouds via continuum mapping in the far-IR.
- Spatial and temporal variability of key gaseous species, including the cloud-forming volatiles (*e.g.*, $CH_4$ and $NH_3$), disequilibrium species ($PH_3$ and para-$H_2$) to serve as tracers of atmospheric motions; and a reassessment of the global helium fraction on all four worlds.
- Isotopic ratios (D/H, $^{13}C/^{12}C$, $^{15}N/^{14}N$) in a variety of species at sufficiently high spectral resolution would enable the team to compare the atmospheric composition of the four worlds to help constrain the origins of their gaseous composition (*e.g.*, accretion from ices or from gases in the protosolar nebula).
- Catalog of exogenic species, such as CO, HCN, and $H_2O$ that can only be accessed using space-based facilities, to understand the ongoing evolution of the giants' stratospheres.

Although snapshots of these parameters have been published previously, attempts to provide long-term, consistent data in the thermal infrared have been hampered by changing conditions on Earth. The team aims to replicate the success of the *Hubble* OPAL program (regular visible-light imaging of all four targets) via regular imaging/spectroscopy observations in the far-infrared. This would require short segments of observations spanning multiple years of *Origins* operations.

**Summary:** Critical enabling *Origins* capabilities for giant planet far-IR science include large FOVs, essential filter choices, and adequate saturation limits for viewing bright, moving, extended, rotating objects. With these tools, *Origins* can become the first facility to provide a regular baseline of atmospheric monitoring, enabling comparative planetology of all four worlds. The planned missions to Jupiter (JUICE and Clipper, scheduled in the 2030s) are not capable of executing similar science, nor are they sensitive to the far-IR (*i.e.,* spectroscopy beyond 30 μm). *Origins* provides the opportunity to overcome the challenges of JWST/MIRI to unlock essential knowledge about our corner of the Universe.

### A.13 Community Participation and Science Proposals

As part of the initial phase of this study, the *Origins* STDT collected 46 two-page science proposals from the astronomical community through an open call made via *Origins*' Science Working Groups (SWGs). These proposals presented science cases and technical requirements, focusing on studies aimed at a far-infrared space-based facility in the mid 2030s. The *Origins* mission definition team read all submitted proposals, which were then ranked through a preferential vote; the proposals and their rankings were used to drive the initial *Origins* Concept 1 (9.1 m telescope) design that was presented to NASA in an interim report (*Origins* Space Telescope Interim Study Report, 2018; see *https://asd.gsfc. nasa.gov/firs/docs/*).

The *Origins* Baseline 5.9-m telescope concept presented as the recommended concept in this report has a corresponding science capability degradation compared to the unconstrained *Origins* Concept 1. Table A-2 presents a summary of the top 25 proposals as ranked by the STDT and the percentage of science in each that can be completed with the *Origins* Baseline design. As summarized in Table A-2, most proposed science cases can be completed with the 5.9 m design , and several additional cases are enabled if the HERO instrument, an upscope option, is adopted.

**Table A-2:** Status of 2016-2017 Community-led Science Proposals

| Proposal Number and Name<br>Available at: https://origins.ipac.caltech.edu/resources<br>(List is ordered based on STDT internal vote rankings) | Origins Baseline Mission | | Reason |
|---|---|---|---|
| | Can be Completed? | % of Science | |
| 1. The Rise of Metals | Theme-1 | 70% | 100% with MISC upgrade |
| 2. Bio-signatures of Transiting Exoplanets | Theme-3 | 100% | |
| 3. The First Dust | Theme-1 | 25% | Aperture size reduction |
| 4. Water Content of Planet-Forming Disks | Theme-2 | 95% | 100% with HERO upscope |
| 5. Connection Between Black Hole Growth and Star Formation Over Cosmic Time | Theme-1 | 100% | |
| 6. Direct Detection of Protoplanetary Disk Masses | Theme-2 | 95% | 100% with HERO upscope |
| 7. Birth of Galaxies During Cosmic Dark Ages | DS 1.4.1 | 10% | Aperture size reduction |
| 8. Galaxy Feedback from SNe and AGN to z~3 | Theme-1 | | |
| 9. Survey of Small Bodies in the Outer Solar System | DS 1.4.8 | 100% | |
| 10. Direct Imaging of Exoplanets | | 0% | Coronagraph not in design for baseline or upscopes |
| 11. Star Formation and Multiphase ISM at Peak of Cosmic Star Formation | Theme-1 | 100% | |
| 12. Thermo-Chemical History of Comets and Water Delivery to Earth | Theme-2 | 80% | 90% with HERO upscope; rest aperture size reduction |
| 13. Galaxy Feedback Mechanisms at z<1 | DS 1.4.3 | 100% | |
| 14. Water Transport to Terrestrial Planetary Zone | Theme-2 | 15% | 100% with HERO upscope |
| 15. Frequency of Kuiper Belt Analogues | DS 1.4.12 | 100% | |
| 16. Feedback on All Scales in the Cosmic Web | Theme-1 | 75% | Aperture size reduction |
| 17. Magnetic Fields and Turbulence - Role in Star Formation | DS 1.4.10 | 50% | 100% with HERO upscope for turbulence |
| 18. The EBL (extra-galactic background light) with Origins | DS 1.4.2 | 50% | 100% with FIP upscope for missing bands |
| 19. Determining the cosmic-ray flux in the Milky Way and nearby galaxies | DS 1.4.15 | 0% | 100% with HERO upscope |
| 20. Episodic Accretion in Protostellar Envelopes and Circumstellar Disks | DS 1.4.7 | 100% | |
| 21. Formation and History of Low-Mass Ice Giant Planets | | 50% | Aperture size reduction |
| 22. Find Planet IX | | 100% | |
| 23. Fundamentals of dust formation around evolved stars | | 0% | 100% with HERO upscope |
| 24. The dynamic interstellar medium as a tracer of galactic evolution | | 0% | 100% with HERO upscope |
| 25. Probing magnetic fields with fine structure lines | | 0% | 100% with HERO upscope |

## APPENDIX B - ORIGINS SPACE TELESCOPE (ORIGINS) MISSION REQUIREMENTS

1. Scope
   This document defines the mission requirements for the *Origins* Space Telescope (*Origins*).

2. Mission Objectives
   The objectives of the *Origins* mission is to perform 2.8–588 μm Mid-IR and Far-IR science observations with unprecedented sensitivity to transform our understanding of the Universe, from the formation of the earliest galaxies, to the formation of habitable worlds, and search for biosignatures on exoplanets.

3. Documents
   3.1 Applicable Documents
       3.1.1 *Origins* Space Telescope (*Origins*) Science Traceability Matrix (STM)
       3.1.2 NPR 8705.4, Risk Classification for NASA Payloads (updated with Change 3)
       3.1.3 GSFC-STD-1000 Rules for the Design, Development, Verification and Operation of Flight Systems*
       3.1.4 GSFC-STD-7000 General Environmental Verification Standard*
       3.1.5 IEST-STD-CC1246D: Product Cleanliness Levels and Contamination Control Program
       3.1.6 *Origins* Space Telescope (*Origins*) Contamination Control Plan
       3.1.7 NPR 8715.6B, NASA Procedural Requirements for Limiting Orbital Debris and Evaluating the Meteoroid and Orbital Debris Environments
       3.1.8 NASA-STD-8719.14A (with Change 1), Process for Limiting Orbital Debris
   3.2 Reference Documents
       3.2.1 TBD

4. Origins Mission Requirements
   4.1 The *Origins* mission shall meet the science requirements documented in the *Origins* Space Telescope (*Origins*) Science Traceability Matrix (STM).

   *Rationale: Origins mission to meet science requirements in the STM*

   4.2 The *Origins* observatory shall be comprised of the payload and the spacecraft bus. The *Origins* spacecraft bus shall be capable of fully accommodating the payload that is comprised of a telescope and the following instruments that meet the Instrument Requirements in the *Origins* STM:
   - The *Origins* Survey Spectrometer (OSS)
   - The Mid-Infrared Imager Spectrometer Camera (MISC)
   - The Far-Infrared Imager Polarimeter (FIP)

   *Rationale: Origins observatory needs to support the instruments listed in the STM*

   *\* The applicable document depends on the NASA center selected to manage the mission*

**Table B-1:** *Origins* Telescope Specifications

| Parameter | Value | Units |
|---|---|---|
| Aperture Size | Circular, 5.9 | meters |
| *f*-number | *f*/14.0 (telescope) <br> *f*/0.63 (primary) | -- |
| Effective focal length | 82.6 | meters |
| Field of view | 46 x 15 | arcmin. |
| Waveband | 2.8 - 600 | microns |
| Operating temp. | $\leq 4.5$ | Kelvin |
| Optical performance | Diffraction limited at $\lambda = 30\mu m$ | |
| Design form | Three mirror anastigmat (TMA), on-axis pupil | |

4.3 The *Origins* telescope shall meet the specifications in **Table B-1**, as well as include a Field Steering Mirror (FSM) with minimum throw of 1 arcmin.

*Rationale:* Telescope specifications defined for meeting STM science requirements (together with the instruments); FSM requirements to enable small area mapping and calibrated measurements with multi-pixel far-IR detector arrays

4.4 The *Origins* orbit shall be a Sun-Earth L2 orbit between 4 degree and 29 degree off the Sun-Earth axis.

*Rationale:* To reduce background noise and enable detection of faint signals

4.5 The *Origins* observatory shall be designed for 5 year mission life with 10 year consumables.

*Rationale:* To support 5 year mission life requirement and allow sufficient comet observation; to allow mission extension to 10 years

4.6 The *Origins* observatory shall be a NASA Class A payload as defined in NPR 8705.4, Risk Classification for NASA Payloads (updated with Change 3).

*Rationale: Origins* mission's priority, national significance, lifetime requirement and budget fit the criteria for Class A mission

4.7 No credible single failure in the *Origins* observatory or single ground operator fault shall permanently prevent meeting the science requirements for the *Origins* mission defined in the STM.

*Rationale:* To better assure success of a Class A mission and meet science requirements

4.8 The *Origins* observatory body coordinate system shall be a right hand orthogonal system defined as follows: The +X axis shall be the telescope boresight, the Z axis shall be perpendicular to the X axis, with the -Z axis intersecting with the longest baffle/barrel/sunshield longitudinal line - extended, the Y axis shall complete the right handed XYZ axes system. The origin shall be at the center of the separation plane. Roll, pitch and yaw shall be defined as rotations about the X, Y and Z axes respectively.

*Rationale:* Observatory coordinate axes defined

4.9 The *Origins* shall be a facility mission for observations selected through a competitive proposal process.

*Rationale:* NASA policy applied to *Origins*

4.10 The *Origins* observatory shall be capable of being launched on at least one of the following two expendable launch vehicles: Space Launch System (SLS) with the 8.4 m fairing and Space X Big Falcon Rocket (BFR) with the 9 m fairing.

*Rationale:* Capable launch vehicles with large fairings to accommodate stowed *Origins* with large, no-deployment telescope

4.11 The *Origins* observatory shall have sun shield(s) to protect the telescope from views of the Sun, Earth and Moon, and the radiated heat from the spacecraft bus.

*Rationale:* To protect the telescope from bright and warm sources, enabling mission science

4.12 The *Origins* telescope and the cold side of the three instruments shall be cooled to, and maintained at < 4.5 kelvin (K) using cryocoolers, sun shields and deep space radiators.

*Rationale:* Cooling needed to achieve required measurement sensitivity

4.13 The *Origins* telescope shall have a baffle to minimize stray light degradation of the telescope image, and to prevent radiation from the innermost sun shield seeing inside the 4.5 K zone.

*Rationale:* Stray light degrades science

4.14 The *Origins* observatory shall be capable of +45 degree/-5 degree pitch rotation, +/- 5 degree roll rotation, and +/- 180 degree yaw rotation for science observations. The Field of Regard (FoR) over mission life shall be $4\pi$ Steradian.

*Rationale:* Wide Field of Regard for observing any target of interest

4.15 The *Origins* observatory shall be 3-axis Attitude Controlled, capable of pointing accuracy of 0.15 arcsec (3 sigma, pitch and yaw), 0.15 arcsec pointing knowledge (3 sigma, pitch and yaw), 50 mas (1 sigma) or less pointing jitter up to 20 Hz, 50 mas or less pointing drift up to10 hours. The telescope FSM shall be capable of accepting feedback from the MISC tip-tilt sensor to reduce these jitter and drift to < 1 mas.
The observatory shall be capable of 60 arcsec/sec scanning rate. After a slew and settle, the pointing accuracy shall be no greater than 2 arcsec (1 sigma) to facilitate target acquisition.

*Rationale:* Attitude control to support science observations, including OSS and FIP surveying, and enabling MISC transit/eclipse spectroscopy measurements.

4.16 The *Origins* observatory onboard clock shall be accurate to 1 milli-second or better relative to the International Atomic Time (TAI).

*Rationale:* Science observation timing accuracy to enable synchronization across multiple observing platforms and to support onboard orbit determination

4.17 The *Origins* observatory design shall not preclude serviceability on orbit.

*Rationale:* Serviceability allows extension of mission life

4.18 The *Origins* observatory design shall be capable of supporting simultaneous survey by the OSS and the FIP and 21 Tbits/day data collection.

*Rationale:* A driving observation scenario for spacecraft bus design to accommodate

4.19 The *Origins* observatory shall achieve a science observation efficiency of at least 80%.

*Rationale:* To enable achieving all driving science requirements within 20% of the 5 year mission life (12 month accumulated observation time)

4.20 At least 99% of the *Origins* science observation data shall be successfully delivered to the *Origins* Science Operations Center, processed and archived.

*Rationale:* To limit worst case science data loss.

4.21 The *Origins* Level Zero production data shall be available 72 hours from the time of data collection onboard the observatory.

*Rationale:* Data latency defined.

4.22 The *Origins* observatory shall not suffer any loss of stored science, engineering, and Health and Safety data in the event of a data downlink outage lasting up to 24 hours (TBD).

*Rationale:* Impact of limited duration downlink outage on data loss defined

4.23 The *Origins* observatory design shall support detection, isolation and recovery capabilities for any single fault in the spacecraft bus. The observatory shall be capable of surviving the occurrence of any single fault in the spacecraft bus without ground intervention for a minimum of 24 (TBD) hours.

*Rationale:* Spacecraft bus fault protection capabilities defined.

4.24 The *Origins* observatory shall be capable of meeting all *Origins* science requirements after exposure to both the natural and induced environments from launch to achieving the mission orbit, and during and after exposure to both natural and induced environments in the mission orbit.

*Rationale:* Natural and self-generated environments shall not prevent achieving mission science.

4.25 The *Origins* observatory design, development, verification and operation shall comply with Applicable Document 3.1.3.

*Rationale:* Adoption of technical guidance distilled from past experiences

4.26 The environmental verification of the *Origins* observatory, its payload and spacecraft bus and their components shall comply with Applicable Document 3.1.4.

*Rationale:* Adoption of environmental verification guidance distilled from past experiences

4.27 The *Origins* observatory shall comply with Applicable Document 3.1.5 and Applicable Document 3.1.6 during handling, fabrication, integration, test, storage and transportation. Contamination induced performance degradation on ground and after launch shall not prevent meeting *Origins* science requirements.

*Rationale:* Cleanliness and contamination requirements

4.28 The *Origins* observatory shall comply with Applicable Document 3.1.7 and Applicable Document 3.1.8.

*Rationale:* Compliance with limiting orbital debris requirements

## APPENDIX C - OBSERVATORY DETAILS

### C.1 Thermal Details

The sunshield and spacecraft shields limit the radiation heat load to the Barrel. The spacing between inner and outer sunshield layers is 1.38 meters. The spacing between the inner sunshield and Barrel is 1.19 meters. This spacing provides a balance between thermal performance through radiation to deep space, and compact design to allow easy ground deployment and stowing within planned fairings. Some overall properties used in the models are shown in Figure C-1.

#### C.1.1 Cryogenic Payload Module Details

The black radiator covers 140 degrees out of the 360-degree circumference of the barrel (Figure C-2). The Vespel® spacers are thin-walled (10 mm outer diameter, 0.25 mm wall thickness) tubes 51- mm long plus end fittings of 6 mm each resulting in a total length of 63 mm. This design will be analyzed for surviving launch loads in Phase A. The 20 K shield covers the total inside area of the barrel, 116 m² in the cylindrical portion and 60 m² at the bottom, for a total area of 176 m². The Vespel® supports are 1 m apart, therefore, there are 176 spacers in parallel conducting heat between the 35 K and 20 K zones. Conductance calculations show a total of 33 mW conducted by the Vespel® spacers between the 35 K barrel and 20 K shield. Lateral thermal conduction in the 4.5 K, 20 K, and 35 K thermal zones exploit high purity aluminum thermal conductance properties to minimize temperature gradients.

The 20 K zone intercepts radiated and conducted heat flowing between the 35 K barrel and the 4.5 K baffle, telescope and instruments. The 20 K shield lines the inside of the 35 K barrel and consists of high thermal conductance, pure aluminum foil sandwiched between 50-micron thick DAK sheets. The 20 K shield construction was used successfully as a thermal shield at 20 K for the Mid InfraRed Instrument (MIRI) instrument on JWST. The DAK/Al/DAK sandwich has a mass of 0.22 kg/m². It is

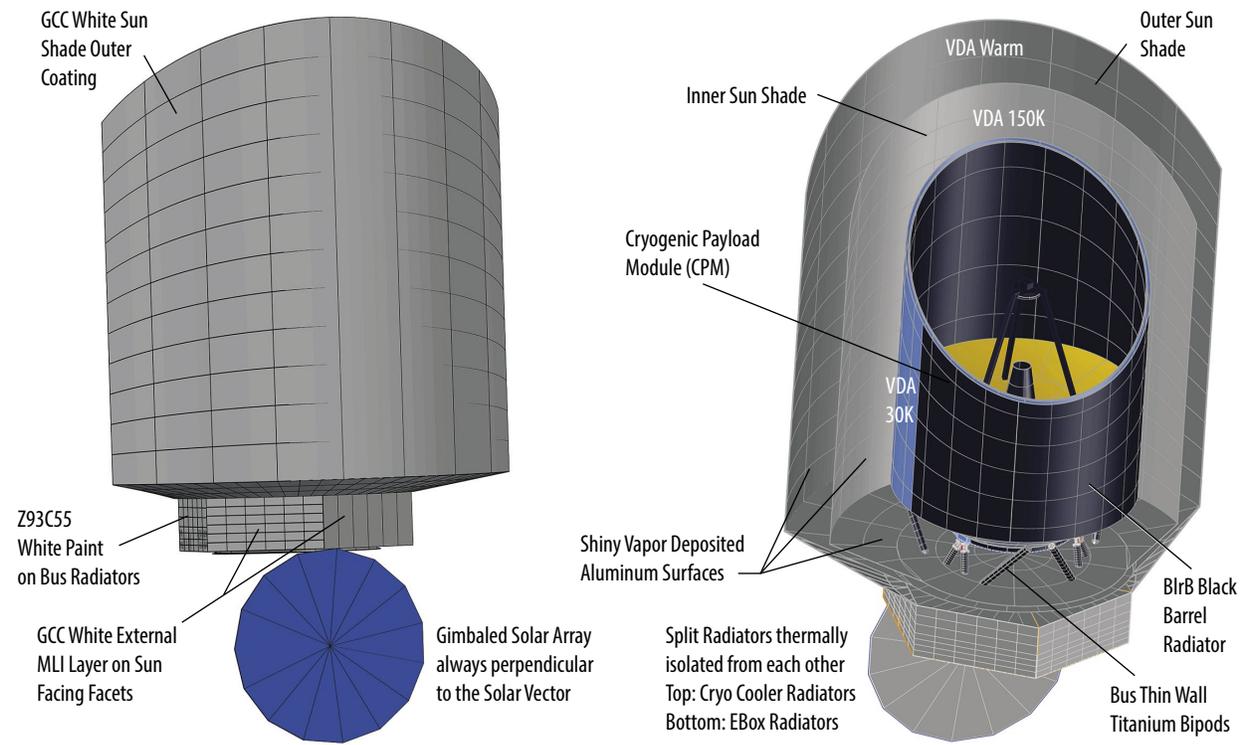

GCC White Sun Shade Outer Coating

VDA Warm

Inner Sun Shade

Outer Sun Shade

VDA 150K

Cryogenic Payload Module (CPM)

VDA 30K

Z93C55 White Paint on Bus Radiators

GCC White External MLI Layer on Sun Facing Facets

Gimbaled Solar Array always perpendicular to the Solar Vector

Shiny Vapor Deposited Aluminum Surfaces

Split Radiators thermally isolated from each other Top: Cryo Cooler Radiators Bottom: EBox Radiators

BlrB Black Barrel Radiator

Bus Thin Wall Titanium Bipods

**Figure C-1:** End to End Thermal modeling of the entire observatory for the full attitude range at SEL2 establishes thermal design feasibility.

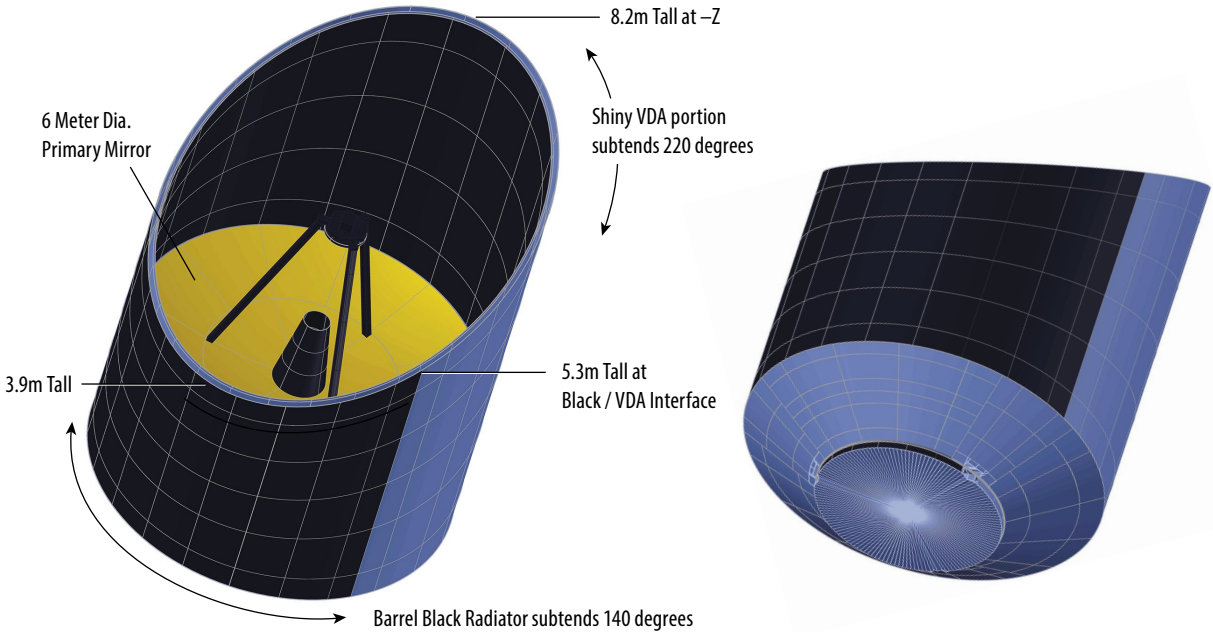

**Figure C-2:** The Barrel radiator spans 140 degrees of the barrel circumference.

mechanically supported off the Barrel surfaces via small Vespel® tubes with an end feature that supports the 20 K material. The 20 K zone provides a radiation barrier and is used to exploit excess refrigeration returning from the 4.5 K cryo lines. The Barrel cylindrical surface is subdivided into two zones. The radiator zone is coated black and the sunshade-facing zone is thermally-coated with VDA.

### Low Temperature VDA Thermal Emittance Properties

Low emittance characteristics of shiny VDA surfaces at cold temperatures are key to satisfying the stringent radiation heat load budget. The lowest emittance was 0.01 for VDA surfaces in the Cold Zone. All VDA surfaces were assumed to provide 98% specular reflections. Figure C-3 shows the low temperature VDA emissivity (Tuttle, 2008) included in the model.

Calculated radiation heat loads are directly dependent upon the thermal emittance value assigned to the various thermal model surfaces. Minimization of parasitic heat loads to the cold zones rely on low-temperature VDA emittance properties. A low-emittance property clearly benefits the engineering solution, as less radiation heat transfer occurs with decreasing emittance. Typical thermal model analysis assumes constant emittance over the temperature range. For *Origins* three different values of VDA emittance were imposed depending on the temperature of the material and wavelength of the dominant thermal radiation upon that surface. See Section 2, Table 2-5. A hundred thousand rays were shot from each surface to calculate radiation exchange couplings. For VDA, the thermal model uses a conservative 98% specular reflectivity and 2% diffuse.

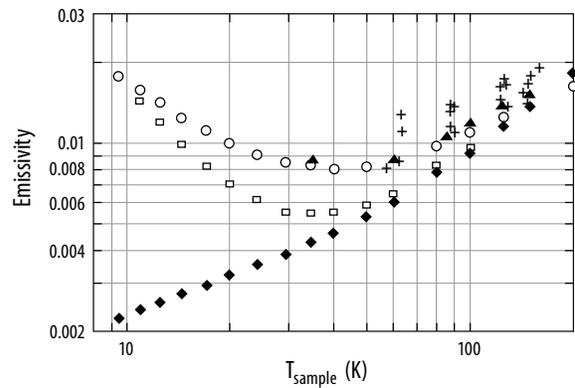

**Figure C-3:** The open circles are data from the work of Tuttle, et al. (2104, 2008). The emissivity of VDA-coated Kapton has been measured accurately from 10 K to room temperature. *Origins* uses these data to accurately account for temperature dependent emissivity.

### Low Temperature Emissivity of Ball IR Black

The JWST team made measurements on candidate black surfaces at low temperature (Tuttle, *et al.* 2014). *Origins* uses BIRB 0.65 kg/m² material (Figure C-4) properties for the black surfaces in its thermal models.

### CPM Thermal Model Elements

The team thermally-modeled the telescope and instruments within the 4.5 K zone. Radiatively, the 4.5 K zone sees a 20 K environment provided by the 20 K shield, allowing a relatively-simple calculation of the small radiative heat input. Conducted heat to the 4.5 K zone is located in discreet locations: the 4.5 K bipod interfaces and the telescope and instrument harnesses as they traverse the 4.5 K bipods. Each instrument also has some dissipation. Each of the three instruments is connected to an instrument mounting plate by a thermal strap, resulting in three simple nodes for heat input to the system. For practical considerations, the team modeled the telescope in detail, while the instruments were treated as single nodes. Figure C-5 shows outer barrel thermal zones.

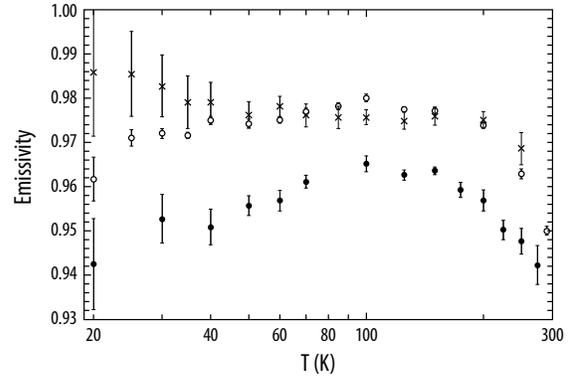

**Figure C-4:** Emissivity measurements made for BIRB show it is very black at low temperatures. The 0.65 kg/m² sample (X) used for JWST was also used in this model. This application of BIRB will be used for *Origins*.

Table C-1 shows thermal model CPM tabular temperature results for Observatory Pitch Angles 90°, 112°, and 150° for hot and cold solar irradiance assumptions. Note that the 4.5 K, 20 K, and 35 K zones are all very isothermal making the entire cryo-thermal system more efficient and easier to analyze.

### Spacecraft Bipod Assembly

The Bus Bipods/Struts are thin wall titanium alloy (Ti6Al4V). The Bus Bipod main cylindrical thermal isolators are 858 mm long, 165 mm in diameter, with a 1.3 mm wall thickness. The Bus Struts are 2438 mm long, 178 mm in diameter, with a 2.3 mm wall thickness. Figure C-5 shows the spacecraft Bipod struts simulated in the thermal model.

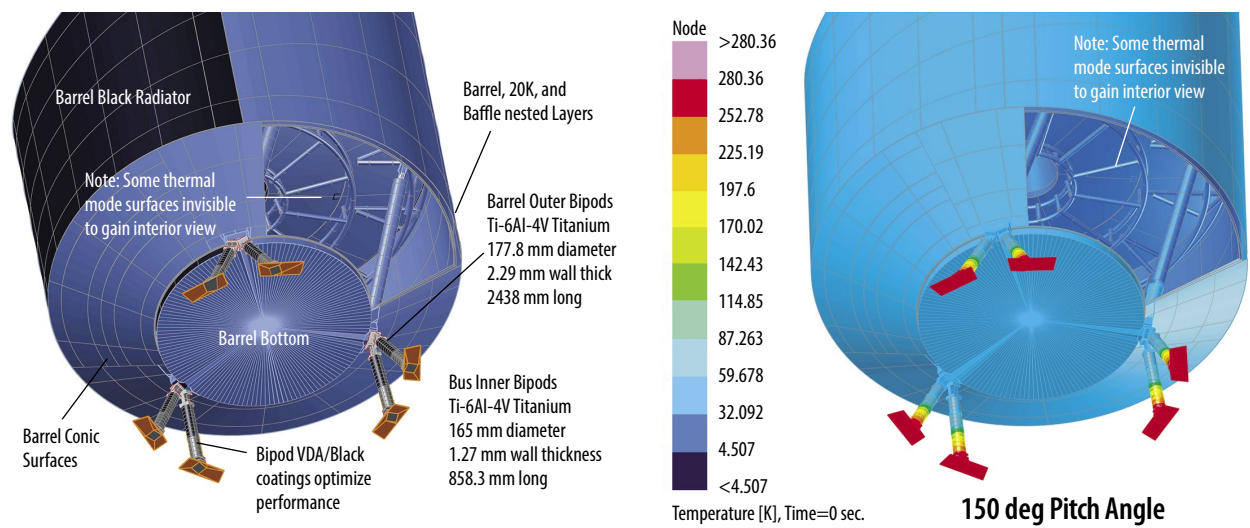

**Figure C-5:** The thermal model shows that the Bus bi-pods effectively thermally isolates the CPM from warmer SBM.

**Table C-1:** Cryogenic payload module temperature results show that the payload is insensitive to observatory attitude.

| Observatory Pitch Angle (deg) | 85 | | | 135 | | | 90 | | | 135 | | | 150 | | |
| Hot or Cold Case | Cold | | | Cold | | | Hot | | | Hot | | | Hot | | |
| | $T_{Avg}$ | $T_{Min}$ | $T_{Max}$ | $T_{Avg}$ | $T_{Min}$ | $T_{Max}$ | $T_{Avg}$ | $T_{Min}$ | $T_{Max}$ | $T_{Avg}$ | $T_{Min}$ | $T_{Max}$ | $T_{Avg}$ | $T_{Min}$ | $T_{Max}$ |
| | (K) | (K) | (K) | (K) | (K) | (K) | (K) | (K) | (K) | (K) | (K) | (K) | (K) | (K) | (K) |
| Cryogenic Payload Module | | | | | | | | | | | | | | | |
| Instrument Nodes | 4.51 | 4.510 | 4.510 | 4.51 | 4.510 | 4.510 | 4.51 | 4.511 | 4.511 | 4.51 | 4.510 | 4.511 | 4.51 | 4.511 | 4.511 |
| Structural Platforms | 4.50 | 4.502 | 4.505 | 4.50 | 4.502 | 4.505 | 4.50 | 4.502 | 4.506 | 4.50 | 4.502 | 4.506 | 4.50 | 4.502 | 4.506 |
| Telescope Components | | | | | | | | | | | | | | | |
| Eighteen Primary Mirror Segments | 4.51 | 4.510 | 4.511 | 4.51 | 4.510 | 4.511 | 4.51 | 4.514 | 4.514 | 4.51 | 4.513 | 4.513 | 4.51 | 4.513 | 4.513 |
| Secondary Mirror Optical Surfaces | 4.51 | 4.510 | 4.510 | 4.51 | 4.510 | 4.510 | 4.51 | 4.513 | 4.513 | 4.51 | 4.512 | 4.512 | 4.51 | 4.512 | 4.512 |
| Secondary Support Struts | 4.51 | 4.510 | 4.510 | 4.51 | 4.510 | 4.510 | 4.51 | 4.513 | 4.513 | 4.51 | 4.512 | 4.512 | 4.51 | 4.512 | 4.512 |
| BAFFLE | | | | | | | | | | | | | | | |
| Cyl Surfaces Above PM | 4.51 | 4.512 | 4.513 | 4.51 | 4.511 | 4.513 | 4.52 | 4.516 | 4.519 | 4.52 | 4.514 | 4.517 | 4.52 | 4.515 | 4.517 |
| Cyl Surfaces Below PM | 4.51 | 4.511 | 4.512 | 4.51 | 4.511 | 4.513 | 4.52 | 4.515 | 4.516 | 4.51 | 4.514 | 4.515 | 4.51 | 4.514 | 4.515 |
| Baffle Conical Surfaces | 4.51 | 4.512 | 4.512 | 4.51 | 4.512 | 4.512 | 4.52 | 4.516 | 4.516 | 4.51 | 4.515 | 4.515 | 4.52 | 4.515 | 4.515 |
| BMIDZONE | | | | | | | | | | | | | | | |
| Mid 20K Cyl Surfaces | 20.00 | 20.000 | 20.006 | 20.00 | 20.000 | 20.008 | 20.00 | 20.000 | 20.009 | 20.01 | 20.000 | 20.010 | 20.01 | 20.000 | 20.011 |
| Mid 20K Zone Conical Surfaces | 20.00 | 20.000 | 20.001 | 20.00 | 20.000 | 20.002 | 20.00 | 20.000 | 20.002 | 20.00 | 20.000 | 20.002 | 20.00 | 20.000 | 20.003 |
| BARREL | | | | | | | | | | | | | | | |
| Space Side Cyl Black Radiator | 31.61 | 31.380 | 31.998 | 32.60 | 32.219 | 33.057 | 33.81 | 33.471 | 34.358 | 34.46 | 34.080 | 35.086 | 34.87 | 34.460 | 35.541 |
| Sun Side Cyl Low E Surfaces | 32.00 | 31.742 | 32.314 | 33.05 | 32.747 | 33.426 | 34.45 | 34.062 | 34.867 | 35.15 | 34.717 | 35.630 | 35.60 | 35.144 | 36.119 |
| Bottom Shield Surfaces Facing Bus | 32.58 | 32.481 | 32.954 | 33.74 | 33.632 | 34.119 | 35.12 | 34.981 | 35.485 | 35.95 | 35.812 | 36.325 | 36.47 | 36.319 | 36.837 |
| Barrel Conic Surfaces | 32.37 | 31.873 | 32.726 | 33.49 | 32.896 | 33.892 | 34.85 | 34.124 | 35.321 | 35.64 | 34.820 | 36.153 | 36.12 | 35.251 | 36.670 |

The 70 K thermal intercepts are located on the Bus Bipods/Struts (Figure C-6). This strategy exploits excess refrigeration capacity in the cryocoolers' upper stages. The team set the thermal model intercept nodes as 70 K boundary nodes. The model was then used to calculate the heat required for these nodes to hold that temperature. The sum of that negative heat load was required to be less than that specified in the cryo-thermal budget. A similar procedure was used to optimally locate the 20 K intercept on the 4 K bipods.

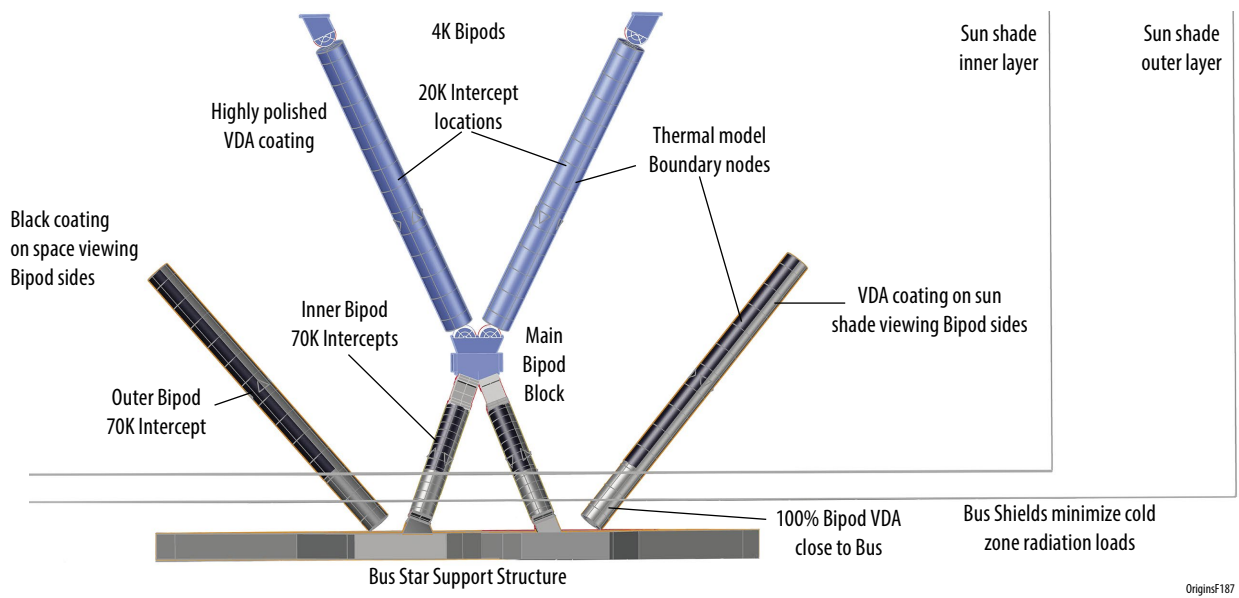

**Figure C-6:** Spacecraft bus bi-pods/strut and 4.5 K Bipods have heat intercepts at 20 and 70K and VDA or Black paint coatings located to minimize heat flow to the Cryogenic Payload Module.

**Table C-2:** Bipod assembly temperatures are listed for the full case set analyzed which shows the effectiveness of the thermal isolation strategy.

| Observatory Pitch Angle (deg) | 85 | | | 135 | | | 90 | | | 135 | | | 150 | | |
| Hot or Cold Case | Cold | | | Cold | | | Hot | | | Hot | | | Hot | | |
| | $T_{Avg}$ (K) | $T_{Min}$ (K) | $T_{Max}$ (K) | $T_{Avg}$ (K) | $T_{Min}$ (K) | $T_{Max}$ (K) | $T_{Avg}$ (K) | $T_{Min}$ (K) | $T_{Max}$ (K) | $T_{Avg}$ (K) | $T_{Min}$ (K) | $T_{Max}$ (K) | $T_{Avg}$ (K) | $T_{Min}$ (K) | $T_{Max}$ (K) |
|---|---|---|---|---|---|---|---|---|---|---|---|---|---|---|---|
| BUS BIPODS | | | | | | | | | | | | | | | |
| Left Lower Bus Mounting Block | 187.2 | 183.6 | 188.9 | 233.4 | 227.9 | 236.1 | 203.7 | 199.6 | 205.7 | 250.9 | 244.7 | 254.0 | 267.5 | 260.4 | 271.2 |
| Left Bus Titanium Tube Iso (Bus Intercept Side) | 131.4 | 84.9 | 186.4 | 157.6 | 92.3 | 232.1 | 141.4 | 87.8 | 202.8 | 168.1 | 95.5 | 249.3 | 177.3 | 98.2 | 265.7 |
| Left Bus Titanium Tube Iso (CPM Intercept Side) | 51.5 | 36.2 | 65.7 | 52.3 | 37.3 | 66.0 | 52.8 | 38.5 | 66.1 | 53.5 | 39.3 | 66.4 | 53.9 | 39.8 | 66.6 |
| Left Flexure Upper Block | 35.2 | 34.0 | 36.2 | 36.3 | 35.1 | 37.3 | 37.5 | 36.4 | 38.5 | 38.3 | 37.2 | 39.3 | 38.8 | 37.7 | 39.8 |
| Right Lower Bus Mounting Block | 186.7 | 183.2 | 188.6 | 233.9 | 228.6 | 236.8 | 203.6 | 199.6 | 205.7 | 251.8 | 245.8 | 255.0 | 268.7 | 261.9 | 272.5 |
| Right Bus Titanium Tube Iso (Bus Intercept Side) | 131.3 | 85.7 | 185.8 | 158.2 | 94.0 | 232.5 | 141.9 | 89.2 | 202.6 | 169.3 | 97.7 | 250.2 | 178.6 | 100.7 | 266.8 |
| Right Bus Titanium Tube Iso (CPM Intercept Side) | 51.7 | 36.4 | 65.8 | 52.6 | 37.6 | 66.2 | 53.2 | 38.7 | 66.2 | 54.0 | 39.6 | 66.6 | 54.4 | 40.1 | 66.8 |
| Right Flexure B Lower Block | 35.3 | 33.9 | 36.4 | 36.4 | 35.0 | 37.6 | 37.6 | 36.3 | 38.7 | 38.5 | 37.2 | 39.6 | 39.0 | 37.7 | 40.1 |
| Barrel Block Main | 33.6 | 32.4 | 34.4 | 34.7 | 33.5 | 35.5 | 35.9 | 34.8 | 36.7 | 36.8 | 35.6 | 37.6 | 37.3 | 36.1 | 38.1 |
| Left IAM M55J Cylinder Isolator Toward Barrel | 29.3 | 24.1 | 33.3 | 30.1 | 24.6 | 34.3 | 31.1 | 25.2 | 35.6 | 31.7 | 25.5 | 36.4 | 32.1 | 25.8 | 36.9 |
| Left IAM M55J Cylinder Isolator Toward Cold Zone | 13.8 | 4.8 | 19.1 | 13.8 | 4.8 | 19.1 | 13.8 | 4.8 | 19.1 | 13.8 | 4.8 | 19.1 | 13.8 | 4.8 | 19.1 |
| Left CPM Attach Block | 4.6 | 4.5 | 4.8 | 4.6 | 4.5 | 4.8 | 4.6 | 4.5 | 4.8 | 4.6 | 4.5 | 4.8 | 4.6 | 4.5 | 4.8 |
| Right IAM M55J Cylinder Isolator Toward Barrel | 29.3 | 24.2 | 33.3 | 30.2 | 24.6 | 34.4 | 31.1 | 25.2 | 35.6 | 31.7 | 25.5 | 36.4 | 32.1 | 25.8 | 36.9 |
| Right IAM M55J Cylinder Isolator Toward Cold Zone | 13.8 | 4.8 | 19.1 | 13.8 | 4.8 | 19.1 | 13.8 | 4.8 | 19.1 | 13.8 | 4.8 | 19.1 | 13.8 | 4.8 | 19.1 |
| Right CPM Attach Block | 4.6 | 4.5 | 4.8 | 4.6 | 4.5 | 4.8 | 4.6 | 4.5 | 4.8 | 4.6 | 4.5 | 4.8 | 4.6 | 4.5 | 4.8 |
| Left Lower Bus Mounting Block | 193.7 | 190.0 | 195.4 | 236.6 | 231.2 | 239.2 | 210.2 | 206.0 | 212.2 | 255.1 | 249.0 | 258.0 | 269.1 | 262.3 | 272.3 |
| Left Bus Titanium Tube Iso (Bus Intercept Side) | 135.3 | 86.8 | 192.9 | 159.8 | 94.5 | 235.4 | 145.6 | 90.3 | 209.3 | 171.2 | 98.3 | 253.6 | 179.0 | 100.8 | 267.4 |
| Left Bus Titanium Tube Iso (CPM Intercept Side) | 51.4 | 35.6 | 65.8 | 52.3 | 36.7 | 66.1 | 52.9 | 38.0 | 66.2 | 53.7 | 38.9 | 66.6 | 54.1 | 39.4 | 66.8 |
| Left Flexure Upper Block | 34.4 | 33.1 | 35.5 | 35.6 | 34.3 | 34.7 | 36.9 | 35.6 | 37.9 | 37.8 | 36.5 | 38.8 | 38.3 | 37.0 | 39.4 |
| Right Lower Bus Mounting Block | 193.5 | 189.7 | 195.6 | 236.1 | 230.6 | 239.3 | 209.8 | 205.5 | 212.3 | 254.3 | 248.1 | 258.0 | 268.3 | 261.3 | 272.4 |
| Right Bus Titanium Tube Iso (Bus Intercept Side) | 134.8 | 85.8 | 192.6 | 159.0 | 92.7 | 234.6 | 144.7 | 88.7 | 208.8 | 169.9 | 96.0 | 252.6 | 177.7 | 98.4 | 266.3 |
| Right Bus Titanium Tube Iso (CPM Intercept Side) | 51.2 | 35.6 | 65.7 | 52.0 | 36.8 | 66.0 | 52.6 | 38.0 | 66.1 | 53.3 | 38.8 | 66.4 | 53.7 | 39.4 | 66.6 |
| Right Flexure B Lower Block | 34.4 | 33.1 | 35.6 | 35.6 | 34.2 | 36.8 | 36.9 | 35.6 | 37.9 | 37.7 | 36.5 | 38.8 | 38.2 | 37.0 | 39.3 |
| Barrel Block Main | 33.0 | 32.4 | 33.4 | 34.1 | 33.5 | 34.6 | 35.5 | 34.8 | 35.9 | 36.3 | 35.6 | 36.8 | 36.8 | 36.0 | 37.3 |
| Left IAM M55J Cylinder Isolator Toward Barrel | 28.9 | 23.9 | 32.7 | 29.8 | 24.4 | 33.9 | 30.8 | 25.0 | 35.2 | 31.4 | 25.3 | 36.0 | 31.8 | 25.6 | 36.5 |
| Left IAM M55J Cylinder Isolator Toward Cold Zone | 13.8 | 4.8 | 19.1 | 13.8 | 4.8 | 19.1 | 13.8 | 4.8 | 19.1 | 13.8 | 4.8 | 19.1 | 13.8 | 4.8 | 19.1 |
| Left CPM Attach Block | 4.6 | 4.5 | 4.8 | 4.6 | 4.5 | 4.8 | 4.6 | 4.5 | 4.8 | 4.6 | 4.5 | 4.8 | 4.6 | 4.5 | 4.8 |
| Right IAM M55J Cylinder Isolator Toward Barrel | 28.9 | 23.9 | 32.7 | 29.8 | 24.4 | 33.9 | 30.8 | 25.0 | 35.2 | 31.4 | 25.3 | 36.0 | 31.8 | 25.6 | 36.4 |
| Right IAM M55J Cylinder Isolator Toward Cold Zone | 13.8 | 4.8 | 19.1 | 13.8 | 4.8 | 19.1 | 13.8 | 4.8 | 19.1 | 13.8 | 4.8 | 19.1 | 13.8 | 4.8 | 19.1 |
| Right CPM Attach Block | 4.6 | 4.5 | 4.8 | 4.6 | 4.5 | 4.8 | 4.6 | 4.5 | 4.8 | 4.6 | 4.5 | 4.8 | 4.6 | 4.5 | 4.8 |

Table C-2 shows the Bipod temperature results for the operational cases analyzed. Temperature results for each Bipod are shown for four temperature zones: the zone closest to the Bus Star Beam structure, the portion of the titanium isolator below the 70 K intercept, the portion of the titanium isolator above the 70 K intercept, and the Barrel connection point. Bus interface temperatures range from 183.6 K to 271.2 K (-90 to 0 C) for the full case set. Figure C-7 shows the temperature contour plot for the Pitch +150 Hot case. Table C-3 shows the required cryocooling for each intercept, as well as for the 4.5 K cold heads mounted to the 4.5 K cold zone structure.

The wires from the spacecraft to the instruments and telescope, including coax carrying sensitive data, are routed along the spacecraft bipods and 4 K bipods and anchored to heat sinks at 70 K, 35 K, along that path. The 20 K heat sink was not used for these cases in order to achieve a better heat load balance. This intercept will be included in future iterations. The parasitic heat flow from 35 K to 4.5 K was calculated separately with simple spreadsheet thermal conductivity curves.

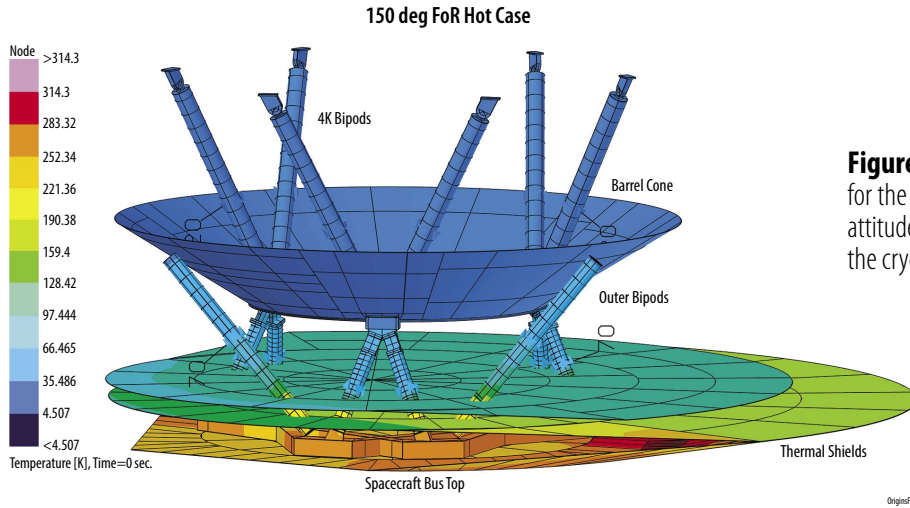

**150 deg FoR Hot Case**

4K Bipods
Barrel Cone
Outer Bipods
Thermal Shields
Spacecraft Bus Top

Node
> 314.3
314.3
283.32
252.34
221.36
190.38
159.4
128.42
97.444
66.465
35.486
4.507
< 4.507
Temperature [K], Time=0 sec.

**Figure C-7:** The temperature contour plot for the Pitch +150° Hot case shows this EOL attitude produces the highest heat loads to the cryocoolers.

**Table C-3:** Cryo cooling requirement for each intercept and the 4.5K cold heads shows that cryo cooler heat loads with 100% margins are satisfied for all cases.

| | | Hot Cases | | | Cold Cases | | |
|---|---|---|---|---|---|---|---|
| **Observatory Pitch Angle (deg)** | | **90** (W) | **135** (W) | **150** (W) | **85** (W) | **90** (W) | **135** (W) |
| CPM Cold Zone Power Dissipation (Watts) | | | | | | | |
| Instrument Power Dissipation | | 0.031 | 0.031 | 0.031 | 0.031 | 0.031 | 0.031 |
| Parasitic Thru Electrical Wires | | 0.014 | 0.014 | 0.014 | 0.014 | 0.014 | 0.014 |
| | Total: | 0.045 | 0.045 | 0.045 | 0.045 | 0.045 | 0.045 |
| Cold Zone 4.5 Cryo Cold Head Nodes (100mW) | | | | | | | |
| Four Cold Heads 4.5 K Refrigeration | | -0.093 | -0.089 | -0.090 | -0.079 | -0.080 | -0.079 |
| Allowable Sums Between 0 and -0.100 W. | Total: | -0.093 | -0.089 | -0.090 | -0.079 | -0.080 | -0.079 |
| Cold Zone 20 K Intercepts (200 mW Available) | | | | | | | |
| Mid Zone 20K Intercept Power | | -0.050 | -0.056 | -0.060 | -0.034 | -0.034 | -0.041 |
| 20 K BiPod Intercepts (Six Total - Once each Bi Pod) | | -0133 | -0.125 | -0.133 | -0.082 | -0.082 | -0.096 |
| Allowable Sums Between 0 and -0.200 W | Total: | -0.163 | -0.181 | -0.192 | -0.115 | -0.116 | -0.137 |
| BUS BiPod 70K Intercept Power (10 Watts Available) | | | | | | | |
| 70 K Central BiPod Intercepts (Six Total - One Each BiPod) | | -4.113 | -6.571 | -7.442 | -2.993 | -3.158 | -5.292 |
| 70 K Outer Strut Intercepts (Six Total - One Each Strut) | | -0.421 | -1.510 | -1.940 | 0.158 | 0.087 | -0.927 |
| Allowable Sums Between 0 and -10.0 W | | -4.533 | -8.081 | -9.382 | -2.835 | -3.072 | -6.219 |

## Sunshield

The team made the initial *Origins* model runs assuming the JWST sunshield layer 1 (sun-facing side) EOL thermal optical properties (alpha=0.47, emittance= 0.72). The initial results were poor. To overcome this, the team used Goddard Composite Coating (GCC) as the sunshield outer layer thermal coating. This coating provides a lower EOL alpha/emittance ratio (alpha=0.17, emittance=0.64) which significantly reduces sunshield temperatures.

GCC has flown on numerous missions. It was developed in the 1970s and variants of it have flown on GOES, GOLD, Triana, SECCHI, and SDO. It was also used on HST as an outer blanket layer replacement on Servicing Mission 3 (SM3). The most recent use is on LCRD as an outer blanket layer. The coating is a thin film sandwich of coatings deposited on a Kapton substrate to provide desired properties at solar wavelengths. Nominally, the coating is ~2 microns thick which enables sufficient flexibility to be used as an Multi-Layer Insulation (MLI) outer blanket layer. The coatings' thin film prescription will be tailored to *Origins* requirements. Temperature results are shown in Table C-4.

Figure C-8 shows *Origins* thermal model results and geometry for FoR cases (85, 112, 135, and 150 degrees). The gimballed solar arrays are always perpendicular to the solar vector.

**Table C-4:** The Sunshield temperature results show that two shields can provide an average temperature of ~120 K on the inner shield.

| Observatory Pitch Angle (deg) | 85 | | | 135 | | | 90 | | | 135 | | | 150 | | |
|---|---|---|---|---|---|---|---|---|---|---|---|---|---|---|---|
| Hot of Cold Case | Cold | | | Cold | | | Hot | | | Hot | | | Hot | | |
| | $T_{Avg}$ (K) | $T_{Min}$ (K) | $T_{Max}$ (K) | $T_{Avg}$ (K) | $T_{Min}$ (K) | $T_{Max}$ (K) | $T_{Avg}$ (K) | $T_{Min}$ (K) | $T_{Max}$ (K) | $T_{Avg}$ (K) | $T_{Min}$ (K) | $T_{Max}$ (K) | $T_{Avg}$ (K) | $T_{Min}$ (K) | $T_{Max}$ (K) |
| Sun Shade Temperatures | | | | | | | | | | | | | | | |
| Sun Shade Outer Layer | 201.9 | 185.6 | 214.2 | 190.1 | 175.1 | 201.4 | 264.1 | 240.3 | 281.4 | 242.3 | 221.1 | 257.7 | 242.2 | 221.1 | 257.6 |
| Sun Shade Inner Layer | 98.9 | 98.4 | 99.2 | 94.4 | 94.0 | 94.8 | 128.3 | 127.3 | 129.2 | 119.3 | 118.5 | 120.0 | 119.4 | 118.6 | 120.2 |
| BUSSHLD | | | | | | | | | | | | | | | |
| Outer Close-Out Surfaces To Top Bus Panels | 189.5 | 136.6 | 202.4 | 255.7 | 189.0 | 273.5 | 228.0 | 150.0 | 256.6 | 310.4 | 220.4 | 332.8 | 311.3 | 221.4 | 334.8 |
| Outer Conical Shield Surfaces Between SS OL & Bus Top | 160.8 | 137.6 | 179.4 | 230.8 | 219.8 | 244.5 | 208.6 | 181.5 | 228.3 | 286.3 | 274.0 | 303.0 | 287.4 | 273.7 | 304.9 |
| Y Facing Central Disk Closer To Bus | 129.0 | 109.6 | 136.0 | 159.9 | 126.5 | 171.2 | 149.3 | 122.4 | 162.1 | 181.8 | 140.9 | 198.5 | 188.0 | 145.4 | 203.6 |
| Y Facing Closeout To Outer Sun Shade | 117.1 | 81.7 | 138.3 | 152.0 | 96.7 | 210.2 | 139.5 | 99.9 | 156.1 | 178.8 | 114.2 | 256.7 | 181.3 | 115.7 | 256.9 |
| Y Facing Central Disk Closer To Cold Zone | 90.4 | 78.4 | 98.6 | 102.2 | 86.8 | 110.7 | 111.1 | 94.2 | 123.1 | 119.5 | 100.0 | 129.9 | 122.2 | 102.2 | 133.0 |
| Y Facing Closeout To Inner Sun Shade | 88.2 | 81.4 | 90.4 | 99.4 | 96.6 | 100.3 | 110.6 | 100.1 | 114.1 | 118.8 | 114.0 | 120.4 | 120.4 | 115.5 | 122.1 |

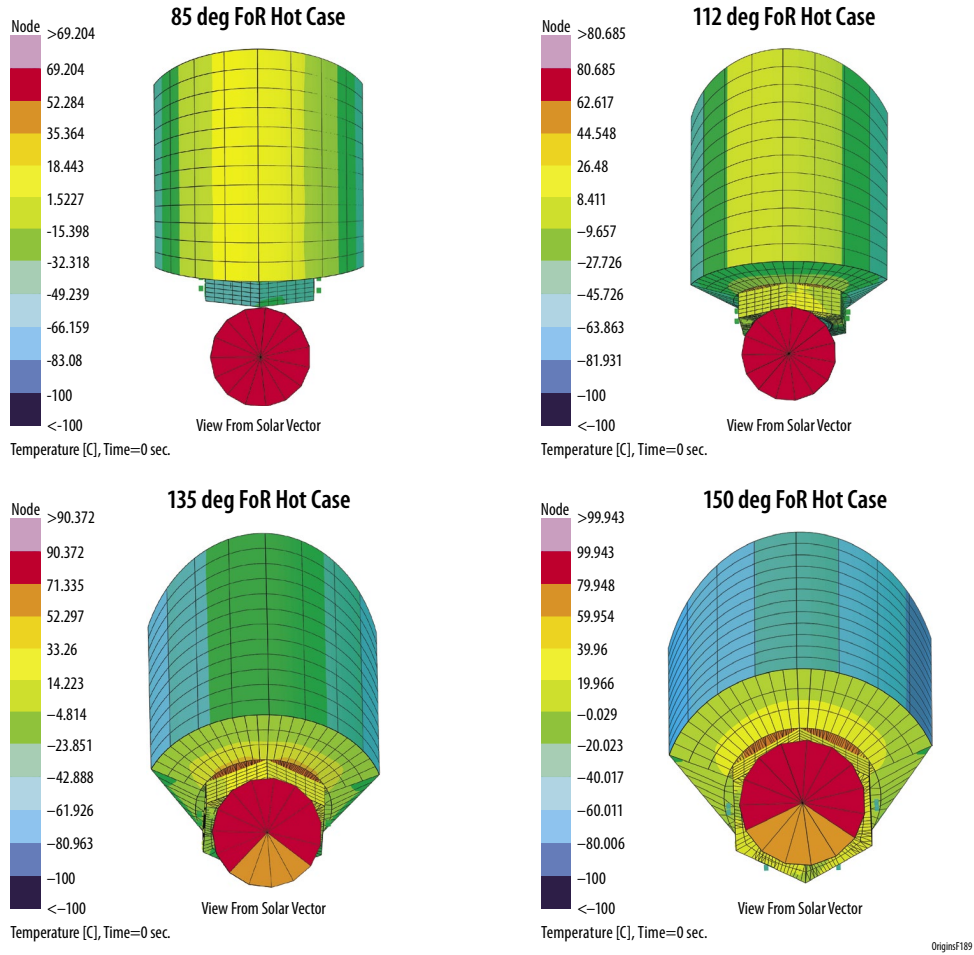

**Figure C-8:** *Origins* pitch angles of 85, 112, 135, and 150 degree (as viewed from the solar vector) shows the full FoR range accommodated by the thermal design. The inner shield and barrel are not visible in these views.

## Spacecraft Thermal System

The spacecraft bus thermal design provides a low-cost solution that satisfies all mission requirements. The overall thermal design strategy cold biases the spacecraft structure to minimize the Cryogenic Pay-load Module (CPM) interface temperature while accommodating over 4 kW component dissipated power. Heaters assure that minimum temperature requirements are not violated and constant conductance ammonia heat pipes are used to transfer high heat loads to exterior radiators and to spread heat within radiators.

The spacecraft bus model is hexagonally-shaped and provides six 1.40 m tall by 3.18 m wide peripheral panels. Additional radiating area is available on the bottom of the spacecraft perimeter outside the payload adaptor ring diameter 4.39 m (173 inches).

The hex vertex to opposing vertex dimension is 6.35 m. Key features are described in Figure C-9.

The thermal design minimizes the absorbed solar input by insulating sun facing sides Bay A and Bay B with MLI. The MLI assumes a GCC outer white outer layer so that incident solar energy is largely reflected back to the space sink.

The Bay C, D, E, and F perimeter panels experience minimal absorbed solar input for the full FoR Bus orientation range which make them ideal surfaces to reject dissipated heat loads. These panels are each split into an upper and lower panel. The 0.9-m tall upper panels are dedicated to the 4.5 K cryocooler heat loads. The 0.5-m tall lower panels are dedicated to various electronics components. Radiator panels are thermally coated with Z93C55 silicate white paint.

The bottom of the spacecraft outside the 4.4 m diameter payload adaptor fitting is also used to reject heat. These radiating surfaces can be exposed to incident solar energy for higher FoR Bus orientations which makes them less attractive for rejecting high heat loads. They are, however, suitable for rejecting smaller loads. It is noted (as shown in Figure C-8) that the gimbaled solar array casts a shadow on the bus bottom that reduces the solar heat load to these radiators. Thermal design performance is therefore dependent upon the solar array location and the shadow it casts.

Cryocoolers are physically mounted to the main structural thrust tube. The M55J composite thrust tube structure and radial panels have poor thermal conductance. Each cryocooler installation is iden-tical which affords cost, verification, and performance advantages. The 450 W cryocooler thermal interface is located at the height coinciding with the bottom of the perimeter cryocooler radiator panel. Constant conductance heat pipes transport the cryo-cooler heat load radially in the X-Z plane to the radiator panel. The aluminum honeycomb radiator panels embed heat pipes within to achieve high radiator efficiency. A U-shaped CCHP pipe at the bottom and along the perimeter edges and straight CCHP's to spread heat to the top of the cryocooler radiator panel are assumed.

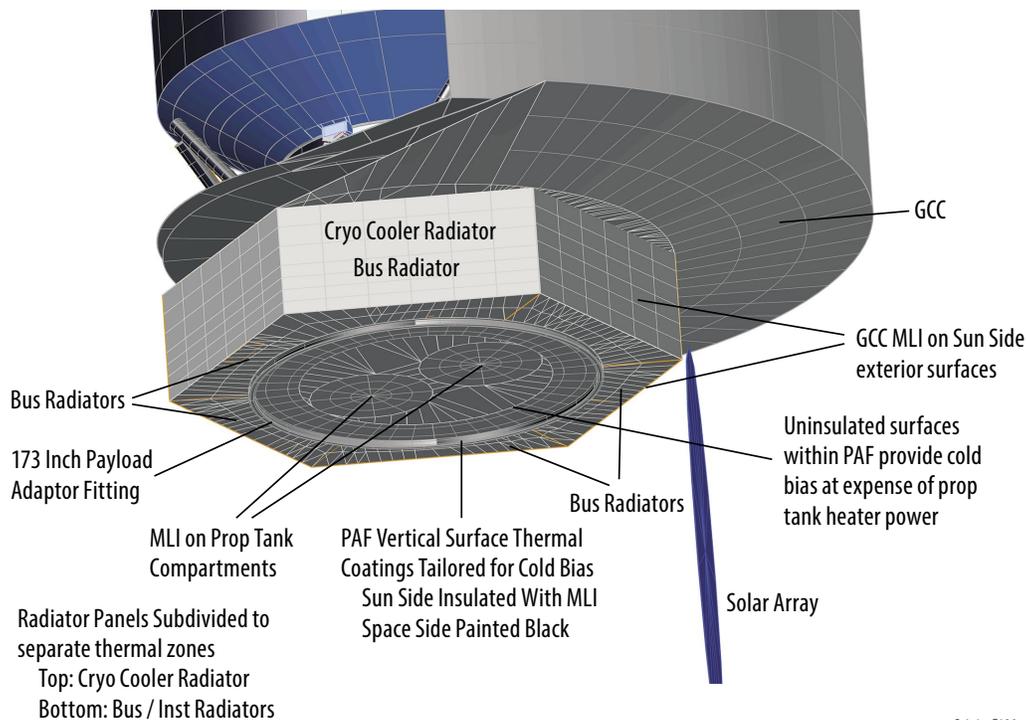

**Figure C-9:** The thermal aspects of the spacecraft bus are designed to minimize CPM interface temperatures while accommodating over 4kW electronics dissipated power.

Hot and cold thermal design cases were established to bound thermal design performance for mission parameters and risk factors associated with early mission design. Solar irradiance was varied from 1286 to 1421 W/m$^2$ for cold and hot cases, respectively. Beginning of Life (BOL) thermal coating properties for the cold case and End-of-Life properties for the hot case. Component power dissipation uncertainty is considered by assuming CBE estimates for the cold case and 30% higher power for the hot cases. Note that the cryocoolers are designed to accept a maximum power of 450 W each, thus no margin need be applied to these items.

**Table C-5:** All box temperatures fall within acceptable limits for hot and cold cases.

| Observatory Pitch Angle (deg) | 85 | | | 135 | | | 90 | | | 135 | | | 150 | | |
|---|---|---|---|---|---|---|---|---|---|---|---|---|---|---|---|
| Hot of Cold Case | Cold | | | Cold | | | Hot | | | Hot | | | Hot | | |
| | $T_{Avg}$ (C) | $T_{Min}$ (C) | $T_{Max}$ (C) | $T_{Avg}$ (C) | $T_{Min}$ (C) | $T_{Max}$ (C) | $T_{Avg}$ (C) | $T_{Min}$ (C) | $T_{Max}$ (C) | $T_{Avg}$ (C) | $T_{Min}$ (C) | $T_{Max}$ (C) | $T_{Avg}$ (C) | $T_{Min}$ (C) | $T_{Max}$ (C) |
| BUS_EBOXES | | | | | | | | | | | | | | | |
| Bay C OSS Main Elextronics #1 (MEB1) | -37.1 | -37.1 | -37.1 | -9.6 | -9.6 | -9.6 | -25.8 | -25.8 | -25.8 | 15.7 | 15.7 | 15.7 | 22.8 | 22.8 | 22.8 |
| Bay C OSS Main Electronics #2 (MEB2) | -43.3 | -43.3 | -43.3 | -10.0 | -10.0 | -10.0 | -30.9 | -30.9 | -30.9 | 15.6 | 15.6 | 15.6 | 23.9 | 23.9 | 23.9 |
| Bay C SIRU Inertial Ref Unit #1 | -32.6 | -32.6 | -32.6 | -7.5 | -7.5 | -7.5 | -22.1 | -22.1 | -22.1 | 17.8 | 17.8 | 17.8 | 24.7 | 24.7 | 24.7 |
| Bay C Transponder #1 | -39.9 | -39.9 | -39.9 | -15.8 | -15.8 | -15.8 | -27.6 | -27.6 | -27.6 | 7.2 | 7.2 | 7.2 | 13.1 | 13.1 | 13.1 |
| Bay C Transponder #2 | -37.5 | -37.4 | -37.4 | -14.5 | -14.5 | -14.5 | -25.2 | -25.1 | -25.1 | 7.9 | 7.9 | 7.9 | 13.4 | 13.4 | 13.4 |
| Bay C FIP BIAS Box #2 (Floor) | -43.8 | -43.8 | -43.8 | -18.4 | -18.4 | -18.4 | -31.8 | -31.8 | -31.8 | 4.4 | 4.4 | 4.4 | 10.8 | 10.8 | 10.8 |
| Bay C FIP BIAS Box #1 (Floor) | -41.7 | -41.7 | -41.7 | -15.9 | -15.9 | -15.9 | -29.6 | -29.6 | -29.6 | 7.7 | 7.6 | 7.6 | 14.1 | 14.1 | 14.1 |
| Bay C FIP RF Sig Proc Box | -30.8 | -30.8 | -30.8 | -7.2 | -7.2 | -7.2 | -16.5 | -16.5 | -16.5 | 18.6 | 18.6 | 18.6 | 24.5 | 24.5 | 24.5 |
| Bay C FIP ADR CTR Assy Box | -35.8 | -35.8 | -35.8 | -11.7 | -11.7 | -11.7 | -22.3 | -22.3 | -22.3 | 12.7 | 12.7 | 12.7 | 18.6 | 18.6 | 18.6 |
| Bay C RF_Translation_RF_Slice_EBox | -29.7 | -29.7 | -29.7 | -6.1 | -6.1 | -6.1 | -18.9 | -18.9 | -18.9 | 20.1 | 20.1 | 20.1 | 26.9 | 26.9 | 26.9 |
| Bay C Accelerometer #1 (Floor) | -45.3 | -45.3 | -45.3 | -7.0 | -7.0 | -7.0 | -31.6 | -31.6 | -31.6 | 20.0 | 20.0 | 20.0 | 29.4 | 29.4 | 29.4 |
| Bay C Accelerometer #2 (Floor) | -45.9 | -45.9 | -45.9 | -8.0 | -8.0 | -8.0 | -32.1 | -32.1 | -32.1 | 18.6 | 18.6 | 18.6 | 28.1 | 28.1 | 28.1 |
| Bay C Stay Tracker EBox (Floor) | -42.2 | -42.2 | -42.2 | -9.4 | -9.4 | -9.4 | -30.2 | -30.2 | -30.2 | 16.6 | 16.6 | 16.6 | 24.9 | 24.9 | 24.9 |
| Bay C Reaction Wheel (Floor) Heat Source | -52.8 | -52.8 | -52.8 | -23.0 | -23.0 | -23.0 | -38.7 | -38.7 | -38.7 | -4.9 | -4.9 | -4.9 | 3.7 | 3.7 | 3.7 |
| Bay D Optical Comm Module #2 | -21.9 | -21.9 | -21.9 | -19.9 | -19.8 | -19.8 | -13.8 | -13.8 | -13.8 | -7.2 | -7.1 | -7.1 | -5.3 | -5.3 | -5.3 |
| Bay D Optical Comm Module #1 | 7.8 | 7.7 | 7.7 | 9.8 | 9.8 | 9.8 | 22.4 | 22.4 | 22.4 | 28.0 | 28.0 | 28.0 | 29.6 | 29.6 | 29.6 |
| Bay D Power Supply Electronics (PSE) | -12.0 | -12.0 | -12.0 | -9.9 | -9.9 | -9.9 | -3.1 | -3.1 | -3.1 | 3.7 | 3.7 | 3.7 | 5.7 | 5.7 | 5.7 |
| Bay D ACE Avionics #1 | -22.8 | -22.8 | -22.8 | -20.3 | -20.3 | -20.3 | -14.5 | -14.5 | -14.5 | -7.4 | -7.4 | -7.4 | -5.4 | -5.4 | -5.4 |
| Bay D ACE Avionics #2 | -23.9 | -23.9 | -23.9 | -21.5 | -21.5 | -21.5 | -16.1 | -16.1 | -16.1 | -9.0 | -9.0 | -9.0 | -7.0 | -7.0 | -7.0 |
| Bay D OST MISC #1 | -46.3 | -46.3 | -46.3 | -36.4 | -36.4 | -36.4 | -37.0 | -37.0 | -37.0 | -23.6 | -23.6 | -23.6 | -19.7 | -19.7 | -19.7 |
| Bay D OST MISC #2 | -28.4 | -28.4 | -28.4 | -25.0 | -25.0 | -25.0 | -20.4 | -20.4 | -20.4 | -12.6 | -12.6 | -12.6 | -10.4 | -10.4 | -10.4 |
| Bay D SSR #1 | -42.6 | -42.6 | -42.6 | -26.8 | -26.8 | -26.8 | -29.5 | -29.5 | -29.5 | -10.3 | -10.3 | -10.3 | -4.7 | -4.7 | -4.7 |
| Bay D SSR #2 | -50.0 | -50.0 | -50.0 | -34.1 | -34.1 | -34.1 | -38.4 | -38.4 | -38.4 | -19.0 | -19.0 | -19.0 | -13.4 | -13.4 | -13.4 |
| Bay D CDH #1 | -46.8 | -46.8 | -46.8 | -22.9 | -22.9 | -22.9 | -34.3 | -34.3 | -34.3 | -2.2 | -2.2 | -2.2 | 4.7 | 4.7 | 4.7 |
| Bay D CDH #2 | -48.3 | -48.3 | -48.3 | -23.5 | -23.5 | -23.5 | -36.4 | -36.4 | -36.4 | -2.5 | -2.5 | -2.5 | 4.4 | 4.4 | 4.4 |
| Bay D OCXO Oven Controlled Crystal Oscillator | -52.7 | -52.7 | -52.7 | -30.2 | -30.2 | -30.2 | -41.2 | -41.2 | -41.2 | -11.7 | -11.7 | -11.7 | -4.9 | -4.9 | -4.9 |
| Bay E OSS BIAS Electronics | -28.2 | -28.2 | -28.2 | -27.3 | -27.3 | -27.3 | -20.1 | -20.1 | -20.1 | -17.0 | -17.0 | -17.0 | -15.9 | -15.9 | -15.9 |
| Bay E ADU And DITCE Electronics | -22.4 | -22.4 | -22.4 | -21.6 | -21.6 | -21.6 | -13.6 | -13.6 | -13.6 | -10.4 | -10.4 | -10.4 | -9.3 | -9.3 | -9.3 |
| Bay E ADR Control Assembly Electronics | -19.0 | -18.9 | -18.9 | -18.2 | -18.2 | -18.2 | -9.5 | -9.5 | -9.5 | -6.3 | -6.3 | -6.3 | -5.2 | -5.2 | -5.2 |
| Bay E OSS MEB #1 | -16.9 | -16.9 | -16.9 | -15.3 | -15.3 | -15.3 | -3.6 | -3.6 | -3.6 | -0.4 | -0.4 | -0.4 | 0.7 | 0.7 | 0.7 |
| Bay E OSS MEB #2 | -15.2 | -15.2 | -15.2 | -13.6 | -13.6 | -13.6 | -1.4 | -1.4 | -1.4 | 1.7 | 1.7 | 1.7 | 2.8 | 2.8 | 2.8 |
| Bay E OSS Translation Electronics | 15.8 | 15.8 | 15.8 | 18.0 | 18.0 | 18.0 | 40.0 | 40.0 | 40.0 | 43.4 | 43.4 | 43.4 | 44.6 | 44.6 | 44.6 |
| Bay E Star Tracker EBox | -6.5 | -6.5 | -6.5 | -4.3 | -4.3 | -4.3 | 11.4 | 11.4 | 11.4 | 14.7 | 14.7 | 14.7 | 15.9 | 15.9 | 15.9 |
| Bay E NavCam Malin ECAM50 (Floor) | -78.4 | -78.4 | -78.4 | -33.1 | -33.1 | -33.1 | -63.9 | -63.9 | -63.9 | -11.1 | -11.1 | -11.1 | -0.3 | -0.3 | -0.3 |
| Bay E Reaction Wheel (Floor) Heat Source | -26.4 | -26.4 | -26.4 | 0.6 | 0.6 | 0.6 | -13.9 | -13.9 | -13.9 | 26.2 | 26.2 | 26.2 | 33.0 | 33.0 | 33.0 |
| Bay F RF Signal Processing EBox | 2.9 | 2.9 | 2.9 | 10.0 | 10.0 | 10.0 | 27.6 | 27.6 | 27.6 | 37.0 | 37.0 | 37.0 | 37.7 | 37.7 | 37.7 |

The SBM thermal design strategically locates heat dissipating components to balance the heat loads to the available radiator panels. Table C-5 shows SBM-located component power dissipation assumptions for the case set considered. Radiator heat flow is localized to each respective panel since the composite structure to which it is structurally joined is a poor thermal conductor. Each electronic box is assumed to be flush mounted to the interior of the radiator panel. Radiator embedded spreader heat pipes are used where necessary in cases where the components base plate area does not provide sufficient radiating area. The internal viewing radiator side and the electronic box chassis are covered with MLI (or low e coating) to minimize the radiative heat flow to the spacecraft interior.

A cold zone within the thrust tube structure is created by employing the central bottom SBM surface as an uninsulated radiator. White paint (Z93C55) thermal optical properties are used for the outside panel. This minimizes the conducted heat load to the Bus Star Beam structure, minimizing its temperature. This is important in reducing the parasitic heat load to the CPM Cold Zone. The thrust tube structure has MLI insulation on the surface facing radially outward. Bare composite thrust tube surfaces face inward which are radiatively coupled to the cold spacecraft bottom central radiator. In addition, selected Star Beam structure surfaces facing the space sink are coated black to minimize its temperature.

Propulsion tanks (see Figure C-10) are insulated with MLI and heaters are applied to maintain allowable propulsion system temperatures. The Propulsion tank structural support conic is M55J composite that thermally-isolates the tanks from the cold structure. External surfaces within the tank structural support circle are insulated. An MLI blanket e* of 0.03 was assumed in the thermal model.

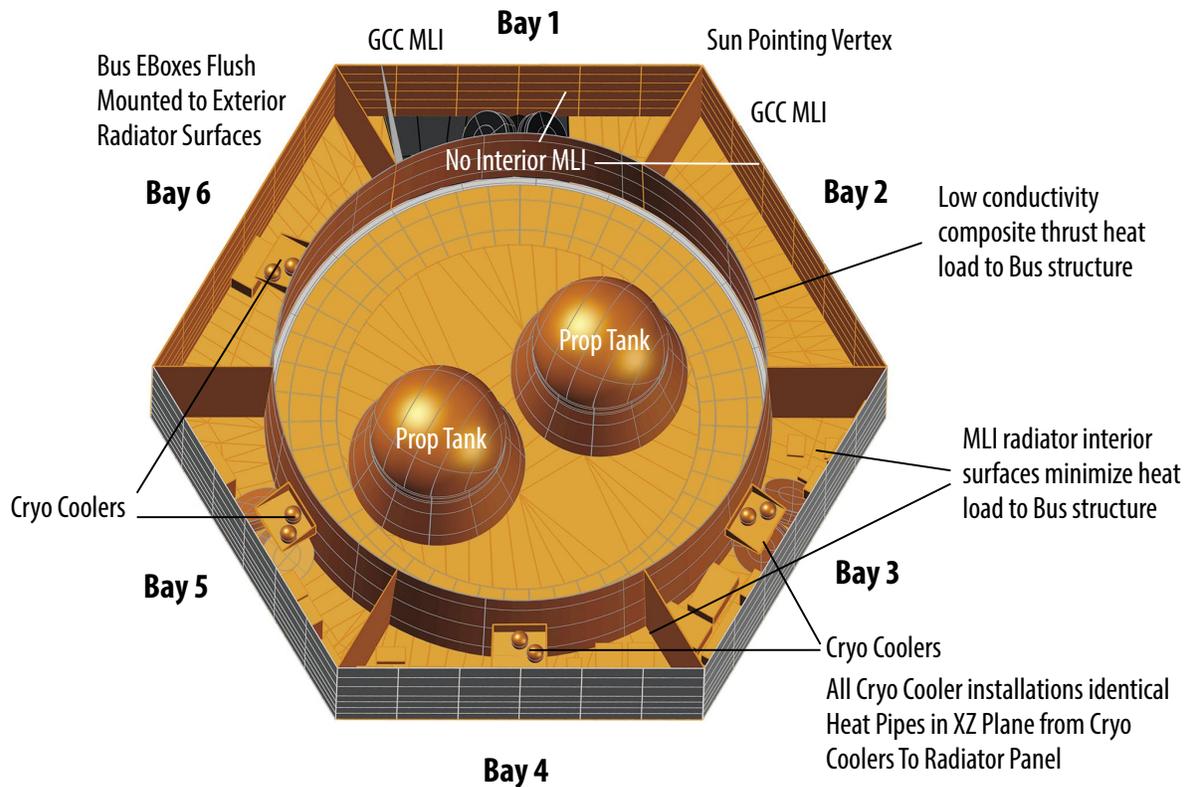

**Figure C-10:** The spacecraft bus thermal model interior view shows how the various heat loads are localized to each radiator panel.

Payload adaptor ring vertical surfaces are tailored to minimize its temperature. For both the sun side and space halves, sides facing the sun are assumed to be insulated with MLI while sides facing anti-sun are painted to maximize heat rejection and achieve cold bias.

The thermal vacuum test orientation directs the telescope to view in the +X direction (up). This orientation places the cryo cooler heat pipes perpendicular to the gravity field in the Y-Z plane. Spreader heat pipes within the perimeter radiator panels are oriented in the +X direction and will operate during thermal vacuum test in the reflux mode.

Figures C-11 to -14 show Bus component locations for Bays C, D, E, and F. No components are located on the Bus A or Bus B perimeter panels.

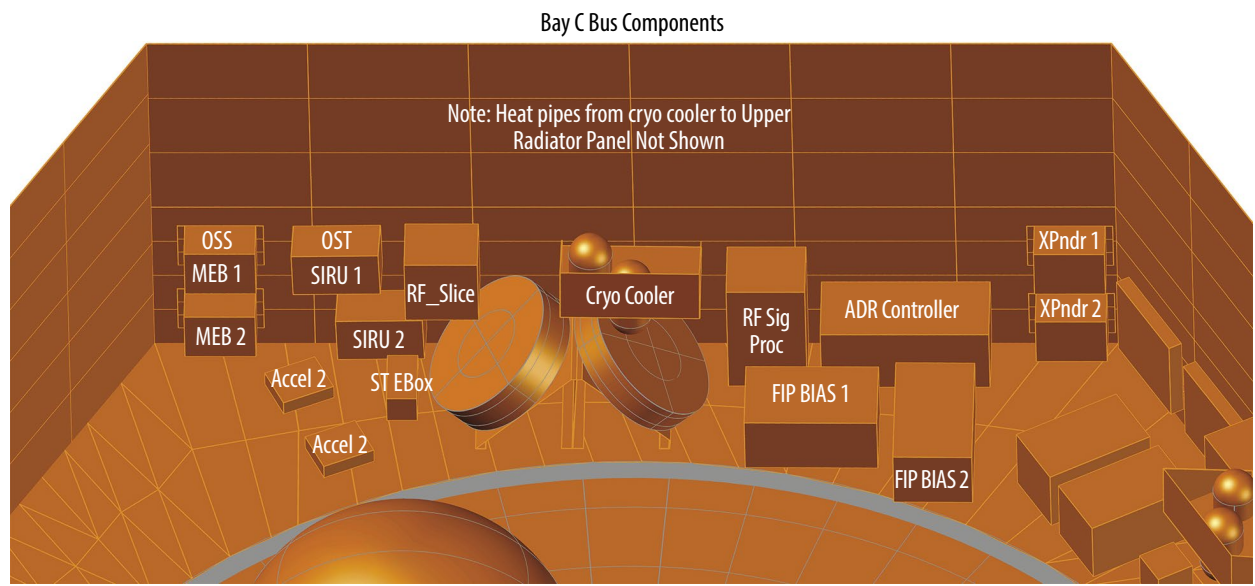

**Figure C-11:** There is plenty of space in each of the SBM bays for electrical boxes and other hardware. Here we show Bus Bay C.

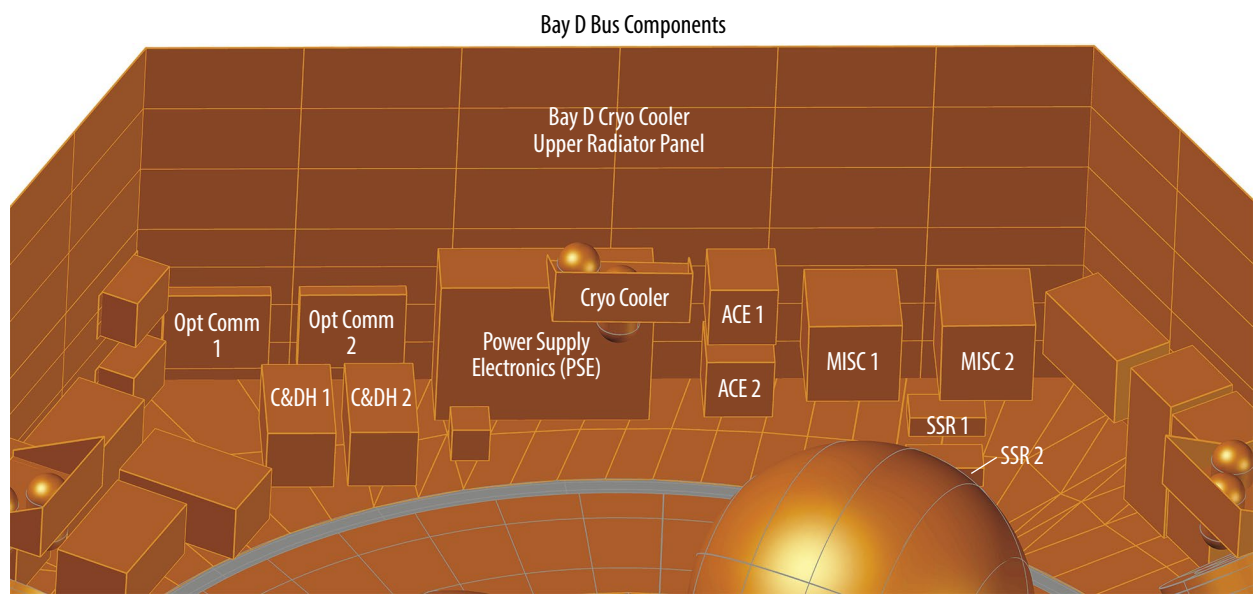

**Figure C-12:** Same as Figure C-11 but for Bus Bay D.

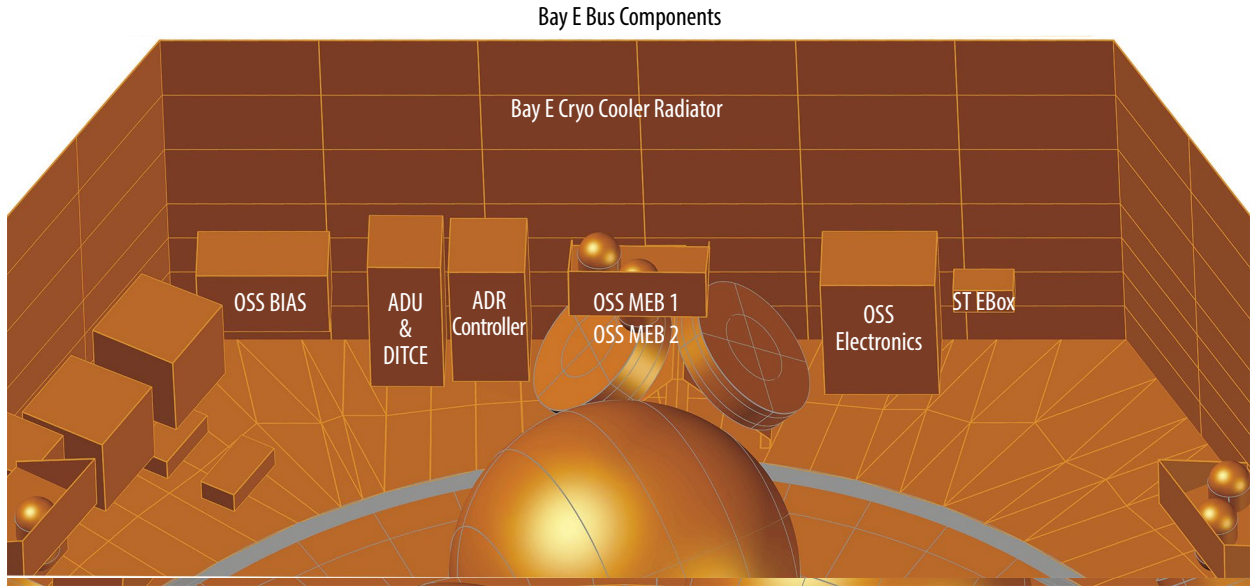

**Figure C-13:** Same as Figure C-11 but for Bus Bay E.

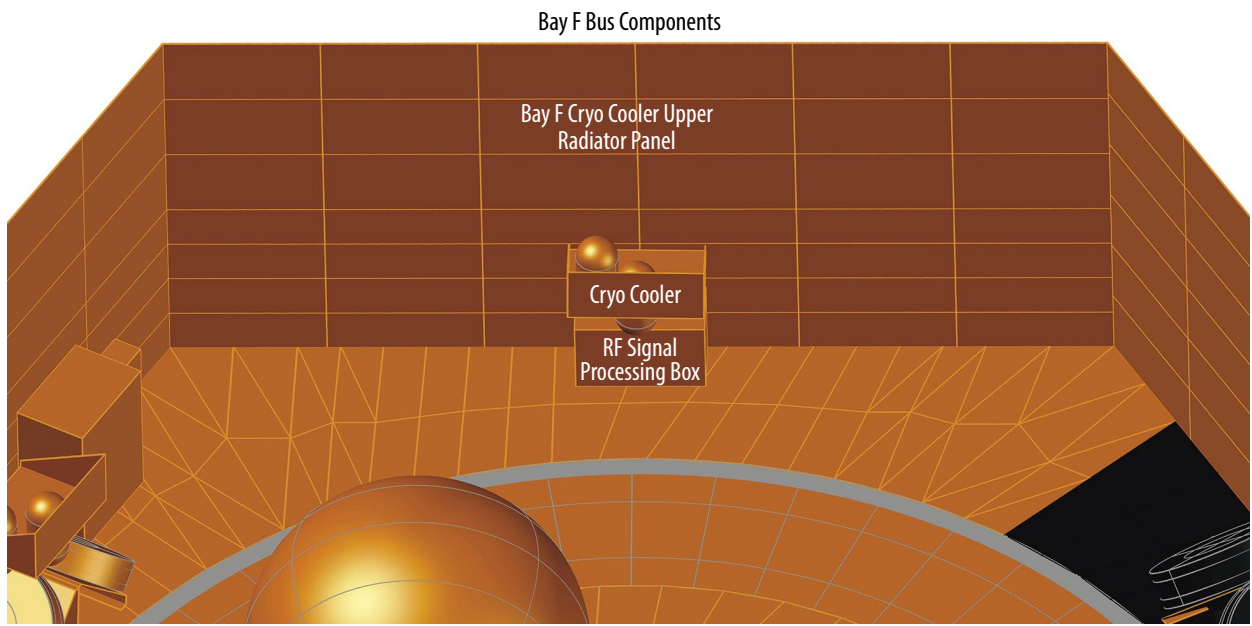

**Figure C-14:** Same as Figure C-11 but for Bus Bay F.

Table C-6 shows other key model heat flows for information purposes.

Figure C-15 plots spacecraft Bus temperature contours for FoR 0 cold case and FoR 150 degree hot cases.

Future study should explore thermal performance sensitivity to solar array location. Translating the array further from the Bus while optimizing its shape for its Bus shading effects might yield thermal performance benefits.

**Table C-6:** Thermal simulation heat flows include heater power and key absorbed solar heat loads.

| Pitch Angle (deg) | | Hot Cases | | | Cold Cases | | |
|---|---|---|---|---|---|---|---|
| | | 90 | 135 | 150 | 85 | 90 | 135 |
| | | (W) | (W) | (W) | (W) | (W) | (W) |
| Heater Power | | | | | | | |
| Propulsion Tank #1 Heater Power | | 86.0 | 47.3 | 23.2 | 94.5 | 88.2 | 49.6 |
| Propulsion Tank #2 Heater Power | | 102.5 | 42.2 | 3.5 | 91.8 | 105.7 | 45.7 |
| Bay C Lower Panel Heater | | 5.8 | 0.0 | 0.0 | 46.0 | 16.9 | 0.0 |
| Bay C Upper Cryo Radiator Panel Heater | | 55.2 | 0.0 | 0.0 | 151.0 | 76.3 | 0.0 |
| Bay D Lower Panel Heater | | 0.0 | 0.0 | 0.0 | 30.1 | 0.0 | 0.0 |
| Bay D Upper Cryo Radiator Panel Heater | | 0.0 | 0.0 | 0.0 | 40.8 | 3.1 | 0.0 |
| Bay E Lower Panel Heater | | 0.0 | 0.0 | 0.0 | 47.2 | 0.0 | 0..0 |
| Bay E Upper Cryo Radiator Panel Heater | | 0.0 | 0.0 | 0.0 | 44.0 | 4.7 | 0.0 |
| Bay F Lower Panel Heater | | 0.0 | 0.0 | 0.0 | 0.0 | 0.0 | 0.0 |
| Bay F Upper Cryo Radiator Panel Heater | | 0.0 | 0.0 | 0.0 | 27.5 | 0.0 | 0.0 |
| | Total: | 249.5 | 89.5 | 26.7 | 572.9 | 294.8 | 95.3 |
| Solar Array Face (Alpha = 0.82) | Total: | 23412.4 | 24342.8 | 24102.6 | 15962.9 | 21138.3 | 23979.5 |
| Sun Shade Absorbed Power | | | | | | | |
| Sun Shade Outer Layer Absorbed Power | | 29369.5 | 20755.7 | 20752.2 | 10897.2 | 10939.7 | 8548.9 |
| Sun Shade Inner Layer Absorbed Power | | 0.0 | 0.3 | 0.3 | 0.5 | 0.1 | 0.3 |
| | Total: | 29369.5 | 20756.0 | 20752.4 | 10897.7 | 10939.8 | 8549.2 |
| BUS Bus Perimeter Panel Absorbed Power | | | | | | | |
| Bay A MLI Exterior Surface | | 1074.3 | 829.5 | 833.5 | 368.7 | 408.0 | 360.2 |
| Bay B MLI Exterior Surface | | 1073.7 | 844.9 | 847.5 | 368.5 | 407.7 | 365.9 |
| Bay C High Cryo Radiator Surface | | 2.4 | 96.6 | 96.3 | 0.2 | 1.4 | 62.3 |
| Bay C Low Bus Radiator Surface | | 1.8 | 38.0 | 37.7 | 0.2 | 0.0 | 24.1 |
| Bay D High Cryo Radiator Surface | | 0.0 | 0.0 | 0.0 | 0.0 | 0.0 | 0.0 |
| Bay D Low Bus Radiator Surface | | 0.0 | 0.0 | 0.0 | 0.0 | 0.0 | 0.0 |
| Bay E High Cryo Radiator Surface | | 0.0 | 0.0 | 0.0 | 0.0 | 0.0 | 0.0 |
| Bay E Low Bus Radiator Surface | | 0.0 | 0.0 | 0.0 | 0.0 | 0.0 | 0.0 |
| Bay F High Cryo Radiator Surface | | 2.4 | 96.5 | 96.3 | 0.2 | 1.4 | 62.5 |
| Bay F Low Bus Radiator Surface | | 1.8 | 37.7 | 37.6 | 0.2 | 1.0 | 24.2 |
| | Total: | 2156.5 | 1943.1 | 1949.1 | 737.9 | 820.6 | 899.3 |
| BUS Bus Bottom Panel Absorbed Power | | | | | | | |
| Bottom Central Region Outside Prop Tank Disks | | 2.5 | 46.4 | 42.8 | 1.3 | 1.9 | 41.8 |
| Bottom Central Disk | | 2.3 | 47.0 | 45.0 | 1.1 | 1.7 | 41.8 |
| Bay A Bottom Surfaces | | 18.0 | 129.3 | 146.6 | 6.0 | 11.4 | 94.8 |
| Bay B Bottom Surfaces | | 18.4 | 165.3 | 179.5 | 5.6 | 11.5 | 114.2 |
| Bay C Bottom Surfaces | | 4.0 | 406.3 | 409.2 | 0.8 | 2.4 | 234.5 |
| Bay D Bottom Surfaces | | 0.1 | 22.9 | 33.3 | 0.1 | 0.1 | 13.8 |
| Bay E Bottom Surfaces | | 0.1 | 46.5 | 56.7 | 0.1 | 0.1 | 27.3 |
| Bay F Bottom Surfaces | | 3.0 | 311.6 | 312.8 | 0.7 | 1.9 | 180.4 |
| | Total: | 48.4 | 1175.1 | 1226.1 | 15.6 | 31.0 | 748.7 |

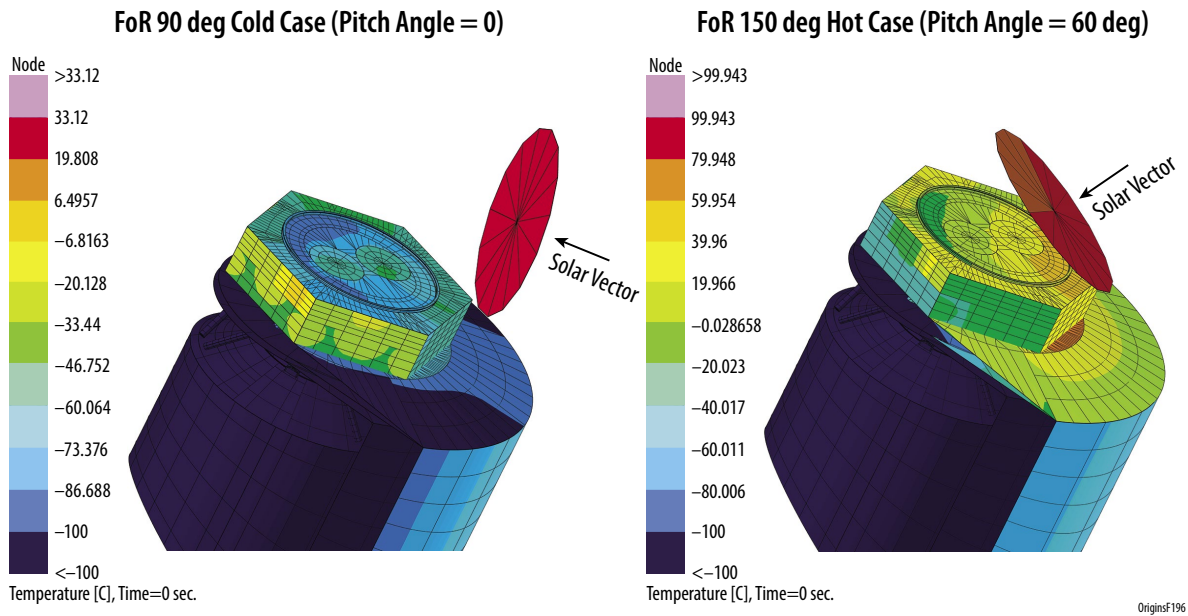

**FoR 90 deg Cold Case (Pitch Angle = 0)** **FoR 150 deg Hot Case (Pitch Angle = 60 deg)**

**Figure C-15:** The external surfaces of the outer Sunshield layer and SBM are all within the required bounds.

## C.2 Assembly and Structural Details

### C.2.1 Assembly Details

#### CPM Assembly

CPM assembly begins at the CPM Base, which is first secured to the OST transportation dolly. The Bus Bipod subassemblies are hinged at the base and require support during the install. Therefore, the 35 K Deck with the Bipod Interface Brackets and Barrel Support Ring are pre-assembled. This 35 K Deck subassembly is then craned over the Star, allowing installation of the Bus Bipods, by installing the hinge pin at the base and bolting the flexure to the Bipod Interface Bracket. This process is repeated for each opposite Bus Bipod until all Bus Bipods are installed.

The PMBSS design is circumferentially symmetric and can be broken down into three sets of two pie-shaped wedge styles (Figure C-16). The size of each wedge subassembly allows the components to be brazed together using brazing ovens known to exist today. The brazing fabricator will have to develop fixtures and jigs to complete the wedge subassembly. When all wedges are completed, they are bolted together to complete the PMBSS assembly.

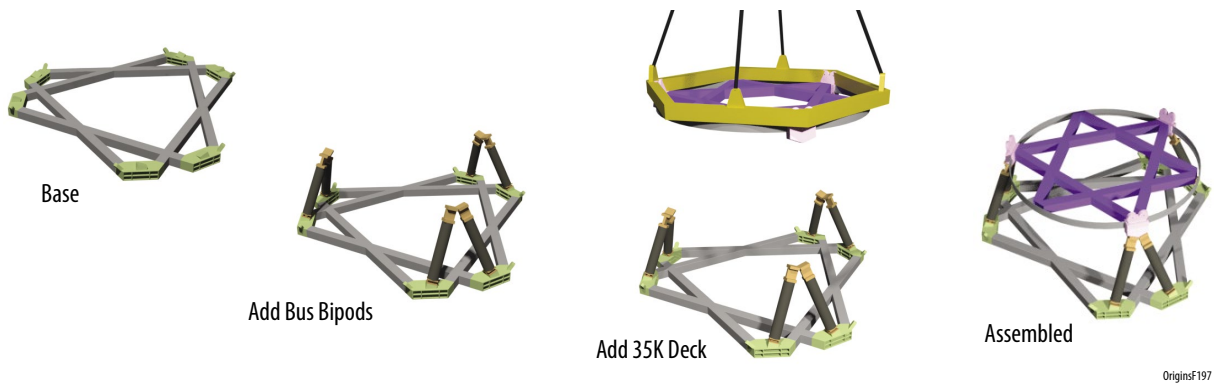

Base    Add Bus Bipods    Add 35K Deck    Assembled

**Figure C-16:** The Bus Bipods are temporarily supported during assembly since they are hinged at the base.

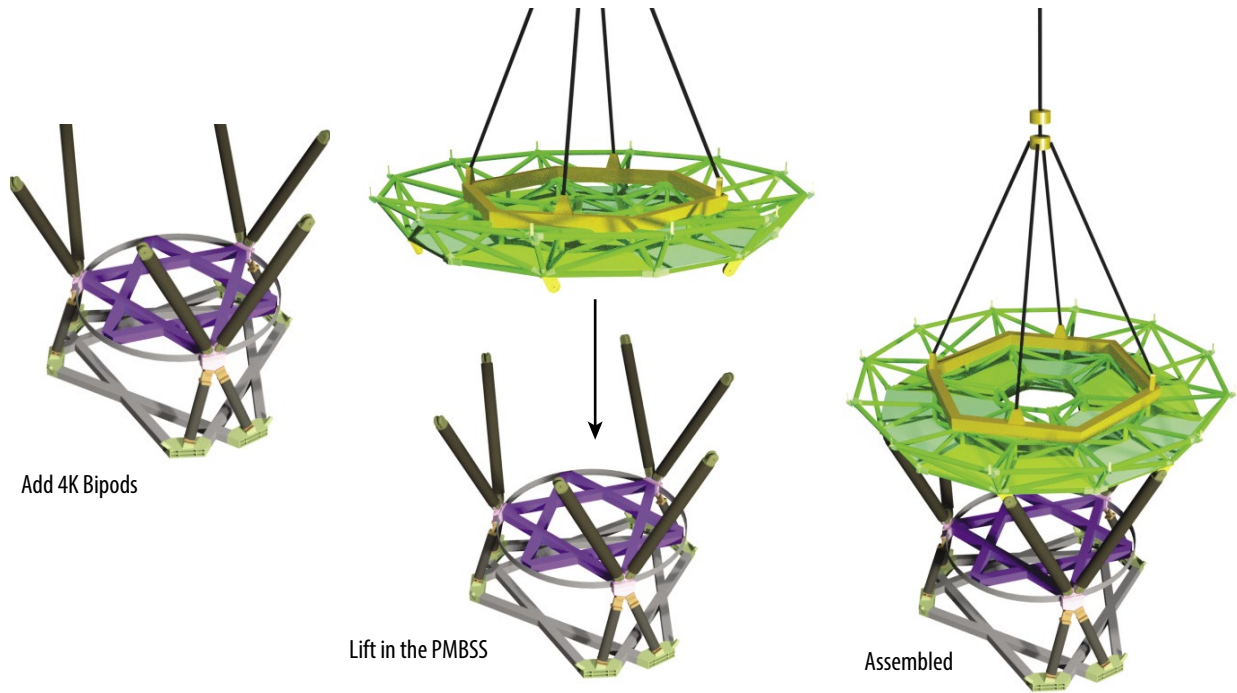

Add 4K Bipods

Lift in the PMBSS

Assembled

**Figure C-17:** The 4.5 K Bipods are supported until the PMBSS is installed.

The interface at each end of the 4.5 K Bipods is a lug-and-clevis joint. At this point in assembly, the lug-side of the upper joint is already attached to the PMBSS. At the opposite end of the 4.5 K Bipod, the lug-side of the joint is an integral part of the Bipod Interface Bracket. The PMBSS is lifted by crane and supported until it is high enough to install the 4.5 K Bipods. Once installed, the PMBSS is lifted further, over the 35 K Deck, to install the lower ends of the 4.5 K Bipods to the Bipod Interface Brackets (Figure C-17).

The Secondary Mirror Tripod, SMSA, and PMSAs can then be installed on the PMBSS. The Baffle Interface Flexures are also installed around the top of the PMBSS (Figure C-18). The main CPM structure is then ready for instrument installation.

The team will construct an Instrument Installation Platform at the height of the 35 K Deck to enable each instrument's GSE to roll its instrument to a point under its mounting location on an IMP. The GSE is required to lift the instrument to enable attachment to the IMP (Figure C-19).

After all electronics, harnessing, and thermal strapping are installed, the Cone sections of the Barrel are attached [*show that the cone comes in several sections*] to the Barrel Support Ring on the

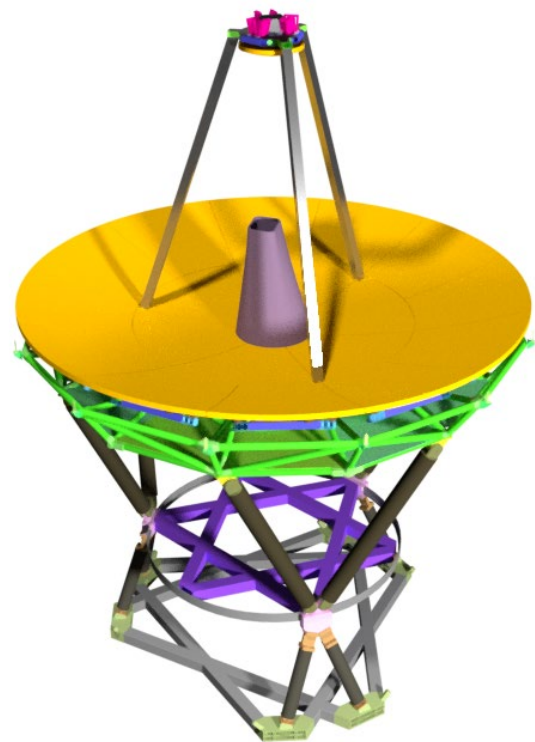

**Figure C-18:** Each Primary Mirror Segment Assembly is installed, then the Tripod, then the Secondary Mirror, and finally the Aft Optics Structure.

35 K Deck. The Baffle is then lowered and attached to the Baffle Interface Flexures at the top of the PMBSS (Figures C-20 and -21). On a parallel path, the DAK-Foil-DAK blanketing and offset brackets are installed on the inside surface of the Barrel. The Barrel is then lifted over the existing subassembly and lowered into the top of the Cone. The pre-assembled Telescope Cover is then installed.

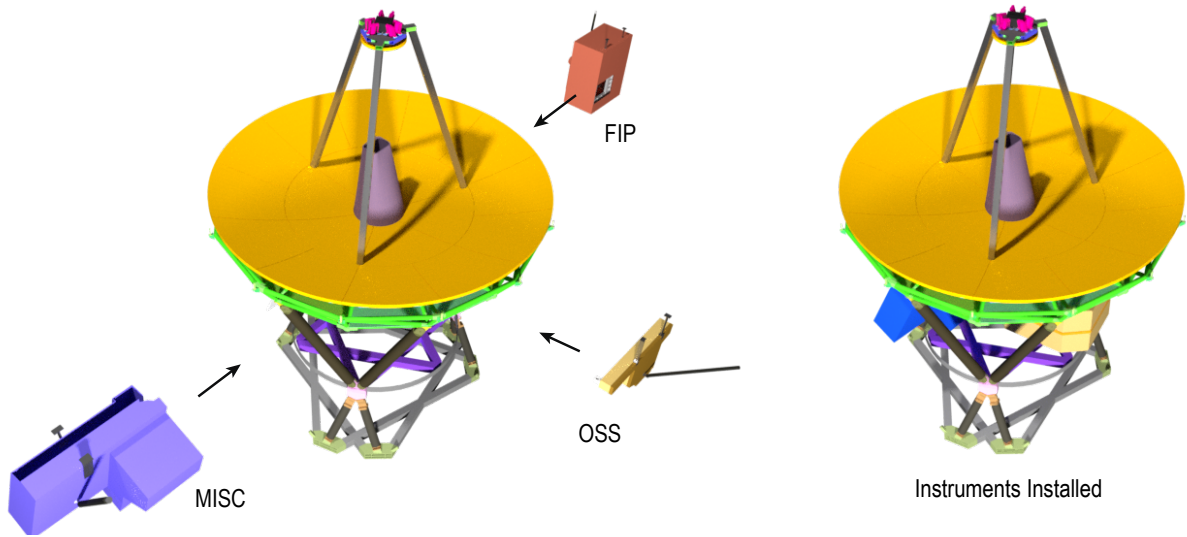

**Figure C-19:** Instrument ground support equipment is used to carry each instrument into area under the PMBSS.

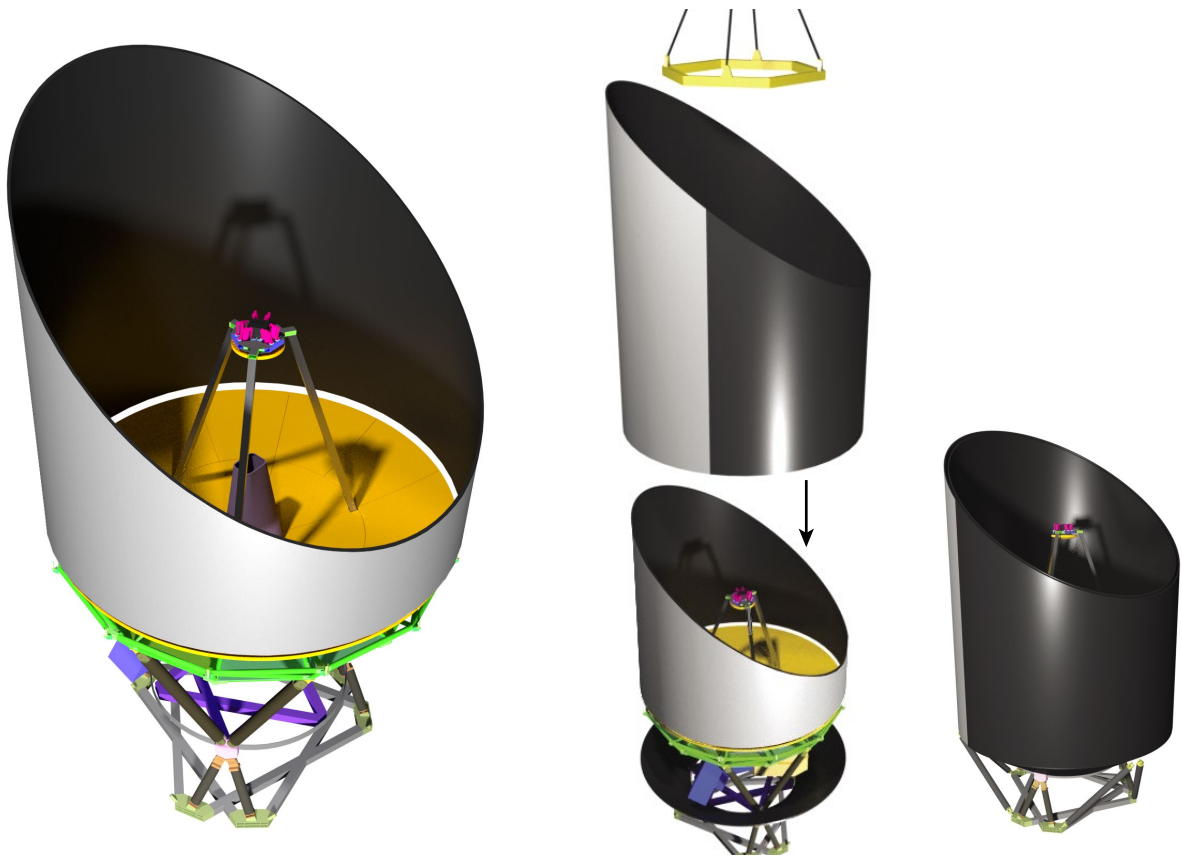

**Figure C-20:** Titanium flexures are installed on the PMBSS, located around the entire top edge of the PMBSS, and then the Baffle is lowered down and attached to the flexures.

**Figure C-21:** The Barrel cone section is installed to the 35 K Deck in multiple subassemblies. The Barrel cylindrical section is then lowered onto the Cone.

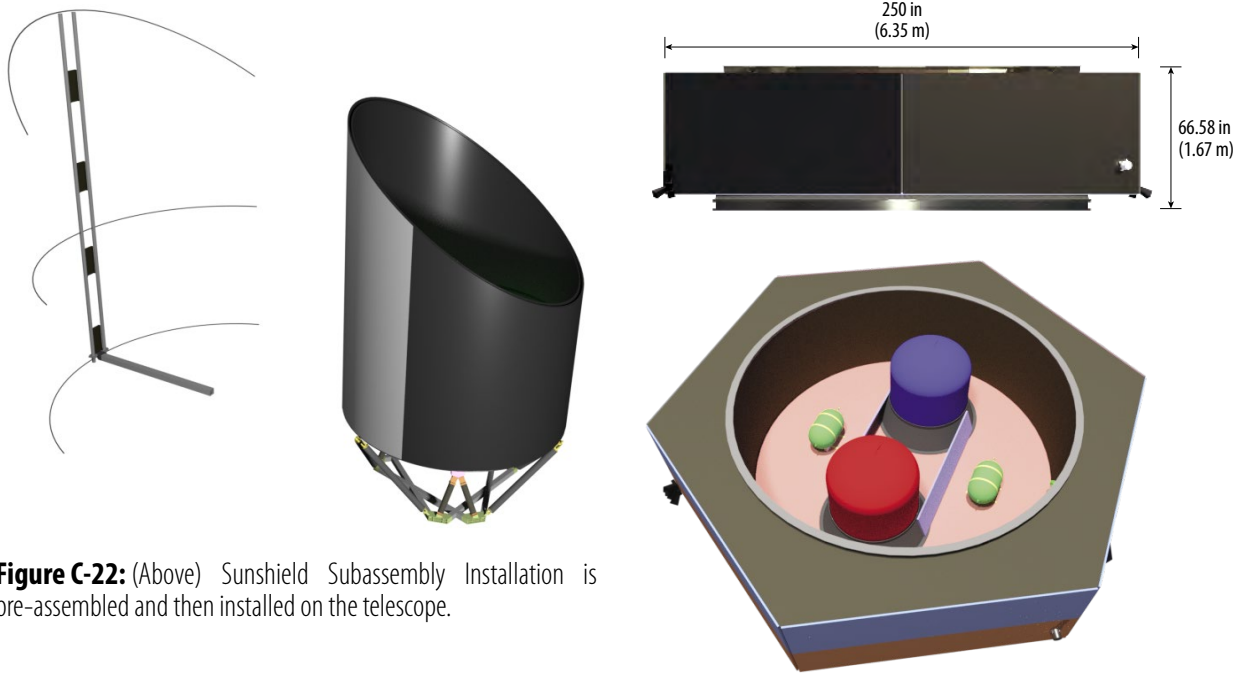

**Figure C-22:** (Above) Sunshield Subassembly Installation is pre-assembled and then installed on the telescope.

Origins F204

**Figure C-23:** (Above/right) The *Origins* spacecraft bus supports the CPM and contains all the warm-side electronics, communication subsystems, and propulsion subsystem.

The Sunshield mechanism arms are pre-assembled; these subassemblies are installed to the Base. The Sunshield masts and horizontal flexible rods are then installed to the arms and masts, respectively. The Sunshield components are then in the deployed configuration. The Inner Fixed Sun Shield and then the Outer Fixed Sunshield are then installed above the Base. The shield material is then ready to be installed to the masts and arms (Figure C-22).

### SBM Assembly

The *Origins* Spacecraft is a 1.67 m tall hexagon-shaped bus with a 6.35 m circumscribed diameter (Figure C-23).

The four non-sun-facing Bus side panels act as heat sink radiators for the cryocoolers, instrument electronics, and Bus electronics. The main load carry component is the centrally located cylindrical Thrust Tube. The six Bus Radial panels are equally spaced and attach on the outside of the Thrust Tube, forming six bays for mounting the Bus equipment (Figure C-24). The Radial panels support the Bus side panels, top close-out panel, and bottom panel.

### Sun-Side

The Thrust Tube attaches to the Payload Adapter Fitting (PAF) Interface Ring (PIR). The PIR flies with the Bus when ejected from the launch

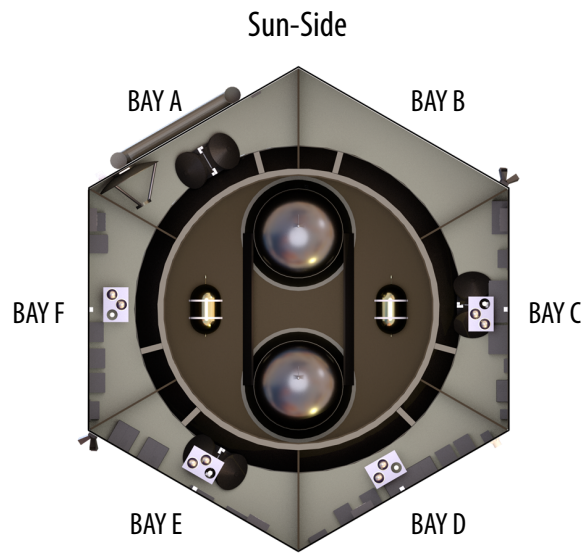

**Figure C-24:** Bus Bay Definitions are used to help describe the locations of the various electronics and other subsystems.

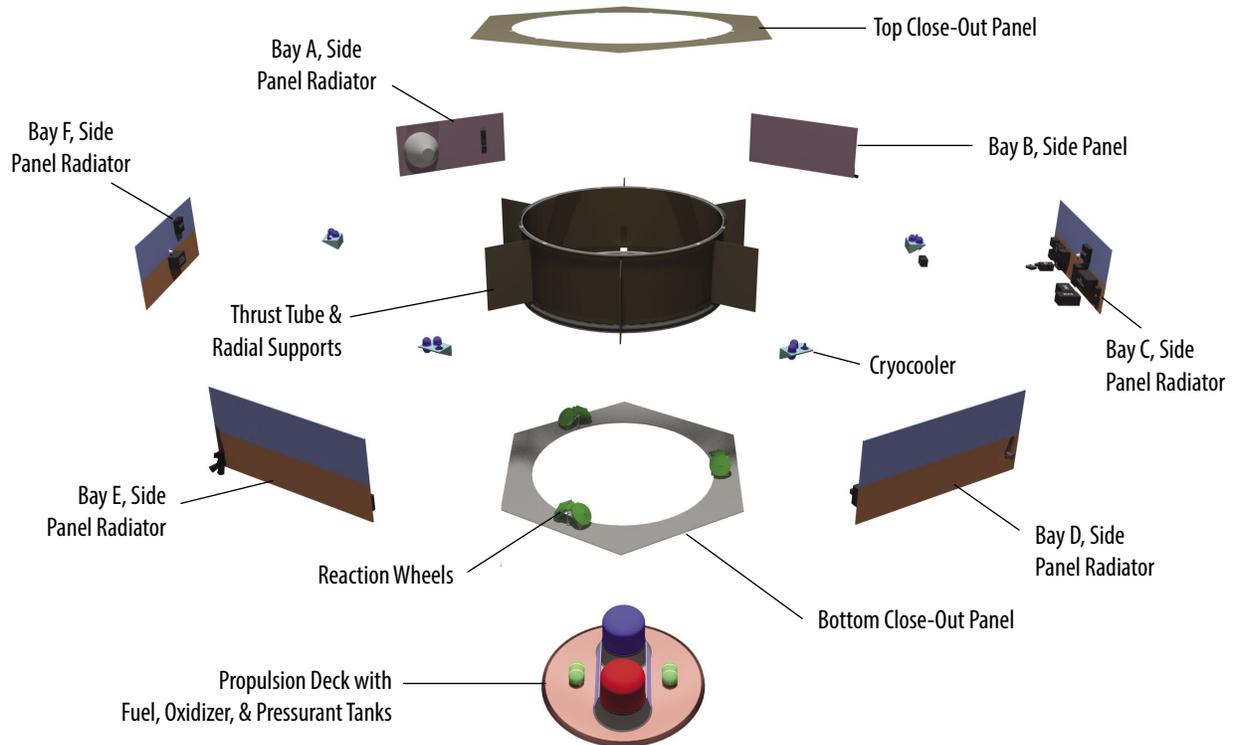

**Figure C-25:** The *Origins* Spacecraft Bus, Exploded View

vehicle second stage. Mounted to the PIR is the Propulsion Deck, which sits inside of the Thrust Tube. On the Propulsion Deck are the Oxidizer and Fuel tanks, pressurant tanks, and propulsion control system panel. Figure C-25 shows the spacecraft bus components.

### Thrust Tube

The Thrust Tube (Figure C-25) is a 1.56 m tall tube, 4.33 m diameter, made from M55J fiber composite (60% FV [*0/45/90/-45*]). The main wall thickness is 6.35 mm, with thicker 12.7 mm by 76.2 mm long bands at each end for a bolted interface to the Upper and Lower Interface Rings. Around the Thrust Tube are three equally-spaced CPM Support brackets that also provide stiffness to the structure. The Upper Interface Ring and CPM Support Brackets enable the CPM to be mounted to the SBM. The Lower Interface Ring attaches the Thrust Tube to the PIR.

### Side Panels

The six Bus Side Panels (Figure C-26) are 25.4 mm thick aluminum face-sheet honeycomb panels, 3.15 mm wide by 1.4 mm tall. Four of the non-sun-facing Bus Side Panels are used as heat sinks, radiating to deep space. These side panels are split length-wise, allowing the upper section to be dedicated to a cryocooler and the lower section to be dedicated to the electronics located in each bay. The radiators contain embedded heat pipes to improve thermal efficiencies.

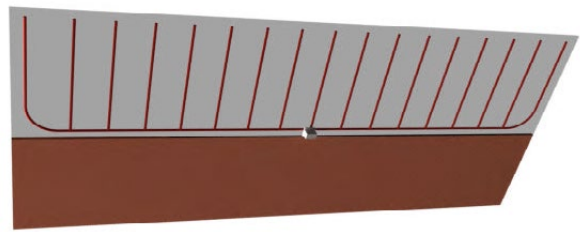

**Figure C-26:** The Spacecraft Bus Side Panel Radiators are split lengthwise to create two separate radiators — the lower acts as a heat sink for electronics and the upper as a radiator for cryocooler heat. The upper section has embedded heat pipes that transfer the heat from the cryocooler.

## Close-out Panels

Closing out the topside of the Bus bays is a lightweight, thin (12.7 mm), aluminum face-sheet honeycomb core composite panel. This panel does not support any hardware and is used only to enclose the bay for thermal and contamination control. The panel does provide in-plane stiffness and structure to the Radials and Side Panels.

The Bottom Close-out panel is a 25.4-mm thick aluminum face-sheet honeycomb panel. These panels are fastened to the Radials and Side Panels. Bus electronics are also mounted on these panels.

## Propulsion Deck

The Propulsion Deck sits inside of the Thrust Tube mounted on the PIR. The Oxidizer and Fuel tanks, pressurant tanks, and propulsion control system panel are on the Propulsion Deck. The tanks are supported by M55J fiber composite skirts. These skirts have a 6.35-mm thick wall and are 438 mm tall. To minimize negative effects on the Deck's first frequency, they are positioned as far as possible from the center of the Deck. The tanks are heated and blanketed. The fiber composite material was selected to provide a low thermal conductance path to the Deck.

The Propulsion Deck is an aluminum face-sheet honeycomb panel with 2-mm thick face-sheets and an overall panel thickness of 101.6 mm. The panel is bolted at the perimeter to the PIR. To help maintain a first natural frequency of 15 Hz or better, two support panels are fastened to the skirts and Deck (Figure C-27).

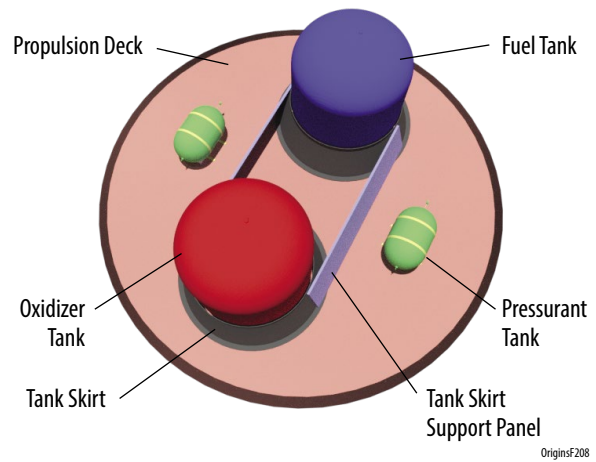

**Figure C-27:** The Propulsion Deck supports the propulsion equipment and has a first natural axial frequency greater than 15 Hz.

## Subsystem Component Layout

Bay A houses three diplexers. The communication antenna subassembly is mounted external to the bus wall. To fit within the launch vehicle, the antenna dish is recessed into Bay A and the dish recess is enclosed (Figure C-28). Bay B is empty. Figure C-28 show the electronic components located in Bays A, & C-F.

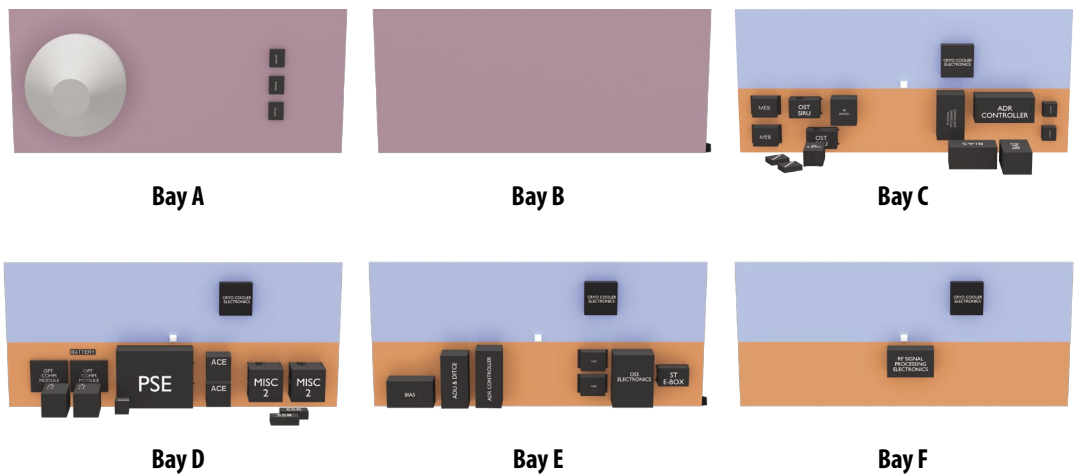

**Figure C-28:** Bay A Components, showing the Communication Antenna Dish's enclosure and three diplexers. Bay C-F Components; the upper (blue) section of the side panel is the cryocooler heatsink and contains embedded heatpipes. Components mounted to the Bottom Close-out panel appear to be floating.

**SBM Strength Test**

The spacecraft Bus will also be strength tested to protoflight levels. For this testing the CPM Mass Simulator will be attached to the spacecraft at the same interface as the CPM. The spacecraft PIR will attach to the vibration table through a test fixture that mimics the flight interface. The spacecraft Bus assembly will undergo protoflight three-axis sine vibration testing, protoflight acoustics testing, and separation shock testing. If vibration testing is deemed inadequate, then a pull test fixture will be attached at the CPM interface and the spacecraft will undergo a pull testing or centrifuge testing program. This testing will verify the workmanship and structural design of the SBM.

**C.2.2 Structural Details**

The structural analysis completed for this report is limited to determining the observatory's primary natural frequencies and a small set of components with the minimum margins of safety. The *Origins* observatory is designed to launch on the Space Launch System (SLS) and SpaceX BFR. The SLS Mission Planner's Guide states that

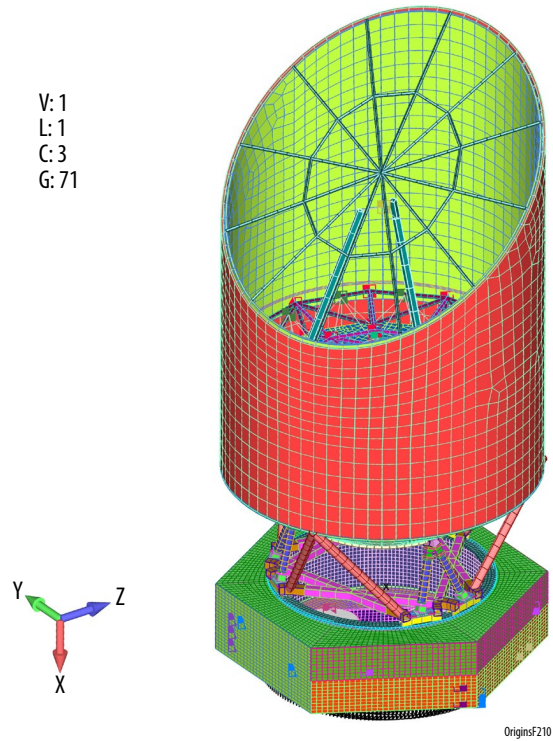

V: 1
L: 1
C: 3
G: 71

OriginsF210

**Figure C-29:** The *Origins* observatory finite element model shows a very detailed design for an accurate first-cut analysis. Colors denote different components.

SLS Block 1B payload design loads can be used for observatories having a minimum lateral primary natural frequency of 8 Hz and a minimum axial primary natural frequency of 15 Hz. Similar information for the BFR has not yet been published. The *Origins* structural model is shown in Figure C-29.

**Normal Modes Analysis**

The *Origins* observatory normal modes analysis predicts primary natural lateral and axial frequencies of 8.09 Hz and 16.7 Hz, respectively, meeting SLS User Guidelines and allowing simpler structural analysis going forward. To reach these frequency values several design improvements were made: the barrel and baffle were stiffened, supported at the top at launch via the aperture cover, extra struts were added from the spacecraft to the barrel, the bottom of the barrel was made in the shape of a cone, and the support of the propellant tanks was stiffened. These improvements added some mass and had thermal impact, but result in a robust structural design without requiring a detailed coupled loads analysis.

The lateral mode shape (Figure C-30) is a rocking mode shape about the spacecraft Payload Interface Ring, primarily driven by the size and shape of the Barrel.

The primary axial frequency mode shape (Figure C-31) is an "oil drum" mode shape of the Propulsion Deck.

**Stress Analysis**

The SLS Mission Planner's Guide Block 1B payload design loads are listed in Table C-7.

The team's conservative analysis uses a fully-enveloping load, as described in the SLS Mission

**Table C-7:** *Origins* design loads. These lower loads are usable since the minimum fundamental frequencies of the payload are in the acceptable range for the SLS.

| Event | | Lateral Acceleration (g's) | Axial Acceleration (g's) |
|---|---|---|---|
| Liftoff | Min | 0 | -1.5 |
| | Max | 2 | -1.5 |
| Ascent - Transonic | Max | 2 | 2.25 |
| Booster Phase - Max G | Max | 0.5 | 3.25 |
| Core Stage Phase - Max G | Max | 0.5 | 4.1 |

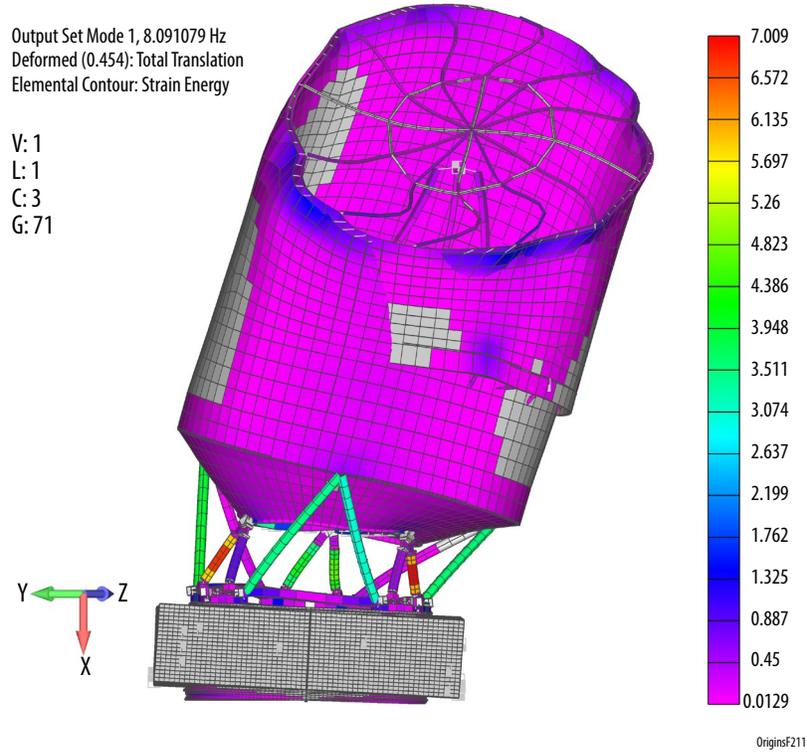

**Figure C-30:** Primary lateral frequency mode shape is driven by the barrel of the observatory. Colors denote strain energy levels.

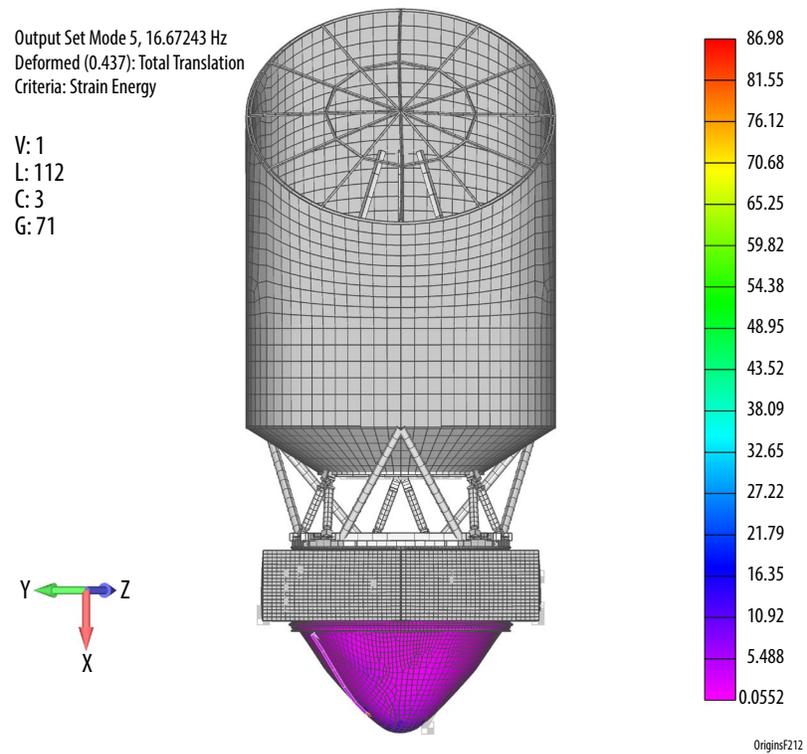

**Figure C-31:** Primary axial frequency mode shape is an oil can mode of the Propulsion Deck. Colors denote strain energy levels.

Planner Guide, which states: "Spacecraft/payload accelerations are estimates from ongoing SLS analysis. Analysis ground rules and assumptions limit the ascent vehicle steady state "axial acceleration to a maximum of 5 g and a lateral acceleration to 3 g." This combined loading exceeds the loads listed in Table C-7.

Figures C-32 to -38 highlight the *Origins* observatory components with the highest stress levels under the defined load condition (Table C-7). Table C-8 provides component structural analysis information.

## C.3 Telescope Details

### C.3.1 Telescope Structure Risk Reduction

To gain confidence in the use of beryllium as a structural element a qualification program involving a section of primary mirror and backplane was conceived.

**Table C-8:** The major component margin of safety data — all designs exceed minimum strength requirements for launch (MS>0).

| Component | Material | FTu (ksi) | FTy (ksi) | Max Stress (ksi) | FoS (u/y) | MS (min) |
|-----------|----------|-----------|-----------|------------------|-----------|----------|
| PIR | Al 6061 | 42 | 35 | 21.05 | 1.25 (y) | 0.33 |
| Barrel | AL 6061 | 37 | 35 | 10.05 | 1.4 (u) | 1.63 |
| PMBSS | Beryllium Tube | 47 | 35 | 20.89 | 1.4 (y) | 0.20 |
| IMP | Beryllium Plate | 40 | 30 | 15.13 | 1.4 (y) | 0.42 |
| Bus Bipods | Ti-6AL-4V | 125 | 119 | 28.49 | 1.4 (u) | 2.13 |
| Thrust Tube | M55J | 75.7 | 40 (cu) | 10.33 | 1.5 (cu) | 1.58 |
| 4.5 K Bipods | M55J | 75.7 | 40 (cu) | 9.06 | 1.5 (cu) | 1.94 |
| Hex Code | Aluminium | 0.17 (su) | 0.7 | 0.053 | 1.5 (su) | 1.15 |

FTu – Ultimate Tensile Material Strength
FTy = Yield Tensile Material Strength
(cu) = Ultimate Compressive Material Strength
(su) = Ultimate Shear Material Strength
FoS (u) = Factor of Safety against FTu
FoS (y) = Factor of Safety against FTy
MS = Margin of Safety. This is the fractional exceedance of the design above the FoS. Anything greater than 0 is OK

The PMBSS beryllium component fabrication and subassembly brazing operations will be further defined in Phase A. Beryllium structural shapes are currently in production at a small number of facilities and should be available for *Origins*. Best practice for joining beryllium components to form subassemblies is high-temperature oven brazing. Ovens of the size that would fit even one 60-degree

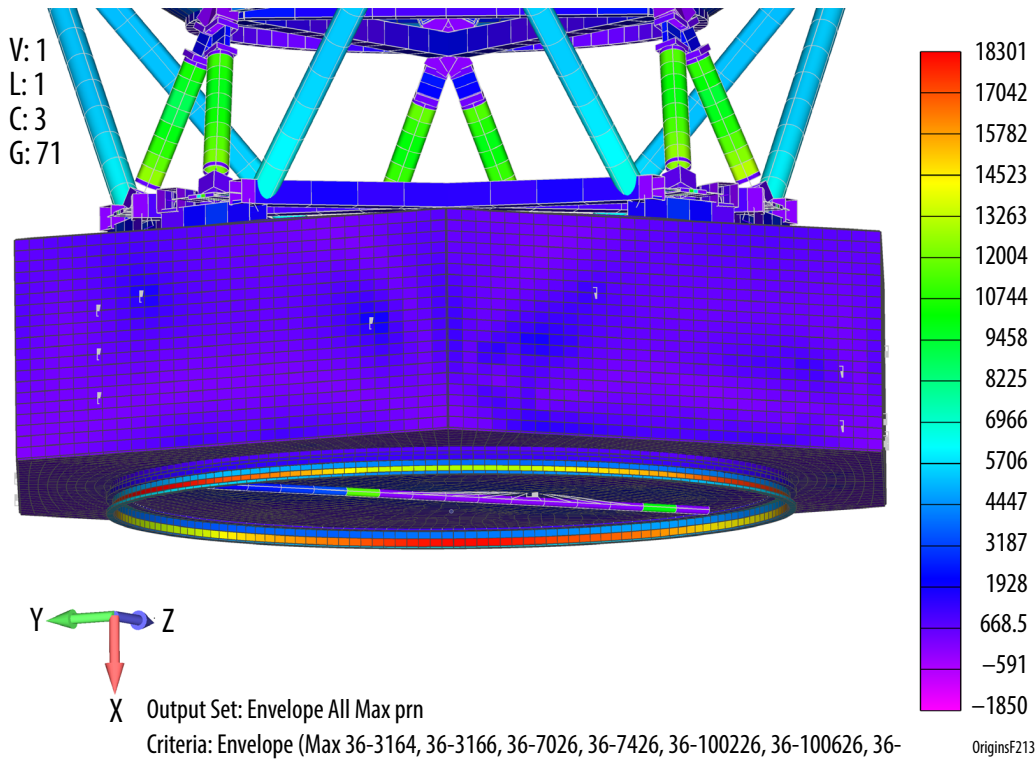

**Figure C-32:** The highest stressed aluminum component is the Lower Interface Ring that interfaces with the Payload Adapter Fitting. Colors denote stress levels in Pounds per Square Inch (PSI).

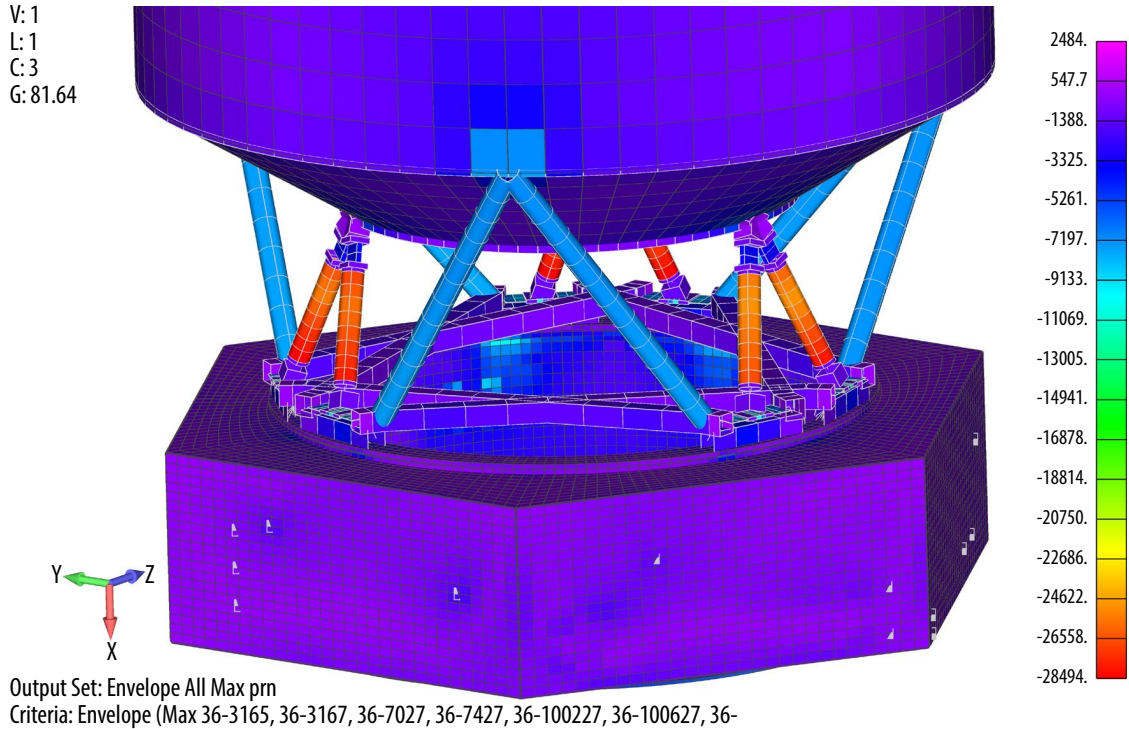

V: 1
L: 1
C: 3
G: 81.64

2484.
547.7
-1388.
-3325.
-5261.
-7197.
-9133.
-11069.
-13005.
-14941.
-16878.
-18814.
-20750.
-22686.
-24622.
-26558.
-28494.

Output Set: Envelope All Max prn
Criteria: Envelope (Max 36-3165, 36-3167, 36-7027, 36-7427, 36-100227, 36-100627, 36-

OriginsF214

**Figure C-33:** The highest stressed titanium component is the Bus Bipod at 28,494 psi. Colors denote stress levels (PSI) .

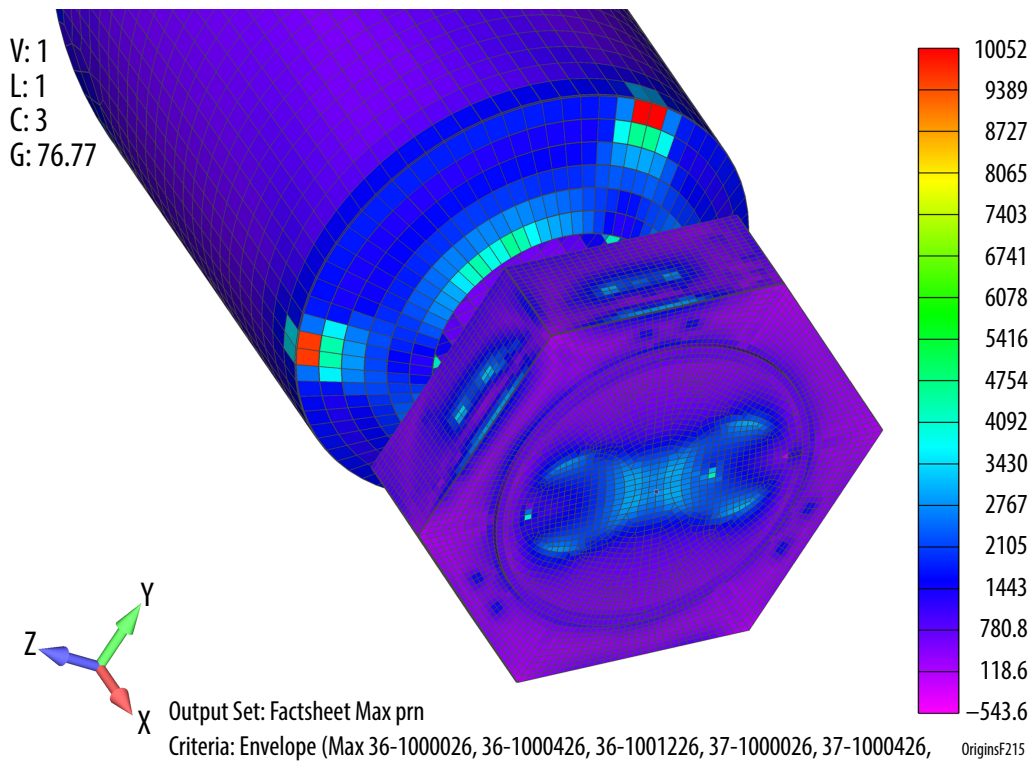

V: 1
L: 1
C: 3
G: 76.77

10052
9389
8727
8065
7403
6741
6078
5416
4754
4092
3430
2767
2105
1443
780.8
118.6
−543.6

Output Set: Factsheet Max prn
Criteria: Envelope (Max 36-1000026, 36-1000426, 36-1001226, 37-1000026, 37-1000426,

OriginsF215

**Figure C-34:** The highest stressed aluminum honeycomb facesheet component is located at the interface of the Barrel cylindrical side panel and conic bottom panel. Colors denote stress levels in PSI.

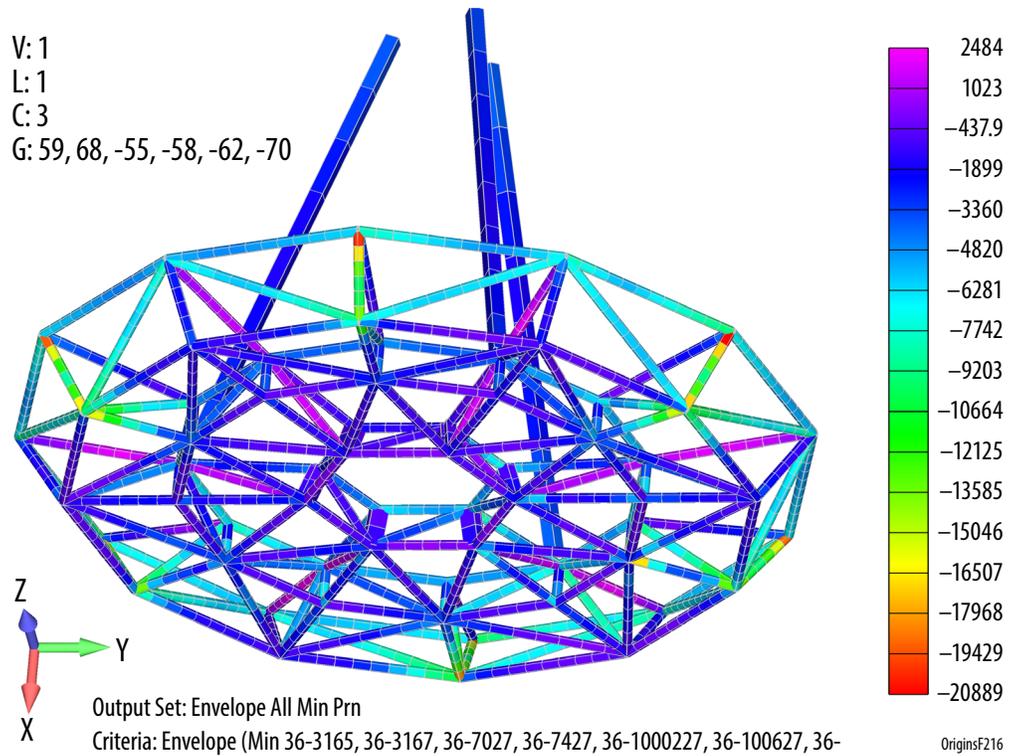

V: 1
L: 1
C: 3
G: 59, 68, -55, -58, -62, -70

Output Set: Envelope All Min Prn
Criteria: Envelope (Min 36-3165, 36-3167, 36-7027, 36-7427, 36-1000227, 36-100627, 36-

OriginsF216

**Figure C-35:** The highest stressed beryllium tube component located at the top end of a vertical tube. Colors denote stress levels in PSI.

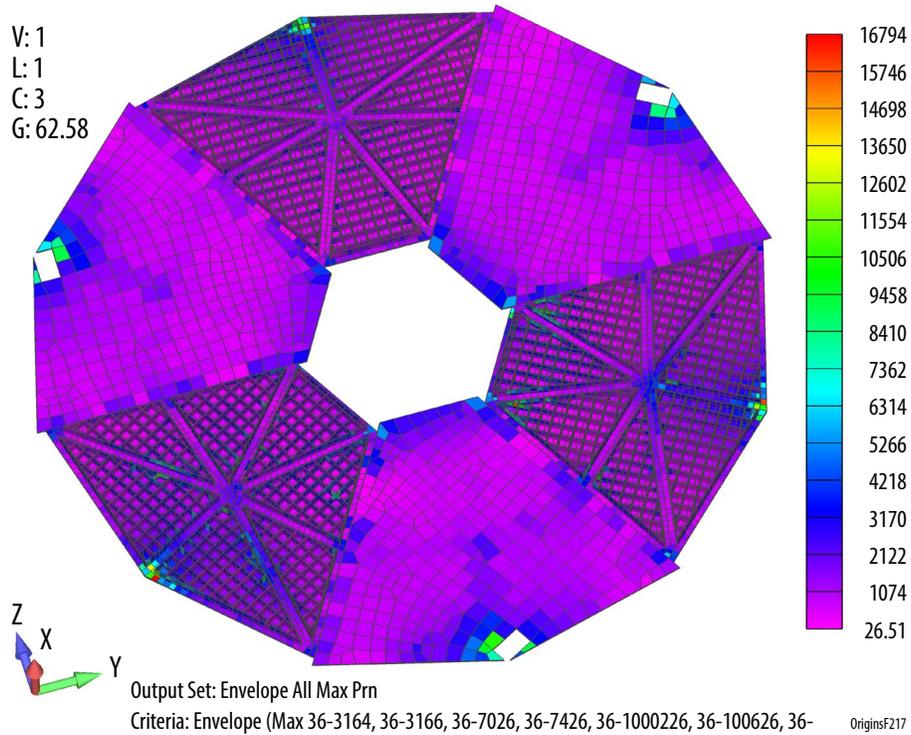

V: 1
L: 1
C: 3
G: 62.58

Output Set: Envelope All Max Prn
Criteria: Envelope (Max 36-3164, 36-3166, 36-7026, 36-7426, 36-1000226, 36-100626, 36-

OriginsF217

**Figure C-36:** The highest stressed beryllium plate component is along the outside edge of an IMP. Color denotes stress levels (PSI).

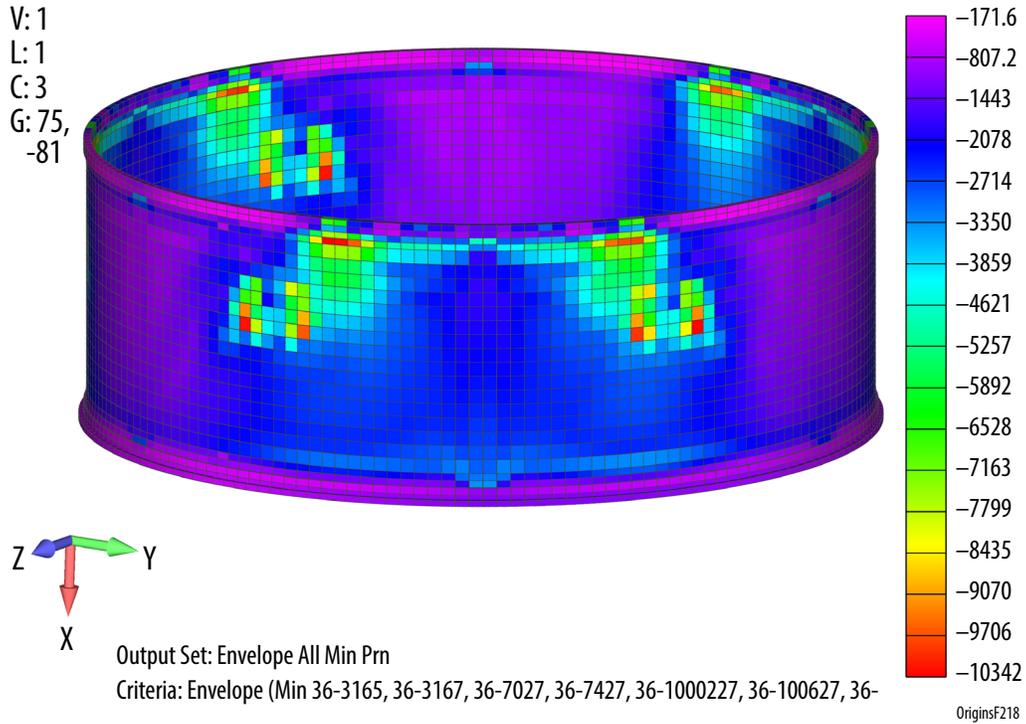

V: 1
L: 1
C: 3
G: 75,
-81

Output Set: Envelope All Min Prn
Criteria: Envelope (Min 36-3165, 36-3167, 36-7027, 36-7427, 36-1000227, 36-100627, 36-

OriginsF218

**Figure C-37:** The highest stressed M55J fiber composite component is shown along the side of the tube and indicated by red elements. Colors represent stress levels (PSI).

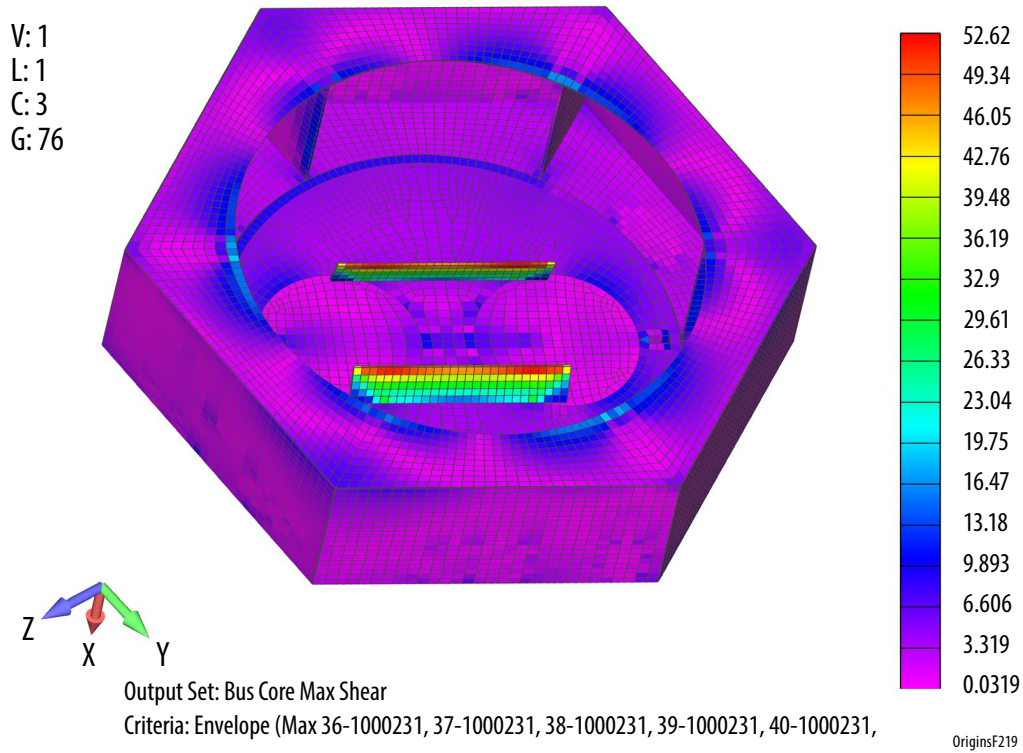

V: 1
L: 1
C: 3
G: 76

Output Set: Bus Core Max Shear
Criteria: Envelope (Max 36-1000231, 37-1000231, 38-1000231, 39-1000231, 40-1000231,

OriginsF219

**Figure C-38:** The highest stressed aluminum hex core component are the fuel tank skirt supports. Colors indicate shear stress levels (PSI).

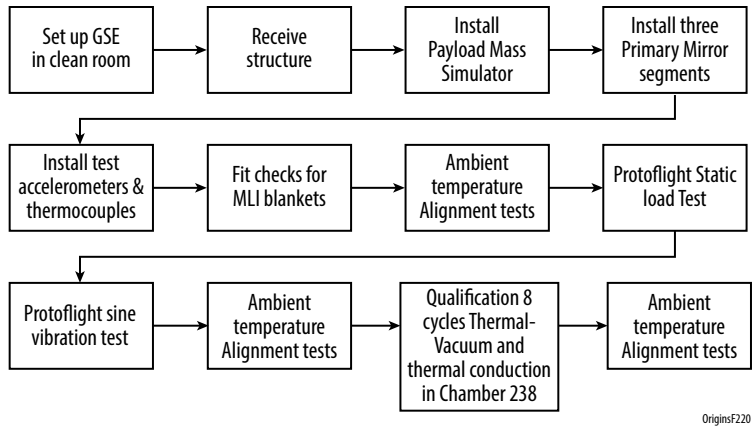

**Origins Telescope Wedge (ETU)**
**Integration and Test for Structural Integrity Summary Flow**

**Figure C-39:** The Telescope Wedge ETU Test Flow shows the rigorous testing performed to ensure good optical performance and robust tolerance to launch loads.

(wedge) section of the *Origins* PMBSS do not currently exist. The PMBSS will therefore be bolted together. To qualify this design, the *Origins* program will invest in structural testing for one PMBSS wedge section. The PMBSS wedge test program flow (Figure C-39) includes optical alignment testing using three Engineering Test Unit (ETU) PMSA segments (two outer segments, one inner segment).

Optical alignment testing at ambient temperature will establish baseline measurements prior to vibration testing. The OSS instrument mass simulator (simulator for the mission's largest and heaviest instrument) is attached to a wedge. PMSAs are also installed on the wedge. For structural testing, the test article is inverted and affixed to a test fixture mounted to a vibration table or the ground, depending on the test performed (Figure C-40). Static load testing and sine vibration environment testing will structurally qualify the PMBSS wedge subassemblies. Optical alignment testing is performed post-vibration and -thermal cycling to verify PMBSS optical performance.

### Component Testing

It is typical to carry out structural tests on honeycomb panels and fiber composite structural components. For honeycomb panels, unique joint designs in areas of expected high stresses and loads will be duplicated in test coupons and tested to failure to establish B-basis (limited quantity statistics) strength values. During panel fabrication, lap shear testing of aluminum facesheet primer adhesion will be tested, along with flatwise tension tests of panel coupons to verify proper panel consolidation. Test coupons that mimic the design of the 4.5 K Bipods fiber composite tubes with bonded titanium end fittings will also be tested to failure to prove design viability.

### C.3.2 Stray Light Analysis

Stray light analysis refers to the process of identifying and preventing light rays from any source other than the object being observed, from reach-ing the detector plane. For this purpose, mechanical structures known as baffles are incorporated into the telescope design to block these unwelcome rays' potential "skip paths." Tight packaging constraints require careful consideration of how to adequately baffle against stray light for open telescope designs. Often the constraints involved make it difficult to achieve well separated paths from image and pupil conjugates which would permit placement of effective baffles. As a result, baffling may be limited to careful use of pupil masks and baffles combined with field masks at or near intermediate images to restrict specular paths from the sky to the focal plane other than from within the field of view. Alignment tolerances can be especially tight if the folded path results in the pupil conjugate being nearly coplanar with the intermediate image conjugate. The stray light analysis of *Origins* was carried out in FRED software.

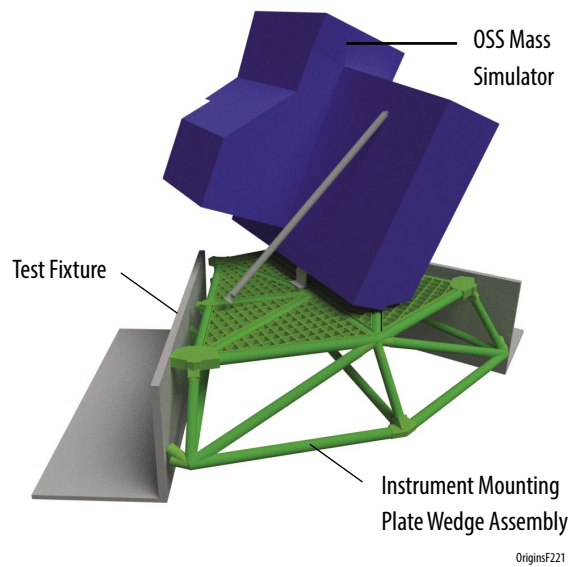

**Figure C-40:** Telescope Wedge ETU Test Setup is inverted during ground-testing, which allows less-robust ground support equipment to support the unit during the tests.

Figure C-41 shows *Origins*' external and internal baffling. The external baffling is shown in red in the image on the left while the image on the right shows the 'snout' baffling and vanes that are coming away from the hole in the primary mirror towards the secondary mirror. Figure C-42 shows the effectiveness of the baffling in blocking stray light. In each case, rays were traced backwards through the system from the instrument FOV footprints at the focal surface towards the primary and onto the sky. In Figure C-43, the rays leaving the focal surface are colored green and remain so until reflecting off of the tertiary mirror, at which point they are colored red. In the top picture (without baffling) there is a significant ray path that allows light to travel directly from the focal surface to the sky without hitting

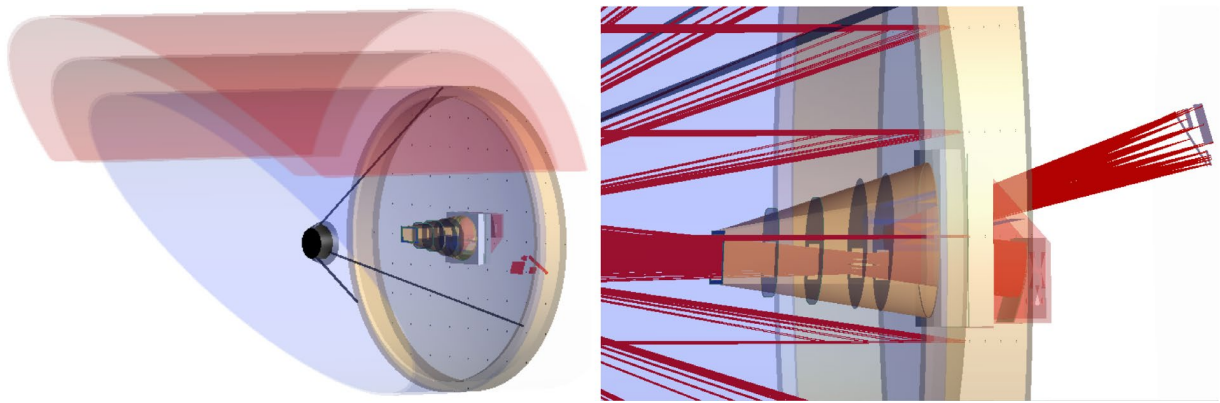

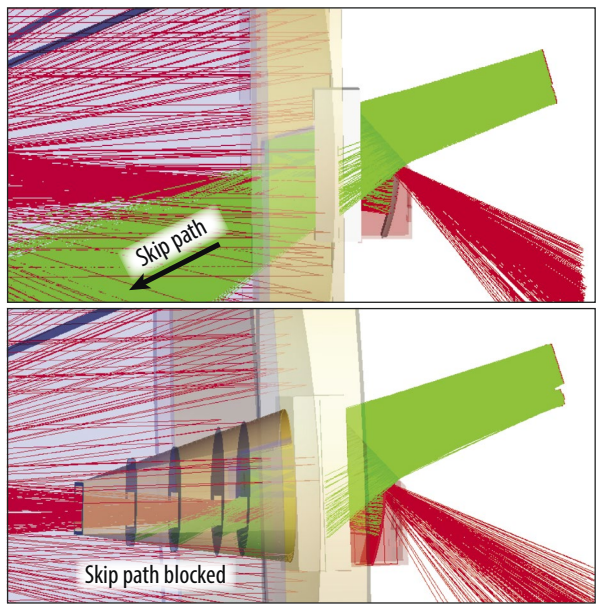

**Figure C-41:** (Above) External (left) and internal (right) baffles. Both sets of baffling are meant to stop the ability of unwanted light sources from reaching the focal surface.

**Figure C-42:** (Left) Images showing the effect of baffling to mitigate stray light. (Top) No baffling allows for skip ray paths and light to travel directly from the sky to focal surface without reflecting off the telescope mirrors versus (bottom) having baffling which prevents this from occurring.

the tertiary (as evident from the fact that the rays remain green in color). In the bottom picture, the baffling blocks this skip path and only the desired FOV (the red rays) make it all the way from the image surface to the sky. Figure C-43 shows the intensity signal to noise ratio between the desired FOV allocated to the instruments and noise due to scatter. Figure C-43 shows that the difference between the two is between about four and five orders of magnitude. As the FSM is located at the exit pupil of the telescope, an image of the primary mirror forms there. To further mitigate stray light, it is useful to place a baffle around the FSM, as was the case for the FSM of JWST.

In addition to blocking skip paths as mentioned above, scatter is another consideration in stray light analysis. Scatter can be reduced by using mirrors that are well polished with low surface roughness as well as clean. In the FRED model, surface roughness is represented by the Harvey-Shack model with particulate contamination modeled by MIL-STD-1246C. Figure C-44 shows the effect of surface roughness in more detail. By increasing the surface roughness of the primary mirror from 1.5 to 15 to 100nm (while leaving the secondary, tertiary, and FSM each with a surface roughness of 1.5 nm in each case) it is apparent that a rougher surface leads to a worsened signal to noise ratio between on and off-axis. This is due to the increased scatter. The bottom plot of Figure C-44 shows the case for which the primary mirror has again a surface roughness of 1.5nm but no longer includes any baffling. The results of Figure C-44 are summarized also in Table C-9 which show relative flux versus scattered power as function of the primary mirror's surface roughness. As the surface roughness increases, more light is lost from the signal to scatter.

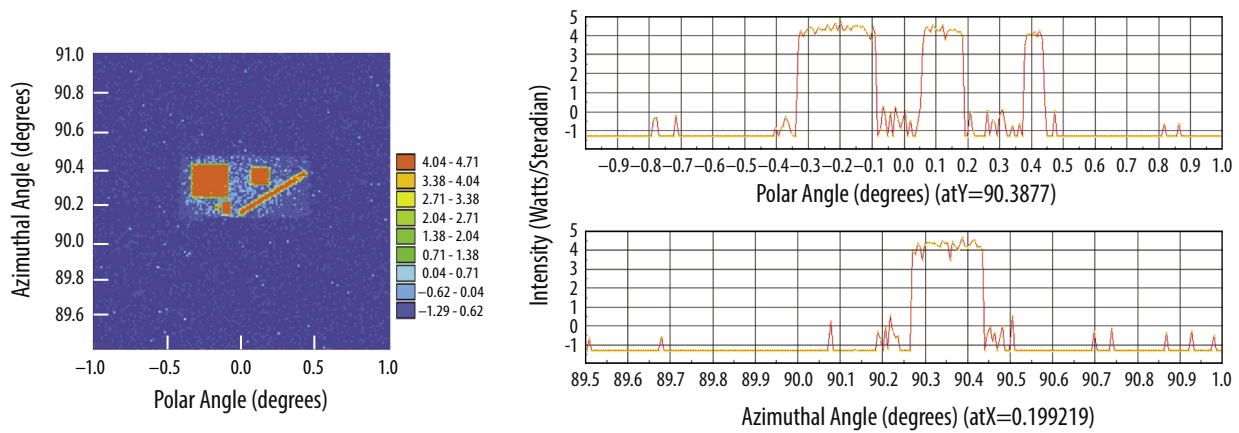

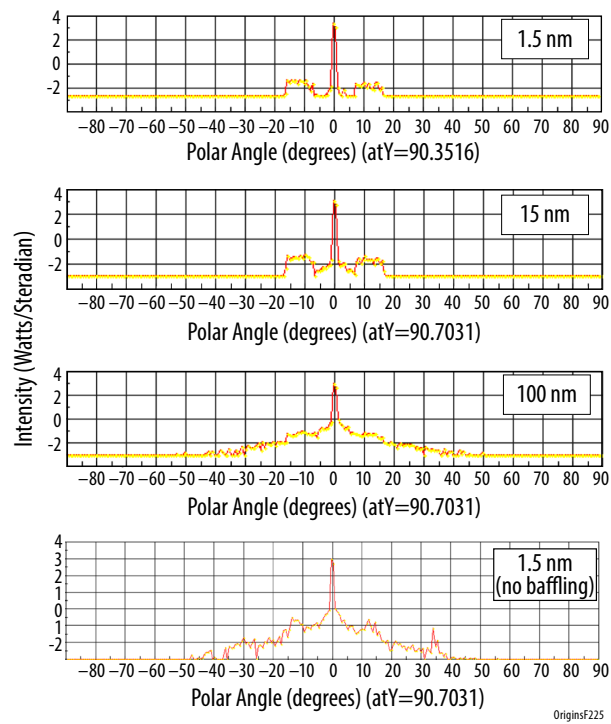

**Figure C-43:** (Above) Analysis showing relative intensity between the desired signal (the instruments' FOV footprints) and noise due to near field scatter. Note that the x and y lines scans are plotted on a logarithmic scale and show between about four and five orders of magnitude difference between signal and noise. Calculations carried out at $\lambda = 30\mu m$.

**Figure C-44:** (Left) The effect of primary mirror surface roughness on intensity as a function of FOV angle. As the primary mirror surface roughness increases, more scatter occurs from the surface, worsening the signal to noise ratio between on and off-axis. Note that the intensity scales are logarithmic. Calculations carried out at $\lambda = 30\mu m$.

**Table C-9:** Relative flux versus scattered power as function of M1 surface roughness

| M1 Surface Roughness (nm RMS) | 1.5 | 15 | 100 |
|---|---|---|---|
| Signal | 99.43% | 99.18% | 87.11% |
| Integrated noise ($2\pi$ solid angle) | 0.55% | 0.63% | 4.30% |
| Integrated noise (1 deg² solid angle) | 0.01% | 0.02% | 0.30% |

### C.3.3 Optical Error Budget

An error budget for this telescope has been developed and is shown in detail below. This document lays out the allocation of wavefront error to design, thermal stability, line of sight/jitter, and reserve for each of the instruments and the telescope itself.

This section presents the optical error budget carried out in conjunction between NASA Goddard and Ball-Aerospace in detail. The purpose of such a budget is to identify different sources that can all contribute to the wavefront error of the telescope and ensure that the diffraction-limited performance specification is still met. Given the design wavelength of 30 µm and a Strehl ratio of 0.8, the goal as-built RMS wavefront error is 2255 nm. Figure C-45 shows that the terms, when root summed squared together, do indeed meet the 2255 nm requirement with the remainder designated as reserve. For

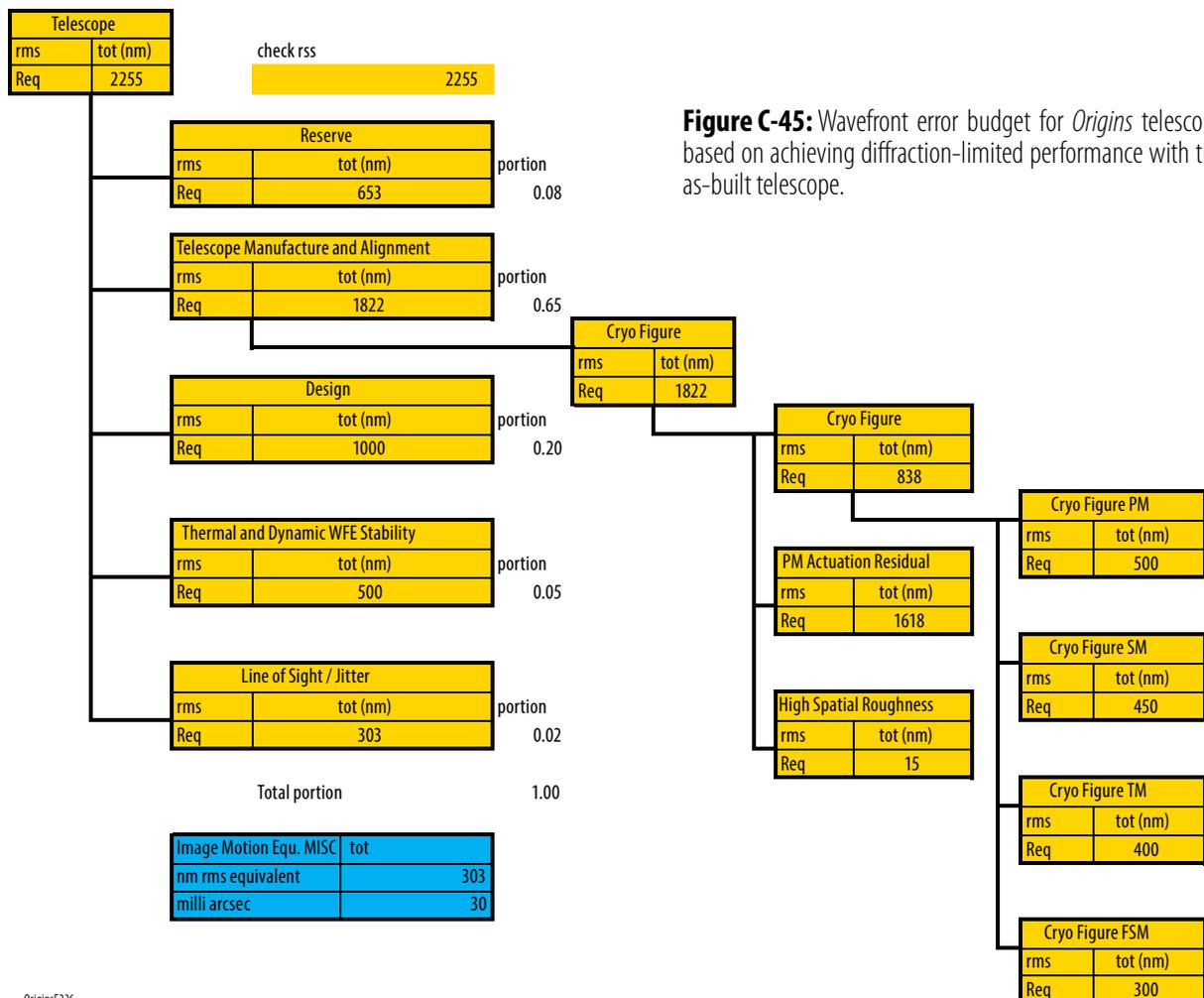

**Figure C-45:** Wavefront error budget for *Origins* telescope based on achieving diffraction-limited performance with the as-built telescope.

OriginsF226

future work this wavefront budget would need to be broken into much greater detail, setting specific wavefront allocations for individual optical elements, field points, etc. Figure C-46 shows a series of budgets combining the allocations to the OSS, FIP, and HERO (studied as upscope option) instruments with that of the telescope. In each case the top-level telescope wavefront budget value (2255 nm) is root summed squared together with the value from each instrument towards the desired value for that individual instrument based on its design wavelength. Any remainder is designated again as reserve. For example, FIP is to be diffraction-limited at a wavelength of 40 μm requiring a total RMS wavefront error 3007 nm. Root sum squaring the 2255 nm from the telescope with the 1612 nm calculated for the FIP instrument leaves over 1165 nm in reserve to total the top-level 3007 nm. For each instrument the goal wavefront performance is met with reserve left over. Note that the MISC-T instrument is omitted from the budget since wavefront error is not the preferred metric for evaluation. MISC-T uses the telescope as a light collector and does not requirement diffraction limited image quality. Rather, it requires light stability over a few hour time span. Using simulated point-spread functions generated in CODEV®,(with a total of 2255 nm of wavefront error) the MISC-T team has concluded that the instrument achieves high stability due to small long-term drift and jitter.

### C.3.4 Optical Testing

*Origins* requires optical testing throughout the integration process of the observatory. Each optical element of both the telescope and the instruments are first verified individually at room temperature. The specific optical tests depend on the element being measured (mirror, lens, grating, etc.). For example, mirrors are typically measured by standard interferometry with a reference sphere (or null corrector or computer-generated hologram (CGH) if the optical surface is more geometrically complex). Due to *Origins* long design wavelength of 30μm, it may be possible to carry out this validation using an optical-probe coordinate measuring machine (CMM). Using a CMM is typically faster and less expensive than interferometry (not requiring a unique test setup for each different mirror to be tested) but does not provide the same resolution which is not as much of a concern for measuring these surfaces. Note that measuring the segments of the telescope's primary mirror is eased by the fact that there are only two different segment prescriptions.

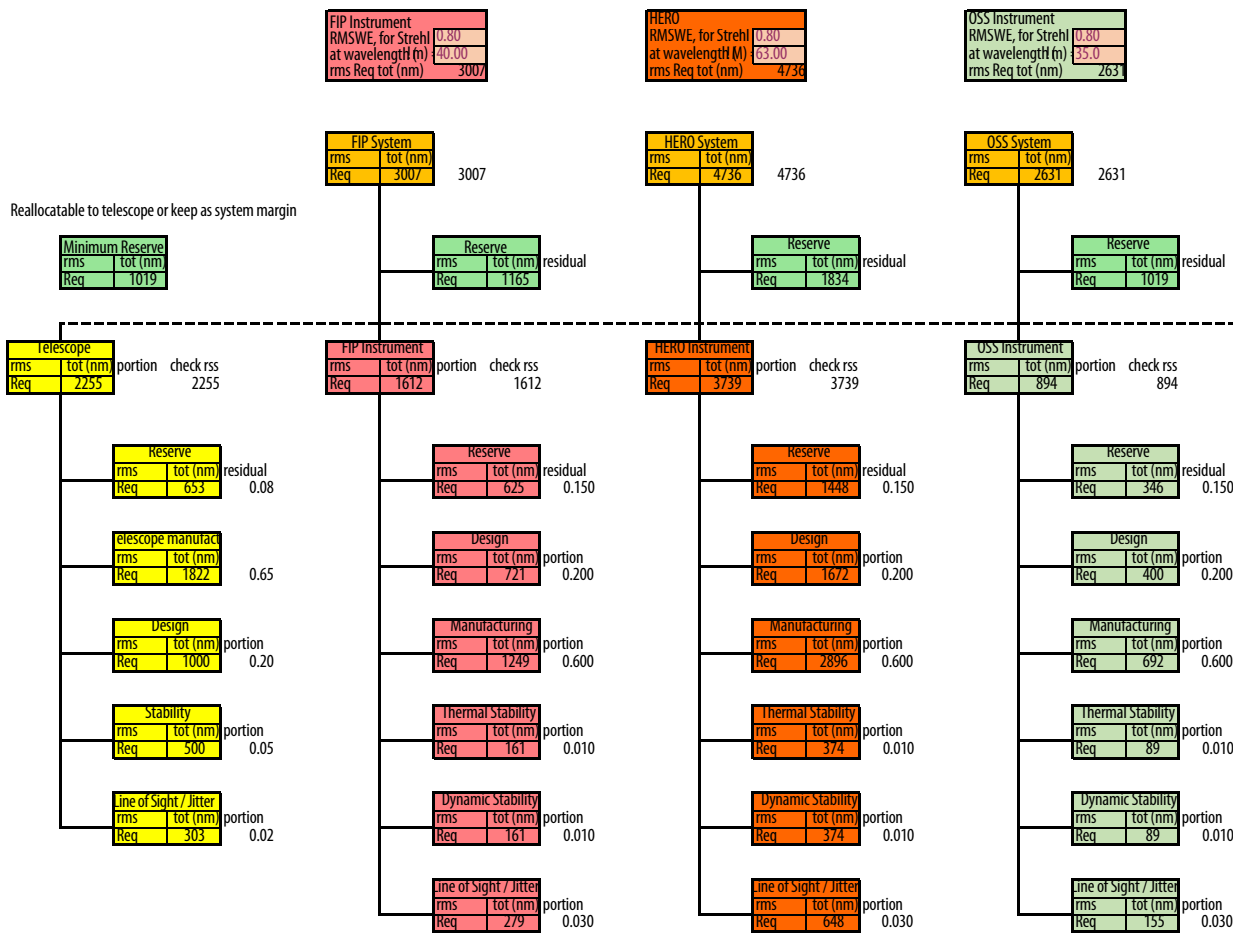

**Figure C-46:** Wavefront error budgets combining the allocations to the OSS, FIP, and HERO instruments with that of the telescope.

Once each instrument is assembled and all are integrated together within one structure they must be tested optically during cryo-vac cycles and post vibration test (to ensure that alignment is not lost in either case). Doing so requires the use of another optical system meant to simulate the output from the telescope. This system is used to evaluate the instruments' response to various optical metrics such as focus shift, wavefront error, boresight, and pupil shear. Instruments with a wavelength dependence (those that are spectroscopic and/or contain refractive optical elements are validated at multiple wavelengths). The segments of the primary mirror are phased together using wavefront sensing algorithms to properly actuate each segment to the required location and orientation. This is the same process used by JWST but requires fewer iterations of the process due to the reduced accuracy with which the segments must be placed.

Once each is independently verified, the telescope and instruments are integrated together and are tested again in cryo-vac. *Origins* uses a similar series of optical alignment tests as JWST. To do so, a series of optical fibers are placed at the internal focus of the telescope (between the secondary and tertiary mirrors) which can be oriented in different directions depending on the test of interest. By directing the fibers towards the tertiary, the optical system is evaluated from the tertiary to the instruments (known as the 'Half-Pass' test). If the light is sent in the opposite direction, the entire system from primary mirror to instruments is tested. This is known as the 'Pass-and-a-Half' test since light starts within the telescope and reflects off of the secondary and primary mirrors before leaving the system collimated and being reflected back into the full optical system by a series of autocollimators.

## C.3.5 Alternative trade studies

Before settling on the final design architecture discussed in the sections above, the *Origins* team explored a number of alternative telescope design forms. These included trade studies between two and three-mirror telescopes, on and off-axis pupils, non-rotationally symmetric pupils, and systems utilizing a primary mirror with a spherically-shaped surface (as opposed to a conical or aspheric one). The results of these studies are cataloged in this report.

The primary trade study examined the four different telescope configurations, A through D, as summarized in Table C-10. The purpose of this study was to identify what combination of number of powered mirrors and pupil bias would be carried forward for the telescope design. Optical layouts of each of these are shown in Figure C-47. For each of these four designs, the primary mirror (PM), secondary mirror (SM) and image surface (IS) are all labeled, along with the tertiary mirror (TM) and fold mirror (FM) as appropriate. In all four cases the image surface is curved to better correct field curvature, an optical aberration caused by a difference in best focus with field. As expected, the curvatures of the image surfaces for both three-mirror designs (C and D) are less than that for both two-mirror designs (A and B). This is due to the additional powered mirror being able to better spread the optical power over a greater number of mirrors and therefore "flatten" the field.

**Table C-10:** Summary of the number of mirrors, pupil bias and field bias for each of the four telescope configurations studied

| Configuration | No. of Mirrors | Pupil Bias | Field Bias |
|---|---|---|---|
| A | Two | On-axis | On-axis |
| B | Two | Off-axis | On-axis |
| C | Three | On-axis | Off-axis |
| D | Three | Off-axis | Off-axis |

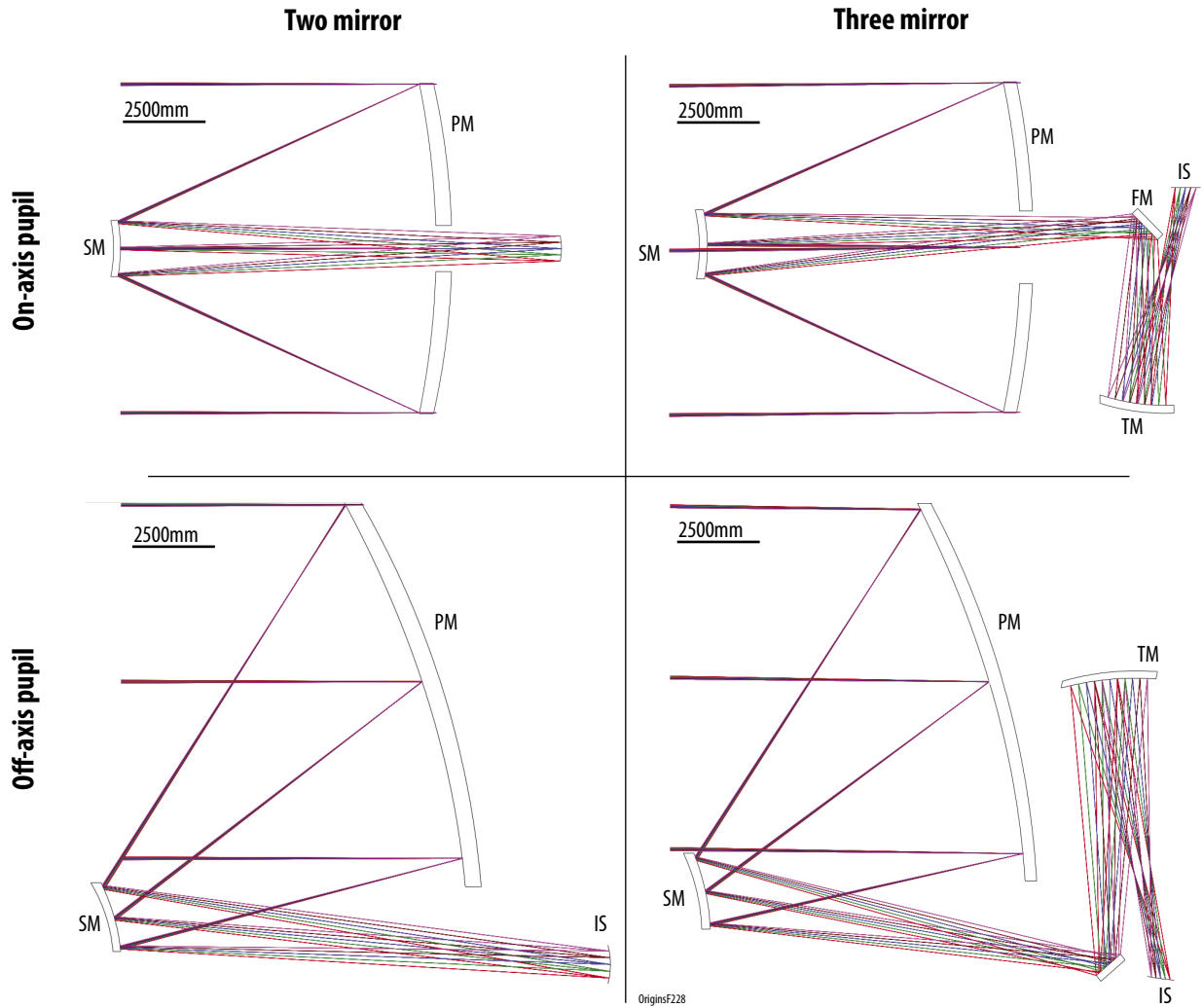

**Figure C-47:** Comparison showing the different optical layouts for each of the four telescopes configurations identified for study. A variant of configuration C was chosen for *Origins*.

## Number of Mirrors

The ability of a telescope to correct optical aberrations depends directly on the shape and number of mirrors that compose that telescope. By providing additional degrees of freedom to the optical designer, a telescope utilizing a greater number of powered mirrors can have superior imaging performance over a telescope with fewer mirrors. Given an object at infinity, a single parabolic mirror is capable of correcting spherical aberration to yield stochastic imaging on axis. Moving to a two-mirror Ritchey-Chrétien configuration (like the HST), spherical and coma can be corrected simultaneously while a third mirror makes it possible to correct astigmatism, resulting in the three-mirror anastigmat (TMA) design (like JWST). With a greater ability to correct optical aberrations, additional mirrors increase the field of view over which the telescope is able to form a high-quality image. This fact is reflected in Figure C-48, which shows Strehl ratio plotted as a function of FOV for each of the four designs introduced in the previous section. For each design and evaluation wavelength, the Strehl ratio is shown over the same 1° x 1° full FOV. As expected, the imaging performance is always superior at longer wavelengths and at the center of the FOV. Based on that, it makes the most sense to allocate the center parts of the telescope's FOV to the instruments with the strictest wavefront performance specifications and/or those operating at the shortest wavelengths (often one in the same).

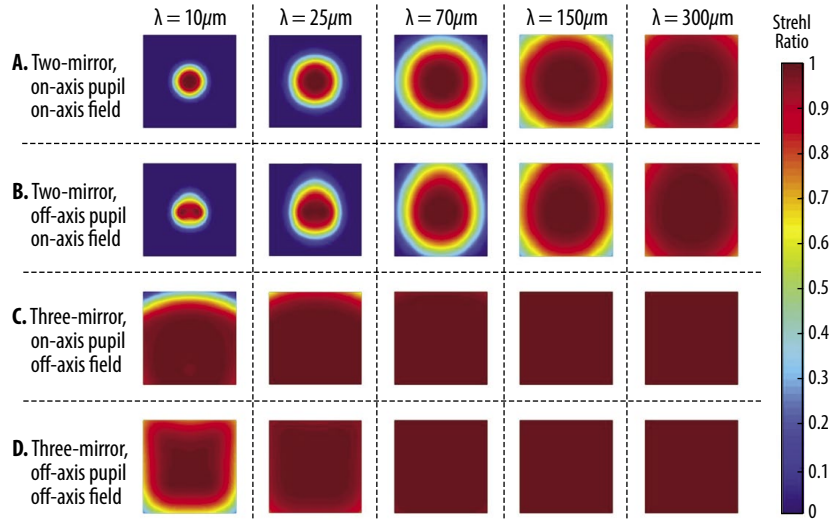

**Figure C-48:** Comparison of Strehl ratio as a function of FOV for four different telescope configurations. In each case the performance is evaluated over the same 1° x 1° full FOV.

While a TMA offers superior imaging over a larger field of view as compared to a two-mirror design, the advantage is not without cost. The TMA often requires the fabrication, integrating, and testing of not just one, but usually two additional mirrors (the powered tertiary as well as a flat fold mirror) as compared to a two-mirror design. The fold flat is often included to ensure the focal surface of the telescope is in a convenient location for interfacing with the instruments. If the telescope designer wishes to include a steering mirror (such as for the case of JWST) for the purpose of pointing stability, this fourth mirror becomes a necessity. It is very difficult to include a steering mirror in a telescope design of two powered mirrors as the steering mirror must be placed at the telescope's exit pupil. TMAs are a useful solution for relaying the pupil to a convenient location for a steering mirror while the same is not true of a two-mirror design. If pointing stability is an issue but the design does not include a steering mirror, pointing can also be carried out in the individual instruments.

The question of pursuing a two versus three-mirror telescope design essentially boils down to one of cost versus performance. Identifying the improved imaging performance over a larger field of view and the appeal of the pointing stability afforded by the steering mirror, the STDT made the decision to pursue a TMA design.

## Aperture Bias

A second consideration in the design of the *Origins* telescope was the question of whether the aperture stop (located on the primary mirror) should be on or off-axis. Figure C-47 shows the difference in layout for on and off-axis telescope designs of two and three mirrors. As seen in Figure C-47, off-axis telescopes are laid out so that the secondary mirror does not at all obstruct the light incident on the primary mirror. This is desirable for a number of reasons, the most readily-apparent of which is that this maximizes the amount of light reaching the detector. While the increased throughput is beneficial from a scientific standpoint, the unobstructed design suffers from being more complicated to fabricate and test, and is less compact as compared to an obstructed design. Segmenting an on-axis mirror is easier than doing so for an off-axis one as it generally requires few specific mirror prescriptions.

Another consideration is how the pupil bias affects the point spread function (PSF) of the telescope. The PSF (also known as impulse response) is one of a number of metrics for determining the image quality of an optical system. One can mathematically calculate the image formed by a given optical system by convolving the object being viewed with that system's PSF (which changes with both location in the FOV and wavelength). For perfect stochastic imaging, the PSF should be a delta function; however, in reality the PSF is not because of the effects of diffraction. For a system with a circular aperture, the PSF is an Airy disk. By increasing the size of the obscuration, more energy ends up in the side lobes of the Airy disk pattern. This is apparent in Figure C-49, which shows a comparison between PSFs for different values of linear obscuration for telescope design A from Figure C-47. The 80% encircled energy diameters (for the on-axis field point at a wavelength of 30 μm) are 0.49, 0.69, 0.90, and 1.05 mm for the no obscuration, 15, 30, and 50% obscuration cases, respectively. This spreading out of the PSF is undesirable as it further spreads out the energy contained in the image of a single field point, deviating the telescope even more from stochastic imaging. In addition to the effect of the size of the obscuration, the mounting structure (spider) of the secondary mirror also causes diffraction of light and reduces the overall throughput to the image surface.

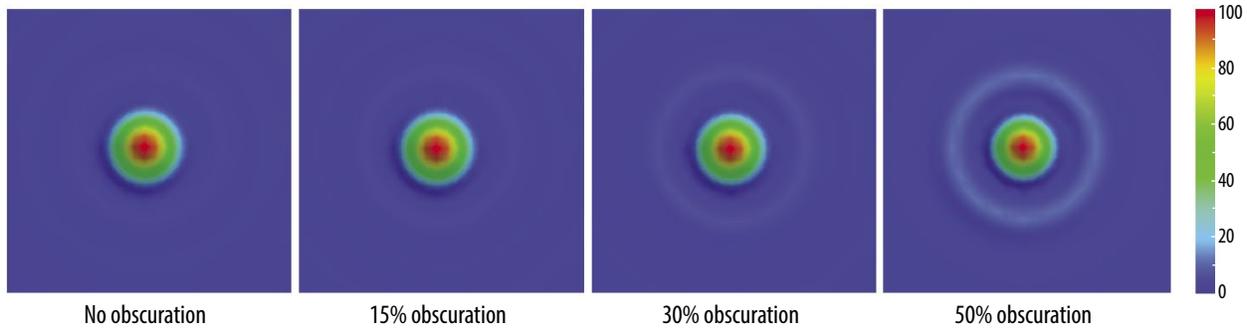

**Figure C-49:** Comparison of PSFs for increasing values of linear obscuration for an on-axis telescope

The team elected to pursue an obstructed, on-axis aperture design for the baseline design for the reasons regarding reduced cost and complexity. Traditionally, the type of decentered mirrors that compose off-axis aperture telescope systems are more complicated and costly to fabricate and test as compared to centered mirrors. Coupling this decision with that of the preference for a three-mirror telescope discussed in the previous section, concept C was identified as the solution space to pursue. *Origins* has the added benefit of operating at such long wavelengths that the polishing requirements for surface figure and surface roughness are much coarser than they would be for an equivalent telescope operating in the visible spectrum, easing the fabrication process and eliminating a cryo-null figuring step.

## Aperture Shape

Most commonly, the aperture of a telescope is circular. In an attempt to design *Origins* to have the largest collecting area possible while fitting in a given fairing size, the engineering team explored a number of alternative mirror shapes, most notably ellipses of differing aspect ratios. Ultimately the team decided opted for a circular primary mirror due to science objectives preferring a symmetric PSF. A non-rotationally symmetric aperture will result in a non-rotationally symmetric PSF. For an elliptical aperture, the effect of this is to essentially have different f-numbers and therefore, different angular resolutions in directions orthogonal to one another. The dependence of the PSF on aperture shape for circular and elliptical apertures of different aspect ratios is shown in Figure C-50. Figure C-51 shows an analogous comparison of PSFs for both square and rectangular primary mirrors. These shapes have the same issue of an asymmetric PSF that the elliptical apertures do but to an even larger degree. Representing these surfaces as a series of hexagonal segments was explored with the expected result that the sharp lines of the hexagons resulted in artifacts in the calculated PSFs determined to be inconsistent with the scientific requirements of *Origins*. As such, a circular aperture was pursued.

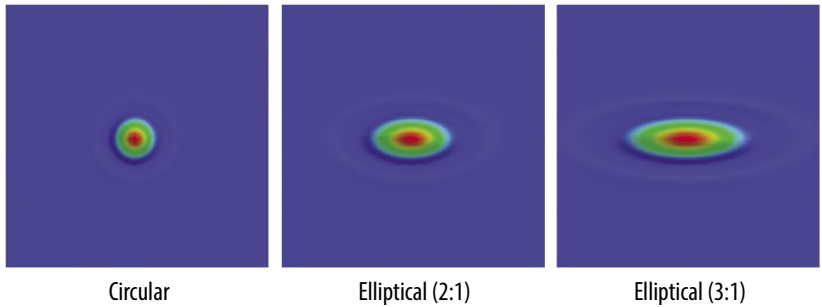

**Figure C-50:** PSFs for circular and elliptical aperture shapes showing that asymmetry in the in the aperture shape leads to undesired asymmetry in the PSF. The circular aperture was chosen for *Origins*.

Circular   Elliptical (2:1)   Elliptical (3:1)

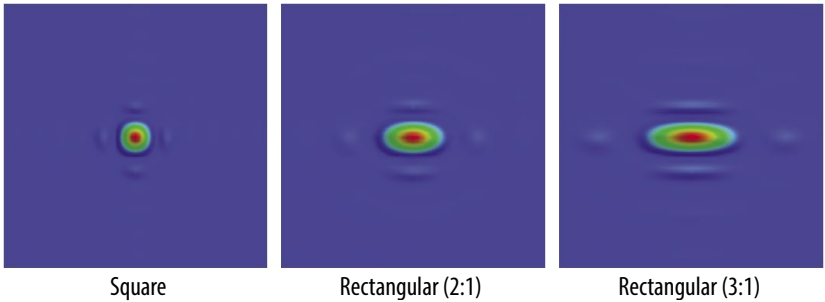

**Figure C-51:** PSFs for square and rectangular aperture shapes. Such apertures were deemed unacceptable for *Origins* due to yielding an asymmetric PSF.

Square   Rectangular (2:1)   Rectangular (3:1)

## Spherical Primary

The final study carried out for the *Origins* telescope design study was that of possibly using a spherical-surface primary mirror (as opposed to an aspheric or conical one). The motivation for this study was the idea of being able to generate segments of identical surface shape to compose the primary mirror. If possible, this would very likely ease the fabrication and testing process as compared to a design requiring each segment or subset of segments to have a unique surface prescription.

Historically, spherical mirrors have been used in telescope designs due to their relative ease of manufacture and test as compared to aspheric surfaces. The major problem with spherical surfaces is that they introduce optical aberration into the telescope, most notably spherical aberration. Spherical aberration is defined as a difference in best focus based on the position of the ray within the aperture. Essentially, the rays striking the center of the optic experience a different optical power than those at the edge of the optic, resulting in a blurred image. The simplest way to correct this error is to utilize a parabolic mirror, which has the property of focusing all the rays from a single on-axis field point to a single point. Unlike a spherical surface, a parabola does not have a constant local curvature across its aperture, which is why it can bring all the rays to the same focus. For the aforementioned reason, spherical mirrors are not often used today in telescope design, especially when image quality is of high concern.

The potential cost-savings of the segmented spherical primary mirror were considered beneficial enough to warrant a design study. For this purpose a 4500 mm-diameter TMA design was attempted, utilizing a 15 x 25 arcmin full FOV and design wavelength of 30 μm. The design is off-axis as the main *Origins* concept being pursued at that time as well. As expected, the optical design proved to be quite challenging due to the greater amount of aberration induced by the primary mirror. The final design is shown in Figure C-52. Note that this design was intended as a first attempt to explore the solution space and does not include all of the constraints included in the final design including the matching of the position of the FSM to the location of the exit pupil. Adding these constraints will only make the design process more challenging. These results of reduced image quality were echoed by Lockheed-Martin's optical designers who attempted the spherical mirror concept in a separate internal study. Based on this, the spherical primary was not pursued for *Origins*. It is expected that this idea presents a much greater benefit as the number of individual segments composing a mirror increases and could still be applied to another mission concept depending on its architecture and needs. As noted above, the primary mirror segments in the *Origins* baseline mission concept have only two distinct optical prescriptions.

### C.3.6 Materials Details

#### Fused Silica

Fused silica has extensive heritage as an optical substrate. It also has the lowest CTE at 4.5 K–nearly indistinguishable from beryllium–the lowest strain at 4.5 K, and relatively low specific stiffness. However, it has poor thermal conductivity and diffusivity, which is critical for *Origins*. In terms of manufacturability, it is roughly equivalent to Ultra Low Expansion (ULE) with boule production, light-weighting, and polishing.

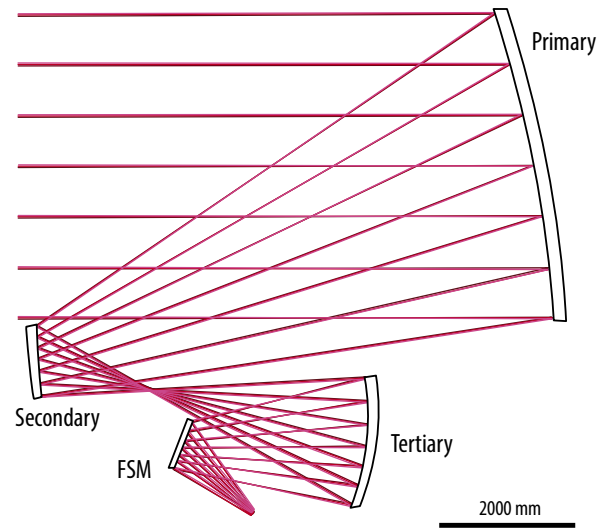

**Figure C-52:** Resulting optical design utilizing a spherical primary mirror. The design concept was not further pursued for *Origins* due to poor optical performance.

#### Silicon Carbide

Silicon carbide (SiC) has spaceflight heritage through the *Herschel* Space Observatory, a large 3.5 meter sintered SiC primary mirror. SiC has high specific stiffness, low strain at 4.5 K–significantly higher than fused silica but lower than others–low CTE at 4.5 K, and excellent thermal conductivity and diffusivity. It also has the potential for many segments to be created from a single mold using a cladding process, but SiC would require subsurface damage management and mitigation.

#### Beryllium

Beryllium has the highest performance because of its high specific stiffness, very low CTE at 4.5 K, and excellent thermal conductivity and diffusivity. It also has the most relevant heritage through the James Webb Space Telescope (JWST) with a large, segmented primary mirror operating at cryogenic temperatures. However, beryllium has relatively high strain at 4.5 K, and it is brittle. It also requires significant schedule lead time to develop segments and imposes higher costs, mainly associated with manufacturing. Even though beryllium is high TRL based on its JWST heritage, it requires extra care in manufacturing due to the human health complexity factor in grinding and light-weighting.

#### Aluminum 6061

Aluminum 6061 has good thermal conductivity and diffusivity. It is also excellent for manufacturability because it can be machined easily, polished, and heat-treated. However, its high density, extremely high strain at 4.5 K–highest of all the materials–and relatively high CTE at 4.5 K make it an overall poor performer for *Origins*. An athermal design, both optically and structurally, would minimize some issues associated with strain and CTE mismatch, but then mass is an issue. Options exist to improve light-weighting for an aluminum mirror but there is also chemical stability to consider because of its highly reactive surface. Spaceflight heritage exists for apertures below 0.5 meter.

## AlBeMet®

AlBeMet® is a metal matrix composite of aluminum and beryllium. Its materials properties fall between beryllium and aluminum and while it has a better performance than aluminum, it still has some of the disadvantages of low manufacturability due the toxicity of beryllium, while also being lower TRL. It also lacks heritage, with little-to-no spaceflight heritage and no meter class heritage. (East, Matthew, 2018)

## Material Performance for *Origins*

The team created material performance plots to better assess the remaining material candidates. Figures C-53 to C-56 show material properties plotted against specific stiffness at room temperature for beryllium, AlBeMet®, silicon carbide, fused silica, and aluminum. Material properties vary at 4.5 K from test to test and alloy to alloy. The properties also vary based on purity, grade, and manufacturer. These figures represent properties from 4.5 K - 30 K based on available data.

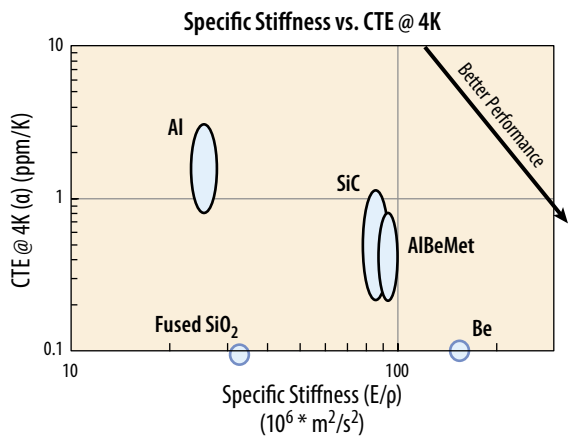

**Figure C-53:** The coefficient of thermal expansion (CTE) at 4.5 K plotted against specific stiffness at room temperature shows beryllium (Be) is the best performer, followed by AlBeMet®, SiC, fused silica, and aluminum (Al).

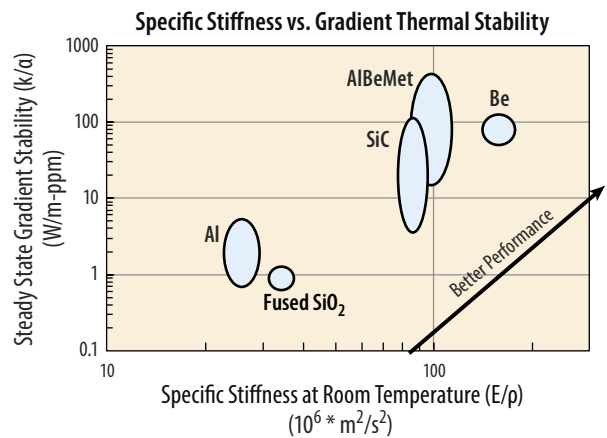

**Figure C-54:** Steady state gradient stability plotted against specific stiffness at room temperature shows beryllium (Be), AlBeMet®, and SiC as the best performers.

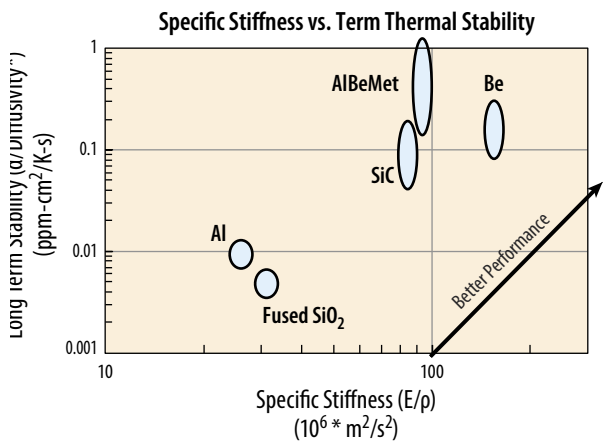

**Figure C-55:** Long-term stability (thermal diffusivity/CTE) plotted against specific stiffness at room temperature shows beryllium (Be), AlBeMet®, and SiC as the best performers.

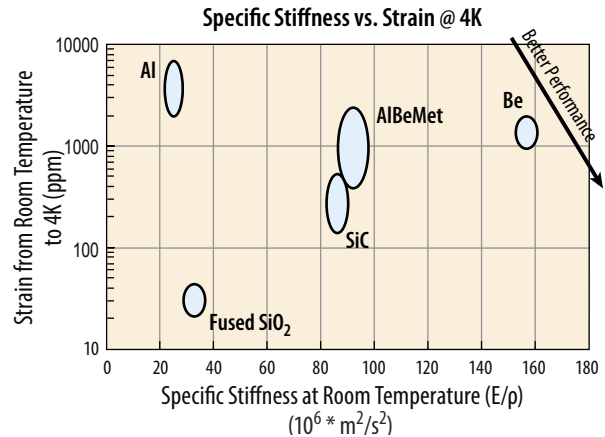

**Figure C-56:** Strain from room temperature down to 4 K against specific stiffness at room temperature shows tradeoffs between beryllium (Be), AlBeMet®, SiC and fused silica with aluminum (Al) as the lowest performer.

**Conclusions and Future Recommendations**

The materials selection is a tradeoff process. The most important parameters for the *Origins* primary mirror are specific stiffness (high stiffness and low density) and thermal performance (high thermal conductivity and low CTE). Beryllium 0-30 has excellent specific stiffness and adequate thermal performance for *Origins*' needs. It is also the highest TRL material and the most relevant cryogenic heritage, which advances the manufacturability and development. SiC is an attractive alternative option, but requires trades on the design and processing. Further assessments to address face sheet thickness, rib thickness, cell size of the mirror segments are recommended for either material option.

For structural materials, an athermal design is an attractive solution for CTE matching, but greatly increases the mass. As a result, the team recommends composites whenever possible. For the backplane and other cold side structures, thermal performance is more important than mass. Therefore, the team recommends these structures be made of the same material as the primary mirror. However, beryllium and SiC are brittle in comparison to most metals and composites, so early work with potential designs and vendors is needed for larger structures made of these materials.

## C.4 Instrument Details

There are no instrument details.

## C.5 ACS Details

Figure C-57 shows the performance of a Kalman filter during inertial hold using two DTU star trackers and a Honeywell Fiber Optic Gyro (FOG). Although it takes several minutes for the estimator to reach steady state, the knowledge requirement of 0.15 arcsec RMS is met after only 10 sec of settling.

During survey operations, the onboard pointing accuracy requirement is relaxed to 2 arcsec RMS. (The definitive attitude knowledge of 0.15 arcsec is achieved via post-processing on the ground.) Figure C-59 shows the performance of a Kalman filter during a representative survey profile using two DTU star trackers and a Honeywell FOG for measurements. The survey profile consists of scanning at a rate of 60 arcsec/sec for 1 min, then reversing the scan direction. The reversal requires ~90 sec. This profile is a conservative version of the expected profile, used to bound the resulting knowledge errors. Figure C-59 shows the estimator exhibits an offset during the scanning portions and is perturbed by the change in scanning direction. The resulting knowledge error for this bounding example is 1.0 arcsec RMS, both while scanning and while changing direction. Since the knowledge error satisfies *Origins*' 2.0 arcsec RMS requirement at all times, no extra settling time is needed after a scan reversal before enabling science data collection.

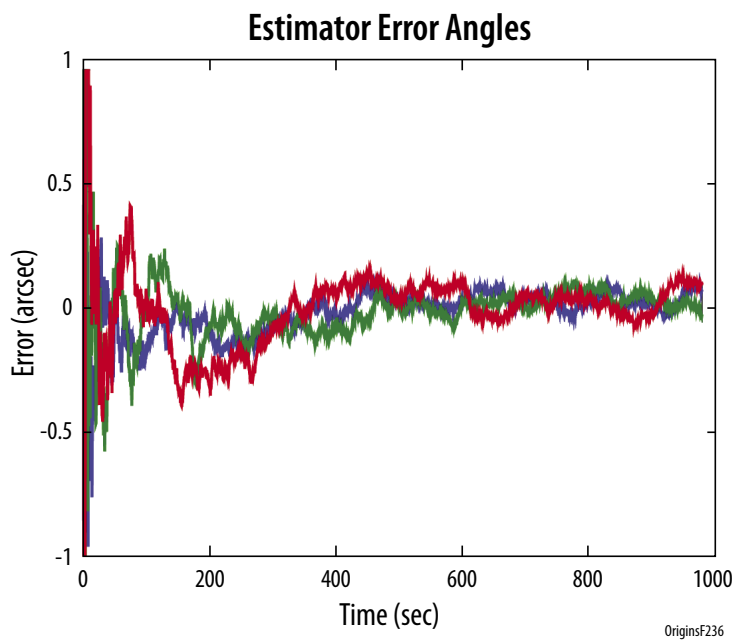

**Figure C-57:** Pointing knowledge during inertial hold is validated by time-domain simulation. Although the Kalman filter requires several minutes to achieve steady-state performance, pointing knowledge is within the 150-mas RMS requirement after only 10 seconds of initialization.

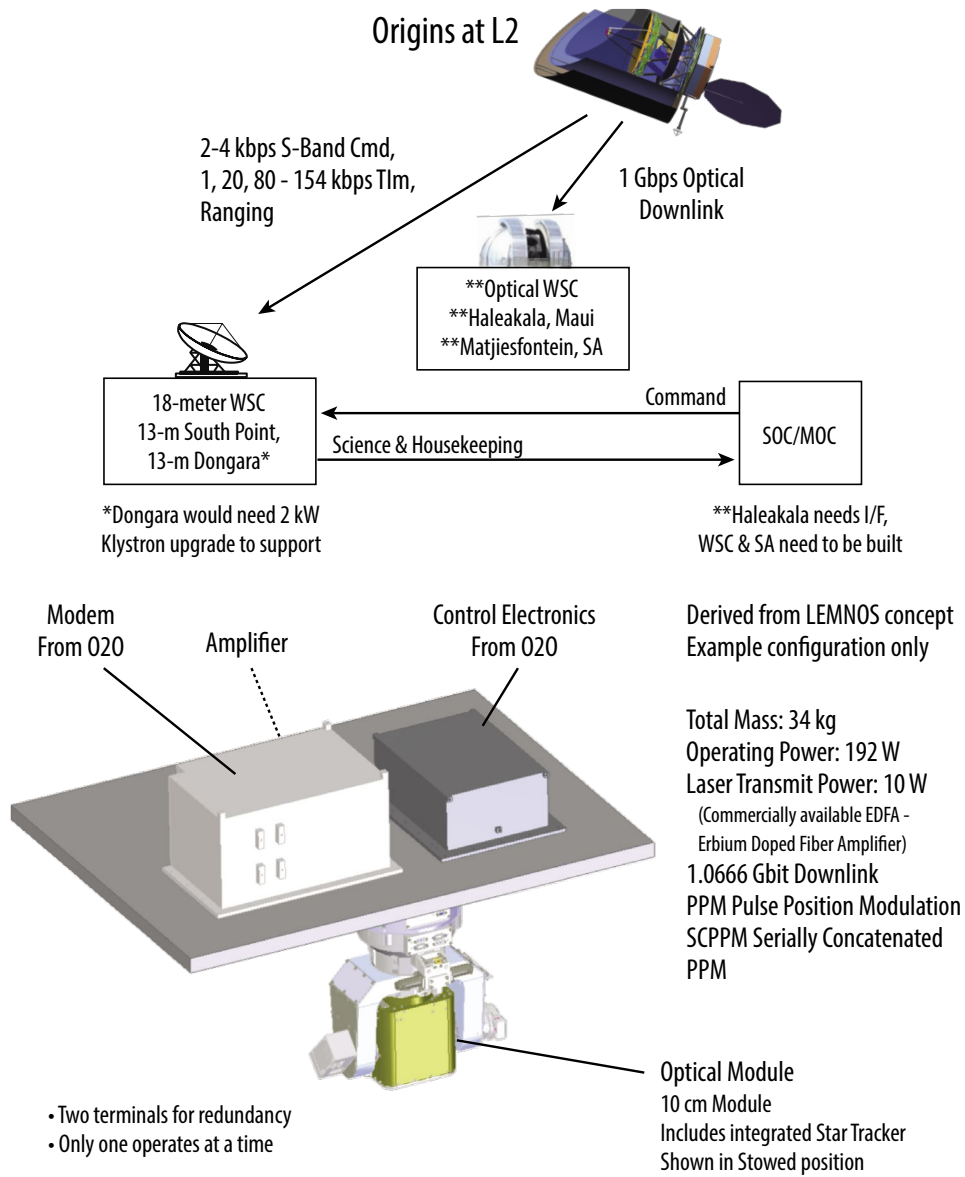

**Figure C-58:** The *Origins* optical communication system is fully redundant.

## C.6 Propulsion Details

There are no propulsion details added.

## C.7 Communications Details

Optical communication was selected to downlink the mission's ~21 Tbits/day science data to the ground (Figures C-58 and -60). The *Origins* optical flight terminal is based on the latest 10 cm design from NASA's **Laser-Enhanced Mission Communications Navigation and Operational Services (LEMNOS) program.** Table C-11 shows flight optical system physical parameters. Flight terminal pointing is guided by ground beacon and the system's internal star tracker.

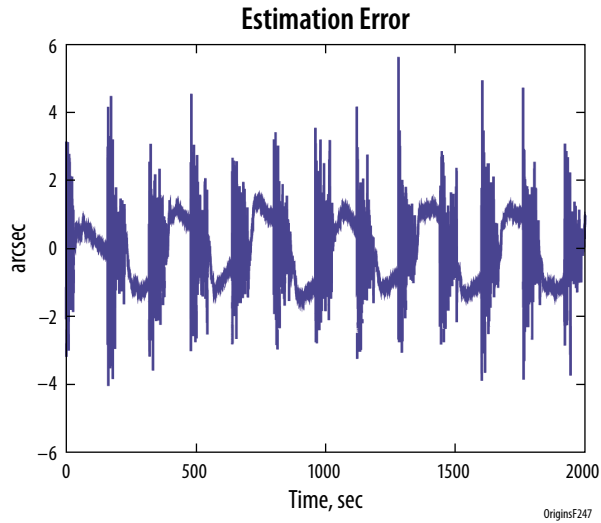

**Figure C-59:** Pointing knowledge during survey operations complies with the 2-arcsec RMS requirement at all times. The square-wave pattern apparent in the plot is due to the raster scan survey pointing profile.

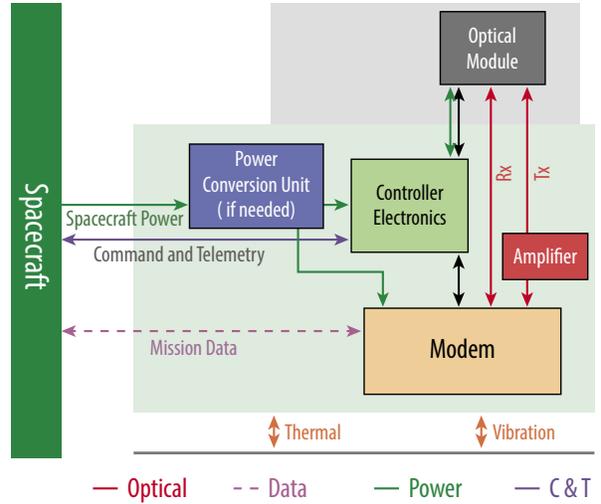

**Figure C-60:** Flight Optical Terminal Block Diagram

**Table C-11:** Flight Optical Terminal Physical Parameters.

| Constraint/ Parameter | Component | Current Estimate | Minimum | Maximum | Comments |
|---|---|---|---|---|---|
| Proximity | • OM / CE<br>• Modem | N/A | TBD | TBD | • OM and CE need to be in close proximity to each other (same panel) due to approx 140 wire connections between them.<br>• Modem can be much farther away from the CE and OM, using fiber to connect. |
| Mass | • OM<br>• CE<br>• Modem<br>• Amplifier | • 12.2 Kg<br>• 6.7 Kg<br>• 11.1 Kg<br>• 4 Kg | N/A | N/A | |
| Power | • OM<br>• CE<br>• Modem<br>• Amplifier | • N/A (included in CE)<br>• <96 Watts, 57 W operating<br>• 85 Watts<br>• 50 Watts | N/A | N/A | |
| Volume | • OM<br>• CE<br>• Modem<br>• Amplifier | • 20 in (H) X 15 in (dia)<br>• 10 in X 8.75 in X 4.5 in<br>• 15.2 in X 16 in X 9 in (TBD)<br>• 6 in X 5 in X 2 in | N/A | N/A | |
| Temperature Range | • CE<br>• Modem | N/A | • Op: -29 deg C<br>• Survival: -40 deg C<br>• Op: +6 deg C<br>• Sur: -24 deg C | • Op: +39 deg C<br>• Survival: +61 deg C<br>• Op: +43 deg C<br>• Sur: +61 deg C | |

Optical communication provides orders of magnitude higher downlink rate than RF via a compact, low-mass, and low-power flight terminal. It is currently the state-of-the-art in communication and is ideally suited for high data volume space missions such as *Origins*. Optical/Ka based European Data Relay System (EDRS) deployment started in 2016 and will be complete in 2020/2021. NASA successfully demonstrated lunar lasercom in 2013-2014. Laser Communication Relay Demonstration (LCRD) is on track for a 2019 launch, and the Deep Space Optical Comm (DSOC) demonstration is slated for 2022 (Figure C-61). A US commercial optical satellite constellation, LeoSat, is planned to complete deployment of 78-108 satellites by 2021, ushering in worldwide service. Based on these developments, optical communication will be well established before the start of *Origins* in 2025.

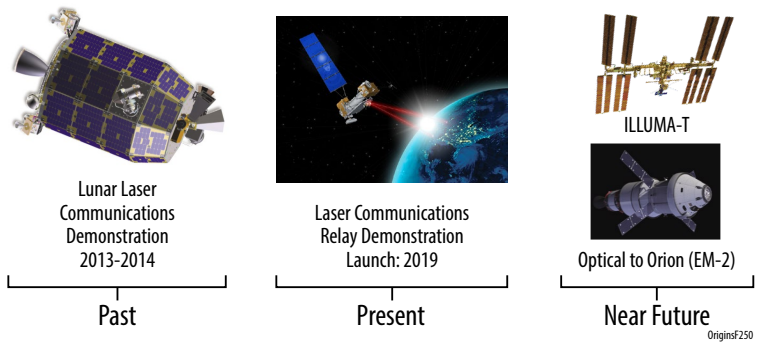

Lunar Laser
Communications
Demonstration
2013-2014

**Past**

Laser Communications
Relay Demonstration
Launch: 2019

**Present**

ILLUMA-T

Optical to Orion (EM-2)

**Near Future**

OriginsF250

**Figure C-61:** Optical communication development started several years ago, and will be well established before the start of *Origins* in 2025.

*Origins* uses S-band for housekeeping telemetry, commands, and tracking. The design is comprised of two omni antennas, redundant transponders with 8 W RF output power and a 0.9-m two-axis gimballed high gain antenna (HGA). The mission's 1 Kbps downlink and 2 Kbps uplink is achieved using an omni antenna and the 18-m NASA Near Earth Network (NEN) ground station antenna at White Sands, NM with 2 kw klystron. The HGA is required for tracking. HGA is also needed for uplink, downlink, and tracking with 13-m NEN antennas (Tables C-12 and -13). Tables C-12 and C-13 show *Origins* communications modes, data rate, and signal margins with and without ranging.

The three NASA Deep Space Network (DSN) stations' 34-m antennas provide S-band emergency backup, closing the links via the OST omni antennas with plenty of margin (Tables C-14 and C-15).

**Table C-12:** *Origins* Optical and S-band via NEN Ground Stations with Comfortable Link Margins

| *Origins* Mode | Data Rate | Margin (dB) | Comments |
|---|---|---|---|
| Optical Comm *Origins* Concept 2 | 1.066 Gbps | 3.22 | <20°, no clouds |
| S-band WSC 18-m CMD 300 Watt Uplink with Ranging via 0.9-m | 2 kbps | 4.5 | 99% availability at 5° |
| S-band WSC 18-m TLM with Ranging via 0.9-m | 79.5 kbps | 3.0 | 99% availability at 5° |
| S-band WSC 18-m Ranging via 0.9-m | | 0/9.0 | 99% availability at 5° |
| S-band WSC 18-m CMD 2 kWatt Uplink with Ranging via 0.9-m | 4 kbps | 10.5 | 99% availability at 5° |
| S-band WSC 18-m TLM with Ranging via 0.9-m | 79.5 kbps | 3.0 | 99% availability at 5° |
| S-band WSC 18-m Ranging via 0.9-m | | 9.0/12.7 | 99% availability at 5° |
| S-band USHI 13-m CMD - 2 kWatt Klystron with Ranging via 0.9-m | 2 kbps | 10.5 | 99% availability at 5° |
| | 4 kbps | 7.5 | |
| S-band USHI 13-m TLM with Ranging via 0.9-m | 20 kbps | 3.0 | 99% availability at 5° |
| S-band USHI 13-m Ranging via 0.9-m | | 6 / 6 | 99% availability at 5° |

\* will need waiver for not having TDRS to cover Near Earth launch critical events in case they are not covered by ground stations.  See Appendix for details.

**Table C-13:** *Origins* S-band via NEN Ground Stations without ranging with comfortable link margins.

| *Origins* Mode | Data Rate | Margin (dB) | Comments |
|---|---|---|---|
| S-band  WSC 18-m 300 W CMD via S/C 0.9 m HGA | 32 kbps | 4.0 | 99% availability at 5° |
| S-band WSC 18-m TLM via S/C 0.9m HGA | 154 kbps | 3.0 | |
| S-band WSC 18-m 2 kW CMD via S/C 0.9 m HGA | 256 kbps | 4.0 | |
| S-band WSC 18-m 2 kW CMD via Omni | 2 kbps | 6.3 | |
| S-band WSC 18-m TLM via Omni | 1 kbps | 2.2 | |
| S-band USHI 130m 2 kW CMD via 0.9 m HGA | 32 kbps | 10.0 | |
| S-band USHI 13-m TLM via 0.9 m HGA | 23 kbps | 3.0 | |

\* will need waiver for not having TSRS to cover Near Earth launch critical events in case they are not covered by ground stations.  See Appendix for details.

**Table C-14:** DSN emergency backup with comfortable link margins.

| *Origins* Mode with DSN (backup) | Data Rate | Margin (dB) | Comments |
|---|---|---|---|
| DSN 34-m BWG Command w Ranging | 2 kbps | 5 | 99% availability at 6° |
| DSN 34-m BWG Command w Ranging | 4 kbps | 2 | |
| DSN 34-m BWG TLM w Ranging | 1 kbps | 9.8 | |
| DSN 34-m BWG Ranging | | 8.4 | |
| DSN 34-m BWG Command w subcarrier | 2 kbps | 13.3 | |
| DSN 34-m BWG Command only | 2 kbps | 15.4 | |
| DSN 34-m BWG Command only | 4 kbps | 12.4 | |
| DSN 34-m BWG Telemetry only | 1 kbps | 12.7 | |

Note: See appendix for details

**Table C-15:** *Origins* Optical Communication Link Analysis with 3.2 dB Margin

| | LASERCOM SPACE-EARTH LINK MARGIN CALCULATION | |
|---|---|---|
| | GSFC C.L.A.S.S. ANALYSIS | DATE & TIME: 08/08/2018 15:19:11 |
| | LINKID: *Origins* Concept 2 Nominal | PREPARED BY: Asoka Dissanayake |
| | STATION: White Sands NM | LATITUDE: 38.5°, LONGITUDE: -106.6°, ALTITUDE: 1449.5m |
| | PATH LENGTH: 1,500,000 km | ELEVATION ANGLE: 20.0° |
| | DATA RATE: 1066.600 Mbps | MODULATION: Pulse Position Modulation |
| | CODE RATE: R1/3 | |

| ITEM# | PARAMETER | VALUE AND REMARKS |
|---|---|---|
| | **GENERAL** | |
| 1 | Wavelength (nm) | 1550.000 User input |
| 2 | Range (km) | 1.50E+06 User input |
| 3 | Data rate (Mbps) | 1066.6 User input |
| 4 | Modulation | PPM User input |
| 4A | PPM order | 4 User input |
| 4B | Guard time (%) | 25.0 User input |
| 5 | Forward error correction type | SCPPM Serially Concatenated PPM |
| 5A | Forward error correction rate | 0.333 User input |
| 6 | Planck constant (Joules sec) | 6.63E-34 |
| 7 | Energy per photon (dBJoule) | -188.92 Planck constant*optical frequency |
| | **ATMOSPHERE AND NOISE** | |
| 9 | Atmospheric attenuation (dB) | 0.68 FASCODE: US Standard, Rural 5km |
| 9 | Cloud attenuation (dB) | 0.00 User input |
| 10 | Scintillation index | 0.12 Vg: 4.0 m/s; Cn²:0.10E-13 m^(-2/3) |
| 11 | Fried parameter (cm) | 13.07 Vg: 4.0 m/s; Cn²:0.10E-13 m^(-2/3) |
| 12 | Sky radiance (W/cm²/μm/sr) | 1.49E-02 MODTRAN: US Standard, Rural 5km |
| 13 | Background object radiance (W/cm²/μm/sr) | 0.00E+00 Background object: None |
| 14 | Stray radiance (W/cm²/μm/sr) | 2.00E-04 User input |
| | **TRANSMITTER** | |
| 15 | Laser signal power (dBW) | 10.00 Input power: 10.0W |
| 16 | Transmitter loss (dB) | 0.00 User input |
| 17 | Telescope aperture diameter (cm) | 10.00 User input |
| 19 | Telescope gain (dBi) | 105.25 Array size: 1 |
| 20 | Telescope optical loss (dB) | 4.00 User input |
| 21 | Pointing loss (dB) | 1.50 User input |
| 22 | EIRP (dBW) | 109.75 15-16+19-20-21 |
| 23 | Free space loss (dB) | 321.70 CLASS computed |
| | **RECEIVER** | |
| 24 | Telescope aperture diameter (cm) | 60.00 User input |
| 26 | Telescope gain (dB) | 127.72 Array size: 4 |
| 27 | Receiver optical loss (dB) | 3.00 User input |
| 27A | Polarization loss (dB) | 0.50 User input |
| 27B | Pointing loss (dB) | 0.50 User input |
| 27C | Fiber coupling loss (dB) | 2.62 Empirical model |
| 27D | Signal power loss due to finite Tx extinction | 0.09 Tx on-off ratio: 23.0dB |
| 28 | Received signal power (dBW) | -91.61 Empirical model-27C-27D |
| 29 | Received signal photons (dBph/bit) | 7.03 CLASS computed |
| 30 | Background object photons (ph/ns) | 0.00E+00 Background object: None |
| 31 | Stray radiance photons (ph/ns) | 2.14E-04 User input |
| 32 | Sky radiance photons (ph/ns) | 1.60E-02 MODTRAN; Sun angle: 1.0° |
| 32A | Unextinct photons (ph/ns) | 2.70E-02 Tx on-off ratio: 23.0dB |
| 33 | External noise photon total (ph/ns) | 4.32E-02 30+31+32+32A |
| 34 | Detector efficiency (dB) | -0.97 Efficiency: 80.0% |
| | **LINK PERFORMANCE** | |
| 35 | Required signal photons at capacity (dBph/bit) | -1.75 CLASS computed |
| 36 | Implementation loss (dB) | 1.50 User input |
| 36A | Dynamic loss (dB) | 0.34 CLASS computed |
| 36B | Detector blocking loss (dB) | 0.71 Array size:128, Dead time: 5ns |
| 36C | Detector jitter loss (dB) | 1.04 Jitter std: 25.0ps |
| 37 | Available signal photons (dBph/bit) | 2.46 29+34-36-36A-36B-36C |
| 38 | Code design from capacity (dB) | 1.00 Default value |
| 39 | Margin (dB) | 3.22 37-35-38 |

*Origins*'s 1 Gbps data downlink will utilize the design of the wide Field Infrared Survey Telescope (WFIRST) SSR with Delay/Disruption Tolerant Networking (DTN) protocols that can handle up to 3.5 Gbps data downlink. DTN is currently being incorporated onto the Plankton, Aerosol, Cloud, ocean Ecosystem (PACE) mission, which will launch 12 years before *Origins*.

## C.8 Flight Software Details

*Origins* has ample (61%+) onboard processing resources margins (Table C-16).

The *Origins* flight software team uses three testbeds for development and testing, initial interface verification, and mission operations preparation. The highest-fidelity testbed is maintained for FSW system testing, as well as sustaining engineering.

The OFSW (Table C-17) will be developed in compliance with NPR 7150.2 and CMMI-compliant processes. A waterfall development model transforms software requirements into tested flight code. Major coding occurs in three incre-mental builds. The first build provides all core FSW and external/internal interfaces drivers, which are mainly inherited from the GSFC cFS library and MUSTANG. The second includes mission-unique functionalities, such as science data acquisition/management, power, thermal, and attitude control. The third build includes mission operations and fault management. All requirements for each build are defined prior to the start of coding. Formal reviews are held before proceeding beyond key phases, including Requirements Analysis, Preliminary Design, Critical Design, and Acceptance Test. Additionally, all new and modified code is inspected by the team and unit tested by the developer prior to build. Once build code has been integrated into the FSW, the software test team performs verification tests of the integrated OFSW requirements. No new testing is planned for heritage flight software components; however, functionality is exercised through system testing. After all builds are completed, a system level acceptance test is performed, during which the entire flight software suite is tested as a total integrated unit. Figure C-63 describes the OFSW integration and test processes. The OFSW team maintains the flight software configuration management and discrepancy reporting systems with full Quality Assurance insight.

**Table C-16:** *Origins* has ample Processors Utilizations Margins

| CSCI | CPU Utilization Estimates | | |
|------|-------|-----|--------|
| | Total | CBE | Margin |
| Command & Data Handling Processor (C&DH) DLEON3 | 100 | 38.57 | 61% |
| Attitude Control Electronics Processor (ACE) LEON3 | 100 | 8.62 | 91% |
| Solid State Recorder Processor (SSR) LEON3 | 100 | 24.14 | 76% |

**Table C-17:** OFSW Modules Commonality

| Components | Descriptions | CDH | ACE | SSR |
|-----------|-------------|-----|-----|-----|
| **cFE Applications** | A set of (5) Mission customizable services that provide critical S/C FSW functionality | | | |
| Event Services | Provides asynchronous telemetry messages to notify ground operators of informational, anomalous, and critical onboard activities | X | X | X |
| Executive Services | Manages startup of cFE services, provides application services to enable Starting, Deletion and Restarting of running FSW tasks | X | X | X |
| Software Bus | Provides routable inter-application CCSDS format messages which form the foundation of all onboard FSW communications | X | X | X |
| Table Services | Provides onboard table management for ground operations | X | X | X |
| Time | Provides time stamping of Mission Telemetry, as well as ancillary, related time functions ("time at tone") | X | X | X |
| **cFS Applications** | The cFS is a set of mission customizable applications that, together with the cFE and other applications provide mission FSW functionality | | | |
| Scheduler | Provides Mission unique scheduling of FSW applications | X | X | X |
| Housekeeping | Provides mechanism for gathering telemetry from FSW applications | X | X | X |
| File Manager | Provides a flight to ground interface for managing onboard files | X | X | X |
| Memory Dwell | Provides Ground with the ability to monitor ad hoc memory locations | X | X | X |
| Memory Manager | Provides a flight to ground interface for managing onboard memory | X | X | X |
| Limit Checker | Provides the capability to monitor onboard telemetry, and provide a rules-based response | X | | |
| Store Command | Provides the capability to store commands for later execution, as well as providing pre-canned sequences of commands | X | | |
| Checksum | Performs data integrity checking of memory, tables and files | X | X | X |
| Data Storage | Records housekeeping, and engineering data onboard for subsequent downlink | X | | X |
| **Mission Applications** | cFS compliant applications that need tailoring to *Origins* mission unique requirements | | | |
| CCSDS File Delivery Protocol | Engineering/science data file uplink and downlink, may need customization from PACE | X | | X |
| Memory Scrub | Scrubs SRAM memory to detect and correct Memory errors, may need customization from PACE | X | | X |
| Telemetry Output | Manages real-time and recorded engineering data downlink, telemetry filtering | X | | |
| Command Ingest | Receives ground commands for distribution to subsystems, support CCSDS, COP-1 uplink protocols provide 1st order command validation and authentication | X | | |
| Power Management | Provide switch services and battery management | X | | |
| Instrument Applications | Provides command, telemetry, and housekeeping support for instrument | X | | |
| Guidance, Navigation and Control | Provides spacecraft navigation, attitude determination and control, SA and HGA pointing. The ACE will have a much reduced set of functions | X | X | |
| Interfaces Manager | Provides interface to custom hardware (1553, SpW, SERDES, etc.) | X | X | X |
| FSM Control | Provides field steering mirror control | X | | |
| Script Processor | Command script process for mission operations similar to JWST | X | | |
| DTN/playback manager | Delay tolerant network, science data playback | | | X |

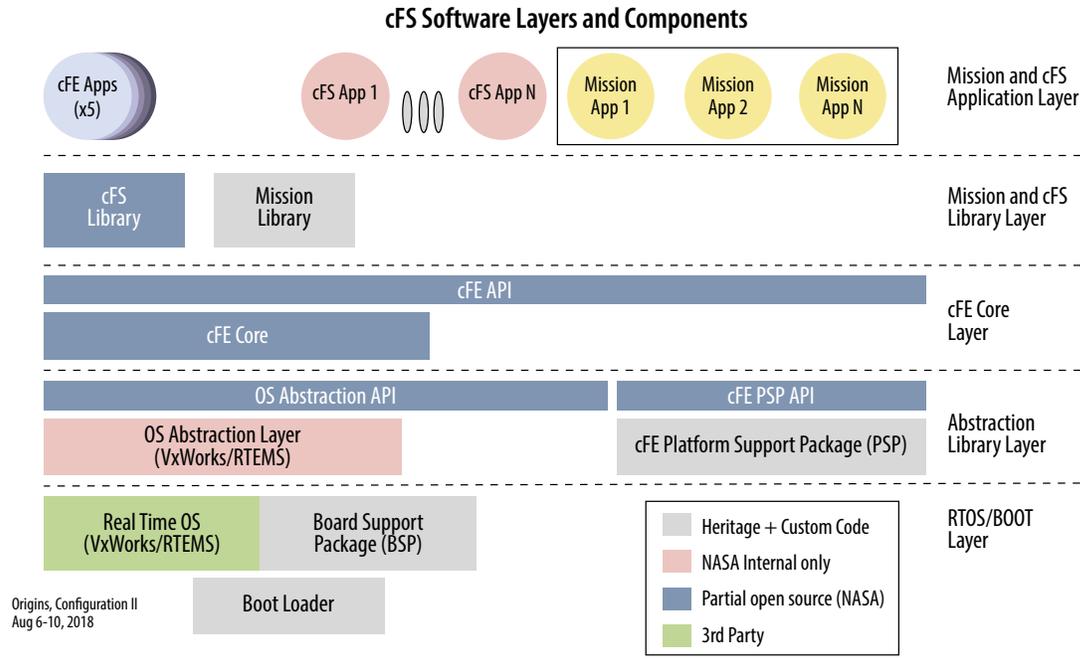

**Figure C-62:** The *Origins* software system builds on heritage and well developed modules.

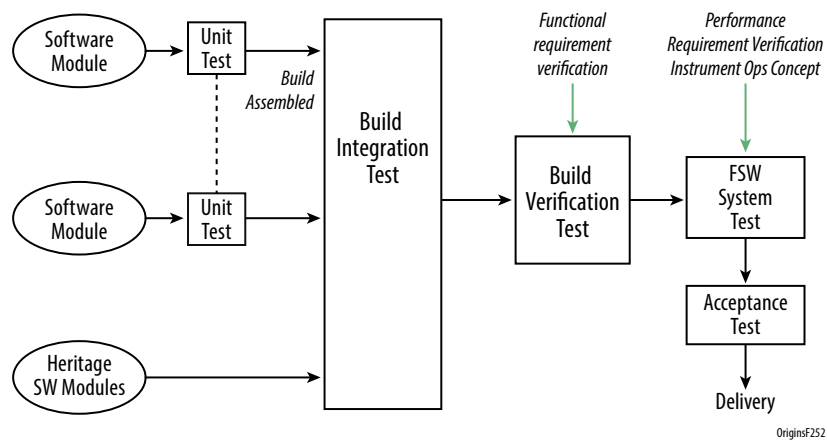

**Figure C-63:** OFSW Integration and Test

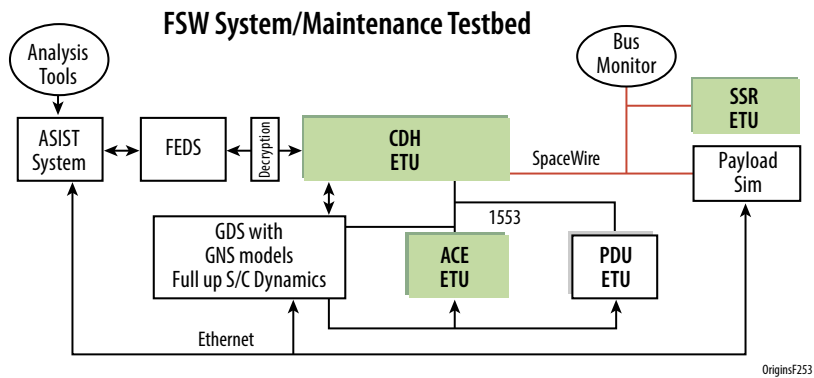

**Figure C-64:** Flight Software System Test Environment

The *Origins* flight software team uses three testbeds for development and testing, initial interface verification, and mission operations preparation. The highest-fidelity testbed is maintained for FSW system testing, as well as sustaining engineering. Figure C-64 shows the fully redundant system configuration for system test and flight software maintenance. Flight software testbeds reduce risk through high-fidelity simulation of the operational interfaces. The development team uses Advance Spacecraft Integration and System Test (ASIST) or commercially available Integrated Test and Operations System (ITOS) ground support equipment (GSE) to command and control the execution of the OFSW. The *Origins* FSW testbed configurations are similar to heritage uses at GSFC or any NASA centers and/or space industry partners, but are dedicated to specific development purposes. 1) The C&DH Testbed supports *Origins* C&DH/SSR FSW development, build integration, and build testing. It includes a payload and dynamic simulator to check out GNC and payload I/Fs. 2) The ACS FSW Testbed supports *Origins* GNC and ACE FSW development, build integration, and build testing. It includes a full dynamic simulator for hardware in-the-loop simulations to verify GNC functional and performance requirements. 3) The FSW System/Maintenance testbed has the highest-fidelity configuration. The team uses this testbed for FSW system testing, mission operations simulations and training, as well as sustaining engineering.

## C.9 Integration and Test Details

### C.9.1 Telescope Testing

The team plans for additional qualification testing during Telescope I&T. After the PMSAs and SMA are installed, instrument mass simulators are attached to the Telescope, along with the Baffle, Barrel, harnessing, Sunshields, and cold electronics. The CPM Base is then attached to a test fixture using the same interface as the Thrust Tube Upper Ring. This assembly will undergo a modal survey test, signature sine sweep vibration testing, protoflight three-axis sine vibration testing, and protoflight acoustics testing.

After Telescope testing is complete, the instrument mass simulators are removed and flight instruments are installed. This structure is now the Cryogenic Payload Module (CPM). The CPM undergoes additional protoflight three-axis sine vibration testing (derived from the launch vehicle specification) and protoflight acoustics testing.

Assembly Integration and Testing of the Telescope can be performed any number of facilities around the US. GSFC was chosen as an example, since the facility costs, sizing and capabilities are well known. The start of Telescope I&T would be with assembly of the structural subassemblies at GSFC in the Spacecraft Systems Development and Integration Facility (SSDIF, Figure C-65), which was also used for JWST telescope assembly.

A rollover fixture is used to orient the structure as required for assembly access. Then optics installation starts. The Aft Optics, all primary mirror assemblies, and the secondary mirror are installed and aligned to the structure. Then the instrument mass simulators, cryocoolers, inner optical baffle, telescope harness, outer barrel, sunshield structure and deployment mechanism, electrical boxes, and protective cover are integrated. The completed telescope assembly is then ready for the mechanical test program (Modal Survey, Signature Sine Sweep Frequency Characterization, Protoflight three-axis Sine Vibration), followed by one more Signature Sine Sweep Frequency Characterization and Protoflight Acoustics. Alignment is performed before and after these tests. The scope of this test program is to verify the stability of the mirrors, structures and mechanisms. The Telescope, which includes the Barrel and stowed sunshield then goes through acoustics testing for the first time. After this, the Telescope is ready for integration with the Instruments. The Telescope I&T summary flow is shown in Figure C-66.

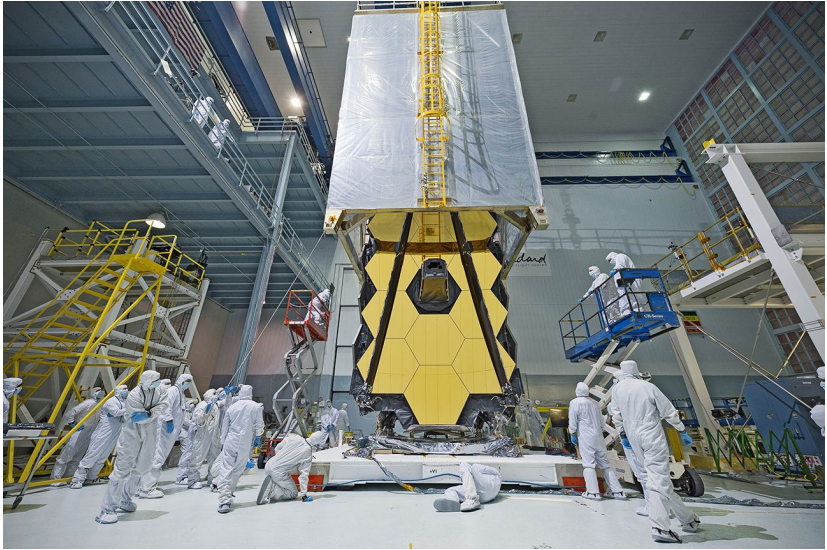

**Figure C-65:** SSDIF High Bay Cleanroom at GSFC was also used to assemble the JWST telescope.

## C.9.2 Cryogenic Payload Module Testing

The primary objective of the integration and test (I&T) program is to successfully deliver a fully-tested and -verified CPM that meets all functional requirements and performs as designed when placed on the spacecraft to create the *Origins* observatory. The CPM I&T summary flow is shown in Figure C-69. While specific locations are given to perform various tests, these are meant to be examples only which show one way that the tests can be accomplished. Other facilities at different locations in the US may also be used for these tests.

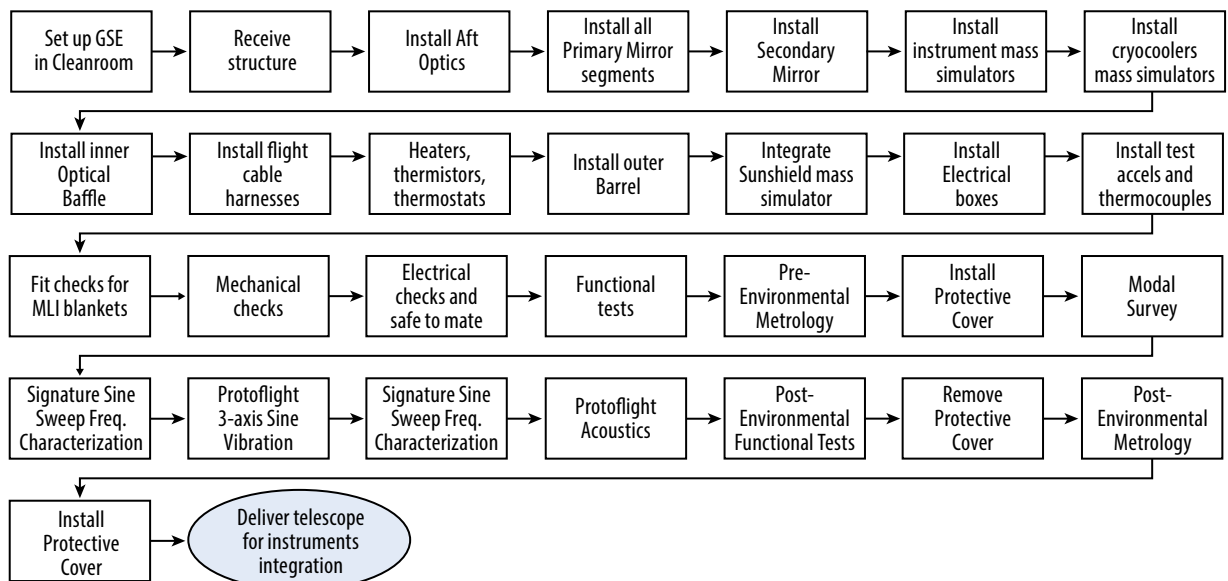

**Figure C-66:** Telescope I&T Summary Flow

The CPM integration can be accommodated at GSFC's Spacecraft Systems Development and Integration Facility (SSDIF, Figure C-65). Specialized Ground Support Equipment (GSE) development is required for the integration of all components. After ambient integration of the fully-assembled telescope with the instruments, the team will perform EMI/EMC testing. This test is followed by Protoflight Vibration and Protoflight Acoustics tests. Aliveness test and ambient metrology are performed pre- and post-mechanical testing to ensure the hardware remains aligned. The environmental test campaign (except for the cryogenic thermal vacuum test) can be accommodated at GSFC. GSFC's Acoustic Chamber is shown in Figure C-67. EMI testing requires using a customized tent.

The team will design and manufacture a specialized In-Plant Transporter (IPT) to move the flight hardware from the cleanroom to the vibration and acoustics facility. The IPT will include a portable clean tent around the CPM to provide the required cleanliness. The sunshield deployment test will be conducted at this point. After the mechanical tests, the team will perform a mass properties test. The hardware is then prepared for transportation and shipped to JSC. Upon arrival at JSC, the flight CPM is removed from the transporter and prepared for installation into the Chamber A (Figure C-68). Preparations include telescope protective cover removal, GSE sensor installation, metrology, and ambient functional testing. Once installed into Chamber A, CPM cryogenic testing begins. The cryogenic vacuum test will verify CPM-level requirements in the conditions of the expected flight environment, with an emphasis on optical measurements that can be performed in this test configuration. The optical tests will verify the CPM system optical workmanship and provide optical test data to support integrated telescope modeling used to predict flight optical performance. After the testing, the CPM is then removed from the chamber and prepared for transportation. The hardware is shipped to GSFC, where it is mated to the spacecraft to create the *Origins* observatory.

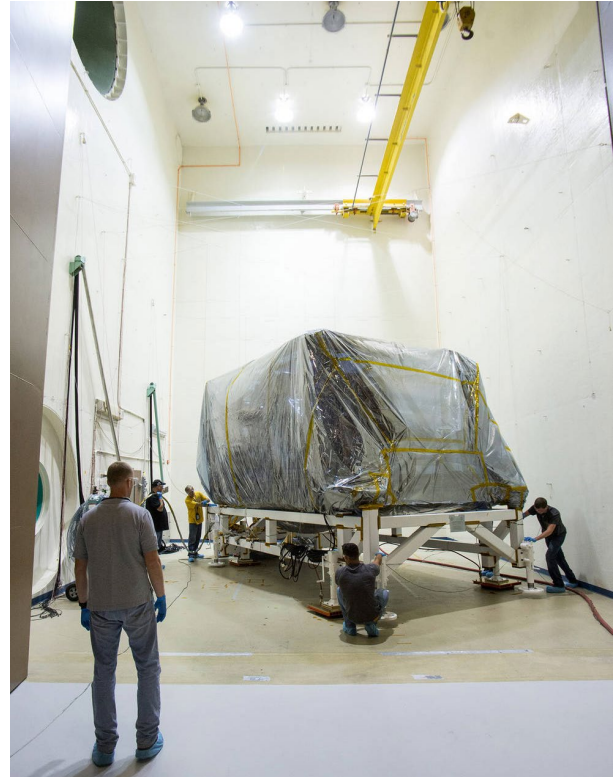

**Figure C-67:** The GSFC Acoustic Chamber can easily accommodate testing the *Origins* CPM.

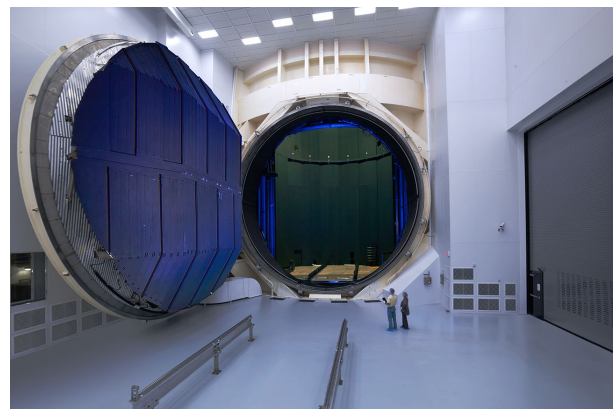

**Figure C-68:** Chamber A at JSC will be used to perform CPM cryogenic testing.

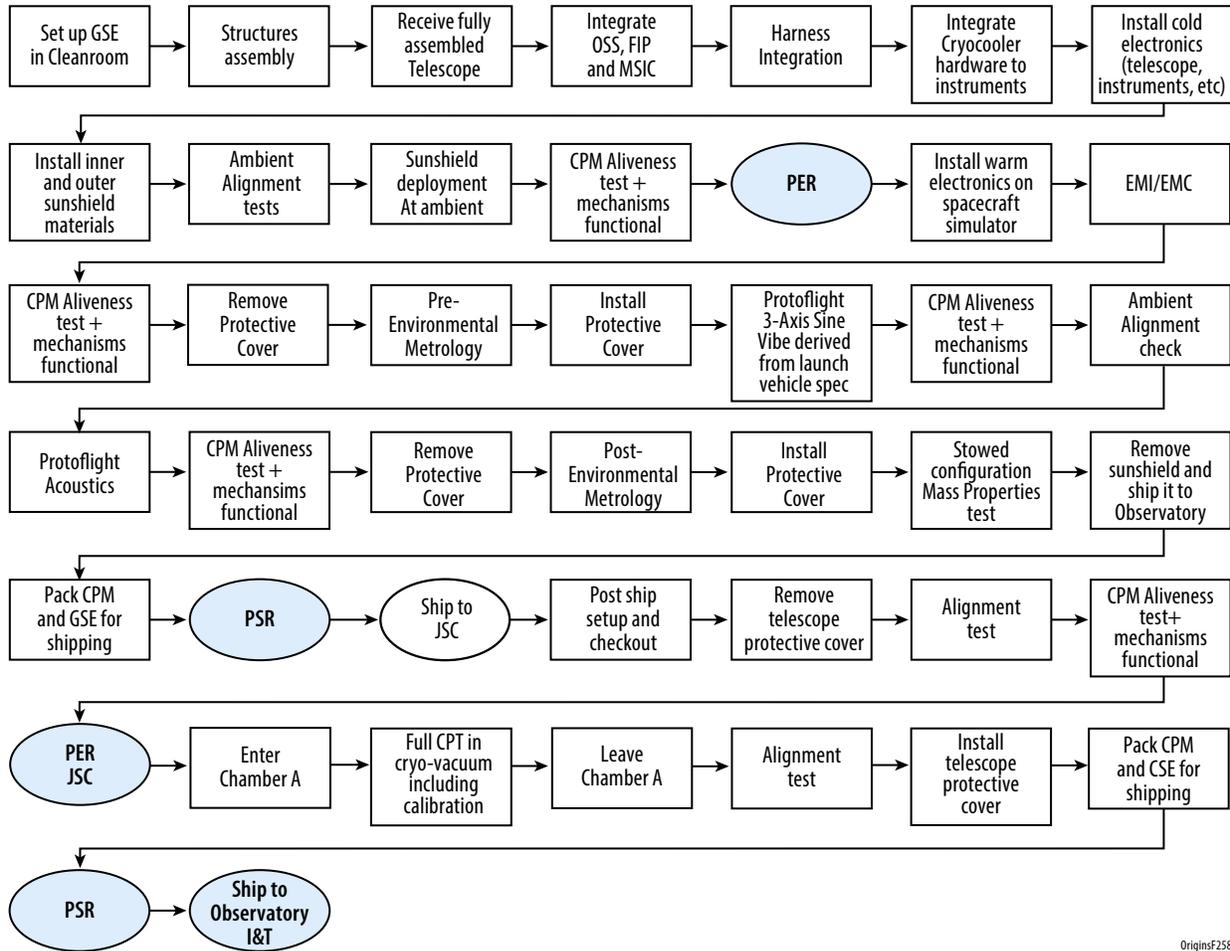

Set up GSE in Cleanroom → Structures assembly → Receive fully assembled Telescope → Integrate OSS, FIP and MSIC → Harness Integration → Integrate Cryocooler hardware to instruments → Install cold electronics (telescope, instruments, etc)

Install inner and outer sunshield materials → Ambient Alignment tests → Sunshield deployment At ambient → CPM Aliveness test + mechanisms functional → PER → Install warm electronics on spacecraft simulator → EMI/EMC

CPM Aliveness test + mechanisms functional → Remove Protective Cover → Pre-Environmental Metrology → Install Protective Cover → Protoflight 3-Axis Sine Vibe derived from launch vehicle spec → CPM Aliveness test + mechanisms functional → Ambient Alignment check

Protoflight Acoustics → CPM Aliveness test + mechanisms functional → Remove Protective Cover → Post-Environmental Metrology → Install Protective Cover → Stowed configuration Mass Properties test → Remove sunshield and ship it to Observatory

Pack CPM and GSE for shipping → PSR → Ship to JSC → Post ship setup and checkout → Remove telescope protective cover → Alignment test → CPM Aliveness test + mechanisms functional

PER JSC → Enter Chamber A → Full CPT in cryo-vacuum including calibration → Leave Chamber A → Alignment test → Install telescope protective cover → Pack CPM and CSE for shipping

PSR → Ship to Observatory I&T

OriginsF258

**Figure C-69:** (Above) The CPM I&T flow shows a simple path to achieve a fully-tested and -verified CPM.

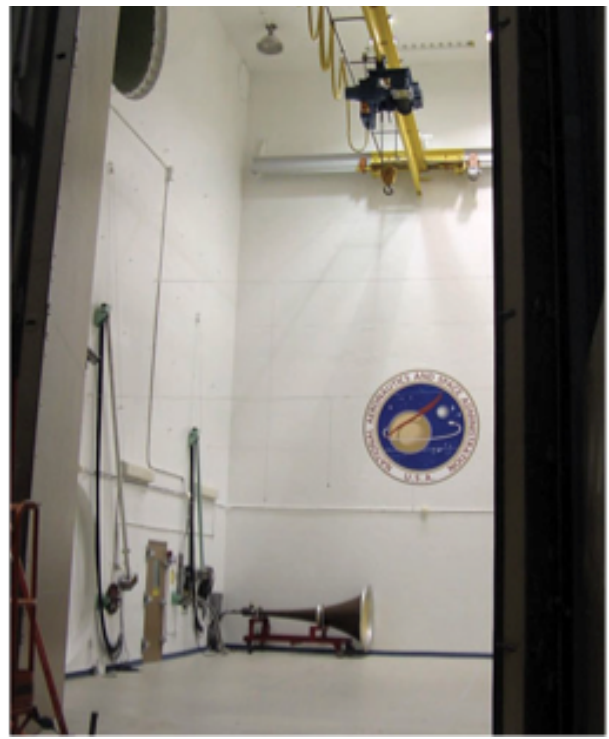

**Figure C-70:** (Right) The Chamber at GSFC. Note that this chamber is not large enough for the fully assembled *Origins* observatory. This test will be performed at JSC or another suitable facility.

The CPM integration could occur at GSFC in the SSDIF. To accomplish the integration of all components, specialized Ground Support Equipment (GSE) needs to be developed. After ambient integration of the fully assembled telescope with instruments, EMI/ EMC testing will be performed. This test will be followed by Protoflight Vibration and Protoflight Acoustics tests. Aliveness test and ambient metrology will also be performed pre- and post- mechanical testing to ensure the hardware remains aligned. The environmental test campaign (except for the cryogenic thermal vacuum test) could be done at GSFC. The Acoustic Chamber is shown in Figure C-70. A customized tent could be used for EMI.

The IPT has a portable clean tent around the CPM that provides the necessary cleanliness required. After the mechanical tests, mass properties test is performed.

## C.9.3 Spacecraft Bus Module I&T

The SBM integration (Figure C-71) begins with the propulsion system integration followed by the harness integration and the assembly of thermal components and electronics. Validated flight software is delivered and utilized throughout the remainder of I&T. A complete electrical integration of all bus mounted electronics is performed in a "flat sat" configuration. Flight electrical equipment that will be mounted to the bus structure is temporarily mounted to a ground support table to enable full flight electrical integration of the entire complement of bus-mounted equipment. Typical electrical system integration issues are discovered during "flat sat" I&T, which is performed independent of the bus structure critical path. Major tests during SBM I&T include an initial ambient baseline electrical Comprehensive Performance Test (CPT), alignments, ambient deployments, RF compatibility/Laser Communications test.

The environmental test campaign includes: EMI/EMC, Protoflight 3-Axis Sine Vibration, Protoflight Acoustics, Protoflight Thermal Balance & Thermal-Vacuum, Propulsion functional and leakage tests. These tests could be done at GSFC using the T4000-3 Shaker shown in Figure C-72, the large EMI chamber (shown in Figure C-73), the Mass Properties Measurement Facility (MPMF), and the Space Environment Simulator (SES, shown in Figure C-74). There are many other facilities in the U.S. where these tests can be done. CPM interfaces are electrically simulated during functional and environmental testing. A fully tested and verified SBM is ready for mate with the CPM in observatory I&T.

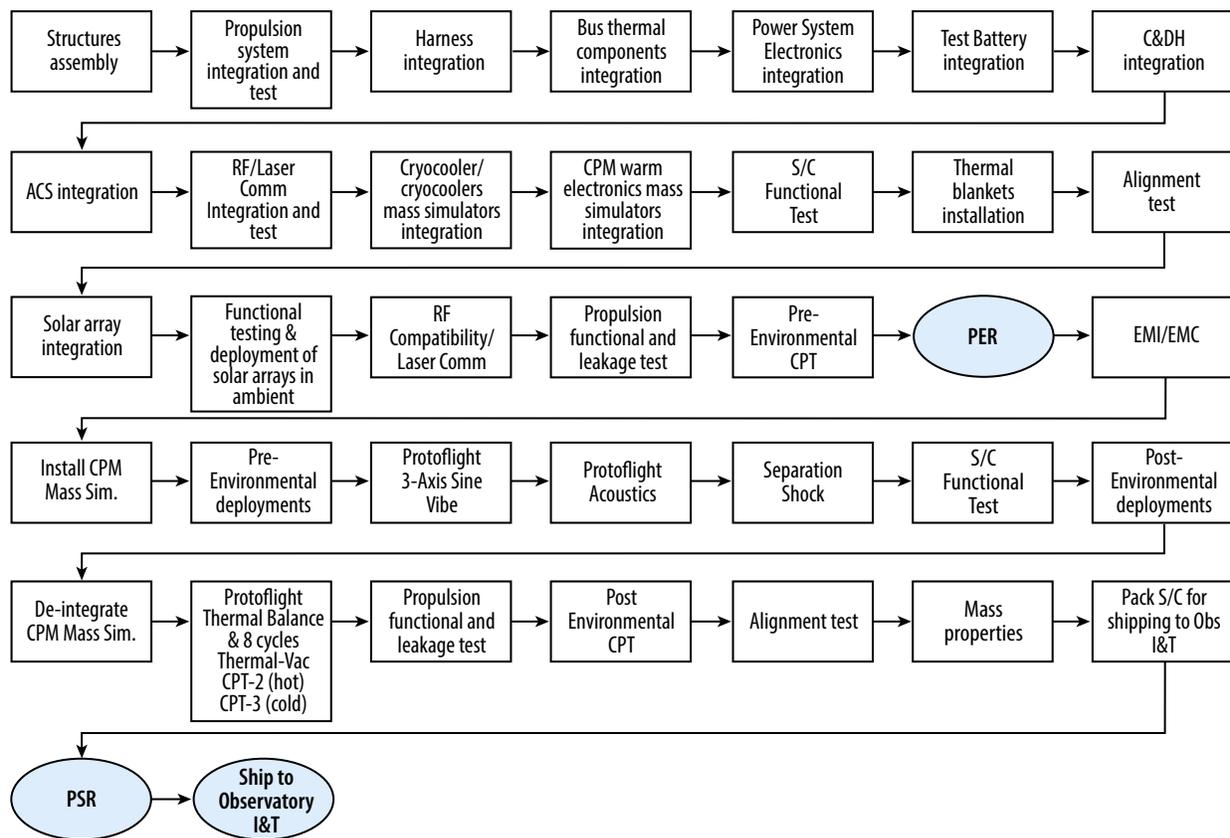

**Figure C-71:** SBM I&T Summary Flow.

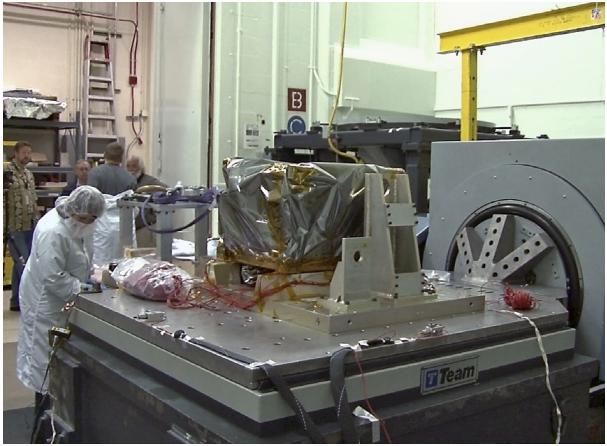

**Figure C-72:** Facility 410 – Unholtz-Dickie T4000-3 Shaker

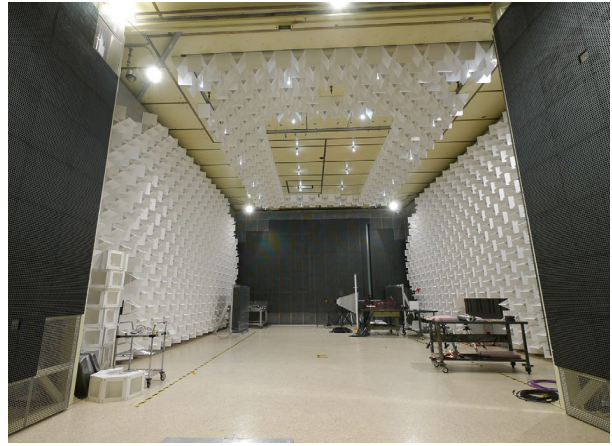

**Figure C-73:** Large EMC Shielded Enclosure

### C.9.4 Origins Observatory I&T
This has been discussed in Section 2. Added here is a detailed flow chart, Figure C-75.

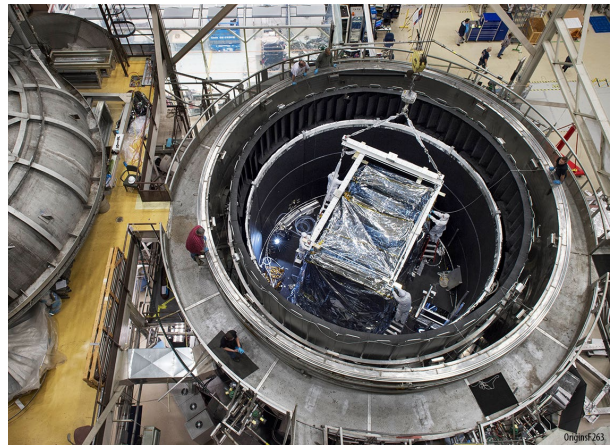

**Figure C-74:** SES Chamber

### C.9.5 Origins Launch Site Operations
Prior to the arrival of the Observatory, the electrical ground support equipment (EGSE) are housed in trailers or equipment shelters that are shipped to the launch site. The EGSE are then tested to verify proper function and prepared to be utilized for final tests. The Observatory is shipped to the launch site in its launch configuration. Functional tests must still be performed to ensure that transportation and handling activities have not adversely affected the observatory. It is important that any problems be identified and corrected before the next level of integration. These standalone activities are performed in the dedicated off-pad processing facility at the launch site. Any additional integration or testing activities are also performed during this phase. These can involve required corrective actions resulting from post shipment test anomalies or tasks otherwise planned to be completed just prior to launch (*e.g.*, alignment, propulsion functional and leakage tests, etc.). *Origins* will follow typical launch site operations that are described in the flow shown in Figure C-76.

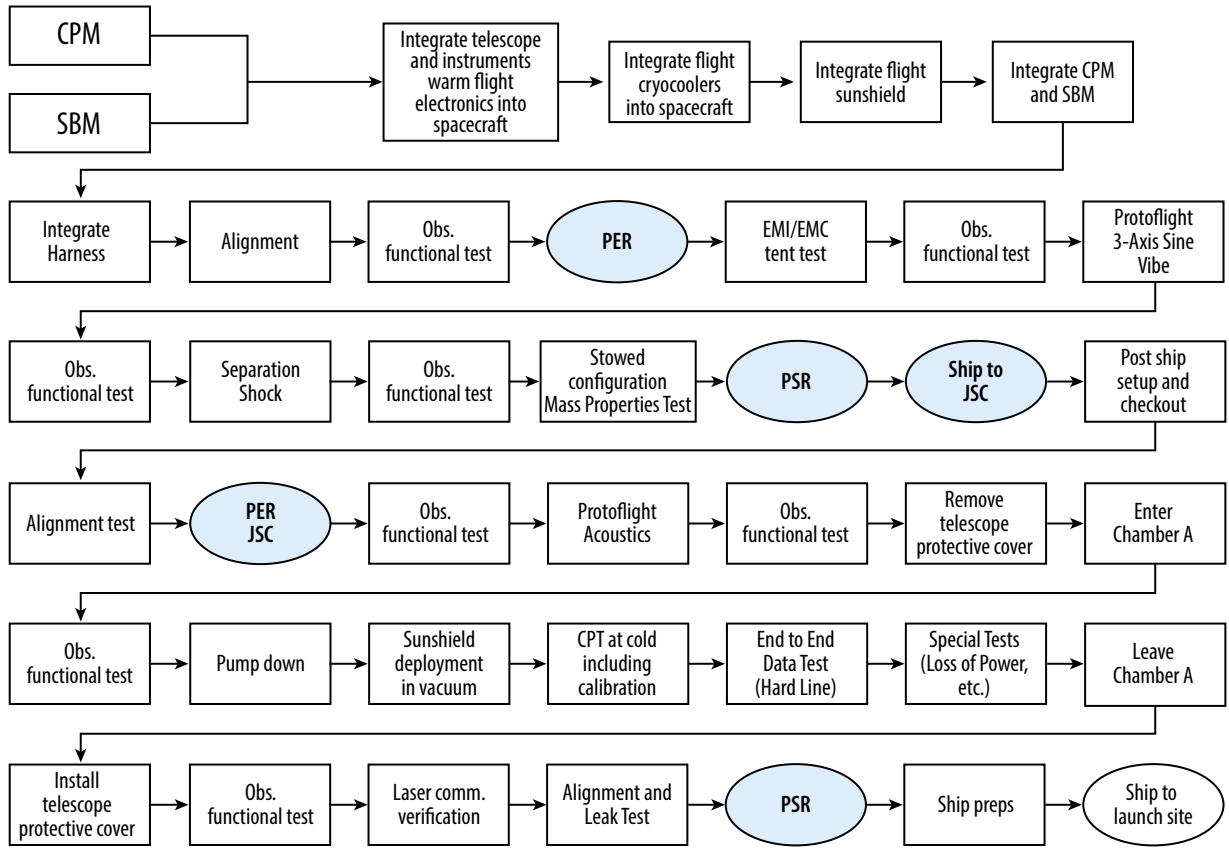

**Figure C-75:** *Observatory I&T Summary Flow.*

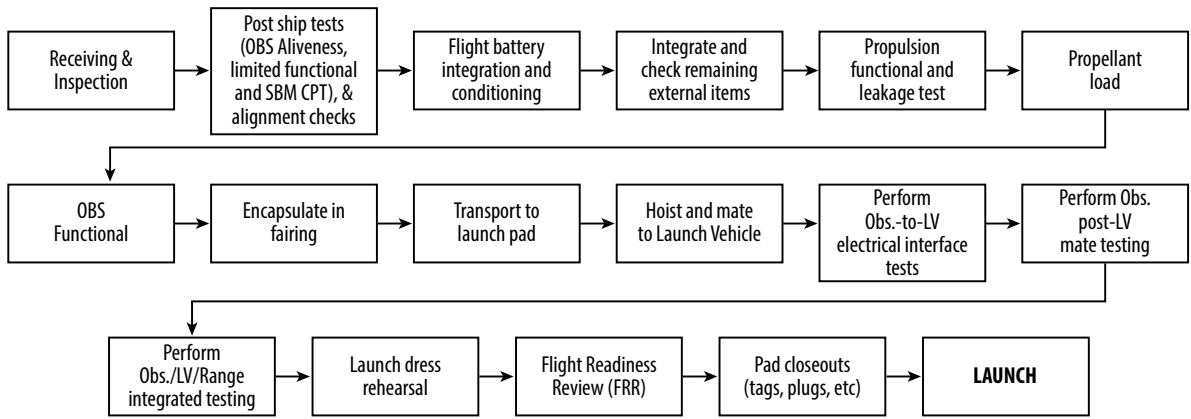

**Figure C-76:** *Origins* Top Level Flow for Launch Site Operations

## C.9.6 Other Notable Facilities

Aside from Chamber A at JSC and SES at GSFC, Marshall Space Flight Center (MSFC) has a large thermal vacuum chamber called the X-Ray Calibration Facility (XRCF) (Table C-18) capable of reaching temperatures below 20 K.

At the Glenn Research Center (GRC) the Space Power Facility (SPF) is a huge vacuum chamber. The SPF is able to simulate a spacecraft's launch environment, as well as in-space environments. NASA has developed these capabilities under one roof to optimize testing of spaceflight hardware while minimizing transportation issues.

**Table C-18:** The XRCF facility has many capabilities that could be utilized for cryogenic mirror testing.

| | |
|---|---|
| **Special Capabilities:**<br>• X-ray calibration and performance testing of large, grazing-incidence x-ray optics, detectors, and telescopes<br>• Cryogenic (< -400F) optical metrology of large direct-incidence optics and structures (± 1 μradian line of sight stability)<br>**Vacuum Volumes:**<br>• 20ft dia x 60ft horizontal cylindrical test volume<br>• 5ft dia x 1728ft illumination beam line<br>**Vacuum Levels:**<br>• $10^{-8}$ Torr via cryogenic and turbomolecular pumps<br>• $10^{-3}$ Torr helium partial pressure to augment heat transfer<br>**Thermal Environments:**<br>• Entire test volume from -200F to +200F (55 zones)<br>• Two enclosures from -424F to +120F<br>   » 10ft x 8ft x 30ft<br>   » 31ft x 16.5ft dia<br>**Clean Rooms:**<br>• 6000 sq ft Class 1,000 (ISO 6)<br>• 2000 sq ft Class 10,000 (ISO 7) | **Helium Refrigeration:**<br>• Two closed-loop gaseous helium expansion cycle refrigerators each capable of 1kW at 20K<br>**X-ray Generators:**<br>• Energy range from .01 to 10 keV<br>• Other ranges available at MSFC<br>**Optical Metrology:**<br>• Leica laser Absolute Distance Meter<br>• 4D PhaseCam, AOA Wavescope, IPI<br>**Optical View Port:**<br>• 10.8 inch dia clear aperture N-BK-7 tiltable window<br>**Metrology Alignment Stages**<br>• Physik Instrumente Hexapod precision motion stage<br>• Aerotech XYZ translation stage<br>**Test Article Alignment Stages:**<br>• 6 Degree of Freedom Stage - cryogenic and vacuum capable<br>• Five Axis Mount - vacuum capable |

## C.10 Orbit Details

Different types of spacecraft orbits can be achieved at the Sun-Earth Libration Points. These orbits can vary from very large amplitude Halos, similar to the one used by JWST, to smaller Quasi-Halos, similar to that used by Wide Field Infrared Survey Telescope (WFIRST), to even smaller Lissajous orbits, similar to the Deep Space Climate Observatory (DSCOVR) orbit. Figure C-77 shows a comparison of JWST and DSCOVR in the XY, XZ, and YZ Rotating Libration Point (RLP) frames. The RLP frame is commonly used when modeling the dynamical motion of the Circular Restricted Three Body Problem (CRTBP). In the case of *Origins*, the RLP frame is defined using a primary body (the Sun) and a secondary gravitating body (the Earth), which is less massive than the primary. The X-axis points from the Sun to the Earth. The Y-axis is orthogonal to the X-axis and lies in the plane of the Earth's motion about the Sun, pointing in that motion's direction. Finally, the Z-axis is orthogonal to the X and Y axes. The origin then lies at the location of the Libration Point of interest. In *Origins*'s case, the RLP origin is at SEL2 like JWST and WFIRST. In DSCOVR's case, the RLP origin is at Sun-Earth L1 (SEL1) since DSCOVR orbits about that Libration Point.

JWST's orbit is much larger in size compared to that of DSCOVR. While the orbits look similar (besides their relative size differences) in Figure C-77 A and B, and C and D, it is not until the orbits are viewed in the Figure C-77 E and F that the true difference between a Halo and Lissajous orbit becomes apparent. JWST orbits in a perfectly periodic Halo, whereas DSCOVR's Lissajous grows/expands for several years and then ultimately shrinks. Although these Halo and Lissajous orbits both have 6-month periods, the larger Halo orbits about the same crossing points, whereas the Lissajous evolves around them and repeats its cycle. This difference is important, because *Origins* utilizes a combination of a Quasi-Halo (a Halo that is not perfectly periodic) and a Lissajous orbit.

*Origins*'s requirements are to maintain an orbit about SEL2 for 5 years (carrying enough consumables for 10 years). To achieve this SEL2 orbit, the Launch Vehicle carrying *Origins* must be capable of providing a $C_3$ of -0.55 to -0.75 km²/sec² to achieve the correct energy on the outbound trajectory to reach SEL2.

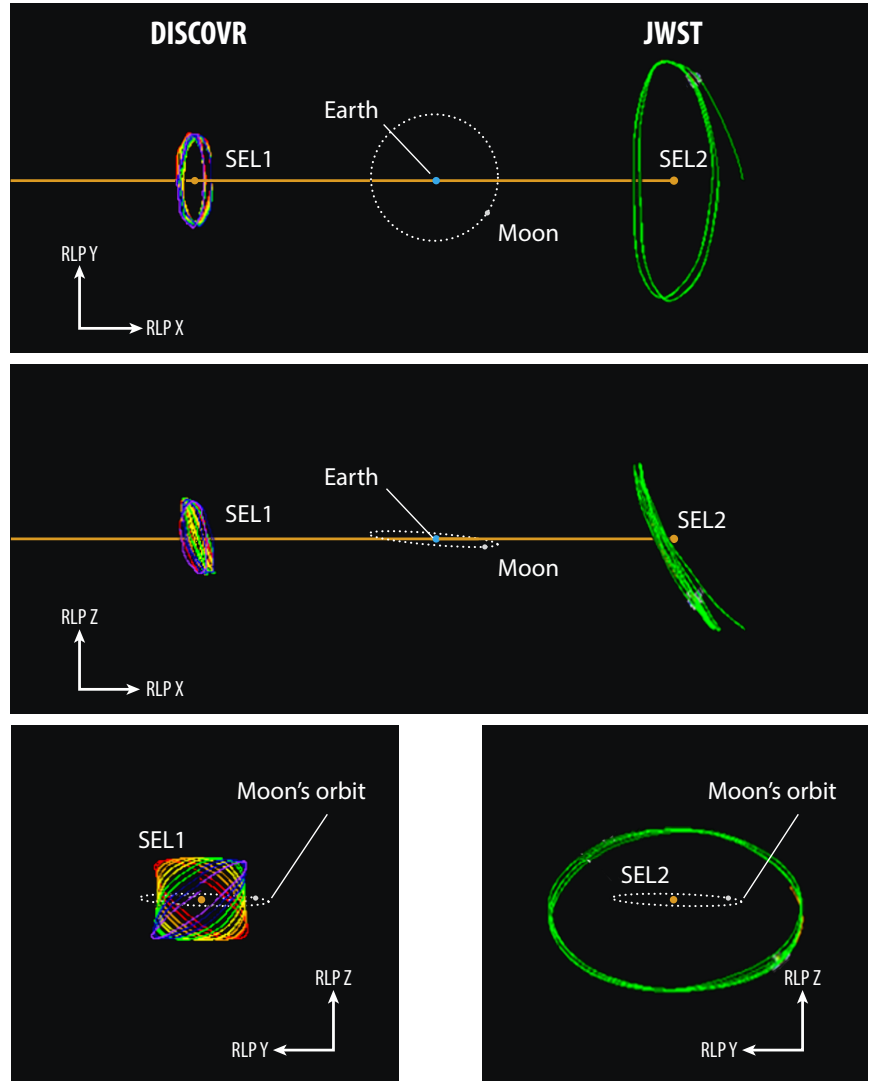

**Figure C-77:** JWST and DSCOVR orbits are shown in the XY, XZ, and YZ Rotating Libration Point Frame show examples of the various orbits (shown to scale) that can be achieved at the Libration Points.

The transfer trajectory from Earth to SEL2 and the mission orbit are shown in the RLP XY Frame with corresponding LEV and Moon Angles in Figure C-78.

For this design, the STK scenario began with a nominal Second Engine Cut-Off 1 (SECO1) state. This SECO1 state represents a typical state provided by the Launch Vehicle (assuming a launch from Kennedy Space Center) used in Flight Dynamics launch window analysis. SECO1 is shown in Figure C-79.

*Origins* will then coast in a Low-Earth Orbit (LEO) of 185 km for less than 10 minutes. After this "Coast Phase", the Launch Vehicle will impart a very large ΔV (3.23 km/sec) onto *Origins* to put it on the correct outbound trajectory, achieving the nominal Transfer to Insertion Point (TIP) state, to ensure *Origins* travels towards SEL₂ on its nominal path. This sequence is shown in Figure C-80.

In most launches, the Launch Vehicle executes a nearly-perfect TIP burn and the spacecraft only needs to execute a very small (cm/s) MCC maneuver 24-31 hours after launch. *Origins* carries enough ΔV to account for worst case 3s Launch Vehicle dispersions using an MCC maneuver.

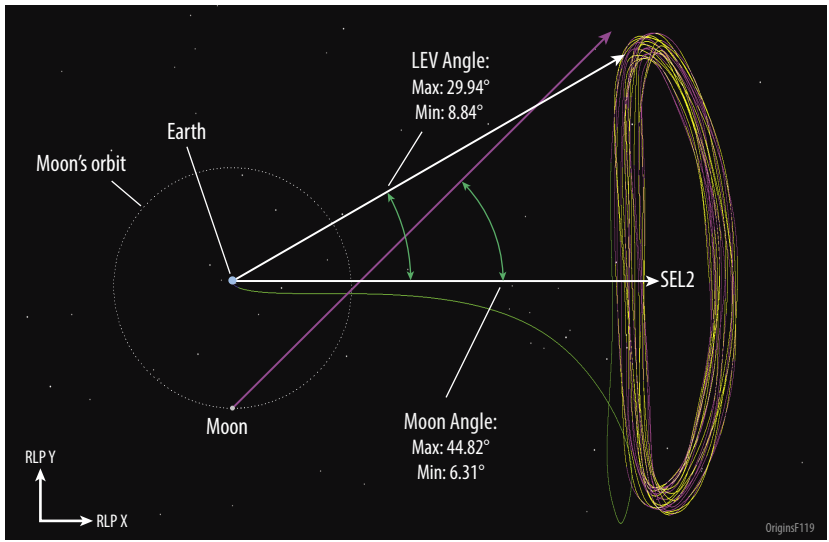

**Figure C-78:** *Origins'* transfer trajectory and mission orbit for 10 years shown in the RLP XY Frame with corresponding LEV and Moon Angles meets current requirements

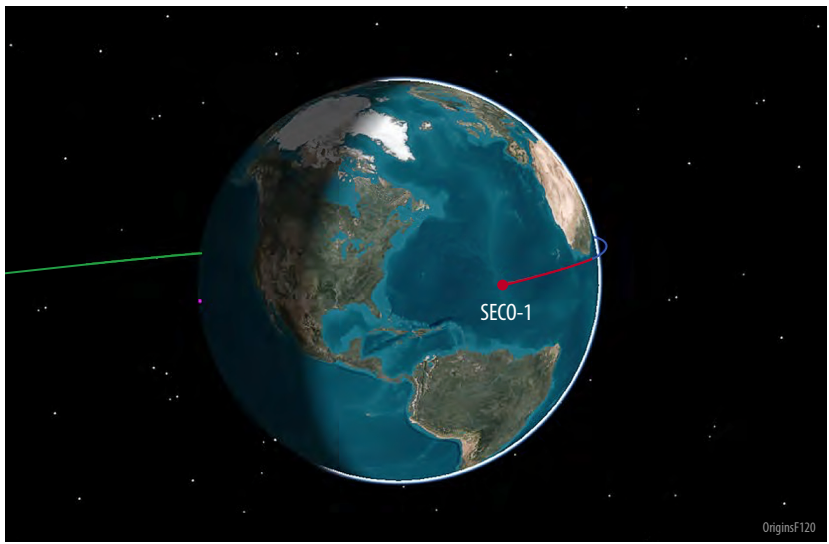

**Figure C-79:** *Origins* will be delivered to a SECO1 state for a notional 2035 launch.

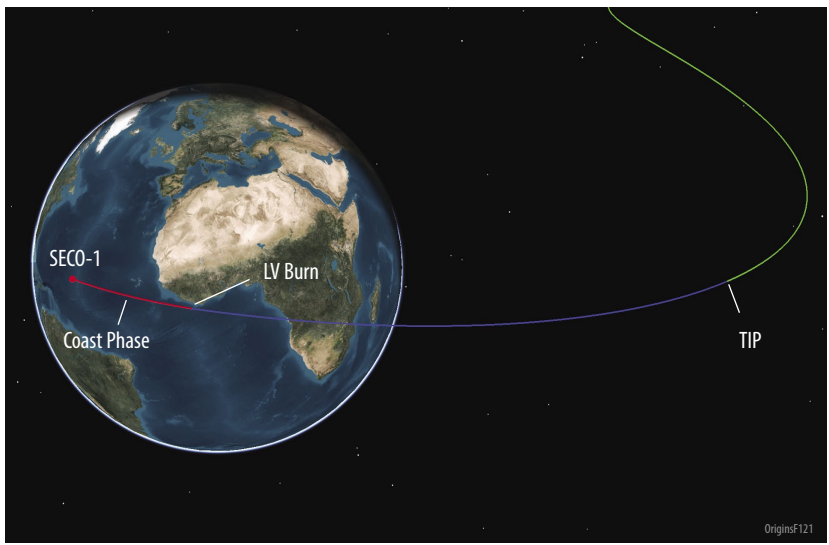

**Figure C-80:** *Origins* will then coast along its parking orbit until a state is reached where the Launch Vehicle will perform a large maneuver that will place *Origins* on the correct outbound trajectory to SEL2.

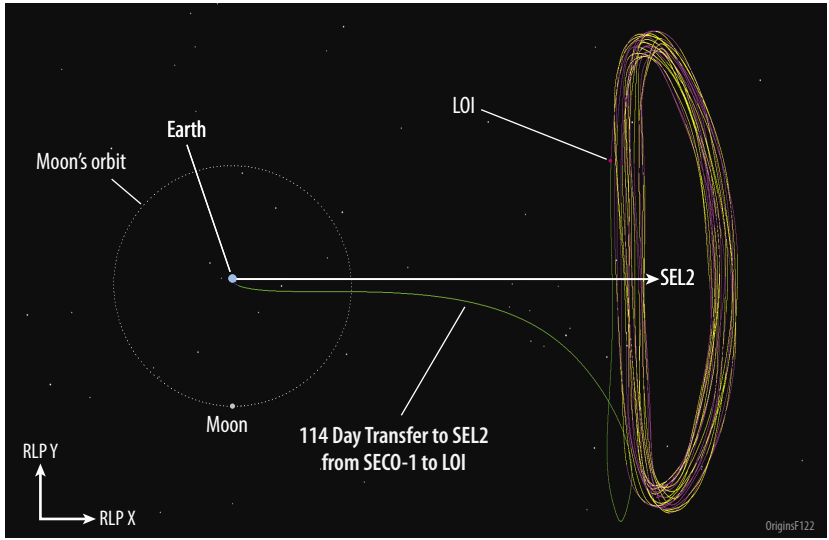

**Figure C-81:** *Origins* transfers to SEL2 over 114 days before executing a LOI maneuver to insert itself into the mission orbit.

After the MCC maneuver, *Origins* performs a small MCC2 maneuver, if required to correct any dispersions from MCC1. Afterward, *Origins* continues on its 114-day transfer to SEL2. After 114 days, *Origins* is within the vicinity of SEL2 and executes a Libration Orbit Insertion (LOI) maneuver to attain the notional 29° LEV mission orbit. The transfer to SEL2 and LOI maneuver are shown in Figure C-81.

Once *Origins* has completed its LOI maneuver, it begins routine station-keeping maneuvers every 21 days to maintain the mission orbit. The total ΔV is provided in Table C-19.

There were several assumptions used in this analysis. For simplicity, the current model assumed thrust could occur in any direction, and all maneuvers were modeled as impulsive. As the design furthers along, and the propulsion system is finalized, the maneuvers should be modeled as finite and the thruster locations should be modeled to account for any thruster pointing errors. There will be a ΔV penalty once these models are implemented. Also, momentum unloads were assumed to occur once a day with a 13.33 mm/sec residual ΔV each time. These momentum unloads were not included in the current orbit determination analysis, but were accounted for in the orbit maintenance ΔV.

**Table C-19:** Orbital Maneuvers and ΔV requirements

| Maneuver | ΔV (m/s) | Mission Time | Notes |
|---|---|---|---|
| MCC1 | 30 | L +24h | Dependent on LV performance reliability |
| MCC2 | 10 | MCC1 + 10 days | Clean up for MCC1 efficiency, if needed |
| LOI | 10 | L + 114 days | Insertion into 4x29° LEV mission orbit |
| Orbit Maintenance | 45 | ~every 3 weeks after LOI | ΔV for 10 years ~3.5 m/s for stationkeeping and 1 m/s for Momentum Unloading per year |
| Additional Momentum Unloading and Shadow Avoidance Maneuvers | 15 | Between orbit-maintenance maneuvers, as needed | Origins needs to perform routine Momentum Unloading maneuvers, which impose a residual ΔV onto the orbit. This can destabilize the orbit between orbit-maintenance maneuvers. Also, if a moon shadow were to be seen in a predicted ephemeris (at least 6 months out) a shadow avoidance maneuver can be performed to raise the amplitude of Origins's orbit so it avoids the shadow. This 15 m/s is accounting for 5 m/s for shadow avoidance and 10 m/s (1 m/s per year for 10 years) for Momentum Unloading disturbances. |
| End-of-Life (EOL) Maneuver | 2 | EOL | Small maneuver to place Origins in Heliocentric orbit (ensures the observatory won't fall back toward Earth) |
| Total | 112 | | Very conservative, needs to be re-evaluated with full thruster models (pointing and efficiency) in simulation |

## APPENDIX D - UPSCOPES

During *Origins'* Concept 2 study, the team designed four highly-capable instruments: OSS (Section 3.1), MISC (MISC-T, Section 3.2), FIP (Section 3.3) and Heterodyne Receiver for *Origins* (HERO). The detailed studies of these instruments included layout and component selection, optical ray trace, mechanics design, MELs and estimations of instrument cost. All four instruments were technologically feasible and offered compelling science. However, they exceeded the cost target even when assuming some international contributions. Thus, the team needed to conduct a descope exercise to trim the total-mission-cost-to-NASA to the target amount. Both OSS and FIP reduced their total pixels in half. In addition, FIP originally had two parallel channels and four colors: 50/250 and 100/500 μm. For a FIP image, one would get two colors simultaneously, *e.g.* 250 and 500 μm. The spacing of the wavelengths 50, 100, 250 and 500 μm was chosen to sample the SED of an object. The 100/500 μm parallel channel of FIP was descoped from FIP leaving just the one channel in which either 50 or 250 μm could be chosen.

This upscopes appendix provides instrument descriptions for the HERO and Mid-Infrared Spectrometer and Camera (MISC) wide field imager (WFI) with associated science cases. Adding the 100/500 μm parallel channel back into FIP is a possible upscope option, but the details of this revised instrument are not described here. The science traceability matrix (STM) for both upscope instruments is shown in Table D-1. This traceability matrix shows some of the key science drivers and the resulting requirements. However, this upscopes STM is not as well developed as the baseline concept STM. During a phase A study of *Origins* with an upscope, predicted performance and requirements will be refined and science impact of failing to meet a requirement will be analyzed.

Figure D-1 shows the ray trace of the upscoped *Origins* design. As compared to the baseline design, this includes the addition of the MISC imager channel, the HERO instrument, and increased FOVs for both FIP and OSS. The optical performance for the fully upscoped *Origins* design is shown in Figure D-2 (RMS wavefront error). In each case the design meets the optical performance specification (< 0.07λ for RMS wavefront error and > 0.8 for Strehl ratio) across each instruments' FOV. Figure D-3 shows the expected spectral resolving power and sensitivity for MISC-WFI and HERO in the context of the *Origins* baseline instrument performance. The estimated cost, shown as percentage of the baseline mission cost, for HERO and MISC-WFI is shown in Table D-2.

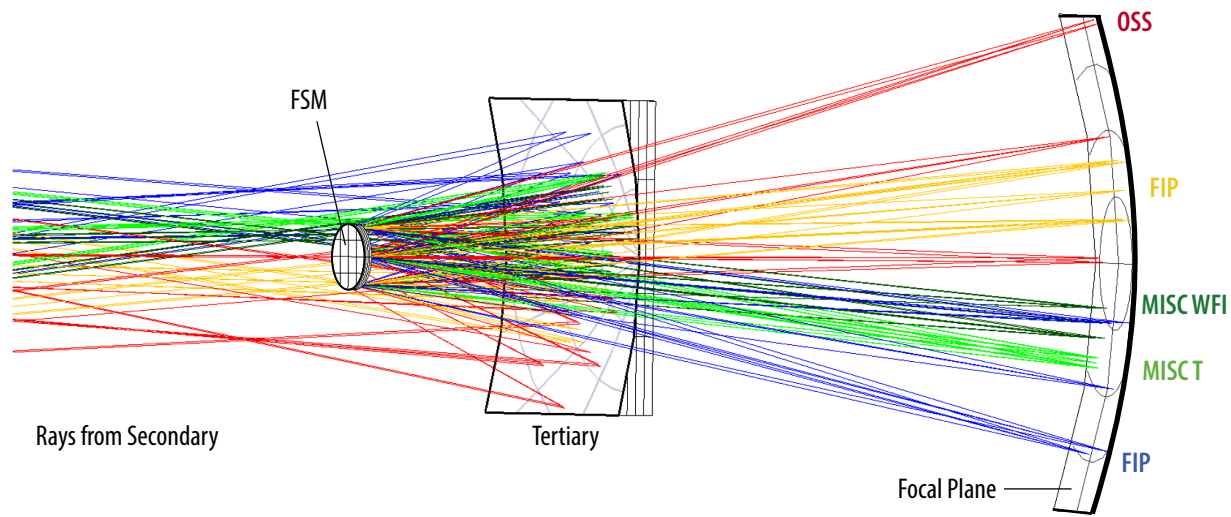

**Figure D-1:** CODEV® layout showing rays for each of the upscoped instruments passing through the telescope onto the image surface. As compared to the baseline design, this includes the addition of the MISC imager channel, the HERO instrument, and increased FOVs for both FIP and OSS.

**Table D-1:** *Origins* Science Traceability Matrix: Discovery Space Sciences

| NASA Science Goals | *Origins* Science Goal/ Question | Science Objectives (Wide ranging. We highlight examples.) | Science Requirements | | Instrument Requirements | | | | *Origins* Mission Requirements | |
|---|---|---|---|---|---|---|---|---|---|---|
| | | | Science Observable | Measurement Requirement | Parameter | Technical Requirement | Ins | CBE Performance | Driver | Parameter |
| How does the Universe work?  How did we get here?  Are we alone? | Ability to perform a wide range of discovery science programs led by the astronomical community, selected through peer-review. | Case #1: Follow-up photometrically selected $5 < z < 10$ galaxy samples from WFIRST deep imaging surveys to measure star-formation, stellar mass, AGN, and other properties using rest-frame optical emission lines. | Rest-frame Hα, Paschen-α, PAH 3.3-μm as tracers of SFR; near-IR continuum for stellar mass | Spectroscopic imaging in the mid-IR covering surveys comparable to WFIRST-deep fields.  Sufficient spectral resolving power to separate Hα from [NII]. | Wavelengths | 5–25 μm | MISC- Wide Field Spectroscopy | 2.8–28 μm | Point source sensitivity, stability and systematic error control | • To enable access to all targets of interest, **the field of regard shall be 4π sr** over the course of the mission. • To meet objectives of Case #1, **add upscope optional MISC wide-field imager and spectroscopy instrument to the instrument suite of *Origins*.** • To meet objectives of Case #3, **add upscope optional heterodyne instrument (HERO) to the instrument suite of *Origins*** |
| | | | | | Spectral resolving power R=λ/Δλ | 250 | | 300 | | |
| | | | | | Spectral line sensitivity | $1 \times 10^{-20}$ W m⁻² at 12 μm to detect Hα from a typical galaxy. (1 hr; 5σ) | | $4 \times 10^{-21}$ W m⁻² at 12 μm (1 hr; 5σ) | | |
| | | Case #3: Measure the water mass at all evolutionary stages of star and planet formation and across the range of stellar masses, tracing water vapor and ice at all temperatures between 10 and 1000 K. (Section B.2; An upscope science case to Theme-2 Objective #1) | Measure water ($H_2O$) content of starless cores in diverse environments. High spectral resolution follow up of young disks in proto-stars to determine the distribution of water within the disks | Line flux density of ortho $H_2O$ 538 μm, para $H_2O$ 269 μm, and other low energy transitions of water and its isotopes to sensitivity limit $1 \times 10^{-20}$ W m⁻² (3 mK) per 0.3 km s⁻¹ velocity channel (5σ) to obtain spectrally resolved line profiles of 10 known starless cores and young disks in 100 h with spatial resolution ≤ 0.1 pc | Wavelength range | 180μm to 550 μm | HERO (Up-scope Instrument) | 111–617 μm | | |
| | | | | | Spectral resolving power R=λ/Δλ | ≥10⁶ for velocity resolution | | up to 10⁷ | | |
| | | | | | Angular resolution | ≤ 25″ at 538 μm to resolve starless cores | | 23″ at 538 μm | | |
| | | | | | Field of view | 1′ x 1′ to map starless cores | | 2′ x 2′ | | |
| | | | | | Spectral line sensitivity | $1 \times 10^{-20}$ W m⁻² per velocity channel (3 mK), 1 h (5σ) | | $6 \times 10^{-21}$ W m⁻² per velocity channel at 538 μm, 1 h (5σ) | | |

**Table D-2:** upscope options from *Origins'* Baseline Mission, their science impact, and cost as percentage of total baseline mission cost.

| upscope Option | Science Impact | Cost Increase as a Percentage (%) of the total mission cost |
|---|---|---|
| HEterodyne Receiver for *Origins* | Enhances Theme 2 (Water and Habitability), Objective 1 (water mass in all evolutionary stages) by improving Doppler tomographic measurements of water in planet-forming disks and measuring water in protostars.  Enables new science, potentially including long baselines for an Event Horizon Telescope system. | 15% |
| MISC-Imager channel | Enhances Theme 1 (Extragalactic), Objective 1 & 2 (SFR, BHAR, metals) by providing imaging survey capability in the mid-infrared and spectroscopy that accesses optical line metallicity and star formation indicators at high redshift. | 14% |

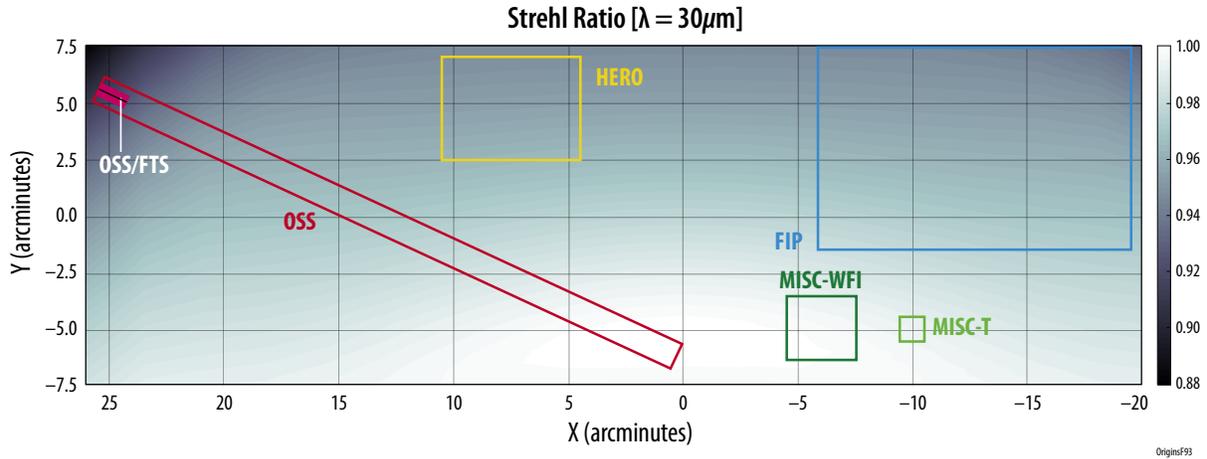

**Figure D-2:** An evaluation of the upscoped *Origins'* RMS wavefront error as a function of FOV shows that the performance requirement (less than 0.07λ) is met across each instruments FOV. Note that the colorbar is set such that light corresponds to better optical performance and dark to poorer optical performance.

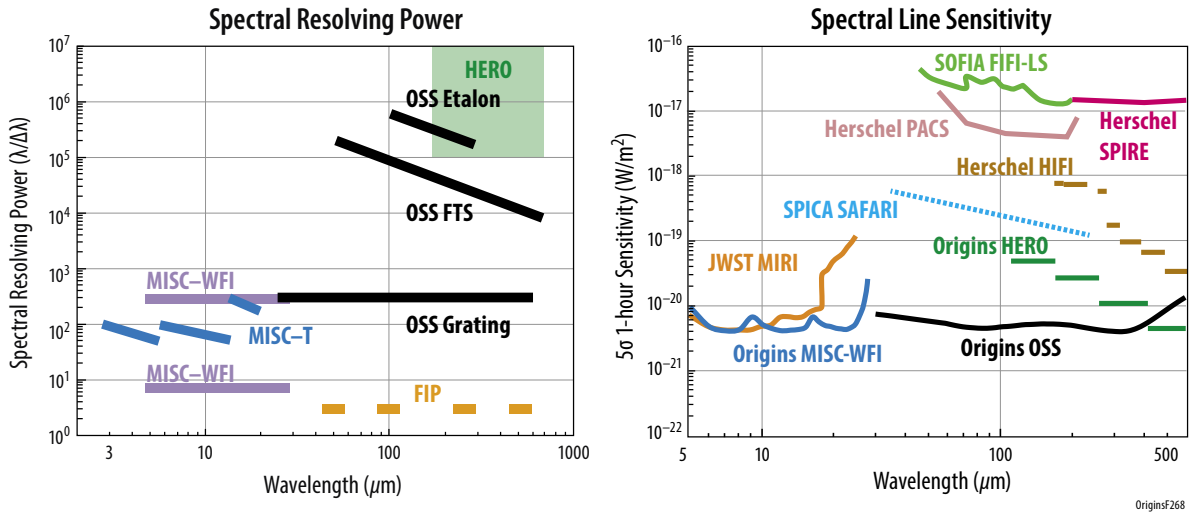

**Figure D-3:** *Origins* upscope option ideally complements the *Origins* baseline configuration by adding the capability of very high spectral resolution observations at high line sensitivity (HERO), by adding an mid-IR MISC imager , adding the 100 and 250 micron windows of FIPS and by increasing the mapping speed of OSS.

## D.1 Heterodyne Receiver for Origins

The Heterodyne Receiver for *Origins* (HERO) complements and extends the baseline instrument suite of *Origins* by adding a 3rd dimension to the observations through line tomography and by allowing detailed kinematic studies. HERO is largely intended for follow-up observations of OSS, going beyond statistical studies and allowing zooming into individual objects to study the physical and chemical processes within them. HERO consists of sensitive, heterodyne focal plane arrays covering a wide wavelength range between 617 and 111 microns and allowing dual-polarization and dual-frequency observations (see Table D-2).

**Table D-3:** HERO Design Parameters

| Band | λmin μm | λmax μm | F min GHz | F max GHz | Pixels | Trx K | Beam " | $T_{rms}$[a] mK | Line Flux[b] Wm$^{-2}$ | Spectral resolv.power |
|------|---------|---------|-----------|-----------|--------|-------|--------|-----------------|------------------------|-----------------------|
| 1 | 617 | 397 | 486 | 756 | 2x9 | 50 | 20.3 | 2.6 | $6.4 \times 10^{-21}$ | up to $10^7$ |
| 2 | 397 | 252 | 756 | 1188 | 2x9 | 100 | 12.9 | 4.2 | $1.6 \times 10^{-20}$ | up to $10^7$ |
| 3 | 252 | 168 | 1188 | 1782 | 2x9 | 200 | 8.5 | 6.8 | $4.0 \times 10^{-20}$ | up to $10^7$ |
| 4 | 168 | 111 | 1782 | 2700 | 2x9 | 300 | 5.6 | 8.4 | $7.3 \times 10^{-20}$ | up to $10^7$ |

[a] Receiver noise for 1h integration at $10^6$ resolution (0.3 km/s) using one polarization.
[b] Detectable line flux at 5σ, for 1h pointed integration (on+off source) in two polarization, with the 5.9 m primary mirror of *Origins*.

### D.1.1 HERO Science Drivers

#### D.1.1.1 Early stages of the Water Trail

The *Origins* baseline concept proposes three extremely powerful instruments: OSS, FIP, and MISC. However, these instruments are incapable of fully investigating the trail of water. With its heterodyne receiver, *Origins* in its upscope configuration will play a critical role in tracing the early path of water from the ISM into young circumstellar disks through its unique access to the lowest-energy rotational transitions of the water molecule and its isotopologues ($H_2^{18}O$, $H_2^{17}O$, HDO) at high spectral resolving power (up to $10^7$), and in synergy with JWST for tracing water ice through its infrared and far infrared bands. With the high sensitivity provided by its large, 5.9-m telescope and HERO's extremely high spectral resolution capabilities, *Origins* will be a transformational tool for following the path of water in the ISM.

The trail of water begins either on dust grains with O+H followed by OH+H to form $H_2O$ that can be released into the gas phase either through FUV photodesorption or thermal heating of the grain or, alternatively, in molecular gas, where water vapor forms in chemical reactions involving molecular hydrogen and different reactants. In warm (shocked) gas, the main water formation route involves atomic oxygen and the hydroxyl radical OH; in cold or diffuse regions, the water chemistry is dominated by a suite of hydrogen abstraction reactions starting from ionized oxygen O+ and leading to the molecular ions OH+, $H_2O+$, and finally the precursor of water vapor $H_3O+$, which produces water by recombining with electrons. Water vapor is destroyed by the ambient far-UV radiation field and can be removed from the gas by various chemical reactions and by freezing onto dust grains.

While some interstellar water is known to be present in diffuse molecular gas and UV-irradiated photodissociation regions, the bulk of water is found in dense molecular clouds as ice mantles on cold (T ~ 10 K) dust grains with tiny traces of water vapor (three orders of magnitude less abundant than water ice). The formation of water ice is an important step in the evolution of the dust, as water ice fosters a more efficient way of sticking grains together (Chokshi *et al.*, 1993; Gundlach & Blum, 2015) and a more efficient chemistry in the ice mantles (Boogert *et al.*, 2015). Water ice forms on grain surfaces by the addition of hydrogen atoms on adsorbed oxygen atoms and by direct freezing of water vapor. The balance between water vapor and water ice is governed by the competition between freezing and desorption induced by energetic radiation (UV and cosmic rays) that can trigger the release of water molecules from the ice. Via a host of observations (*e.g.*, Whittet *et al.*, 1983; Öberg *et al.*, 2011; Boogert *et al.*, 2015), it is now known that the water ice mantle first forms in pre-stellar cores. This is the water that is provided to the young disk and sets the stage for all that follows.

State-of-the-art chemo-dynamical models of the prestellar core evolution that include water ice and cosmic ray-induced production of water vapor predict that, overall, for a typical prestellar core of 1 solar mass ($2 \times 10^{33}$ g), the total mass of water vapor can be estimated to lie between 20 and $2 \times 10^3$ Earth ocean mass, while the total mass including water ice would be up to a few millions Earth ocean mass. The spread in these estimates is not only due to the individual variation between cores related to their environment, but also to the very limited knowledge about this part of the water trail. Our understanding of the physical processes controlling the water abundance has not been quantitatively tested because of the absence of high-quality observations; hence, the relative fractions of water vapor and ice in the inner regions of cores remain very poorly known (Keto *et al.*, 2014; Schmalzl *et al.*, 2014).

While *Herschel* has brought a decisive confirmation of the water formation pathways in diffuse UV irradiated gas and in shocks of all kinds, the sensitivity and imaging capabilities of its heterodyne receiver, Heterodyne Instrument for the Far-Infrared (HIFI), operating with high resolving power (R=$10^7$) had just begun the exploration of water in cold prestellar cores, the birth places of new stellar and planetary systems. The only published detection of water vapor in a starless core (Caselli *et al.*, 2012) shows that water vapor is not completely frozen despite the very low (6K) temperature reached there. There are two exciting aspects to this detection: (1) it highlights the role of cosmic rays in water chemistry and shows that cosmic ray-induced water desorption, which also operates in the disk mid plane, must be finely tuned regarding the efficiency of water vapor release from the ice and the cosmic ray propagation to obtain the amount of detected water vapor; (2) the velocity profile of the 110-101 ground state line of ortho-water at 538 μm (557 GHz) (Figure D-4) presents an inverse P-Cygni profile characteristic of infalling gas, providing definite evidence that this core is gravitationally collapsing and will soon form a new star. A static core would show a pure absorption spectrum. The infall speed can be deduced from the difference of the velocities between maximum emission and absorption, while detailed information about the radial variations of the water fractional abundance along the line of sight can be extracted from the overall line profile using a sophisticated radiative transfer code.

Overall, the water vapor content of this core amounts to about $2.2 \times 10^3$ Earth ocean mass (the mass of the Earth's oceans is $1.35 \times 10^{24}$ g), while the ice content would be two thousand times larger. This observation illustrates the unique information carried by the water ground state lines, probing the total water content, its radial distribution, and dynamics. Because of their sensitivity to hydrogen densities higher than $10^7$ cm$^{-3}$, the ground state water line profile uniquely reveals the dynamics of the inner regions of the collapsing core, enabling a clear and accurate test of star formation theories and an accurate measurement of the amount of water that is delivered to the forming protostar/protoplanetary system (Keto, Rawlings, & Caselli, 2014, 2015).

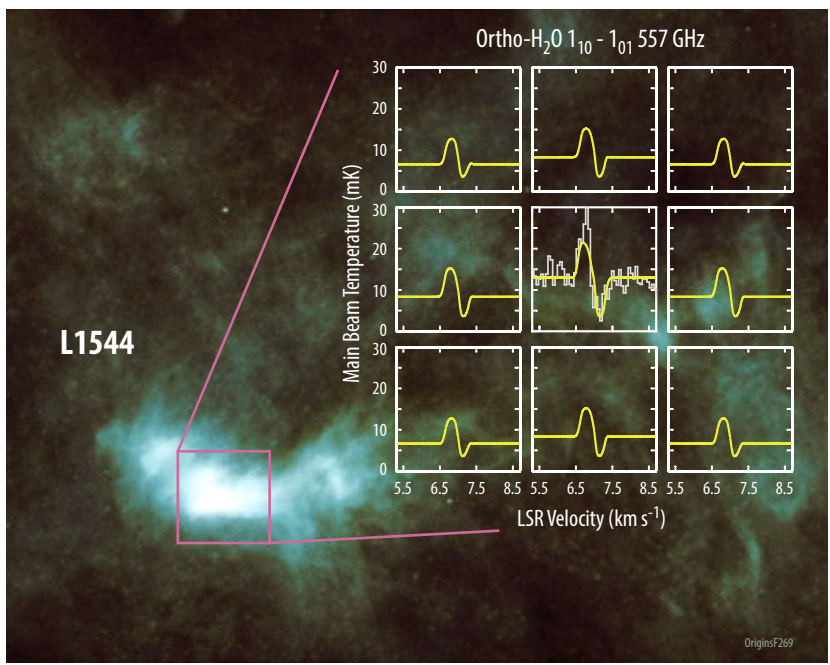

**Figure D-4:** High spectral resolution observations reveal complex line shapes of the ground state water line that allow us to determine the dynamics of the inner cores. The L1544 prestellar core showing the dust continuum emission. The insert is the modeled line profiles of the H2O $1_{10}$–$1_{01}$ transition at 538 μm (557 GHz) in the 25″ FWHM beam of Origins using the MOLLIE radiative transfer code (Keto et al., 2014). Each panel is separated by 25″. The central panel includes the HIFI spectrum with a 40″ beam. Note the higher continuum level and deeper absorption in the central panel, which shows the forming core.

However, it is not possible to draw general conclusions about the water trail from a single object. L1544 is the best case studied by *Herschel*. Only a few other cores have been detected, but not fully modeled. It is currently unknown if this core is representative in its water content, vapor, and ice relative abundances and infall speed. Because *Herschel* observations have only probed the ortho spin symmetry state, it is not known whether a larger quantity of the more stable spin symmetry state (para water) is present, which could significantly increase the total water content, potentially by an order of magnitude. This is crucial to understanding the chemistry of water's beginning.

Besides the difficulty in obtaining sensitive high spectral resolution spectra of very faint signals, which explains the very small number of detections, *Herschel* used single pixel receivers with a limited angular resolution of 40" FWHM at 538 µm, leading to difficulties and ambiguities in the interpretation of the water line profiles. As illustrated in Figure D-4, radiative transfer models of collapsing cores predict that the water line profile changes across the spatial extent of the core getting narrower and more dominated by the blue-shifted emission peak away from the core center. Such spatial information is necessary to reconstruct with a high confidence level the spatial variation of the infall speed and determine the accretion rate of water on the future protostar and its protoplanetary disk. This dynamical information can only be provided by velocity resolved profiles of gas-phase water ground state lines. The ice content can be estimated from chemical-dynamical models knowing the water vapor content. Water ice can also be directly detected in the near infrared by JWST and in the far infrared with *Origins*/OSS. The main gas phase precursors, OH and $H_3O+$, will be accurately measured with HERO, leading to a complete account of the chemical network of water. With its upgraded instrument suit *Origins* can fully determine the origin and distribution of water and its precursor molecules and is uniquely suited to perform a deep study of a core sample at different stages of evolution and in different star-forming environments.

*Origins* will provide unparalleled information about how common water is and how it originated. Water is the most abundant ice. JWST can study line-of-sight water-ice in a couple near-infrared absorption features (*e.g.*, at 3.05 microns), but can't study extended water-ice in emission, which requires observations at 43, 47, and 63 microns. OSS observations of the ice and HERO's water vapor observations at high spectral resolution are a powerful tool of *Origins*. ALMA has great sensitivity but even ALMA has difficulty with pre-stellar cores because of their large size. Thus ALMA is not a competitive facility for this science case. Other single dish facilities have made significant contributions and have characterized several species in different spatial regimes. However, water, as the most abundant and best characterized ice, provides the central background for any interpretation. The water vapor above the ice is proportional to the ice reservoir and therefore traces it directly. Thus observations of the water vapor lines are crucial in order to understand the chemistry that precedes stellar birth. Furthermore, these lines can only be detected and characterized with the spectral resolution offered by HERO.

The HERO instrument has been designed with a spatial footprint of 3x3 pixels in two polarizations covering a FOV of 2'x 2' at 538 µm that perfectly matches the sizes of prestellar cores. The optic design allows simultaneous observations of o-$H_2O$ at 538 µm and p-$H_2O$ at 269 µm ground state lines, for an efficient and complete characterization of the gas-phase water. State of the art radiative transfer calculations based on chemical-dynamical models and with the minimal assumption of equal abundances for ortho- and para-$H_2O$, predict that the p-$H_2O$ line is three times fainter than the o-$H_2O$ transition. *Origins* will therefore be able to detect both lines down to the predicted minimum water content of a few tens of Earth ocean masses, and then establish for the first time a complete and accurate census of the gas phase water content in cold prestellar cores. OSS will provide the ice content.

Once the amount of water delivered to the forming disks is determined, the next step along the water path is the evolution of the bulk gas and water content of protoplanetary disks. OSS will provide a nearly complete view of the mass and condition of water throughout most regions of typical disks. The lowest-lying water line available to OSS at high spectral resolving power is the 179.5 micron line of ortho-water with ($\Delta E/k = 80K$). However, even this line may struggle detecting the coldest water at temperatures as low as 10 K. This very cold water is sequestered in the coldest outer regions, and deep midplanes of large, massive disks, and is best observed and characterized using the 538 micron ground-state line of ortho-water. For the coldest water, the high spectral resolving power at 538 micron, as offered by HERO, will be used on well-studied, nearby disks from the OSS survey to accurately measure the profile of the 538 micron water line thus allowing a determination of the radial distribution of gas-phase water within disks. As discussed for CO lines by Yu *et al.* (2017) information on the underlying radial distribution of the molecular carrier as well as the thermal and dynamical (turbulence) structures can be inferred from a detailed analysis of the line profile. The thermal and dynamical structures control the vapor/ice equilibrium and the transport of volatiles in the outer disk (>50 AU), which can be accurately determined over a representative sample of protoplanetary disks.

HERO can zoom into the disk with line-tomography and significantly contribute to the disk case, because:

- The high spectral resolving power of 30 m/s allows much more detailed line tomography than OSS, especially of the coldest gas in large disks, which will have Keplerian velocities as low as a 1-2 km/s.
- The high spectral resolving power is available over the whole frequency range of HERO, including the lowest-lying ortho-water line at 538 microns.
- The 538 micron o-$H_2O$ traces the coldest gas ($\Delta E/k = 27K$), whereas the lines targeted by OSS trace slightly warmer water ($\Delta E/k > 80K$).
- The 538 micron o-$H_2O$ line (and its comparable isotopologues) is significantly less effected by dust attenuation than the 179.5 micron line, giving access to deeper layers, even in massive disks.
- A single setting of HERO observes both the ortho-line at 538 micron, and the para-line at 269 micron, and includes the entire line with at least 1000 resolution elements without the need for scanning. OSS provides simultaneous spectral coverage as well.

### D.1.1.2 Water in our Solar System - Final Stages of the Trail of Water

High spectral resolution observations of water in our solar system allows us to understand the end of the water trail and indicate possible environments for the emergence of life. Understanding in detail how water was transported to Earth and where it is found in our solar system allows us to apply that knowledge to other planetary systems and to direct the search for life.

Comets are one possible origin of water on Earth, where the D/H ratio gives a clue to whether comets delivered the water to Earth. Complementary to the large OSS comet survey, HERO carries out follow up observations of the brighter comets at very high spectral resolution (100 m/s) in order to determine the origin of the outgassing (nucleus versus dust around nucleus) for different gases (by analyzing the line profiles), to get a refined D/H ratio, to determine the excitation mechanisms of the gases, and to determine the gas coma structure. HERO contributes to questions about the origin of the solar system by providing isotopic ratios (*e.g.* D/H, $^{16}O/^{17}O$ and $^{16}O/^{18}O$) for a large number of comets and linking the age of these objects to other primitive bodies (*e.g.* FUN-CAIs - Calcium-aluminum-rich inclusions with isotopic mass fractionation).

Water does not only exist on Earth, but also on other planets, such as Mars, which may have supported life. HERO observations constrain the water cycle, hydrogen/oxygen chemistry and origin of the Martian atmosphere by very sensitive and highly spectrally resolved observations of molecules and their isotopologues (*e.g.* $H_2O$, $H_2O_2$, OH, O, $O_2$, $O_3$) (the retrieval of their vertical distribution requires very high spectral resolution) and provide upper limits for a large number of molecules so far not detected (*e.g.* HCl). HERO constrains the origin of water in the stratospheres of the gas and ice giants, determines the D/H ratio in hydrogen and water (constrains isotopic equilibration processes and ice mass fraction on the surface of ice giants) and isotopic ratios of at least C, S and O with high precision. Moons of these planets also contain water. HERO is capable of monitoring the composition, physical conditions and variability of the Enceladus torus, the atmospheres of the Galilean satellites (including detection of plumes), Titan and the dwarf-planet Ceres (Figure D-5).

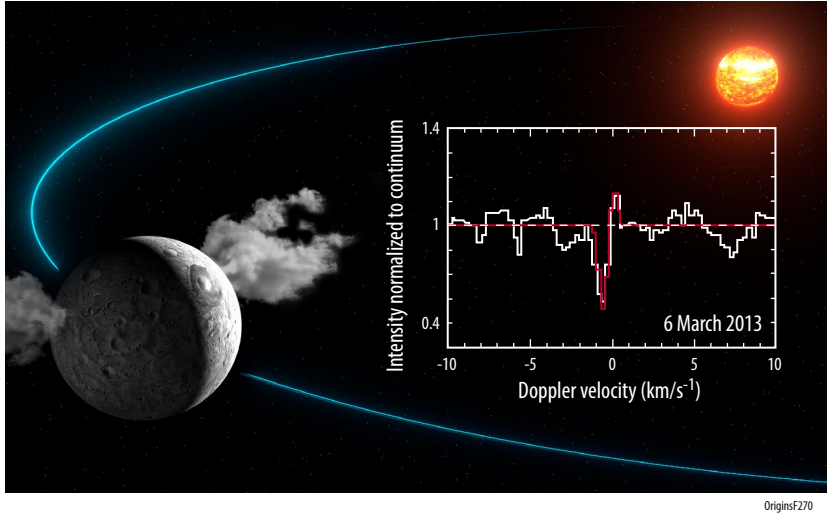

**Figure D-5:** Water is ubiquitous in our solar system. Detection of water vapour around the dwarf planet Ceres. The narrow 557 GHz $H_2O$ line was detected with HIFI in October 2012 and 2013 (Kueppers et al, 2014, Nature).

### D.1.1.3 Discovery Science

HERO was designed for the trail of water case, but it has large potential for discovery science. While the astronomical community proposes the science to be carried out, the *Origins* team gives two examples.

### D.1.1.3.1 Discovery Science: Determining Cosmic Ray Flux in the Milky Way and Nearby Galaxies

High-resolution absorption-line spectroscopy (with $\lambda/\Delta\lambda \geq 10^6$) of key interstellar molecules and molecular ions will allow the cosmic ray flux to be determined within the Milky Way and nearby galaxies. A 100-hour survey will obtain estimates of the cosmic ray ionization rate along ~40 sightlines across the Galaxy and will address key questions about the origin and distribution of cosmic rays.

**Scientific Importance:** Low-energy cosmic rays control the heating, ionization, and chemistry of dense molecular clouds. Therefore, cosmic rays set the initial conditions for star formation, while their production – *e.g.*, in supernova remnants (SNR) – is associated with stellar death (Grenier *et al.*, 2015). This program addresses the key questions: (1) What is the typical cosmic ray ionization rate as a function of Galactocentric distance? (2) How much does the cosmic ray ionization rate vary from one molecular cloud to another? (3) To what extent are cosmic rays excluded from dense molecular clouds? (4) What are the sources (*e.g.*, SNR) of low-energy cosmic rays? While high-energy (E > 280 MeV) cosmic-rays can be probed using gamma-ray observations, submillimeter observations from *Herschel* of molecular ions (OH+, $H_2O+$ & $H_3O+$) provide a unique probe of the low-energy cosmic rays that control the heating and ionization of star-forming molecular clouds thanks to the simple chemical processes forming them (Gerin *et al.* 2016, Neufeld & Wolfire, 2017). Whereas extinction in the Galactic disk severely limits near-IR measurements of the cosmic ray ionization rate probe $H_3+$, and the interpretation of millimeter molecular line emission for cosmic ray ionization rate determination requires ancillary data to remove ambiguities, submillimeter observations of hydride absorption can be performed toward background sources at large distances within the disk and are relatively straightforward to interpret. All absorption lines of $H_3O+$ and $H_2O$ are completely inaccessible from the ground, as are the strongest transitions of OH+ and H$_2$O+. SOFIA can observe some of the required transitions, but *Origins* is required to obtain all the necessary lines and assemble a meaningful sample.

**Proposed Observations:** High-resolution absorption-line spectroscopy of specific molecular transitions in the 0.5 – 2 THz spectral range is needed along sight-lines toward a large sample of submillimeter continuum sources along the Galactic plane and in nearby galaxies. Such measurements will provide the column densities of OH+, H$_2$O+, H$_2$O, and H$_3$O+ – all of which are produced via reaction sequences initiated by the cosmic ray ionization of H or H$_2$ – together with the column densities of CH and HF as probes of the H$_2$ column. In addition, *Origins*/HERO will determine the relative populations of several metastable states of H$_3$O+, providing additional constraints on the H$_3$O+ formation rate (Lis *et al.*, 2014), and thus the cosmic ray ionization rate. With these observations, *Origins*/HERO will perform the most comprehensive survey of the cosmic-ray density in the Milky Way and nearby galaxies.

### D.1.1.3.2 Discovery Science: The Birth Path of Dust

Generations of stars have created the enrichment of the interstellar medium of galaxies, which the "Rise of Metals" baseline science case observes. For a better understanding, it is desirable to understand the formation process of dust and the stellar yields. For low- to intermediate-mass stars ($M_i$<10$M^{\odot}$), the largest fraction of all stars, we cannot yet describe these processes from first principles. A critical missing ingredient is a prescription for how dust forms in the upper atmospheres of red giants. Radiation pressure on dust close to the star, through absorption and scattering of stellar light, is essential to drive the outflows and remove material from the star. In conjunction with OSS measurements of molecular clusters close to the star, HERO observations of H$_2$O and hydrides, molecules essential for the formation of molecular clusters from the "simple" molecular gas, sets strong constraints on the chemical birth path of dust. The simultaneous tracing at high-spectral resolution (sub-km/s) of these light molecules and the high-excitation transitions of heavy molecules, such as AlO and TiO$_2$, is critical to disentangle the complex velocity field in the upper atmosphere, which is caused by infall and outflow of material as a consequence of shocks. Existing facilities like ALMA provide high-angular resolution observations but miss H$_2$O, hydrides, and the high-excitation transitions of heavier molecules. SOFIA can study some of these, but lacks the sensitivity (1 day *Origins* observations correspond to 1 year SOFIA observations assuming a 10 time higher sensitivity with HERO) and the access to H$_2$O. Without a representative sample and the water lines, we cannot constrain the chemistry at the basis of this fundamental issue in astrophysics: how is dust formed from the gas phase?

The idea of using the *Origins* to study black hole physics on event horizon scales has recently been raised as a potential extension of the HERO science case. This topic is addressed separately in Section D.1.12.

### D.1.2 HERO Science Traceability and Instrument Requirements

See upscopes introduction for the HERO Science Traceability Matrix.

### D.1.3 HERO Instrument Description

HERO (see Figure D-6) uses the heterodyne principle and is based on the successful HIFI (Heterodyne Instrument for the Far-Infrared, de Grauuw *et al.* 2009) instrument on *Herschel*, as well as the upGREAT (upgraded German REceiver at Terahertz, Risacher *et al.* 2016 and 2017) instrument on Stratospheric Observatory For Infrared Astronomy, but exceeds both in sensitivity and in frequency coverage. HERO is the first heterodyne array receiver ever designed for a satellite mission, which requires special components with low power consumption.

**Mixers:** HERO uses the most sensitive cryogenic mixers that exist: Superconducting Insulating Superconducting (SIS) mixers (*e.g.* Kerr *et al.* 2015) for the two lower frequency bands and Hot Electron Bolometers (HEB) mixers (*e.g.* Cherednichenko *et al.* 2008) for the upper two bands. The mixers are ar-

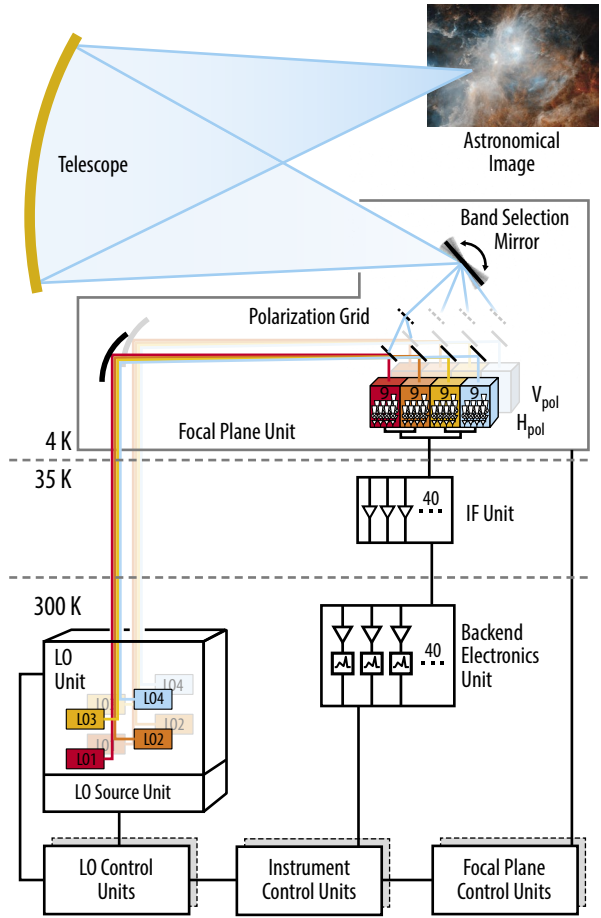

| Sub-system | Component | HERO |
|---|---|---|
| Local Oscillator | Synthesizer Technology | CMOS or YIG-based W-band synthesizer + GaN. amps |
| | Multiplied LO | Cascaded Multipl. + On-chip. Power Combining +a 3D integ. |
| | DC power/pixel | 2 W |
| | Fractional Bandwidth | 45% |
| Cryogenic Detectors | Mixer Technology | SIS, HEB |
| | LNA Technology | Low-power SiGe HBT |
| | Number of pixels | 2 (polarizations) x3x3 |
| | DC power/pixel | 0.5 mW |
| | Mixer. Assembly | Waveguide |
| Backend | IF Processing | GaAs HEMT ampl |
| | Spectrometer Tech. | CMOS based SoC |
| | DC power/pixel | 2 W |
| | IF Bandwidth | 8 GHz |
| Total DC power per pixel | | 3.5 W |

OriginsF271

**Figure D-6:** The HERO is the first heterodyne array receiver designed for space. The schematic diagram shows a relatively simple design that covers the widest frequency coverage and the highest sensitivity of any heterodyne receiver. HERO uses sensitive state-of-the-art components with low power consumption and weight.

ranged in 3x3 arrays with a 10x10mm² size for the SIS and a 5x5mm² size for the HEB mixers. There are two identical arrays for each frequency band, one for each linear polarization. HERO has balanced mixers, and one mixer per array is sideband separating to allow sideband ratio calibration. To avoid losses in beam splitters LO and sky are fed to the mixers in orthogonal linear polarization and separated by an orthomode transducer at the entrance of the mixer.

**Local Oscillators (LO):** The LO provides the artificial monochromatic references signal. HERO uses extremely wideband, continuously tuneable, multiplier amplifier LOs with up to 45% fractional bandwidth. The LO signal is split in waveguide into 9 beams to match the mixer arrays (see Figure D-7).

**Signal Amplification and Processing:** After mixing the LO and the sky signal the inter-mediate frequency signal (IF) between 0.5 and

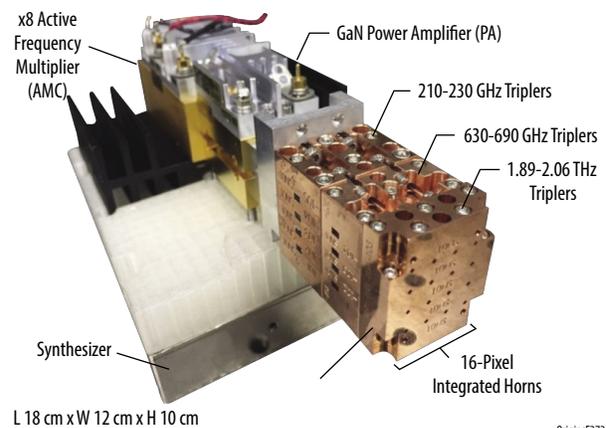

L 18 cm x W 12 cm x H 10 cm

OriginsF272

**Figure D-7:** Compact multi-pixel electronically tunable LO systems have been demonstrated to 1.9 THz.

6.5 GHz (baseline, 8.5 GHz goal) is created. Extremely low dissipation (< 0.5mW), low noise (< 4K) SiGe (see *e.g.*, Montazeri *et al.* 2016) (or InPh) amplifiers enhance the signal strength in three stages at 4.5K, at 35K and in the warm space bus. HERO uses miniaturized IF circuits (*e.g.* Zailer *et al.* 2016) in the space craft bus.

**Backend:** HERO has compact, low power (<2W), configurable backends with at least 1000 channels, either CMOS-based spectrometers (Kim *et al.* 2018) or autocorrelation spectrometers (Vogt *et al.* 2010).

**Instrument control:** Similar to HIFI/*Herschel* HERO has three standard instrument control units: one for the cold focal plane unit, one for the LO unit and one for the backends.

### D.1.4 HERO Mechanical and Thermal Design and Resource Requirements

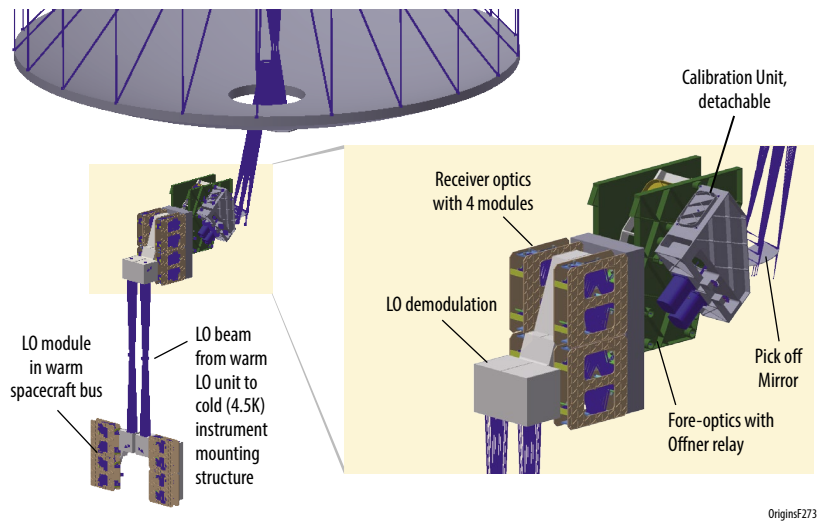

**Figure D-8:** The mechanics and ray trace of HERO's focal plane unit show a very compact design.

**Table D-4:** The HERO instrument requirements fit easily within the spacecraft allocations.

| HERO | Initial Allocation | Current best estimate | Margin |
|------|--------------------|-----------------------|--------|
| Mass | < 200kg | 183 kg | 9.2% |
| Volume | <2 m³ | 1.28 m³ | 56% |
| Power | < 250 W | 205 W | 22% |
| Heat Dissipation @ 4.5K | < 35 mW | 20 mW | 75% |
| Heat Dissipation at 35K | < 70 mW | 40 mW | 75% |
| Data Rate | — | 5.1 kbit/s | — |
| TRL in 2018 | ≥ 3 | ≥ 4 | — |
| TRL in 2027 | ≥ 6 | ≥ 6 | — |

**Table D-5:** HERO uses two proven mechanisms and two shutters.

| Mechanism | Function | Accuracy | Heritage |
|-----------|----------|----------|----------|
| Offner Relay | Select frequency band, point at calibration load | 1'30" Accuracy 15 Deg Range | Herschel/SPIRE (Paine et al. 2003) |
| Tip-Tilt for LO | Mirror allowing to adjust LO beam for mechanical drift between spacecraft bus and cold instrument unit | 1" Accuracy | Herschel/HIFI (Focal Plane Chopper, Huisman et al. 2011) |
| Calibration Load shutter | Cover calibration load to minimize heat load on | 1mm @ 50 mm base Rotational cover | Herschel/HIFI |
| LO door/shutter | Close optical connection between LO units in warm spacecraft bus and telescope | 1mm @ 150 mm base Rotational cover | Herschel/HIFI |

### D.1.5 HERO Risk Management Approach

**Table D-6:** HERO Risk management

|  | Element |
|---|---|
| Single Element | Fore optics, Offner Relay, LO door |
| Internally redundant | Mixers and LO of both polarizations are identical, Mixer pixels within an array are identical, hot and cold calibration loads can be exchanged if one heater fails, Tip-Tilt mechanism |
| Externally redundant | Control electronics |

### D.1.6 HERO Test, Integration, Alignment and Calibration

The *Herschel*/HIFI testing, integration, alignment and calibration procedures are used for HERO. The HERO instrument group tests, calibrates and internally aligns the instrument in existing cryostats prior to delivery. *Origins* has a modular design and the NASA team can integrate the instruments in any order.

### D.1.7 HERO Descope Options

HERO described above represents the best compromise between simplicity and science capabilities. HERO has a modular design and can easily be down-scoped by reducing the number of pixels, frequency bands, polarizations, or the dual frequency operation. Table D-7 gives two representative descope options.

**Table D-7:** There are several possibilities to descope HERO, either to lower power, mass and cooling budgets (Option 1) or alternatively to use Herschel/HIFI components with TRL 8/9 (Option 2). The descope has some impact on the science.

| Instrument Description | HERO | Descope Option 1 (Reduced power, mass, dissipation) | Descope Option 2 (High TRL components) |
|---|---|---|---|
| Frequency bands | 617–397 µm | 589–364 µm | 617–397 µm |
|  | 397–252 µm | 364–219 µm | 397–252 µm |
|  | 252–168 µm | — | 252–168 µm |
|  | 168–111µm | — | 168–111µm |
| Polarizations | 2 | 2 | 2 |
| Pixels | 3 x 3 | 1 x 5 | 1 x 4 |
| Simultaneous observing frequency. | Dual - frequency | Dual -frequency | Single frequency |
| # backends | 40 | 24 | 10 |
| Technology | Innovative, low power TRL 4–9 | Innovative, low power TRL 4–9 | Herschel/HIFI TRL 8/9 but wider RF BW and shorter wavelength |
| Heat dissipation at 4.5 K | 20mW | 12mW | 50mW |
| Required power | 205W | ~ 190W | ~ 530W |
| **Science impact** | | | |
| Trail of water | — | Some $H_2O$ lines not available, HD not observable, slower mapping | Ok, all lines, slower mapping only single frequency |
| GO - projects | — | Less lines: Lose discovery space, less molecules for birth of dust | Only single frequency observations, i.e. slower |

### D.1.8 HERO Partnership Opportunities

The HERO study was carried out as an international collaboration led by France (Figure D-9). There are several groups including US laboratories that could build HERO for *Origins*.

### D.1.9 Spacecraft Modifications Required for HERO upscope

Nearly no spacecraft modifications are needed to integrate HERO into the *Origins* baseline spacecraft design, as it is relatively small and light and fits between OSS and FIPS. HERO requires slightly lower cooling power, electrical power, and data rates than the other instruments. As HERO does not observe at the same time as other instruments, no extra cryocoolers, solar panels, or data links are required. However, HERO requires a free line of sight to the spacecraft bus, where its local oscillators are located. To minimize any heat transfer to the cryogenic unit, IR blocking filters are added as well as a door that can be shut when HERO is not in operation (Figure D-10).

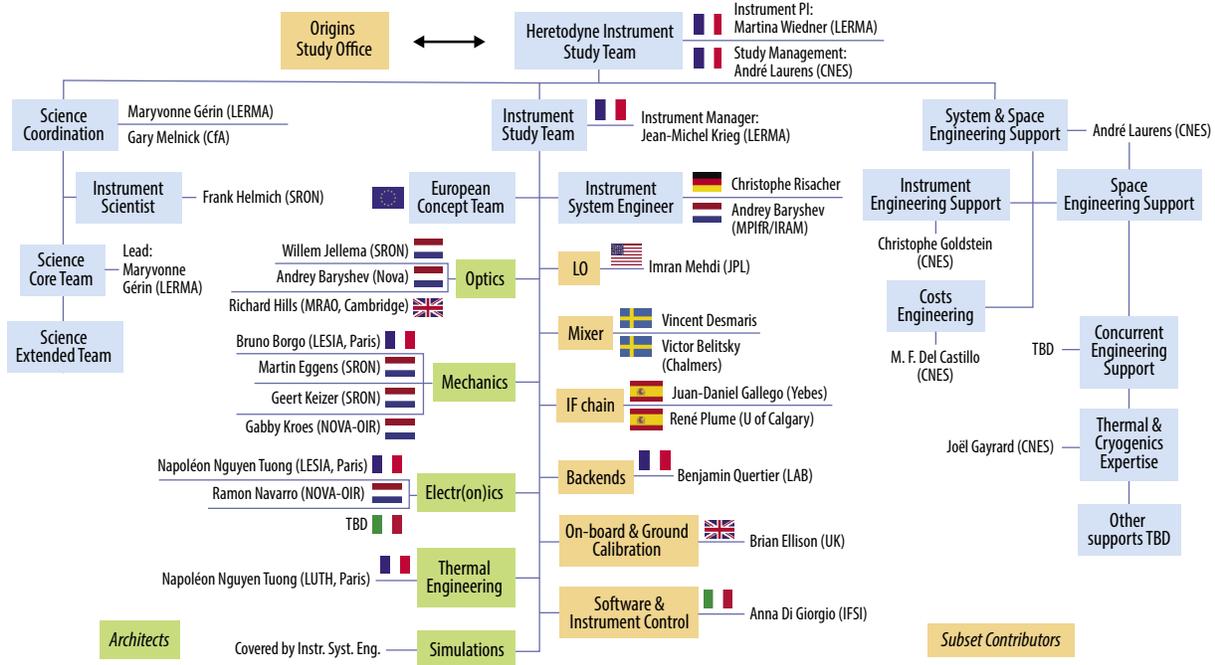

**Figure D-9:** An international consortium team carried out the HERO study, showing the worldwide interest in a heterodyne receiver.

### D.1.10 Technology Roadmap for HERO

HERO builds on the current state-of-the-art receiver but surpasses them. Some development work is required, however there is a very high TRL fallback option (see Table D-7, Figure D-11) of a smaller HERO using *Herschel*/HIFI components.

### D.1.11 HERO Performance

The Heterodyne Receiver for *Origins* (HERO) builds on a strong heritage, has near quantum-limited performance and has unrivaled sensitivity over a very large frequency range (see Figure D-12). It consists of small mixer arrays, observes in both polarizations simultaneously and allows dual frequency observations. It uses very little power per pixel and is the first heterodyne array receiver designed for a satellite. It is ten times more sensitive to point sources then *Herschel*/HIFI, i.e. can observe a factor of 100 faster and it can map a field of non-extended sources 8000 times faster, due to its arrays. Being on a satellite the full far-IR spectrum becomes available to HERO, whereas ground based telescopes and even SOFIA are hidden behind a more or less opaque curtain of atmospheric

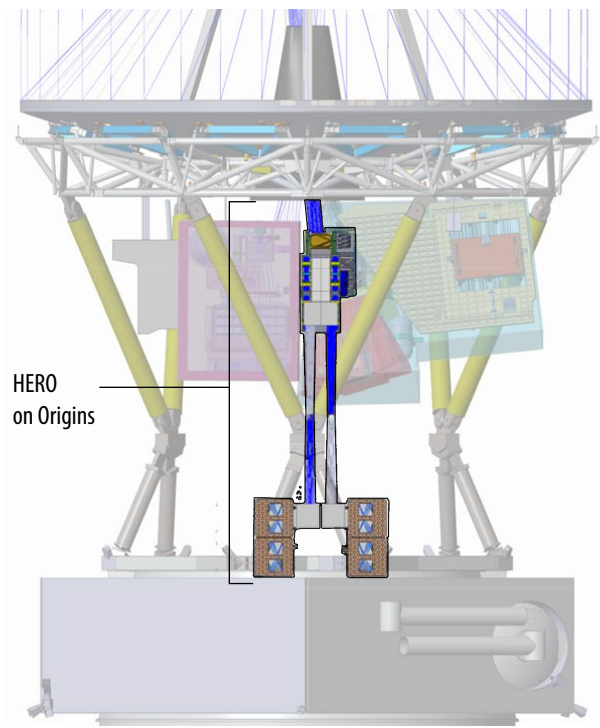

HERO on Origins

**Figure D-10:** HERO is a relatively small instrument that fits easily into the spacecraft. To minimize the cryogenic heat load the reference signal (Local Oscillator) is located in the spacecraft bus and optically linked to the cold unit (blue ray traces).

**Table D-8:** Current State-of-the-art and HERO requirements

| Sub-system | Component | HIFI on Herschel | STO-2 on balloon | GREAT/4GREAT/upGREAT on SOFIA | HERO on Origins | TRL HERO | TRL Descope 2 see Tab D.1-6 |
|---|---|---|---|---|---|---|---|
| Local Oscillator | Synthesizer Technology | YIG-based Ka-Band+Active Frequency Multipliers + GaAs W-band power amps | | VCO-based oscillators from VDI | CMOS or YIG-based W-band synthesizer + GaN. amps | 5 | 8 |
| | Multiplied LO | Cascaded GaAs frequency multipliers | | Multipliers/ QCL for 4.7 THz | Cascaded Multipl. + On-chip. Power Combining + 3D integ. | | |
| | Number of Pixels per array | 1 | 2 | 7 | 9 | | |
| | DC power/pixel | 25 W | 20 W | AMC: 14W QCL: 71 W | ~ 1.5 W | | |
| | Fractional Bandwidth | ~12 % | ~1 % | 10 - 33 % | ~45 % | | |
| Cryogenic Detectors | Mixer Technology | SIS, HEB | HEB | SIS, NbN HEB | SIS, HEB | — | — |
| | SIS sensitivity | 2–6 hf/k | — | 2–6 hf/k | 2hf/k | 5 | 7/8 |
| | HEB sensitivity | 13–18 hf/k | 3,5 hf/k lab only | 4–10 hf/k | 3hf/k | 4 | 6/7 |
| | Mixer. Assembly | Quasi-optical (QO) | QO | Waveguide | Waveguide | — | — |
| | Number of pixels | 1 | 2 | 2x7 | 2x9 | — | — |
| | LNA Technology | InP HEMT | SiGe HBT | SiGe, InP | Low-power SiGe HBT | 4 | 8 |
| | DC power/pixel | 10 mW | 4 mW | 5 mW | 0.5 mW | | |
| Backend | IF Processing | GaAs HEMT amplifiers | | | | 8 | 8 |
| | Spectrometer Tech. | FFT FPG | FFT | FFT | CMOS based SoC | 4/5 | 8 |
| | DC Power/pixel | 10 W | 10 W | 35 W | 2 W | | |
| | IF Bandwidth | 1.5 GHz | 1.5 GHz | 0.2–4 GHz | 6 GHz goal 8GHz | | |
| **Total DC power per pixel** | | 35 W | 35 W | 49–106 W | 3.5 W | — | — |

**Table D-9:** Development Goals, Risks and Mitigations

| Item | Goal | Risk | Mitigation | Consequence |
|---|---|---|---|---|
| LO | 45% fractional bandwidth with 10 micro W/pixel output using < 0.5 W/pixel | BW not obtained with enough output power | i) more power input or ii) more LO bands | More satellite resources required: i) power ii) weight and power |
| | | | Two kinds of LOs will be developed. QCLs have more output power | For QCL need frequency selection, PLL and extra cooler → more engineering effort |
| Mixers | 2–3 hf/k with 6 (goal 8 GHz IF BW) | i) Sensitivity not reached | Two mixer types are developed SIS and HEB | Longer observing time |
| | | ii) BW not reached for HEB mixers | Use SIS mixers as much as possible | Use two LO settings to cover wide lines |
| | Compact arrays | Difficulty miniaturizing | Use individual mixers | Slightly more volume required |
| Spectrometers | 6 GHz (goal 8 GHz BW), at 1 W/pixel, 8192 channels | SoC ASIC not radiation hard | Put in radiation shielding box | Equipment becomes heavier |
| | | Stability or BW selection cannot be reached | There are two spectrometer types developed: Autocorrelators can reach these specs | Autocorrelators are likely to be heavier and consume more power, might need to have fewer backends, in worst case no dual frequency observations → increase in observing time |
| Low Noise Amplifiers | 2K noise, 6 (goal 8GHz BW), 0.5 mW power dissipation | Lower BW and low power dissipation cannot be obtained at the same time | Two kinds of LNAs are developed: i) SiGe which have low power dissipation, but currently not the bandwidth ii) InP have BW but at higher power | Use InP LNAs, require more cooling or reduce backends, in worst case no dual frequency observations → increase in observing time in some cases |
| Optics | Wide bandwidth at excellent performance | Performance degrades over BW | i) increase number of frequency bands ii) Accept slightly lower sensitivity | i) Increase in mass and volume of receiver ii) Slightly longer observing time |

| | | 2019 | 2020 | 2021 | 2022 | 2023 | 2024 | 2025 | 2026 | 2027 |
|---|---|---|---|---|---|---|---|---|---|---|
| | | | | | | | | TRL 5 | EM | TRL 6 |
| Optics | Lenslet arrays, filters, calib source, etc | | | | Develop broadband design | → | Envir./Qual. testing | Characterization in relevant environment | Subsystem level integration | System level testing |
| Mixers | SIS and Hot Electron Bolometer | Single pixel with required sensitivity and frequency coverage, search for materials for high freq. SIS → | | Single pixel with sensitivity and IF BW | Array proof of concept, balanced mixers | Envir. test | Qual. test | | | |
| Local Oscillator | Schottky Multipliers | Single pixel with fractional BW → | | | Array proof of concept | | | | | |
| Local Oscillator | QCL | | Powerful and efficient source, large and continuous coverage, high operating temp → | | Development of mode selection optics, PLL, beam divider | Efficient and long distance (few m) LO and mixer coupling scheme | Down select. Envir./Qual. testing | | | |
| Local Oscillator | Parametric multipliers | | Single pixel with output power | Single pixel with BW | Array proof of concept | | | | | |
| IF | LNAs | | | | Low noise, wideband and Low DC power → | | Down select. Envir./Qual. testing | | | |
| Backends | SoC ASIC | | Bandwidth, DC power, calibration | | | Envir./Qual. testing and Down selection → | | | | |
| Backends | Autocorrelators | | Bandwidth, DC power → | | | | | | | |

OriginsF276

**Figure D-11:** The development schedule ensures a low-risk technology development path for HERO.

**Table D-10:** HERO Instrument Characteristics

| HERO Instrument Characteristics | |
|---|---|
| **Parameter** | **Value** |
| Operating Mode | Dual Frequency<br>Dual Polarization |
| Sensitivity | $6.4 \times 10^{-21}$ W/m$^2$ at 480 μm<br>$1.6 \times 10^{-20}$ W/m$^2$ at 300 μm<br>$4.0 \times 10^{-20}$ W/m$^2$ at 200 μm<br>$7.3 \times 10^{-20}$ W/m$^2$ at 130 μm |
| Resolving Power | $10^5$ to $10^7$ |
| Wavelength coverage | 617 – 111 μm |
| Field of View | 2.1′ x 2.1′ at 480 μm,<br>1.3′ x 1.3′ at 300 μm,<br>0.8′ x 0.8′ at 200 μm,<br>0.6′ x 0.6′ at 130 μm |
| Array size | 3x3 |
| Saturation | None |
| Polarization | 2 |

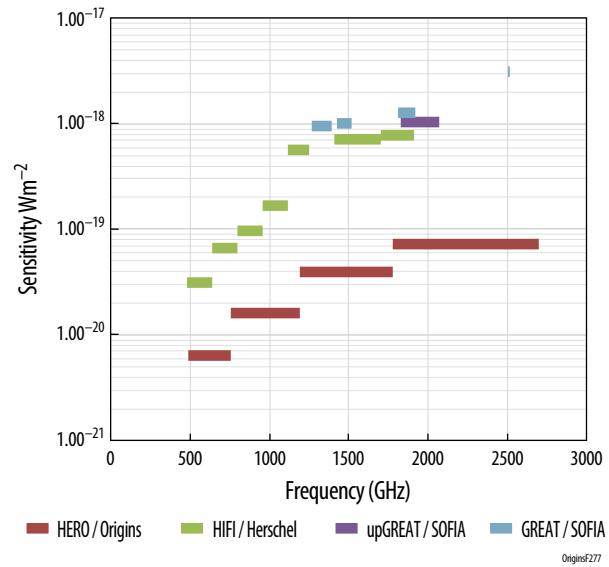

OriginsF277

**Figure D-12:** HERO is more sensitive by an order of magnitude than any prior heterodyne instrument and has the widest frequency coverage of any previous instrument. (The sensitivity calculations for SOFIA assume a perfectly transparent atmosphere, which is not true close to the water lines.)

absorption. HERO is a powerful but low-risk upscope option that has the potential to open a new era of high spectral resolution observations in the far-IR.

### D.1.12 Black Hole Demographics Across the Universe and New Tests of General Relativity Made Possible Using Extremely Long-Baseline Interferometry with the Origins Space Telescope and HERO

The concept of using the *Origins* Space Telescope (*Origins*) to study black hole physics on event horizon scales has recently been raised as a potential extension of the HERO science case. Preliminary science concept exploration shows much promise, and merits further study. The unprecedented angular resolution resulting from the combination of *Origins* with existing ground-based submillimeter/millimeter telescope arrays would increase the number of spatially resolvable black holes by a factor of $10^6$, permit the study of these black holes across all cosmic history, and enable new tests of General Relativity by unveiling the photon ring substructure in the nearest black holes. Expanding the HERO instrument (Wiedner *et al.* 2018) to be an interferometric station will require several technology enhancements. Top-level discussions have not revealed any showstoppers, the necessary additional requirements are presented here but the design considerations warrant more focused attention.

### D.1.12.1 Key Science Goals and Objectives

A leap forward in our understanding of black holes came in April 2019, when the Event Horizon Telescope (EHT) collaboration revealed the first horizon-scale image of the active galactic nucleus of M87 (EHT Collaboration *et al.* 2019a). This feat was accomplished using a ground-based very long baseline interferometry (VLBI) network with baselines extending across the planet, reaching the highest angular resolution currently achievable from the surface of the Earth. We could soon have an opportunity to make another leap by utilizing the extremely long baselines between Earth and the *Origins* Space Telescope (*Origins*). The corresponding ~120 × improvement in angular resolution would increase the expected number of spatially resolvable black hole shadows from $\approx 1$ to $\geq 10^6$, enabling new studies of black hole demographics across cosmic history. For the closest supermassive black holes – such as the one in M87 – the unprecedented angular resolution would yield access to photon ring substructure, providing a new tool for making precise black hole spin measurements and testing the validity of General Relativity (GR).

Dramatically improving the angular resolution of (sub)mm VLBI is an exciting but daunting prospect that requires either increasing the observing frequency, extending the baseline lengths, or both. Not many sites on Earth offer atmospheric conditions that are good enough for observations at (sub) mm wavelengths to be routinely viable, and the current longest baselines are already nearly equal to one Earth diameter. Sizable angular resolution improvements will thus inevitably require stations in space. The L2 orbit planned for *Origins* provides a unique opportunity for extending VLBI to extremely long baselines by observing in tandem with sensitive ground-based stations (such as ALMA, LMT, NOEMA, GBT, or ngVLA). Such observations would achieve an unprecedented angular resolution. The EHT is currently the highest-frequency ground-based VLBI network, operating at a frequency of 230 GHz and attaining a resolution of ~20 μs; by comparison, an Earth-L2 baseline would have typical fringe spacings well under a micro-arcsecond at observing frequencies of 86, 230, 345, and 690 GHz.

### SMBH Demographics and Cosmic Evolution

For a black hole viewed by an observer at infinity, the locus of event horizon-grazing photon trajectories forms a nearly circular closed curve on the sky (Bardeen *et al.* 1973). This boundary defines the inner edge of the "photon ring," and for astrophysical black holes emission from the black hole "shadow" region interior to the photon ring is expected to be substantially depressed (see EHT Collaboration *et al.* 2019a and references therein).

Given a uniform distribution of supermassive black holes (SMBHs) in flat space, we expect the number of sources, $N$, with spatially resolved black hole shadows to increase as the cube of the maximum baseline length. Current Earth-based arrays, such as the EHT, are able to spatially resolve black hole shadows for $N \approx 1$ source (EHT Collaboration *et al.* 2019a; 2019b; 2019c). Extending a baseline from Earth to L2, at a distance of ~120 Earth diameters, would increase the expected number of spatially resolvable black hole shadows from $N \approx 1$ to $N \approx 120^3 > 10^6$. Each black hole with a resolved shadow would have a corresponding black hole mass estimate (or, more specifically, an estimate of the mass-to-distance ratio M/D), enabling studies of SMBH mass demographics with access to an unprecedented statistical sample.

In our Universe, the angular diameter distance reaches a maximum value at a redshift of $z \approx 2$; sources of a given physical size thus have a minimum possible angular size, and if a source can be spatially resolved at $z \approx 2$ then it can be spatially resolved at any redshift. The left panel of Figure D-13 shows that on the extremely long baselines between Earth and L2, with fringe spacings $\theta < 0.1$ µas, this minimum angular size ensures that SMBHs with masses $M \geq 10^9 M_\odot$ have shadow diameters that can be spatially resolved across cosmic history. Measurements of the SMBH mass distribution as a function of redshift would shed light on SMBH-galaxy coevolution and inform cosmological models of structure formation, helping to understand how supermassive black holes formed in the early Universe (see *e.g.* Latif *et al.* 2013).

**Measuring SMBH Spin and Testing GR Using Photon Ring Substructure**

For the nearest SMBHs, such as M87, a baseline between Earth and L2 opens up the possibility of spatially resolving the substructure of the photon ring itself, enabling new, stringent tests of GR. As detailed by Johnson *et al.* (2019), the self-similar nature of the photon ring is naturally decomposed by interferometers, with successive "windings" of photon orbits dominating the signal in discrete baseline intervals (see right panel of Figure D-13). The period of the visibility signal along a particular orien-tation is a function of the ring diameter along that orientation. The orbit of L2 around the Sun over the course of a year ensures that all orientations can in principle be sampled, enabling precise measurements of the photon ring size and shape. The shape of the photon ring around a Kerr black hole is uniquely defined by its spin and inclination angle, with mass acting exclusively as an overall scaling factor. Measuring the ratio between shadow diameters at different orientations thus provides a sensitive probe of the SMBH spin as well as a way to test the validity of GR itself (Johannsen & Psaltis 2010; Broderick *et al.* 2014).

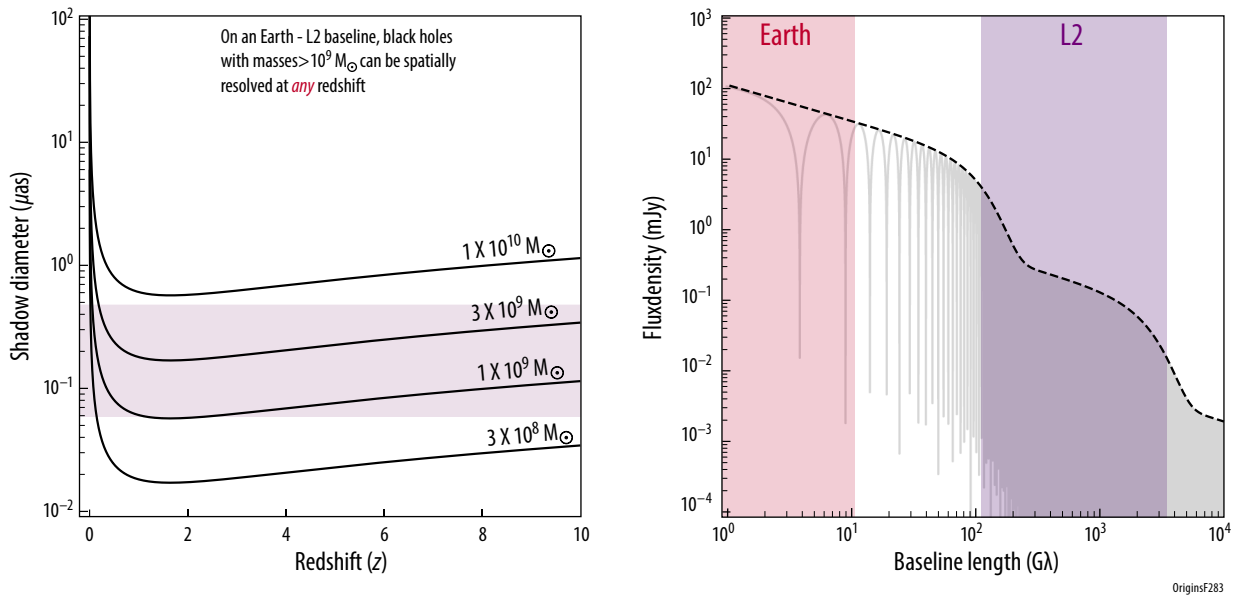

**Figure D-13:** (Left): Black hole shadow diameter vs. redshift for SMBHs of varying mass; the range of sizes accessible with a maximal Earth-L2 baseline operating at frequencies of 86 – 690 GHz is shaded in purple. We can see that as shadows reach a minimum size at $z \approx 2$, SMBHs with masses greater than $\sim 10^9 M_\odot$ are resolvable at all redshifts. (Right): Flux density as a function of baseline length is plotted in gray for a simple model of the M87 photon ring structure (Johnson et al. 2019) with the envelope shown as a dashed black line; note that the periodic structure evident at short baselines continues at longer baselines with the same period. The purple shaded region shows the range of baselines accessible over the course of a year on an Earth–L2 baseline operating at frequencies of 86 – 690 GHz; the long-baseline end of this range corresponds to a maximal baseline at 690 GHz, while the short-baseline end of the range corresponds to a minimal projected baseline (for M87, the minimal projected baseline is a factor of ~ 4 shorter than the maximal baseline) at 86 GHz. For comparison, the range of base-lines accessible from the ground at 230 GHz observing frequency is shown in red. The Earth-L2 baseline coverage samples many more periods of the visibility structure than accessible from the ground, enabling correspondingly more precise measurements of the photon ring diameter.

### D.1.12.2 Technology Overview

All extremely long baseline observations will face sensitivity challenges related to the physical brightness temperature limits of synchrotron radiation imposed by self-absorption and inverse- Compton scattering (Kellermann & Pauliny-Toth *et al.* 1969; see left panel of Figure D-14). On a maximal Earth-L2 baseline, no source is expected to have a flux density exceeding ~ 1 mJy. This strict sensitivity requirement drives the technology considerations.

The sensitivity of an interferometric baseline depends on: (1) the geometric mean of the system equivalent flux densities of the individual telescopes; (2) the averaged bandwidth; and, (3) the coherent integration time. The first property allows telescopes such as *Origins* (at 5.9-meter diameter) to form sensitive baselines when paired with a large ground-based telescope (*e.g.*, ALMA or ngVLA). The second allows digital enhancements (*e.g.*, wider recorded bandwidths) to offset limitations in telescope sensitivity. The third ties sensitivity to phase stability, which is limited by the atmosphere and reference frequency.

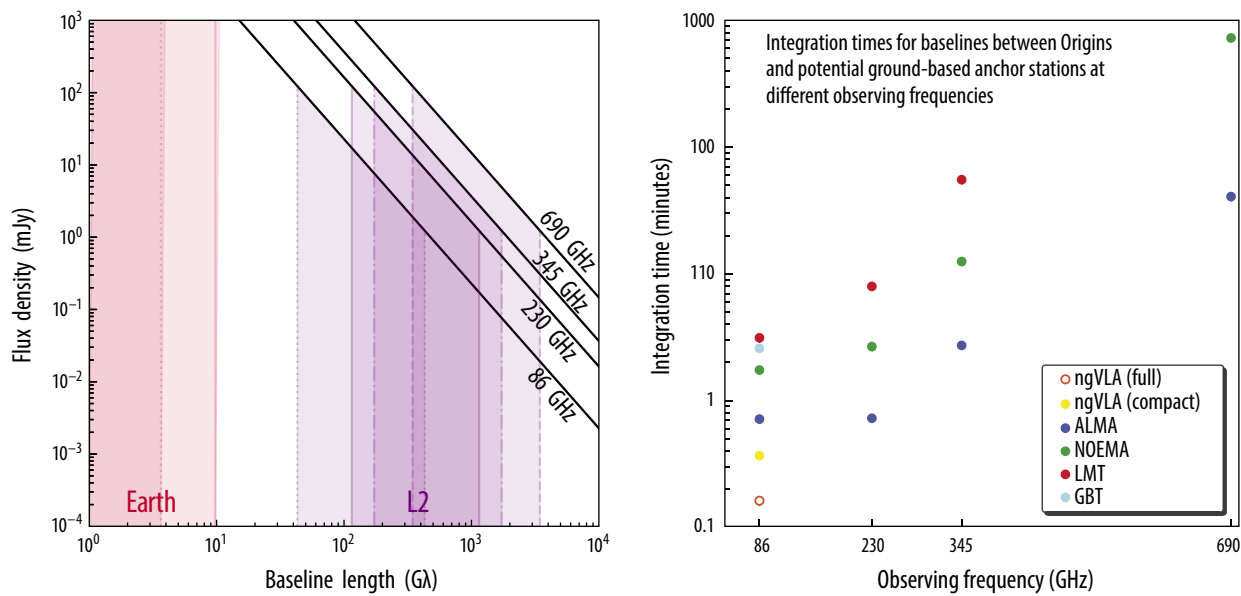

**Figure D-14:** (Left): For baselines to L2, synchrotron self-absorption and inverse-Compton scattering limit the brightness temperature to $T_b \leq 10^{12}$ K (Kellermann & Pauliny-Toth 1969); the black lines mark this limit for each of the labeled observing frequencies. Below each of these lines, the region shaded purple indicates physically allowed flux densities observable on Earth-L2 baselines ranging between $10 - 100\%$ of the maximum projected baseline length; overlapping regions indicate where sources could be observed with matched resolution at two or more frequencies (albeit at different times). The red shaded regions are analogous to the purple ones, but for Earth-Earth baselines at 86 and 230 GHz only. (Right): Estimated integration times required to achieve a $5\sigma$ detection of a source with brightness temperature $T_b = 10^{12}$ K on a maximal baseline between Origins and various potential Earth-based anchor stations as a function of observing frequency. For the sensitivity estimates, we have assumed a total bandwidth (across all sidebands and polarizations) of 32 GHz for the 86 GHz observations, and 64 GHz for all other frequencies. All ground stations are assumed to be observing at 45 degrees elevation, with zenith opacity of 0.05 (ALMA), 0.09 (NOEMA), 0.13 (LMT) at 230 GHz, and 0.05 (GBT, ngVLA) at 86 GHz. An aperture efficiency of 0.7 was assumed for all sites. Receiver temperatures are taken from station specifications or projections.

The following component recommendations are based on first-pass analyses and discussions, and are meant to give a general idea of the technology enhancements *Origins* would need to perform as an interferometric VLBI station.

### D.1.12.3 Technology Drivers

#### Receiver

Atmospheric conditions above ground-based stations limit the data collection frequencies to windows around 86 GHz, 230 GHz, 345 GHz, and 690 GHz. Band 1 of the HERO instrument already includes 690 GHz. The remaining frequencies could be covered using two additional bands, one that spans the 86 GHz and 230 GHz windows, and another that covers 230 GHz through 345 GHz. Simultaneous multifrequency observations will be useful as the ground-based array expands its capabilities to include multifrequency phase transfer.

#### Data Processing/Storage/Downlink

To achieve the sensitivities required for an Earth-L2 baseline, a combination of wide bandwidths and long integration times are needed. Though an improvement of one relaxes the requirement of the other, a first-pass operational configuration suggests that a total bandwidth of 64 GHz (16 GHz per sideband) for an integration time of two hours could achieve the requisite sensitivity on baselines to ALMA at any frequency (see right panel of Figure D-14). These requirements inform the data processing, storage/downlink, and timing reference components.

#### Data Processing

Onboard data processing requires high-speed analog-to-digital converters (ADC) and a high- performance Field Programmable Gate Array (FPGA). The FPGA quantizes and packetizes the digitized data for storage. Each of the four 8 GHz channels sampled at Nyquist requires a 16 Gbps ADC. The FPGA receives multiple streams and quantizes the data from the 4-bit sample to 2 bits, giving a total data rate of 256 Gbps. Current FPGA transceiver technology meets the speeds required, and both Xilinx (Ultrascale Kintex) and Microsemi (RTG-4) are currently working to make these high-speed interfaces available in space-qualified components.

#### Storage/Downlink

With an observation time of 6 hours, a 1/3 duty cycle, and a rate of 256 Gbps, a total of approximately 230 TB of data will be captured. A tradeoff between onboard storage and downlink speed will need to be conducted to determine the most optimal design. Storage continues to increase in both capacity and speed, with 1TB currently available in a small (22 mm x 80 mm) package. Downlink capabilities, pushed by the telecommunications industry, are also getting faster with laser communication speeds currently at 200 Gbps from LEO to small ground-based receivers (Robinson *et al.* 2018).

#### Timing Reference

To achieve phase stability on a single baseline for integration times of ~hours requires either that both stations are individually equipped with extremely stable timing references or that they share a common reference. Individual clocks would need to have Allen deviations better than ~ $5 \times 10^{-18}$ over the integration time, which is roughly three orders of magnitude more stable than current ground-based technology and likely unachievable on a 1–2-decade timescale. Instead, a shared reference would relax the more stringent stability requirement almost entirely and will likely prove to be the more feasible option. A shared ground-space timing reference has been demonstrated at lower observing frequencies by RadioAstron (Kardashev *et al.* 2013), though additional work would be required to extend such capabilities to higher observing frequencies and to a station at L2.

**Positional/Velocity Accuracy**

Antenna position and velocity (and possibly higher order derivatives) must be known in order to coherently average the correlated signal across finite bandwidth and over time. Initial searches can be conducted with wide search windows in the associated delay, delay-rate, and acceleration parameters, with residual values being used to refine the orbit determination, as is currently done for RadioAstron (Zakhvatkin *et al.* 2018). To create a baseline of position, velocity, and acceleration requirements, we take the RadioAstron specifications and increase the requirement by a factor of 10 for velocity and acceleration:

- Position error less than 600 m
- Velocity error less than 2 mm s$^{-1}$
- Acceleration error less than $10^{-9}$ m s$^{-2}$

These requirements can be potentially relaxed with improvements in computational delay and rate searching routines at the correlation stage.

## D.2 Mid-Infrared Spectrometer and Camera: Wide Field Imager

The Mid-Infrared Spectrometer and Camera (MISC) instrument UpScope configuration consists of a second separate module, the MISC Wide Field Imager (WFI) module. The MISC Wide Field Imager (WFI) module which offers a wide field imaging (3x3 arcmin) and low-resolution spectroscopic capability with filters and grisms for 5-28 microns. As needed, the MISC WFI is also used for focal plane guidance for itself and the other *Origins* science instruments.

The MISC instrument studied for the *Origins* report had two major components: a transit spectrometer described in the baseline, MISC-T, (Section 3.2) and a camera with low-resolution grism spectroscopy (this Section). The camera became an upscope option during the team's descope process (Section 2.1). This appendix describes the science drivers for the MISC camera and the full MISC instrument design.

### D.2.1 MISC Science Drivers

### D.2.1.1 MISC Follow-up of WFIRST-Deep and WFIRST-Wide Galaxies at 5 < z < 10

Using *Origins*/MISC to study the galaxies detected at 5 < z < 10 in WFIRST-Deep and WFIRST-Wide surveys will enable critical measurements of the star formation rate (SFR), stellar mass (M$_\star$), and dust attenuation. Two surveys using *Origins*/MISC WFI photometric and spectroscopic capabilities will provide unique data that broaden our understanding of the evolution of the mass function (cosmic mass assembly), star formation rate density, and average dust attenuation for a representative sample of galaxies at 5 < z < 10.

**Introduction:** Stars emit photons over the entire wavelength range, but their main emissions fall in the rest-frame UV, optical, and near-IR (~0.25 to 3.5 µm range). Several SFR tracers are available in this wavelength range (Hα, Paα, PAH 3.3 µm). Measuring the stellar mass of galaxies in the early universe at 5 < z < 10 will translate into wavelengths that uniquely match *Origins*/MISC capabilities (*i.e.,* 5 to 30 µm). The rest-frame UV spectrum will provide an access to the young stars likely to be predominant at z > 6; but, to perform a complete census, including potential older stars, requires measurements in the rest-frame optical and near-IR.

To follow up on WFIRST's objects at z > 5 after JWST, requires *Origins* equipped with MISC. *Origins* is the only existing or planned facility able to efficiently measure these objects in the rest-frame optical+near-IR ranges. Extremely Large Telescope (ELT) may provide some data, but mainly below 2.5 µm, which is not sufficient to determine accurate stellar masses of galaxies during reionization. An estimate of important parameters from the WFIRST-Deep and WFIRST-Wide surveys is required to inform this *Origins* study. The planned WFIRST-Deep and WFIRST-Wide surveys are expected to result in appreciable samples of galaxies as Lyman-break dropouts for which physical properties will only be known partially due to the wavelength range of those observations. *Origins*/MISC-WFI is uniquely ideal to making additional observations in this spectral range.

**Scientific Importance:** In 2030 and beyond, a number of facilities will have surveyed the sky to identify galaxies during the epoch of reionization. JWST will open up this era, but galaxies are expected to be very rare and faint at 5 < z < 10. For galaxy samples during reionization from WFIRST Deep and Wide surveys, crucial parameters, such as stellar mass and dust attenuation (mandatory to compute the total star formation density), will be missing, because tracers are not observable from the ground and no other planned mission will be able to observe the sky at these redshifts and faint levels (*e.g.*, SPICA's 2.5-m diameter barely reaches z ~ 5).

**Proposed Observations:** To solve this outstanding question the team uses the *Origins*/MISC instrument for two kinds of observations: Photometry and Spectroscopy. Projections for this *Origins*/MISC science program were built using the J-band luminosity function from Ma, Hopkins, and Garrison-Kimmel (2018) and the SED modeling code CIGALE (Noll *et al.*, 2009).

**Photometry:** To build the Mass Functions (MFs), SFR vs. $M_\star$, and other physical parameters (specific SFR, specific $M_{dust}$), *Origins*/MISC could be used to observe a minimum of ten objects per redshift bin and per magnitude bin (constrained from WFIRST's sample). Using the SPIE paper from Sakon, *et al.* (2018) and CIGALE's estimated flux densities, the total exposure time needed is about 280 hours for 120 galaxies with S/N > 5 measurements. Adding the rest-frame UV data from WFIRST and, possibly the rest-frame far-IR data from *Origins*/FIP, would enable an even more detailed analysis.

**Spectroscopy:** To estimate the SFR and dust attenuation for these galaxies, the team will observe Paα, and PAH 3.3 μm features with MISC/WFI at R = 300. Paα will be observed over the entire redshift range, whereas Hα is observed only at z > 6, and PAH 3.3 μm only at 0.8<z < 8. A minimum of ten objects per redshift bin and per magnitude bin (starting from WFIRST's sample) is required. The Paα/Hα ratio constrains the dust attenuation (nominal case B = 8.46). With this information, SFR tracers (Hα, Paα, PAH 3.3 μm) and far-UV (incorporating ancillary data from WFIRST could be used to determine stellar initial mass function of these galaxies during reionization and address if stars are made up of PopII or PopIII stars, for example. If the same objects are observed with *Origins*/FIP, it also possible to determine the amount of dust attenuation needed to estimate the total star formation rate.

This program aimed at detailed characterization of 100-150 galaxies during reionization can be carried out using *Origins*/MISC's capability to observe two wavelength ranges (and two objects per slit, on average) simultaneously in 250 hours.

### D.2.1.2 Giant Planet Atmospheres: Templates For Brown Dwarfs and Exoplanets

> The four giant planets of our Solar System represent the closest and best example of a whole class of gaseous, substellar objects that are commonplace in our universe. Their ever-changing atmospheres provide the interface between their dynamic interiors and the external magnetospheric environment. *Origins* mid-infrared observations (5-30 μm) provide the capability for comparative planetology of the four giants, as well as long-term monitoring of atmospheric cycles to connect the JWST and *Origins* eras.

**Introduction:** Brown Dwarfs and directly-imaged exoplanets demonstrate rotational variability of their light curves, related to poorly-understood cloud formation and thermal contrasts on these distant, unresolved worlds. The four giants of our Solar System provide an ideal template for studying the sources of spatial and temporal variability, in an effort to understand how dynamic activity (banded structures, discrete vortices, vertical mixing) varies as a function of planetary metallicity, irradiation, and other driving properties. Although the giants have been studied for decades at visible and near-IR wavelengths, the required spatial resolutions in the mid-infrared for robust meteorological studies have been realized only recently (particularly for the ice giants). This wavelength range is key, providing the temperatures, humidity, wind shears, atmospheric stability, gaseous composition, and aerosol structure within these planetary atmospheres. In short, the mid-infrared reveals the environmental conditions underpinning the color and cloud changes observed in the visible. JWST/MIRI observations (Figure D-15) are expected to significantly improve the quality of spatially-resolved mid-infrared spectroscopy of all four targets. However, the low saturation limits and small fields-of-view of the integral field units limits the capabilities of this observatory for the giant planets. Furthermore, a key requirement for a step-change in giant planet atmospheric characterization is the need for high-temporal-cadence observations under invariant conditions. This time-domain science is required over multiple timescales: 1) short-term to identify small-scale changes in the atmospheric properties due to moist convective processes and waves; 2) intermediate-term to understand large-scale changes to the belt/zone structure; and 3) long-term to monitor seasonal and non-seasonal changes to these worlds.

**Scientific Importance:** Understanding these temporal variations, and specifically the thermal, gaseous, and aerosol changes that underpin them, will provide a ground-truth catalog of giant planet variability as a resource to the exoplanetary community. Combined with the proposed decade-long observing record from JWST in the same 5-30 μm spectral range, *Origins* would add to an unparalleled record of atmospheric variability on all four worlds.

**Proposed Observations:** This proposal program would be complementary to any *Origins* observations of the giant planets performed in the far-IR beyond 30 μm, although this focuses specifically on the mid-IR. This science case requires regular, global scale mapping of the four giants in narrowband imaging, low-resolution (R~300-500) spectroscopy, and high-resolution (R~3000) spectroscopy. *Origins* must be able to track the four giant planets, and to fit the full discs (45" for Jupiter) into the FOV (or with efficient mosaicking built into the observing strategy). Saturation limits must be sufficient to observe the full spectrum (unlike JWST/MIRI), potentially via the use of neutral density filters. Sensitivity limits should be improved beyond MIRI to permit rapid imaging of the ice giants Uranus and Neptune.

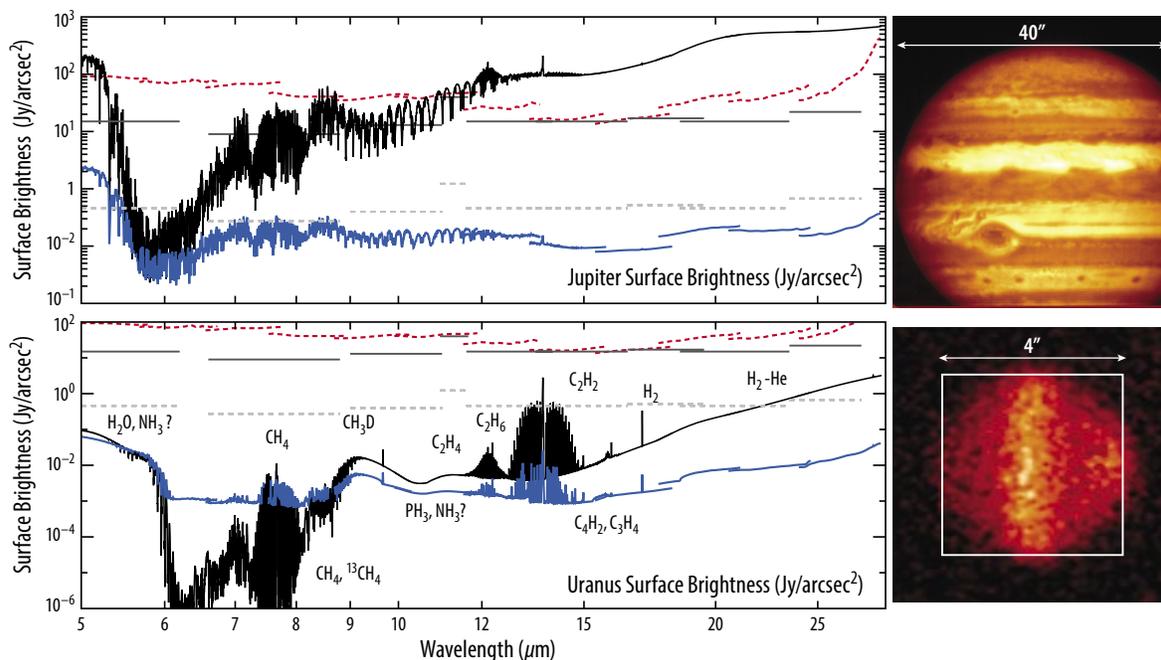

**Figure D-15:** Imaging capabilities for Jupiter and Uranus from an 8-m observatory (VLT) compared to synthetic MIRI spectra in the 5-30 μm range (black). Gaseous features are labelled in the lower plot. The blue line shows MIRI sensitivity to extended objects in 5 minutes and the challenge of observing Uranus. The red dotted lines show MIRI saturation and the challenge of observing Jupiter. The yellow boxes indicate the small fields-of-view of MIRI. It is hoped these challenges can be overcome by Origins in the mid-IR, so these targets can be observed with a regular cadence throughout the Origins lifetime

If all of these conditions are met, the scientific products would include:

- Global temperature maps from the troposphere (via $H_2$-He collision induced continuum at 15-30 μm) and stratosphere (from $CH_4$ at 7.7 μm), including emissions from regions heated by planetary aurorae. Temperatures are used to derive 3D wind patterns and key meteorological tracers such as the vorticity distributions, as well as assessing the global energy balance of all four worlds.
- Distributions of tropospheric aerosols and condensation clouds via mapping at 5 and 10 μm.
- Spatial and temporal variability of key gaseous species, including the cloud-forming volatiles (*e.g.*, $CH_4$, $NH_3$), disequilibrium species ($PH_3$ and para-$H_2$), and stratospheric photochemical products (*e.g.*, ethane and acetylene).
- Isotopic ratios (D/H, $^{13}C/^{12}C$, $^{15}N/^{14}N$) in a variety of species provided sufficiently-high spectral resolution, allowing the team to compare the atmospheric composition of the four worlds to help constrain the origins of their gaseous composition (*e.g.*, accretion from ices or from gases in the protosolar nebula).
- Exogenic species, such as CO, $CO_2$, and $H_2O$, to understand the ongoing evolution of their stratospheres, which can only be accessed from space-based facilities.

Although snapshots of these parameters have been published previously, attempts to provide long-term, consistent data in the thermal infrared have been hampered by changing conditions on Earth. The goal of MISC upscope is to replicate the success of HST's Outer Planet Atmospheres Legacy (OPAL) program (regular visible-light imaging of all four targets) via regular imaging/spectroscopy observations utilizing the mid-infrared and legacy from JWST/MIRI. This requires short segments of observations spanning multiple years of *Origins* operations.

### D.2.2 MISC Science Traceability

The imaging capability of the MISC WFI will be used for general science objectives. The spectroscopic capability with a resolving power of a few hundreds of the MISC WFI will be used to measure the mid-infrared dust features and ionic lines at z up to ~1 in Rise of Metals and Black Hole Feedback programs. The MISC WFI also serves as the focal plane pointing and guiding for the observatory, including when the MISC-T channel is performing exoplanet spectroscopy observations. Table D-11 summarizes the basic measurement capabilities of the upscope MISC instrument.

### D.2.3 MISC Instrument Description

#### D.2.3.1 General MISC Operation Principle

The MISC WFI covers the wavelength from 5 to 28 μm with two channels; WFI-short (WFI-S) covers the wavelength from 5 to 9 μm and WFI-long (WFI-L) covers from 9 to 28 μm. MISC WFI-S serves as the focal plane pointing and guiding for the observatory and has two redundant sets of 2k×2k Si:As detector arrays. During the normal operation, one of the two Si:As detector arrays in the WFI-S is used and the mirror changer switches the optical path if needed to use the other detector in case the performance of the first detector is degraded for some reason. It should be noted that between the two channels, the WFI-L channel was prioritized for science promise over WFI-S by the STDT, in case a trade choice was had to be made between the two.

**Table D-11:** The MISC instrument fact sheet (upscope) shows the additional capabilities needed to perform the science described in the sections above.

| Parameter | MISC Transit Spectrometer | MISC Wide Field Imager |
|---|---|---|
| Observing modes | [1] MISC Ultra Stable Spectroscopy | [2] MIR Imaging<br>[3] MIR low-resolution spectroscopy (slit)<br>[4] MIR low-resolution spectroscopy (slitless)<br>[5] MIR scan mapping |
| Spectral Range | 2.8–20 µm<br>MISC-T-S: 2.8–5.5 µm<br>MISC-T-M: 5.5–11 µm<br>MISC-T-L: 11–20 µm | 9–28 µm<br>WFI-S: 5-9 µm (SG1: 5-9 µm)<br>WFI-L: 9-28 µm (LG1: 9-16.2 µm, LG2: 16.2-28 µm) |
| Resolving power | R=50–100 in 2.8–5.5 µm (T-S)<br>R=50–100 in 5.5–11 µm (T-M)<br>R=165–295 in 11–20 µm (T-L) | R=5-10 for MIR Imaaging<br>R=300 for MIR low-resolution spectroscopy |
| Angular resolution | Cannot attain spatially resolved information within the FOV | 0.21 arcsec at 5 µm, 0.38arcsec at 9 µm<br>0.68 arcsec at 16 µm, 0.98 arcsec at 23 µm,<br>1.18 arcsec at 27.6 µm<br>(pixel scale; 0.088 arcsec/pix) |
| Field-of-View | Determined by the field stop size<br>3."0 in radius (MISC-T-S)<br>3."0 in radius (MISC-T-M)<br>2."0 in radius (MISC-T-L) | [Imager] 3 arcmin x 3 arcmin<br>[Slit for spectroscopy] Length; 3 arcmin, Width; 0.38 arcsec<br>(WFI-S/SG1), 0.68 arcsec (WFI-L/LG1), 1.18 arcsec (WFI-L/LG2) |
| Detectors | A 2kx2k HgCdTe detector array (30 K) for MISC-T-S<br>A 2kx2k HgCdTe detector array (30 K) for MISC-T-M<br>A 2kx2k Si:As detector array (~8 K) with a calibration source for MISC-T-L | A 2kx2k Si:As detector array (~8 K) for WFI-S1<br>A 2kx2k Si:As detector array (~8 K) for WFI-S2<br>A 2kx2k Si:As detector array (~8 K) for WFI-L |
| Sensitivity (Current best estimates) | SNR/sqrt(hr) = 12952 at 3.3 µm<br>SNR/sqrt(hr) = 13339 at 4.2 µm<br>SNR/sqrt(hr) = 9726 at 5 µm<br>SNR/sqrt(hr) = 9873 at 6.3 µm<br>SNR/sqrt(hr) = 8552 at 7.6 µm<br>SNR/sqrt(hr) = 8373 at 8 µm<br>SNR/sqrt(hr) = 7084 at 9.6 µm<br>SNR/sqrt(hr) = 6948 at 10 µm<br>SNR/sqrt(hr) = 4570 at 14 µm<br>SNR/sqrt(hr) = 3064 at 20 µm<br>assuming a R=50 with a 10.8 K-mag star | [Imager]<br>1-hr 5σ Continuum Sensitivity for a point source<br>0.06 µJy at 5µm, 0.25 µJy at 9 µm,<br>0.64 µJy at 16 µm, 0.96 µJy at 23 µm,<br>1.93 µJy at 25 µm<br>[Low-resolution Spectroscopy; R=300]<br>1-hr 5σ Line Sensitivity for a point source<br>$5.0\times10^{-21}$ W/m$^2$ at 6 µm, $4.5\times10^{-21}$ W/m$^2$ at 8µm,<br>$5.3\times10^{-21}$ W/m$^2$ at 10 µm, $4.3\times10^{-21}$ W/m$^2$ at 12 µm,<br>$5.2\times10^{-21}$ W/m$^2$ at 18 µm, $5.4\times10^{-21}$ W/m$^2$ at 24 µm,<br>$1.1\times10^{-20}$ W/m$^2$ at 26 µm, $5.4\times10^{-19}$ W/m$^2$ at 28 µm |
| Saturation limit | 29.8 Jy at 3.3 µm, 27.5 Jy at 6.3 µm, 4.4 Jy at 14 µm<br>calculated for the shortest readout time (assuming partial readout, 10 µsec per pixel per read, two reads per pixel to sample up the ramp) | |

## D.2.3.2 MISC Instrument Overview

The block diagrams of MISC wide field imager (WFI) in the upscope design is shown in Figure D-16.

## D.2.3.3 MISC Optical Design

In the upscope MISC, MISC-T does not include a Lyot-Coronagraph based tip-tilt sensor – this pointing correction function is now provided by the MISC WFI module. Other than that, the optical design (Figure D-17) of the MISC-T is the same as that described in Section 3.2. The MISC Wide Field Imager (WFI) offers a wide field imaging (3 arcmin x 3 arcmin) and low-resolution (R~300) spectroscopic capability with filters and grisms covering from 5 to 28 µm. MISC WFI-S serves as the focal plane pointing and guiding for the observatory and has two redundant sets of 2k×2k Si:As detector arrays. The MISC WFI-L has a 2kx2k Si:As detector array. WFI-S and WFI-L share the same 3 arcmin by 3 arcmin FOV by means of the beam splitter and the diffraction-limited image quality over the entire FOV is achieved at any wavelength from 5 to 28 µm with the help of a deformable mirror and a tip tilt mirror in the fore optics of the MISC WFI. The reflected beam (*i.e.,* 5-9 µm) by the beam splitter is diverted to WFI-S and the transmitted beam (*i.e.,* 9-28 µm) is sent through to WFI-L.

**Origins/MISC Wide Field Imager (WFI)**

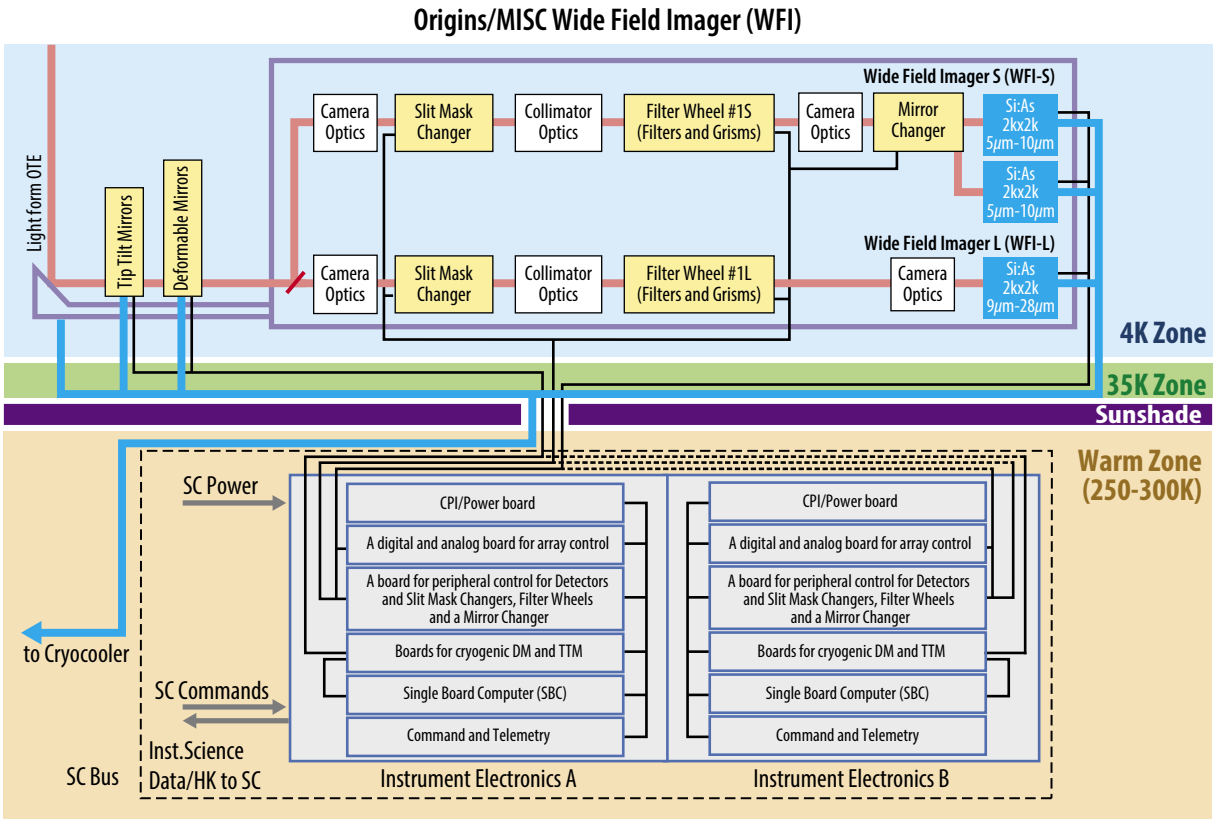

**Figure D-16:** The MISC WFI block diagram indicates the second module that was added to the upscoped MISC instrument.

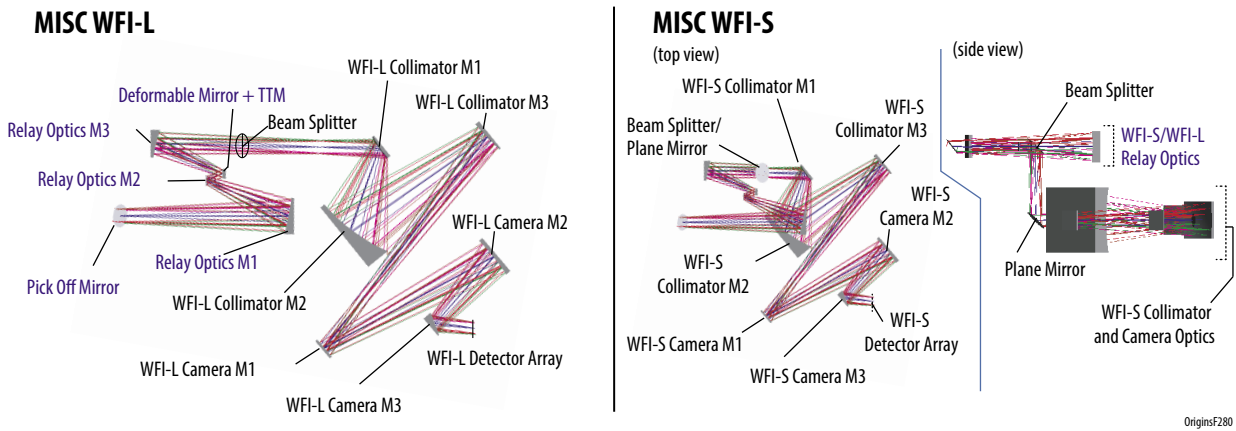

**Figure D-17:** The MISC WFI optical design includes a A) wave front error correction system in the fore optics and B) the imager optics.

### D.2.3.4 MISC WFI Detection Subsystem

The MISC WFI-S optical path contains two 2k x 2k Si:As detector arrays and the MISC WFI-L optical path contains a 2k x 2k Si:As detector array, each of which is bonded to a Si readout multiplexer operating at ~8 K to provide good detective quantum efficiency in the 5-28 μm wavelength range. This design has been used extensively in previous space missions such as *Spitzer*, AKARI, and JWST, as well as in SOFIA instruments. The detector development work needed for the MISC-WFI upscope comprises a modest factor of 2 increase in the detector dimension format for the WFI module.

### D.2.3.5 MISC WFI Signal Amplification

Amplification of the signals from the detectors is a straightforward reuse of the technology used in previous space applications of Si:As BIB detectors such as *Spitzer*, JWST, and WISE, and will employ dedicated satellite chips that are located in close proximity to the detectors and also operate at cryogenic temperatures.

### D.2.3.6 MISC WFI Read-out Electronics

Again, the readout electronics design of the MISC WFI detectors is very similar to that used in previous space missions, and will incorporate flexible readout patterns and a variety of readout strategies, including double-correlated and Fowler sampling techniques.

### D.2.3.7 MISC Mechanical Design

Based on the results of the MISC WFI instrument mechanical design, a 3D solid model of the MISC WFI instrument is shown in Figure D-18. To reduce MISC WFI instrument mass, the team assumed Beryllium as the baseline material for the mirrors, mirror support structures, and base plate.

### D.2.3.8 MISC Mechanisms

The MISC WFI module has the following mechanisms noted with the following numbers in Figure D-18: [1] a deformable mirror (DM) assembly in the fore optics of the WFI module, [2] a tip-tilt mirror (TTM) assembly in the fore optics of the WFI module, [3] a slit mask changer in WFI-S, [4] a slit mask changer in WFI-L, [5] a filter wheel for WFI-S, [6] a filter wheel for WFI-L, and [7] a mirror changer in WFI-S. Since the telescope optics is limited to diffraction-limited image quality performance only at 30 µm and longer wavelengths, the MISC WFI module needs to have its own internal wave front error correction optics. The MISC WFI module includes a DM and a TTM in its fore-optics and achieves diffraction limited image performance at >5µm for sources within a wide 3'x3' FOV. The slit mask changer used in MISC WFI-S has two 0.4-inch slots (a slit mask and a hole) and that in MISC WFI-L has three 0.4-inch slots (two slit masks and a hole). The filter wheel assembly used in each of the MISC WFI-S and WFI-L has triple wheels. Each wheel has six 0.3-inch positions for band-pass filters and/or grisms for low-resolution spectroscopy. The mirror changer is placed just before the two Si:As detector arrays in the WFI-S to switches the optical path when needed to do so.

### D.2.3.9 MISC Instrument Control

The MISC WFI has redundant dual-string warm electronics boxes (WEB A and WEB B for the MISC WFI module). Each electronics box contains a single board computer (SBC) and four boards: (i) a CPI/ Power board, (ii) a digital and analog board for array control (two 2kx2k Si:As arrays for WFI-S, one 2kx2k Si:As array for WFI-S), (iii) a board for peripheral control for detectors (two 2kx2k Si:As arrays for WFI-S, one 2kx2k Si:As array for WFI-L), two slit mask wheels, two filter wheels, and one mirror changer, and (iv) a control board for the cryogenic DM and TTM. A SBC is used to analyze the shape of the point-spread function (PSF) and to give feedback to the DM and TTM to achieve the diffraction-limited image quality at 5µm. This SBC also provides reference star centroid information that is sent to the *Origins* telescope pointing control.

### D.2.3.10 MISC Observing Modes and Data Rates

The MISC offers seven observing modes (Table D-12). In addition to the MISC-T mode, MISC has guiding, imaging, low resolution spectroscopy and scan mapping.

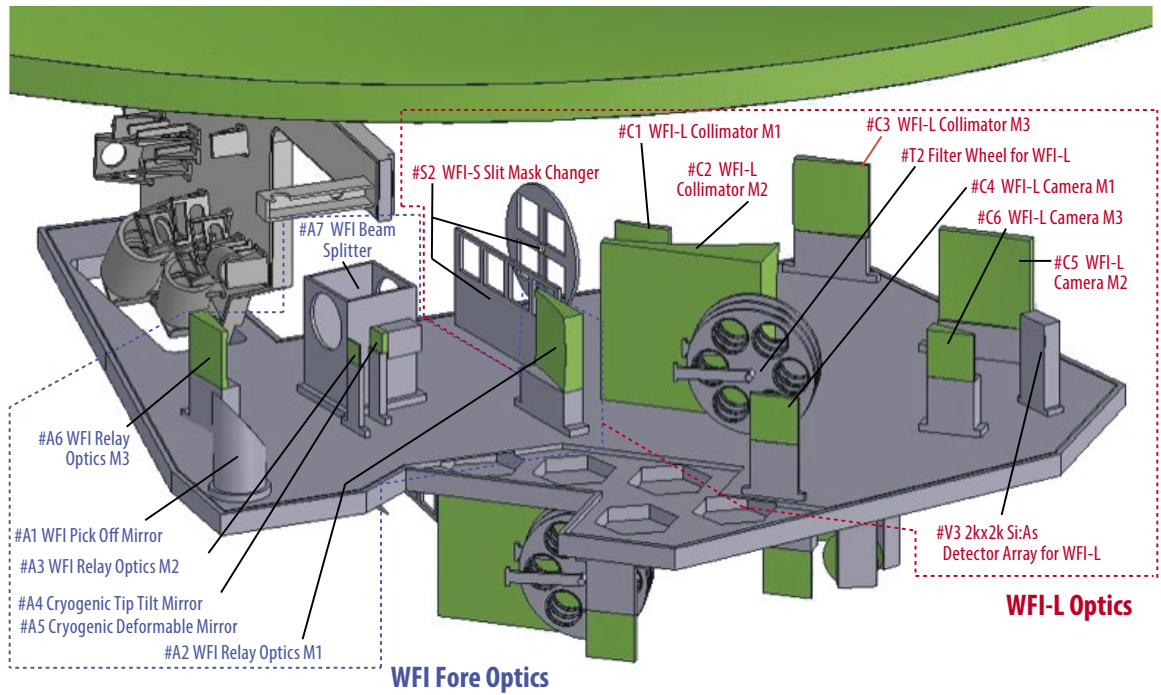

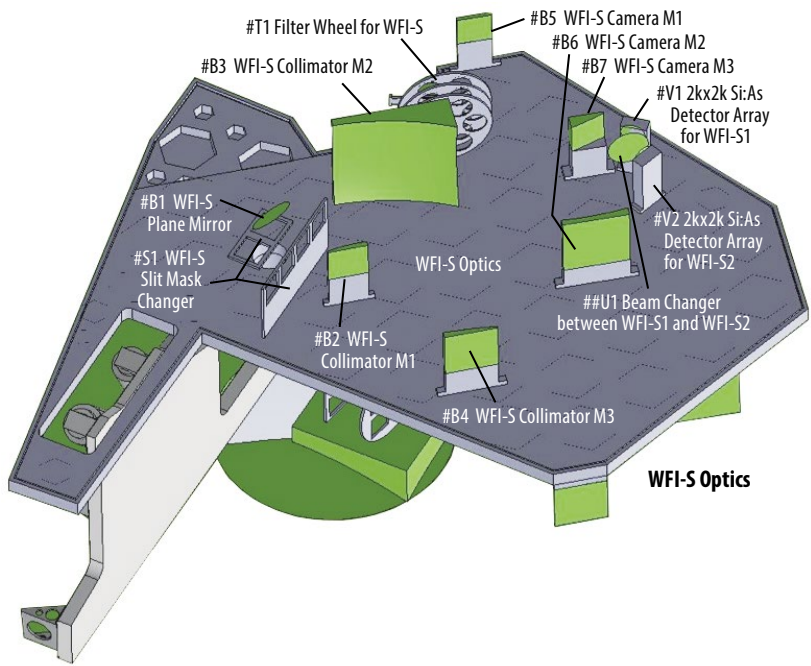

**Figure D-18:** The MISC WFI 3D solid model shows the instrument's A) WFI Fore optics, WFI-L optics, and B) WFI-S optics.

## D.2.4 MISC Predicted Performance

The following assumptions are made for the sensitivity estimate of the MISC WFI module. The 5σ 1-hour line sensitivity of MISC WFI (R=300) is shown in Figure D-19.

**Exposure Time:** The longest exposure time is determined as 300 sec for the Si:As 2k × 2k detectors (30 μm/pix; Raytheon) taking account of the number of pixels affected by cosmic ray hit events during an exposure and assuming the cosmic ray hit event rate at L2 as $5 \times 10^4$ m$^{-2}$ sec$^{-1}$ (Swinyard *et al.*, 2004).

**Table D-12:** The MISC observing modes (upscope) include the baseline MISC-T modes as well as additional modes used in the imaging and spectroscopy module.

| AOT | Mode | Data rate |
|-----|------|-----------|
| AOT00 | Guiding | 2.63 Mbps (max/average) |
| AOT01 | MISC Transit Spectroscopy | 3.73 Mbps (max)[2]−1.59Mbps (average)[3] [TRA] 2.63 Mbps (max/average) [Guiding with WFI-S] |
| AOT02 | MISC Imaging | 12.6 Mbps (max)[4]−0.45 Mbps (average)[5] [WFI] |
| AOT03 | MISC Low-Resolution Spectroscopy (slit) | 12.6 Mbps (max)[4]−0.45 Mbps (average)[5] [WFI] |
| AOT04 | MISC Low-Resolution Spectroscopy (slitless) | 12.6 Mbps (max)[4]−4.5 Mbps (average)[6] [WFI] |
| AOT05A | MIR Scan Mapping (AKARI/IRC type) | 3.745 Mbps (max/average)[7] |
| AOT05B | MIR Scan Mapping (to use TTM moving in a "freeze frame") | 16.78 Mbps (max/average)[8] |

Note 1: up to 8 windows of 32 x 32 pixels each of the MISC WFI-S module are read out at 20 Hz
Note 2: Max data rate of AOT01 is calculated for the shortest exposure $t_{exp}$=4s
Note 3: Average data rate is calculated for the longest exposure $t_{exp}$=10s
Note 4: Max data rate is calculated by assuming shortest exposure $t_{exp}$=2s and co-adding 6 exposures
Note 5: Average data rate is calculated for the longest exposure time $t_{exp}$=300s
Note 6: Average data rate is calculated for the longest exposure time $t_{exp}$=30s
Note 7: using double lines (2 x 1 pix x 2048 pix) for the purpose of mili-second confirmation. Neighboring 4 lines are read to avoid unstable behavior of the detector. Consequently 2 x (4+1+4) =18 lines are read. 2 lines are down linked and other 16 lines are discarded. Assuming 4 sec for full (2048 rows) readout, 4 x 18 / 2048 = 35 msec per double lines. The data rate becomes 16 bits x 2 x 1 pix x 2048 pix x 2 ch / 35 msec = 3.745 Mbps.
Note 8: readout half of the array at 0.25Hz. The data rate becomes 16 bits/pix x 2048 x 1024 pix x 2 ch x 0.25Hz = 16.78 Mbps.

**Detector Performance:** Detector Dark Current is assumed to be 0.06 e⁻sec⁻¹pixel⁻¹ for the Si:As 2k × 2k detector. The readout noise (reduced up to ¼ by means of Fowler-16 sampling) is assumed to be 10 e⁻ for the Si:As 2k × 2k detectors. Saturation Full Well is assumed as 2.5 x 10⁵ e⁻ for the Si:As 2k × 2k.

**Background:** The high background case assumes zodiacal emission modeled as a greybody at 268.5 K, normalized to 80 MJy/sr at 25 µm for the high background case and 274.0 K, normalized to 15 MJy/sr at 25 µm for the low background case.

### D.2.5 MISC Alignment, Integration, and Test

MISC instrument alignment and integration are straightforward. Dividing the instrument into two separate modules, each accessing its own re-gion of the telescope focal plane, does not require particularly stringent confocality of the entrance apertures, and as one of the modules has deformable mirrors, a small degree of accommodation can be realized after integration. However, since the MISC WFI module is also used by the observatory for pointing, this instrument must meet confocality requirements with all other *Origins* instruments. As these other instruments all operate in the far-infrared, meeting this requirement should not be difficult. The two-module configuration also greatly simplifies and speeds up instrument testing, as the modules can be tested independently and in parallel, if desired. Specialized test hardware development is required, but although the scope of this development will be significant, it is straightforward, with no new technology development required. The integration and test of the MISC instrument is expected to take 12 months, once all external test support hardware and software is ready.

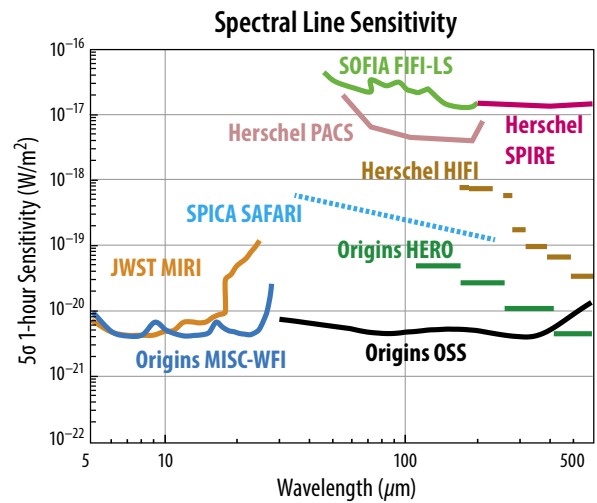

**Figure D-19:** The 5σ 1-hour line sensitivity of the Origins /MISC WFI (R=300) module.

### D.2.6 MISC Heritage

The wavelength coverage and the capabilities of the MISC instrument are partly overlapping with those of the JWST/MIRI, SOFIA/FORCAST, SOFIA/EXES, SPICA/MCS, SPICA/SCI, SPICA/SMI, TAO/MIMIZUKU, *Spitzer*/IRS, and TMT/MICHI instruments.

### D.2.7 MISC Upscope Enabling Technology

In addition to the enabling technologies identified in the baseline MISC (Section 4.4) there are two new technologies that will need to be further developed to include the MISC upscope capabilities: larger format Si:As arrays and a deformable mirror capable of operation at ~8K temperatures. The increase in Si:As array format size is expected to be relatively straight-forward, since as other than the array size the current performance of the detectors used in the JWST MIRI instrument will be sufficient for the MISC WFI. A cryogenic deformable mirror has been demonstrated, although not at quite as low as the temperature of the MISC WFI (Table D-14).

**Table D-13:** MISC instrument (upscope) resource requirements are a modest extension from the baseline values.

| | | | MISC (upscope) | |
|---|---|---|---|---|
| | | | **MISC-T** | **MISC WFI** |
| Information related to the Electrical Subsystem | Total Max Data Rate | Mbps | See Table D-11 | See Table D-11 |
| | Total Average Data Rate | Mbps | See Table D-11 | See Table D-11 |
| Information related to the Mechanical model | Volume (cold component) | [m³] (m x m x m) | 0.15 (2.0 x 0.25 x 0.30) [foreoptics] 0.12 (1.0x 0.8 x 0.15) [Collimator + Densified Pupil Spectrographs] | 3.23 (1.7 x 1.9 x 1.0) |
| Information related to the Thermal Model | Mass (cold component) | [kg] | 68.87 | 126.51 |
| | Mass (warm component) | [kg] | 16.00 | 16.00 |
| | Mass (Harnessing, bipods, etc) | [kg] | 16.36 | 27.7 |
| | Total Mass | [kg] | 101.23 | 170.21 |
| | Total Peak Power (cold part) | [W] | 0.159 | 0.197 |
| | Total Peak Power (warm part) | [W] | 10 | 120 |
| | Total Average Power (cold part) | [W] | 0.009 | 0.050 |
| | Total Average Power (warm part) | [W] | 10 | 46 |
| | Total Standby Power (cold part) | [W] | 0.009 | 0.009 |
| | Total Standby Power (warm part) | [W] | 10 | 46 |
| | Average Power Dissipation (detectors) | [W] | 0.008 | 0.008 |
| | Average Power Dissipation (heater) | [W] | 0.009 | 0.009 |

**Table D-14:** MISC WFI Enabling Technology and Component Heritage

| Enabling Technology | Specification | Maturity of the technology | Heritage, References |
|---|---|---|---|
| kilo-DM (BOSTON micro-machines corporation) | 32x32 MEMS Deformable Mirror | Tested at crygenic temperature [1] | Based on heritages of SPICA Coroagraph Instrument (SCI; Enya et al. 2011, "A high dynamic-range instrument for SPICA for coronagraphic observation of exoplanets", SPIE, 8146, 81460Q) [1] Aoi Takahaski et al. "Laboratory demonstration of a cryogenic deformable mirror for wavefront correction of space-borne infrared telescopes," Appl. Opt. 56, 6694-6708 (2017) |
| Cryogenic DM studied in NAOJ and Kyocera | 35 element PZT bimorph 45mm diameter 1mm thickness | In an early prototype and R&D phase. Not yet tested at cryogenic temperature. | Based on the AO technology for Subaru telescope as well as for the TMT. Shin Oya et al. "Characterization of vibrating shape of a bimorph deformable mirror", Proc. SPIE, vol. 7015:3R (8pp) (2008). Shin Oya et al. "Deformable mirror design of Subaru LGSAO system", Proc. SPIE, vol. 5490, pp. 1546-1555 (2004). |
| Cryogenic DM studied in SRON | No continuous power dissipation (set and forget) in cryogenic part. High reliability because of minimal harness and electronics. | In an early prototype and R&D phase. Not yet tested at cryogenic temperature. | R. Huisman et al., "Deformable Mirror concept utilizing piezoelectric hysteresis for stable shape configurations", in preparation, 2019. |
| DM studied in Laboratoire d'Astrophysique de Marseille, CNRS | TBD | In an early prototype and R&D phase. Not yet tested at cryogenic temperature. | Marie Laslandes, Emmanuel Hugot, Marc Ferrari, Claire Hourtoule (Aix Marseille Universite, CNRS, LAM) "Mirror actively deformed and regulated for applications in space: design and performance" https://arxiv.org/abs/1305.0476 |

## APPENDIX E - SUPPORTING SCIENCE CALCULATIONS

This appendix contains descriptions of detailed science calculations that support the science requirements summarized in the science traceability matrix in Section 1.4.

### E.1 Extragalactic source simulations and spectral extraction

### E.1.1 Extracting mid-IR and far-IR sources from 3D data cubes

To achieve the *Origins* extragalactic scientific objectives described in Section 1.1, it will be necessary to retrieve the spectral line intensities of millions of galaxies. Although deep *Origins* extragalactic surveys will be spatially "confused," with multiple unresolved sources contributing to the signal detected in each beam, the planned spectroscopic surveys envisaged in Section 1.1.5 will yield spatial-spectral data cubes with fewer than 1 spectral line per 15 spatial-spectral resolution elements, or "spaxels". Hence, the spectral lines are not confused (see Figure 1-21), but how accurately can the line strengths be measured? Answering this question requires a high-fidelity model of the far-infrared sky and extensive analysis, along the lines of the effort reported by Raymond *et al.* (2010) for SPICA.

One basic idea is to search for the long-wavelength counterparts of sources cataloged at much higher resolution in another waveband (*e.g.*, optical or near-IR for typical galaxies, and X-ray or radio for AGN), as exemplified in the XID method for constructing *Herschel*-SPIRE catalogs (Roseboom *et al.*, 2010). XID was further developed into the XID+ algorithm (Hurley, 2017) and applied in several papers (*e.g.*, Pearson *et al.*, 2017). This method was extended to deep data with wide wavelength coverage in the so-called "super-deblending" approach (Liu *et al.*, 2018). The SEDeblend approach (MacKenzie *et al.*, 2016) deblends sources while fitting continuum spectral energy distributions (SEDs).

It should be possible to combine these ideas with existing optical spectral extraction codes. The ideal retrieval method would use all available prior information. The priors would include:

- deep high-resolution data, which would already exist in well-known extragalactic survey fields;
- the positions and types (*e.g.*, star-forming galaxies or AGN) of any known sources;
- statistical information about expected line strengths based on galaxy types, and perhaps parameterized spectral templates (SEDs and lines);
- photometric redshifts; and
- calibrated telescope and instrument response functions

All such available information will help to reduce the search volumes so that solutions are tractable with reasonable computing power. A Markov chain Monte Carlo approach or an iterative "cleaning" approach could be used to fit the entire 3D data set. Raymond *et al.* (2010) investigated an approach like this for SPICA. With deep prior catalogs (including photo-z's, which we will already have), we expect to do considerably better. Yet exactly how well we will be able to recover broad SED shapes remains to be determined.

Below we report on the preliminary results of a blind "stress test" designed to locate extragalactic sources and retrieve their spectra for comparison with an input source and line catalog unknown to the person (Mr. Alex Griffiths, PhD candidate, University of Nottingham) who kindly volunteered his time to conduct the analysis. The software used in the "stress testing" conducted to date is sub-optimal relative to software that will ultimately be developed to retrieve spectra from *Origins* data. More appropriate methods outlined above and will be explored during Pre-Phase A.

### E.1.2 Far-IR Sky Model

Our Far-IR Sky Simulator (FIRSS) includes extragalactic and Galactic sources of emission, as well as emission from the interplanetary dust in the solar system. The COBE Diffuse Infrared Background Experiment (DIRBE) "zodi" model (Kelsall *et al.* 1998) is used to estimate the zodiacal emission inten-sity at the chosen wavelength, sky coordinates, and solar elongation angle. Galactic interstellar dust in "infrared cirrus" clouds dominates the high Galactic latitude far-IR sky. Cirrus clouds exhibit structure on all spatial scales measured to date and are thus a source of confusion noise. The FIRSS uses the dust maps of Schlegel *et al.* (1998), derived from absolutely calibrated DIRBE data in combination with IRAS data, to set the mean brightness of the cirrus emission to a value appropriate for the chosen sky coordinates. The Zubko, Dwek, & Arendt (2004) dust spectrum is used to estimate the cirrus brightness at unobserved wavelengths. The FIRSS extrapolates the power-law cirrus structure found on large angular scales by IRAS (Gautier 1986) to the smaller scales not yet observed (Kiss *et al.*, 2003). The structure function is matched to the observed power spectrum at coarser scales, maintaining the mean brightness as a function of sky position and wavelength.

The FIRSS statistically represents the three-dimensional distribution of extragalactic sources (*i.e.,* their projected density in the sky and their distribution in luminosity and redshift) according to the source count model described by Bonato *et al.* (2019). The FIRSS uses cosmological parameters based on Wilkinson Microwave Anisotropy Probe (WMAP) observations (Bennett et al 2003; Spergel *et al.* 2003) to determine the geometry and predict the structure formation history of the universe. The FIRSS employs a dated model of galaxy clustering (Kashlinsky, 1998), but this is of little consequence in the present context.

Finally, the FIRSS includes a simple model for galaxy morphology. All star formation-dominated galaxies are represented as disks, which follow Freeman's Law for the galactocentric radial distribution of surface brightness. (Galaxy size is assumed not to evolve, and no radiative transfer calculations are performed.) Each disk galaxy is assigned random inclination and position angles. Active Galactic Nuclei (AGN) are treated as point sources with no surrounding disk. The FIRSS assigns a spectrum to each galaxy according to whether it is dominated by star formation activity or an AGN (Bonato *et al.*, 2019). A typical spectrum includes thermal dust emission; the polycyclic aromatic hydrocarbon (PAH) features seen at mid-IR wavelengths; and the many ionic, atomic, and molecular emission lines cataloged by Bonato *et al.* (2019).

We used the FIRSS to generate a spatial-spectral data cube centered in the COSMOS field at Galactic coordinates (l, b) = 236.822, 42.1216, respectively, and we adopted a solar elongation angle of 90 deg for the simulated observation. We assumed a cirrus power-law index of -2.5, which reasonably fits observed angular scales. The simulated data cube comprises 2048 x 2048 spatial pixels and 2301 spectral channels. The pixel size is 0.14 arcsec. Each wavelength slice of the data cube was simulated as if observed with R = 300 spectral resolving power. At this resolution, galactic spectral lines are not resolved, but molecular PAH features are mostly resolved. The spectral channels run from 25 to 600 μm in uneven steps, which approximately Nyquist sample the spectral resolution.

The simulated data cube contains 5724 star-forming galaxies, which range in redshift from 0.05 to 5.19 and in luminosity from $5.6 \times 10^5$ to $3.3 \times 10^{12}$ L$_\odot$. The data cube also contains 753 AGN, which range in redshift from 0.15 to 3.93 and in luminosity from $7.2 \times 10^5$ to $1.6 \times 10^{11}$ L$_\odot$. Figure E-1 shows two wavelength slices of the simulated data cube, each depicting ¼th of the simulated field (1024 x 1024 pixels).

To prepare for the "stress test," each slice of the data cube was convolved with the wavelength-dependent *Origins* point spread function at a field position in the middle of the OSS slit (Figure 2-24). The "convolved" data cube was turned over to Mr. Griffiths for analysis. The only metadata provided were the attributes of the data cube, such as sky coordinates, pixel dimensions, spectral resolving power assumed, and the wavelengths of the image slices. The testing was blind in the sense that the contents of the cube (extragalactic sources and their individual characteristics) were unknown to Mr. Griffiths.

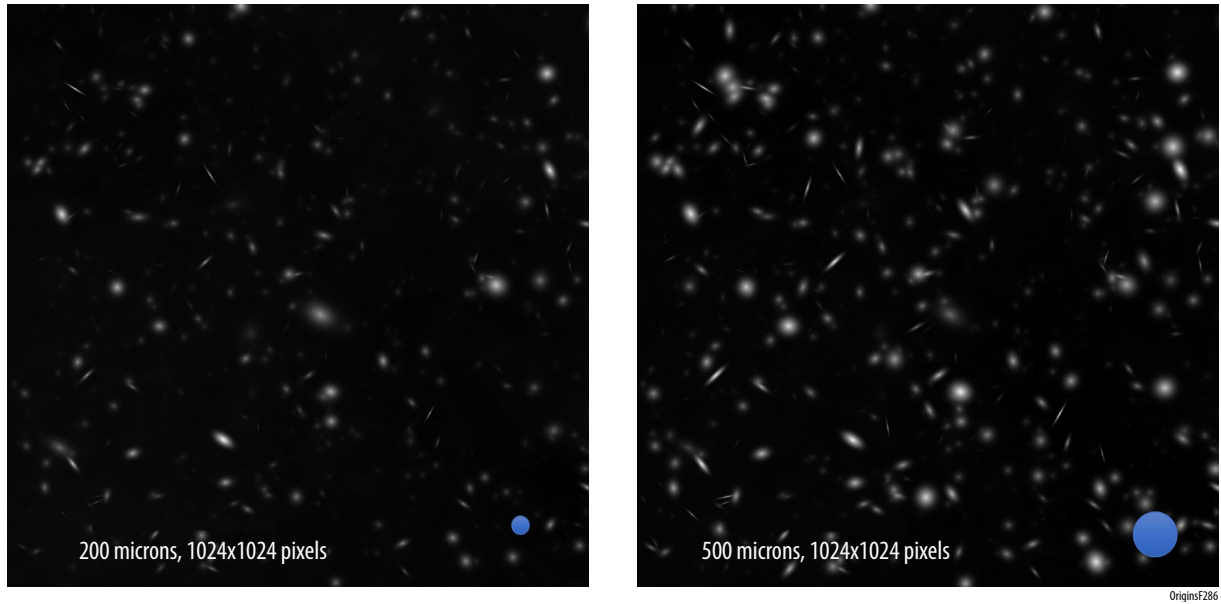

**Figure E-1:** Wavelength slices of the simulated far-IR sky at 200 μm (left) and 500 μm (right) showing the simulated sky at resolutions unattainable even with the Origins Space Telescope. These are two of the 2301 wavelengths simulated, and each image shows only 1/4th of the simulated field. After convolving the model sky with the Origins PSF (approximate beam size shown), the convolved data cube was used to conduct a "stress test" to see how many galaxies could be identified and to derive their spectra.

### E.1.3 Source detection

To conduct the stress test, we adopted methods and software commonly used to extract overlapping spectra from optical survey data, as our volunteer research assistant is experienced in that domain. Below we discuss some ideas for improving on this approach. For test purposes, we added Gaussian noise to the simulated sky data cube, with standard deviation 0.3 MJy/sr (same at all wavelengths).

Due to the increasing popularity of three-dimensional data cubes produced by integral-field units (IFUs) at optical wavelengths, with instruments such as MUSE (Bacon *et al.*, 2010) and upcoming facilities like JWST, numerous source-detection methods have been developed. Here we investigate some of these methods and discuss how they may be applied to the simulated data. We apply these techniques to a "sub-cube" – a 100 x 100 pixel region extracted from the full simulated data cube – to reduce the software run time and enable multiple iterations.

We considered two well-established methods of optical data-cube source detection. First, traditional source-detection methods, such as SExtractor (Bertin and Arnouts, 1996) can be used on 2D images created directly from a data cube. These images are created by collapsing or "flattening" (*i.e.,* taking the median or sum) the data cube along the spectral axis. This can be done over the whole wavelength range to obtain what is often referred to as a "white-light" image, or alternatively, one can create narrow- or broad-band images by limiting the wavelength range. The severe spatial blending that occurs when there are many sources per beam is a significant issue. We ran SExtractor on various 2D images created from the simulated sub-cube and recovered around 40% of sources compared to the input catalog (our measure of "truth"), which was divulged at the end of the stress-testing period. This could potentially be improved by fine tuning the SExtractor parameters, along with careful selection of the wavelength range, but this method is unlikely to recover 100% of the sources due to blending.

The second approach we explored is the blind detection of emission-line sources. Specifically, we considered two of the most popular python packages available: the MUSE Line Emission Tracker (MUSELET; Bacon *et al.* 2016); and the Line Source Detection and Cataloging Tool (LSDCat;

Herenz and Wisotzki, 2017). MUSELET builds on the former method by splitting the data cube into line-weighted pseudo-narrow-band images across the full wavelength range. The continuum is estimated from spectral medians on either side of the narrow-band region. SExtractor is then used to detect line emission in the individual narrow-band images and a composite catalogue of objects is created. Our tests show that this method is efficient at detecting a large fraction of the emission lines within the sub-cube; however, the compilation of the final source catalog is far from complete, at only about the 50% level relative to "truth."

The solid angle subtended by the *Origins* OSS beam grows by a factor of 553 from the shortest to the longest accessible wavelength, a much greater range than customarily seen in optical data cubes, such as those from MUSE. We attribute the relatively low source recovery fraction to the fact that the software was not designed to handle a wavelength-dependent PSF. Emission line sources can be assigned positions far from the true position of the source due to positional error, particularly at longer wavelengths. In the MUSELET tests, we find that almost all objects show multiple line detections. This is a promising indication that all of the sources might be identifiable.

We also experimented with the python package LSDCat, which cross-correlates a continuum-subtracted data cube with a 3D emission-line template in order to blindly identify sources in a data cube. The algorithm runs a median filter along the wavelength axis in order to remove source continua. A 3D Gaussian template is then used to identify emission lines, from which a catalog is created by analyzing the spatial locations of the identified line emitters. The results are similar to those seen with MUSELET. The vast majority of the emission lines are detected (see Figure E-2), but the method finds only about 70% of the "truth" sources.

In summary, we find that the techniques explored here provide a good starting point for the detection of extragalactic sources in the simulated OSS data cube. However, due to the wavelength-dependent PSF, the resulting source catalog is incomplete. Because the software is sub-optimal for this application, we regard the results as conservative. A customized algorithm that takes the PSF into

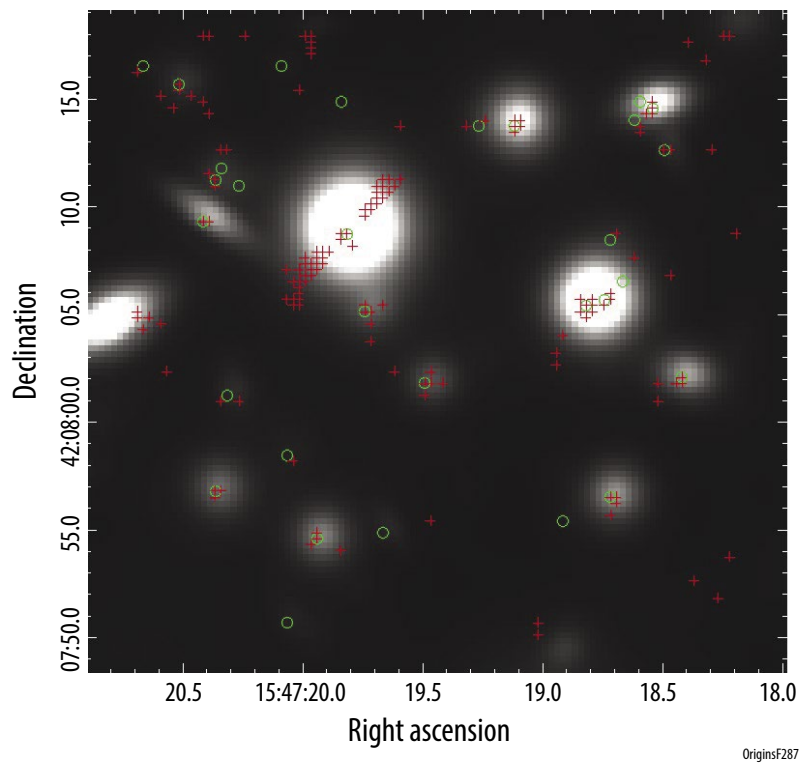

**Figure E-2:** All emission-line detections for a LSDCat run on a 100 x 100 pixel sub-cube. Red crosses show the locations of detected emission line sources and green circles show all input sources that are detected above the flux threshold. The simulation is based on the COSMOS field simulation and used to test the various software packages. The image is a sum over all wavelengths in the PSF-convolved data cube.

OriginsF287

account and combines line detection with source identification techniques commonly used at longer wavelengths is expected to yield better results, as discussed below.

### E.1.4 Spectral extraction

Isolation and extraction of a source's far-IR spectrum is particularly challenging and should rely on knowledge of the wavelength-dependent PSF. We explored the use of AutoSpec (Griffiths and Conselice, 2018) for this purpose. By utilizing the wealth of information available within a visible-wavelength IFU data cube, AutoSpec has been shown to isolate sources and can lead to increased signal-to-noise ratio spectra. To test the applicability of AutoSpec to *Origins* data, we examined a number of spatially blended sources identified in the SExtractor output.

Figure E-3 shows an example of this procedure and illustrates AutoSpec's ability to distinguish spatially confused sources and derive their spectra. However, our exploratory tests on the *Origins* sky simulations have not resolved all of the known issues. Similarly to MUSELET and LSDCat, the software was designed for optical data, in which the PSF is wavelength-independent over the wavelength range of the data cube. Since this is not the case for *Origins* spectroscopic survey data, we find that source continuum fluxes and line ratios are not accurately reproduced. We expect that more accurate source spectra could be recovered through a combination of the approaches applied here and with far-IR deblending techniques and source modeling.

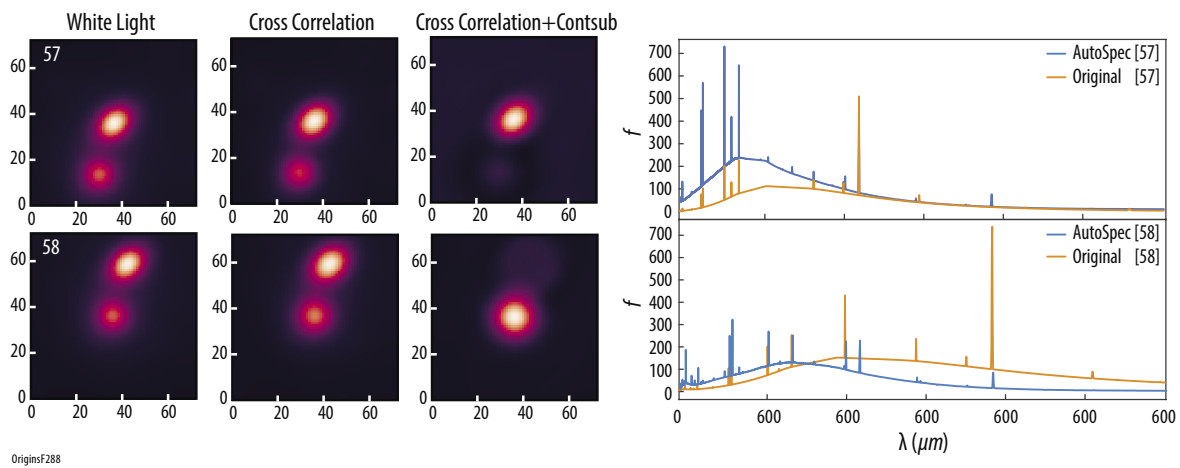

**Figure E-3:** AutoSpec software retrieves the spectra of a pair of simulated galaxies whose emission would be spatially blended when viewed with the Origins Space Telescope. The first two rows show images centered on the galaxy in question. The left panels show the "white-light" image constructed by summing the data cube along the wavelength axis. Central and right panels show the cross-correlation weight maps produced by the Autospec code, run without and with continuum subtraction, respectively. The overall similarity of the continuum will bias cross-correlation results when the two objects are this close to each other. The power of the cross-correlation algorithm (combined with continuum subtraction) can be seen by comparing the central and right panels; each source is successfully isolated from its neighbor. The resulting spectra for the two objects are shown in the bottom panel and compared with "truth" spectra from the original (prior to convolution) data cube. Autospec retrieves lines and continua, but the retrieved spectra don't accurately match "truth."

## E.2 Planetary system formation technical calculations

### E.2.1 Why Do We Have to Spectrally Resolve Lines?

*Spitzer* and *Herschel*/PACS did not spectrally resolve water lines in disks, and this greatly limited the confidence with which water abundance and its spatial distribution could be derived. Indeed, despite extensive efforts, *Herschel*/PACS did not unambiguously detect the 179.5 μm ground-state line in any disk (Meeus *et al.*, 2012; Dent *et al.*, 2013), almost certainly due to its low spectral resolving power. In contrast, cold water detections were made by *Herschel*/HIFI at much greater spectral resolving power for a few disks, and these spectra resulted in the best estimates to date of the mass of the cold-water reservoir (Hogerheijde *et al.*, 2011). High spectral resolving power is important for two reasons:

1. **Sensitivity:** Disks cool primarily via dust emission in the mid- to far-infrared wavelength range. As a result, all optically thick, planet-forming disks are characterized by a strong far-infrared continuum. Any observed lines suffer photon noise from this spectral dust continuum. This effect is important when astronomical background radiation, rather than the telescope, sets the noise floor. For instance, at full spectral resolving power, typical far-infrared water lines have line-to-continuum ratios of 10-100%. Under-resolving the lines leads to line-to-continuum ratios of 1-10%, negating much of the sensitivity advantage of a cold telescope (**Figures 1-29** and **E-4**).

2. **Tomography:** The scientific objective of measuring water lines is to retrieve the mass distribution of water vapor in each disk. When the line emission cannot be spatially resolved, velocity-resolved line spectra can be used to derive the radial distribution of line emission (**Figure E-4**, and **Section E.2.3**). Line tomographic imaging can also be used on any other disk line, such as the 112 μm HD line and the atomic fine-structure lines discussed above.

### E.2.2 Line Detection Yields and Design Reference Program

The goal for *Origins* is to conduct a survey of a large number of disks around stars with masses down to the brown dwarf limit. To determine if the requirements for sample size and line sensitivity can be met, the team developed a design reference program for observing water with *Origins*/OSS. Since the nearest star-forming regions within 150 pc (such as Taurus or Ophiuchus) do not contain 1000 planet-forming disks, it is necessary to explore more-distant clusters.

**The 1000 disk survey:** There are many young clusters within 1 kpc that can be used for a survey of water and HD in disks. The team selected the Orion cluster (Megeath *et al.*, 2012), which consists of ~3000 infrared-identified protoplanetary disks located within a combined area on the sky of 16 deg². The Orion cluster is known to be young, with an estimated age of 1-3 Myr (Hillenbrand, 1997), ensuring access to large numbers of gas-rich planet-forming disks. Optical spectroscopic follow-up of the L1630 and L1641 subclusters confirmed 399 young stars and showed the median spectral energy distribution of the disks is similar to that found in Taurus and similar nearby star-forming regions (Fang *et al.*, 2009).

The stellar luminosities and stellar masses of the disk sample (~3000 disks) are estimated using their H-band magnitudes, which are corrected for extinction using the average measured extinction law for the region ($A_H/A_K$=1.55) and the 3 Myr isochrone from Siess *et al.* (2000). The line flux for each disk is then predicted using the reference disk model based on the disk RNO 90, as fitted to *Spitzer* and *Herschel* spectra by Blevins *et al.* (2016) and scaling by the stellar luminosity. Since RNO 90 has some of the strongest water lines observed by *Spitzer*, the reference model has decreased water abundance by a factor 2 to approximate a more typical planet-forming disk (*e.g.*, Pontoppidan *et al.*, 2010; Carr *et al.*, 2011).

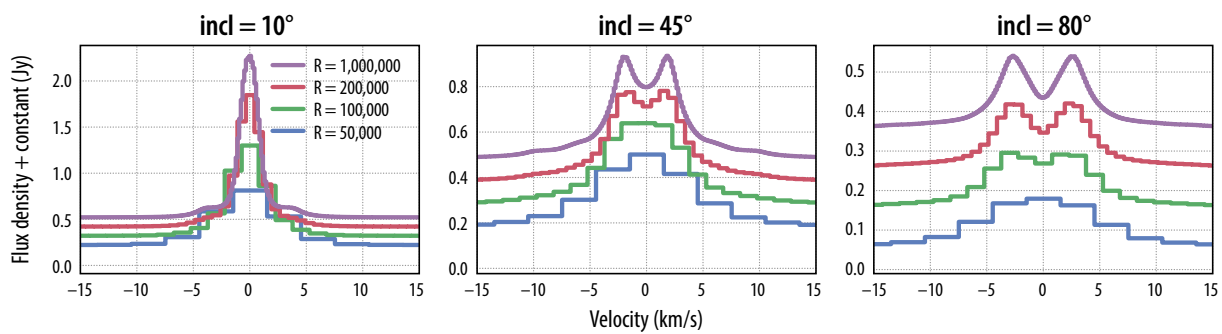

**Figure E-4:** The 179.5 μm water ground-state line simulated for different inclinations and resolving powers shows that the double peak at intermediate inclinations is separated only at resolving powers of at least R=200,000. Origins resolves these lines and is capable of exploiting the kinematic information.

For the purposes of a yield calculation, the ground-state $H_2O$ $2_{1\,2} - 1_{0\,1}$ line at 179.5 µm is used to define when water vapor is detected. The predicted flux of this line, along with *Origins*/OSS sensitivity in FTS mode, is used to estimate how many disks can be observed per 1000 hours, assuming a luminosity-limited sample and randomly selected disks. Many other water lines at shorter wavelengths tracing warmer gas are also essential, but these tend to be brighter than the ground-state line (Table 1-11) and will be observed concurrently. This results in a sample brightness distribution with the same shape as that of the parent population. For comparison, the same exercise is repeated for SPICA/SAFARI. The resulting yield is shown in Figure E-5, which demonstrates that *Origins* is able to obtain complete, luminosity-limited measurements of water vapor masses at the distance of Orion down to stellar masses of 0.25 M⊙. Crucially, this mass range overlaps with the expected mass range of the host stars of habitable planets to be targeted by *Origins*/MISC-T, enabling *Origins* to link the trail of water from the youngest disks all the way to planetary surfaces in the same class of stellar system.

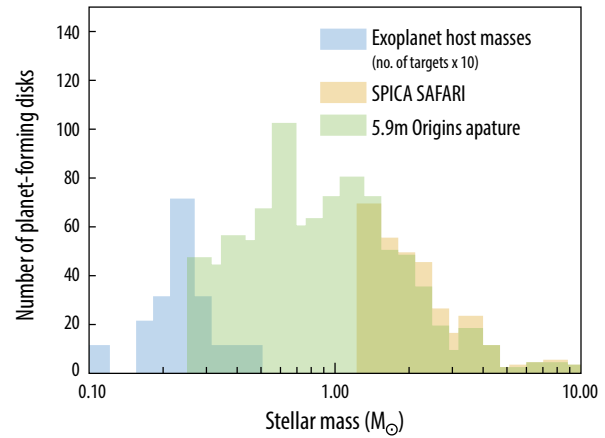

**Figure E-5:** Stellar mass yield of OST/OSS detections of water vapor in the Orion cluster compared to to the expectation for SPICA/SAFARI in the same time (1000 hours). Also shown is the mass distribution of exoplanet host stars with transiting exoplanets targeted by OSS/MISC.

### E.2.3 Line Tomographic Imaging

In the absence of velocity information in the water lines, it is necessary to fit a model to many lines to infer the water distribution. Therefore, *Origins* also deploys the line tomography method, enabling it to obtain this information more directly (Bast *et al.*, 2013; Manser *et al.*, 2016). In this method, lines that are spectrally well resolved can be inverted to produce radial images of the line emission. Under the assumption of a Keplerian velocity field, the relation between line shape and radial intensity is one-to-one over a wide range of disk radii (Figure E-6; Bast *et al.*, 2013). The *Origins* design reference mission aims to obtain high-resolution spectra of the ground-state water line at 179.5 µm and the HD J=1-0 line using the OSS Etalon.

Line tomography works best if the double peak of the Keplerian line profile is well separated, with at least five spectral resolution elements across the line. Figure 1-32 demonstrates that a resolving power of R~200,000 is required for tomographic imaging of the ground state water line. Since the FTS resolving power is higher at shorter wavelengths, and warmer lines are wider (because they trace smaller disk radii), the FTS can be used with the addition of the etalon for tomographic imaging of warm disk gas.

Line tomographic imaging requires high signal-to-noise ratio (SNR). Specifically, the SNR per resolution element must be high enough to distinguish the trough between the double-peaks of a Keplerian line profile. This is the most important detail, as it determines the line emission intensity at the greatest distance from the central star. For lines tracing warm gas, this central line component is the most important for characterizing snow lines (*e.g.*, Notsu *et al.*, 2016). The trough is typically 20% of the peak line flux density. With a minimum SNR of 20 per resolution element, the trough can be detected with 99% confidence. A minimum sensitivity requirement for line tomography is then defined as that which yields SNR=20 per resolution element for the 179.5 μm water line in a nearby solar-mass disk at 125 pc. At R=200,000, this line is resolved with ~5 resolution elements, and has a total strength of $3 \times 10^{-18}$ W/m². Tomography of this line can be achieved with a 1σ sensitivity of $3 \times 10^{-20}$ W/m² per resolution element.

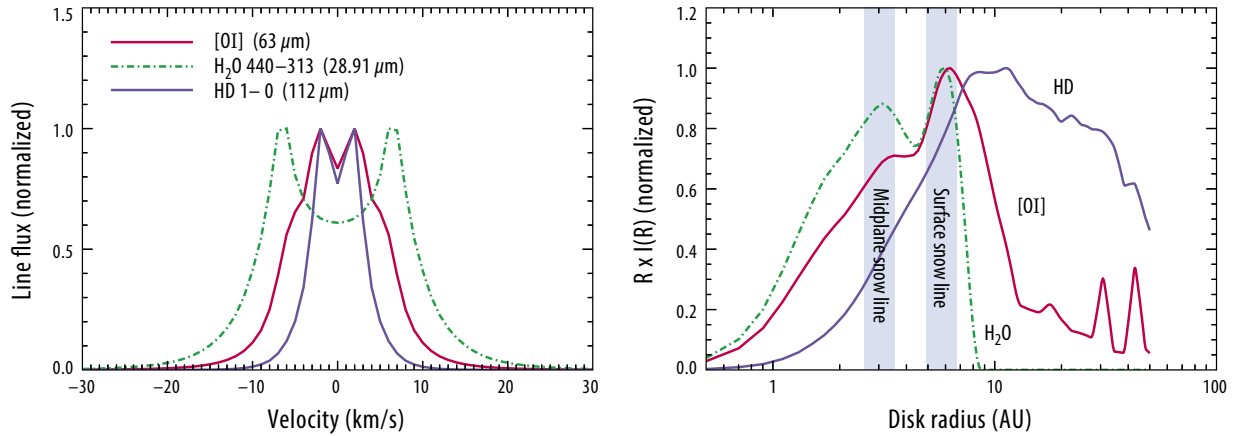

**Figure E-6:** Example line-tomographic mapping of mid- to far-infrared emission lines from planet-forming disks. Left: The observed line profiles from water, HD, and [OI] 63 μm. Right: The retrieved radial intensity using the line profiles, assuming a Keplerian velocity profile.

### E.2.4 Technical Note On Estimated Sensitvity to Water

For pre-stellar cores, which are not included in baseline *Origins* program, uncertainty is set by the range of ortho/para ratios of water vapor and overall mass. For protostars, the range is set by calculations of the water vapor mass with T > 150 K (*i.e.*, evaporated water) by Harsono *et al.* (2015), along with an assumption that the proper probe is $H_2^{18}O$, which requires the higher sensitivity offered by *Origins*' cooled far-infrared telescope. The inner disk refers to hot water emission inside the midplane snowline and assumes a gas rich disk, such as those modeled by Du and Bergin (2014). The maximum content of water and the lower limit are set by assuming the mass could be lower by 2 orders of magnitude, and the water abundance is set by the fundamental chemical limit of pure gas phase chemistry in gas with little UV exposure (~$10^{-6}$ photons per $H_2$ molecule). Because of its higher spectral resolving power and corresponding ability to detect sources with lower line-to-continuum flux ratios, *Origins* is more sensitive than JWST. For the outer disk (beyond the water snowline), the expected range is motivated by the *Herschel* water line survey water mass limits in Du *et al.* (2017), with an assumed factor of 100 range down to gas/dust mass ratios of unity. *Origins* is more sensitive than *Herschel* due to its larger, cooled aperture and sensitive detectors. *Origins* is also more sensitive than SPICA due to its larger aperture and higher spectral resolving power (Section 1.2.5).

Disk ice mass upper-limits are taken from the existing *Herschel* detection by McClure *et al.* (2015). The lower bound is uncertain, as it depends on the growth of bodies to larger sizes and gaseous disk dissipation. The team therefore assumed a three order-of-magnitude range. As a result of its 5.6x larger aperture, *Origins* is more sensitive than SPICA in this range.

For gas in debris disks, the team draws upon the comprehensive work of Kral *et al.* (2017), using the full range of expectations for debris systems at 100 pc. For such detections, *Origins* is more sensitive than *Herschel* due to its cooled aperture and SPICA due to its higher spectral resolving power.

For comets, we use the range of the D/H ratio seen in water in the solar system compiled by Hallis (2017). The decisive advantage of *Origins* lies in its superlative sensitivity and high spectral resolving power, which enable detection of the narrow $H_2^{18}O$ and HDO lines. *Origins* sees weaker comets with production rates as low as $10^{27}$ mol/s, which is 2 orders of magnitude below the level detectable from *Herschel* for comets at 1 AU. With an improved ability to determine the D/H to higher accuracy *Origins* can cover a broad D/H range, measure the water D/H in more comets than ever before and characterize the heterogeneity in the population.

The pre-stellar cores science requires an instrument with very high spectral resolving power ($R > 10^6$). Although this is not included in the *Origins* baseline, an instrument upscope, HERO, that will fulfill the requirements of this additional work is detailed in Appendix D.

### E.2.5 Technical Note on HD Flux Calculations

The HD J=1-0 fluxes and line-to-continuum ratios are provided in Table 1-13 of Section 1.2 of this report. Trapman *et al.* (2017) has predicted line-to-continuum contrasts of ~10-20% for *Herschel*/PACS. However, the line-to-continuum ratios seem to be inconsistent with the measurements and upper limits of McClure *et al.* (2016), who reported contrasts of <3% for four of six disks, one detection of 8%, and one upper limit of <10%. *Origins'* disk models for estimating HD line fluxes and line-to-continuum contrast are therefore consistent in terms of line flux, but are significantly more conservative in terms of line-to-continuum contrast, and consistent with the range of *Herschel* observations. There is a critical need for high spectral resolving power to consistently detect the line over a range of disk masses and sizes.

### E.3 Exoplanet calculations

In order to determine the ideal wavelength range, number of transits/eclipses, resolving power and aperture size to characterize terrestrial M-dwarf planets and develop well-justified science requirements for *Origins*, members of the *Origins* Exoplanet Science Working Group conducted a comprehensive exoplanet trade-space study (Tremblay *et al.*, in prep), simulating transmission and emission spectra with realistic uncertainties and performing full atmospheric retrievals. While there are likely nearly limitless choices and plausible arguments for particular climates and compositions for potentially habitable planets, for this trade study the team assumed a modern Earth composition (Robinson *et al.*, 2011). We simulate our spectra using the physical parameters (*e.g.*, size) of TRAPPIST-1e (Gillon *et al.*, 2016), orbiting an M8 star with a K-band magnitude that represents the median of our observational sample, Kmag = 9.85. This magnitude is based on TESS and SPECULOOS simulated yields.

*Origins* ultimately aims to detect biosignatures and habitability indicators in the atmospheres of temperate M-dwarf planets—notably robust detections of ozone and methane. Standard atmospheric spectral retrieval models and Bayesian approaches (Line *et al.*, 2013a,b; Greene *et al.*, 2016; Line & Parmentier, 2016; Kreidberg *et al.*, 2015; Line *et al.*, 2016; Benneke & Seager, 2012; MacDonald & Madhusudhan, 2017; Swain, Line, & Deroo, 2014; Trotta, 2008; Feroz & Hobson, 2008) adapted for terrestrial atmospheres are used to derive atmospheric parameter constraints and molecular detection significances. We compute detection significances using a fully Bayesian evidence-based method (Trotta, 2009) where a significant detection is considered to be $>3.6\sigma$ according to the Jeffery's scale. These quantities are used to assess the impact of each instrument trade assumption under these particular atmospheric scenarios. The uncertainties derive from using realistic instrument throughputs and adding (in quadrature) the photon-limited performance to an assumed 5 ppm noise floor. While the primary trades investigated include wavelength coverage, spectral resolution and mirror diameter, in this appendix we focus on the former two trades, as they were critical in setting the observational program—namely, the number of transits/eclipses in each observational tier. (We note that the figure constraining aperture size is already shown in Figure 1-52). Table E-2 presents a summary of the two trades. In the following two subsections, we detail the major conclusions for each of these trades.

**Table E-1:** Key terrestrial molecular bands between 3 - 22 μm. Having access to multiple bands of the same feature is critical to breaking degeneracies in molecular abundances.

| Absorber | Wavelength (μm) |
|----------|-----------------|
| $CO_2$ | 4.3, 15 |
| $H_2O$ | 6.3, >17 |
| $CH_4$ | 3.3, 7.7 |
| $N_2O$ | 4.5, 7.8, 8.6, 17 |
| $O_3$ | 9.7, 14.3 |

## Wavelength

We evaluate five bandpasses in our trade study as shown in Table E-2: 3-5, 3-11, 3-30, 5-11, and 5-30 microns. The primary habitability indicator and biosignature molecules contain their strongest molecular bands between 3 and 22 $\mu m$, making this range optimal for detection and abundance constraints. It is critical to have at least two bands per each molecular species, as degeneracies due to overlapping spectral signatures result in ambiguities if only one molecular band is observed. The 3-22 $\mu m$ range ensures the presence of at least two bands per each of the key molecules, which can break these degeneracies, resulting in robust molecular detections. The trade study shows 3-5 $\mu m$ is a critical wavelength region for the detection of $CO_2$; for bandpasses inclusive of the 3-5 $\mu m$ range, only ~6 transits are required for a significant detection of $CO_2$. Therefore, the detection of $CO_2$ forms the first (bottom) tier of Origin's 4000-hr exoplanet observational program, relying on this wavelength range. The detection of $N_2O$ in < 60 transits requires a bandpass that includes 3-11 $\mu m$. While $CH_4$ is not included in the table, as >100 transits are required to detect the individual molecule at high significance given our median K-band magnitude, the combined detection of $CH_4$ and $N_2O$ will be easier to accomplish in a reasonable number of transits (see Figure 1-55), and it would be possible to detect $CH_4$ to high significance for brighter targets. Lastly, we find longer wavelengths (10-22 $\mu m$) are key to assessing the effective surface temperature, as a large amount of flux at habitable terrestrial planet temperatures is radiated at these wavelengths (see Figure 1-49).

## Resolution

It is a common misconception that high-resolution is required to detect molecules. While stellar atomic lines are relatively narrow and require high resolutions, molecular signatures commonly present in planetary atmospheres present broad spectral features that are readily resolved with resolutions as low as ~50. Indeed, Table E-2 shows that there is not a large reduction in transits required to detect $CO_2$, $O_3$ and $N_2O$ to high significance as you increase spectral resolution from R = 50 to R = 100. Only $H_2O$ shows a significant reduction in transits, as R=100 is necessary to resolve the broad $H_2O$ bands. However, resolutions lower than 50 can cause features from different molecules to blend, resulting in greater degeneracies between molecular absorbers. These enhanced degeneracies inhibit robust molecular detections. This is particularly true for the detection of $O_3$.

## Trade Summary

Given our wavelength and resolution trade space study of a TRAPPIST-1e like planet with an Earth-like atmospheric composition around a star of K-band magnitude 9.85, we conclude that in order for *Origins* to detect habitability indicators and biosignatures $CO_2$, $O_3$, $N_2O/CH_4$ to high significance (>3.6σ) and measure effective surface temperature within a 5-year mission lifetime, it must support a wavelength range of approximately 3-30 microns and a spectral resolution of at least 50. Further simulations and retrievals on transmission and emission spectra showed that the 3-22 micron wavelength range was best for detecting biosignatures/habitability indicators and constraining surface temperature, while simultaneously falling within optimal design specifications for MISC-T.

**Table E-2:** Number of transits required to detect each molecule to 3.6σ confidence for each of our test cases. A dash indicates that achieving a 3.6σ detection was not possible within 100 transits. Note the detection significance values for not calculated for molecules other than $CH_4$ due to the excessive computational expense associated with retrieving a broad wavelength, high resolution spectrum.

| Bandpass | Resolution | $H_2O$ | $CO_2$ | $O_3$ | $N_2O$ |
|---|---|---|---|---|---|
| 3-5 microns | 300 | - | 5.7 | - | 61.5 |
| | 100 | - | 5.9 | - | 79.1 |
| | 50 | - | 6.2 | - | - |
| 3-11 microns | 100 | 84.6 | 5.9 | 57.7 | 43.4 |
| | 50 | - | 6.1 | 70.9 | 58.3 |
| 3-30 microns | 100 | 79.1 | 5.5 | 56.1 | 42.3 |
| | 50 | 95.6 | 5.5 | 64.8 | 52.8 |
| | 30 | - | 6.3 | 80.8 | 67 |
| 5-11 microns | 100 | - | - | 81.9 | - |
| | 50 | - | - | 100 | - |
| 5-30 microns | 100 | - | 37.1 | 70.9 | - |
| | 50 | - | 37.8 | 83 | - |
| | 30 | - | 38.2 | 100 | - |

## APPENDIX F - INDEPENDENT STUDY OF LARGE COLD (≤4.5K) TELESCOPE

An independent study of a cooling a large space telescope to 4.5 K or less is described in this paper and supports the feasibility of cooling the *Origins* Space Telescope. Arenberg *et al.* (2019) has a comprehensive discussion of a 4.5 K JWST architecture.

## An Alternate Architecture for the *Origins* Space Telescope

Jonathan W. Arenberg, John Pohner, Michael Petach, Ryan Hall and Jeffrey Bautista
Northrop Grumman Aerospace Systems, Redondo Beach CA 90278

### Abstract

This paper our investigation into adapting the design of the James Webb Space Telescope to the needs and requirements of the *Origins* Space Telescope (*Origins*). *Origins* is mission study being performed for the Astro 2020 decadal review and is a Far-IR mission, ~6-600 μm with optics at 4K. This low operating temperature is not achievable by passive means, requiring active cooling. We find this cooling to be within the state of the art and discuss modifications needed to the JWST design necessary to reach the needed temperatures. The thermal architecture and the heats loads at 18K and 4K are discusses as are the modification existing coolers necessary. Our ultimate conclusion is that 4K telescopes are possible and in-family designs such as JWST can be modified to fulfill the needs of this mission.

**Keywords:** *Origins* Space Telescope

### F.1 INTRODUCTION

The *Origins* Space Telescope is a mission study being prepared for the Astro 2020 decadal. [*1*] Our paper reports on the results of a study to re-use the design of the James Webb Space Telescope to the extent possible. The motivation is quite simple reuse the JWST to the extent possible, resulting in a 6.5 m primary (approximately 23.4 m² collecting area) and avoidance of much of the non-recurring expenses involved in the design of the space telescope. Our proposed architecture is called *Origins*' alternate architecture, to avoid confusion of this work the study's baseline design

In some respects this study is deja vu. A previous decadal study called, Single Aperture Far Infra-Red (SAFIR) pursued the same path of adapting the JWST design for the far-IR mission with the addition of active cooling. [*2*] [*3*] These studies were performed from the vantage point of the early 2000's made two basic assumptions, the technical maturation of the planned cryo-coolers for JWST's MIRI instrument and the JWST design. At the time of this writing, 2019, both of these events have been realized. This enables our modern study of this question based on validated flight JWST thermal models.

This quick study also presents an argument showing the truth in two seminal tenets. The first is that 4K class telescopes are viable and possible. These systems are a problem in design and manufacture, they do not require new generations of cooling technology. The second tenet is that space observatory architecture have mission flexibility and new missions can leverage previous designs to achieve higher engineering productivity.

Because we are adapting a design for a mission it was not originally intended, we have taken a step by step approach illustrated in **Figure F-1**.

### F.2 Architecture and Loads

The alternate architecture is derived from the validated flight models used in the design and verification of JWST. **Figure F-2** shows a picture of the proposed configuration and identifies some of the major changes to the Webb architecture.

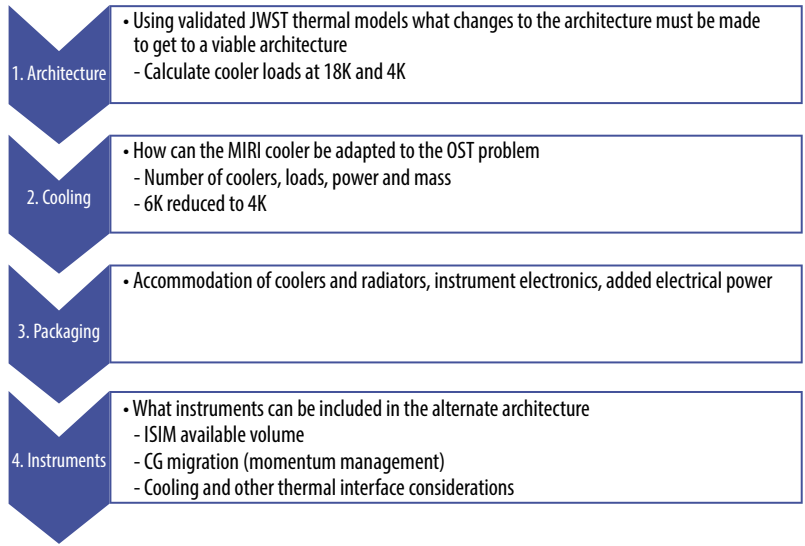

| | |
|---|---|
| **1. Architecture** | • Using validated JWST thermal models what changes to the architecture must be made to get to a viable architecture<br>  - Calculate cooler loads at 18K and 4K |
| **2. Cooling** | • How can the MIRI cooler be adapted to the OST problem<br>  - Number of coolers, loads, power and mass<br>  - 6K reduced to 4K |
| **3. Packaging** | • Accommodation of coolers and radiators, instrument electronics, added electrical power |
| **4. Instruments** | • What instruments can be included in the alternate architecture<br>  - ISIM available volume<br>  - CG migration (momentum management)<br>  - Cooling and other thermal interface considerations |

**Figure F-1:** Steps in the definition of *Origins'* Alternate Architecture

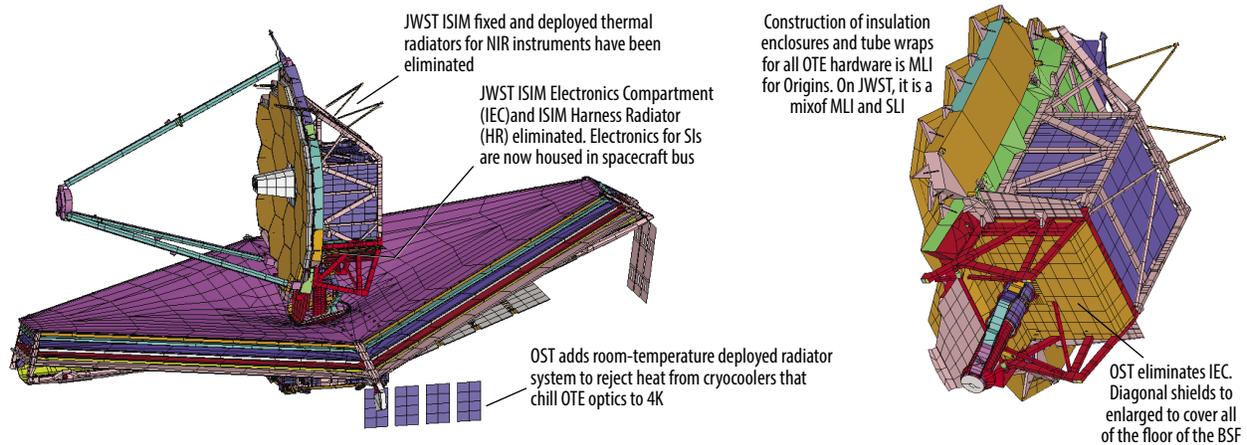

JWST ISIM fixed and deployed thermal radiators for NIR instruments have been eliminated

JWST ISIM Electronics Compartment (IEC) and ISIM Harness Radiator (HR) eliminated. Electronics for SIs are now housed in spacecraft bus

OST adds room-temperature deployed radiator system to reject heat from cryocoolers that chill OTE optics to 4K

Construction of insulation enclosures and tube wraps for all OTE hardware is MLI for Origins. On JWST, it is a mix of MLI and SLI

OST eliminates IEC. Diagonal shields to enlarged to cover all of the floor of the BSF

**Figure F-2:** View of the JWST derived *Origins* Alternate Architecture

**Table F-1:** Modifications to the Webb architecture leading to *Origins* Alternate Architecture

| Subsystem` | JWST | Origins |
|---|---|---|
| Cooling | Passive thermal architecture for mirror and all instruments except MIRI, which is actively cooled to ~6K | Active cooling of telescope optics and instruments to 4K |
| Insulation | Some sides of ISIM enclosure – as well as insulation for backplane and BSF - covered with mixture of MLI and black Kapton SLI for added radiative cooling. | Completely cover instrument volume and with high-performance MLI to minimize refrigerant loads for instruments. Backplane and BSF structure are covered with MLI with low-ε surfaces facing mirrors |
| Harness Design | Design employs copper harness running along the backplane and BSF, with OTE's use of PhBr restricted to the transition harness running from bottom to top of the DTA (where harness temperature gradients are the largest). | Electrical harness will employ only PhBr or other low-k material in harnesses from DTA base upward. |
| Aft optics subsystem (AOS) | AOS covered with black Kapton SLI with Kevlar mesh | AOS insulated with MLI to better isolate AOS |
| AOS | JWST cools FSM to <40K using two radiator panels and heat straps. | Origins cools Fine Steering Mirror (FSM) and Tertiary Mirror (TM) using heat straps connected to 4K cold finger. |
| Deployable Tower Assembly | Leaves almost all DTA tube surfaces bare. | design makes extensive use of MLI and low- e surfaces to prevent heat from 300K vibration isolator assembly near base of DTA from radiating into cold side |

A more detailed listing of the differences between the Webb architecture and *Origins* alternate architecture is given in **Table F-1**.

### F.2.1 18K Thermal Loads

The *Origins* Alternate Architecture has two temperature intercepts, 18K and 4K. At each intercept the heat is removed by active refrigeration. The JWST thermal model was modified as described above and used to calculate the loads on the cooler at various locations.

The 18K heat loads are considered first. The surfaces shown in yellow on the backplane support fixture (BSF) in **Figure F-3** can be cooled to 18K to reduce the required cooling load for the Primary Mirror (4K), Tertiary Mirror (4K), Fine Steering Mirror (4K), and science instruments (detectors potentially <1K). This is done transferring heat from these locations to a limited number of cold fingers using high-purity aluminum straps. The unmargined total heat from all locations at 18 K is 203.9 mW.

Several other locations were selected to cooled to 18 K. On Webb, all thermally-significant harnesses that run from warm spacecraft bus to cold OTE must pass through a specific interconnect panel, ICP6. This harness includes those for controlling PMSAs, SM, and FSM, as well as OTE deployments mechanisms and heaters, temperature sensors, etc. Cooling this ICP is an unimagined load of 339 mW. The other item selected as for cooling is the Cold Junction Box (CJB) located near ICP6 on underside of BSF floor. The harnesses for controlling PMSAs, SM, and FSM pass through the CJB. Cooling CJB to 18K helps augment cooling at 18K and further reduces the heat load on mirrors operating at 4K The predicted 18K heat load on the CJB is 38 mW. This gives a total calculated 18K heat load of 581 mW.

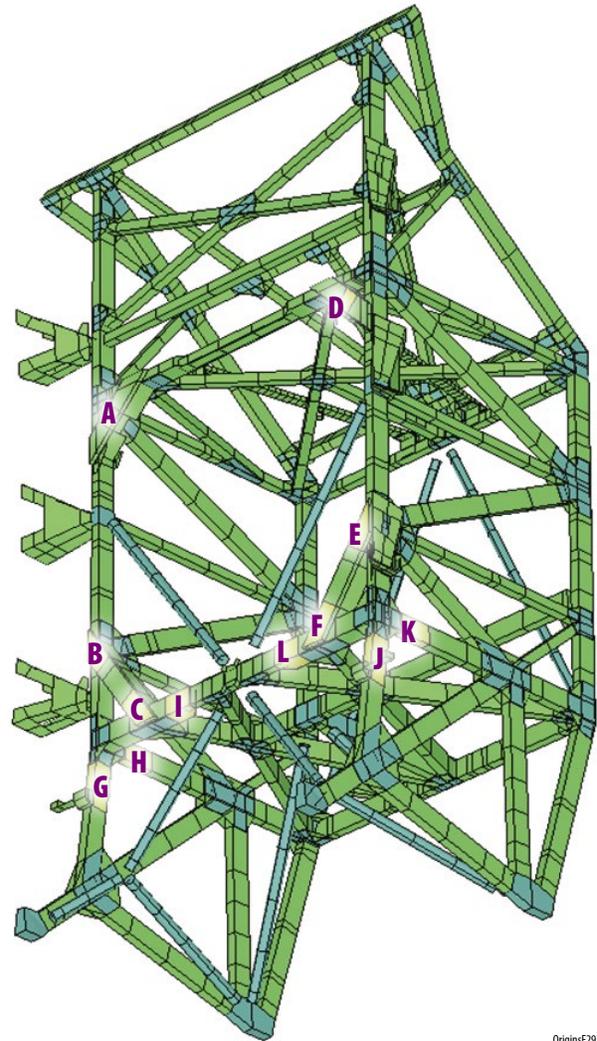

**Figure F-3:** BSF 18K stage cooling locations are indicated by the lettered points

### F.2.2 4K Thermal Loads

The thermal loads on the optics are also calculated directly from the thermal model. **Figure F-4** shows the heat loads on the primary mirror.

The total 4K load from the primary mirror is 118.8 mW. The very noticeable "hot spot" in the mirror is due to an asymmetry in the design of the Webb harness, is clearly an area that can be investigated for further improvement.

The 4K loads on the other optics have also been determined and are 9.9 mW for the secondary mirror, 1.7 mW for the tertiary mirror and 4.7 mW for the fine steering mirror. This gives an unmargined total of 143.3 mW at 4K.

Lightsey et al provides an assessment of the stray light performance of this configuration. [*4*]

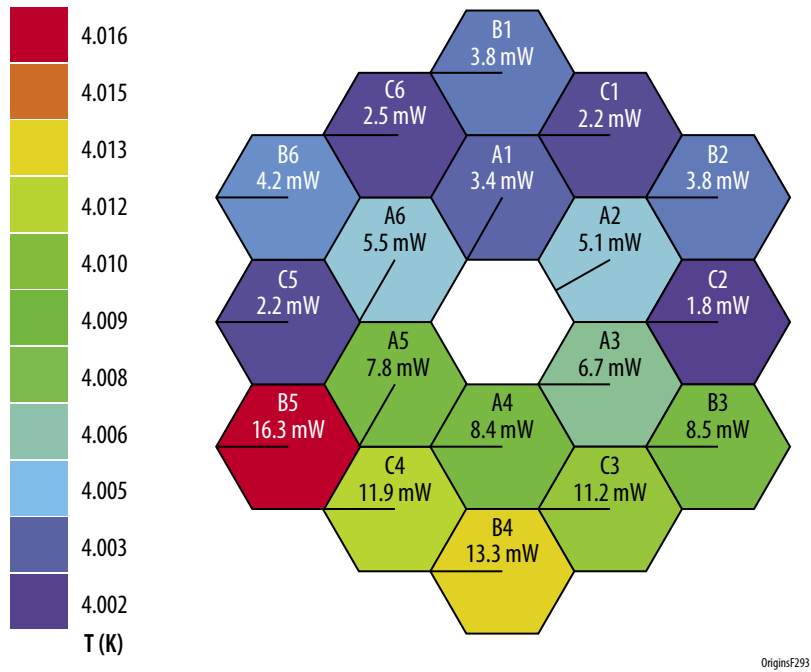

**Figure F-4:** Primary Mirror Loads

## F.3 Cooler

From the preceding section the unmargined cooler loads have been calculated and are; 134 mW at 4K and 581 mW at 18K. As has been mentioned above, these loads are no margin applied to them and are thus an optimistic estimate. To provide a more sober assessment of the alternate architecture thermal challenge we adopted a 33% margin as the loads required. Other investigators use larger margins on heat loads. The fact that validated models are used in this study justifies the smaller, but still significant, margins chose. Use of this margin gives the following loads: 178 mW at 4K and 1.005 W at 18K.

Reuse of the Webb architecture suggests that t cooler that will be used for the MIRI instrument will be re-employed for *Origins*, as such no new cooler development is needed for *Origins*. The Webb MIRI cooler has the requirements of 6.25K: 55 mW at 6.25K and 232 mW at 18K. It is our assessment that an unmodified MIRI cooler is unsuitable for use in the *Origins* application. However, a high TRL modifications can be made that will allow for sufficient heat removal at 4K. [5] The cooler modification adds a second JT compressor in series with MIRI cooler JT compressor. The additional compressor stage is a modified JT compressor with 4X area pistons. These larger pistons are being used on other programs and are this not new. The 1$^{st}$ stage recuperator tube diameters will be increased compared to MIRI. It should be noted that this increase is small, and only the "trained eye" will notice the difference, without the drawing to indicate dimensions. It is also remarked that this increase in tube diameter is local and does not ripple through the system. **Figure F-5** shows a diagram of the modified cooler.

With the margined loads above and using the anchored MIRI cooler mode modified to the *Origins* cooler configuration, it is found that 3 coolers will be needed for the *Origins* mission. The estimated electrical power needed is about 1.5 kW total, approximately 500 W/cooler.

Initial assessment is that $^3$He and $^4$He are viable options as the refrigerant gas. This selection will be subject of a future trade.

## F.4 Packaging

As part of this study an initial look at packaging the instrument and cooler electronics was performed. One objective of this packaging study was to find room within the existing Webb spacecraft for the relocated instrument electronics, which in the Webb configuration are located on the cold side of the sunshield. The second objective is to find a means to reject the additional 1.5 kW coming from the additional cryo-coolers. In the Webb design heat form the single MIRI cooler is rejected on a radiator panel forming a side of the bus. This is the location planned for the instrument electronics.

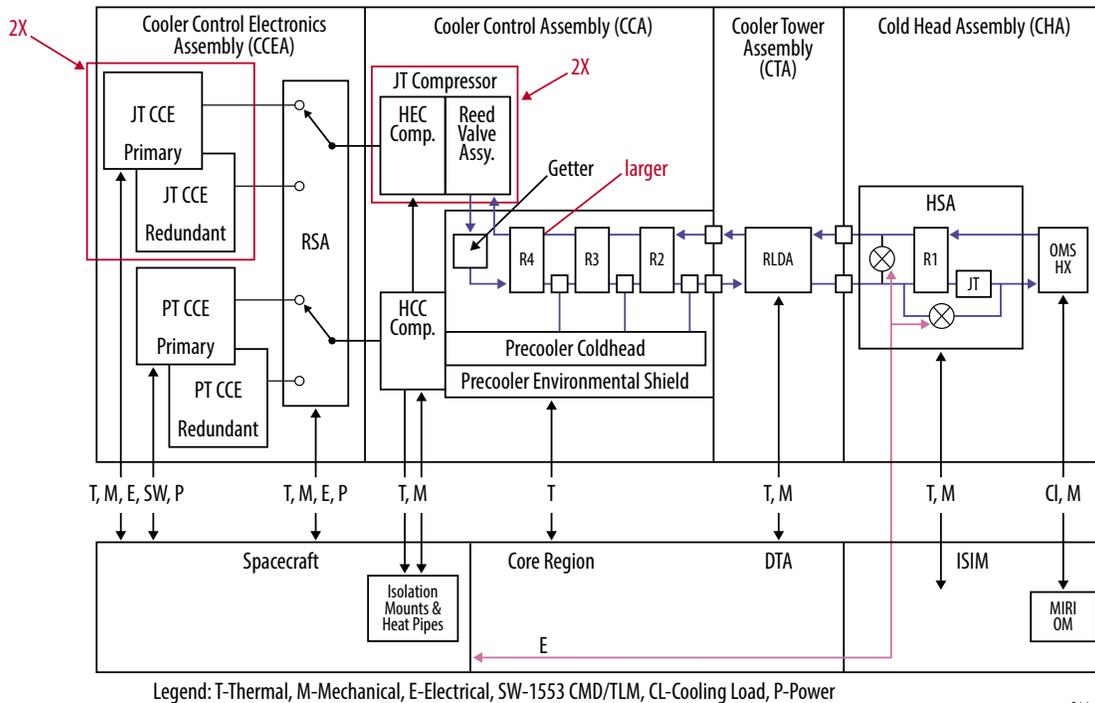

**Figure F-5:** 2 stage JTC compressor version of MIRI cooler modified for use on Origins

The solution for *Origins* to reject the heat from the 3 cryocoolers, is to have a deployable room temperature radiator made of panels. Each panel measuring 1.25m by 0.7 m, with one panel per cooler. The ensemble of radiations is capable of rejecting nearly 1.6 kW at 290K. This frees up the space now used for MIRI and this is the initial location for the relocated instrument electronics. A sketch of this layout is given as **Figure F-6**.

Another key consideration in this architecture is the ability to accommodate the science instruments. To assess this the volume footprints of the existing *Origins* instruments were fitted to the available volume in the JWST ISIM. Since the electronics for the instrument have been moved into the space craft volume, there is additional volume available for instrument accommodation. The results of the instrumentation

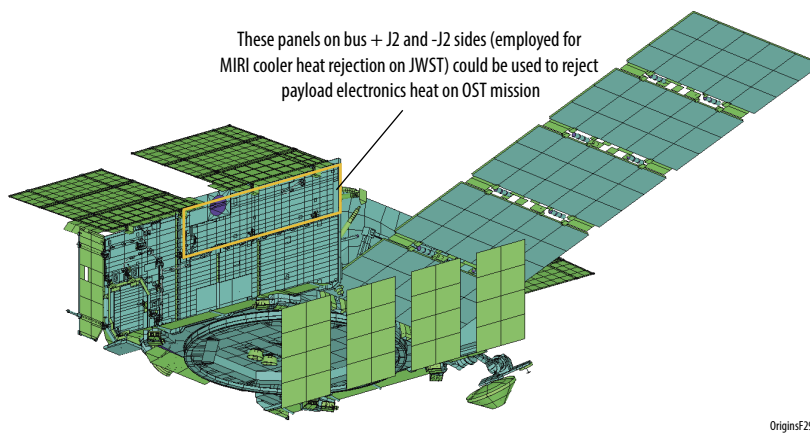

These panels on bus + J2 and -J2 sides (employed for MIRI cooler heat rejection on JWST) could be used to reject payload electronics heat on OST mission packaging study is shown in **Figure F-7**. To first order, all of the instrument envelopes can be accommodated. There are few areas of interference with current backplane struts and the *Origins* instrument envelopes. These small interferences are not viewed as "show stoppers" as the instruments are represented by their envelopes and these envelopes were developed for the *Origins* baseline not this architecture. Given the small size of the interferences, we are quite confident a deliberate effort to accommodate the *Origins* instruments would be successful.

**Figure F-6:** Instrument and Cooler electronics packaging

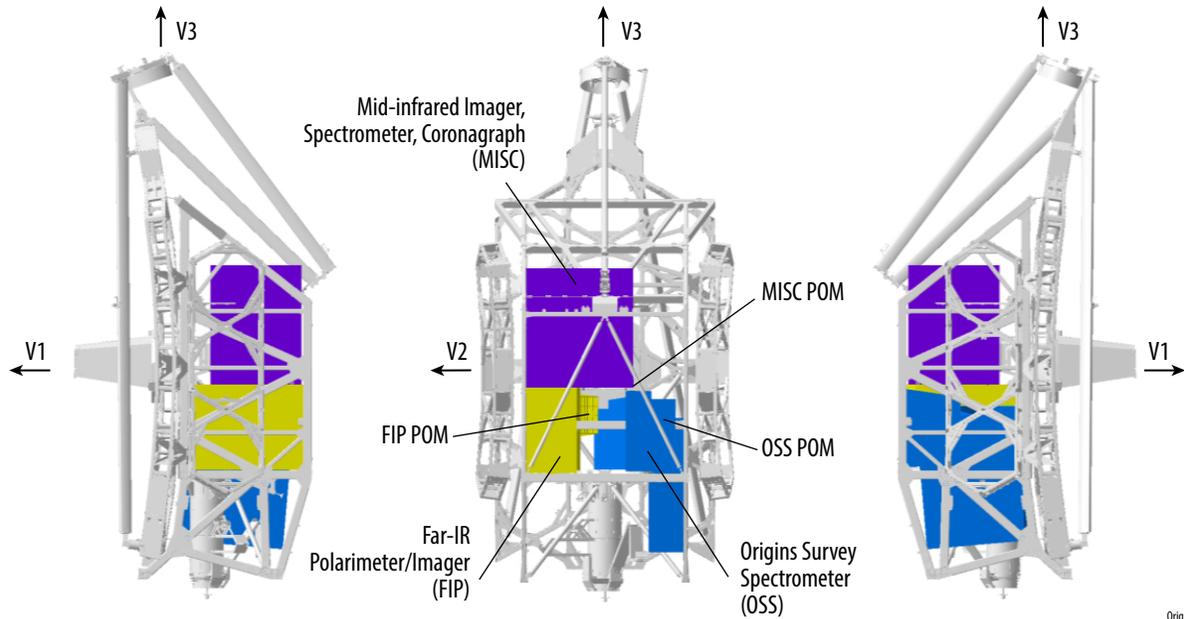

**Figure F-7:** *Origins* instruments packaged in the ISIM and IEC volumes within the JWST structure.

We also examined the electrical power budget, which is higher for *Origins* than for JWST, requiring the addition of two extra panels to the current 5 panel array. This array packaged for launch is shown in **Figure F-8**. This thicker stack does encroach of the keep put zones defined by the Ariane 5 launch vehicle. This small envelope violation is not a "show stopper" for two reasons. The first is that that envelope violations are small and in family with others that we know of that have been waived for JWST. Second is that when this JWST derived *Origins* design might launch the Ariane 5 will be retired, so a small violation is no reason to reject a concept at this level of development.

We also investigated the longer length of the solar array for it potential interference with deployment envelopes. In this case, the envelopes are inviolate. To accommodate the longer solar array the angle of deployed solar array had to be adjusted down 10 degrees as show in **Figure F-9**.

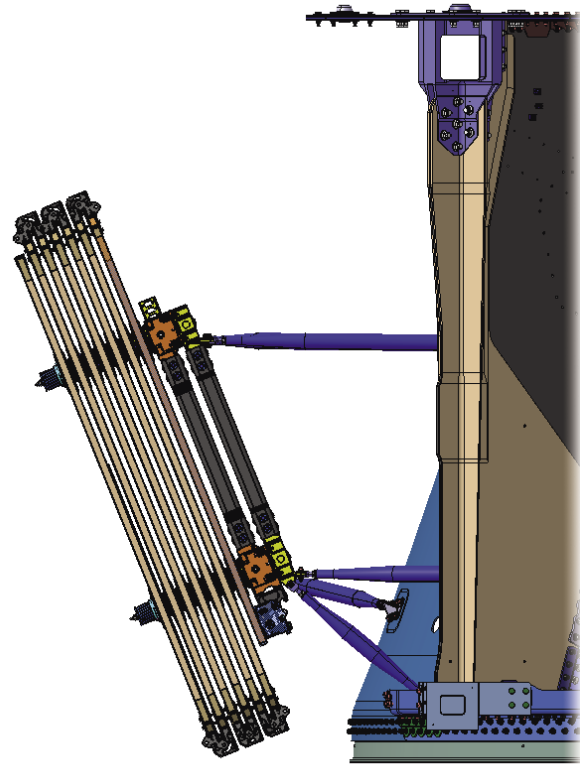

**Figure F-8:** Stowed 7 panel solar array on Webb Bus

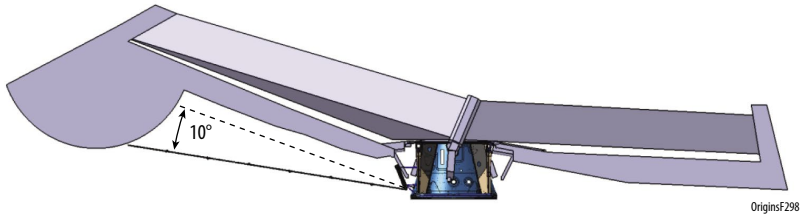

**Figure F-9:** Deployed envelope for Origins length array and adjustment to avoid interference with the aft flap.

Finally, a mass assessment was made based on the largely JWST mass properties and study values for the instruments. All told the Observatory wet mass is approximately 6,505 kg, with small mass margins on the JWST legacy hardware and the same mass growth allowance used by the *Origins* study and reported there.[1] Our study also indicates that the CG of the alternate architecture is compliant to the requirements of JWST momentum management solution. Allowing us to conclude at the conceptual level that once the alternate architecture has reached operating temperature, it can operated in the same efficient manner as will JWST, allowing reuse of all the apparatus and software of Webb planning and operations.

### F.5 Summary

This brief report has shown that the concept of using a modified Webb architecture can meet the thermal needs of *Origins* and is a viable concept. This alternate architecture employs 3 MIRI coolers with well understood and demonstrated modifications to achieve the 4K required for *Origins*. The important conclusion here is that realizing a 4K space telescope is within the current state of technology with no major developments required. A seminal argument the authors have been making in many forums. [6]

It is also worth noting that baseline *Origins* design requires a similar number of state of the art cooler to our findings. [1] The *Origins* team approached the design problem in a very different way and came to the same conclusion. Namely, *Origins* requires only a manageable number of existing coolers to achieve its mission. In short, such observatories are viable and possible. The challenge to realizing *Origins* lies in design and workmanship, not technology.

We have also demonstrated that there are possible observatory architectures that can accommodate telescopes of various spectral ranges and operating temperatures. This exciting possibility opens the way for planned reuse of architecture, design and hardware, providing an increase in productivity and launch tempo that may well enable a sustainable future for such flagship missions. [7]

## Acronyms and Definitions

| Acronym | Definition |
|---|---|
| ACE | Advanced Composition Explorer |
| ACEIT | Automated Cost Estimating Integrated Tools |
| AC-Modulated | Alternating-Current Modulated |
| ACS | Attitude Control System |
| ACTDP | Advanced Cryocooler Technology Development Program |
| ADC | Analog digital Converter |
| ADR | Adiabatic Demagnetization Refrigerator |
| ADRC | Adiabatic Demagnetization Refrigerator Controller |
| AGN | Active Galactic Nucleus or Nuclei |
| ALMA | Atacama Large Millimeter/submillimeter Array |
| AO | Adaptive Optics |
| AO | Announcement of Opportunity |
| AOS | Aft Optics Structure |
| APD | Astrophysics Division |
| ARC | AMES Research Center |
| ARIEL | Atmospheric Remote-sensing Infrared Exoplanet Large-survey |
| ASIC | Application-Specific Integrated Circuits |
| ASIST | Advance Spacecraft Integration and System Test |
| ATP | Authority/Authorization to Proceed |
| AU | Astronomical Unit |
| B | Billion |
| BIB | blocked impurity band |
| BIRB | Ball infrared Black |
| BFR | Big Falcon Rocket |
| BHAR | Black-Hole Accretion Rate |
| BHARD | Black-Hole Accretion-Rate Density |
| BLASTPol | Balloon-borne LargeAperture Submillimeter Telescope for Polarimetry |
| BOL | Beginning of Life |
| BW | Bandwidth |
| BY | Base Year (is defined equivalent to Fiscal Year (FY)) |
| C | Cooper Pair |
| CADR | Continuous Adiabatic Demagnetization Refrigerator |
| CBE | Current Best Estimate |
| CCA | Cryocooler Assembly |
| CCHP | Constant Conduction Heat Pipes |
| CDF | Cummulative Density Function |
| C&DH | Command and Data Handling |
| CDM | Code Division Multiplexing |
| CDR | Critical Design Review |
| CEA | French Alternative Energies and Atomic Energy Commission |
| CEMA | Cost Estimating and Modeling Analysis |
| cFE | core Flight Executive |
| CFRP | Carbon Fiber Reinforced Polymer |
| cFS | core Flight Software |
| CGH | computer-generated hologram |
| CGM | Circumgalactic Medium |
| CIDC | interdigitated capacitor |
| CL | confidence levels |
| CMB | Cosmic Microwave Background |

| Acronym | Definition |
|---|---|
| CMM | Coordinate Measuring Machine |
| CMMI | Capability Maturity Model Integration |
| CNES | Centre National d'Études Spatiales (National Center for Space Studies) |
| CM&O | Center Management and Operations |
| CMOS | Complementary metal oxide semi-conductor |
| COBE | Cosmic Background Explorer |
| COTS | Commercial off the shelf |
| CPA | chrome-potassium-alum |
| CPI | Cloud Particle Imager |
| CPM | Cryogenic Payload Module |
| CPT | Comprehensive Performance Test |
| CPU | Computer Programmable Unit |
| CRM | Continuous Risk Management |
| CRTBP | Circular Restricted Three Body Problem |
| CSO | Chief Safety Officer |
| CTE | Coefficient of Thermal Expansion |
| CTIA | capacitive trans-impedance amplifier |
| CVZ | Continuous Viewing Zone |
| DAC | Distributed Monte Carlo Method Analysis Code |
| DAK | Double Aluminized Kapton |
| DC | Direct Current |
| DDL | Detector Development Lab |
| DES | Dark Energy Survey |
| DET | Direct Energy Transfer |
| DIT | Differential Impedance Transducer |
| DIT | Dublin Institute of Technology |
| DM | Deformable Mirror |
| DR | dilution refrigerator |
| DRO | Distant Retrograde Orbit |
| DSCOVR | Deep Space Climate Observatory |
| DSN | Deep Space Network |
| DSOC | Deep Space Optical Comm |
| DTU | Data Transfer Unit |
| DTN | Delay/Disruption Tolerant Network |
| EAR | Export Administration Regulations |
| EDRS | European Data Relay System |
| EDU | Engineering Demonstration Unit |
| EGSE | Electrical Ground Support Equipment |
| ELT | Extremely Large Telescope |
| EM | Electromagnetic |
| EM | Engineering Model |
| EMC | Electromagnetic Compatibility |
| EMI | Electro-Magnetic Interference |
| EOL | End-of-Life |
| EoR | Epoch of Reionization |
| EPS | Electrical Power System |
| ESA | European Space Agency |
| ETU | Engineering Test Unit |
| EQW | Equivalent Width |
| F&A | Facilities and Administrative |
| Far-IR SIG | Far-IR Science Interest Group |

| Acronym | Definition |
| --- | --- |
| FDM | frequency-division-multiplexed |
| FIP | Far-infrared Imager Polarimeter |
| FIR | Far-InfraRed (~30-300µm) |
| FM | Flight Model |
| FM | Fold Mirror |
| FOG | Fiber Optic Gyro |
| FoR | Field of Regard |
| FOR | Flight Operations Review |
| FOV | Field of View |
| FPA | Focal Plane Array |
| FPGA | Field Programmable Gate Array |
| FPI | Fabry-Perot interferometer |
| FRR | Flight Readiness Review |
| FS | Fine Structure |
| FSM | Field Steering Mirror |
| FSR | Final Systems Review |
| FSW | Flight Software |
| FTE | full-time-equivalent |
| FTS | Fourier Transform Spectrometer |
| FUN CAIs | Calcium-aluminium-rich inclusions with isotopic mass fractionation effects and unidentified nuclear isotopic anomalies |
| FWHM | Spectral Line Full Width at Half-Maximum |
| GCC | Goddard Composite Coating |
| GI | Guest Investigator |
| GLF | gadolinium-lithium-fluoride |
| GMT | Giant Magellan Telescope |
| GNC | Guidance Navigation and Control |
| GO | General Observer |
| GOES | Geostationary Operational Environmental Satellite |
| GOLD | Goddard Open Learning Design |
| GRC | Glenn Research Center |
| GSE | Ground Support Equipment |
| GSFC | Goddard Space Flight Center |
| GW | gravitational wave |
| HAWC+ | High-resolution Airborne Wideband Camera-plus |
| HEB | Hot Electron Bolometer mixer |
| HEMT | High-electron-mobility transistor |
| HERO | HEterodyne Receiver for Origins |
| HFI | Planck High Frequency Instrument |
| HGA | High Gain Antenna |
| HIFI | Heterodyne Instrument for the Far-Infrared |
| HIRMES | HIgh Resolution Mid-infrarEd Spectrometer |
| HQ | NASA Headquarters |
| HST | Hubble Space Telescope |
| HWP | Half-wave Plate |
| IBC | impurity band conductor |
| IC | integrated circuit |
| IDE | Integrated Development Environment |
| IDL | Instrument Design lab |
| IF | Intermediate Frequency |
| IFU | Integral Field Unit |
| IGM | Intergalactic Medium |
| IMF | Initial Mass Function |

| Acronym | Definition |
| --- | --- |
| IMP | Instrument Mounting Plate |
| IMS | Integrated Master Schedule |
| IMS | Inner Mirror Segments |
| IMU | Inertial Measurement Unit |
| IPT | In-Plant Transporter |
| IPAC | Infrared Processing and Analysis Center |
| IPC | interpixel capacitance |
| IR | InfraRed |
| IRAC | InfraRed Array Camera |
| IRAS | InfraRed Astronomy Satellite |
| IRS | InfraRed Spectrometer (on Spitzer) |
| IRSA | IPAC Infrared Science Archive |
| IS | Image Surface |
| ISE | Instrument Systems Engineer |
| ISM | Interstellar Medium |
| ISO | Infrared Space Observatory |
| I&T | Integration and Test |
| ITAR | International Traffic in Arms Regulations |
| ITOS | Integrated Test and Operations System |
| JAXA | Japan Aerospace and eXploration Agency |
| JCMT | James Clerk Maxwell Telescope |
| JHU | Johns Hopkins University |
| JPL | Jet Propulsion Laboratory |
| JSC | Johnson Space Center |
| JWST | James Web Space Telescope |
| KB | Kuiper-belt |
| KDP | Key Decision Point |
| KIDs | Kinetic Inductance Detectors |
| L | inductor |
| LCRD | Laser Communication Relay Demonstration |
| LEMNOS | Laser-Enhanced Mission Communications Navigation and Operational Services |
| LEO | Low Earth Orbit |
| LEV | SEL2 Earth-Vehicle |
| LIGO | Laser Interferometer Gravitational Wave Observatory |
| LIR | Infrared Luminosity |
| LIRG | Luminous Infrared Galaxies |
| LISA | Laser Interferometer Space Antenna |
| LM | Lockheed Martin |
| LO | Local Oscillator |
| LOFAR | Low Frequency Array |
| LOI | Libration Orbit Insertion |
| LOS | Line Of Sight |
| LRS | Low-Resolution Spectrometry |
| LSST | Large Synoptic Survey Telescope |
| LTE | Local Thermal Equilibrium |
| LV | Launch Vehicle |
| LWS | Long Wavelength Spectrometer (on infrared space observatory) |
| M | Million |
| MAST | Mikulski Archive for Space Telescopes |
| MAM | Mission Assurance Manager |
| MCC | Mid-Course Correction |
| MDL | microdevices lab |

| Acronym | Definition |
|---------|------------|
| MDL | Mission Design Lab |
| MEB | Main Electronics Box |
| MEL | Master Equipment List |
| MEMS | Micro-Electro-Mechanical Systems |
| MEV | Maximum Expected Value |
| MHD | magneto-hydrodynamical |
| MIR | Mid-Infrared (~3-30μm) |
| MIRI | Mid InfraRed Instrument |
| MIRI-IFU | Mid InfraRed Instrument  Integral Field Unit |
| MISC | Mid-Infrared Spectrometer and Camera |
| MISC-T | Mid-Infrared Spectrometer and Camera Transit Spectrometer Module |
| MKID | Microwave Kinetic Inductance Device |
| MLI | Multi-Layer Insulation |
| MOC | Mission Operations Center |
| MOR | Mission Operations Readiness Review |
| MPC | Minor Planet Center |
| MPMF | Mass Properties Measurement Facility |
| MPSoC | multiprocessor system-on-chip |
| MPV | Maximum Possible Value |
| MSE | Mission Systems Engineer |
| MSFC | Marshall Space Flight Center |
| MSX | Mid-course Space eXperiment |
| NASA | National Aeronautics and Space Administration |
| NEA | Noise Equivalent Angle |
| NEN | Near Earth Network |
| NEOCam | Near-Earth Object Camera mission |
| NEP | Noise Equivalent Power |
| NIRCam | Near-Infrared Camera |
| NIRSpec | Near InfraRed Spectrograph |
| NOEMA | Northern Extended Millimeter Array |
| NRHO | Near Rectilinear Halo Orbit |
| OFCO | Office of the Chief Financial Office |
| OFSW | Origins Flight SoftWare |
| OMS | Outer Mirror Segments |
| OPD | Optical Path Difference |
| Origins | Origins Space Telescope |
| ORR | Operations Readiness Review |
| OSS | Origins Survey Spectrometer |
| OTIS | Optical Telescope Element and Integrated Science Instrument Module |
| PACE | Plankton, Aerosol, Cloud, ocean Ecosystem |
| PACS | Photodetector Array Camera and Spectrometer |
| PAF | Payload Adaptor Fitting |
| PAH | Polycyclic Aromatic Hydrocarbon |
| PAH-SFR | PAH Star Formation Rate |
| PCA | Pressure Control Assembly |
| PDR | Preliminary Design Review |
| PDR | Photon-Dominated Region or Photo-Dissociation Region |
| PEEK | PolyEthylEtherKetone |
| PER | Pre-Environmental Review |
| PES | PRICE Estimating Suite |
| PHOENIX | A radiative-transfer atmosphere code (not an acronym) |
| PM | Primary Mirror |

| Acronym | Definition |
|---------|------------|
| PMBSS | Primary Mirror Backplane Support Structure |
| PMSA | Primary Mirror Segment Assembly |
| PIA | Propellant Isolation Assembly |
| PIR | PAF Interface Ring |
| POM | Pick-off-Mirror |
| PM | Primary Mirror |
| PMD | Propellant Management Device |
| PPM | Pulse Position Modulation |
| PS | Propulsion System |
| PSE | Power Supply Electronics |
| PSF | Point Spread Function |
| PSI | Pounds per Square Inch |
| PSR | Pre-Ship Review |
| QCDs | Quantum Capacitance Detectors |
| QE | quantum efficiency |
| QML | Quality Management Plan |
| QSO | Quasi-Stellar Object |
| R | Resolving Power |
| RA | Resource Analyst |
| RAO | Resource Analysis Office |
| RF | Radio Frequency |
| RFSoC | RF-system on chip |
| RLP | Rotating Libration Point |
| RM | Risk Manager |
| RMS | Root Mean Squared |
| RMSWE | Root Mean Square Wavefront Error |
| ROICs | Read Out Integrated Circuits |
| RV | Radial Velocity |
| RXTE | Rossi X-ray Timing Explorer |
| RY | Real Year |
| S/A | Solar Array |
| SAFARI | SpicA FAR-infrared Instrument |
| SAM | System Assurance Manager – check this…"Safety" |
| SAMPEX | Solar, Anomalous, and Magnetospheric Particle Explorer |
| SAT | Strategic Astrophysics Technology |
| SB | Spectroscopic Binary |
| SBC | Single Board Computer |
| SBIR | Small Business Innovation Research |
| SBM | Spacecraft Bus Module |
| SC | Spacecraft |
| SCPPM | Serially Concatenated Pulse Position Modulation |
| SC TILT | Spacecraft Top-Level Integration and Test |
| SDO | Scattered-Disk Object |
| SDO | Solar Dynamics Observatory |
| SDSS | Sloan Digital Sky Survey |
| SECCHI | Sun Earth Connection Coronal and Heliospheric Investigation |
| SECO1 | Second Engine Cut-Off 1 |
| SED | Spectral Energy Distribution |
| SED | Sciences and Exploration Directorate |
| SEL1 | Sun-Earth Libration Point 1 |
| SEL2 | Sun-Earth Libration Point 2 |
| SEM | Scanning Electron Microscope |

| Acronym | Definition |
|---------|------------|
| SES | Space Environment Simulator |
| SEU | single event functional upsets |
| SF | Star Formation |
| SFD | source follower per detector |
| SFR | Star Formation Rate |
| SFRD | Star Formation Rate Density |
| SHI | Sumitomo Heavy Industries |
| SiC | Silicon Carbide |
| SIR | System Integration Review |
| SIS | Superconducting Insulating Superconducting mixer |
| SKA | Square Kilometre Array |
| SLS | Space Launch System |
| SM | Secondary Mirror |
| SM3 | Servicing Mission 3 |
| SMA | Secondary Mirror Assembly |
| SMA | Safety and Mission Assurance |
| SMBH | Supermassive Black Hole |
| SMSS | Secondary Mirror Support Structure |
| S/N | See SNR (if we're using both, we should pick one!) |
| SNe | Supernovae |
| SNR | Signal-to-Noise Ratio |
| SNR | Supernova remnants |
| SOA | state-of-the-art |
| SOC | Science Operations Center |
| SOFIA | Stratospheric Observatory for Infrared Astronomy |
| SPECULOOS | Search for habitable Planets EClipsing ULtra-cOOl Stars |
| SPF | Space Power Facility |
| SPICA | SPace Infrared telescope for Cosmology and Astrophysics |
| SPIFI | South Pole Imaging Fabry-Perot Interferometer |
| SPIRE | Spectral and Photometric Imaging Receiver |
| SQUIDs | Superconducting Quantum-Interference Devices |
| SRR | System Requirement Review |
| SSDIF | Spacecraft Systems Development and Integration Facility |
| SSPD | Satellite Servicing Projects Division |
| SSR | Solid State Recorder |
| SRON | Netherlands Institute for Space Research |
| SS | Study Scientist |
| SSA | Sun Shield Assembly |
| STDT | Science and technology Definition Team |
| STM | Science Traceability Matrix |
| STScI | Space Telescope Science Institute |
| STTARS | Space Telescope Transportation Air, Road and Sea |
| SV | Servicing Vehicle |
| SWAS | Submillimeter Wave Astronomy Satellite |
| SWS | Short Wave Spectrometer (on Infrared Space Observatory) |
| T | Transit Spectrometer |
| TAA | Technical Assistance Agreements |
| TAC | Time Allocation Committee |
| TAI | International Atomic Time |
| TBD | To Be Determined |
| TDM | Time-Frequency Division Multiplexing |
| TDRSS | Tracking and Data Relay Satellite System |
| TES | Transition-edged Sensors |

| Acronym | Definition |
|---------|------------|
| TESS | Transiting Exoplanet Survey Satellite |
| TFSM | Telescope Fine Steering Mirror |
| TID | Total Ionizing Dose |
| TIM | Technical Interchange Meeting |
| TIP | Transfer to Insertion Point |
| TIS | Teledyne Imaging Sensors |
| T-L | MISC Transit Spectrometer Long Wavelength Channel |
| TLS | two-level-system (fluctuators) |
| TM | Tertiary Mirror |
| T-M | MISC Transit Spectrometer Mid Wavelength Channel |
| TMA | Triple Mirror Assembly |
| TMA | Three Mirror Anastigmat |
| TMT | Thirty Meter Telescope |
| TNO | Trans Neptunian Object |
| TOO | Targets of Opportunity |
| TRAPPIST | Transiting Planets and Planetesimals Small Telescope |
| TRL | Technology Readiness Level |
| TRMM | Tropical Rainfall Mapping Mission |
| T-S | MISC Transit Spectrometer Short Wavelength Channel |
| TTM | Tip-Tilt Mirror |
| TTS | MISC Tip-Tilt Sensor |
| ULE | Ultra Low Expansion |
| ULIRG | Ultra-Luminous Infrared Galaxy |
| upGREAT | upgraded German REceiver at Terahertz |
| UR | University of Rochester |
| VDA | Vapor Deposited Aluminum |
| VELLOs | Very Low Luminosity Objects |
| VNA | Vector Network Analyzer |
| VSMOW | Vienna Standard Mean Ocean Water (standard used for D/H) |
| WBS | Work Breakdown Structure |
| WFI | MISC Wide Field Imager Module |
| WFIRST | Wide Field InfraRed Survey Telescope |
| WFIRST/HLS | WFIRST High Latitude Survey |
| WISE | Wide-Field Infrared Explorer |
| WMAP | Wilkinson Microwave Anisotropy Probe |
| WSC | White Sands Complex |
| XRCF | X-Ray Calibration Facility |
| XUV | X-ray and Ultraviolet |
| YSO | Young Stellar Object |
| ZEUS | redshift (Z) and Early Universe Spectrometer |